\pdfoutput=1

\newcommand{\thesisTitleFrontmatter}{ENABLING EFFECTIVE ERROR MITIGATION\\IN MEMORY CHIPS\\THAT USE ON-DIE ERROR-CORRECTING CODES}

\newcommand{\thesisTitlePlain}{Enabling Effective Error Mitigation in Memory Chips That Use On-Die Error-Correcting Codes}
\newcommand{\thesisDissNumber}{27894}
\newcommand{\thesisAuthor}{Minesh Patel}
\newcommand{\thesisUni}{\protect{ETH Z\"urich}}

\newcommand{\thesisYear}{2021}
\newcommand{\thesisVersion}[0]{1.0}

\documentclass[12pt,oneside,a4paper]{ethzthesis}

\newcommand*\circled[1]{\tikz[baseline=(char.base)]{
    \node[shape=circle,fill,inner sep=1pt] (char) {\textcolor{white}{\textbf{#1}}};}}

\usepackage{siunitx}
\usepackage{algorithm2e}

\usepackage{multirow}
\usepackage{float}
\usepackage{bm}
\usepackage{amsmath}
\DeclareMathOperator*{\argmax}{argmax} 

\newcommand{\xmo}[1]{#1}

\definecolor{blue}{rgb}{0.0, 0.0, 0.0}
\definecolor{red}{rgb}{0.0, 0.0, 0.0}
\definecolor{blue}{rgb}{0.0, 0.0, 0.0}
\definecolor{green}{rgb}{0.0, 0.0, 0.0}
\definecolor{yellow}{rgb}{0.0, 0.0, 0.0}

\newcommand{\dmpa}[1]{#1}

\newcommand{\dmpc}[1]{#1}
\newcommand{\dmpd}[1]{#1}

\newcommand{\mpo}[1]{#1}
\newcommand{\mpt}[1]{#1}
\newcommand{\mpf}[1]{#1}

\newcommand{\jk}[1]{#1}
\newcommand{\mps}[1]{#1}
\newcommand{\revmp}[1]{#1}

\newcommand{\revmpf}[1]{#1}
\newcommand{\cro}[1]{#1}
\newcommand{\mpi}[1]{#1}
\newcommand{\mpii}[1]{#1}
\newcommand{\mpiii}[1]{#1}
\newcommand{\mpiv}[1]{#1}
\newcommand{\mpv}[1]{#1}
\newcommand{\mpvi}[1]{#1}
\newcommand{\mpvii}[1]{#1}

\newcommand{\hht}[1]{#1}
\newcommand{\jkz}[1]{#1}
\newcommand{\jkx}[1]{#1}

\newcommand{\mpg}[1]{#1}
\newcommand{\mph}[1]{#1}
\newcommand{\mpj}[1]{#1}
\newcommand{\mpk}[1]{#1}
\newcommand{\mpl}[1]{#1}
\newcommand{\mpm}[1]{#1}
\newcommand{\mpn}[1]{#1}
\newcommand{\mpp}[1]{#1}
\newcommand{\mpq}[1]{#1}
\newcommand{\mpr}[1]{#1}
\newcommand{\mpu}[1]{#1}

\newcommand{\gfi}[1]{#1}
\newcommand{\gfii}[1]{#1}

\newcommand{\rev}[1]{#1}

\usepackage{cleveref}
\crefformat{section}{Section #2#1#3}
\crefformat{subsection}{Section #2#1#3}
\crefformat{subsubsection}{Section #2#1#3}
\crefformat{appendix}{Appendix #2#1#3}

\makeatletter
\newcommand\requiredelimiter[2][########]{%
  \ifdefined#2%
    \def\@temp{\def#2#1}%
    \expandafter\@temp\expandafter{#2}%
  \else
    \@latex@error{\noexpand#2undefined}\@ehc
  \fi
}
\@onlypreamble\requiredelimiter
\makeatother

\newcommand\mechanism{\mbox{EIN}}
\newcommand\simulator{\mbox{EINSim}}
\requiredelimiter[/]{\mechanism}
\requiredelimiter[/]{\simulator}

\newcommand{\cmark}{\ding{51}}%
\newcites{S}{Survey Sources}

\begin{document}
\frenchspacing
\raggedbottom
\selectlanguage{english}
\pagenumbering{roman}
\pagestyle{plain}

\bstctlcite{IEEEexample:BSTcontrol}
\bstctlcite[@auxoutS]{IEEEexample:BSTcontrol}
\setbiblabelwidth{1000} 

\bstctlcite{IEEEexample:BSTcontrol}

\begin{titlepage}
    \large
    \begin{center}
        \begingroup
        \MakeUppercase{Diss. ETH No. \thesisDissNumber{}}
        \endgroup
    
        \hfill

        \vfill

        \begingroup
            \textbf{\thesisTitleFrontmatter}
        \endgroup

        \vfill

        \begingroup
            A thesis submitted to attain the degree of\\
            \vspace{0.5em}
            \MakeUppercase{Doctor of Sciences} of \MakeUppercase{\thesisUni} \\
            \vspace{0.5em}
            (Dr. sc. \thesisUni) \\
            
        \endgroup

        \vfill

        \begingroup
            presented by\\
            \vspace{0.5em}
            \MakeUppercase{\thesisAuthor}\\
            \vspace{0.5em}
            born on 22 October 1992\\
            \vspace{0.5em}
            citizen of the United States of America
        \endgroup

        \vfill

        \begingroup
            accepted on the recommendation of\\
            \vspace{0.5em}
            Prof.\ Dr.\ Onur Mutlu, examiner\\
            \vspace{0.5em}
            Prof. Dr. Mattan Erez, co-examiner \\
            \vspace{0.5em}
            Prof. Dr. Moinuddin Qureshi, co-examiner \\
            \vspace{0.5em}
            Dr. Vilas Sridharan, co-examiner \\
            \vspace{0.5em}
            Dr. Christian Weis, co-examiner
        \endgroup

        \vfill

        \thesisYear%

        \vfill
    \end{center}
\end{titlepage}

\thispagestyle{empty}

\hfill

\vfill

\noindent\thesisAuthor: \textit{\thesisTitlePlain,}
\textcopyright\ \thesisYear

\cleardoublepage
\thispagestyle{empty}

\vspace*{3cm}

\begin{center}
    \textit{To my loving parents, Alpa and Hamen, and my sister, Shreya.}
\end{center}

\medskip

\clearpage
\chapter*{Acknowledgments}
\addcontentsline{toc}{chapter}{Acknowledgments}

This thesis is the culmination of six years of effort throughout my graduate
studies, and a vast number of individuals contributed either directly or
indirectly. These acknowledgments comprise my reflections on their influence
herein.

\xmo{First and foremost}, I thank my adviser, Onur Mutlu, for shaping me into
the individual I am today. Through his supervision, the environment he has built
in SAFARI, and the trust he has put in me, I have acquired the expertise, tools,
and mindset I need to better understand myself, my goals, and my work. Although
I cannot know how or where I will apply these skills in the future, I am certain
that they will be central to my continued development as a researcher,
scientist, and engineer.

I thank my committee members, Mattan Erez, Moinuddin Qureshi, Vilas Sridharan,
and Christian Weis, for supervising this thesis. Your feedback is integral to
improving both the thesis and its constituent works, and I look forward to
seeing the impact this work has due to our collective efforts.

\xmo{I thank the many anonymous peer reviewers who vetted my work across many
conference submissions to ISCA, MICRO, HPCA, ASPLOS, and DSN. They
provided the external expert feedback necessary to identify and address the
weaknesses of the works, allowing us to improve their scientific value and
resulting impact.}

I am grateful for the SAFARI group members who helped and supported me along the
way, both in a working and social capacity. At its foundation, the Ph.D. has
been a lonely journey. However, you all minimized the impact of that isolation,
turning an otherwise independent process into something much, much more. In
particular, I thank Jeremie Kim for his time and energy as a friend, sounding
board, confidant, roommate, and generally someone who was always willing to
listen. I thank Amirali Boroumand and Saugata Ghose, who helped guide me through
my first days in the Ph.D. with their friendship and support. I thank the others
I interacted with during my time at CMU, including Nandita Vijaykumar, Vivek
Seshadri, Samira Khan, Rachata Ausavarungnirun, Hongyi Xin, Yang Li, Kevin
Chang, Kevin Hsieh, Yixin Luo, and Donghyuk Lee for their advice and support
during my early years. Similarly, I thank those I got to know at ETH, including
Hasan Hassan, Giray Ya\u{g}l\i{}k\c{c}\i{}, Geraldo Franciso de Oliveira, Taha
Shahroodi, Nastaran Hajinazar, Roknoddin Azizi, Can F\i{}rt\i{}na, Nika
Mansouri-Ghiasi, Konstantinos Kanellopoulos, Rahul Bera, Jisung Park, Arash
Tavakkol, Lois Orosa, Juan Gomez-Luna, and many others for their support, both
in the context of friendship and research. I would also like to extend a special
thank you to Max Rumpf and Arvid Gollwitzer for helping me with translating my
abstract into German.

Durign my time at CMU and ETH, I met many wonderful people who helped me adjust
and enjoy my time outside of lab. At CMU, I thank my roommate, Dipanjan Saha,
whose friendship was essential to surviving my sojourn into the winter
wonderland that is Pittsburgh; Vignesh Balaji, with whom I shared the experience
of the Ph.D. life like no other; and Madhav Iyengar, who supported me with his
constant friendship. At ETH, I thank my roommate, Dominik Christe, and the
Systems Group members, including Johannes de Fine Licht, Kaan Kara, David
Sidler, and Ghislain Fourny, who helped me adjust to life in Z\"{u}rich and ETH.

I thank my internship mentors, including Tyler Huberty, Stephan Meier, Jared
Zerbe, Heonjae Ha, Seung Lee, Griffin Branham, Taehyun Kim from Apple and Stefan
Saroiu and Alec Wolman from Microsoft Research for guiding me in both my work at
the company and my research in general. I am glad to have worked with them and
have my perspectives broadened.

I am eternally grateful for my family and friends from back home, whose
uncompromising support and encouragement was essential for me to make it this
far in my own journey. I thank Philip Li, Thejas Prasad, Rohan Mutalik, Chirag
Sakhuja, Nikhil Joglekar, Karthik Sethuraman, Alex Hsu, Kevin Pham, Kevin Chen,
Nikhil Dixit, Chris Roberts, Sean Shen, Nathan Kwan, Sudha Verma, and the many
others who helped me through the more difficult times. In many ways, I believe
that these individuals took the same journey with me, side-by-side, in spirit.

\xmo{I thank all the industrial partners who provided funding to support this
work: Microsoft, Intel, Google, IBM, Facebook, Qualcomm, VMware, NVIDIA,
Samsung, Alibaba, Huawei.}

\xmo{Finally, my most important thank you goes to my parents}, Alpa and Hamen
Patel, and my sister, Shreya Patel, for their love, support, and encouragement
throughout my journey to advance the state-of-the-art. Scientific contribution
is difficult, and its success is grounded on the foundation you provided me with
through your own efforts and dedication. I am endlessly grateful for the
opportunities I have had, and I thank you all, in addition to my cousins, aunts,
uncles, and other family members who stood by me, for making this possible.

\clearpage
\chapter*{Abstract}
\addcontentsline{toc}{chapter}{Abstract}

Improvements in main memory storage density are primarily driven by process
technology shrinkage (i.e., technology scaling), which negatively impacts
reliability by exacerbating various circuit-level error mechanisms. To
compensate for growing error rates, both memory manufacturers and consumers
develop and incorporate error-mitigation mechanisms that improve manufacturing
yield and allow system designers to meet reliability targets. Developing
effective error mitigation techniques requires understanding the errors'
characteristics (e.g., worst-case behavior, statistical properties).
Unfortunately, we observe that proprietary on-die Error-Correcting Codes (ECC)
used in modern memory chips introduce new challenges to efficient error
mitigation by obfuscating CPU-visible error characteristics in an unpredictable,
ECC-dependent manner.

In this dissertation, we experimentally study memory errors, examine how on-die
ECC obfuscates their statistical characteristics, and develop new testing
techniques to overcome the obfuscation through four key steps. First, we
experimentally study DRAM data-retention error characteristics to understand the
challenges inherent in understanding and mitigating memory errors that are
related to technology scaling. Second, we study how on-die ECC affects these
characteristics to develop Error Inference (EIN), a new statistical inference
methodology for inferring key details of the on-die ECC mechanism and the raw
errors that it obfuscates. Third, we examine the on-die ECC mechanism in detail
to understand exactly how on-die ECC obfuscates raw bit error patterns. Using
this knowledge, we introduce Bit Exact ECC Recovery (BEER), a new testing
methodology that exploits uncorrectable error patterns to (1) reverse-engineer
the exact on-die ECC implementation used in a given memory chip and (2) identify
the bit-exact locations of the raw bit errors responsible for a set of errors
that are observed after on-die ECC correction. Fourth, we study how on-die ECC
impacts error profiling and show that on-die ECC introduces three key challenges
that negatively impact profiling practicality and effectiveness. To overcome
these challenges, we introduce Hybrid Active-Reactive Profiling (HARP), a new
error profiling strategy that uses simple modifications to the on-die ECC
mechanism to quickly and effectively identify bits at risk of error. Finally, we
conclude by discussing the critical need for transparency in DRAM reliability
characteristics in order to enable DRAM consumers to better understand and adapt
commodity DRAM chips to their system-specific needs. 

This dissertation builds a detailed understanding of how on-die ECC obfuscates
the statistical properties of main memory error mechanisms using a combination
of real-chip experiments and statistical analyses. Our results show that the
error characteristics that on-die ECC obfuscates can be recovered using new
memory testing techniques that exploit the interaction between on-die ECC and
the statistical characteristics of memory error mechanisms to expose physical
cell behavior. We hope and believe that the analysis, techniques, and results we
present in this dissertation will enable the community to better understand and
tackle current and future reliability challenges as well as adapt commodity
memory to new advantageous applications.

\clearpage
\chapter*{Zusammenfassung}
\addcontentsline{toc}{chapter}{Zusammenfassung}

Verbesserungen in der Arbeitsspeicherdichte werden in erster Linie durch die
Miniaturisierung der Prozesstechnologie (d.h. Technologieskalierung)
vorangetrieben. Diese Verkleinerung wirkt sich jedoch negativ auf die
Zuverlässigkeit aus, da sie verschiedene Fehlerarten auf Schaltkreisebene
verstärkt. Um diesen steigenden Fehlerraten entgegenzuwirken, entwickeln sowohl
die Speicherhersteller als auch die Verbraucher Fehlerminderungstechniken, die
die Fertigungsausbeute verbessern und es den Systementwicklern ermöglichen, ihre
Zuverlässigkeitsziele zu erreichen. Die Entwicklung von effektiven
Fehlerminderungstechniken erfordert jedoch ein genaues Verständnis der
Fehlereigenschaften (z. B. Worst-Case-Verhalten, statistische Eigenschaften).
Leider stellen wir fest, dass die in modernen Speicherchips eingesetzten
proprietären On-Die Error-Correcting Codes (ECC) die Fehlereigenschaften auf
unvorhersehbare und ECC-abhängige Weise verschleiern und somit effiziente
Fehlerminderung erschweren.

In dieser Dissertation untersuchen wir Speicherfehler experimentell und
betrachten, wie On-Die ECC ihre statistischen Eigenschaften verschleiert und
entwickeln neue Testverfahren, um diese Verschleierung in vier Schritten zu
überwinden. Um die Herausforderungen zu verstehen, die mit der Miniaturisierung
zusammenhängen, untersuchen wir zunächst die
DRAM-Datenerhaltungsfehlereigenschaften experimentell. Zweitens untersuchen wir,
wie sich On-Die ECC auf diese Eigenschaften auswirkt und entwickeln Error
Inference (EIN), eine neue statistische Inferenzmethode, mit der sich wichtige
Details des On-Die ECC-Mechanismus und der von ihm verschleierten Bit-Fehler
ableiten lassen. Drittens untersuchen wir den On-Die-ECC-Mechanismus, um zu
verstehen, wie er Bit-Fehlermuster verschleiert. Mit diesem Wissen stellen wir
Bit Exact ECC Recovery (BEER) vor, eine neue Testmethode, die unkorrigierbare
Fehlermuster ausnutzt, um (1) die exakte On-Die ECC-Implementierung eines
Speicherchips und (2) die genauen Stellen der Bitfehler zu bestimmen. Viertens
untersuchen wir, wie On-Die ECC sich auf die Fehlerprofilierung auswirkt, und
zeigen, dass On-Die ECC drei zentrale Herausforderungen birgt, die sich negativ
auf die Praktikabilität und Effektivität der Profilerstellung auswirken. Zur
Überwindung dieser Herausforderungen, führen wir Hybrid Active-Reactive
Profiling (HARP) ein, eine neue Fehlerprofilierungsstrategie, die einfache
Modifikationen am On-Die ECC-Mechanismus nutzt, um schnell und effektiv
fehlergefährdete Bits zu identifizieren. Abschließend erörtern wir die kritische
Notwendigkeit von Transparenz bei DRAM-Zuverlässigkeitsmerkmalen, um es
DRAM-Kunden zu erlauben, handelsübliche DRAM-Chips besser zu verstehen und an
ihre systemspezifischen Anforderungen anzupassen.

In dieser Dissertation wird ein detailliertes Verständnis entwickelt, wie
On-Die-ECC die statistischen Eigenschaften von Arbeitsspeicherfehlern
verschleiert, wozu wir Experimente auf echten Chips mit statistischen Analysen
kombinieren. Unsere Ergebnisse zeigen, dass die Fehlercharakteristiken, die
durch On-Die ECC verschleiert werden, mit unseren neuen Speichertesttechniken
wiederhergestellt werden können. Wir hoffen und glauben, dass die Analyse, die
Techniken und die Ergebnisse dieser Dissertation es Entwicklern erlauben wird,
Zuverlässigkeitsprobleme besser zu verstehen und zu bewältigen sowie mit
handelsüblichem Arbeitsspeicher neue, vorteilhafte Anwendungen zu erschliessen.

\pagestyle{headings}
\cleardoublepage
\tableofcontents
\newpage
\listoffigures
\newpage
\listoftables

\cleardoublepage
\pagenumbering{arabic}%
\setstretch{1.3}
\chapter{Introduction}

Memory has been an integral component of computing systems for over half a
century~\cite{ibm2021dram} and has evolved alongside the rest of the system to
achieve high capacity, performance, reliability, and energy-efficiency at low
cost. In particular, dynamic random access memory (DRAM), \xmo{first introduced
by Robert Dennard at IBM in the late 1960s~\cite{ibm2021dram, dennard1968field,
dennard1974design, markoff2019ibm, nature2018memory}, has served as the de-facto
standard main memory technology across a broad range of computing systems for
decades. This is primarily due to its large, reliable storage capacity at low
cost relative to other memory technologies (e.g., SRAM,
Flash)~\dmpd{\cite{hennessy2011computer, meena2014overview, mutlu2013memory,
mutlu2015main}}.}

To remain competitive within the memory market \xmo{while satisfying growing
demands brought about by continual advancements in computing systems,} DRAM
manufacturers have relentlessly pursued improvements in storage density and
bandwidth while minimizing the \xmo{overall cost-per-bit. Manufacturers have
achieved this through improving both their circuit designs and manufacturing
processes (e.g., process technology scaling), enabling denser, faster memory
devices~\cite{ibm2021dram, kang2014co, mandelman2002challenges,
childers2015achieving}.}

Unfortunately, these improvements exacerbate various unwanted circuit-level
phenomena (e.g., charge leakage) that can lead to \emph{errors}, i.e., memory
cells that do not store data correctly~\cite{nair2013archshield, gong2017dram,
mutlu2013memory, liu2013experimental, childers2015achieving}. \xmo{Therefore, as
manufacturers continue to improve DRAM technology, they face growing error rates
that threaten both (1) DRAM manufacturers' own device reliability and factory
yield~\cite{micron2017whitepaper, kang2014co, park2015technology, son2015cidra,
cha2017defect, flamm2010impact}; and (2) system designers' ability to build
robust systems that their customers can rely
upon~\dmpd{\cite{meza2015revisiting, kim2020revisiting, kim2014flipping,
nair2016xed, son2015cidra}}.}

\begin{figure}[h]
    \centering
    \includegraphics[width=0.6\linewidth]{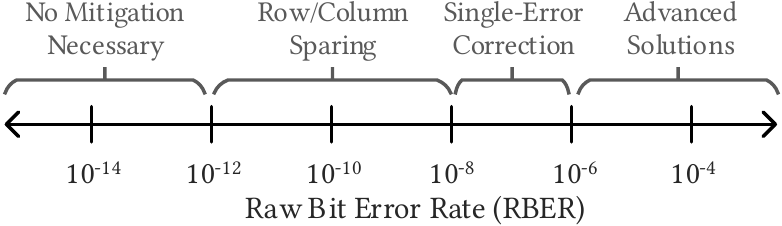}
    \caption{Error-mitigation mechanisms used to address different ranges of raw
    bit error rates (RBERs). Reproduced from prior
    work~\cite{nair2013archshield, kline2020flower}.}
    \label{fig:rber_ranges}
\end{figure}

To compensate for these errors, DRAM manufacturers incorporate
\emph{error-mitigation mechanisms}, which are circuits designed to prevent
memory errors from causing software-visible bit flips. DRAM manufacturers choose
error-mitigation mechanisms that are \xmo{appropriate for the types of errors
that need to be dealt with, e.g., depending on when and how the errors occur.}
Figure~\ref{fig:rber_ranges} illustrates different error-mitigation mechanisms
that are typically used to address different ranges of raw bit error rates
(RBERs), i.e., the proportion of cells that exhibit errors before mitigation.

Error rates below $\approx 10^{-12}$ are generally considered to be safe for
normal operation and do not require mitigation~\cite{nair2013archshield,
kline2020flower}. Relatively low error rates up to $10^{-8}$ are addressed using
\emph{row and column sparing}, where memory designers provision extra rows
and/or columns of storage cells at design-time to serve as replacements for
other rows and columns that contain erroneous
cells~\cite{kang2014co,mandelman2002challenges, nair2013archshield,
son2015cidra, horiguchi2011nanoscale}. Higher error rates up to $10^{-6}$ may be
treated using single-error correcting (SEC) error-correcting codes
(ECC)~\cite{costello1982error} (e.g., Hamming codes~\cite{hamming1950error})
that can detect and correct a limited number of errors throughout the memory
device. Finally, even higher error rates of $>10^{-6}$ require more advanced
error-mitigation solutions, such as stronger, more complex ECC (e.g.,
BCH~\cite{bose1960class, hocquenghem1959codes} or RS~\cite{reed1960polynomial}
codes), \xmo{fine-granularity repair techniques~\cite{nair2013archshield,
kline2020flower, longofono2021predicting, lin2012secret}, or solutions that
address specific types of errors (e.g., higher refresh rates to address
refresh-~\dmpd{\cite{liu2012raidr, venkatesan2006retention}} or
RowHammer-related~\dmpd{\cite{kim2014flipping, apple2015about,
mutlu2019rowhammer}} errors).} Section~\ref{sec:bg_error_mitigations} discusses
error-mitigation mechanisms in greater detail.

Recently, DRAM manufacturers have begun using single-error correcting ECC
(on-die ECC)~\cite{kang2014co, micron2017whitepaper, oh2014a, oh20153,
kwak2017a, kwon2017an} to manage \xmo{increasingly prevalent single-bit errors,
which are generally the predominant type of error that manufacturers must
contend with once their process technology is mature~\cite{micron2017whitepaper,
cha2017defect, son2015cidra, lee2001new}. On-die ECC operates entirely within
the memory device to detect and correct errors, so its operations are invisible
outside of the memory device. Internally, on-die ECC subdivides the physical
memory into \emph{ECC words} (typically 64 or 128 bits wide~\cite{nair2016xed,
gong2018duo, im2016im}) and is capable of correcting one error within each
word.}

\begin{figure}[h]
    \centering
    \includegraphics[width=0.75\linewidth]{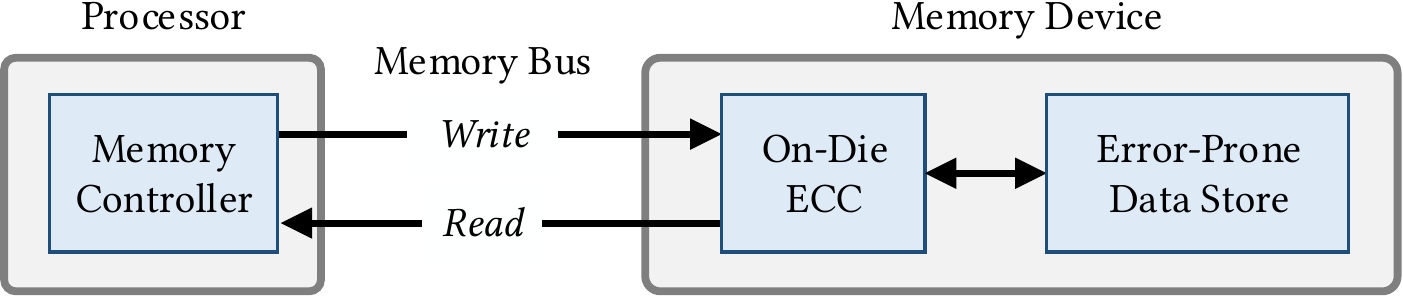}
    \caption{Block diagram of a typical system that uses a memory device with on-die ECC.}
    \label{fig:intro_sys_ondieecc}
\end{figure}

\xmo{Figure~\ref{fig:intro_sys_ondieecc} illustrates the high-level architecture
of a system that uses a memory device equipped with on-die ECC. A \emph{memory
controller} within the processor interacts with the memory device over a
\emph{memory bus}. The memory controller issues \emph{read} and \emph{write}
operations to load and store data to the memory. Within the memory device, we
see that the load and store operations first interact with the on-die ECC logic,
which then interacts with the physical data store that is susceptible to errors.
To perform error detection and correction, on-die ECC \emph{encodes} the data
provided by the memory controller using an algorithm (e.g., Hamming
encoding~\cite{hamming1950error}) such that a decoding algorithm (e.g., syndrome
decoding~\cite{moon2005error, richardson2008modern}) can tell when one (or more,
given a more complex coding algorithm) error(s) have occurred.}

\xmo{In this way, on-die ECC enables DRAM manufacturers to tolerate errors that
are randomly distributed throughout the memory device, providing consumers with
the appearance of a more reliable memory device without exposing the underlying
reliability and error characteristics of the memory chips. In particular,
manufacturers can preserve acceptable manufacturing yield at relatively low cost
(i.e., 6.25-12.5\% storage capacity~\cite{gong2018duo} with low area,
performance, and energy overheads~\cite{micron2017whitepaper, kang2014co,
cha2017defect}) using a simple single-error correcting ECC
today~\cite{micron2017whitepaper, jedec2020ddr5}. However, because on-die ECC is
a key tool for DRAM manufacturers to manipulate factory yield, it is inherently
tied to profitability and business interests~\cite{nair2013archshield}.
Therefore, DRAM manufacturers treat the details of on-die ECC and the error
rates it masks as proprietary, not disclosing them in publicly-available
documentation (e.g., device datasheets).}

\section{Problem Discussion}

Unfortunately, on-die ECC introduces new challenges for scientists and engineers
\xmo{(both researchers and practitioners in the field, in academia as well as
industry)} who need to understand memory reliability characteristics in the
course of their work. This is because on-die ECC has a limited correction
capability: although it can fully correct all single-bit error patterns, its
behavior, when faced with multi-bit error patterns (i.e., \emph{uncorrectable}
error patterns), is undefined based on the \xmo{particular ECC function used
inside a given memory chip to correct the errors~\cite{son2015cidra,
pae2021minimal, jeong2020pair, gong2018duo}. As a result, on-die ECC obfuscates
raw error patterns (i.e., the \emph{pre-correction} errors) into unpredictable,
implementation-dependent error patterns after on-die ECC correction (i.e., the
\emph{post-correction} errors).}

\xmo{Figure~\ref{fig:intro_sys_ondieecc} explains why this is the case. Although
the memory controller reads from and writes to the memory device, it has no
visibility into the encoded data that is stored into the physical memory cells.
Instead, it only has access to data that has already been modified by the on-die
ECC mechanism. Therefore, the memory controller is entirely unaware when errors
occur and/or are corrected by the on-die ECC logic. Similarly, when \dmpd{the
on-die ECC logic faces uncorrectable errors, the memory controller simply
receives the corrupted data returned by the DRAM chip without any indication
that the on-die ECC logic was unable to correct the errors (if they were even
detected at all) or even exacerbated the total number of errors through
miscorrection~\cite{jeong2020pair, criss2020improving, son2015cidra}. As a
result, the memory controller cannot easily reason about the errors that it
observes}, even if the physical processes that cause the errors to begin with
(e.g., circuit-level models) are well understood.\footnote{Note that, even if
the on-die ECC circuitry were not proprietary (i.e., its details were publicly
known), the memory controller would still lack the visibility into the on-die
ECC mechanism's error-correction process \dmpd{because the mechanism provides no
feedback to the memory controller when errors are detected and/or corrected.}}}

\xmo{On-die ECC's error-correction properties are desirable during normal
operation when uncorrectable errors are infrequent. However, its obfuscation of
raw error characteristics severely impacts studies that seek to understand how
errors occur in order to improve the system's overall \emph{robustness} (e.g.,
in terms of reliability, performance, security, safety, etc). We identify and
discuss three concrete example consumer use-cases that are negatively impacted
by the use of on-die ECC.}

\textbf{System design.} System designers who integrate commodity memory chips
into their designs must understand the chips' reliability characteristics in
order to \xmo{create a system that meets} their own design goals. Unfortunately,
on-die ECC alters \xmo{the memory chip's} reliability characteristics in a way
that makes predicting the worst-case behavior difficult to anticipate: when
faced with an uncorrectable error pattern, on-die ECC may act to
\emph{exacerbate} the error pattern into an even worse pattern, depending on the
particular ECC function chosen by the memory manufacturer~\cite{son2015cidra,
gong2018duo, jeong2020pair, pae2021minimal, patel2020bit, cha2017defect}. This
forces the system designer to make conservative assumptions about the worst-case
reliability characteristics, often overprovisioning their own designs relative
to what would be necessary if on-die ECC were not present~\cite{gong2017dram,
cha2017defect}.

\textbf{Third-party test and qualification.} Test and qualification engineers
exercise various operating conditions to determine whether a given memory chip
satisfies design requirements~\cite{kingston2012whitepaper,
kingston2021kingston, alliance2014reliability, alars2005dram, atp2018stringent}. If
and when errors are observed, the engineer must often understand (i.e., debug)
\emph{why} the error occurred, a process known as \emph{root-cause
analysis}~\cite{alars2005dram, baseman2016improving}. Unfortunately, on-die ECC
obfuscates the raw errors in a way that is opaque to the engineer. Upon
observing an error, the engineer cannot easily determine the underlying cause
for the error since the details of the error-correction process are contained
within the memory chip and not visible to the engineer.

\textbf{Memory error characterization.} \dmpd{Scientists monitor and/or
deliberately induce memory errors (e.g., by violating manufacturer-recommended
operating timings~\cite{lee2015adaptive, lee2017design, kim2018solar,
kim2018dram, kim2019d, talukder2018exploiting, talukder2018ldpuf,
chang2016understanding, kim2014flipping, liu2013experimental,
kim2020revisiting}, using extreme operating environments~\cite{ditali2007x,
bacchini2014characterization, wang2018dram, tannu2017cryogenic,
goiffon2019radiation, lee2019evaluation}, monitoring large-scale
systems~\cite{bautista2016unprotected, baseman2016improving, cai2017error,
meza2015revisiting, siddiqua2013analysis}) to understand how errors behave}
across a broad range of operating conditions. This process is known as
\emph{error characterization} and is generally performed with the goal of
developing models and insights that can be used to build smarter systems and
improve the technology itself. Unfortunately, on-die ECC conflates the
technology-specific characteristics with the particular ECC circuitry used in a
given memory chip~\cite{patel2019understanding}. Therefore, the results of an
error characterization study no longer reliably reflect properties of the memory
technology alone and are difficult to put into context alongside results from a
memory device that uses a different on-die ECC mechanism.

All three use-cases suffer from the lack of transparency into DRAM error
characteristics. In each case, not being able to understand and reason about the
observed behavior limits the DRAM consumer's ability to work with and improve
upon DRAM chips that use on-die ECC. In Chapter~\ref{chap:position}, we take
this observation further and argue that the more general lack of transparency
into DRAM operation and reliability discourages system designers from innovating
upon the DRAM substrate and adapting commodity DRAM chips to \xmo{the unique
needs of their particular systems.}

\xmo{The obfuscation that on-die ECC causes is expected to worsen as newer
memory chips with higher error rates incorporate even stronger error-mitigation
mechanisms that, in turn,} may further obfuscate error characteristics. Thus, it
is crucial to develop techniques that overcome this problem not only for today's
DRAM chips that incorporate on-die ECC, but also \xmo{other memory technologies
that are prevalent (e.g., NAND flash~\dmpd{\cite{micheloni2010inside,
cai2012error, cai2017error, cai2018errors})} and emerging \dmpd{(e.g., phase-change
memory~\cite{lee2009architecting, lee2010phasecacm, zhao2014improving, lee2010phase,
qureshi2009scalable, lee2009study, wong2010phase}, magnetoresistive
memories~\cite{everspin2021sttmram, wong2012metal, chun2012scaling,
huai2008spin, apalkov2013spin})} that will, or already do, require similar error
mitigations.}

\section{Solution Direction}

In this dissertation, we seek to (1) \emph{understand} how on-die ECC obfuscates
the underlying memory error characteristics; and (2) develop new testing
techniques that scientists and engineers can use to \emph{overcome} the
obfuscation.

\section{Thesis Statement}

Our approach is encompassed by the following thesis statement:

\begin{center}
\parbox{12.5cm}{\textit{The error characteristics that on-die ECC obfuscates can be recovered
using new memory testing techniques that exploit the interaction between on-die
ECC and the statistical characteristics of memory error mechanisms to expose
physical cell behavior, thereby enabling scientists and engineers to make
informed decisions towards building \xmo{smarter and more robust} systems.}}
\end{center}

\section{Our Approach}

To identify and address the challenges that on-die ECC introduces, we
\textit{(i)} introduce a series of new testing techniques that enable scientists
and engineers to make sense of the post-correction error characteristics that
are visible outside of the memory chip in terms of the raw pre-correction errors
that are hidden behind on-die ECC; and \textit{(ii)} build a detailed
understanding of how on-die ECC obfuscates raw error characteristics thorough a
combination of experimental and analytical studies. We briefly review our
\xmo{new} mechanisms and studies in the remainder of this section.

\subsection{Understanding DRAM Data-Retention Error Profiling}

Volatile DRAM cell storage capacitors naturally lose charge over time, resulting
in data loss (i.e., \emph{data-retention errors}) if the charge is not
periodically restored (i.e., \emph{refreshed})~\dmpd{\cite{hamamoto1995well,
liu2013experimental,
hamamoto1998retention}}.\footnote{Section~\ref{bg:subsec:data_retention}
discusses the relevant circuit-level error mechanisms in greater detail.}
Unfortunately, periodically refreshing all DRAM cells (e.g., performed every 32
or 64 ms for modern DRAM~\cite{jedec2008ddr3, jedec2012ddr4, jedec2014lpddr4})
incurs significant performance and energy overheads~\dmpd{\cite{liu2012raidr}}.
To overcome these overheads, prior works~\dmpd{\cite{liu2012raidr,
ohsawa1998optimizing, wang2014proactivedram, venkatesan2006retention,
lin2012secret, nair2013archshield, ghosh2007smart, qureshi2015avatar,
jafri2020refresh, kim2000dynamic, kim2003block, katayama1999fault,
patel2017reach, mathew2017using}} develop a variety of techniques to eliminate
unnecessary refresh operations by exploiting the observation that only a small
subset of all cells actually require frequent refreshing. 

However, identifying these few cells is a difficult problem known as
\emph{data-retention error profiling}. We observe that prior works generally
rely on a slow and expensive approach to profiling that we refer to as
\emph{brute-force} profiling. Brute-force profiling consists of iteratively
writing data to DRAM, waiting for errors to occur, and reading the data back to
identify possible errors. Unfortunately, this approach is too slow and is
detrimental to system execution, especially when profiling may be continuously
performed during run time.

To develop a faster, more efficient profiling algorithm, we perform a thorough
experimental data-retention error characterization study of 368 LPDDR4 DRAM
chips from across three major DRAM manufacturers. We closely examine how
data-retention errors occur, both collectively and at an individual cell
granularity. Based on our study, we identify three key metrics that define a
complex tradeoff space surrounding data-retention error profiling:
\emph{coverage}, \emph{false positive rate}, and \emph{runtime}. We show that
making concessions in one or more dimensions can enable benefits in the others.

We use this principle to propose \emph{reach profiling}, a new profiling
algorithm that searches for errors at a longer refresh interval and/or higher
temperature relative to the desired operating conditions. In doing so, reach
profiling identifies errors under conditions at which they are \emph{more
likely} to be observed, thereby both coverage and runtime at the cost of
identifying false positives. Using experimental data, we show that our first
implementation of reach profiling (called REAPER) can achieve on average of over
99\% coverage with less than a 50\% false positive rate while running
2.5$\times$ faster than the brute-force algorithm by profiling at 250 ms above
the target refresh interval. Furthermore, our end-to-end evaluations show that
REAPER enables significant system performance and DRAM power improvements,
outperforming the brute-force approach and enabling larger reductions of the
refresh overhead that were previously unreasonable to the profiling overheads.

\subsection{Inferring On-Die ECC and Raw Error Characteristics}

To understand how on-die ECC impacts data-retention error characterization
studies, we study how the on-die ECC error correction process affects the
statistical properties of data-retention errors. These properties are well
understood based on extensive experimental studies performed throughout decades
of literature~\dmpd{\cite{hamamoto1995well, hamamoto1998retention,
jin2004modeling, jin2005prediction, kim2009new, kong2008analysis,
liu2013experimental, kim2003block, venkatesan2006retention, qureshi2015avatar,
katayama1999fault, patel2017reach, mathew2017using, patel2019understanding,
patel2020bit, weis2015retention, shirley2014copula, weis2015thermal,
khan2014efficacy, khan2016case, khan2016parbor, khan2017detecting,
kraft2018improving, restle1992dram, saino2000impact, ditali2007x, bacchini2014characterization, wang2018dram,
tannu2017cryogenic, goiffon2019radiation, lee2019evaluation, hassan2017softmc}}.
Unfortunately, as we show in this work, the errors observed after ECC correction
no longer exhibit the expected statistical characteristics but rather depend on
the particular on-die ECC implementation used in a given DRAM chip.

To better understand both on-die ECC and the raw error characteristics, we
develop Error Inference (EIN), a new statistical inference methodology capable
of inferring details of both aspects using only the processor-visible
post-correction errors. EIN uses maximum a posteriori (MAP) estimation over
statistical models that we develop to represent ECC operation to: (1)
reverse-engineer the type and correction capability of on-die ECC; and (2) infer
high-level statistical properties of the pre-correction errors (e.g., the raw
bit error rate). As part of our work, we develop and release
EINSim~\cite{eccsimgithub}, a flexible open-source simulator that facilitates
applying EIN to a wide variety of DRAM devices and standards.

We evaluate EIN through the first experimental data-retention error
characterization study of DRAM devices with on-die ECC in open literature. We
test 232 (82) LPDDR4 devices with (without) on-die ECC to demonstrate that EIN
enables (1) reverse-engineering the type and strength of the on-die ECC
implementation, which we find to be a single-error correcting Hamming code with
\{136 code, 128 data\} bits; (2) inferring pre-correction data-retention error
rates given only the post-correction errors; and (3) recovering the well-studied
raw bit error distributions that on-die ECC obfuscates.

\subsection{Determining On-Die ECC Implementation Details}

To determine exactly how on-die ECC acts upon specific raw bit error patterns,
we observe that data-retention errors are data-dependent (i.e., they only occur
in cells programmed with specific data values). By programming specific cells to
specific values, we can prevent those cells from experiencing errors. We exploit
this capability to develop a new testing methodology, Bit-Exact Error Recovery
(BEER), that systematically determines the exact mathematical function used for
a given on-die ECC implementation (i.e., its parity-check matrix) without
hardware tools, prerequisite knowledge about the DRAM chip or the on-die ECC
mechanism, or access to ECC metadata (e.g., error syndromes, parity
information). BEER extracts this information by using carefully-crafted test
patterns that reveal behavior unique to a given ECC function when inducing
data-retention errors.

We use BEER to identify the ECC functions of 80 real LPDDR4 DRAM chips that use
on-die ECC from across three major DRAM manufacturers. We then evaluate BEER's
correctness in simulation and performance on a real system to show that BEER is
both effective and practical when applied to a wide range of ECC functions. To
show that BEER is useful in various settings, we introduce and discuss several
concrete ways that DRAM consumers can use BEER to improve upon their design and
test practices. In particular, we introduce Bit-Exact Error Profiling (BEEP),
the first error profiling algorithm that uses the known on-die ECC function
(e.g., via BEER) to recover the number and bit-exact locations of the
unobservable raw bit errors responsible for a given observed uncorrectable error
pattern.

\subsection{Enabling Error Profiling in Memory That Uses On-Die ECC}

Efficiently mitigating errors at high error rates (e.g., $>10^{-4}$) typically
requires fine-grained repair strategies that focusing error-mitigation resources
only on those bits that are susceptible to error (i.e., \emph{at-risk bits}).
However, these strategies require a fast and effective way to identify the
at-risk bits. Unfortunately, on-die ECC obfuscates how the memory controller
perceives errors, thereby complicating the process of identifying at-risk bits
(i.e., error profiling).

To understand how on-die ECC impacts error profiling, we analytically study how
on-die ECC operations alter the system's view of memory errors. Based on our
study, we find that on-die ECC introduces statistical dependence between errors
that occur in different bits that would otherwise be uncorrelated. We observe
that this dependence introduces three key challenges for error profiling: (1) a
small set of probabilistic raw bit errors can give raise to a combinatorially
large set of at-risk bits; (2) because the profiler cannot observe raw bit
errors, it wastes significant testing time to explore various combinations of
raw bit errors to determine whether a given bit is at-risk; and (3) inducing
worst-case testing conditions requires identifying complex multi-bit worst-case
test patterns, which are significantly harder to develop than single-bit
worst-case test patterns.

To address these three challenges, we introduce Hybrid Active-Reactive Profiling
(HARP), a new bit-granularity error profiling strategy that operates from within
the memory controller to identify at-risk bits. HARP rapidly achieves high
coverage by exploiting two key insights. First, uncorrectable errors have two
key sources: (1) raw bit errors that pass through on-die ECC unmodified,
referred to as \emph{direct errors}; and (2) mistakenly ``corrected'' bits as a
result of implementation-dependent undefined on-die ECC behavior, called
\emph{indirect errors}. Second, the number of indirect errors can never exceed
the correction capability of the on-die ECC code. Therefore, the key idea of
HARP is to separately identify bits that are at risk of direct and indirect
errors: with small modifications to the on-die ECC logic, the memory controller
can exploit an on-die ECC read bypass path to quickly and efficiently identify
all bits at risk of direct errors. Then, the memory controller can safely
capture any remaining bits at risk of indirect errors using a secondary ECC
within the memory controller.  

We evaluate HARP relative to two baseline profiling strategies that rely only on
post-correction errors to identify at-risk bits. We find that HARP achieves 99th
percentile coverage of at-risk bits in 20.6\%/36.4\%/52.9\%/62.1\% of the number
of profiling rounds required by the best baseline strategy given 2/3/4/5
pre-correction errors. Based on our evaluations, we conclude that HARP
effectively overcomes the three aforementioned profiling challenges introduced
by the presence of on-die ECC.

\subsection{Advocating for Transparent DRAM Reliability Characteristics}

Finally, based on the insights we developed throughout our studies, we argue the
importance of having transparency into basic commodity DRAM reliability
characteristics. We observe that, although \dmpd{techniques exist for system
designers to adapt commodity DRAM chips to their systems' particular
performance, energy, reliability, security, etc. needs (e.g., by implementing
error-mitigation mechanisms~\cite{kim2015bamboo, cardarilli2000development,
yoon2010virtualized, udipi2012lot, jian2013low, kim2015frugal, nair2016xed,
jian2013adaptive, chen2015ecc, chen2013e3cc, manzhosov2021muse, patil2021dve,
wang2018content}, reducing refresh overheads~\cite{liu2012raidr,
ohsawa1998optimizing, wang2014proactivedram, venkatesan2006retention,
lin2012secret, nair2013archshield, ghosh2007smart, qureshi2015avatar,
khan2014efficacy, khan2016case, khan2016parbor, khan2017detecting,
jafri2020refresh, kim2000dynamic, kim2003block, katayama1999fault,
patel2017reach, mathew2017using}, optimizing access
latency~\cite{chandrasekar2014exploiting, chang2016understanding, kim2018solar,
lee2015adaptive, lee2017design, wang2018reducing, zhang2016restore,
hassan2016chargecache, koppula2019eden, mathew2017using}), implementing these
techniques requires knowing key information related to DRAM reliability (e.g.,
details of the DRAM microarchitecture, testing best-practices). Unfortunately,
this information is not available today through official channels. Therefore,
system designers must rely on assuming or inferring this information through
unofficial channels (e.g., reverse-engineering techniques), which poses a
serious barrier to practical adoption of such
optimizations~\cite{khan2017detecting, qureshi2021rethinking, saroiu2022price}.}

We substantiate this argument through \dmpd{four concrete case studies: (1)
improving system-level memory reliability; (2) reducing DRAM refresh overheads;
(2) improving the DRAM access latency; and (3) implementing defense mechanisms
against the RowHammer vulnerability. In each study, we find that implementing a
solution based on previously-proposed mechanisms requires making restrictive,
potentially incorrect assumptions about a given DRAM chip.} We identify DRAM
standards as the root of the problem: current standards rigidly enforce a fixed
operating point with no specifications for how a system designer might explore
alternative operating points. 

\dmpd{To overcome this problem, we introduce a two-step approach that
reevaluates DRAM standards with a focus on transparency of reliability
characteristics. In the short term, we ask for information release through a
combination of both (1) crowdsourced testing from DRAM consumers; and (2)
official information from DRAM manufacturers, possibly standardized by
extensions to DRAM standards. In the long term, we propose extending DRAM
standards with explicit DRAM reliability standards that provide
industry-standard guarantees, tools, and/or information that enable DRAM
consumers to perform their own reliability analyses and understand DRAM
reliability at different operating points.}

\section{Contributions}

This dissertation makes the following contributions:

\begin{enumerate}
\item We build a detailed understanding of how on-die ECC obfuscates the
statistical characteristics of main memory error mechanisms. To our knowledge,
ours is the first study to explore this problem in detail. In doing so, we
introduce new statistical models to represent how pre-correction errors are
transformed by on-die ECC and draw new insights concerning the challenges that
on-die ECC introduces for understanding and mitigating errors.
\dmpd{Chapters~\ref{chap:reaper}--\ref{chap:harp} describe how we build this
understanding through studying basic data-retention error characteristics and
how on-die ECC affects those characteristics from the perspective of the memory
controller.}

\item We perform the first experimental characterization study of data-retention
error characteristics in LPDDR4 DRAM chips in order to understand the complex
tradeoffs inherent to data-retention error profiling. Using the results of our
study, we introduce and evaluate reach profiling, a new error profiling strategy
that improves coverage and performance relative to prior data-retention error
profiling strategies. Chapter~\ref{chap:reaper} describes our experimental study
and reach profiling in detail.

\item We perform the first experimental characterization study of DRAM chips
with on-die ECC in order to understand how on-die ECC affects the memory
controller's perception of data-retention errors. We show that on-die ECC
obfuscates expected statistical characteristics in a way that depends on the
particular on-die ECC implementation. Based on these observations, we introduce
Error Inference (EIN), a new statistical inference methodology for identifying
key properties of the on-die ECC implementation (e.g., correction capability,
word length) and pre-correction errors (e.g., error rate) by comparing the
post-correction errors' expected and observed statistical characteristics.
Chapter~\ref{chap:ein} describes our experimental study and EIN in detail.

\item We introduce BEER, the first testing methodology that identifies the full
on-die ECC function (i.e., its parity-check matrix) without requiring hardware
tools, prerequisite knowledge about the DRAM chip or on-die ECC mechanism, or
access to ECC metadata (e.g., error syndromes, parity information). We use BEER
to identify the ECC functions of real LPDDR4 DRAM chips and evaluate BEER both
in simulation and on a real system to show that it is practical and effective.
Chapter~\ref{chap:beer} describes BEER and its evaluations in detail.

\item We introduce BEEP, a new DRAM data-retention error profiling methodology
that leverages the information provided by BEER to infer the precise counts and
bit-exact locations of raw bit errors that are responsible for an observed
uncorrectable error pattern. We show that BEEP enables characterizing
pre-correction error locations across different ECC functions, codeword lengths,
error patterns, and error rates. Chapter~\ref{chap:beer} describes BEEP and its
evaluations in detail.

\item We analytically study how on-die ECC affects the systems' view of
uncorrectable errors in order to understand the effects that on-die ECC has on
error profiling. Based on our study, we identify three key challenges that
on-die ECC introduces for error profiling (discussed in detail in
Chapter~\ref{chap:harp}). We then introduce HARP, a new profiling strategy that
overcomes the three challenges by leveraging the key insight that only a limited
number of at-risk bits need to be identified in order to guarantee that all
remaining at-risk bits can be safely identified. We evaluate HARP to show that
it effectively overcomes the three challenges as compared with existing
profiling strategies that rely only on observing post-correction errors.
Chapter~\ref{chap:harp} describes our analysis and HARP in detail.

\item Based on the insights developed in Contributions 1-6, we \dmpd{advocate
for increased transparency into commodity DRAM reliability so that system
designers can make informed design decisions when integrating DRAM chips. We
support our argument with four concrete case studies: (1) improving overall DRAM
reliability; (2) reducing DRAM refresh overheads; (3) improving DRAM access
latency; and (4) integrating RowHammer defenses. Based on these studies, we
identify DRAM standards as the root of the problem and propose a two-step plan
for improving transparency going forward.} Chapter~\ref{chap:position} describes
the importance of DRAM transparency and provides our recommendations for current
and future commodity DRAM-based systems.

\item \dmpd{We describe remaining challenges in overcoming the problems caused
by hidden ECC mechanisms, and in a broader sense, the lack of transparency of
DRAM reliability. Chapter~\ref{chap:conc} discusses these challenges, including
extending the proposed techniques to other devices and systems, leveraging the
techniques to enable more robust system design, designing alternative
error-mitigation mechanisms to replace on-die ECC, and improving transparency
into DRAM reliability and operation.}

\end{enumerate}

\section{Outline}

This dissertation is organized into 9 chapters. Chapter~\ref{chap:bg} gives
relevant background information about DRAM operation, failure modes, and error
mitigation. Chapter~\ref{chap:related} discusses related works that address
relevant problems in error profiling, characterization, and mitigation.
Chapter~\ref{chap:reaper} presents our experimental study of DRAM data-retention
errors in chips without on-die ECC in addition to introducing and evaluating
reach profiling. Chapter~\ref{chap:ein} introduces EIN and its experimental
evaluation using DRAM chips that use on-die ECC. Chapter~\ref{chap:beer}
introduces BEER and BEEP and their respective evaluations.
Chapter~\ref{chap:harp} introduces HARP and its evaluations.
Chapter~\ref{chap:position} discusses the value of having transparent DRAM
reliability characteristics and argues for improved transparency going forward.
Finally, Chapter~\ref{chap:conc} provides a summary of this dissertation as well
as future research direction and concluding remarks.
\chapter{Background}
\label{chap:bg}

\dmpa{In this chapter, we provide the necessary overview of the background
material necessary to understand our discussions, analyses and contributions.
Section~\ref{bg:sec:dram_overview} reviews DRAM organization and operation.
Section~\ref{bg:sec:dram_errors} reviews key DRAM failure modes that are
relevant to our work. Section~\ref{sec:bg_error_mitigations} reviews memory
error mitigation techniques. Section~\ref{bg:sec:block_coding} provides an
overview of block coding, the algorithm used for typical memory error-correcting
codes. Finally, Section~\ref{bg:sec:sat_solver} introduces boolean
satisfiability (SAT) solvers as used in our work in Chapters~\ref{chap:beer}
and~\ref{chap:harp}.}

\section{DRAM Overview}
\label{bg:sec:dram_overview}

This section describes DRAM organization and operation as relevant to this
dissertation. 

\subsection{DRAM Organization}
\label{subsec:dram_organization}

\begin{figure}[h]
    \centering
    \includegraphics[width=0.6\linewidth]{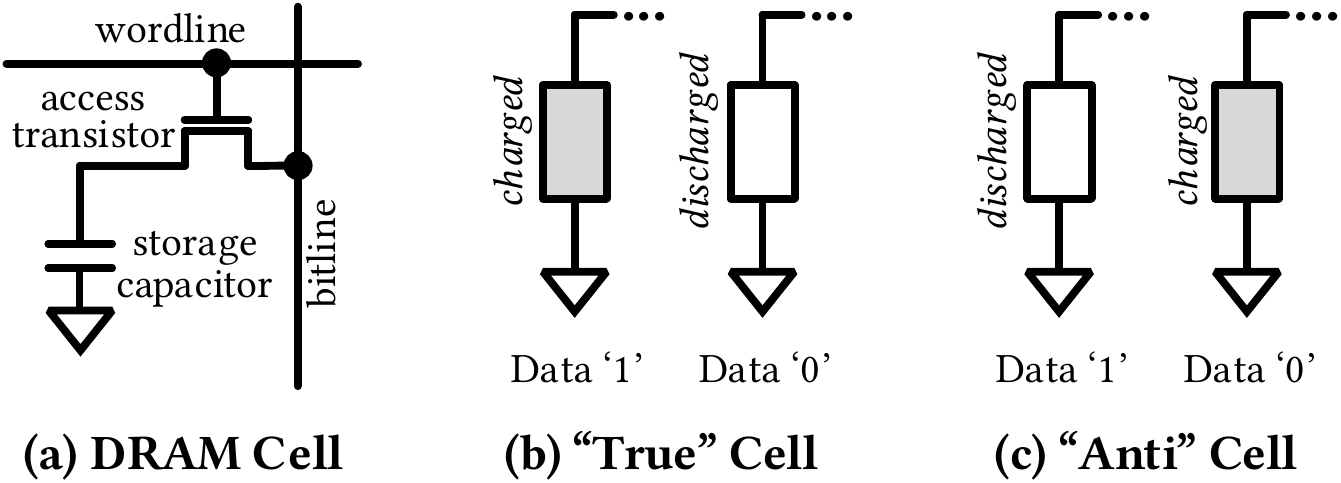}
    \caption{Overview of a DRAM cell and the two different charge encoding conventions.}
    \label{fig:dram_cell}
\end{figure}

DRAM stores each bit of data in its own DRAM cell, as illustrated in
Figure~\ref{fig:dram_cell}. The cell comprises a \emph{storage capacitor} that
encodes the binary data value using the charge level of the capacitor and an
\emph{access transistor} that is used to read and modify the stored data. The
access transistor's gate is manipulated by a control signal known as the
\emph{wordline}. When the wordline is enabled, the storage capacitor is
connected to the \emph{bitline}, so the charge stored in the storage capacitor
equalizes with that in the bitline. Whether the capacitor encodes data `1' or
data `0' as the \texttt{charged} state is an arbitrary convention and depends on
the manufacturer's particular circuit design. A cell that encodes data `1' using
the charged state is referred to as a `true cell' and the discharged state as an
`anti cell'~\cite{kraft2018improving, liu2013experimental, wu2019protecting,
kim2020charge}.

\begin{figure}[h]
    \centering
    \includegraphics[width=0.3\linewidth]{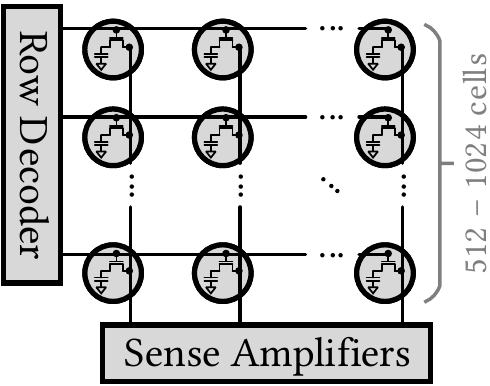}
    \caption{Two-dimensional grid of DRAM cells organized as a DRAM mat.}
    \label{fig:dram_mat}
\end{figure}

A DRAM \emph{mat} stores multiple bits of data by organizing DRAM cells into a
two-dimensional grid of rows and columns (typically 512-1024 cells per
dimension). Figure~\ref{fig:dram_mat} shows the architecture of a typical DRAM
mat. A common wordline connects the gates of all access transistors of the cells
along a \emph{row}. Therefore, when a wordline is asserted (i.e., driven to
logic `1'), all cells along the row are \emph{enabled}. The drains of all access
transistors along a \emph{column} are connected by a common bitline. Each
bitline is connected to a \emph{sense amplifier}, whose circuits detect a small
voltage shift on the bitline (i.e., as a result of charge sharing between the
storage capacitor and the bitline) and amplify it to a CMOS-readable value. In
this way, when a wordline is asserted, each cell along the row shares its
capacitor's charge with its respective bitline. Each bitline's sense amplifier
then detects and amplifies this shift, effectively reading the value stored in
each cell along the row. Note that this process is destructive: once the value
stored in a cell capacitor is read out, the storage capacitor no longer stores
the original data. Therefore, the sense amplifier must recharge the cell
following the read process in order to restore the correct data value before
attempting a subsequent read. The sense amplifier can be thought of as
temporarily caching the row's data during the read operation, so the sense
amplifiers are collectively referred to as a \emph{local row buffer}.

\begin{figure}[h]
    \centering
    \includegraphics[width=0.7\linewidth]{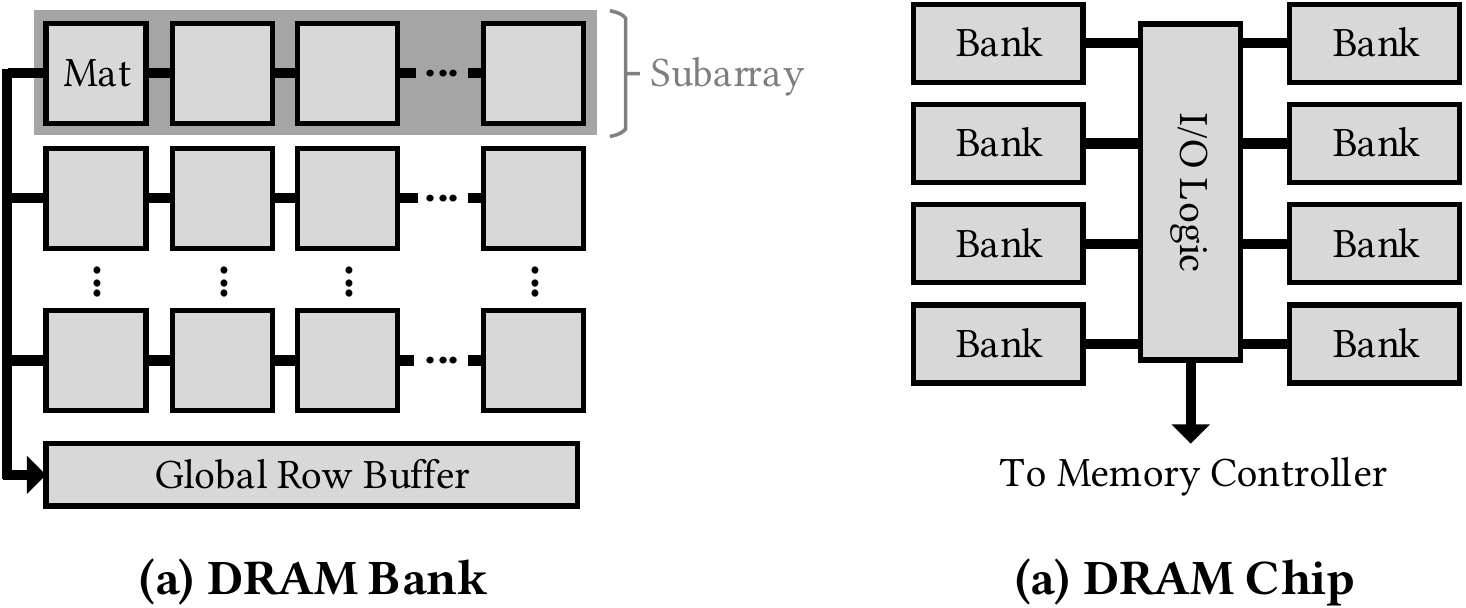}
    \caption{DRAM mats organized in a two-dimensional array to form a DRAM bank, and banks connected by common I/O logic to form a DRAM chip.}
    \label{fig:dram_bank_and_chip}
\end{figure}

Figure~\ref{fig:dram_bank_and_chip}a illustrates how mats are in turn organized into a
grid to form a DRAM \emph{bank}. Each bank typically comprises 32K-256K rows and
512-2K columns. Note that a row of mats is sometimes referred to as a
\emph{subarray}. Each bank contains a single \emph{global row buffer} that is
shared among all mats for I/O operations when the bank is read from or written
to. Figure~\ref{fig:dram_bank_and_chip}b shows how multiple DRAM banks are then combined
to form a DRAM \emph{chip}. Depending on the DRAM standard, each chip may
comprise 8~\cite{jedec2008ddr3} or 16~\cite{jedec2012ddr4} banks.

\begin{figure}[h]
    \centering
    \includegraphics[width=0.8\linewidth]{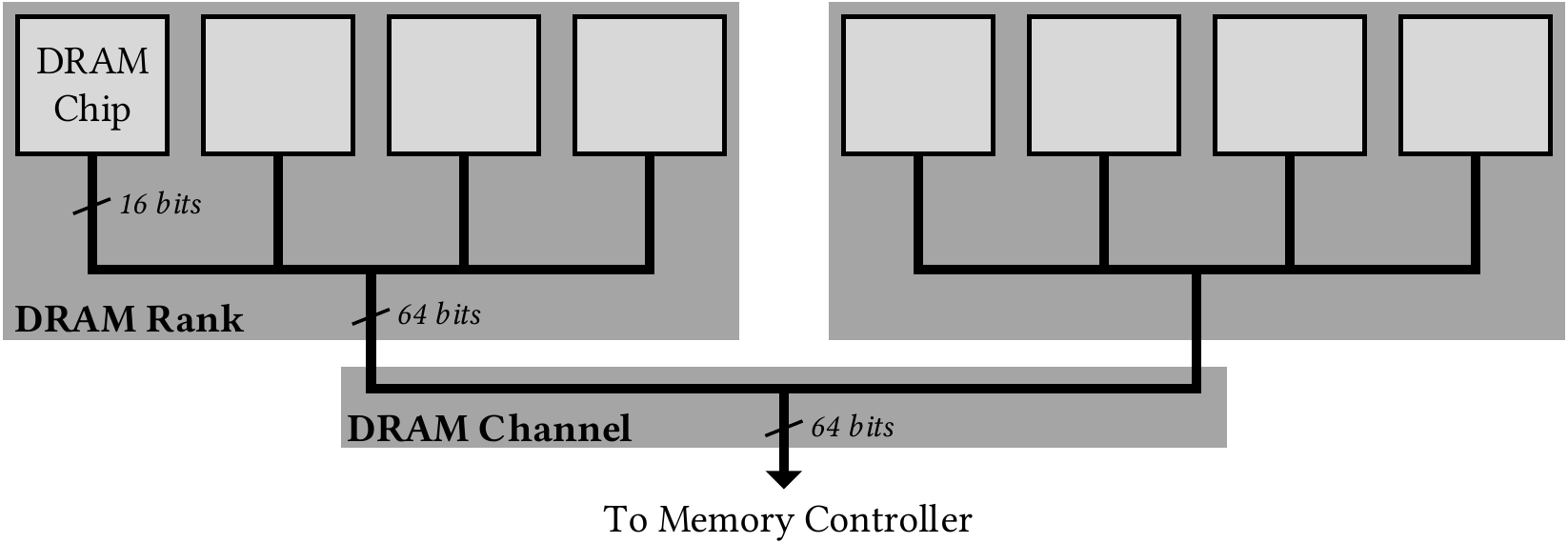}
    \caption{Multiple DRAM chips organized to form two ranks that in turn communicate over the same 64-bit channel.}
    \label{fig:dram_rank_channel}
\end{figure}

Similarly, depending on the DRAM standard, multiple chips may be combined to
form a single interface, known as a \emph{channel}, with the memory controller.
Mobile DRAM (e.g., LPDDR\emph{n}~\cite{jedec2014lpddr4}) typically uses a single
DRAM chip per channel, while desktop DRAM (e.g.,
DDR\emph{n}~\cite{jedec2008ddr3, jedec2012ddr4}) typically aggregates 4-16 DRAM
chips to form a \emph{rank} that connects to the channel. In general, up to
eight ranks may time-multiplex the same channel.
Figure~\ref{fig:dram_rank_channel} illustrates an example of a typical dual-rank
DDR4 system with four DRAM chips per rank. The figure illustrates how each chip
provides a 16-bit interface that is concatenated with all other chips in the
rank to form a 64-bit rank interface. The two 64-bit rank interfaces then
communicate with the memory controller by time-multiplexing the channel. Note
that the DRAM channel also transfers command and address information to each
chip in addition to the 64-bit data (not illustrated for simplicity). 

\subsection{DRAM Addressing}

Accessing data at a given DRAM location is done by specifying its \emph{rank},
\emph{bank}, \emph{row}, and \emph{column} address. The rank address specifies
which DRAM rank the channel should communicate with in order to access the data.
The bank, row, and column addresses are broadcast to all chips within the target
rank. The resulting data is then concatenated across all chips and returned to
the memory controller over the channel.

DRAM manufacturers are not required to maintain a direct mapping between the row
or column addresses provided on the bus and the physical location of the
corresponding cell.\footnote{This is unrelated to the virtual and physical
addresses that the system software deals with. The DRAM bus address is created
by the memory controller as a function of the physical address that is
determined by the system's virtual-to-physical memory address translation
process. Therefore, the system's physical address is orthogonal to the DRAM's
internal address mapping for locating physical cells in the DRAM device.} This
means that manufacturers are free to shuffle rows and columns as desired to
optimize their circuit designs. This mapping is generally irrelevant to the
system software since there is no perceived difference between physical cells at
different locations in the memory. Unfortunately, the mapping becomes relevant
in the presence of reliability issues such as RowHammer~\cite{kim2014flipping,
kim2020revisiting, redeker2002investigation} because the physical locations of
errors depend on the physical locations that have been accessed. If the
bus-to-physical address mapping is not known, it is difficult to determine where
errors might have occurred.

Prior works~\dmpd{\cite{lee2017design, wang2020dramdig, barenghi2018software,
jung2016reverse, kim2020revisiting, tatar2018defeating, kim2018solar}} develop
techniques to reverse-engineer the DRAM internal row and/or column mappings by
deliberately inducing errors and observing their bus address locations. Knowing
that the induced errors should follow well-understood characteristics based on
the physical chip design (e.g., RowHammer errors should occur in rows that are
physically adjacent to accessed rows), these works are able to infer the
internal row and/or column organization. In Chapter~\ref{chap:position}, we
argue that this information should be made public given that (1) it is helpful
for researchers and engineers to know in order to design DRAM optimizations and
security defenses; and (2) it can already be reverse-engineered, as shown by the
aforementioned works.

\subsection{DRAM Operation}

The memory controller is responsible for orchestrating all DRAM operations,
including DRAM initialization, access, and maintenance (e.g., periodic refresh
and calibration). These operations are invoked through the use of DRAM commands,
and DRAM manufacturers clearly specify the timing constraints surrounding their
usage~\cite{jedec2008ddr3, jedec2012ddr4, jedec2014lpddr4, jedec2020ddr5,
jedec2020lpddr5}. In this section, we briefly introduce the history of DRAM
specifications in order to provide context for how they came about. Then, we
discuss the primary commands and timings concerning DRAM access and refresh as
mandated by modern JEDEC standards.

\subsubsection{DRAM Specifications}

Following the conception of single-transistor DRAM by IBM in the late
1960s~\cite{markoff2019ibm, nature2018memory, ibm2021dram, dennard1968field,
dennard1974design}, the first commodity DRAM chips (e.g., Intel
1103~\cite{intel1975data}, Mostek MK4096~\cite{mostek1973mk4096}) quickly gained
popularity due to their cost-per-capacity advantages relative to other viable
main memory technologies~\cite{flamm1993measurement}. Through the following
decades, various manufacturers developed their own DRAM chips, resulting in a
proliferation of different DRAM types and access protocols (e.g., FPM, EDO) that
steadily improved storage density through continued improvements in both design
and manufacturing. Several competing DRAM standards arose in the 1990s,
including Intel's PCx~\cite{intel1998pc}, Rambus' RDRAM~\cite{rambus2000direct},
and JEDEC's SDRAM~\cite{jedec1994synchronous} standards. However, by the late
1990s, JEDEC thoroughly secured market dominance, effectively crowding out
competing standards by the mid 2000s~\cite{cavicchi2010patent}. By the advent of
modern DDR\emph{n} memory, JEDEC had established significant barriers to entry
for all non-JEDEC-compliant chips~\cite{cavicchi2010patent,
european2010competition}.

Today, all major commodity DRAM manufacturers comply with JEDEC standards for
DRAM interfaces and operating characteristics because JEDEC-compliance
guarantees interoperability between DRAM chips and other JEDEC-compliant
systems~\cite{jedec2021jc42}. JEDEC DRAM standards are collectively developed by
committees comprising both DRAM manufacturers and customers. Each specification
discusses requirements for customer-visible aspects of a DRAM chip's design,
including features, electrical characteristics, access protocols, recommended
operating conditions (e.g., timings, voltage, temperature). This enables a
\emph{separation of concerns} between DRAM manufacturers and consumers. In the
ideal case, \emph{DRAM specifications} are synonymous with \emph{customer
requirements}: DRAM manufacturers can focus on manufacturing the best possible
DRAM chips that conform to the specification. 

\subsubsection{DRAM Access}

\newcommand{\cmdact}[0]{\texttt{ACT}}
\newcommand{\cmdrd}[0]{\texttt{RD}}
\newcommand{\cmdwr}[0]{\texttt{WR}}
\newcommand{\cmdpre}[0]{\texttt{PRE}}
\newcommand{\cmdref}[0]{\texttt{REF}}

DRAM specifications outline the precise command sequence used for each DRAM
access. As described in Section~\ref{subsec:dram_organization}, each bank
contains only one global row buffer. Therefore, only one row may be read from
any given bank at a time. To meet this constraint, each DRAM access requires up
to \emph{three} operations:
\begin{enumerate}
    \item \emph{Row activation (\cmdact{}):} The \cmdact{} command
    \emph{activates} a row within a bank by asserting the row's wordline and
    allowing the sense amplifiers to develop to CMOS-readable voltage values.
    This is referred to as \emph{opening} the row.

    \item \emph{Column access (\cmdrd{} and \cmdwr{}):} \cmdrd{} and \cmdwr{}
    commands read and write to a given column within the open row. The
    definition of a column depends on the DRAM geometry and standard, but
    typically refers to a 16-64-byte aligned region of the open row. While a
    given row is open, the memory controller is free to issue as many \cmdrd{}
    and \cmdwr{} commands as necessary to columns within the open row.
    
    \item \emph{Bank precharge (\cmdpre{}):} The \cmdpre{} command \emph{precharges}
    the currently open row within a bank by de-asserting the wordline and
    resetting all bitlines to their quiescent states (typically
    $\frac{1}{2}V_{dd}$). This prepares the bank for the next row activation.
\end{enumerate}

\newcommand{\trcd}[0]{\texttt{tRCD}}
\newcommand{\trp}[0]{\texttt{tRP}}
\newcommand{\tccd}[0]{\texttt{tCCD}}
\newcommand{\tras}[0]{\texttt{tRAS}}
\newcommand{\trefw}[0]{\texttt{tREFW}}

\begin{figure}[h]
    \centering
    \includegraphics[width=\linewidth]{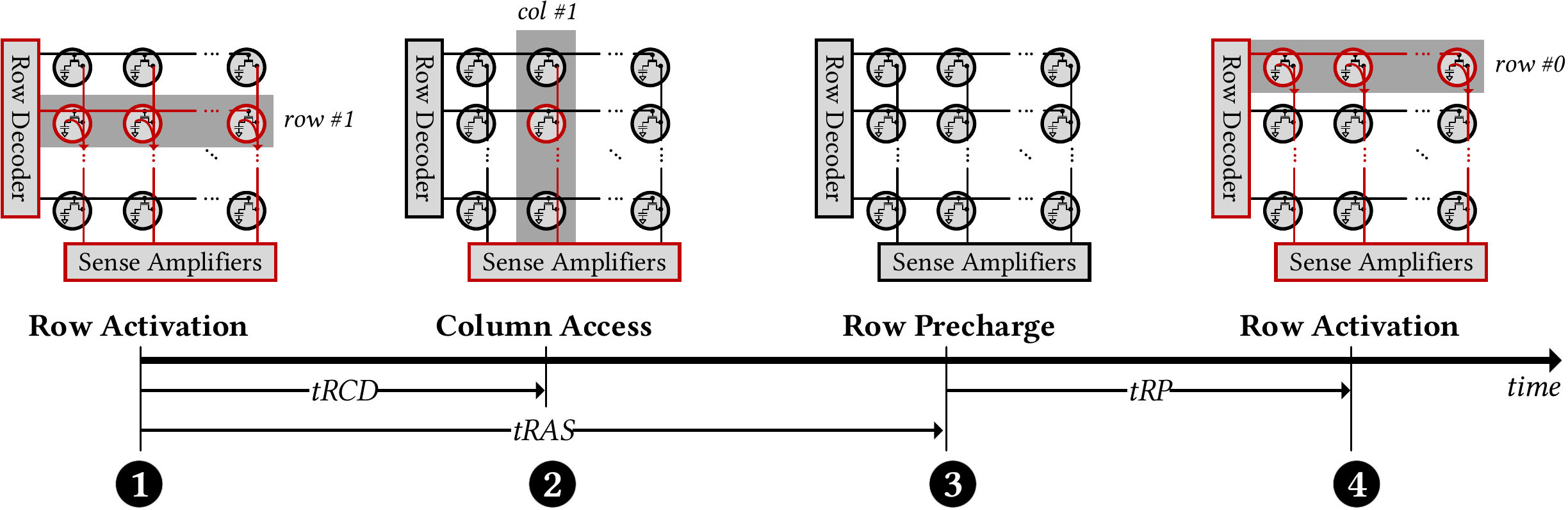}
    \caption{Core DRAM access timings for the \cmdact{}, column access (i.e.,
    \cmdrd{} and \cmdwr{}), and \cmdpre{} commands.}
    \label{fig:dram_access_timings}
\end{figure}

Figure~\ref{fig:dram_access_timings} summarizes the timing constraints that the
memory controller must obey when issuing ACT, RD and WR, and PRE commands in the
context of a DRAM read access. DRAM chip datasheets provide
manufacturer-recommended values for each of these timings, in addition to many
other timings for other commands.\footnote{We refer the reader to the JEDEC DRAM
specifications for a more detailed overview of standardized DRAM commands and
timings~\cite{jedec2008ddr3, jedec2012ddr4, jedec2014lpddr4, jedec2020ddr5,
jedec2020lpddr5}.} First, \circled{1} a new row (e.g., row address 1) in the
bank is activated. This causes the row decoder to activate the row's wordline,
allowing the sense amplifiers to detect and amplify the row's stored data
values. Next, \circled{2} a particular column (e.g., column address 1) in the
newly-opened row is accessed following a time delay of \trcd{}. Peripheral I/O
logic (not illustrated) will transfer the corresponding data to the chip's
external interface. Then, \circled{3} the bank is precharged in order to close
the open row and prepare the bank for the next row access. The \cmdpre{} command
may not be issued until at least \tras{} has elapsed since the row activation.
Finally, \circled{4} another row may be opened (e.g., row address 0) following
\trp{} after the precharge operation.

\subsubsection{DRAM Refresh}

For as long as commodity DRAM has existed, the DRAM refresh algorithm has
favored simplicity and correctness: the memory controller issues a fixed number
of refresh commands every \emph{refresh window} (\trefw{}), which corresponds to
the retention time of the worst-case cell in the memory chip. The refresh
algorithm guarantees that every DRAM cell is refreshed \emph{at least} as often
as required to prevent data loss,\footnote{This is true whether each refresh
operation operates on all DRAM rows (e.g., burst-mode
refresh~\cite{micron1999tn}) or on a subset of all rows (e.g., distributed-mode
refresh~\cite{micron1999tn}, fine-grained refresh~\cite{jedec2012ddr4}).} and
its simplicity means that both DRAM manufacturers and system designers can treat
the refresh algorithm as a fixed design constraint without needing further
specification. 

Current DRAM specifications~\cite{jedec2008ddr3, jedec2012ddr4, jedec2020ddr5,
jedec2014lpddr4, jedec2020lpddr5} mandate that the memory controller issue
\cmdref{} commands every 3.9-7.8$\mathrm{\mu}$s. Each refresh operation invokes
dedicated refresh circuitry within the DRAM chip that refreshes a small number
(e.g., 32) of rows. Throughout each \trefw{} period (i.e., 32-64ms), all rows in
the DRAM chip are refreshed once.\footnote{Note that, while this is the common
understanding, DRAM specifications do not mandate that internal refresh
operations refresh all rows exactly once within \trefw{}. By adjusting the
number of rows refreshed with each \cmdref{} command, DRAM manufacturers may
adjust the effective refresh rate for each row according to their needs in a
manner that is invisible to the memory controller.} However, it is important to
note that different cells can withstand different refresh rates, and refreshing
all cells at a fixed rate is a conservative strategy that accounts for the
\emph{worst-case} cell under \emph{worst-case} operating conditions. As we will
see in Chapter~\ref{chap:reaper}, this implies that there is significant room
for improvement in the DRAM refresh algorithm.

\subsection{Violating Recommended Timings}
\label{bg:subsec:violating_recommended}

We can induce errors in real DRAM chips by deliberately violating
manufacturer-recommended timings. The resulting error distributions allow us to:
1) reverse-engineer various proprietary DRAM microarchitectural
characteristics~\cite{jung2016reverse, lee2015adaptive, patel2019understanding,
chang2016understanding, mukhanov2020dstress, kim2018dram, kraft2018improving,
kim2018solar, orosa2021deeper, hassan2021uncovering} and 2) understand the
behavior of different DRAM errors (e.g., data
retention~\cite{hamamoto1998retention, liu2013experimental,
venkatesan2006retention, liu2012raidr, qureshi2015avatar, khan2014efficacy},
access latency-related errors~\cite{kim2018dram, kim2019d, talukder2018ldpuf,
talukder2019prelatpuf, gao2019computedram, olgun2021quac}).

By increasing $t_{REFW}$, we observe \emph{data-retention errors} in certain
cells with higher charge leakage rates~\cite{venkatesan2006retention,
hamamoto1998retention, liu2012raidr, kang2014co, kim2009new, hamamoto1995well,
liu2013experimental, nair2013archshield, khan2014efficacy, khan2016parbor,
khan2017detecting, weis2015retention, jung2017platform,jin2005prediction,
kim2009new, kong2008analysis, lieneweg1998assesment}. The quantity and locations
of these errors depend on \emph{i}) the data pattern programmed into cells,
\emph{ii}) the layout of true- and anti-cells in DRAM~\cite{liu2013experimental,
kraft2018improving}, and \emph{iii}) environmental factors such as operating
temperature and voltage~\cite{hamamoto1998retention, liu2012raidr,
liu2013experimental, khan2014efficacy, patel2017reach, chang2017understanding,
khan2016parbor, khan2016case, kim2014flipping, lee2010mechanism, li2011dram,
khan2017detecting}. Section~\ref{bg:sec:dram_errors} discusses the statistical
characteristics of data-retention errors in greater depth.

\section{DRAM Errors and Failure Modes}
\label{bg:sec:dram_errors}

In this section, we first clarify the terminology used throughout this
dissertation to refer to memory errors. Then, we review three key DRAM failure
modes that are relevant to our work: (1) DRAM data retention; (2) DRAM access
timing violations; and (3) RowHammer.

\subsection{Reliability-Related Terminology}

We observe that different works use different terms to refer to related
reliability concepts. Therefore, we clarify the reliability-related terms that
we use throughout this dissertation.

\subsubsection{Faults, Errors, and Failures}

We use the terms \emph{fault}, \emph{error}, and \emph{failure} according to the
ISO/IEC/IEEE 24765 standard~\cite{ieee2010iso} as follows:

\begin{itemize}
    \item \emph{Fault}: A defect in a hardware device or component.
    \item \emph{Error}: A discrepancy between the computed, observed, or measured value and the theoretically correct value.
    \item \emph{Failure}: The termination of the ability of a product to perform its required function.
\end{itemize}

Therefore, we use the term \emph{error} to refer to bit-flips observed during
testing and \emph{fault} or \emph{error mechanism} to refer to the underlying
reason for an observed error (e.g., excessive charge leakage, electrical
coupling between circuit components). Finally, we use the term \emph{failure
mode} when describing the manifestation of an error from the perspective of the
rest of the system. 

\subsubsection{Types of Faults and Errors}

Following from prior work~\cite{gong2018duo, jeong2020pair}, \dmpa{we classify}
DRAM faults as \emph{inherent} or \emph{operational}. Inherent faults are caused
by aspects of the DRAM technology and broadly encompass scaling-related faults
such as data-retention and RowHammer. Operational faults are those that can
cause errors during normal operation and include factors such as aging
effects~\cite{jedec2016failure} and particle strikes~\cite{may1979alpha}.
Section~\ref{subsec:bg:key_failure_modes} discusses both classes of faults in
greater detail.

We refer to errors that occur within the physical memory device as \emph{raw
errors} or \emph{pre-correction errors} to distinguish that they have not yet
gone through error-correction logic. Similarly, we refer to errors that have
gone through error correction (e.g., through on-die ECC) as
\emph{post-correction} errors. Note that all post-correction errors are the
result of the error-correction mechanism being unable to fully correct an
uncorrectable raw error pattern. \dmpd{Therefore, post-correction errors are
either detectable-but-uncorrectable (DUE) errors or silent data corruption (SDC)
errors.} Given that on-die ECC is a single-error correcting
code~\dmpd{\cite{oh2014a, oh20153, kwak2017a, kwon2014understanding,
micron2017whitepaper, im2016im, jedec2020ddr5, kang2014co, patel2020bit} that}
has no additional error-detection capability and does not report error detection
or correction events to the memory controller, all post-correction errors we
work with in this dissertation can be considered SDCs.

\subsection{Key DRAM Failure Modes}
\label{subsec:bg:key_failure_modes}

DRAM suffers from a broad range of error mechanisms that can lead to different
types of errors. Our work focuses primarily on errors that result from inherent
faults due to DRAM technology scaling and deliberate violations of recommended
operating timings and conditions (e.g., data-retention
errors~\cite{hamamoto1995well, hamamoto1998retention, venkatesan2006retention,
liu2013experimental, bacchini2014characterization, weber2005data,
yamaguchi2000theoretical}, errors due to access timing
violations~\cite{kim2018solar, lee2013tiered, lee2015adaptive, lee2017design,
chang2016understanding, olgun2021quac, koppula2019eden,
chandrasekar2014exploiting}). In this section, we review key DRAM failure modes
to provide context for our studies.

\subsection{Manufacturing-Time Faults}

Manufacturing-time faults include any defects incurred at manufacturing time
that cause one or more DRAM cells to not meet design requirements. These faults
can cause errors occur due to a variety of DRAM-specific (e.g.,
RowHammer~\dmpd{\cite{kim2014flipping, mutlu2017rowhammer, mutlu2019rowhammer,
walker2021on, yang2019trap, park2016statistical, kim2020revisiting}},
data-retention~\dmpd{\cite{hamamoto1995well, hamamoto1998retention,
venkatesan2006retention, liu2013experimental, bacchini2014characterization,
weber2005data, yamaguchi2000theoretical}}, variable-retention
time~\cite{bacchini2014characterization, kim2015avert, qureshi2015avatar,
yaney1987meta, restle1992dram, shirley2014copula, kim2011characterization,
kim2011study, kumar2014detection, mori2005origin, ohyu2006quantitative,
khan2014efficacy, kang2014co, liu2013experimental}) and non-specific (e.g.,
stuck-at faults due to manufacturing defects~\cite{alars2005dram,
horiguchi2011nanoscale, mrozek2019multi}) error mechanisms and are the primary
reliability concern with DRAM technology
scaling~\dmpd{\cite{micron2017whitepaper, kang2014co, gong2018duo,
cha2017defect, nair2013archshield, park2015technology, mutlu2013memory}}. These
errors are dealt with by manufacturers using on-die error-mitigation mechanisms,
including row and column sparing~\dmpd{\cite{horiguchi2011nanoscale,
cha2017defect, kang2014co, nair2013archshield, keeth2007dram, son2015cidra,
kwon2014understanding}}, on-die ECC~\dmpd{\cite{oh2014a, oh20153, kwak2017a,
kwon2014understanding, micron2017whitepaper, im2016im, jedec2020ddr5,
kang2014co, patel2020bit}}, and RowHammer-prevention logic (e.g., on-die target
row refresh~\dmpd{\cite{hassan2021uncovering, frigo2020trrespass,
jattke2022blacksmith, marazzi2022protrr}}). Therefore, assuming the error
mitigation mechanisms work as intended, these errors are normally unknown to the
end user and are considered proprietary information by DRAM
manufacturers~\cite{gong2017dram, nair2013archshield, saroiu2022price,
qureshi2021rethinking, loughlin2021stop, farmani2021rhat}.

Prior works~\dmpd{\cite{gong2017dram, nair2013archshield, kim2020revisiting}}
explain that these errors become worse with continued process technology scaling
because manufacturing variations between components become more significant as
the components themselves shrink. For example, shrinking cell storage capacitors
means that they will hold less charge and therefore be more sensitive to various
charge leakage mechanisms~\cite{kang2014co, nair2013archshield}. Therefore, as
DRAM technology continues to scale, a larger proportion of cells fail to meet
design targets and require error mitigation. 

When deliberately changing access and refresh timings (e.g., during testing or
custom operation), the user effectively chooses to operate under conditions that
DRAM manufacturers did not design for. At these conditions, the on-die error
mitigations that manufacturers provisioned to address manufacturing-time errors
(e.g., errors related to data retention and access timings) no longer hide all
errors, allowing the user to study their characteristics. Many prior
works~\dmpd{\cite{jung2014optimized, hamamoto1995well, hamamoto1998retention,
yaney1987meta, shirley2014copula, weis2015retention, jung2015omitting,
weis2015thermal, baek2014refresh, khan2014efficacy, liu2013experimental,
qureshi2015avatar, venkatesan2006retention, hassan2017softmc,
chang2017understanding, lee2017design, schroeder2009dram, sridharan2012study,
jung2016reverse, khan2016case, khan2016parbor, qureshi2015avatar, sutar2016d,
chang2017thesis, lee2016reducing, chandrasekar2014exploiting,
chang2016understanding, lee2015adaptive, kim2018dram, kim2018solar, kim2019d,
talukder2018exploiting, talukder2018ldpuf, kim2014flipping,
kim2015architectural, gao2019computedram, kim2009new, kim2020revisiting}} use
this approach to observe scaling-related errors and understand their
characteristics in order to develop more robust systems
Section~\ref{sec:rel:experimental_studies} reviews these studies in more detail.

\subsection{Operational Failures}

DRAM suffers from various faults that are prevalent among general semiconductor
devices~\cite{jedec2016failure}, including possible circuit-level aging
effects~\dmpd{\cite{jedec2016failure, meza2015revisiting}} and random external
events (e.g., particle strikes~\cite{may1979alpha}). Unfortunately, the details
of commodity DRAM failure rates are proprietary~\cite{gong2017dram,
nair2013archshield, saroiu2022price, qureshi2021rethinking, loughlin2021stop,
farmani2021rhat} and prior studies~\dmpd{\cite{schroeder2009dram,
hwang2012cosmic, sridharan2012study, sridharan2015memory, sridharan2013feng,
meza2015revisiting, bautista2016unprotected, siddiqua2013analysis}} have had to
conduct their own large-scale experimental analyses to estimate failure rates.
These studies do \emph{not} know the underlying faults responsible for an
observed failure; instead, they infer reasons for the failures based on how the
failures appear to the system (e.g., correlations between memory
addresses)~\cite{baseman2016improving}. Therefore, the failure rates observed by
these studies represent the cumulative effect of whichever circuit-level error
mechanisms are dominant for the systems' particular configurations. 

Systems that need long-term resilience against operational failures typically
incorporate additional error-mitigation mechanisms \emph{outside} the DRAM chip
(e.g., rank-level ECC within the memory controller~\cite{cha2017defect,
kim2015bamboo}). These mechanisms are better equipped to deal with operational
failures because they are not as resource-constrained as the DRAM die.
Therefore, they can provision resources to mitigate a wide variety of errors,
including multi-bit errors and even failures of entire DRAM
chips~\cite{kim2015bamboo}. Section~\ref{subsec:rank_lvel_ecc} discusses these
mechanisms in more detail.

\subsection{DRAM Data Retention Errors}
\label{bg:subsec:data_retention}

\begin{figure}[h]
    \centering
    \includegraphics[width=0.6\linewidth]{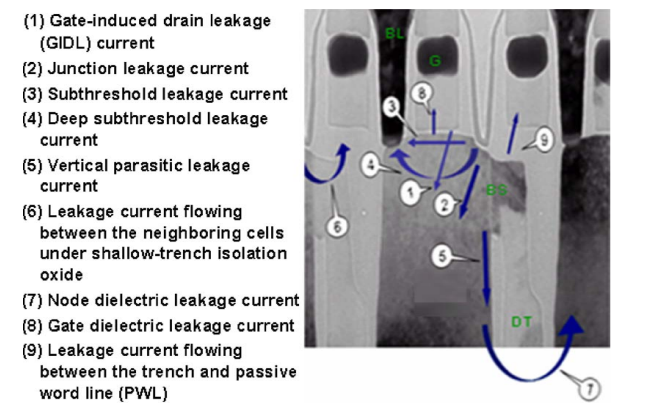}
    \caption{Major DRAM cell charge leakage paths illustrated in a
    cross-sectional view of a DRAM cell, showing the bitline (BL), access
    transistor gate (G), and storage capacitor (DT). Image taken
    from~\cite{lee2012dram}.}
    \label{fig:dram_cell_leakage_paths}
\end{figure}

DRAM cell storage capacitors are susceptible to a large number of charge leakage
mechanisms that vary in strength depending on the data values stored in cells,
switching activity nearby a given cell, and with environmental factors such as
temperature and voltage~\cite{chatterjee1979leakage, restle1992dram,
hamamoto1998retention, iniewski2011nano, lee2012dram}.
Figure~\ref{fig:dram_cell_leakage_paths} shows a cross-sectional view of a DRAM
cell capacitor (illustrated as ``DT'') alongside various charge leakage paths
that it is susceptible to. The dominant leakage path(s) vary depending on the
particular DRAM circuit design and operating conditions; however, all DRAM cells
are susceptible to charge leakage and rely on periodic DRAM refresh operations
to prevent data loss. At manufacturer-recommended operating conditions, DRAM
refresh operations occur frequently enough to prevent data-retention errors from
appearing.

However, when the memory controller \emph{deliberately reduces} the frequency of
refresh operations (e.g., in order to improve system performance and
energy-efficiency~\cite{venkatesan2006retention, liu2012raidr}) certain cells
with worse leakage characteristics exhibit errors. Prior
work~\dmpd{\cite{venkatesan2006retention, hamamoto1995well,
ohsawa1998optimizing, hamamoto1998retention, liu2012raidr, wyns1988temperature,
liu2013experimental, khan2014efficacy}} has extensively studied how long
different DRAM cells can correctly retain data (i.e., their \emph{retention
times}) and shown that only a small set of worst-case cells actually require
frequent refreshing (i.e., at the JEDEC-standardized rate of 32-64
ms~\cite{jedec2008ddr3, jedec2012ddr4, jedec2014lpddr4}). In contrast, the vast
majority of cells have much longer retention times (e.g., $>1$s). This high
cell-to-cell variation is a result of manufacturing process variation and is
becoming worse with continued technology shrinkage~\cite{nair2013archshield,
cha2017defect, hong2010memory}.

Through their experimental studies, prior works~\cite{hamamoto1995well,
hamamoto1998retention, liu2013experimental, venkatesan2006retention, liu2012raidr,
jin2005prediction, weis2015retention, kong2008analysis} have identified several
key characteristics of DRAM data-retention errors that are essential to our work
in this dissertation. 

First, data-retention errors are strongly dependent on and can be easily induced
by manipulating the refresh window (e.g., using configuration registers within
the memory controller~\cite{amd2018amd}). Except for certain cells
that are susceptible to a phenomenon known as variable retention time (discussed
later in this section), \dmpd{most strong} DRAM cells exhibit relatively
\dmpd{stable} retention times and fail repeatably when the refresh window
exceeds a cell's retention time. 

Second, data-retention errors are uncorrelated and occur uniform-randomly
throughout the memory device as a result of random process variation. Prior
works demonstrate this property through extensive error-characterization
studies, showing that data-retention errors exhibit no discernible spatial
patterns~\cite{hamamoto1998retention, liu2012raidr, baek2014refresh,
venkatesan2006retention, shirley2014copula} and can be realistically modeled as
uniform-randomly distributed~\cite{kim2018dram, shirley2014copula,
baek2014refresh} independent events~\cite{shirley2014copula}.

Third, data-retention errors are data-dependent, occurring only in cells whose
storage capacitors are in the charged state. This means that only true-cells
programmed to data `1' are likely to fail, and vice-versa for anti-cells. This
behavior has been observed across DRAM chips from a broad range of manufacturers
and standards~\cite{kraft2018improving, liu2013experimental, wu2019protecting,
kim2020charge}, and we hypothesize this is due to the dominant leakage modes
causing charge to drain the storage capacitor rather than to charge it.

\subsubsection{Variable Retention Time}

Certain cells are vulnerable to an error mechanism known as \emph{variable
retention time (VRT)}, where a cell unpredictably switches between two or more
distinct retention times states~\cite{bacchini2014characterization,
kim2015avert, qureshi2015avatar, yaney1987meta, restle1992dram,
shirley2014copula, kim2011characterization, kim2011study, kumar2014detection,
mori2005origin, ohyu2006quantitative, khan2014efficacy, kang2014co,
liu2013experimental}. This affects the cell's data-retention error
characteristics because the likelihood that a cell fails changes over time
depending on its retention time state. Prior works~\cite{micron2017whitepaper,
kang2014co, cha2017defect} identify VRT errors as a key DRAM scaling challenge,
and on-die ECC is partially motivated by their presence. Although techniques for
identifying and mitigating VRT-related errors have been
proposed~\cite{qureshi2015avatar, sharifi2017online}, efficiently identifying
and mitigating VRT errors remains a difficult problem.

\subsection{DRAM Access Timing Violations}

The memory controller may \emph{deliberately choose} to violate the
manufacturer-recommended inter-command timings in order to improve DRAM access
performance~\dmpd{\cite{chandrasekar2014exploiting, chang2016understanding,
kim2018solar, lee2015adaptive, lee2017design, wang2018reducing,
zhang2016restore, hassan2016chargecache, koppula2019eden, mathew2017using,
zhang2021quantifying, lee2016reducing, chang2017thesis}} or exploit beneficial
side effects of the resulting undefined behavior (e.g., to generate random
values~\cite{talukder2018exploiting, kim2019d, olgun2021quac} and device
fingerprints~\cite{talukder2018ldpuf, kim2018dram}, perform logic
operations~\dmpd{\cite{gao2019computedram, olgun2021pidram, seshadri2019dram,
seshadri2017ambit, seshadri2015fast, seshadri2013rowclone}}). However, if
timings are violated beyond what a given DRAM chip can withstand, errors may
occur due to internal DRAM circuitry (e.g., sense amplifiers, peripheral logic)
having insufficient time to complete necessary operations. We briefly discuss
activation- and precharge-related errors because they have been extensively
studied and exploited by prior works.

\subsubsection{Activation-Related Errors}

Reducing the time given for DRAM row activation (i.e., \trcd{}) can cause errors
due to reading out data from the row buffer before the sense amplifiers are able
to fully develop to CMOS-readable values~\cite{chang2016understanding}. Several
works study the effects of reducing \trcd{} and find that (1) more errors occur
with greater timing violation~\cite{chandrasekar2014exploiting,
chang2016understanding, kim2018solar, kim2018dram, lee2015adaptive,
hassan2017softmc}; (2) many of the resulting errors tend to cluster along
bitlines~\cite{chang2016understanding, kim2018solar, kim2019d}; and (3) errors
are more likely to occur closer to one edge of a given
subarray~\cite{lee2017design}. These observations are all consistent with sense
amplifier behavior: with lower \trcd{}, the sense amplifier has less time to
develop, thereby becoming more likely to yield incorrect data. In particular,
due to manufacturing process variation, certain sense amplifiers develop more
slowly than others, causing them to exhibit errors for all cells along their
respective bitline. Furthermore, cells that are physically further away from the
sense amplifier (e.g., on the other side of the bitline) experience a longer
signal propagation delay after wordline activation before their bitline
perturbation arrives at the sense amplifier, introducing asymmetry between
different cells along the same bitline.

\subsubsection{Precharge-Related Errors}

Reducing the time given for a bank to precharge a row (i.e., \trp{}) can cause
errors because all bitlines may not be fully precharged to their quiescent
state, thereby disturbing the sensing process for the following row activation.
Several works~\cite{chandrasekar2014exploiting, lee2015adaptive,
chang2016understanding, lee2017design, talukder2018ldpuf,
talukder2018exploiting} study how precharge-related errors manifest, finding
that (1) overall error rates increase sharply as \trp{} is
reduced~\cite{lee2015adaptive, chang2016understanding, talukder2018ldpuf}; (2)
precharge-related errors are more likely to occur near the boundaries of
mats~\cite{lee2017design}; and (3) at extremely high error rates that are
observed when aggressively reducing \trp{}, errors appear to occur roughly
uniform-randomly~\cite{talukder2018ldpuf}. Similar to the case with
activation-related errors, these observations can be explained by a combination
of design-dependence and manufacturing process variation. When errors begin to
appear, they cluster near the edges of mats, where the signal propagation delay
from the precharge signal drivers and row decoders to the cells is greatest. In
the extreme case, when \trp{} is very short, the design-dependence is subsumed
by cell-to-cell variation causing random cells to be more susceptible to error
than others.

\subsection{RowHammer}

RowHammer~\dmpd{\cite{kim2014flipping, bains2014row, mutlu2017rowhammer,
mutlu2019rowhammer}} is a failure mode in which repeated accesses to a given
DRAM row (known as an \emph{aggressor} row) can induce bit-flips in physically
nearby rows (known as \emph{victim} rows). RowHammer can be classified as a
\emph{read-disturb} fault, and prior works~\dmpd{\cite{yang2019trap,
walker2021dram, jiang2021quantifying, kim2014architectural, baeg2022estimation,
redeker2002investigation}} identify two circuit-level charge leakage mechanisms
that contribute towards RowHammer errors: (1) electron migration; and (2)
capacitative crosstalk. As with data-retention errors, the strengths of each of
these error mechanisms varies greatly between DRAM chip designs and with
operating conditions, and significant research
effort~\dmpd{\cite{kim2014flipping, kim2020revisiting, mutlu2017rowhammer,
mutlu2019rowhammer, park2016experiments, park2016statistical, yang2019trap,
cojocar2020are, walker2021dram, hassan2021uncovering, yaglikci2022understanding,
baeg2022estimation}} has been expended to understand how RowHammer errors
behave. \dmpd{Based on this understanding, many works~\cite{aichinger2015ddr,
apple2015about, aweke2016anvil, kim2014flipping,
kim2014architectural,son2017making, lee2019twice, you2019mrloc,
seyedzadeh2018cbt, van2018guardion, konoth2018zebram, park2020graphene,
yaglikci2021blockhammer, kang2020cattwo, bains2015row, bains2016distributed,
bains2016row, brasser2017can, gomez2016dummy, hassan2019crow, devaux2021method,
ryu2017overcoming, yang2016suppression, yang2017scanning, gautam2019row,
yaglikci2021security, greenfield2012throttling, marazzi2022protrr,
kim2015architectural, jedec2012ddr4} have proposed defense mechanisms ranging
from the device- to system-level.}
\section{Memory Error-Mitigation Mechanisms}
\label{sec:bg_error_mitigations}

This section provides an overview of error-mitigation mechanisms that are
commonly used in memory hardware design, including those that operate within the
memory die and in the memory controller. 

\subsection{Row and Column Sparing}

DRAM manufacturers have long since provisioned extra rows and columns within
storage arrays in order to provide replacements in the event that some are
defective~\cite{horiguchi2011nanoscale, cha2017defect, kang2014co,
nair2013archshield, keeth2007dram, son2015cidra, kwon2014understanding}. This is
known as \emph{row and column sparing} and helps DRAM manufacturers tolerate
imperfections during manufacturing, thereby improving manufacturing yield at a
modest expense of chip area resources. Each manufacturer evaluates the tradeoffs
involved based on their own designs, manufacturing processes, and business goals
in order to determine how many spare rows and columns to provision. Several
works~\cite{cha2017defect, kang2014co, nair2013archshield, son2015cidra} argue
that current inherent DRAM error rates have surpassed the correction capability
of row and column sparing alone, necessitating solutions such as on-die ECC, or
even stronger mechanisms. In general, the exact error rates and spare element
counts are proprietary and vary between manufacturers~\cite{nair2013archshield,
gong2017dram}.

\subsection{On-Die ECC}

For single-bit error rates that are too expensive to mitigate using row and
column sparing, DRAM manufacturers have turned to on-die ECC as a
solution~\cite{nair2016xed, micron2017whitepaper, kang2014co, oh20153,
gong2017dram, son2015cidra, oh2014a, kwak2017a, kwon2014understanding,
patel2019understanding, patel2020bit, patel2021enabling}.
Because on-die ECC operations happen within a DRAM die that is not optimized for
general logic operations, keeping on-die ECC simple and efficient is a primary
design goal~\cite{cha2017defect, kwon2014understanding}. Therefore, DRAM
manufacturers today use simple single-error correcting Hamming
codes~\cite{hamming1950error} that operate on data chunks of 64~\cite{im2016im}
or 128~\cite{oh2014a, oh20153, kwak2017a, kwon2017an} bits.

\begin{figure}[h]
    \centering
    \includegraphics[width=0.9\linewidth]{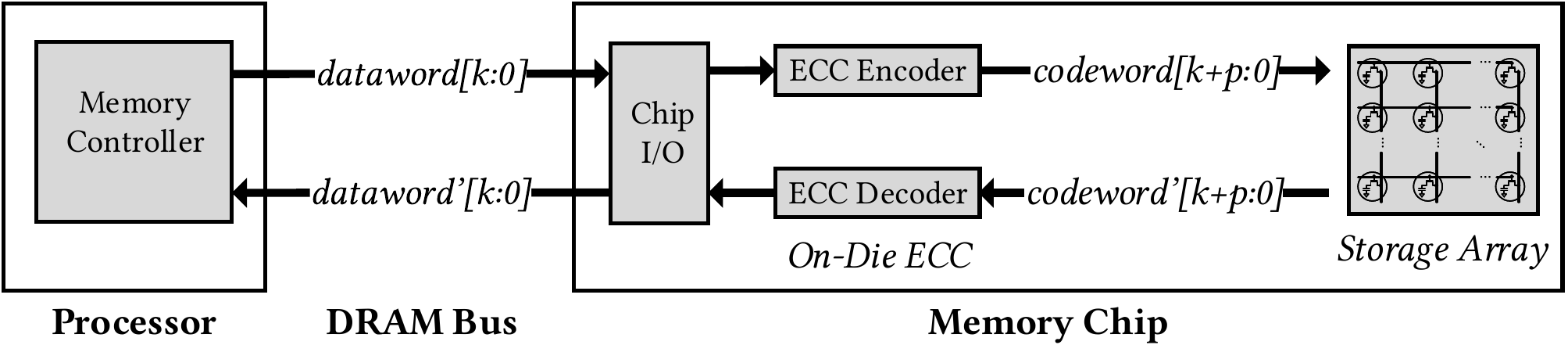}
    \caption{Memory controller interfacing with a memory chip that uses on-die ECC.}
    \label{fig:ondie_ecc_bg}
\end{figure}

Figure~\ref{fig:ondie_ecc_bg} illustrates a simple example where a memory
controller interfaces with a single memory chip that uses on-die ECC. On a write
operation, the memory controller writes a $k$-bit \emph{dataword} to the memory
device. Internally, the memory device encodes the data into a ${k+p}$-bit
\emph{codeword} before writing it to the physical storage array. On a read
operation, the data that the memory controller receives is first decoded by the
ECC logic from a ${k+p}$-bit \emph{codeword'} (where the ''` indicates that the
codeword may contain errors) into a a $k$-bit \emph{dataword'}. In this way, the
memory controller observes only the post-correction errors and has no visibility
whatsoever into how the ECC encoding and decoding mechanism operates.
Section~\ref{subsec:ecc_bg} provides an overview of the mathematical operation
of on-die ECC.

The details of the on-die ECC implementation are considered proprietary secrets
because they, like row and column sparing, are tied to a manufacturer's factory
yield rates~\cite{cost1997yield, childers2015achieving}, which are in turn
directly connected to business interests, potential legal concerns, and
competitiveness in a USD 45+ billion DRAM market~\cite{qy2019global,
verified2019global}. Therefore, DRAM manufacturers do not disclose these details
in public documentation. Furthermore, in our experience, they are unwilling to
reveal on-die ECC details under confidentiality agreements, even for
\dmpd{high-volume customers} for whom knowing the details can be mutually
beneficial.\footnote{Even if such agreements were possible, they would likely be
possible only for parties with large stakes in the DRAM industry. The majority
of DRAM consumers (e.g., smaller industry teams, academics) who do not have
close relations with DRAM manufacturers will be unlikely to forge such
agreements.}

\subsection{Rank-Level ECC}
\label{subsec:rank_lvel_ecc}

Row and column sparing and on-die ECC are unsuitable for addressing high error
rates (e.g., $>10^{-6}$) and multi-bit errors (e.g., burst errors,
component-level failures)~\dmpd{\cite{nair2013archshield,
longofono2021predicting, kang2014co}}. For high-reliability systems (e.g.,
large-scale clusters) that need to be resilient against such errors, system
designers typically integrate \dmpd{an} ECC mechanism within the memory
controller, referred to as \emph{rank-level} ECC~\dmpd{\cite{kim2015bamboo,
gong2018duo, nair2016xed}}. Rank-level ECC uses data blocks at the granularity
of the DRAM interface (or larger) to provide stronger error correction than
on-die ECC can feasibly provide. \dmpd{This section reviews major types of
rank-level ECC designs} that provide varying degrees of reliability.

\subsubsection{SECDED ECC}

Single error correcting, double error detecting (SECDED) ECC is a simple but
prevalent rank-level ECC solution~\cite{kim2015bamboo}. This scheme interfaces
with DRAM using 72 total bits per transfer, 8 of which are metadata used for
error detection and correction (i.e., a 72,64 extended Hamming code). SECDED ECC
is capable of correction one error and detecting two errors within each 72-bit
transfer. 

\subsubsection{Chipkill ECC}

Chipkill ECC maps ECC words to DRAM chips in a way that allows tolerating the
failure of one or more entire DRAM chips within a rank. Several approaches to
chipkill ECC exist, including interleaving SECDED codes~\cite{dell1997white},
using Reed-Solomon (RS) codes~\cite{reed1960polynomial} with symbol sizes that
are aligned with entire DRAM chips (e.g., 4 bits/symbol when using DRAM chips
with 4-bit interfaces~\cite{amd2009bkdg}, 8 bits/symbol mapped across two
transfers from DRAM chips with 4-bit interfaces~\cite{amd2013bkdg}). However,
many of these schemes require datawords that are longer than the 64-bit
interface, which incurs additional reliability challenges due to some
interface-related errors occurring multiple times at the same interface bit
positions~\cite{kim2015bamboo}.

\subsubsection{Other System-Level ECC Designs}

Other rank-level ECC designs have been proposed~\cite{kim2015bamboo,
yoon2010virtualized, udipi2012lot, jian2013low, kim2015frugal, gong2015clean,
chen2016ratt, kim2016all, malek2017odd, chen2015ecc, zhang2018exploring,
yeleswarapu2018sscmsd, chen2018configurable, manzhosov2021muse,
zhang2019boosting, chen2021care} to both enhance reliability and overcome
limitations of existing schemes. In general, each design provides a different
tradeoff between reliability and resource consumption, and it is ultimately up
to the system designer to decide which solution is most appropriate for their
design goals.

\subsection{Repair Mechanisms}
\label{bg:subsec:repair_mech}

As memory manufacturers continue to improve storage density, they exacerbate
various technology-specific error mechanisms that then increase error rates. A
broad class of techniques referred to as \emph{repair mechanisms} identify and
repair bits that are at risk of error from within the memory controller in order
to mitigate high error rates (e.g., $>10^{-4}$) that are infeasible to mitigate
using traditional ECC-based
\dmpa{methods}~\cite{kline2017sustainable,longofono2021predicting,
kline2020flower,nair2013archshield, lin2012secret, qureshi2015avatar}. Prior
works~\cite{lin2012secret, nair2013archshield, kline2020flower,
longofono2021predicting, kline2017sustainable, schechter2010use, nair2019sudoku,
zhang2017dynamic, wang2017architecting, kim2016relaxfault} propose a variety of
repair mechanisms for both DRAM and emerging memory technologies.

\subsubsection{Repair Granularity}

Repair mechanisms~\cite{horiguchi2011nanoscale, yoon2011free,
nair2013archshield, ipek2010dynamically, schechter2010use, lin2012secret,
tavana2017remap, kline2017sustainable, longofono2021predicting, kline2020flower,
seong2010safer, wilkerson2008trading, mandelman2002challenges, son2015cidra,
kim2018solar, lee2017design} perform repair at granularities ranging from
kilobytes to single bits. The granularity at which \mpr{a repair mechanism
identifies} at-risk locations \mpr{is its} \emph{profiling granularity}. For
example, on-die row and column sparing~\cite{jedec2012ddr4, jedec2020ddr5,
kang2014co, mandelman2002challenges, nair2013archshield, son2015cidra,
horiguchi2011nanoscale, cha2017defect} requires identifying at-risk locations at
(or finer than) the granularity of a single memory row.
Table~\ref{tab:categorization} categorizes key repair mechanisms based their
profiling granularities. In general, coarse-grained repair requires less
intrusive changes to the system datapath because repair operations can align
with data blocks in the datapath (e.g., DRAM rows, cache lines, processor
words). However, this means that the repair mechanism suffers from more internal
fragmentation because each repair operation sacrifices more memory capacity
regardless of how \mpq{few} bits are actually at risk of error.

\begin{table}[H]
  \centering
 \scriptsize
 \setlength{\tabcolsep}{2pt}
  \begin{tabular}{llL{10.5cm}}
    \textbf{Profiling Granularity} & \textbf{Size (Bits)} & \textbf{Examples} \\\hline
    System page & 32 K & RAPID~\cite{venkatesan2006retention}, RIO~\cite{baek2014refresh}, Page~retirement~\cite{mcelog2021bad, nvidia2020dynamic, venkatesan2006retention, baek2014refresh, hwang2012cosmic} \\\hline
    \mpp{DRAM external row} & 2-64 K & PPR~\cite{jedec2012ddr4, jedec2020ddr5, kang2014co, mandelman2002challenges,
    nair2013archshield, son2015cidra, horiguchi2011nanoscale, cha2017defect}, Agnos~\cite{qureshi2015avatar}, RAIDR~\cite{liu2012raidr}, DIVA~\cite{lee2017design} \\\hline
    \mpp{DRAM internal row/col} & 512-1024 & Row/col sparing~\cite{kang2014co,
    mandelman2002challenges, nair2013archshield, son2015cidra,
    horiguchi2011nanoscale, cha2017defect}, Solar~\cite{kim2018solar}, \dmpd{CROW~\cite{hassan2019crow}} \\\hline
    Cache block  & 256-512  & FREE-p~\cite{yoon2011free}, CiDRA~\cite{son2015cidra} \\\hline
    Processor word  & 32-64  & ArchShield~\cite{nair2013archshield} \\\hline
    Byte & 8 & DRM~\cite{ipek2010dynamically} \\\hline
    Single bit & 1 &  ECP\footnotemark~\cite{schechter2010use},
    SECRET~\cite{lin2012secret}, REMAP~\cite{tavana2017remap},
    SFaultMap~\cite{kline2017sustainable}, HOTH~\cite{longofono2021predicting},
    FLOWER~\cite{kline2020flower}, SAFER~\cite{seong2010safer}, Bit-fix~\cite{wilkerson2008trading} \\\hline
\end{tabular}
  \caption{Survey of \mpq{prevalent} memory repair mechanisms.}
  \label{tab:categorization}
\end{table}
\footnotetext{\rev{ECP corrects individual bits, but its pointer size can be
adjusted to different granularities as required.}}

\begin{figure}[H]
  \centering
  \includegraphics[width=0.7\linewidth]{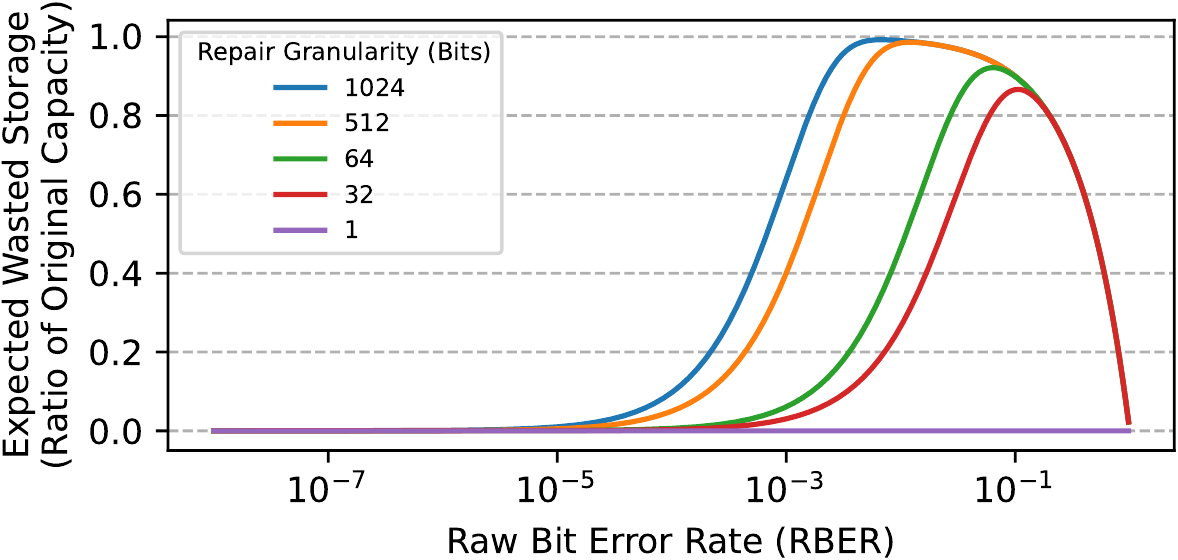}
  \caption{Expected amount of \mpq{wasted storage capacity} when \mpq{repairing} single-bit errors at various \mpq{repair} granularities.}
  \label{fig:wasted_cap}
\end{figure}

Because of this tradeoff between intrusiveness and fragmentation, finer repair
granularities \mpj{are more efficient at higher error
rates~\cite{nair2013archshield, son2015cidra, cha2017defect}}.
Fig.~\ref{fig:wasted_cap} \mpj{illustrates this by showing} the expected
proportion of unnecessarily repaired bits (i.e., the amount of non-erroneous
memory capacity that is sacrificed alongside truly erroneous bits due to
internal fragmentation \dmpd{caused by the repair granularity}) (y-axis) at
various raw bit error rates (x-axis) when mitigating uniform-random single-bit
errors at different repair granularities. We see that coarse-grained repair
becomes extremely wasteful as errors become more frequent, e.g., wasting over
99\% of total memory capacity in the worst case for a 1024-bit granularity
\mpq{at a raw bit error rate of $6.8\times 10^{-3}$}. Note that the expected
wasted storage decreases once the error rate is sufficiently high because an
increasing proportion of bits become truly erroneous, \mpq{which reduces the
wasted bits} for each repair operation. In contrast, bit-granularity repair
\mpq{(denoted with the line for `1')} does \emph{not} suffer from internal
fragmentation. For this reason, repair mechanisms \mpq{designed for} higher
error rates generally employ finer-granularity profiling and
repair~\cite{lin2012secret, kline2017sustainable, longofono2021predicting,
kline2020flower}.

\subsection{Fault-Specific Error Mitigations}

Prior works develop error mitigation mechanisms that are specific to particular
types of faults. This is because, if the fault is well understood, its errors
can be mitigated more efficiently relative to using a general-purpose mitigation
mechanism such as ECC. 

\subsubsection{Mitigating DRAM Data-Retention Errors}
\label{background:subsec:ret_fail_mitigation} 

Prior works~\dmpd{\cite{liu2012raidr, ohsawa1998optimizing,
wang2014proactivedram, venkatesan2006retention, lin2012secret,
nair2013archshield, ghosh2007smart, qureshi2015avatar, khan2016case,
khan2017detecting}} develop error mitigation mechanisms that
specifically target DRAM data-retention errors. This is because data-retention
errors can be mitigated by coordinating with the DRAM refresh algorithm to
refresh cells only when they need to be refreshed, safely enabling reduction of
the refresh rate without requiring expensive general-purpose error mitigation
hardware. In general, these works seek to reduce unnecessary refresh operations
by extending the default refresh interval while mitigating the handful of
resulting retention failures. 

RIO~\cite{baek2014refresh} works with the system software to identify memory
pages with short retention times and prevent software from allocating memory to
them. RAIDR~\cite{liu2012raidr} refreshes DRAM rows at different intervals
according to the retention time of the worst-case cell in each row. Ohsawa et
al.~\cite{ohsawa1998optimizing} propose storing the retention time of each row's
weakest cell within DRAM structures and varying the refresh interval based on
this information. ProactiveDRAM~\cite{wang2014proactivedram} extends the default
refresh interval and issues additional refreshes to rows that cannot sustain the
longer refresh interval. RAPID~\cite{venkatesan2006retention} prioritizes
allocating data to rows with longer retention times and chooses the refresh
interval based on the retention time of the allocated row with the highest
leakage rate. SECRET~\cite{lin2012secret} identifies the set of failing cells at
a longer refresh interval and remaps such cells to known-good cells.
ArchShield~\cite{nair2013archshield} maintains a data structure of known-failing
cells at an extended refresh interval and replicates these cells using a portion
of DRAM at an architectural level. 

These works all enable significant improvements in overall system performance
and energy consumption due to reduction in unnecessary refresh operations (e.g.,
35\% average performance improvement~\cite{qureshi2015avatar}). Unfortunately,
all of these works depend on \emph{accurate identification} of failing cells at
a longer refresh interval (i.e., error profiling).
Section~\ref{related:sec:error_profiling} discusses prior approaches to error
profiling.

\subsubsection{Mitigating RowHammer-Related Errors}

Prior works~\dmpd{\cite{aichinger2015ddr, apple2015about, aweke2016anvil,
kim2014flipping, kim2014architectural,son2017making, lee2019twice, you2019mrloc,
seyedzadeh2018cbt, van2018guardion, konoth2018zebram, park2020graphene,
yaglikci2021blockhammer, kang2020cattwo, bains2015row, bains2016distributed,
bains2016row, brasser2017can, gomez2016dummy, hassan2019crow, devaux2021method,
ryu2017overcoming, yang2016suppression, yang2017scanning, gautam2019row,
yaglikci2021security, greenfield2012throttling, yaglikci2021blockhammer,
marazzi2022protrr}} develop error mitigation mechanisms that identify
error-prone locations (e.g., based on memory access patterns or performance
counters) and take action to mitigate possible RowHammer errors (e.g., by
issuing additional row refresh operations). Similar to data-retention error
mitigation mechanisms, these mechanisms are highly efficient compared to using
general-purpose error mitigations to address RowHammer errors. 

\subsection{Software-Driven Error-Mitigation Techniques}

Effective main memory error management is a \mpp{large} research space with
solutions spanning the entire hardware-software stack. There are many promising
solution directions, including software-driven repair techniques such as
\mpp{page retirement~\cite{mcelog2021bad, baek2014refresh, nvidia2020dynamic,
venkatesan2006retention, meza2015revisiting, hwang2012cosmic}} and
software-assisted techniques such as post-package repair
(PPR)~\cite{jedec2012ddr4, jedec2020ddr5, kang2014co, mandelman2002challenges,
nair2013archshield, son2015cidra, horiguchi2011nanoscale, cha2017defect}.
Unfortunately, these mechanisms have limitations that make them ill-suited to
address the high error rates that we target. For example, page \mpp{retirement}
operates at a coarse (i.e., system memory page) granularity, so it both wastes
significant capacity to repair the many errors at high error rates and cannot
easily repair in-use pages~\cite{meza2015revisiting, lee2019exploiting,
mcelog2021bad}. PPR provides only a few spare rows (e.g., one per bank in
DDR4~\cite{micron2020tn, kim2016relaxfault}) and suffers \mpp{from similar
drawbacks as page retirement due to operating at a coarse granularity
\mpq{(i.e., DRAM row)}}. In contrast, hardware-based repair mechanisms represent
the state-of-the-art in addressing scaling-related main memory errors.

\section{Overview of Block Coding}
\label{bg:sec:block_coding}

In this section, we provide an overview of block error-correction codes that are
used for on-die ECC.

\subsection{Block Error-Correction Codes}
\label{subsec:ecc_bg}

Block coding~\dmpd{\cite{moon2005error, richardson2008modern,
roth2006introduction, clark2013error, costello1982error, costello2004ecc}}
enables data communication over a noisy channel by breaking the data stream into
\emph{datawords} of length $k$ symbols, where each \emph{symbol} \dmpc{can take
on $q$ different values (e.g., $q=2$ for single-bit symbols).} During
\emph{encoding}, the \emph{ECC encoder} maps each dataword to a single
\emph{codeword} of length $n$ using $n-k$ \emph{redundant symbols}. Each symbol
is a function (e.g., xor-reduce) of a subset of the data symbols such that an
error will cause one or more of these functions to evaluate incorrectly.
Encoding results in $q^k$ \emph{valid codewords} out of $q^n$ possible
$n$-symbol words. Upon receiving an $n$-symbol word that may contain erroneous
symbol(s), the \emph{ECC decoder} attempts to determine the originally
transmitted dataword using a \emph{decoding algorithm}.

As a demonstrative example, we consider a common decoding algorithm for binary
(i.e., $q=2$) block codes known as \emph{maximum-likelihood decoding}, which
uses Hamming distance as a metric to find the closest valid codeword to a
received word. Using this approach, the \emph{error-correction capability}, or
$t$, is defined by the \emph{minimum Hamming distance}, or $d$, between any
two valid codewords in the space of all valid codewords. With $d = 2$, a
single-symbol error can always be detected but not always corrected since
there may exist two valid codewords equidistant from the received word. In
general, the error-correction capability can be computed using the
relationship $t = \lfloor\frac{d - 1}{2}\rfloor$, which shows that a minimum
Hamming distance of at least 3 is necessary for single-symbol correction and 5
for double-symbol correction.

When faced with more errors than the code can correct, the decoding result is
\emph{implementation-defined} based on the exact circuitry used to implement
the encoding and decoding algorithms. This is because a code designer has
complete freedom to choose the precise functions that map data symbols to each
redundant symbol, and the same errors induced in two different code
implementations can result in two different post-correction words. In each
implementation, the decoding logic may \emph{i}) manage to correct one or more
actual errors, \emph{ii}) mistakenly do nothing, or \emph{iii}) ``miscorrect''
a symbol that did \emph{not} have an error, effectively \emph{exacerbating}
the number of errors in the decoding result.

\dmpd{Throughout this work, we focus on an important class of block codes that
are used in main memory systems, which are known as \emph{linear block codes}
because their encoding and decoding operations can be expressed as linear
functions of the data.} We follow a commonly used notation for linear block
codes, in which a tuple $(n, k, d)$ describes the length of the codeword ($n$),
the length of the dataword ($k$), and the minimum Hamming distance ($d$),
respectively. This allows us to concisely express the type and strength of a
block code. However, certain codes are also well-known by name (e.g., Repetition
(REP)~\cite{cover1994elements}, Hamming Single-Error Correction
(HSC)~\cite{hamming1950error}, Bose-Chaudhuri-Hocquenghem
(BCH)~\cite{hocquenghem1959codes, bose1960class}), and we will use these names
where appropriate.

\subsection{On-Die ECC and Hamming Codes}
\label{bg:subsec:ondieecc_hamming_code}

\begin{figure}[h]
    \centering
    \includegraphics[width=0.7\linewidth]{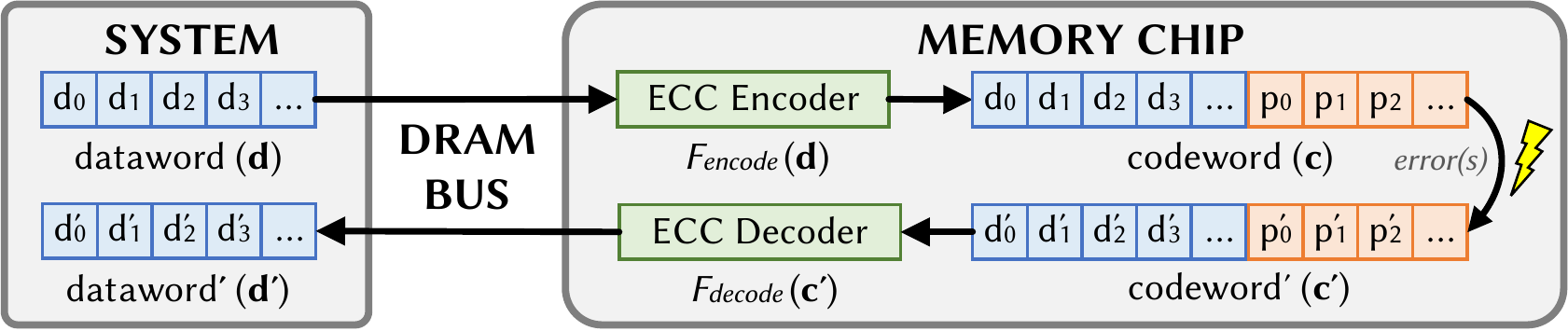}
    \caption{\mpii{Interfacing a memory chip that uses} on-die ECC.}
    \label{fig:ondieecc}
\end{figure}

Single-error correcting Hamming codes are \dmpd{a class of linear block code}
that are used for DRAM on-die ECC~\cite{nair2016xed, micron2017whitepaper,
kang2014co, oh20153, gong2017dram, son2015cidra}. Figure~\ref{fig:ondieecc}
shows how a system might interface with a memory chip that uses on-die ECC. The
system writes $k$-bit \emph{datawords} ($\mathbf{d}$) to the chip, which
internally maintains an expanded $n$-bit representation of the data called a
\emph{codeword} ($\mathbf{c}$), \jkz{created by the ECC encoding of d}. The
stored codeword may experience errors, resulting in a potentially erroneous
codeword ($\mathbf{c'}$). If more errors occur than ECC can correct, e.g., two
errors in a single-error correction (SEC) code, the final dataword read out
\jkz{after ECC decoding} ($\mathbf{d'}$) may also contain errors. The encoding
and decoding functions are labeled $F_{encode}$ and $F_{decode}$.

For all linear codes (e.g., SEC Hamming
codes~\cite{hamming1950error}), $F_{encode}$ and $F_{decode}$ can be
represented using matrix transformations. \jk{As a demonstrative example
throughout} this paper, we use the (7, 4, 3) Hamming
\jkz{code~\cite{hamming1950error}} shown in Equation~\ref{eqn:hg}. $F_{encode}$
represents a generator matrix $\mathbf{G}$ such that the codeword $\mathbf{c}$
is computed from the dataword
$\mathbf{d}$ as $\mathbf{c} = \mathbf{G} \cdot \mathbf{d}$. 
\vspace{-0.25\baselineskip}
\begin{equation}
\vspace{-0.5\baselineskip}
\arraycolsep=0.9pt
\def\arraystretch{0.5}
\small
F_{encode}=\mathbf{G^T}=\left[
  \begin{array}{cccc|ccc}
    1 & 0 & 0 & 0 & 1 & 1 & 1 \\
    0 & 1 & 0 & 0 & 1 & 1 & 0 \\
    0 & 0 & 1 & 0 & 1 & 0 & 1 \\
    0 & 0 & 0 & 1 & 0 & 1 & 1
  \end{array}
\right]
\quad
F_{decode}=\mathbf{H}=\left[
  \begin{array}{cccc|ccc}
    1 & 1 & 1 & 0 & 1 & 0 & 0 \\
    1 & 1 & 0 & 1 & 0 & 1 & 0 \\
    1 & 0 & 1 & 1 & 0 & 0 & 1 
  \end{array}
\right]
\label{eqn:hg}
\end{equation}

\noindent
\textbf{Decoding.}
The most common decoding algorithm is known as \emph{syndrome decoding}, which
simply computes an \emph{error syndrome} $\mathbf{s} = \mathbf{H} \cdot
\mathbf{c'}$ that describes if and where an error exists:
\begin{itemize}
\item $\mathbf{s}=\mathbf{0}$: no error detected.
\item $\mathbf{s}\neq\mathbf{0}$: error detected, and $\mathbf{s}$ describes its bit-exact location.
\end{itemize}
\noindent
Note that the error syndrome computation is \emph{unaware} of the true error
count; it blindly computes the error syndrome(s) assuming a low probability of
uncorrectable errors. If, however, \jk{an} uncorrectable error is present (e.g.,
deliberately induced during testing), one of three possibilities may occur:
\begin{itemize}
\item \emph{Silent data corruption:} syndrome is zero; \mpt{no error}.
\item \emph{Partial correction:} syndrome points to one of the errors.
\item \emph{Miscorrection:} syndrome points to a non-erroneous bit.
\end{itemize}
\noindent
\mpi{When a nonzero error syndrome occurs, the ECC decoding logic simply flips
the bit pointed to by the error syndrome, potentially \emph{exacerbating} the
overall number of errors.}

\noindent
\textbf{Design Space.} \jkz{Each manufacturer can freely select $F_{encode}$ and
$F_{decode}$ functions, whose implementations can help to meet a set of design
constraints (e.g., circuit area, reliability, power consumption). The space of
functions that a designer can choose from is quantified by the number of
arrangements of columns of $\mathbf{H}$. This means that for an $n$-bit code
with $k$ data bits, there are ${2^{n-k} - 1 \choose n}$ possible ECC functions.
Section~\ref{subsubsec:determining_func_from_mcp} formalizes this space of
possible functions in the context of our work.} 

Recent work~\cite{pae2021minimal} shows that some column arrangements can lead
to more miscorrections than others for certain uncorrectable error patterns.
However, designers may choose column arrangements based on circuit latency,
energy, or area concerns, regardless of the resulting behavior with
uncorrectable errors. In our work, we only assume that the code uses a
\emph{systematic encoding}, which requires that $H$ and $G$ do not modify data
bits during encoding (note the identity submatrices in Equation~\ref{eqn:hg}).
This encoding greatly simplifies the hardware encoding and decoding circuitry
and is a realistic assumption for low-latency main memory
chips~\cite{zhang2015vlsichapter41}.

\section{Boolean Satisfiability (SAT) Solvers}
\label{bg:sec:sat_solver}

Satisfiability (SAT) solvers~\cite{de2008z3, bjorner2015nuz,
cimatti2010satisfiability, dillig2012minimum, gomes2008satisfiability,
prasad2005survey} find possible solutions to logic equation(s) with one or more
unknown Boolean variables. A SAT solver accepts one or more such equations as
inputs, which effectively act as \emph{constraints} over the unknown variables.
The SAT solver then attempts to determine a set of values for the unknown
variables such that the equations are satisfied (i.e., the constraints are met).
The SAT solver will return either (1) one (of possibly many) solutions or (2) no
solution if the Boolean equation is unsolvable.

\section{\dmpd{Summary and Further Reading}}
\label{bg:sec:summary_and_pointers}

\dmpd{In summary, significant research effort is necessary throughout the entire
computing stack to realize robust main memory systems. In particular,
understanding memory reliability and how to most efficiently improve it is
central to the design and manufacturing process. We refer the interested reader
to other works on DRAM circuit-level design~\cite{keeth2007dram,
dennard1968field, dennard1974design, choi2015multiple, zhang2014half,
luo2020clr, son2013reducing, lee2013tiered, takemura2007long, kim2006an,
hsu2001intertwin, kim2012case, hassan2019crow, chang2016low, hidaka1990cache,
lee2015adaptive, zhang2016restore, hassan2016chargecache,
chang2016understanding, lee2016reducing, kim2018solar, wang2018reducing,
chang2017understanding, lee2016simultaneous, nguyen2021obet}, system-level
design (e.g., processing-using-DRAM~\cite{seshadri2017ambit,
seshadri2013rowclone, olgun2021pidram, seshadri2015fast, gao2019computedram,
kim2019d, kim2018dram, seshadri2016buddy, seshadri2019dram, seshadri2017simple,
hajinazar2021simdram}, near-data processing~\cite{ahn2016scalable, ahn2015pim,
li2016pinatubo, mutlu2021primer, ghose2019processing, mutlu2019processing,
ghose2018enabling, oliveira2021damov, juan2021benchmarking, boroumand2019conda,
boroumand2018google, boroumand2021mitigating, fromm1997energy, stone1970logic,
kang1999flexram, gokhale1995processing}), testing~\cite{jha2003testing,
kingston2012whitepaper, kingston2021kingston, mrozek2010analysis, nasa2016nasa,
sfikas2016testing, van1991testing, wang2015backward, adams2002high,
atp2018stringent, avl2019testingfaq, cheng2002neighborhood, mrozek2019multi},
reliability (e.g., repair~\cite{horiguchi2011nanoscale, yoon2011free,
nair2013archshield, ipek2010dynamically, schechter2010use, lin2012secret,
tavana2017remap, kline2017sustainable, longofono2021predicting, kline2020flower,
seong2010safer, wilkerson2008trading, mandelman2002challenges, son2015cidra,
kim2018solar, lee2017design}, error-correction coding~\cite{kim2015bamboo,
yoon2010virtualized, udipi2012lot, jian2013low, kim2015frugal, gong2015clean,
chen2016ratt, kim2016all, malek2017odd, chen2015ecc, zhang2018exploring,
yeleswarapu2018sscmsd, chen2018configurable, manzhosov2021muse,
zhang2019boosting, chen2021care}) , and security (e.g.,
RowHammer~\cite{kim2014flipping, bains2014row, mutlu2017rowhammer,
mutlu2019rowhammer, redeker2002investigation, yang2019trap, walker2021dram,
jiang2021quantifying, kim2014architectural, baeg2022estimation,
kim2020revisiting, park2016experiments, park2016statistical}, cold-boot
attacks~\cite{yitbarek2017cold, halderman2008lest, simmons2011security,
gruhn2013practicability, bauer2016lest,muller2010aesse,
villanueva2019cold,mcgregor2008braving, lindenlauf2015cold, lee2011correcting,
hilgers2014post}) for additional information.}
\chapter{Related Work}
\label{chap:related}

Many prior works study DRAM errors and develop techniques for understanding and
mitigating memory errors. This section provides an overview of closely-related
works.

\section{Studying and Improving On-Die ECC}

Developing an on-die ECC design that provides the benefits of highly-efficient
on-die error correction without introducing additional challenges for the rest
of the system is a complex problem involving the interests of both DRAM
manufacturers and consumers. This section reviews general techniques for
exposing ECC mechanisms and recent works that approach the on-die ECC design
problem in different ways.

\subsection{Reverse-Engineering ECC Mechanisms}
\label{related:subsec:rev_engr_ecc}

Both EIN \dmpa{(Chapter~\ref{chap:ein})} and BEER
\dmpa{(Chapter~\ref{chap:beer})} are capable of reverse-engineering details of a
given on-die ECC implementation. Several works provide techniques to
reverse-engineer details of the ECC mechanisms used in NAND flash
memory~\cite{van2015mathematical, van2017bit, wise2018reverse} and DRAM
rank-level ECC~\cite{cojocar2019exploiting}. However, each of these techniques
require some amount of visibility into the ECC mechanism. These works rely on
either directly observing the encoded data (i.e., the codewords) using a
side-channel~\cite{cojocar2019exploiting}, bypassing ECC and directly probing the
raw memory~\cite{van2015mathematical, van2017bit, wise2018reverse,
cojocar2019exploiting}, or knowing when ECC performs an error-correction operation
(e.g., by observing the change in memory access latency when ECC performs a
correction operation~\cite{cojocar2019exploiting, gottscho2016measuring}, or using
custom driver software~\cite{cojocar2019exploiting}). 

However, on-die ECC is fully contained within the memory die and provides
visibility into neither the error-correction operations nor the underlying raw
memory. Therefore, these techniques are not directly applicable to on-die ECC.
In contrast, EIN and BEER use only the post-correction information visible at
the chip interface to extract details of the on-die ECC implementation. Both
techniques achieve this without resorting to special tools or hardware
intrusion. To our knowledge, \dmpa{EIN and BEER are the first (and only)
techniques} that enable reverse-engineering the details of on-die ECC without
doing so.

\subsection{Rearchitecting On-Die ECC}

Several works~\cite{nair2016xed, gong2018duo, pae2021minimal, jeong2020pair}
propose redesigning DRAM on-die ECC in order to improve system-level reliability
by modifying the way in which errors are handled. XED~\cite{nair2016xed} and
DUO~\cite{gong2018duo} expose errors to the memory controller, where stronger
error-correction mechanisms can supplement on-die ECC. Pae et
al.~\cite{pae2021minimal} and PAIR~\cite{jeong2020pair} reorganize the on-die
ECC symbols to improve compatibility with other error-mitigation mechanisms
within the memory controller. 

In contrast, all of our \dmpa{new techniques} except for HARP
\dmpa{(Chapter~\ref{chap:harp})} focus on understanding the behavior of errors
and on-die ECC independently of the on-die ECC design (e.g., whether or not it
provides visibility into its internal operations). Therefore, our works are
applicable to both devices in the field today and future devices that may or may
not use a different on-die ECC design. HARP proposes modifying on-die ECC to
expose the raw encoded data similar to how DUO proposes, but without
transferring the parity-check bits, which DUO accomplishes through an extra read
operation. HARP and DUO use the raw encoded data for different purposes. While
DUO uses it to perform strong error correction operations, HARP uses it to
identify bits that are at risk of post-correction errors (i.e., for error
profiling) in order to support an arbitrary repair mechanism. We note that HARP
and DUO can potentially be used synergistically, but we leave exploring this
direction to future work.

\section{Error Profiling}
\label{related:sec:error_profiling}

We introduce three new error profiling methodologies, i.e., reach profiling
\dmpa{(Chapter~\ref{chap:reaper})}, BEEP \dmpa{(Chapter~\ref{chap:beer})}, and
HARP \dmpa{(Chapter~\ref{chap:harp})}, throughout the course of this
dissertation. This section reviews error profiling algorithms proposed by prior
works.

\subsection{DRAM Data-Retention Error Profiling}

Prior works present various error profiling algorithms to identify DRAM
data-retention errors when reducing the refresh rate to alleviate DRAM refresh
overheads~\dmpd{\cite{venkatesan2006retention, liu2012raidr, khan2014efficacy,
khan2016case, khan2016parbor, khan2017detecting, qureshi2015avatar,
choi2020reducing, sharifi2017online, bacchini2014characterization,
liu2013experimental}}. We roughly categorize these algorithms into two groups: 1)
\emph{active} profiling, which performs multiple rounds of trial-and-error
search using different test patterns and operating conditions that maximize the
chance of observing errors (i.e., that induce worst-case testing conditions);
and (2) \emph{reactive} profiling, which relies on an error-detection mechanism
such as ECC to errors as they occur during runtime. Unfortunately, each of these
proposals has shortcomings that limit their effectiveness in DRAM retention
error profiling.

\subsubsection{Active Profilers}

In order to maximize coverage of bits at risk of error, active profilers use
multiple \emph{rounds} of testing using different \emph{data patterns} to induce
worst-case testing conditions. We call this approach \emph{brute-force}
profiling or \emph{na\"ive} profiling since it blindly runs a testing loop that
1) writes data patterns to DRAM, 2) waits for the target refresh window
($t_{REF}$), and 3) checks for data-retention errors, as shown in
Algorithm~\ref{alg:bf}. Prior works identify effective data patterns useful for
exacerbating data-retention errors, including solid 1s and 0s, checkerboards,
row/column stripes, walking 1s/0s, random data, and their
inverses~\cite{liu2013experimental, khan2014efficacy, kim2014flipping,
mukhanov2020dstress}. By testing sufficiently many data patterns, this approach
approximates testing with the worst-possible data pattern and therefore can find
a significant fraction of all errors at the target refresh window. In
addition, multiple iterations of testing are required in order to account for
the probabilistic nature of data-retention errors (as we show in
Section~\ref{reaper:subsubsection:individual_bit_fail_prob}). However, as we
experimentally demonstrate in
Sections~\ref{reaper:subsubsec:vrt_anal}-\ref{reaper:subsubsection:individual_bit_fail_prob},
brute-force profiling requires many iterations to achieve a high coverage of
errors at a given refresh window. This causes brute-force profiling to have a
high performance overhead, which is only exacerbated by circumstances that
require frequent profiling (Section~\ref{reaper:subsec:perf_energy_eval}).

\RestyleAlgo{ruled}
\SetAlgoVlined
\LinesNumbered
\begin{algorithm}\footnotesize
    \SetAlgoNlRelativeSize{0.7}
	\SetAlgoNoLine
	\DontPrintSemicolon
	\caption{Basic Active Profiling Algorithm} 
    \label{alg:bf}
    {\ttfamily
    {\textbf{PROFILE}($target\_t_{REF}$, num\_iterations):} \par
	\Indp
    failed\_{cells} = [] \par
	\textbf{for}~{$\mathrm{it}\gets\left\{1~\textbf{to}~\mathrm{num\_iterations}\right\}$}: \par
	\Indp
        \textbf{for}~{$\mathrm{dp}$ $\in$ $\mathrm{data\_patterns}$}: \par 
        \Indp
			write\_DRAM(dp) \par
			disable\_refresh() \par
			wait($target\_t_{REF}$) \par
			enable\_refresh() \par
            this\_iteration\_failures $\gets$ get\_DRAM\_errors() \par 
			failed\_{cells}.add(this\_iteration\_failures) \par
	    \Indm\Indm
        \textbf{return} failed\_{cells} \par
    }
\end{algorithm} 

\subsubsection{Reactive Profilers}

Reactive profilers (e.g., AVATAR~\cite{han2014data, qureshi2015avatar})
periodically check an error-detection mechanism (e.g., ECC) during normal
operation and record the set of memory locations that are observed to fail.
Reactive profilers have far lower performance impact than active profilers
because reactive profilers do not need to take exclusive control of the memory
chip for testing purposes. However, as a consequence, reactive profilers do
\emph{not} account for worst-case testing conditions, so they can potentially
miss errors that occur in between monitoring operations. Therefore, reactive
profiling is most often used to mitigate relatively infrequent errors that
appear unpredictably~\cite{qureshi2015avatar}.

\subsubsection{\dmpc{Synergy Between Active and Reactive Profilers}}

\dmpc{Prior work~\cite{qureshi2015avatar} shows that active and reactive
profiling can be combined in a complementary way: using periodic active
profiling to identify the majority of error-prone cells and ongoing reactive
profiling to provide resilience against any errors that active profiling may
have missed. This combined approach is suitable (even
necessary~\cite{liu2013experimental, khan2014efficacy}) to provide resilience
against hard-to-find errors (e.g., variable-retention time errors).
Chapter~\ref{chap:reaper} shows that combined active and reactive profiling may
be necessary to reliably identify data-retention errors, and
Chapter~\ref{chap:harp} exploits the advantages of each profiling approach to
develop a new error profiling methodology suitable for memory chips that use
on-die ECC.}

\subsubsection{Accounting for On-Die ECC}

Unfortunately, none of these algorithms account for or address the effects of
on-die ECC. Furthermore, the insights and observations exploited by these works
(e.g., ways to identify low-probability errors~\cite{qureshi2015avatar,
khan2016parbor, sharifi2017online, qureshi2011pay}) are complementary to the
techniques that we propose.

\subsection{Profiling for Other Error Mechanisms}

Prior works present various error profiling algorithms to identify errors
related to DRAM access latency~\dmpd{\cite{lee2017design,
chang2016understanding, kim2018solar, kim2018dram, kim2019d, lee2015adaptive}},
RowHammer~\dmpd{\cite{kim2020revisiting, cojocar2019exploiting,
frigo2020trrespass, kim2014flipping, cojocar2020are, hassan2021uncovering}}, and
various error mechanisms in emerging memory technologies such as PCM and
STT-RAM~\cite{zhang2012memory, hamdioui2017test, tavana2017remap,
qureshi2011pay}. However, similar to profiling algorithms developed for DRAM
data retention, none of these works consider or address the effects of on-die
ECC. 

\section{Experimental Studies of Memory Devices}
\label{sec:rel:experimental_studies}

Significant work performs experimental error-characterization studies using real
memory devices in order to understand the error mechanisms involved. This
section reviews these works in order to differentiate our studies.

\subsection{Studies of DRAM Chips}

To understand the challenges inherent in error profiling and those that on-die
ECC introduces, we perform several experimental DRAM error-characterization
studies, including the first study to look at data-retention characteristics of
(1) LPDDR4 DRAM, in Chapter~\ref{chap:reaper} and (2) DRAM with on-die ECC, in
Chapter~\ref{chap:ein}. Many other works perform their own experimental studies
of real DRAM chips that focus on various areas of interest, including
data-retention~\dmpd{\cite{jung2014optimized, hamamoto1995well,
hamamoto1998retention, yaney1987meta, shirley2014copula, weis2015retention,
jung2015omitting, weis2015thermal, baek2014refresh, khan2014efficacy,
liu2013experimental, venkatesan2006retention, hassan2017softmc,
weis2017dramspec, khan2016case, khan2016parbor, qureshi2015avatar, sutar2016d,
kim2009new, kong2008analysis, lieneweg1998assesment}}, access
latency~\dmpd{\cite{chang2017thesis, lee2016reducing,
chandrasekar2014exploiting, chang2016understanding, lee2015adaptive,
kim2018dram, kim2018solar, kim2019d, talukder2018exploiting, talukder2018ldpuf,
mukhanov2020dstress, hassan2017softmc}}, read
disturbance~\dmpd{\cite{kim2014flipping, kim2015architectural,
kim2020revisiting, frigo2020trrespass, jung2016reverse, park2016experiments,
park2016statistical, hassan2021uncovering, kim2020improving,
yaglikci2022understanding}}, power consumption~\cite{ghose2018your,
david2011memory} \dmpd{and voltage~\cite{chang2017understanding}}, and the
effects of issuing non-standard command sequences~\cite{gao2019computedram,
olgun2021quac, olgun2021pidram}. Other
studies~\dmpd{\cite{schroeder2009dram, hwang2012cosmic, sridharan2012study,
sridharan2015memory, sridharan2013feng, meza2015revisiting,
bautista2016unprotected, siddiqua2013analysis, meza2018large,
zhang2021quantifying}} examine failures observed in large-scale systems. Both
types of studies are complementary to those we perform.

\subsection{Studies of Other Memory Devices}

Significant work has \dmpd{examined} other memory technologies, including
SRAM~\cite{maiz2003characterization, autran2009altitude,
radaelli2005investigation}, NAND flash memories~\dmpd{\cite{meza2015revisiting,
cai2011fpga, cai2015read, luo2015warm, cai2015data, cai2014neighbor,
cai2013program, cai2013error, cai2013threshold, cai2012flash, cai2012error,
meza2015large, schroeder2016flash, luo2016enabling, vaid2016ssd,
fukami2017improving, cai2017error, cai2017vulnerabilities, luo2018heatwatch,
luo2018improving, kim2020evanesco, park2021reducing}}, hard
disks~\cite{bairavasundaram2008analysis, bairavasundaram2007analysis,
pinheiro2007failure, schroeder2007understanding} and emerging memories such as
phase-change memory~\cite{pirovano2004reliability, zhang2012memory}. These works
are also complementary to the experimental studies that we perform.

\chapter[Reach Profiling for DRAM Data Retention Errors]{Understanding and Improving\\DRAM Data-Retention Error Profiling}

\label{chap:reaper}

\dmpa{In this chapter, we study the DRAM data-retention error characteristics of
DRAM chips without on-die ECC in order to understand the tradeoffs inherent in
data-retention error profiling. Using what we learn from our studies, we
introduce \emph{reach profiling}, a new error profiling strategy that improves
coverage and runtime relative to prior approaches to data-retention error
profiling.}

\section{Background and Motivation}

DRAM stores data in volatile capacitors that constantly leak charge and
therefore requires periodic charge restoration to maintain data correctness. As
cell capacitor sizes decrease with process scaling and the total number of
cells per chip increases each device generation~\cite{itrs}, the total amount
of time and energy required to restore all cells to their correct value, a
process known as DRAM {\it refresh}, scales unfavorably~\cite{liu2012raidr,
kang2014co, chang2014improving}. The periodic refresh of DRAM cell capacitors
consumes up to 50\% of total DRAM power~\cite{liu2012raidr} and incurs large
performance penalties as DRAM cells are unavailable during
refresh~\cite{chang2014improving, liu2012raidr, mukundan2013understanding,
nair2014refresh}.

The DRAM refresh rate, dictated by the {\it refresh interval} or $t_{REFI}$, is
a standardized constant to ensure interoperability of devices from different
vendors and across product generations. DRAM chips are designed to operate
reliably at this refresh rate under worst-case operating conditions. However,
it is \textcolor{red}{well known} that DRAM cells exhibit large variations in
charge retention time~\cite{kim2009new, lee2015adaptive,
liu2013experimental, liu2012raidr, qureshi2015avatar, khan2014efficacy}. Therefore, a fixed
refresh interval causes all DRAM cells to be refreshed at the worst-case rate
even though most DRAM cells can hold data for much longer. Prior works explore
increasing the default refresh interval to what we call a {\it target refresh
interval} while maintaining correctness of DRAM operation. These works assume
that a \emph{small finite set} of \emph{failing cells} (due to the extended
refresh interval) are handled with various \emph{retention failure mitigation
mechanisms} (e.g., ECC, more frequent refreshes, bit repair
mechanisms)~\cite{baek2014refresh, bhati2016dram,
cui2014dtail, khan2014efficacy, lin2012secret, liu2013experimental, liu2012raidr,
qureshi2015avatar, venkatesan2006retention, wang2015radar, nair2013archshield,
wang2014proactivedram}.

\subsection{Motivation: Need for Efficient Error Profiling}

Many of these works that extend the refresh interval assume that the set of
failing cells can be quickly and efficiently identified using a
\emph{brute-force} retention failure \emph{profiling mechanism}, which involves
writing known data to DRAM, waiting for the duration of the required target
refresh interval, and reading the data back out to check for errors, for every
row of cells in a DRAM chip. However, these works do \emph{not} rigorously
explore the efficacy and reliability of such a mechanism. Effects such as
variable retention time (VRT)~\cite{khan2014efficacy, liu2013experimental,
qureshi2015avatar, yaney1987meta, restle1992dram, kang2014co, mori2005origin}
and data pattern dependence (DPD)~\cite{khan2014efficacy, khan2016case,
kim2014flipping, li2011dram, liu2013experimental, khan2016parbor, lee2010mechanism}
invalidate the assumption that a small \emph{finite} set of failing cells
exists and therefore complicate the brute-force approach, likely requiring
\emph{efficient online profiling} mechanisms to discover a
\textcolor{red}{\emph{continuously-changing}} set of failing cells. To this
end, \textbf{our goal} is to $(1)$ thoroughly analyze the different tradeoffs
inherent to retention failure profiling via experimental analysis of a large
number of real state-of-the-art LPDDR4 DRAM chips and $(2)$ use our
observations to develop a robust retention failure profiling mechanism that
identifies an overwhelming majority of all possible failing cells at a given
target refresh interval within a short time.

\subsection{Overview of Experimental Studies}
\label{reaper:sec:overview_of_exp_studies}

We identify three key properties that any effective retention failure profiling
mechanism should have. First, for a given target refresh interval, profiling
should achieve high {\it coverage}, i.e., the ratio of the number of failing
cells discovered by the profiling mechanism to the number of all possible
failing cells at the target refresh interval. This is required to minimize the
cost of the mitigation mechanism for the failing cells at the target refresh
interval. Second, profiling should result in only a small number of {\it false
positives}, i.e., cells that are observed to fail during profiling but never
during actual operation at the target refresh interval. This is beneficial to
minimize the necessary work done by, and thus the overhead of, the retention
failure mitigation mechanism. Third, the profiling {\it runtime} should be as
short as possible to minimize the performance impact of profiling on the
system. 

To understand the complex tradeoffs inherent in DRAM retention failure
profiling with a focus on these three metrics, we conduct a thorough
experimental analysis of 368 state-of-the-art LPDDR4 DRAM chips across three
major DRAM vendors. Our rigorous experimental analysis yields three new major
observations.  First, the brute-force approach to profiling is inefficient and
does not attain high coverage within a short time.  Second, the cells that fail
at a given refresh interval fail more reliably at a higher refresh interval and
at a higher temperature. Third, retention failure profiling has a complex
tradeoff space between three key parameters (coverage, \textcolor{red}{false
positive rate, and} runtime). 

\subsection{Overview of Reach Profiling}

Based on our observations from this rigorous experimental characterization of
the complex tradeoff space, we propose a novel low-overhead and high-coverage
DRAM retention failure profiling methodology called \emph{reach profiling}. The
key idea of reach profiling is to profile at \emph{reach conditions}, which
consist of either a larger refresh interval and/or a higher temperature
relative to the target refresh interval/temperature to quickly discover an
overwhelming majority of all possible failing cells at the target conditions.
Intuitively, given that failing cells are more likely to fail at a higher
refresh interval and at a higher temperature, we can quickly obtain all failing
cells for a target condition by profiling at reach conditions, at the cost of
also identifying some false positive cells as failing. The profiling parameters
can be adjusted to \emph{maximize} coverage while \emph{minimizing} the false
positive rate and runtime. 

We find that on average, across 368 different DRAM chips from three vendors,
reach profiling attains a speedup of 2.5x over brute-force profiling while
providing over 99\% coverage of failing cells with less than a 50\% false
positive rate.  Further speedups of up to 3.5x can be obtained at the expense
of significantly greater false positive rates. By manipulating the profiling
conditions relative to the target conditions, we can select an appropriate
tradeoff between coverage, \textcolor{red}{false positive rate}, and runtime
that suits the desired system configuration.

Reach profiling results in a set of failing \textcolor{blue}{cells} that can be
handled by various retention failure mitigation mechanisms (e.g., error
correcting codes, higher refresh rates, and remapping mechanisms for failing
cells) to reliably operate a system at a target refresh rate.  As an example,
consider a scheme where the DRAM memory controller maps addresses with failing
cells out of the system address space.  For any target refresh interval, reach
profiling can be used to quickly determine the set of failing cells with high
coverage. The memory controller can then map out the addresses containing the
failed \textcolor{blue}{cells} from the system address space in order to
maintain system reliability at a higher refresh interval.  Alternatively, the
system can employ any other error mitigation mechanism proposed by prior work
to handle the failing cells (e.g.,~\cite{baek2014refresh,
bhati2016dram, cui2014dtail, khan2014efficacy,
lin2012secret, liu2013experimental, liu2012raidr, qureshi2015avatar, venkatesan2006retention,
wang2015radar, nair2013archshield}). For example, when used with
ArchShield~\cite{nair2013archshield}, REAPER reliably enables an average
end-to-end system performance improvement of 12.5\% (maximum 23.7\%)
\textcolor{red}{on our workloads (Section~\ref{reaper:subsubsec:e2eresults})}. 

\section{Experimental Methodology}
\label{reaper:section:methodology} 

In order to develop an effective and efficient retention failure profiling
mechanism for modern DRAM chips, we need to first better understand their data
retention characteristics. To this end, we developed a thermally-controlled DRAM
testing infrastructure to characterize state-of-the-art LPDDR4 DRAM chips. Our
infrastructure provides precise control over DRAM commands, which we verified
via a logic analyzer by probing the DRAM command bus. 

All tests, unless specified otherwise, were performed using 368 2y-nm LPDDR4
DRAM chips~\cite{jedec2014lpddr4} from three major vendors in a thermally-controlled
chamber at 45$^{\circ}$C ambient temperature. In our infrastructure, ambient
temperature is maintained using heaters and fans controlled via a
microcontroller-based PID loop to within an accuracy of 0.25$^{\circ}$C, with a
reliable range of 40$^{\circ}$C to 55$^{\circ}$C. DRAM temperature is held at
15$^{\circ}$C above ambient using a separate local heating source and
temperature sensors to smooth out temperature variations due to self-heating. 

\section{LPDDR4 Data-Retention Error Characterization}
\label{reaper:section:new_observations} 

We experimentally study the retention characteristics of 368 LPDDR4 chips and
compare our findings with prior works, which examine DDR3
DRAM~\cite{liu2013experimental, khan2014efficacy, qureshi2015avatar, kim2014flipping}.
We present data showing failure discovery rates, data pattern dependence (DPD)
effects and variable retention time (VRT) effects to demonstrate the
difficulties involved in retention failure profiling. We then extensively
analyze single-\textcolor{red}{cell} failure probabilities to motivate
profiling at a longer refresh interval and/or a higher temperature to quickly
discover failing cells.  We present four key observations resulting from our
experimentation and describe the implications of each observation on retention
failure profiling.

\subsection{Temperature Dependence}

We observe an exponential dependence of failure rate on temperature for refresh
intervals below 4096ms, as consistent with prior work~\cite{hamamoto1998retention,
liu2013experimental, hou2013fpga}. We find and make use of the following temperature
relationships throughout the rest of this paper:
\begin{equation}
R_A \propto e^{0.22 {\Delta}T} \quad\quad\quad R_B \propto e^{0.20 {\Delta}T} \quad\quad\quad R_C \propto e^{0.26 {\Delta}T}
\end{equation}
\noindent
where $R_X$ represents the proportion of failures for Vendor $X$ and
${\Delta}T$ is the change in ambient temperature.  These relationships
approximately translate to scaling the retention failure rate by a factor of 10
for every 10$^\circ$C increase in temperature.

\subsection{Aggregate Retention Time Distributions}
\label{reaper:subsubsection:agg_ret_dist}

\begin{figure}[h]
    \centering
    \includegraphics[width=0.7\linewidth]{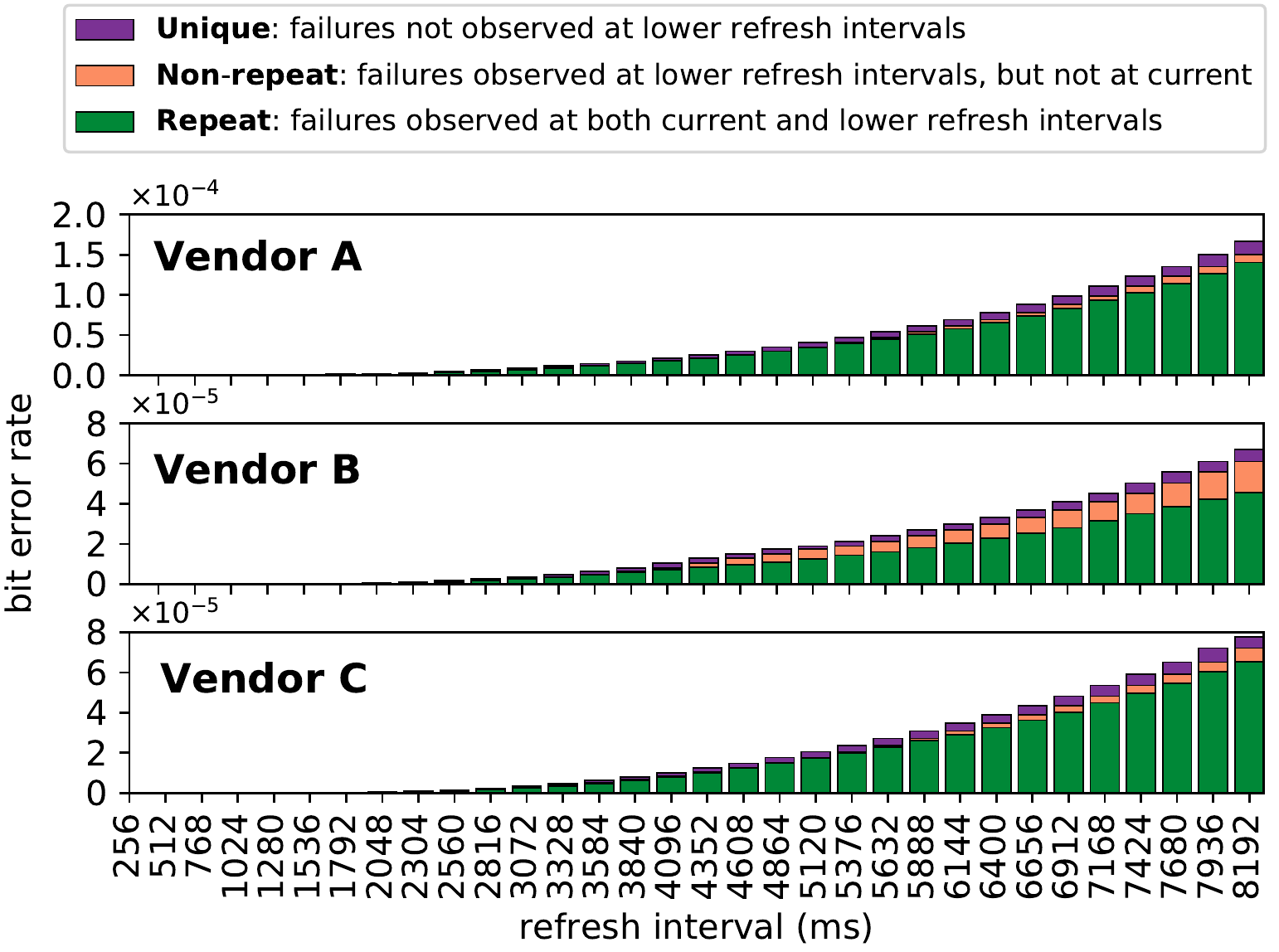}%
\caption{Retention failure rates for different refresh intervals.
\textcolor{blue}{Cells} are categorized as unique, repeat, or non-repeat based
on whether or not they are observed at the given interval \emph{and} at lower
refresh intervals.} \label{fig:retention_time_distribution} 
\end{figure}

Figure~\ref{fig:retention_time_distribution} shows the effect of increasing the
refresh interval on the bit error rates (BER) averaged across 368 total chips
from three different vendors. For each refresh interval, we compare the
population of failing \textcolor{red}{cells} to the population of failing
\textcolor{red}{cells} at all \emph{lower} refresh intervals. Failures that are
1) \emph{also} observed at lower refresh intervals are shown in green (called
\emph{repeat}), 2) not observed at lower refresh intervals are shown in orange
(called \emph{non-repeat}), and 3) observed at lower refresh intervals but
\emph{not} at the given interval are shown in purple (called \emph{unique}). We
make one key observation. 

\textbf{Observation 1}: A large proportion of \textcolor{red}{cells} that are
observed to fail at a given refresh interval are likely to \emph{fail again at
a higher refresh interval}.

\textbf{Corollary 1}: Determining the set of failing cells at a given refresh
interval provides a large proportion of the failures found at \emph{lower}
refresh intervals. 
\subsection{Variable Retention Time Effects}
\label{reaper:subsubsec:vrt_anal}
\begin{figure}[h]
    \centering
    \includegraphics[width=0.7\linewidth]{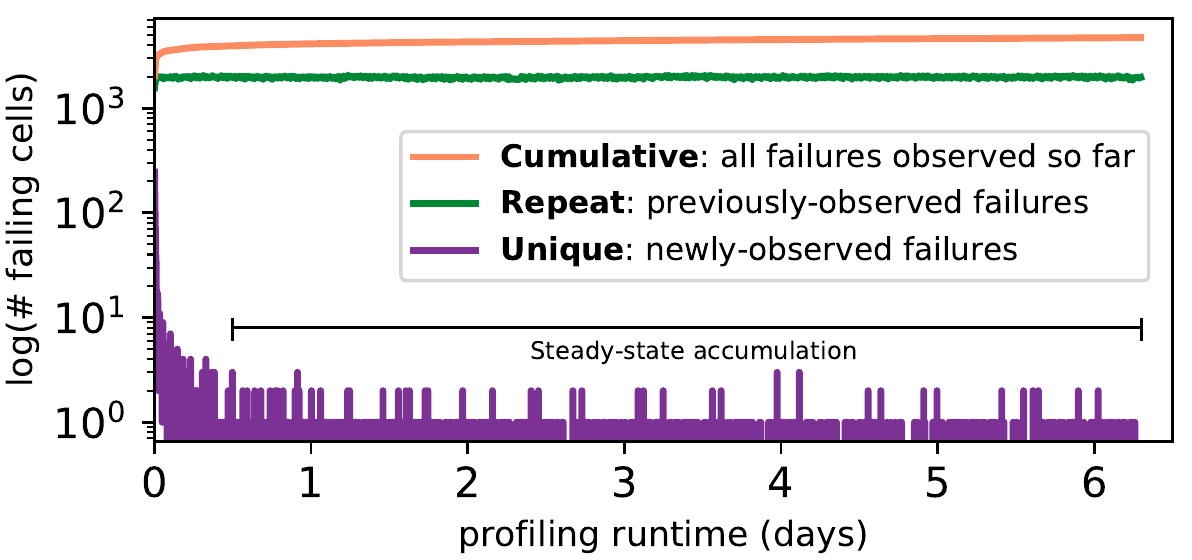} 
    \caption{Number of failing \textcolor{red}{cells} discovered using
brute-force profiling at a refresh interval of 2048ms at 45$^{\circ}$C ambient
temperature using a single representative chip (from Vendor B).} 
    \label{fig:slow_brute_force} 
\end{figure}

We observe variable retention time (VRT) effects that prevent reliable
detection of failing cells. Figure~\ref{fig:slow_brute_force} shows the number
of failing cells discovered over six days of data collection using brute-force
profiling at a refresh interval of 2048ms at 45$^\circ$C ambient temperature.
The data is shown for a single representative chip from Vendor B, but we find a
similar trend across all chips from all vendors. Profiling follows the procedure
described in Algorithm~\ref{alg:bf}, using six data patterns and their inverses
(Section~\ref{related:sec:error_profiling}) per iteration over 800 iterations
throughout the length of the test. The cumulative set of discovered failures is
shown in orange and the set of failures discovered each iteration is broken up
into unique (i.e., newly-discovered) failures and repeat failures, shown in
purple and green, respectively.

We find that after about 10 hours of testing, brute-force profiling enters the
\emph{steady-state accumulation} phase, in which new failures continue to
accumulate with a rate of approximately one \textcolor{red}{cell} every 20
seconds. In other words, it takes about 10 hours to find the base set of
failures for a given refresh interval using the brute-force approach. We
attribute the continual discovery of new failures to the VRT phenomenon, which
can cause a cell to shift from a retention time greater than 2048ms to one that
is lower.  However, we also observe that the total set of failures (unique +
repeat) found in each iteration is nearly constant in size, implying that the
rate of failure accumulation is very close to the rate of failures leaving the
failing set.  This finding is consistent with prior work on DDR3
chips~\cite{liu2013experimental,qureshi2015avatar}, showing that newer DRAM
generations face similar profiling difficulties to older generations.

\begin{figure}[h]
    \centering
    \includegraphics[width=0.7\linewidth]{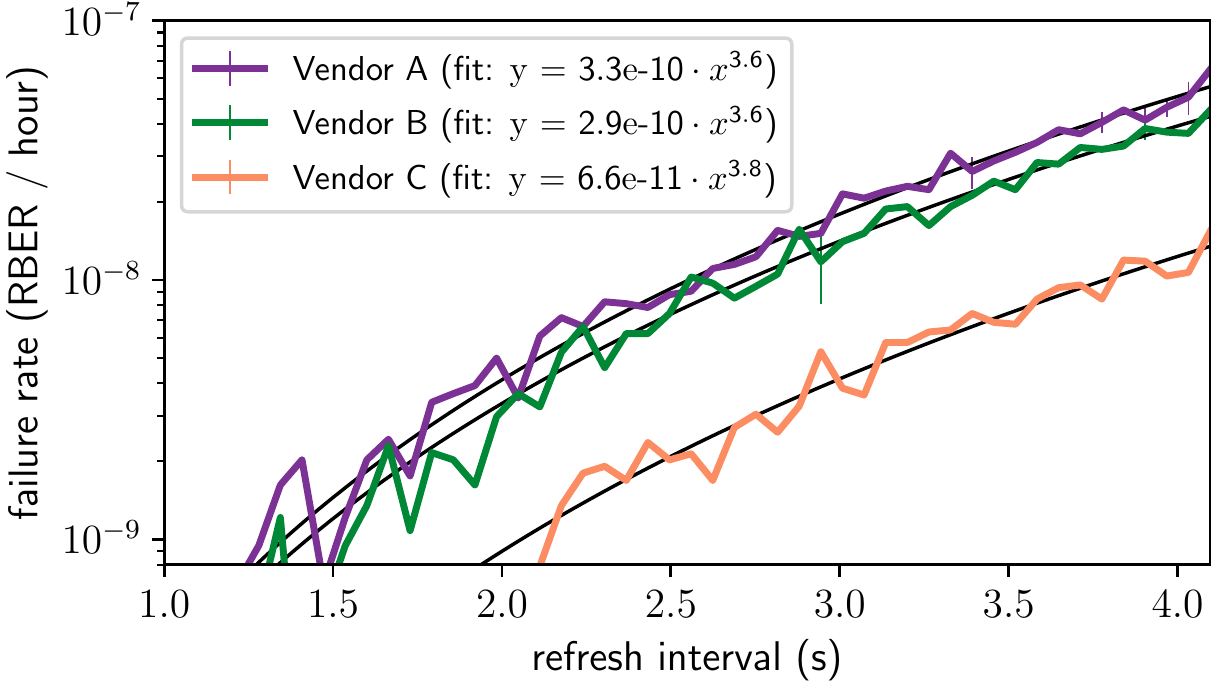} 
    \caption{Steady-state failure accumulation rates vs.$~$refresh interval for
\textcolor{red}{368 chips across the three DRAM vendors} at 45$^\circ$C.
\textcolor{yellow}{The best-fit curve for each vendor is shown in black.}} 
    \label{fig:VRT_per_trefi} 
\end{figure}

We conduct this same analysis for \emph{different refresh intervals}
\textcolor{red}{on 368 chips from across the three DRAM vendors} at 45$^\circ$C.
Figure~\ref{fig:VRT_per_trefi} shows the \emph{steady-state new failure
accumulation} rates for different refresh intervals aggregated across all chips
of each vendor. The \textcolor{red}{y-axis} represents the steady-state new
failure accumulation rate, and the \textcolor{red}{x-axis} shows the refresh
interval. Each data point is drawn at the average value across all chips from
the vendor with error bars representing the respective standard deviation. We
find that the steady-state failure accumulation rate grows at a
\emph{polynomial} rate with respect to the refresh interval.
Figure~\ref{fig:VRT_per_trefi} overlays the data with well-fitting polynomial
regressions of the form $y = a * x^b$, and the exact fit equations are provided
in the figure itself.

\textbf{Observation 2}: No matter how comprehensive the set of failures
discovered is, the population of failing cells continues to change due to
VRT effects. 

\textbf{Corollary 2}: The retention failure profile inevitably needs to be
re-generated after some period of time. In other words, \emph{online} profiling
is required.

\subsection{Data Pattern Dependence Effects}
\label{reaper:subsubsection:dpd_effects}
\begin{figure}[h]
    \centering
    \includegraphics[width=0.7\linewidth]{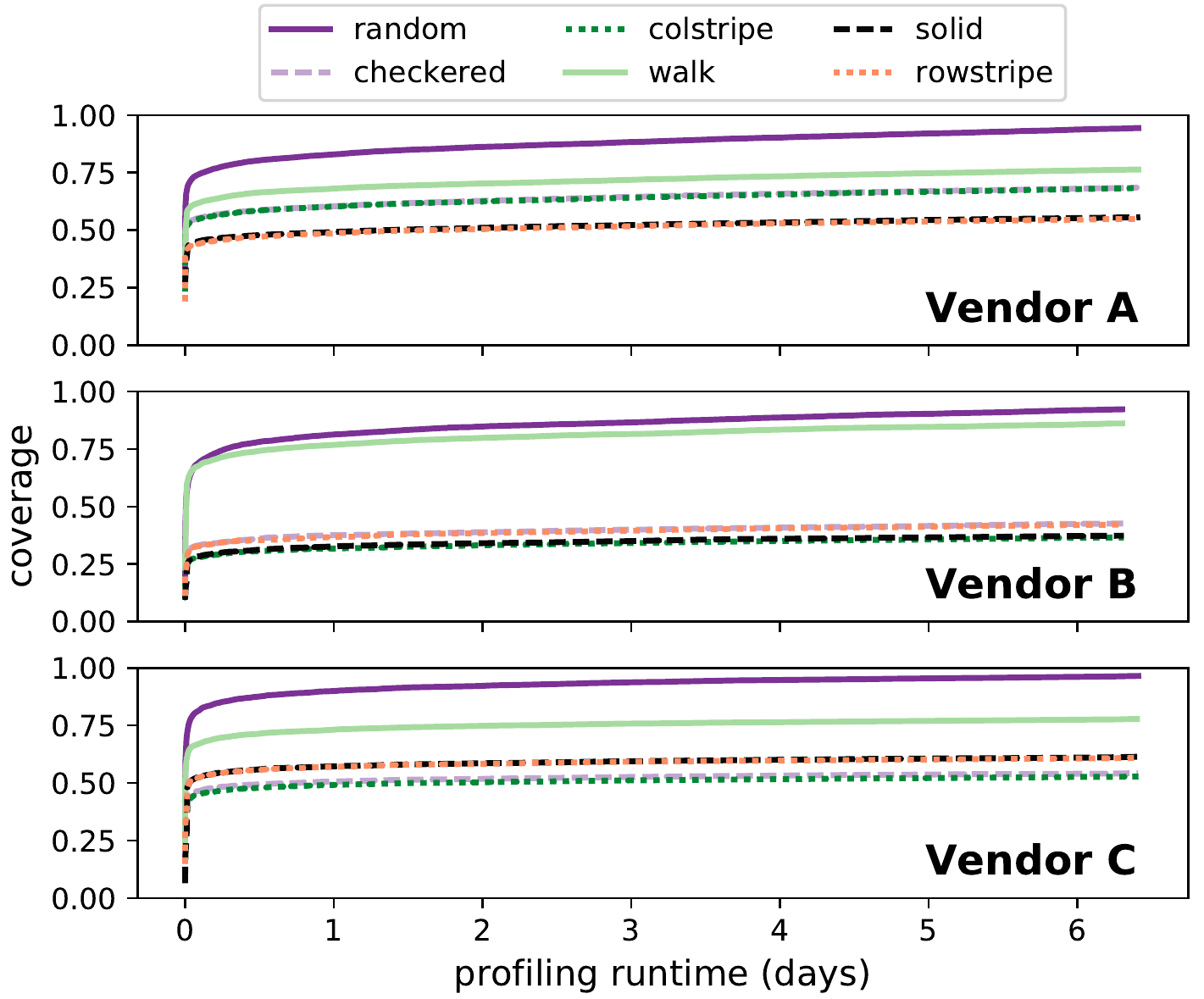} 
    \caption{Coverage of unique retention failures \textcolor{green}{discovered
by different data patterns using brute-force profiling} across 800 iterations
spanning 6 days.} 
    \label{fig:DPD_exists} 
\end{figure}

Figure~\ref{fig:DPD_exists} shows the cumulative number of retention failures
discovered over 800 iterations spanning 6 days of continuous brute-force
profiling with different data patterns across the three vendors at
45$^{\circ}$C. Profiling uses Algorithm~\ref{alg:bf}, in which each iteration
consists of writing a data pattern to DRAM, pausing refresh for 2048ms, and
then checking for retention failures. New failures found by each data pattern
each iteration are added to a running set representing the total number of
failures discovered so far. The ratio of failures discovered by each data
pattern to the \emph{total} number of failures discovered by \emph{all data
patterns together} is plotted against time. We see that failures continue to
accumulate over time, as expected from
\textcolor{red}{Section~\ref{reaper:subsubsec:vrt_anal}}, and different data patterns
uncover different fractions of failures. Of the six data patterns tested, the
random pattern (solid dark purple) discovers the most failing cells across all
three vendors. 
\textbf{Observation 3}: Unlike DDR3 DRAM~\cite{liu2013experimental}, the random data
pattern most closely approaches full coverage after 800 iterations spanning 6
days, but it still \emph{cannot} detect \emph{every} failure on its own. 

\textbf{Corollary 3}: A robust profiling mechanism should \textcolor{green}{use} 
\emph{multiple data patterns} to attain a high coverage of the set of failing
cells. 

\subsection{Individual Cell Failure Probability}
\label{reaper:subsubsection:individual_bit_fail_prob}
\begin{figure}[h]
    \centering
    \includegraphics[width=0.8\linewidth]{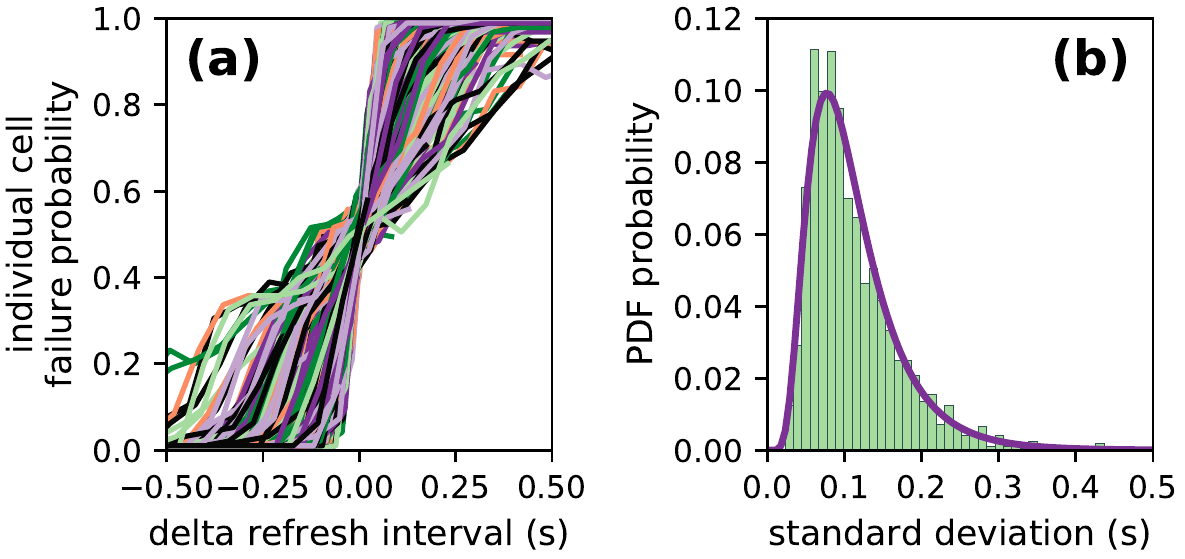} 
    \caption{(a) Individual \textcolor{red}{cells} fail with a
normally-distributed cumulative distribution function with respect to refresh
interval (\textcolor{yellow}{CDF means are} normalized to
\textcolor{blue}{$\mathbf{x=0.00}$s}) and (b) their standard deviations follow
a lognormal distribution (right).  This data is taken from a single
representative chip of Vendor B at 40$^\circ$C and base refresh intervals
ranging from 64ms to 4096ms.}
    \label{fig:individual_bit_failure_distro} 
\end{figure}

We experimentally study the retention characteristics of individual DRAM cells.
We find that each cell's probability of retention failure follows a normal
distribution with respect to \textcolor{red}{the} refresh interval.
Figure~\ref{fig:individual_bit_failure_distro}(a) shows the effect of changing
the refresh interval (shown normalized from between 64ms and 4096ms to
\textcolor{blue}{$x=0.00$s}) on the probability of retention failure for almost
all of the failing \textcolor{red}{cells} from a representative DRAM chip of Vendor
B.\footnote{Cells exhibiting VRT behavior ($\sim$2\% of all cells for these
conditions) are excluded from this plot, for ease of explanation.} Each curve
represents the failure cumulative distribution function (CDF) of a single
\textcolor{red}{cell}, with failure probabilities of 0.0 and 1.0 representing
0\% and 100\% incorrect reads out of 16 total test iterations, respectively.
All distributions' means are normalized to \textcolor{blue}{$x=0.00$s} in order
to highlight the similarity in failure pattern between different
\textcolor{red}{cells}. We see that due to the
\textcolor{red}{normally-distributed failure probability of each cell},
\emph{every \textcolor{red}{cell} becomes more likely to fail at longer refresh
intervals}.

Figure~\ref{fig:individual_bit_failure_distro}(b) shows a histogram of the
standard deviations of each cell's unique failure distribution. We find that
the standard deviations follow a tight lognormal distribution, with the
majority of cells having a standard deviation of less than 200ms at these
conditions.\footnote{Note that this is \emph{not} the same result as the
well-known lognormal distribution of the number of retention failures with
respect to refresh interval~\cite{hamamoto1998retention,li2011dram}. Our observation is that
the spread of \emph{each individual \textcolor{red}{cell} failure distribution} is
lognormally-distributed as opposed to the \emph{overall number of failing
cells} at different refresh intervals.} This means that \emph{slightly
extending the refresh interval results in significantly increasing the failure
probability of almost all \textcolor{red}{cells}, which makes
\textcolor{red}{cells} with low failure probabilities easier to detect}.

\begin{figure}[h]%
    \centering
    \begin{subfigure}{0.7\textwidth}
        \includegraphics[width=\linewidth]{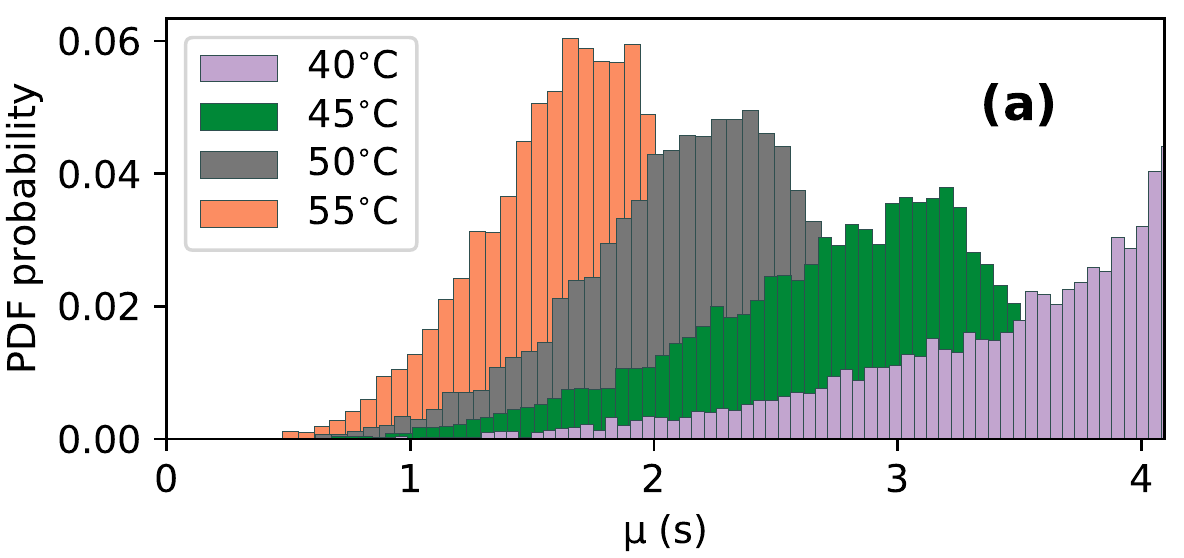}%
    \end{subfigure}
    \qquad 
    \begin{subfigure}{0.7\textwidth}
        \includegraphics[width=\linewidth]{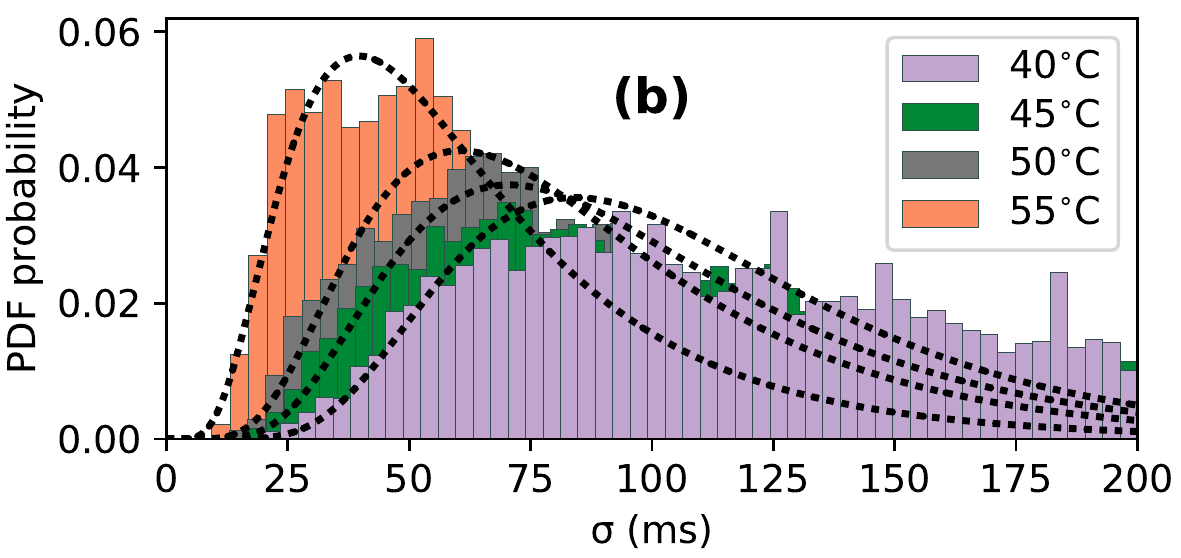}%
    \end{subfigure}
    \caption{Distributions of individual \textcolor{red}{cells'} normal
distribution parameters ($\mu$, $\sigma$) over different temperatures taken
from a representative chip from Vendor B. We see that both distributions shift
left with increasing temperature, which means that \textcolor{red}{an increase
in temperature causes individual \textcolor{red}{cells} to both fail at a lower
refresh interval and also exhibit a narrower failure probability
distribution.}} 
    \label{fig:failure_distributions_agg} 
\end{figure}

We can explain the shape of these distributions using circuit-level DRAM
characteristics. As described in Section~\ref{bg:sec:dram_overview}, the DRAM
cell read process requires sense amplifiers to resolve the direction of a shift
in bitline voltage. This shift is known as the \emph{sense amplifier voltage
offset}, and due to various manufacturing variations, the magnitude of this
shift across different cells is found to be normally
distributed~\cite{li2011dram,wicht2004yield}. While using an extended refresh
interval, DRAM cell leakage causes some cells' voltage offsets to be too small
to resolve correctly, causing the sense amplifier to probabilistically read an
incorrect value. Therefore, our observation of
\textcolor{red}{normally-distributed} failure probabilities \textcolor{red}{for
all cells} (Figure~\ref{fig:failure_distributions_agg}(a)) is a direct
consequence of the normal distribution of the sense amplifier voltage offsets.
Furthermore, Li et al.~\cite{li2011dram} find that leakage components in DRAM
cells follow lognormal distributions, which we hypothesize is responsible for
our observation of lognormally-distributed standard deviations
(Figure~\ref{fig:failure_distributions_agg}(b)).  Given that different cells
leak at lognormally-distributed rates, we would expect the length of time that a
cell's charge is within the noise margin of the sense amplifier to also be
\textcolor{red}{lognormally distributed}.

We extend this analysis to different temperatures, and we aggregate the normal
distribution fit parameters ($\mu$, $\sigma$) for each failing
\textcolor{red}{cell} into the distributions shown in
Figure~\ref{fig:failure_distributions_agg} (a) and (b), respectively. We see
that at higher temperatures, the distributions for both the means and standard
deviations shift left. This means that on average, cell retention times shift
to lower values and their failure distribution becomes narrower around that
retention time. We can take advantage of this observation by profiling not only
at a longer refresh interval, but also at a \emph{higher temperature}, thus
ensuring that a large majority of the cells observed to \emph{inconsistently}
fail at the target refresh interval and temperature are very likely to fail at
the (longer) profiling refresh interval and the (higher) profiling temperature.
\vspace{8pt}
\textbf{Observation 4}: \textcolor{red}{Cells} that inconsistently fail at the
target conditions may be missed with brute-force profiling, but are more likely
to be found when profiling at a longer refresh interval and/or a higher
temperature.

\textbf{Corollary 4}: In order to find failures quickly and consistently, we
should profile at \emph{a longer refresh interval} or \emph{a higher
temperature}. 

\begin{figure}[h]
    \vspace{5pt}
    \centering
    \includegraphics[width=0.7\linewidth]{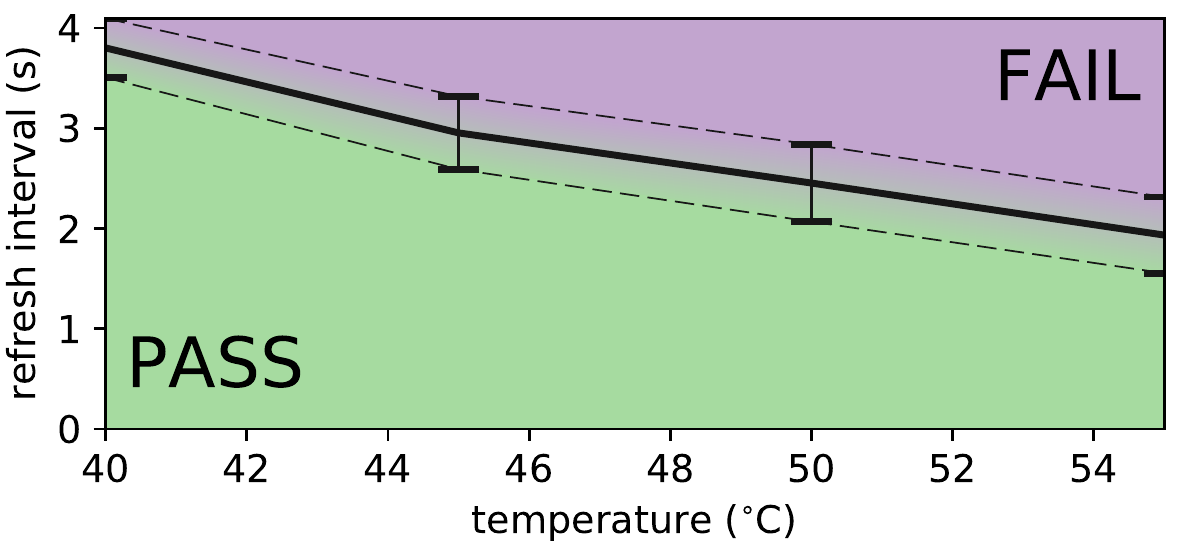} 
	\caption{Effect of manipulating temperature/refresh interval on the
combined normal distribution of failing cells from a representative chip from
Vendor B. Dashed regions represent the combined standard deviation and the
central line represents the combined mean \textcolor{red}{for each tested
temperature}.} 
    \label{fig:individ_bit_ref_and_temp} 
    \vspace{5pt}
\end{figure}

By combining the normal distributions of individual \textcolor{red}{cell}
failures from a representative chip of Vendor B, we obtain the data in
Figure~\ref{fig:individ_bit_ref_and_temp}, which shows the failure probability
for the combined distribution over different temperatures and refresh
intervals. The dashed regions represent the combined standard deviation for
each tested temperature, and the solid black curve in-between represents the
combined mean. From this data, we draw two major conclusions: 1) at a higher
temperature or a longer refresh interval, the typical \textcolor{red}{cell} is
more likely to fail, and 2) raising the temperature and extending the refresh
interval have similar effects \textcolor{red}{on cell failure probability}
(e.g., at 45$^\circ$C, a 1s change in refresh interval has a similar effect to
a 10$^\circ$C change in temperature).

\section{Reach Profiling}
\label{reaper:section:mechanism} 
Reach profiling accelerates the process of finding the set of DRAM cells that
experience retention failures by profiling DRAM under conditions that increase
the likelihood of failure, called \emph{reach conditions}. This enables the
profiler to discover the failures much more quickly, leading to a significant
reduction in overall profiling runtime. Given a set of desired \emph{target
conditions} consisting of a target refresh interval and a target operating
temperature, the \emph{key idea} of reach profiling is to profile at
\emph{reach conditions}, a combination of a \emph{longer refresh interval} and
a \emph{higher temperature} relative to the target conditions. As
experimentally demonstrated in
Section~\ref{reaper:subsubsection:individual_bit_fail_prob}, each DRAM cell is
\emph{more likely} to fail at such reach conditions than at the target
conditions.  While this enables reach profiling to attain high failure
\emph{coverage} (i.e., the ratio of correctly identified failing cells over the
total number of failing cells at the target conditions), it also causes reach
profiling to have \emph{false positives} (i.e., failing cells at the reach
conditions that do \emph{not} fail at the target conditions), leading to more
work for a retention failure mitigation mechanism that is used to correct the
retention failures (including false positives). More work done
by the mitigation mechanism leads to higher overhead in terms of performance,
energy, or area. This results in a \emph{complex tradeoff space} between the
choice of reach conditions and the resulting profiling
\textcolor{red}{coverage, false positive rate, and runtime}.

Given this complex tradeoff space, we need to address three \textbf{key
questions} to develop an effective implementation of reach profiling: 

\begin{enumerate}
	\item What are the desirable \emph{reach conditions}?
	\item Given the continuous accumulation of new failures (e.g., due to VRT),
	how long does a retention failure profile remain useful (i.e., how often
	must we reprofile)? 
	\item What information does the profiler need in order to determine 
	desirable reach conditions for a given system?
\end{enumerate}
We explore the answers to these questions in the rest of this
section. Then, in Section~\ref{reaper:section:e2e}, we use our exploration to develop
REAPER, a robust implementation of reach profiling.
\subsection{Desirable Reach Conditions}

The three key metrics of profiling, \emph{coverage}, \emph{false positive rate},
and \emph{runtime} (Section~\ref{reaper:sec:overview_of_exp_studies}), lead to
contradictory optimization goals. While an ideal configuration would achieve
high coverage, \textcolor{red}{low false positive rate, and low runtime,} we
find that there is a large tradeoff space involving these three goals. Using
experimental data, we demonstrate the large scale of the tradeoff space inherent
in reach profiling and show the effects of changing the refresh interval and the
temperature on the three key profiling metrics. We present data for interesting
choices of reach conditions and analyze the results to show how the profiling
system can make a reasonable choice.
\subsubsection{Manipulating Refresh Interval and Temperature}

\begin{figure}[h]
    \centering
    \begin{subfigure}{\textwidth}
        \centering
        \includegraphics[width=0.7\linewidth]{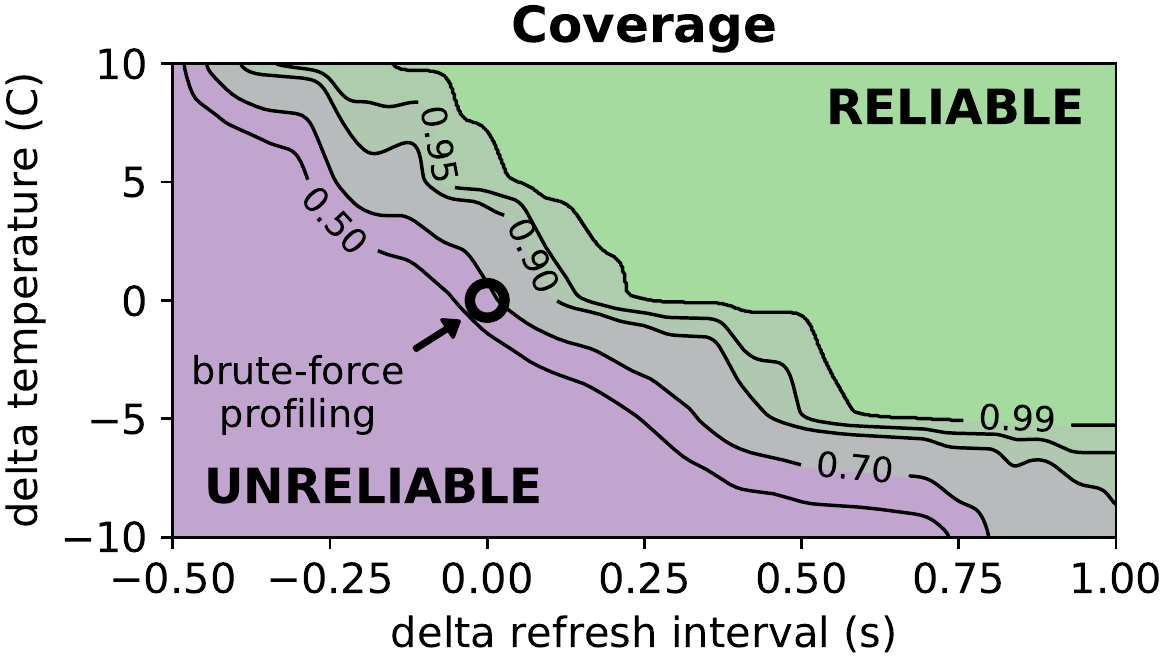}%
    \end{subfigure}
    \qquad 
    \begin{subfigure}{\textwidth}
        \centering
        \includegraphics[width=0.7\linewidth]{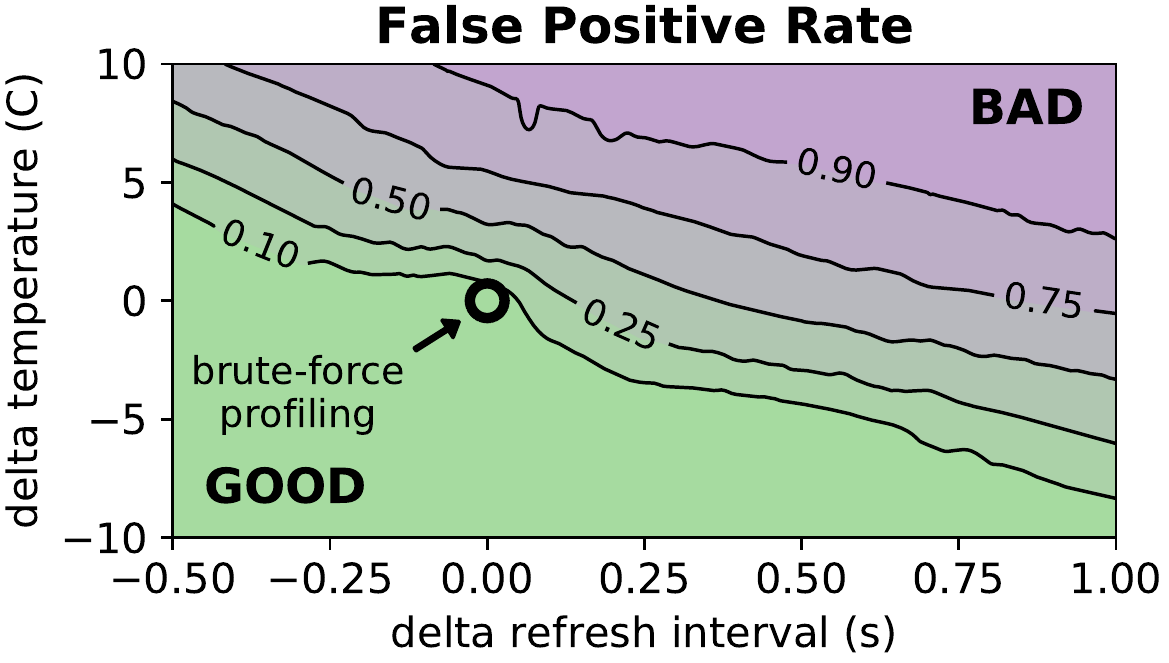}%
    \end{subfigure}
    \caption{Failure coverage (top) and false positive rates (bottom) as a
result of different choices of reach profiling conditions for a representative
chip from Vendor B. Contours represent coverage \textcolor{red}{(top)} and
false positive rate \textcolor{red}{(bottom)} obtained relative to brute-force
profiling at \textcolor{green}{$\mathbf{(x,y)=(0.00,0)}$}.} 
    \label{fig:covandfpr} 
\vspace*{10pt}
\end{figure}

Figure~\ref{fig:covandfpr} demonstrates the effect of choosing different reach
profiling conditions on failure coverage and false positive rate for a
representative chip. In order to perform this analysis, brute-force profiling
is conducted at regularly spaced points throughout the graphs in
Figure~\ref{fig:covandfpr} using 16 iterations of 6 different data patterns and
their inverses as per Algorithm~\ref{alg:bf}. Each point in the
graph space is then treated as a target condition with all
other points as its reach conditions.  This results in a distribution of
coverage/false positive rates for \emph{every} delta temperature/refresh
interval combination.

We find that these distributions are highly similar for each
delta temperature/refresh interval combination, with standard deviations of
less than 10\% of the data range for both coverage and false positive rate.
This allows us to use the means of these distributions to reasonably
demonstrate the overall effect of manipulating the temperature and/or refresh
interval \emph{independently} of a particular target refresh interval. We show
these means in Figure~\ref{fig:covandfpr}, with \textcolor{green}{$(x,y)=(0.00,0)$}
effectively representing \emph{any} target refresh interval and all other
points in the figure representing its reach conditions. The contours represent
the coverage (top graph) and false positive rate (bottom graph).\footnote{The
contours are not perfectly smooth due to variations in the data resulting from
small shifts ($<$0.25$^\circ$C) in temperature throughout testing and the
probabilistic nature of retention failures, including VRT effects.} As these
two graphs show, by increasing the refresh interval and/or temperature, we can
obtain a higher failure coverage \textcolor{red}{(Figure~\ref{fig:covandfpr},
top)}, as expected given the individual \textcolor{red}{cell} failure
probability analysis in Section~\ref{reaper:subsubsection:individual_bit_fail_prob}.
However, this also results in an increase in the false positive rate
\textcolor{red}{(Figure~\ref{fig:covandfpr}, bottom). Thus,} there is a direct
tradeoff between coverage and false positive rate.

\begin{figure}[h]
    \centering
    \includegraphics[width=0.7\linewidth]{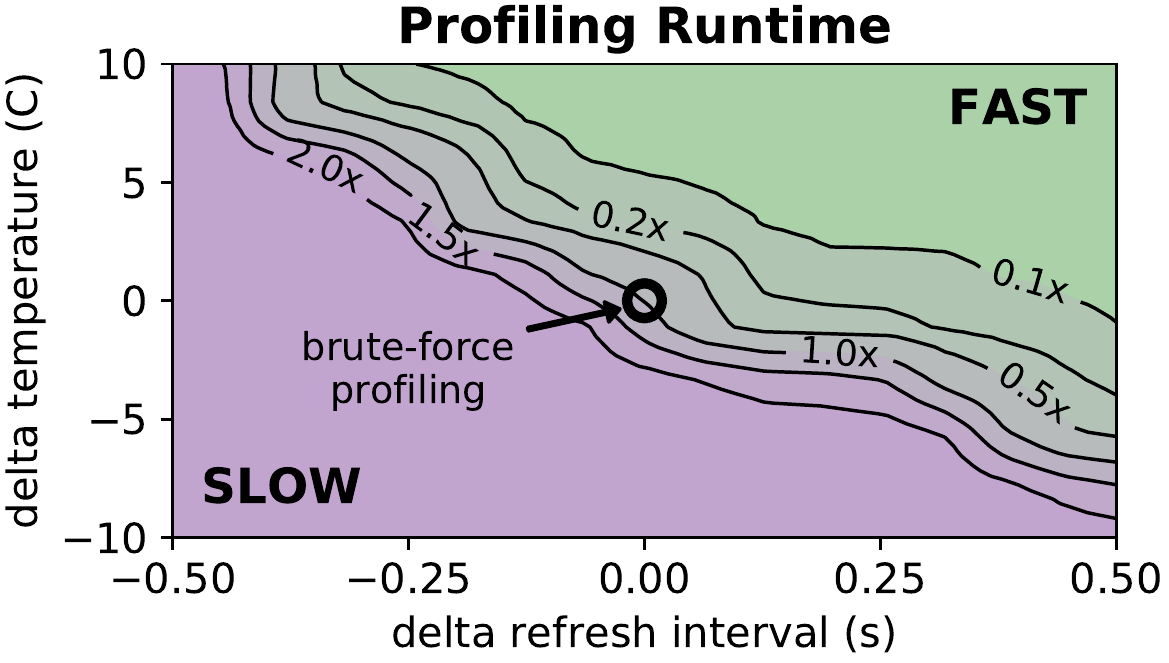} 
	\caption{Profiling runtime as a result of different choices of reach
conditions for a representative chip from Vendor B. Contours represent
profiling runtime relative to brute-force profiling at
\textcolor{green}{$\mathbf{(x,y)=(0.00,0)}$}.} 
    \label{fig:runtime} 
\end{figure}

Profiling runtime is more difficult to evaluate since it depends on $1)$ the
profiling refresh interval, $2)$ number of profiling iterations (see
Algorithm~\ref{alg:bf}), and $3)$ overheads of reading and writing the data
patterns to all of DRAM. We experimentally find that the amount of time taken
to read/write data to all DRAM channels and check for errors is slightly
less than 250ms for our evaluated chips\textcolor{red}{. So,} we assume a fixed overhead of
250ms per profiling iteration per data pattern tested. Figure~\ref{fig:runtime}
shows the results of such an analysis for a representative chip where \textcolor{green}{$(x,y)=(0.00,0)$}
represents profiling at the target conditions (i.e., brute-force profiling)
and all other points show the results of reach profiling with the same
analysis as in Figure~\ref{fig:covandfpr}. Here, each contour
curve shows the profiling runtime at different reach conditions, all normalized
to the runtime at the target refresh interval (i.e., brute-force profiling
runtime). \textcolor{red}{Profiling runtime is} determined by the number of
profiling iterations required to achieve over 90\% coverage.

We see that we can obtain drastic profiling \textcolor{red}{runtime} speedups
by aggressively increasing the reach conditions (e.g., +0.5s, +5$^\circ$C), but
from Figure~\ref{fig:covandfpr}, we know that this will result in a very high
number of false positives (e.g., over 90\% false positive rate). Although the
runtime numbers in Figure~\ref{fig:runtime} assume a fixed coverage, we observe
the same trends across different coverage requirements. Therefore, we conclude
that high coverage and low runtime are generally \emph{competing goals} to
low false positive rates, and choosing a good set of reach conditions at which
to profile depends on the user's and the system's requirements and failure
mitigation capabilities. 

We repeat this analysis for all 368 of our DRAM chips and find that each chip
demonstrates the same trends, showing that the same tradeoff analysis
\textcolor{red}{we have done for} Figures~\ref{fig:covandfpr}
and~\ref{fig:runtime} also applies to every other chip we tested. In the
following section, we analyze the data presented here and determine how to
select a desirable set of reach conditions.

\subsubsection{Choice of Desirable Parameters}
\label{reaper:subsubsec:desirable_params}

The exact choice of reach conditions depends on the overall system design and
its particular optimization goals. For example, in a highly latency-sensitive
system, the primary goal may be to minimize profiling runtime, in which case
\textcolor{red}{giving up a low false positive rate to obtain a low profiling
runtime} may be necessary.  On the other hand, if the retention failure
mitigation mechanism in use is intolerant to high false positive rates (e.g.,
discarding all DRAM rows containing failing cells), a low false positive rate
may be the primary design goal. In general, the system designer should take
these factors into account and make the best tradeoff for the target system.

However, although the exact choice of reach conditions depends on the system
goals, we observe from Figures~\ref{fig:covandfpr} and \ref{fig:runtime} that
higher coverage and higher runtime are directly and positively correlated, and
both come at the expense of false positives. This means that when selecting
suitable reach profiling conditions, the system designer can feasibly select as
high a refresh interval/temperature as possible that keeps the resulting amount
of false positives tractable. This approach primarily relies on identifying the
cost of false positives for the particular system in question,
\textcolor{red}{possibly sacrificing the ability to handle a large number of false
positives in favor of low profiling runtime.}

As a concrete example, from studying data similar to those presented in
Figures~\ref{fig:covandfpr} and \ref{fig:runtime} averaged across all 368
chips, we find that for 99\% coverage at a modest 50\% false positive rate, we
attain a profiling runtime speedup of 2.5x over the brute-force mechanism when
we extend the refresh interval by 250ms. We find that we can push the speedup
to over 3.5x while maintaining the same level of coverage by either increasing
the temperature or lengthening the refresh interval even further, but such
aggressive reach profiling conditions result in greater than 75\% false
positive rates.

\vspace{-5pt}
\subsection{Online Profiling Frequency}
\label{reaper:subsec:online_profiling}
In order to enable a longer refresh interval, we not only have to provide a way
to profile for retention failures, \textcolor{red}{but also have to} guarantee
that retention failure profiling provides sufficient coverage to prevent
erroneous system operation. Given failures that are missed by profiling due to
\textcolor{red}{less-than-100\%} coverage and non-zero new failure accumulation
rates (Section~\ref{reaper:subsubsec:vrt_anal}), we need a method by which to estimate
\emph{profile longevity, i.e.,} the amount of time before a profile is no
longer correct (i.e., when reprofiling becomes necessary). In this section, we
analyze the types of errors that can be missed by profiling and present a
theoretical model for allowable error rates and profile longevity.
\subsubsection{Failures Missed by Profiling}
\label{reaper:subsubsection:missed}

Retention failure profiling can only capture \textcolor{blue}{cells} that fail
at the profiling conditions. \textcolor{blue}{Cells} missed during profiling
due to DPD effects (Section~\ref{reaper:subsubsection:dpd_effects}) and newly-failing
\textcolor{blue}{cells} that did not fail during profiling due to environmental
factors (e.g., temperature shifts, soft errors) or due to VRT effects
(Section~\ref{reaper:subsubsec:vrt_anal}) cannot be observed by profiling. Prior works
acknowledge that there will inevitably be failures that are missed by profiling
and argue that \textcolor{red}{\emph{some form of ECC is necessary to allow
safe operation with a longer refresh
interval}}~\cite{liu2013experimental,khan2014efficacy, qureshi2015avatar}. Our
observation of continuously-accumulating new failures in LPDDR4 DRAM chips
(Section~\ref{reaper:section:new_observations}) \textcolor{red}{leads} us to agree
with this argument \textcolor{red}{for the use of ECC}. Thus, in order to
enable online profiling, we need to determine 1) what error rates can be
tolerated by an ECC-based system and 2) at what point the accumulated failures
will exceed the correction capability of ECC and require reprofiling.
\subsubsection{Allowable Errors and Tolerable Error Rates}
\label{reaper:subsection:allowable_errors}

In order to estimate the probability of system failure given DRAM retention
errors in the presence of various types of error correction codes (ECC), we
estimate the error rate as observed by the system, called the uncorrectable bit
error rate (UBER), as a function of the raw bit error rate (RBER), defined as
the ratio of failing DRAM cells. We define \emph{system failure} as exceeding
an UBER of 1) $10^{-15}$ for consumer
applications~\cite{micheloni2015apparatus} and 2) $10^{-17}$ for enterprise
applications~\cite{samsungbrochure}.

Given a general system using $k$-bit ECC (with $k=0$ defined as no ECC, $k=1$
as SECDED ECC~\cite{luo2014characterizing}, etc.) and an ECC word size of
$w$, we define the UBER as the probability of observing an error in a
single DRAM ECC word, normalized to the number of bits per ECC word: 
\begin{equation}
\label{eqn:uber1}
\mathrm{UBER} = \frac{1}{w}\mathrm{P}[\text{uncorrectable~error~in~a~}w\text{-}\mathrm{bit~ECC~word]} \\
\end{equation}
\noindent
We obtain an uncorrectable error in an ECC word when we have an $n$-bit error
$\forall~n>k$. This means that we can expand Equation~\ref{eqn:uber1} in terms
of $n$:
\begin{equation}
\label{eqn:uber2}
\mathrm{UBER} = \frac{1}{w}\sum\limits_{n=k + 1}^{w}\mathrm{P}[n\text{-bit}\mathrm{~failure~in~a~}w\text{-}\mathrm{bit~ECC~word}]
\\
\end{equation}

\noindent
representing the sum of the probabilities of all possible $k$-bit ECC
uncorrectable errors. In particular, for \textcolor{red}{$k=0$ and $k=1$}, we
have 0 and 8 additional bits per 64-bit data word, respectively. For these cases,
Equation~\ref{eqn:uber2} takes the form:
\begin{equation}
\begin{aligned}
\mathrm{UBER}(k=0) &= \frac{1}{64}\sum\limits_{n=1}^{64}\mathrm{P}[n\text{-bit}\mathrm{~failure~in~a~\text{64-bit}~ECC~word}] \\
\mathrm{UBER}(k=1) &= \frac{1}{72}\sum\limits_{n=2}^{72}\mathrm{P}[n\text{-bit}\mathrm{~failure~in~a~\text{72-bit}~ECC~word}]
\end{aligned}
\end{equation}

The probability of an $n$-bit error in a $k$-bit word can be modeled by a
binomial distribution, assuming that DRAM retention failures are independent
and randomly distributed, as has been shown in previous
work~\cite{baek2014refresh, venkatesan2006retention}. Given that these
assumptions hold, we can use the RBER $R$ as the probability of a single-bit
failure within DRAM, and expand the probability of an $n$-bit failure in a
$k$-bit ECC word using a binomial distribution in terms of $R$: 
\begin{equation}
\mathrm{P}[n\text{-bit}\mathrm{~failure~in~a~w\text{-bit}~ECC~word]} = {w\choose
n}R^{n}(1-R)^{w-n} \\
\end{equation}
\noindent
Putting the pieces together, we arrive at:
\begin{equation}
\mathrm{UBER} = \frac{1}{w}\sum\limits_{n=k + 1}^{w}{w\choose n}R^{n}(1-R)^{w-n}
\\
\end{equation}
\noindent 
Table~\ref{tab:ubervals} summarizes the maximum tolerable \emph{RBER} for a
target UBER of $10^{-15}$, for various choices of ECC strength. In order to
provide comparison points, the table also translates the tolerable RBER into
actual \emph{number of tolerable bit errors} for various DRAM sizes.
\vspace{1pt}
\begin{table}[h]
\centering
\begin{tabular}{ |cl|c|c|c|c|c| }
\cline{3-5}
        \multicolumn{2}{c|}{} & \multicolumn{3}{ c| }{\bf{ECC Strength}} \\ \cline{3-5}
        \multicolumn{2}{c|}{} & \bf{No ECC} & \bf{SECDED} & \bf{ECC-2} \\ \hline
        \multicolumn{2}{|c|}{\bf{Tolerable RBER}} 
                & $\num{1.0e-15}$ & $\num{3.8e-9}$ & $\num{6.9e-7}$ \\ \hline \hline

        & 512MB & $\num{4.3e-6}$ & $\num{16.3}$ & $\num{3.0e3}$ \\ \cline{2-5}
        & 1GB   & $\num{8.6e-06}$ & $\num{32.6}$ & $\num{5.9e+03}$ \\ \cline{2-5}
        & 2GB   & $\num{1.7e-05}$ & $\num{65.3}$ & $\num{1.2e+04}$ \\ \cline{2-5}
        & 4GB   & $\num{3.4e-05}$ & $\num{130.6}$ & $\num{2.4e+04}$ \\ \cline{2-5}
        \rotatebox{90}{\rlap{\shortstack[c]{\bf{\# Tolerable} \\ \bf{Bit Errors}}}}
        & 8GB   & $\num{6.9e-05}$ & $\num{261.1}$ & $\num{4.7e+04}$ \\ \hline
\end{tabular}
\caption{Tolerable RBER and tolerable number of bit errors for UBER =
$10^{-15}$ across different ECC strengths for selected DRAM sizes}
\label{tab:ubervals}
\end{table}
\vspace{-10pt}
Thus, for any target UBER, we can compute the \emph{tolerable RBER}, and hence
the \emph{maximum} number of \textcolor{blue}{cells} that can be allowed to
\emph{escape} (i.e., not be detected by) our retention failure profiling
mechanism \textcolor{red}{while} still \textcolor{red}{maintaining} correct
DRAM operation. By applying the maximum tolerable RBER to the RBER at any
desired target refresh interval, we can directly compute the minimum coverage
required from a profiling mechanism in order for the retention failure
mitigation mechanism to guarantee correct system operation.
\vspace*{-10pt}
\subsubsection{Profile Longevity}
\label{reaper:subsubsection:logngevity}

Given the maximum tolerable number of retention failures $N$ provided by
Table~\ref{tab:ubervals}, the number of failures $C$ missed by profiling due to
imperfect profiling coverage, and the accumulation rate $A$ of new failures as
measured in Section~\ref{reaper:subsubsec:vrt_anal}, we can estimate $T$, the length
of time before we need to reprofile, as:
\begin{equation}
\label{eqn:longevity}
T = \frac{N - C}{A}
\vspace*{10pt}
\end{equation}

For example, with a 2GB DRAM and SECDED ECC as the failure mitigation
mechanism, we can afford up to $N=65$ failures while still maintaining correct
DRAM operation. Assuming an aggressive target refresh interval of 1024ms at
45$^\circ$C, we empirically observe 2464 retention failures
(Figure~\ref{fig:retention_time_distribution}) and a new failure accumulation
rate of $A=0.73$ \textcolor{blue}{cells} / hour
(Figure~\ref{fig:VRT_per_trefi}).  With 99\% coverage, we
\textcolor{red}{compute} $C=24.6 \approx 25$ \textcolor{blue}{cells}, and
applying Equation~\ref{eqn:longevity}, we obtain $T=2.3$ days. We can use a
similar analysis to determine the profile longevity for any desired
configuration and choose an appropriate tradeoff point between the different
reach profiling parameters.
\subsection{Enabling Reliable Relaxed-Refresh Operation}
\label{reaper:subsection:enabling}
In order to determine good reach conditions at which to profile for a real
system running at a longer refresh interval than the default, we need to know
two key pieces of information: 1) the particular retention failure mitigation
mechanism in use, so that we can constrain the reach profiling tradeoff space,
and 2) detailed chip characterization data, in order to make reliable estimates
of reach profiling benefits and profile longevity, allowing us to determine
which point in the tradeoff space will provide the best overall system
performance/energy improvement.

The choice of \emph{retention failure mitigation mechanism} determines the
hardware/energy/performance overhead of managing failing cells, which determines
how aggressively we can push the target/reach conditions before the system can
no longer cope with the number of failures and false positives. The mitigation
mechanism therefore constrains both the target conditions (i.e., the resultant
RBER \emph{without} mitigation) and the maximum number of false positives,
which in turn restricts the range of the reach conditions. 

\emph{Detailed chip characterization data} (as in Figures~\ref{fig:covandfpr}
and \ref{fig:runtime}) is necessary to produce accurate estimates of all of the
parameters of the actual system, including the expected RBER, the profile
longevity, the required coverage, and even the reach conditions themselves.
While these parameters can be estimated from general trends across many chips,
the variations among different chips mean that truly reliable relaxed refresh
operation requires estimating profiling parameters based on data from the
actual chip. 

Currently, DRAM vendors do not provide this data, but it would be reasonable
for vendors to provide this data in the on-DIMM serial presence detect (SPD) as
done in~\cite{kim2012case}. Otherwise, the user would have to characterize
his/her own device. Even though a detailed characterization may take
prohibitive amounts of time, a few sample points around the tradeoff space
could provide enough information in conjunction with the general trends across
many devices to develop accurate estimations. However, the general problem of
efficiently obtaining per-chip characterization data is itself an open research
direction and is beyond the scope of this work. 
Once these two critical pieces of information are available, the system
designer can decide what the best tradeoff is between \textcolor{red}{profile
longevity, coverage, false positive rate, and runtime} according to his/her own
system configuration. Despite the overall trend similarities we observe between
the DRAM chips we evaluate in Section~\ref{reaper:section:new_observations}, the
\emph{optimal} choice of reach conditions depends on the particular system and
its tradeoffs between cost, performance, power, and complexity, and it is up to
the user to determine the most suitable configuration for his/her own needs.

\section{REAPER: Example End-to-End Implementation} 
\label{reaper:section:e2e}

Reach profiling provides a general-purpose retention failure profiling
methodology whose exact use and design parameters
(Section~\ref{reaper:subsection:enabling}) depend on the error mitigation mechanism
used for \textcolor{red}{reduced-refresh-rate} operation. A system designer
should utilize reach profiling uniquely depending on desired performance/energy
targets and ease/overhead of implementation. For example, the designer is free
to choose where to implement the core profiling algorithm (e.g., at the memory
controller, the operating system, etc.), how to save and restore the state of
DRAM before and after a profiling \emph{round} (i.e., each individual instance
of online profiling consisting of all iterations and data patterns), how to
efficiently profile large portions of DRAM without significant performance
loss, etc. However, a thorough exploration of the design space regarding the
implementation of reach profiling is beyond the scope of this paper.  

In order to effectively present the insight behind the tradeoffs involved in
reach profiling, we provide and evaluate REAPER, a na\"ive (yet robust) example
implementation of reach profiling that assumes a full-system
\textcolor{red}{pause} each time retention failures are profiled. REAPER
enables a variety of retention failure mitigation mechanisms~\cite{liu2012raidr,
ohsawa1998optimizing, wang2014proactivedram, venkatesan2006retention,
lin2012secret, nair2013archshield} for reliable operation at longer refresh
intervals. We experimentally validate that, even with these worst-case
assumptions, reach profiling preserves a significant amount of the benefit of
an ideal profiling mechanism \textcolor{red}{that has zero overhead} while
outperforming brute-force profiling in terms of both performance and power
consumption.

\subsection{REAPER Implementation}
Reach profiling requires the ability to manipulate the DRAM refresh interval
and/or the DRAM temperature, both of which can be achieved with commodity
off-the-shelf DRAM. REAPER implements reach profiling in firmware running
directly in the memory controller. Each time the set of retention failures must
be updated, profiling is initiated by gaining exclusive access to DRAM. REAPER
has the ability to manipulate the refresh interval directly, but for
simplicity, we assume that \emph{temperature} is not adjustable (as shown in
Section~\ref{reaper:subsubsection:individual_bit_fail_prob}, manipulating either
\textcolor{red}{the} temperature or the refresh interval achieves the same
effect).  REAPER conducts reach profiling and stores \textcolor{blue}{the addresses
of} the failing \textcolor{blue}{cells} it finds at locations dictated by the
retention failure mitigation mechanism of choice.\footnote{Profiling involves
modifying the contents of DRAM. Data must be saved and restored before and
after profiling is run, respectively.  While a na\"ive implementation may flush
all DRAM data to secondary storage (e.g., hard disk drive, SSD, NVM) or other
parts of DRAM, a more efficient implementation could hide most or all of this
latency.  Efficiently implementing DRAM data save and restore is an orthogonal
problem to this work.  Therefore, we do not take DRAM data save and restore
overheads into account in our performance and energy evaluations.} At the
completion of profiling, REAPER releases exclusive DRAM access and the system
resumes normal operation.  In the remainder of this section, we describe how
this implementation of reach profiling can be combined with two example
retention failure mitigation mechanisms from prior work to reduce refresh
operations. 
\subsubsection{REAPER Supporting ArchShield}
As a demonstrative example, we combine REAPER with
ArchShield~\cite{nair2013archshield} to enable refresh rate reduction.
ArchShield requires a small amount of additional logic in the memory controller
for detecting accesses to known-faulty addresses, which are stored in 4\% of
DRAM in a reserved segment known as the FaultMap. The REAPER firmware
provides these faulty addresses via periodically profiling DRAM for
retention failures following the methodology in
\textcolor{red}{Section~\ref{related:sec:error_profiling}}.
All detected failures are stored into the FaultMap, which ArchShield
accesses to determine which failing addresses must be remapped. 
\subsubsection{REAPER Supporting Multi-Rate Refresh}
In order to show REAPER's flexibility for integration with other retention
failure mitigation mechanisms, we describe REAPER working with
RAIDR~\cite{liu2012raidr}. RAIDR groups DRAM rows into different
bins, according to the rows' retention times, and applies different
refresh rates to each bin. REAPER enables RAIDR by periodically updating the
Bloom filters using the set of failures discovered each time profiling is
conducted.  This enables the system to reduce the refresh interval for
\textcolor{red}{most of the} rows within DRAM, which results in overall system
performance increase and energy reduction. 

\subsection{Evaluation Methodology}
\label{reaper:subsubsec:eval_method}
Given that profiling rounds require on the order of seconds or minutes
(Section~\ref{reaper:subsection:prof_overhead}) and the time between online profiling
rounds can be on the order of hours (Section~\ref{reaper:subsubsection:logngevity}),
we cannot feasibly simulate workloads in order to demonstrate latency effects
due to online profiling. Instead, we simulate our workloads without the
profiling overhead and report performance results using throughput in terms of
instructions-per-cycle (IPC). We then use the following model to compute
overall system performance accounting for online profiling overhead:  
\begin{equation}
\label{eq:perf_calc}
\mathrm{IPC\_real} = \mathrm{IPC\_ideal} \times (1 - \mathrm{profiling\_overhead})
\end{equation}
where $IPC\_ideal$ is the measured throughput of simulated workloads, and
$profiling\_overhead$ is the proportion of real system time spent in profiling.
Our profiling overhead model assumes worst-case configurations, pessimistically
assuming that applications make zero forward progress while
profiling. Our full system performance model roughly estimates the worst-case
performance degradation expected from an implementation of reach profiling.

We use Ramulator~\cite{kim2016ramulator, ramulatorgithub} to evaluate
performance and DRAMPower~\cite{drampowergithub} to evaluate DRAM power
consumption\textcolor{red}{. We simulate 20 multiprogrammed heterogeneous
workload mixes, each of which is constructed by randomly selecting 4 benchmarks
from the SPEC CPU2006 benchmark suite~\cite{spec2006}. Multi-core system}
performance is measured in terms of the weighted speedup
metric~\cite{snavely2000symbiotic,eyerman2008system}. We provide our evaluated
system configuration in Table~\ref{table:system_config}.

\begin{table}[h!] 
    \small
    \centering 
    \renewcommand{\arraystretch}{1.0} 
    \begin{tabular}{| m{4cm} || m{8cm} |}
		\hline Processor & 4 cores, 4GHz clock frequency,
			3-wide issue, 8 MSHRs/core, 128-entry instruction window \\ \hline 
		Last-level Cache & 64B cache line, 16-way,
			8MB cache size \\ \hline 
		Memory Controller & 64-entry read/write
			request queues, FR-FCFS scheduling policy~\cite{rixner2000memory,
			zuravleff1997controller}, open/closed row
			policy~\cite{kim2010thread, kim2010atlas} for single/multi-core \\
			\hline
        DRAM & LPDDR4-3200~\cite{jedec2014lpddr4}, 4 channels,
        1 rank, 8 banks/rank, 32K-256k rows/bank, 2KB row buffer \\
        \hline 
    \end{tabular}
    \caption{Evaluated system configuration}
    \label{table:system_config} 
\end{table} 
\subsection{Performance and Energy Evaluation}
\label{reaper:subsec:perf_energy_eval} 

We develop a detailed model for profiling overhead, taking into account extra
DRAM accesses required for profiling, latencies involved in reading/writing
data patterns to DRAM, and time consumed waiting for \textcolor{red}{the
extended refresh intervals}. We use this model in conjunction with
Equation~\ref{eq:perf_calc} to reliably estimate worst-case system performance
impact and additional DRAM power consumption.  In this section, we present our
1) model \textcolor{red}{for} profiling performance and power consumption
overhead for brute-force profiling and reach profiling, and 2) results for
end-to-end system performance and DRAM power consumption using brute-force
profiling, reach profiling, and the ideal profiling mechanism that does not
impact system performance or power consumption. We find that while profiling
power consumption overheads are very low across a wide range of DRAM sizes and
refresh intervals, profiling performance overheads become significant for large
DRAM chips and high online profiling frequencies. In \textcolor{red}{such}
cases where profiling performance significantly impacts overall system
performance, reach profiling maintains a large fraction of the benefits of
using a longer refresh interval (e.g., providing 14.8\% higher performance than
brute-force profiling on average for a 64Gb chip running at a 1280ms refresh
interval at 45$^\circ$C), as we show in Section~\ref{reaper:subsubsec:e2eresults}.

\subsubsection{Profiling Overhead}
\label{reaper:subsection:prof_overhead}

\renewcommand*{\thefootnote}{\fnsymbol{footnote}}
We model the overhead of online profiling by accumulating the latencies of
individual operations required to run a single \emph{round} of profiling. We
assume that profiling requires a full system \textcolor{red}{pause,} with no
useful work being done while the profiler is running. One round of profiling
consists of $N_\mathrm{it}$ iterations, each of which consists of
reading/writing DRAM with $N_\mathrm{dp}$ different data patterns. Our
end-to-end runtime model is as follows:
\begin{equation}
\begin{aligned}
T_\mathrm{profile} &= 
(t_{REFI} + T_\mathrm{wr~DRAM} + T_\mathrm{rd~DRAM}) * N_\mathrm{dp}*N_\mathrm{it}
\end{aligned}
\end{equation}
\noindent
where $T_{REFI}$ is the profiling refresh interval that DRAM must wait with
refresh disabled in order to accumulate errors; $T_{wr~DRAM}$ is the time to
write a data pattern into DRAM; and $T_{rd~DRAM}$ is the time to read DRAM and
compare against the original data pattern. For 32 8Gb DRAM chips with
$t_{REFI}=$ 1024ms, $T_\mathrm{rd/wr~DRAM}=$ 0.125s\footnote[2]{This value is
based on empirical measurements from our infrastructure using 2GB of LPDDR4
DRAM. We scale this number according to DRAM size throughout our evaluation to
account for the \textcolor{red}{larger} number of accesses.}, $N_\mathrm{dp}=$ 6, and
$N_\mathrm{it}=$ 6, we find that $T_\mathrm{profile}\approx$ 3.01 minutes, and
for 32 64Gb chips, $T_\mathrm{profile}\approx$ 19.8 minutes.
\renewcommand*{\thefootnote}{\arabic{footnote}}

\begin{figure}[h]
    \centering
    \includegraphics[width=0.7\linewidth]{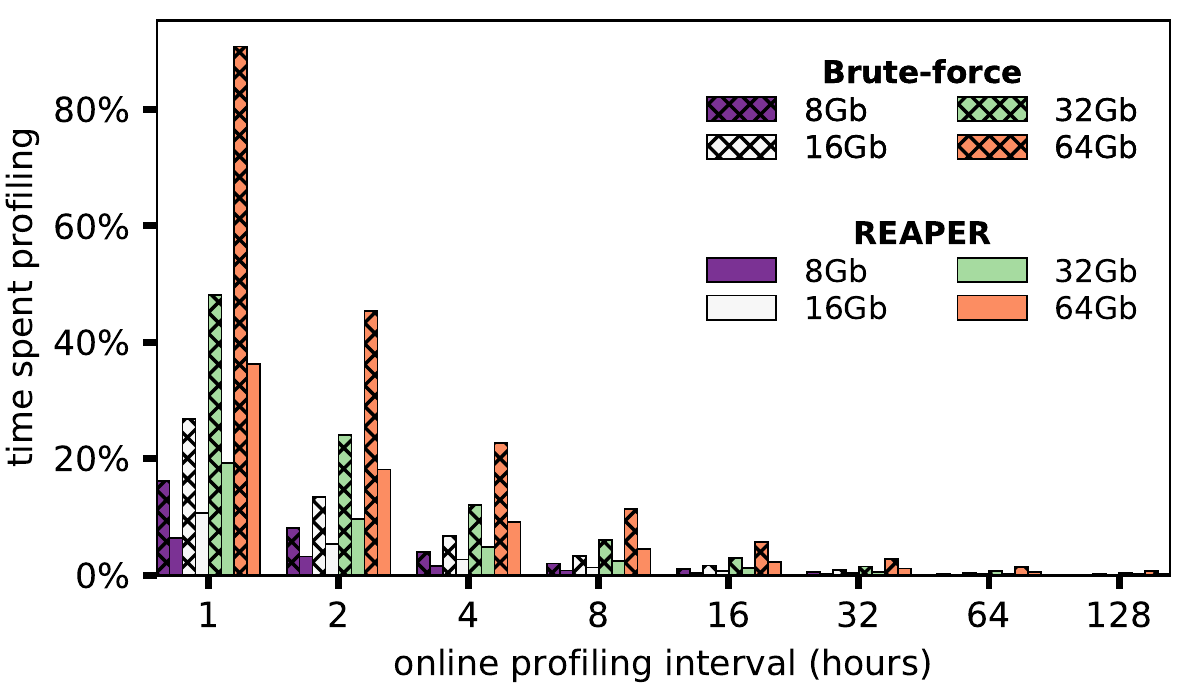} 
	\caption{Total system time spent profiling with REAPER and brute-force
profiling for different online profiling intervals using 32 \textcolor{red}{DRAM} chips, \textcolor{red}{for different chip sizes.}}
    \label{fig:perf_degradation} 
\end{figure}

In order to demonstrate the overall system performance degradation from
profiling, Figure~\ref{fig:perf_degradation} plots the proportion of overall
system time spent profiling for a variety of different profiling intervals and
DRAM sizes assuming 16 iterations of brute-force profiling at 1024ms using 6
data patterns and their inverses. The x-axis represents the
\textcolor{red}{online profiling frequency} in hours, and the y-axis shows the
proportion of total system-time spent profiling. To compare the brute-force
method against REAPER, we plot the performance overheads with hashed
(brute-force) and solid (REAPER) bars. We show results for the 2.5x profiling
runtime speedup of reach profiling over brute-force profiling,
\textcolor{red}{found experimentally} in
Section~\ref{reaper:subsubsec:desirable_params}. The
\textcolor{red}{differently-colored} bars represent DRAM modules consisting of
32 individual DRAM chips, with each chip ranging from 8Gb to 64Gb in size. We
observe that at \textcolor{red}{shorter} profiling intervals, the performance
degradation is prohibitively high and is exacerbated by larger DRAM chip sizes.
For example, we find that for a profiling interval of 4 hours and a 64Gb chip
size, 22.7\% of total system time is spent profiling with brute-force profiling
while 9.1\% of time is spent profiling with REAPER.

\begin{figure}[h]
    \centering
    \includegraphics[width=0.7\linewidth]{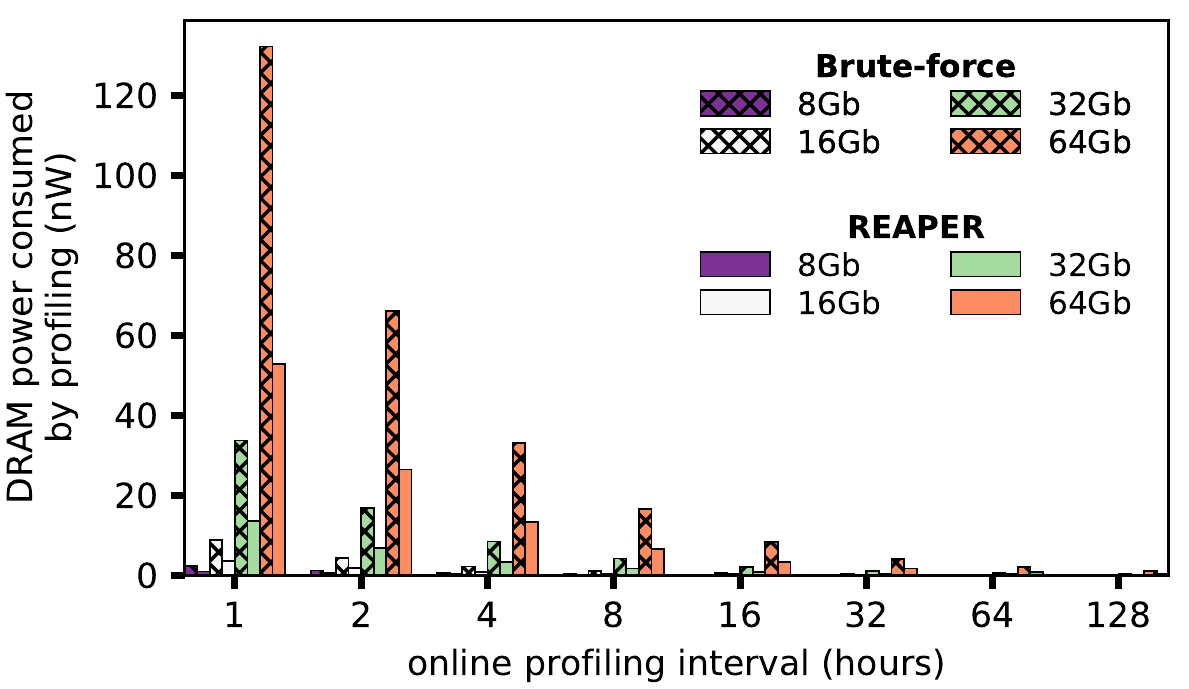} 
	\caption{DRAM power consumption of REAPER and brute-force profiling for
	different online profiling intervals using 32 \textcolor{red}{DRAM} chips, \textcolor{red}{for different chip sizes}}. 
    \label{fig:energy_degradation} 
\end{figure}

In order to demonstrate the overall DRAM power overhead associated with
profiling, Figure~\ref{fig:energy_degradation} shows the total DRAM power
required for profiling across the same sweep of configurations as
Figure~\ref{fig:perf_degradation}. Power consumption overhead is calculated
using the average energy-per-bit costs for different DDR commands obtained from
our LPDDR4 DRAMPower model~\cite{drampowergithub}. We estimate the number of
extra commands required for profiling and show the total DRAM energy consumed
for one round of profiling divided by the online profiling interval. The x-axis
represents the online profiling frequency in hours, and the y-axis shows the
DRAM power consumption in nanowatts. We see that power consumption has a strong
dependence on DRAM size and shows a similar scaling trend as profiling
performance overhead with respect to \textcolor{red}{the} profiling interval.
However, the absolute power consumption is very low (on the order of
nanowatts), because the majority of profiling runtime is spent waiting for
retention failures to occur (i.e., the profiling refresh interval) rather than
actively accessing DRAM. We show in Section~\ref{reaper:subsubsec:e2eresults} how
profiling itself results in negligible additional power \textcolor{red}{in the
DRAM system}, even in the extreme case of very frequent profiling and large
DRAM sizes.

\subsubsection{End-to-end System Evaluation Results}
\label{reaper:subsubsec:e2eresults}
We evaluate the overall effects of REAPER on system performance and DRAM power
consumption by applying our profiling overhead model to
Equation~\ref{eq:perf_calc}. \textcolor{red}{We compare REAPER} with both
online brute-force profiling and the ideal profiling mechanism that incurs no
performance or energy overhead. The ideal mechanism mirrors the approach taken
by prior works on refresh rate reduction~\cite{liu2012raidr, ohsawa1998optimizing,
wang2014proactivedram, venkatesan2006retention, lin2012secret,
nair2013archshield}, which assume that \emph{offline} profiling is sufficient.
Considerable evidence against the sufficiency of offline profiling is provided
by both prior work on DRAM
characterization~\cite{liu2013experimental, khan2014efficacy} and our own observations
\textcolor{red}{(Sections~\ref{reaper:subsubsec:vrt_anal}
and~\ref{reaper:subsubsection:missed})}. We exclude ECC-scrubbing based mechanisms
from our evaluations due to their passive approach to failure profiling, which
cannot guarantee failure coverage in the same way as an active profiling
mechanism, as discussed in Section~\ref{related:sec:error_profiling}.

\begin{figure*}[h]

    \includegraphics[width=\textwidth]{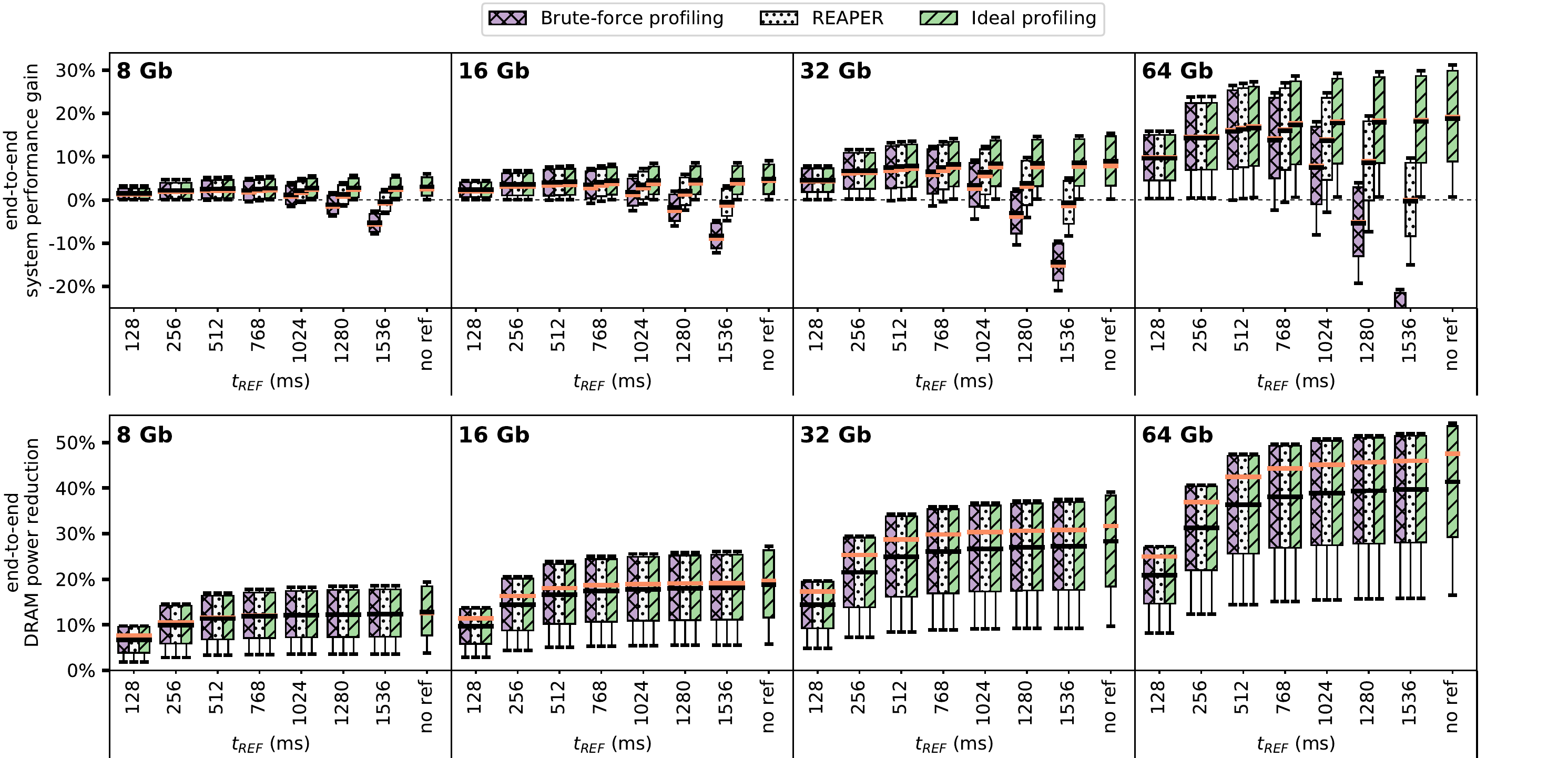}
	\caption{Simulated end-to-end system performance improvement (top) and
    DRAM power reduction (bottom) over 20 heterogeneous 4-core
    workloads for different refresh intervals at 45$^\circ$C,
    taking into account online profiling frequency and profiling overhead.} 
    \label{fig:e2ee} 
\end{figure*}

Figure~\ref{fig:e2ee} shows the results of this analysis, demonstrating the
overall system performance improvement (top) and DRAM power consumption
reduction (bottom) due to operating with a longer refresh interval across all
of our simulated workloads for a variety of DRAM sizes each consisting of 32
individual DRAM chips \textcolor{red}{of the specified capacity (8-64Gb)}. Each
triplet of boxes presents results for online brute-force profiling, REAPER, and
the ideal profiling mechanism, in order from left to right. Results are shown
for various lengthened refresh intervals from 128ms to 1536ms and for the case
of no refresh, which is shown as a single box. For each refresh interval, the
profiling parameters used to compute overhead are obtained using the
experimental data for failure rates at 45$^\circ$C as shown in
Section~\ref{reaper:section:new_observations}. Profile longevity is estimated in the
best case for each configuration, assuming that the profilers achieve
\emph{full coverage} each time they are run. This assumption is reasonable for
profilers achieving high (e.g., 99\%) coverage and allows us to decouple the
results from failure rates of specific DRAM chips. Results are all normalized
to the base performance or power consumption at the default refresh interval of
64ms without profiling. The boxes represent the distribution of benchmarks from
the 25th to the 75th percentiles, the whiskers show the data range, and the
orange and black lines represent medians and means, respectively. We make four
major observations based on these results.

\begin{sloppypar}
First, REAPER enables a high performance improvement and a high DRAM power
reduction by reliably increasing the refresh interval. REAPER enables 512ms as
the \emph{best overall operating point} across all \textcolor{red}{evaluated}
DRAM chip sizes, providing an average of 16.3\% (maximum 27.0\%) performance
improvement and an average of 36.4\% (maximum 47.4\%) power reduction for 64Gb
chips. This is very close to the average performance gain of 18.8\% (31.2\%
maximum) and average power reduction of 41.3\% (54.1\% maximum) that comes with
eliminating all refreshes (rightmost ``no ref'' bars).
\end{sloppypar}

Second, REAPER significantly outperforms the brute-force
mechanism, especially at high refresh intervals ($>=$ 1024ms). While the ideal
profiling mechanism enables increasing gains with longer refresh intervals,
both REAPER and brute-force profiling overheads become more significant for
refresh intervals beyond 1024ms. This is due to the high VRT failure
accumulation rate (Section~\ref{reaper:subsubsec:vrt_anal}) at long refresh intervals,
which requires a high online profiling frequency. However, at such high refresh
intervals, REAPER preserves much more of the ideal profiling benefit than does
brute-force profiling. For example, with 64Gb DRAM chips, at a refresh interval
of 1024ms, using REAPER provides 13.5\% average (24.7\%) maximum performance
improvement while using brute-force profiling provides only
7.5\% average (18\% maximum) performance improvement. REAPER's
performance benefit over brute-force profiling increases with longer refresh
intervals because REAPER can sustain higher online profiling frequencies
better, showing that REAPER is superior to brute-force profiling.
For refresh intervals below 512ms, both REAPER and brute-force
profiling provide benefits that are very close to that of ideal profiling since
the performance overhead of both mechanisms is low. This is due to the low VRT
failure accumulation rate observed in Section~\ref{reaper:subsubsec:vrt_anal} for
short refresh intervals, resulting in a high profile longevity. However, short
refresh intervals do \emph{not} provide the full benefits of employing longer
refresh intervals, so we would like to enable as high a refresh interval as
possible with as low overhead as possible.

Third, REAPER enables high performance operation at very long refresh intervals
that were previously unreasonable to operate \textcolor{red}{the system} at.
For refresh intervals longer than 512ms, the high online profiling frequency
means that profiling overhead becomes significant (as supported by
Figure~\ref{fig:perf_degradation}), and \textcolor{red}{this} overhead
eventually results in overall performance degradation. At a refresh interval of
1280ms, the performance impact of profiling becomes clearly
visible\textcolor{red}{:} using brute-force profiling leads to overall system
performance \emph{degradation} (-5.4\% on average). However, REAPER still
maintains significant performance benefit (8.6\% on average) at 1280ms,
demonstrating that REAPER enables longer refresh intervals that were
\textcolor{red}{previously unreasonable} to operate the system at.

Fourth, both REAPER and brute-force profiling have negligible
impact on overall DRAM power consumption across all refresh intervals and DRAM
chip sizes. While Figure~\ref{fig:energy_degradation} shows that REAPER
consumes significantly less power than brute-force profiling at the same
configuration, profiling power consumption is a very small fraction of total
DRAM power consumption to begin with. This means that although the profiling
process has a large effect on overall system performance, it does
\textcolor{red}{\emph{not}} contribute significantly to DRAM power consumption.
Thus, both REAPER and brute-force profiling are effective at greatly reducing
DRAM power consumption by enabling a longer refresh interval.

We can use the results of Figure~\ref{fig:e2ee} to estimate the benefits
obtained by using \emph{any} retention failure mitigation mechanism.  For
example, ArchShield~\cite{nair2013archshield} extends the default refresh
interval up to 1024ms at the cost of approximately 1\% overall system
performance (Section 5.1 of~\cite{nair2013archshield}). The overall system
performance improvement can be estimated by subtracting ArchShield's reported
performance cost (in~\cite{nair2013archshield}) from the ideal profiling
performance gain (in Figure~\ref{fig:e2ee}), resulting in an overall
performance improvement of 15.7\% on average (28.2\% maximum) using 64Gb DRAM
chips at a refresh interval of 1024ms.  However, as we show the need for an
online profiling mechanism (Section~\ref{reaper:subsec:online_profiling}), the actual
overall performance benefit must be adjusted to account for profiling overhead.
We observe performance improvements of 6.5\% on average (17\% maximum) when
ArchShield is combined with brute-force profiling, and 12.5\% on average
(23.7\% maximum) when it is combined with REAPER. Thus, we find that using
REAPER leads to an average performance improvement of 5.6\% over using
brute-force profiling. A similar analysis can be conducted using other
state-of-the-art retention failure mitigation mechanisms to estimate overall
system speedup and DRAM power consumption.

We conclude that REAPER is an effective and low-overhead profiling mechanism
that enables high-performance operation at very long refresh intervals (beyond
1024ms) that were previously not reasonable to employ due to the high
associated profiling overhead. As a caveat to the data presented in this
section, it is important to note that
Figures~\ref{fig:perf_degradation},~\ref{fig:energy_degradation},
and~\ref{fig:e2ee} are all based on specific assumptions (e.g., 45$^\circ$C
temperature, 2.5x performance improvement of reach profiling vs. online
brute-force profiling, 32 chips per DRAM module, 100\% coverage by profiling,
\textcolor{red}{20 randomly-formed heterogeneous workload mixes}) and we are not
covering the entire design space with these results. This means that there
likely exist other conditions under which REAPER may perform even better,
including both different temperatures, at which we would expect failure rates
and therefore profile longevity to change, and varying system requirements
(e.g., choice of retention failure mitigation mechanism, target profiling
coverage and false positive rate, target profile longevity, target UBER, etc.).
These conditions may result in different choices of reach profiling parameters,
which can lead to a higher-than-2.5x performance improvement for reach
profiling over brute-force profiling.  This increase in reach profiling
performance translates directly to reduction in profiling overhead, which in
turn translates to even greater end-to-end system performance improvement and
greater DRAM power reduction than presented
here.

\section{Summary}

In this chapter, we rigorously explore the complex tradeoff space associated
with DRAM retention profiling mechanisms using new experimental data from 368
modern LPDDR4 DRAM chips. In an effort to develop a rigorous understanding of
how retention failure profiling can be made viable for increasing the refresh
interval, we experimentally characterize DRAM chips at various temperatures and
refresh intervals and analyze the probability of failure of different cells in
DRAM as well as tradeoffs of retention time profiling. Following rigorous
analysis of the collected data, we propose \emph{reach profiling}, a technique
whose key idea is to profile DRAM at a longer refresh interval and/or a higher
temperature relative to the target refresh interval/temperature in order to
quickly discover an overwhelming majority of all possible failing cells at the
target conditions. We show that reach profiling enables significant system
performance improvement and DRAM power reduction, outperforming the brute-force
approach and enabling high-performance operation at longer refresh intervals
\textcolor{red}{that} were previously unreasonable to employ due to the high
associated profiling overhead. We conclude that reach profiling can be an
enabler for many past and future DRAM refresh reduction mechanisms.  We also
hope that the new experimental characterization and the analysis of the data
retention characteristics of modern LPDDR4 DRAM devices presented herein will
serve as an enabler for others to develop new techniques to tackle the difficult
yet critical problem of DRAM refresh.
\newcommand{\mpvone}[1]{{\color{black}#1}}
\newcommand{\mpvtwo}[1]{{\color{black}#1}}
\newcommand{\mpvthree}[1]{{\color{black}#1}}
\newcommand{\mpvfour}[1]{{\color{black}#1}}
\newcommand{\mpvfive}[1]{{\color{black}#1}}
\newcommand{\mpvsix}[1]{{\color{black}#1}}
\newcommand{\mpvseven}[1]{{\color{black}#1}}
\newcommand{\mpveight}[1]{{\color{black}#1}}
\newcommand{\mpvnine}[1]{{\color{black}#1}}
\newcommand{\minp}[1]{{\color{black}#1}}
\newcommand{\minptwo}[1]{{\color{black}#1}}

\chapter[EIN: Understanding and Modeling DRAM On-Die ECC]{Understanding and Modeling\\DRAM On-Die ECC}

\label{chap:ein}

In the previous chapter, we present out experimental study of DRAM
data-retention error characteristics in DRAM chips \emph{without} on-die ECC. In
this chapter, we extend such studies to DRAM chips \emph{with} on-die ECC.
\textbf{Our goal} in this chapter is to understand the relationship between pre-
and post-correction errors in a way that enables us to understand overcome the
barrier that on-die ECC poses for understanding DRAM pre-correction error
characteristics.

\section{Background and Motivation}

DRAM has long since been a crucial component in computing systems primarily
due to its low cost-per-bit relative to alternative memory
technologies~\cite{lee2013tiered, mutlu2013memory, nair2013archshield,
mutlu2014research}. However, while subsequent technology generations have
substantially increased overall DRAM capacity, they have not achieved
comparable improvements in performance, energy efficiency, and
reliability~\cite{chang2016understanding, mutlu2014research, lee2013tiered,
hassan2016chargecache}. This has made DRAM a significant performance
and energy bottleneck in modern systems~\cite{mutlu2013memory,
mutlu2014research}.

To address this challenge, researchers propose a wide variety of solutions
based on insights and understanding about DRAM behavior gleaned from
\emph{system-level DRAM error characterization} studies~\cite{baek2014refresh,
chandrasekar2014exploiting, chang2016understanding, chang2017understanding,
ditali2007x, ghose2018your, hassan2017softmc, henkels19914, hwang2012cosmic,
jung2014optimized, jung2015omitting, jung2016reverse, keller2014dynamic,
khan2014efficacy, khan2016case, khan2016parbor, kim2009new, kim2014flipping,
kim2018dram, kim2019d, kim2018solar, lee2015adaptive, lee2017design,
liu2012raidr, liu2013experimental, liu2014trustworthy, meza2015revisiting,
ohsawa1998optimizing, park2016experiments, park2016statistical,
patel2017reach, qureshi2015avatar, rahmati2015probable,
schaller2017intrinsic, schroeder2009dram, sridharan2012study,
sridharan2015memory, sutar2018d, talukder2018exploiting, talukder2018ldpuf,
tang2017dram, tehranipoor2016robust, tehranipoor2017investigation,
venkatesan2006retention, wang2018dram, weis2015thermal, xiong2016run}. These
studies deliberately induce errors in a DRAM device by experimentally
testing the device at conditions that exacerbate physical DRAM error mechanisms (e.g.,
charge leakage, circuit interference). The resulting errors directly reflect
the effects of the error mechanisms, providing researchers with insight into
the physical properties that underlie DRAM operation (e.g., data-retention,
circuit timings, data-pattern sensitivity). Researchers can then exploit these
insights to develop new mechanisms that improve DRAM and overall system
efficiency.

Unfortunately, continued DRAM technology scaling heralds grave reliability
concerns going forward primarily due to increasing single-bit error rates that
reduce manufacturing yield~\cite{lee2013tiered, nair2013archshield,
nair2016xed, micron2017whitepaper, kang2014co, hong2010memory,
gu2003challenges, meza2015revisiting, sridharan2012study, sridharan2015memory,
luo2014characterizing, schroeder2009dram}. While manufacturers traditionally
use redundant circuit elements \mpvseven{(e.g., rows, columns)} to repair
manufacturing faults~\cite{nair2013archshield, son2015cidra, kang2014co,
horiguchi2011nanoscale, mandelman2002challenges, gu2003challenges}, mitigating
growing single-cell error rates is no longer tractable using
\mpvseven{circuit-level redundancy} alone~\cite{micron2017whitepaper}.

To maintain desired yield targets, DRAM manufacturers have recently
supplemented circuit-level redundancy with \emph{on-die error correction codes
(on-die ECC)}\footnote{Also known as \emph{in-DRAM ECC} and \emph{integrated
ECC.}}~\cite{nair2013archshield, nair2016xed, micron2017whitepaper,
kang2014co, oh20153}. On-die ECC is \emph{completely invisible to the
system}~\cite{kang2014co, nair2016xed}: its implementation, encoding/decoding
algorithms, and metadata are all fully contained within the DRAM device and
provide no feedback about error detection and/or correction to the rest of the
system. On-die ECC is independent of any particular DRAM standard, and JEDEC
specifications do not constrain how the on-die ECC mechanism may be
designed~\cite{jedec2014lpddr4}. Since DRAM manufacturers primarily employ
on-die ECC to transparently improve yield, they do not publicly release the
ECC implementation details. Therefore, on-die ECC is typically not described
in DRAM device datasheets, and neither publications~\cite{kang2014co,
kalter199050, oh20153, kwak2017a, kwon2017an, chun201816gb} nor
whitepapers~\cite{micron2017whitepaper, im2016im} provide details of the ECC
mechanism for a given product.
\subsection{Motivation: On-Die ECC's Impact on Studies}
\label{ein:bgmot:ecc_impact_on_studies}

Unfortunately, on-die ECC has dire implications for DRAM error
characterization studies since it censors the true errors that result from
physical error mechanisms inherent to DRAM technology. For a device with
on-die ECC, we observe only \emph{post-correction} errors, which do not
manifest until \emph{pre-correction} error rates exceed the ECC's correction
capability. However, the way in which the ECC mechanism transforms a specific
uncorrectable error pattern is implementation-defined based on the mechanism's
design, which is \emph{undocumented}, \emph{proprietary}, \emph{opaque}, and
possibly \emph{unique} per product. Thus, on-die ECC effectively obfuscates
the pre-correction errors such that they \emph{cannot} be measured simply by
studying post-correction errors without knowing the ECC scheme.

\begin{figure}[h]
    \centering
    \includegraphics[width=0.7\linewidth]{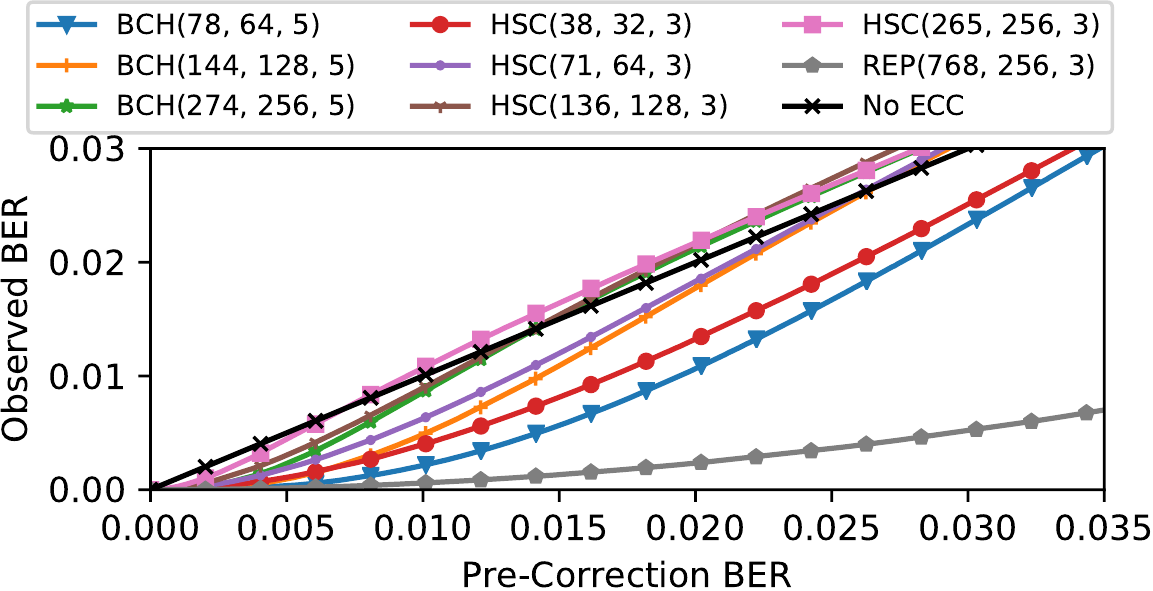}
    \caption{Observed vs. pre-correction data-retention bit error rate (BER)
    for various ECC schemes (color) and no ECC (black) assuming 256 data bits
    written with \texttt{RANDOM} data \mpvseven{(simulated)}.}
    \label{fig:simulated_ber}
\end{figure}

Figure~\ref{fig:simulated_ber} demonstrates the differences in the
\emph{observed} data-retention bit error rate (BER) (y-axis) for different
on-die ECC schemes (explained in Section~\ref{subsec:ecc_bg}) given the same
\emph{pre-correction} BER (x-axis). We generate this data in simulation using
\mpvsix{\simulator/}, which is described in detail in
Section~\ref{sec:simulation}. We see that the observed error rates are
\emph{dependent} on the particular ECC scheme used, and without knowledge of
which ECC scheme is used in a given device, there is no easy way to tie the
\emph{observed} error rates to the \emph{pre-correction} error rates.

This means that post-correction errors may not follow expected, well-studied
distributions based on physical error mechanisms (e.g., exponential
temperature dependence of charge leakage rates~\cite{hamamoto1998retention,
liu2013experimental, bacchini2014characterization}, lognormal retention-time
distributions~\cite{hamamoto1998retention, li2011dram, liu2012raidr,
patel2017reach}) but rather device-architecture-specific shapes that cannot
be reliably compared with those from a device with a different ECC scheme. We
discuss and experimentally demonstrate the implications of this observation in
Sections~\ref{sec:motivation} and~\ref{sec:proof_of_concept}, respectively.

\subsection{Overview of Error Inference (EIN)}

Thus, on-die ECC effectively \emph{precludes} studying DRAM error mechanisms,
motivating the need for a DRAM error characterization methodology that
isolates the effects of intrinsic DRAM behavior from those of the ECC
mechanism used in a particular device. To this end, \textbf{our goal in this
work} is to overcome the barrier that on-die ECC presents against
understanding DRAM behavior in modern devices with on-die ECC. To achieve this
goal, we develop \underline{E}rror-correction \underline{IN}ference
(\mechanism/), a statistical inference methodology that uses maximum a
posteriori (MAP) estimation to 1) reverse-engineer the ECC scheme and 2) infer
the \emph{pre-correction} error rates from only the observed
\emph{post-correction} errors. We follow a methodical four-step process:

\textbf{First,} we tackle the unique reverse-engineering problem of
determining the on-die ECC scheme \emph{without} any visibility into the
error-correction algorithm, the redundant data, or the locations of
pre-correction errors. Our approach is based on the \textbf{key idea} that
even though ECC obfuscates the exact locations of the pre-correction errors,
we can leverage known statistical properties of pre-correction error
distributions (e.g., \mpveight{uniform-randomness}~\cite{shirley2014copula,
baek2014refresh, patel2017reach, kim2018dram}) in order to disambiguate the
effects of different ECC schemes (Section~\ref{sec:prob_anal}).

We develop statistical models to represent how a given pre-correction error
distribution will be transformed by an arbitrary ECC scheme
(Section~\ref{sec:prob_model_ecc}). Our models are parameterized by \emph{i})
the \mpvseven{desired ECC scheme} and \emph{ii}) statistical properties of the
pre-correction error distribution. We then formulate the reverse-engineering
problem as a maximum a posteriori (MAP) estimation of the most likely model
given experimental data from real devices (Section~\ref{sec:bayesian_theory}).

\textbf{Second,} in order to compute several expressions in our statistical
models that are difficult to evaluate analytically, we develop
\simulator/\cite{eccsimgithub}, a flexible open-source simulator that
numerically estimates the error-detection, \mbox{-correction}, and
\mbox{-miscorrection} effects of arbitrary ECC schemes for
different pre-correction error distributions (Section~\ref{sec:simulation}).
\simulator/ models the lifetime of a given ECC dataword through the encoding,
error injection, and decoding processes faithful to how these steps would
occur in a real device (Section~\ref{subsec:high_level_sim_arch}).
\mpveight{To ensure that \simulator/ is applicable to a wide range of DRAM
devices and standards, we design \simulator/ to be modular and easily
extensible to additional error mechanisms and distributions.}
\textbf{Third,} we perform the \emph{first} experimental study of DRAM devices
with on-die ECC in open literature and demonstrate how
\mechanism/ infers both: \emph{i}) the on-die ECC scheme and
\emph{ii}) the pre-correction error rates. We study the data-retention
characteristics of 232 (82) state-of-the-art LPDDR4 DRAM devices with
(without) on-die ECC from one (three) major DRAM manufacturers across a wide
variety of temperatures, refresh rates, and test patterns. To accurately model
pre-correction errors in \simulator/, we first reverse-engineer:
\begin{itemize}
	\item The layout and dimensions of internal DRAM cell arrays.
	\item The locations and frequency distribution of redundant DRAM rows used
	for post-manufacturing repair.
\end{itemize}
Applying \mechanism/ to data from devices with on-die ECC, we:

\begin{itemize}
	\item Find that the on-die ECC scheme is a \mpveight{single-error
	correction Hamming code~\cite{hamming1950error} with ($n$ = 136, $k$ =
	128, $d$ = 3).}	
	\item Show that \mechanism/ can \mpvone{infer} \emph{pre-correction}
	error rates given only \emph{post-correction} errors.	

\end{itemize}

\textbf{Fourth,} we demonstrate \mechanism/'s usefulness by providing a
proof-of-concept experimental characterization study of the data-retention
error rates for the DRAM devices with on-die ECC. We test across different
refresh intervals and temperatures \mpveight{to} show that \mechanism/
effectively enables inferring the pre-correction error rates, which, unlike
the ECC-obfuscated post-correction error rates, follow \mpveight{known shapes
that result from well-studied device-independent error mechanisms.}

\subsection{Applicability of EIN}

\mechanism/ allows researchers to more \mpvsix{holistically study the
reliability characteristics of DRAM devices with on-die ECC by exposing the
pre-correction error rates beneath the observed post-correction errors. This
enables researchers to propose new ideas based on a more general understanding
of DRAM devices. To demonstrate how \mechanism/ may be useful, we provide: 1)
several examples of studies and mechanisms that \mechanism/ enables and 2) a
discussion about the implications of continued technology scaling for future
error characterization studies.}

\subsection{\mpvseven{Example Applications}}

\label{subsec:error_char}

We provide several examples of potential studies and mechanisms that are
enabled by knowing pre-correction error rates, which on-die ECC masks and
\mechanism/ reveals:
\begin{itemize}
\item \textit{Runtime Error Rate Optimization}: A mechanism that intelligently
adjusts operating timings/voltage/frequency to \mpveight{meet dynamically
changing}  system performance/energy/reliability targets (e.g.,
Voltron~\cite{chang2017understanding}, AL-DRAM~\cite{lee2015adaptive},
\mpvseven{AVA-DRAM~\cite{lee2016reducing}, DIVA-DRAM~\cite{lee2017design}})
\mpveight{typically} needs to profile the error characteristics of a device
for runtime decision-making. If the particular ECC scheme is known (e.g.,
using \mechanism/), such a mechanism can leverage device-independent DRAM
error models for decision-making and quickly interpolating or extrapolating
``safe'' operating points rather than having to: (1) use complex, likely
non-parametric, device-specific models for each supported ECC scheme or (2)
characterize each device across its entire region of operation.
\item \mpvone{\textit{Device Comparison Studies}: \mechanism/ enables fair
comparisons of DRAM error characteristics between devices with different (and
without) on-die ECC mechanisms. This is useful for studying the evolution of
error characteristics over time, which provides insight into the state of the
industry and future technology trends. With DRAM error rates continuing to
worsen (discussed in Section~\ref{subsec:future_of_characterization}), such
studies can help predict how much worse future devices may be and how well
current/future error-mitigation mechanisms will cope. }

\item \textit{Reverse-Engineering Other ECCs}: As \mpvtwo{we discuss in}
Section~\ref{subsec:application_to_other_tech}, \mechanism/ \mpveight{is
applicable} to \mpvsix{other systems (e.g., rank-level ECC, Flash memory)
whose ECC schemes are typically also proprietary. Reverse-engineering their
ECC schemes can be useful for various reasons~\cite{courbon2016reverse,
cai2017error, fukami2017improving, wise2018reverse, cojocar2019exploiting},
including failure analysis, security evaluation, forensic analysis, patent
infringement, and competitive analysis. For these systems, \mechanism/ may
provide a way to reverse-engineer the ECC scheme without requiring hardware
intrusion or internal access to the ECC mechanism as typically required by
previous approaches~\cite{van2015mathematical, van2017bit, wise2018reverse,
cojocar2019exploiting} (discussed in Section~\ref{related:subsec:rev_engr_ecc}).}
\end{itemize}
\mpvone{We hope that future work will use \mechanism/ well beyond these
use cases} and will develop new characterization-driven understanding of
devices with on-die ECC.
\subsection{\mpvone{Applicability to Future Devices}}
\label{subsec:future_of_characterization}

\mpvfour{Despite its energy and reliability benefits, on-die ECC} \mpvtwo{does
not fundamentally prevent error rates from increasing. Therefore, future DRAM
devices \mpvsix{may} require \emph{stronger} error-mitigation solutions,
\emph{further obfuscating} pre-correction error rates and making error
characterization studies even more difficult.}

\mpvnine{Similarly}, other memory technologies (e.g.,
Flash~\cite{cai2012error, cai2017error}, STT-MRAM~\cite{ishigaki2010multi,
zhang2012multi, kultursay2013evaluating}, PCM~\cite{qureshi2009scalable,
lee2009architecting, lee2010phasecacm, wong2010phase, seong2013tri},
Racetrack~\cite{zhang2015hi}, RRAM~\cite{wong2012metal}) suffer from
\mpvnine{ongoing} reliability concerns, and characterizing their error
mechanisms requires surmounting any error-mitigation techniques they use.
\mechanism/ takes a first step towards enabling a holistic understanding of
devices whose error characteristics are not directly visible, and we hope that
future work leverages this opportunity to develop new mechanisms to tackle the
reliability challenges that lie ahead.

\section{\mpvseven{Statistically Modeling DRAM and ECC}}
\label{sec:prob_anal}

We begin by formalizing the relationship between pre- and post-correction
\mpvtwo{error distributions and expressing reverse-engineering as a
\mpvfive{maximum a posteriori (MAP)} estimation problem.} Our approach is
grounded on the key idea that \mpvtwo{pre-correction errors arise from
physical error mechanisms with known statistical properties, and because
different ECC schemes transform these distributions in different ways, we can
use what we know about both the pre- and post-correction error distributions
to disambiguate different ECC schemes. \mpvfive{This section provides a
step-by-step derivation of \mechanism/, the statistical inference methodology
we propose in this work}.}

\subsection{\minptwo{Statistically Modeling Error Correction Codes}}
\label{sec:prob_model_ecc}

\begin{figure}[h]
    \centering
    \includegraphics[width=0.7\linewidth]{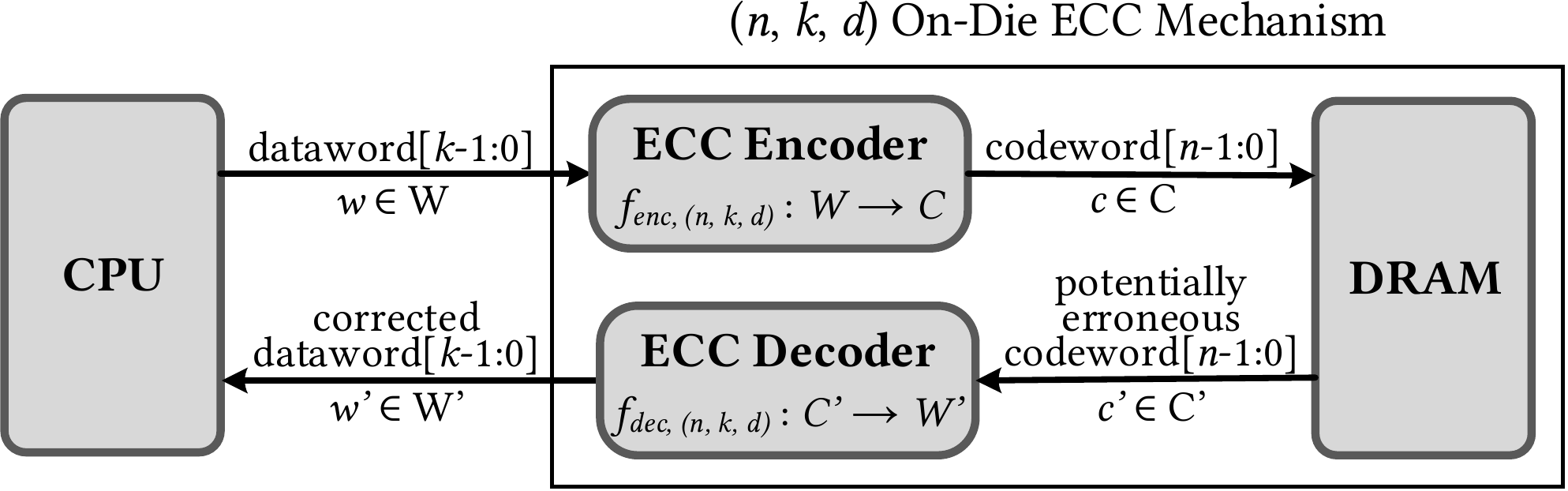}
    \caption{Illustration of an on-die ECC mechanism implementing an $(n, k,
    d)$ binary block code.}
    \label{fig:ecc_impl} 
\end{figure}

Consider an ECC mechanism implementing an $(n, k, d)$ binary block code as
illustrated in Figure~\ref{fig:ecc_impl}. The ECC \emph{encoding} algorithm
$f_{enc,~ ECC}$ transforms a dataword $w$ out of the set of all possible
datawords $\mathcal{W} = \mathbb{Z}_2^k$ into a valid codeword $c$ out of the
set of all possible valid codewords $\mathcal{C} \subset \mathbb{Z}_2^n$.
Likewise, the \emph{decoding} algorithm $f_{dec,~ECC}$ transforms a codeword
$c'$ \mpvsix{(potentially invalid due to errors)} out of the set of all
possible codewords $\mathcal{C'} = \mathbb{Z}_2^n$ into a corrected dataword
$w'$ out of the set of all possible corrected datawords $\mathcal{W'} =
\mathbb{Z}_2^k$.

$f_{dec,~ECC}$ can be thought of as a deterministic mapping\footnote{While
non-deterministic encoding/decoding algorithms exist, they are
\mpvone{typically not} used with the simple ECCs found in DRAM. If more
complex ECCs must be considered (e.g., LDPC~\cite{costello1982error,
cai2017error}), our models can be extended to treat the encoding/decoding
functions as probabilistic transformations themselves.} from the finite set of
inputs $\mathcal{C'}$ to a finite set of outputs $\mathcal{W'}$:
\begin{equation}
f_{dec,~ECC}~:~\mathcal{C'} \mapsto \mathcal{W'}
\end{equation}
\noindent
This means that for a particular ECC scheme $f_i$, the probability of
observing output $w'$ is determined by the probabilities that its
corresponding inputs $\{c_j' \in
\mathcal{C'},~\forall j : f_{dec,~i}(c_j') = w'\}$ occur:
\begin{equation}
\label{eqn:relative_frequency_distributions}
\mathrm{P}_{f_{i}}[w'] = \sum_{\forall j~:~{f_{dec,~i}(c_j') = w'}}\mathrm{P}[c_j']
\end{equation}
\noindent
From this perspective, if we know both 1) the ECC scheme $f_i$ and 2) the
frequency distribution of all possible \emph{input} values $c_{k}$, we can
calculate the corresponding distribution of all possible \emph{output}
values.\footnote{Note that since ECC decoding is generally \emph{not}
injective (i.e., multiple codewords may map to a single decoded dataword), we
cannot determine exactly which input produced an observed output.} Inverting
this relationship, if we \emph{experimentally measure} the frequency
distribution of output values from a real device, we can determine the
probability of having made such observations given 1) a suspected ECC scheme
and 2) an expected frequency distribution of all possible inputs.
Section~\ref{subsec:prob_model} \mpvseven{describes} what we know about
pre-correction error distributions and how we leverage this knowledge to
disambiguate different suspected ECC schemes.

\subsection{Experimental Observables}
\label{sec:observables}
Solving Equation~\ref{eqn:relative_frequency_distributions} requires measuring
the relative frequency distribution of post-correction datawords (i.e.,
$w_i'$). For example, if we use 64-bit datawords, we have $2^{64}$
\emph{unique} datawords. Unfortunately, a single DRAM device has on the order
of \emph{millions} of datawords, which is nowhere near enough to obtain a
representative sample of the full distribution.

Instead, we divide $\mathcal{W'}$ into $N + 1$ subsets, $\mathcal{W}_n'$,
which each comprise all possible datawords with $n \in [0, N]$ errors.
\mpveight{Using this approach}, a relative frequency distribution of the
$\mathcal{W}_n'$ contains only $N + 1$ categories, and even a single DRAM
device contains more than enough samples to obtain a representative
distribution. Experimentally, measuring the number of errors in each dataword
simply requires counting errors.\footnote{There is no fundamental reason for
this choice beyond experimental convenience; if another choice is made, our
analysis still holds but will need to be modified to accommodate the new
choice of $\mathcal{W}_n'$.} We can then rewrite
Equation~\ref{eqn:relative_frequency_distributions} in terms of the subsets
$\mathcal{W}_n'$:
\begin{equation}
\label{eqn:observables}
\mathrm{P}_{f_i}[w' \in \mathcal{W}_n'] = \sum_{\forall
j~:~f_{dec,~i}(c_j') \in \mathcal{W}_n'}\mathrm{P}_{f_i}[c_j']
\end{equation}
\noindent

Unfortunately, this approach requires knowing the exact layout of ECC words in
memory. This may be difficult since multiple bits are read/written together at
the granularity of a \emph{burst} (Section~\ref{subsec:dram_organization}),
and each burst may contain one or more ECC words with an arbitrary
bit-to-ECC-word mapping.

To circumvent this problem, we instead consider the probability of observing
$n$ errors \emph{per burst}, where each burst comprises dataword(s) from one
or more ECC schemes. Mathematically, the total number of \mpveight{errors in a
burst is the sum of the} individual per-dataword error counts and is
\mpveight{computed by convolving the per-dataword error-count distributions.
Counting errors at burst-granularity} is independent of the layout of ECC
words within a burst assuming that ECC words are contained within burst
boundaries so that bursts can be read/written independently. However, if a
different design is suspected, even longer words (e.g., multiple bursts) may
be used as necessary.
\subsection{Statistically Modeling DRAM Errors}
\label{subsec:prob_model}

To estimate the relative frequencies of the \emph{pre-correction} codewords
$c_j' \in \mathcal{C'}$, we exploit the fact that errors arise from physical
phenomena that follow well-behaved statistical distributions. Throughout this
work, we focus on \mpvseven{data-retention} error distributions since they are
well-studied and are easily reproduced in experiment. However, \mechanism/
\mpveight{is applicable} to \emph{any} \mpveight{experimentally-reproducible
error distribution whose statistical properties are well-understood}
(e.g., reduced activation-latency~\cite{chang2016understanding, lee2017design,
kim2018solar, lee2015adaptive, kim2018dram, kim2019d}, reduced
precharge-latency~\cite{chang2016understanding, talukder2018ldpuf,
talukder2018exploiting}, reduced voltage~\cite{chang2017understanding},
RowHammer~\cite{kim2014flipping, mutlu2017rowhammer, mutlu2019rowhammer}).

As described in Section~\ref{bg:subsec:data_retention}, data-retention errors
occur when a charged cell capacitor leaks enough charge to lose its stored
value. \mpvseven{This represents a ``1'' to ``0'' error for a charged true-cell
(i.e., programmed with data ``1'')}, and vice-versa for an
anti-cell~\cite{liu2013experimental, kim2014flipping, cojocar2019exploiting}. Due
to random manufacturing-time variations~\cite{desai2012process,
zhang2015exploiting, li2011dram, kim2009new, lee2015adaptive, lee2016reducing,
kim2014flipping}, certain cells are more prone to data-retention errors than
others~\cite{liu2013experimental, venkatesan2006retention,
hamamoto1998retention, liu2012raidr, kang2014co, kim2009new}. Furthermore,
absolute data-retention error rates depend on operating conditions such as
refresh timings, data patterns, ambient temperature, and supply voltage. Through
extensive error characterization studies, prior works find that, for a fixed set
of testing conditions (e.g., $t_{REFW}$, temperature), data-retention errors
show no discernible spatial patterns~\cite{hamamoto1998retention, liu2012raidr,
baek2014refresh, venkatesan2006retention, shirley2014copula} and can be
realistically modeled as uniform-randomly distributed~\cite{kim2018dram,
shirley2014copula, baek2014refresh} independent events~\cite{shirley2014copula}.

To model an arbitrary pre-correction error distribution in our
\mpvtwo{analysis}, we introduce an abstract model parameter $\theta$ that
encapsulates all state necessary to describe the distribution. In general,
$\theta$ is a \emph{set} of two key types of parameters: \emph{i})
experimental testing parameters (e.g., data pattern, timing parameters,
temperature) and \emph{ii}) device microarchitectural characteristics (e.g.,
spatial layout of true- and anti-cells). We incorporate $\theta$ into our
analysis as a dependency to the terms in Equation~\ref{eqn:observables}:
\begin{equation}
\label{eqn:observables_parameterized}
\mathrm{P}_{f_i,\theta}[w' \in \mathcal{W}_n'] = \sum_{\forall
j~:~f_{dec,~i}(c_j') \in \mathcal{W}_n'}\mathrm{P}_{f_i, \theta}[c_j']
\end{equation}
\noindent
Ideally, \emph{all} of the parameters that comprise $\theta$ are known at
testing time. Unfortunately, experiments are often imperfect, and internal
device characteristics are difficult to obtain without proprietary knowledge
or laborious reverse-engineering. If such parameters are unknown, \mpvtwo{we
can infer them alongside} the unknown ECC scheme.\footnote{\mpt{While we
could lump the unknown ECC scheme into $\theta$ as an unknown
microarchitectural characteristic, we keep it logically separate since
$\theta$ represents what we already understand about DRAM devices, and the
unknown ECC scheme represents what we do not.}}

In this work, \mpvfour{we model \mpvsix{data-retention} errors as
uniform-random, independent events among cells programmed to the ``charged''
state with a fixed probability determined by testing conditions.}
\mpvone{$\theta$ then encapsulates \emph{i}) the single-bit error probability,
called the \emph{raw bit error rate (RBER)}, \emph{ii}) the programmed data
pattern, and \emph{iii}) the spatial layout of true-/anti-cells.}

Unfortunately, evaluating Equation~\ref{eqn:observables_parameterized}
analytically is difficult even for \mpvsix{data-retention} errors due to the
complexity of the interactions between the ECC scheme and the parameters
encompassed by $\theta$. Instead, we numerically estimate the solution to
Equation~\ref{eqn:observables_parameterized} using Monte-Carlo simulation as
described in Section~\ref{sec:simulation}. This approach allows our analysis
to flexibly take into account arbitrarily complex model parameters (e.g.,
detailed microarchitectural characteristics, nontrivial error models).

\subsection{\mpvseven{Inferring the Model Parameters}}
\label{sec:bayesian_theory}
We now formulate the reverse-engineering task as a maximum a posteriori (MAP)
estimation problem over a set $\mathcal{F}$ of hand-selected ECC schemes that
are either directly mentioned in context with on-die ECC (HSC(71, 64,
3)~\cite{im2016im, issi2018ddr3, nair2016xed} and HSC(136, 128,
3)~\cite{kalter199050, micron2017whitepaper, kwak2017a, kwon2017an,
chun201816gb}) or are used as demonstrative examples of applying our
methodology to devices with stronger and/or more complicated codes (e.g.,
BCH($n$, $k$, $d$), HSC($n$, $k$, $d$), REP(3, 1, 3)). Note that we also take
into account implementation details of each of these schemes (e.g., systematic
vs. non-systematic encodings) using our simulation infrastructure as we
describe in Section~\ref{subsec:estimating_relative_freqs}.
To reverse-engineer the unknown ECC scheme $f_{\mathrm{unknown}}$, we start by
expressing it as the \emph{most likely} ECC scheme out of all possible schemes
$f_i \in \mathcal{F}$ given a set of observations $\mathcal{O}$:
\begin{equation}
\label{eqn:high_level_problem}
f_{\mathrm{unknown}} = \argmax_{f_i}(\mathrm{P}[f_i~|~\mathcal{O}])
\end{equation}
\noindent
\mpvone{Unfortunately, we cannot directly evaluate
Equation~\ref{eqn:high_level_problem} since our observations $\mathcal{O}$ are
measured from a device with a \emph{fixed} ECC scheme. Instead, we use
\mpvfour{the} Bayes theorem to express Equation~\ref{eqn:high_level_problem}
in terms of the probability of obtaining measurements $\mathcal{O}$ given an
\emph{arbitrary} ECC scheme $f_i$, which we can calculate using the
relationship in Equation~\ref{eqn:observables}. This yields:}
\begin{equation}
\begin{split}
\label{eqn:bayes}
f_{\mathrm{unknown}} &= \argmax_{f_i}\left(\frac{\mathrm{P}[\mathcal{O}~|~f_i]
            \mathrm{P}[f_i]}{\mathrm{P}[\mathcal{O}]}\right) \\
            &= \argmax_{f_i}\left(\mathrm{P}[\mathcal{O}~|~f_i]
            \mathrm{P}[f_i]\right)
\end{split}
\end{equation}
\noindent
Note that we ignore the denominator (i.e., the marginal likelihood) in
Equation~\ref{eqn:bayes} because \mpvone{it is a fixed scale factor
independent of $f_i$ and does not affect the maximization result.}

We assume a \mpvseven{uniformly-distributed} prior (i.e., $\mathrm{P}[f_i]$)
given that we cannot guarantee anything about the on-die ECC implementation.
By restricting our analysis to only the aforementioned ECC schemes, we already
exclude any schemes that we consider to be unrealistic. In principle, we could
assign greater or lower probability mass to schemes that have been mentioned
in prior work or that are exceedingly expensive, respectively, but we choose
not to do so because \emph{i}) we cannot guarantee that the devices we test
are similar to those mentioned in prior work, and \emph{ii}) we want to
demonstrate the power of our methodology without biasing the results towards
any particular ECC schemes.
The likelihood function (i.e., $\mathrm{P}[\mathcal{O}~|~f_i]$) incorporates
the experimental data we obtain from real devices. As we show in
Section~\ref{sec:observables}, our measurements provide us with the
probability of observing an $n$-bit error in each of $j$ independent DRAM
bursts. Defining $N$ as a random variable representing the number of erroneous
bits observed in a single burst and assuming observations are independent
events (validated in Section~\ref{subsec:prob_model}), we rewrite the
likelihood function as:
\begin{equation}
\label{eqn:likelihood}
\mathrm{P}[\mathcal{O}~|~f_i] = \mathrm{P}_{f_i,\theta}\left[\bigcap_{j = 0}^{j_{max}} N =n_j \right] = \prod_{j = 0}^{j_{max}} \mathrm{P}_{f_i,\theta}[N =n_j ] \\
\end{equation}
\noindent
\mpvone{This is essentially a multinomial probability mass function (PMF)
evaluated at $\mathcal{O}$, where each probability mass is computed using
Equation~\ref{eqn:observables_parameterized}. Unfortunately, as described in
Section~\ref{subsec:prob_model}, the model parameter $\theta$ encapsulates the
\emph{pre-correction} error rate, which we do not know and cannot measure
post-correction. Therefore, for each ECC scheme $f_i$, we first
maximize the likelihood distribution over $\theta$:}
\begin{equation}
\label{eqn:best_fit}
\mathrm{P}[\mathcal{O}~|~f_i] = \max_{\theta}\left(\prod_{j = 0}^{j_{max}} \mathrm{P}_{f_i,\theta}[N =n_j ]\right)
\end{equation}

Inserting the result of Equation~\ref{eqn:best_fit} into our original
optimization objective (Equation~\ref{eqn:bayes}), we obtain the final
objective function to optimize in order to reverse-engineer the ECC scheme
$f_{\mathrm{unknown}}$ used in our devices:
\begin{equation}
\label{eqn:final_opt_problem}
f_{\mathrm{unknown}} = \argmax_{f_i}\left(\max_{\theta}\left(\prod_{j = 0}^{j_{max}} \mathrm{P}_{f_i,\theta}[N =n_j ]\right)
            \mathrm{P}[f_i]\right)
\end{equation}
\noindent
where the inner product term is calculated using
Equation~\ref{eqn:observables_parameterized}.

\mpvone{After the ECC scheme is reverse-engineered, we can repeatedly} apply
Equation~\ref{eqn:best_fit} to solve for $\theta$ across many different
experiments (i.e., observations). Since $\theta$ represents all parameters
necessary to describe the pre-correction error distribution (described in
Section~\ref{subsec:prob_model}), this is equivalent to reverse-engineering
the pre-correction error rate. With the ECC scheme known as $f_\mathrm{known}$,
Equation~\ref{eqn:best_fit} simplifies to:
\begin{equation}
\label{eqn:final_underlying_theta}
 \theta_{\mathrm{unknown}} = \argmax_{\theta}\left(\prod_{j = 0}^{j_{max}} \mathrm{P}_{f_\mathrm{known},\theta}[N =n_j ]\right)
\end{equation}
With Equations~\ref{eqn:final_opt_problem}
and~\ref{eqn:final_underlying_theta}, we can reverse-engineer both \emph{i})
the ECC scheme and \emph{ii}) the pre-correction error rates from observed
post-correction errors for \emph{any} DRAM device whose error distributions
are obscured \mpveight{by} ECC. In Section~\ref{subsec:rev_engr_ecc_scheme},
we experimentally demonstrate how to apply
Equation~\ref{eqn:final_opt_problem} to real devices with on-die ECC.

\section{Simulation Methodology}
\label{sec:simulation}

To apply \mechanism/ to data from real devices, we develop and publicly
release \simulator/~\cite{eccsimgithub}, a flexible open-source C++-based
simulator that models the life of a dataword through the entire ECC
encoding/decoding process. \mpveight{\simulator/ accounts for different ECC
implementations and pre-correction error characteristics to ensure that
\mechanism/ is applicable to a wide variety of DRAM devices and standards.
This section describes \simulator/'s extensible design and explains how
\simulator/ can be used to solve the optimization problems formulated in
Section~\ref{sec:prob_anal}.}
\subsection{\simulator/ High-Level Architecture}
\label{subsec:high_level_sim_arch}

\begin{figure}[h]
    \centering
    \includegraphics[width=0.7\linewidth]{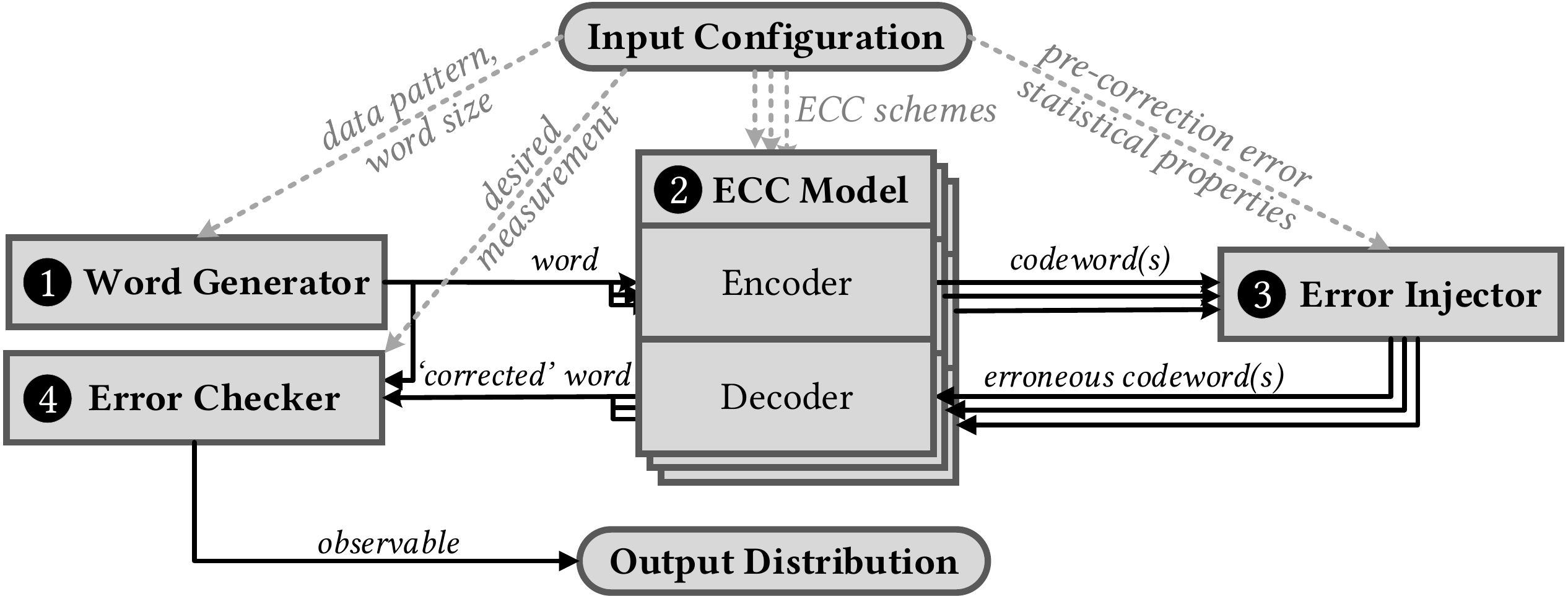}
    \caption{High-level block diagram showing the logical flow of data through
    the different components in our simulator.}
    \label{fig:sim_block_diagram} 
\end{figure}

Figure~\ref{fig:sim_block_diagram} shows a high-level diagram of the logical
flow of data through \simulator/'s different components. To model a DRAM
experiment, we simulate many individual burst-length accesses that each access
a different group of cells. Each burst simulates an experimental measurement,
yielding a distribution of measured values across all simulated bursts. We
describe the function of each simulator component.

\noindent
\circled{1} \textbf{Word generator} constructs a bitvector using commonly tested data
patterns (e.g., \texttt{0xFF}, \texttt{0xAA},
\texttt{RANDOM)}~\cite{adams2002high, liu2013experimental, khan2014efficacy,
khan2016parbor, weis2015retention, lee2017design, patel2017reach}, simulating
the data written to DRAM.

\noindent
\circled{2} \textbf{ECC model} encompasses an ECC implementation, including the
encoding/decoding algorithms and implementation details such as systematic vs.
non-systematic encodings. \mpvseven{Because a single word from the word
generator may comprise multiple ECC datawords, \simulator/ provides a
configurable mapping for decomposing the word into ECC datawords.}

\noindent
\circled{3} \textbf{Error injector} injects errors into a given codeword according to a
configurable error distribution. We implement support for data-retention
errors as described in Section~\ref{subsec:prob_model}. We provide
configurable parameters for the spatial distribution of true-/anti-cells
(e.g., alternating \mpvone{per row}) and the single-bit probability of failure
(i.e., RBER). Errors are injected uniform-randomly across each bit that can
fail (i.e., each ``charged'' cell per the chosen true-/anti-cell layout and
data pattern) using a Bernoulli distribution with $p$ equal to the desired
RBER normalized by the ratio of all cells that can fail, which ensures that
the simulated error rate meets the target RBER on average.

\noindent
\circled{4} \textbf{Error checker} takes the \mpt{pre- and post-correction words as inputs
and calculates a user-defined measurement (e.g., total number of bit-flips).
This corresponds to an experimental observable as explained in
Section~\ref{sec:observables}.}

\subsection{\simulator/ Validation}
\label{subsec:sim_validation}
We validate \simulator/ using a combination of manual and automatic unit
tests. For the ECC model, we 1) provide tests for detecting/correcting the
right amount of errors (exhaustively/sample-based for short/long codes); 2)
hand-verify the inputs/outputs of encoders/decoders where reasonable; 3)
hand-validate the generator/parity-check matrices and/or code generator
polynomials against tables of known values (e.g.,~\cite{clark2013error}); and
4) validate the minimum distance and weight distributions of codewords. Due to
the simplicity of how we model the true-/anti-cell layout and data-retention
errors, we validate the error-injection correctness by 1) manual inspection
and 2) using summary statistics (e.g., distribution of errors across many
simulated bursts).
\subsection{\mpvseven{Applying \simulator/ to Experimental Data}}
\label{subsec:estimating_relative_freqs}
\mpvnine{To analyze data taken from a real experiment, we configure the
simulation parameters to match the experiment and} simulate enough read
accesses (e.g., \textgreater$10^5$) to allow the distribution of simulated
measurements to numerically estimate the real experimental measurements. This
approach effectively solves Equation~\ref{eqn:observables_parameterized}
through Monte-Carlo simulation \mpvnine{for any model parameters \{$f_i$,
$\theta$\} that can be simulated using \simulator/}.
Figure~\ref{fig:simulated_ber} in Section~\ref{ein:bgmot:ecc_impact_on_studies}
provides several examples of evaluating
Equation~\ref{eqn:observables_parameterized} across a wide range of model
parameters \{$f_i$, $\theta$\} using a 256-bit input word programmed with a
\texttt{RANDOM} data pattern. The X-axis shows the \emph{pre-correction} bit
error rate (BER), i.e., the RBER component of $\theta$, and the Y-axis shows the
\emph{observed} BER, which is computed by taking an expectation value over the
distribution resulting from solving
Equation~\ref{eqn:observables_parameterized}. Curves represent different ECC
schemes $f_i$, and each data point represents one simulation of $10^6$ words,
subdividing each word into multiple ECC datawords as necessary.

We see that each ECC scheme transforms the pre-correction error rate
differently. For example, stronger codes (e.g., REP(768, 256, 3), BCH(78, 64,
5)) dramatically decrease the observed BER, whereas weaker codes (e.g.,
HSC(265, 256, 3)) have a relatively small effect. Interestingly, we see that
many of the codes actually \emph{exacerbate} the error rate at high enough
pre-correction error rates \mpvone{because, on average, the decoder mistakenly
``corrects'' bits without errors more often than not}. These examples
demonstrate that different ECC schemes have different effects on the
pre-correction error distribution, and Equation~\ref{eqn:final_opt_problem}
exploits these differences to disambiguate schemes.
\subsection{Inferring the Model Parameters}
\label{subsec:determining_map_params}
To infer the model parameters $f$ and $\theta$, which represent the ECC scheme
and pre-correction error distribution characteristics, respectively, we need
to perform the optimization given by Equation~\ref{eqn:final_opt_problem}. We
do this using a grid search across $f$ and $\theta$, simulating $10^4$
uniformly-spaced error rates for each of several different ECC schemes, data
patterns and true-/anti-cell layouts. While a denser grid \mpveight{may
improve} precision, this configuration sufficiently differentiates the models
we analyze (Section~\ref{subsec:rev_engr_ecc_scheme}).

\mpvone{The solutions to Equation~\ref{eqn:final_opt_problem} are the inferred
ECC scheme and pre-correction error distribution characteristics} that best
explain the experimental observations. From there, we can use
Equation~\ref{eqn:final_underlying_theta} evaluated with the known ECC scheme
in order to determine $\theta$ for any \emph{additional} experiments that we
run (e.g., different error rates).
\subsection{Inference Accuracy}
\label{subsec:accuracy}

MAP estimation rigorously selects between known models and inherently can
neither confirm nor deny whether the MAP estimate is the ``real'' answer. We
identify this as a limitation of \mechanism/ in
Section~\ref{subsec:limitations}. However, in the event that a device uses a
scheme that is not considered in the \mpvsix{MAP estimation}, it would be
evident when testing across different experimental conditions and error rates
since it is unlikely that any of the chosen ECC schemes would be the single
maximum-a-posteriori scheme (i.e., best explaining the observed data) across
all \mpvseven{experiments}.

We can also use confidence intervals to gauge the error in each MAP estimate.
This requires repeating the MAP estimation over $N$ bootstrap
samples~\cite{efron1992bootstrap} taken from the observed data $\mathcal{O}$.
The min/max or 5$^\mathrm{th}$/95$^\mathrm{th}$ percentiles are typically
taken to be the confidence bounds.
\subsection{\mpveight{Applying \mechanism/} to Other Systems}
\label{subsec:application_to_other_tech}

\mechanism/ can be extended to any ECC-protected communication channel
provided that we can induce uncorrectable errors
\mpvfour{whose pre-correction spatial distribution follows some known}
property (e.g., uniform-randomness). Examples include, but are not limited to,
DRAM rank-level (i.e., DRAM-controller-level) ECC and other memory
technologies (e.g., SRAM, Flash, Phase-Change Memory).

\subsection{\mpveight{Applying \mechanism/} to Data from Prior Studies}
\label{subsec:impact}

\mechanism/ \mpveight{is applicable} to data presented in a prior study if the
study supplies enough information to solve
Equation~\ref{eqn:final_underlying_theta}. This requires that the study
provides both: 1) the pre-correction error characteristics, either directly as
statistical distributions or implicitly through \mpvnine{the experimental
methodology} (e.g., device model number, tested data patterns) and 2) the
distribution of errors amongst post-correction words as discussed in
Section~\ref{sec:observables}. If these are known, \mechanism/ \mpveight{can}
infer both the ECC scheme and the pre-correction error rates from the given
data.
\subsection{Limitations of \mechanism/}
\label{subsec:limitations}

\mechanism/ has three \mpvone{main} limitations. However, in practice, these
limitations do not hurt its usability since both DRAM and ECC design are
mature and well-studied topics. We discuss each limitation individually:
\begin{enumerate}
\item \emph{Cannot guarantee success or failure.} As described in
Section~\ref{subsec:accuracy}, MAP estimation cannot guarantee whether the
correct solution has (not) been found. However, Section~\ref{subsec:accuracy}
describes how testing across different operating conditions and using
confidence intervals helps mitigate this limitation.

\item \emph{Requires knowledge and control of errors.} \mpvone{Using
\mechanism/ requires \mbox{\emph{i}) knowing} statistical properties of the
spatial distribution of pre-correction errors, and \emph{ii}) the ability to
induce uncorrectable errors. Fortunately, \mechanism/ can use \mpvnine{any one
of the many} well-studied, easily-manipulated error mechanisms that are
fundamental to DRAM technology (e.g., data retention, RowHammer,
reduced-latency access; see Section~\ref{subsec:prob_model}). Such mechanisms
are unlikely to change dramatically for future devices (e.g., retention errors
are modeled similarly across decades of DRAM
technologies~\cite{hamamoto1995well, hiraiwa1996statistical,
lieneweg1998assesment, hamamoto1998retention, mori2001new, weber2005data,
kong2008analysis, cho2017novel}), which means that
\mechanism/ will likely continue to be applicable.}

\item \emph{Cannot identify bit-exact error locations.} \mpvone{While
\mechanism/ infers pre-correction error \emph{rates}, it cannot determine the
\emph{bit-exact locations} of pre-correction errors.} Unfortunately, since
multiple erroneous codewords may map to each visible dataword, we are not
aware of a way to infer error \mpvone{locations} without insight into the
exact ECC implementation (e.g., algorithms, redundant data). However,
inferring error rates is sufficient to study aggregate distributions, and we
leave error \mpvone{localization} to future work.
\end{enumerate}

\section{Experimental Methodology} 
\label{sec:methodology}

We experimentally characterize 232 LPDDR4~\cite{jedec2014lpddr4} DRAM devices
\emph{with} on-die ECC from a single major DRAM manufacturer that we cannot
disclose for confidentiality reasons. For comparison purposes, we test 82
LPDDR4 DRAM devices of the previous technology generation \emph{without}
on-die ECC from across three major DRAM manufacturers. Given that DRAM
manufacturers provide \mpvseven{neither:} \emph{i}) non-ECC counterparts of
devices with on-die ECC nor \emph{ii}) a mechanism by which to disable on-die
ECC, the older-generation devices provide our closest point of comparison.
We perform all testing using a home-grown infrastructure that provides precise
control over DRAM timing parameters, bus commands, and bus addresses. Our
infrastructure provides reliable ambient temperature control between
40$^{\circ}$C - 55$^{\circ}$C with a tolerance of of $\pm 1^{\circ}$C. To
improve local temperature stability for each DRAM device throughout testing, a
local heating source maintains DRAM at 15$^\circ$C above the ambient
temperature at all times, providing an effective DRAM temperature testing
range of 55$^{\circ}$C - 70$^{\circ}$C.

\section{Experimentally Inferring On-Die ECC \\ and Pre-Correction
Error Rates Using \mechanism/}
\label{sec:app_to_our_data}

In this section, we apply \mechanism/ to infer the \emph{i}) on-die ECC scheme
and \emph{ii}) pre-correction error rates of real devices with on-die ECC.
Before doing so, we validate our uniform-random statistical model for
pre-correction errors and determine the layout of true-/anti-cells to
accurately model the pre-correction error distribution of the devices that we
test.

\subsection{\mpvseven{Validating Uniform-Random Retention Errors}}
\label{subsec:validate_ret_err_dist}

Our model for data-retention errors (Section~\ref{subsec:prob_model}) treats
errors as independent, uniform-randomly distributed events based on
observations made in several prior works~\cite{hamamoto1998retention,
liu2012raidr, baek2014refresh, venkatesan2006retention, shirley2014copula,
kim2018dram}. \mpvseven{For such errors}, the total number of errors $X$
in each fixed-length $n$-bit region of DRAM follows a binomial
distribution~\cite{patel2017reach, baek2014refresh, venkatesan2006retention}
parameterized by the RBER $R$:
\begin{equation}
\mathrm{P}[X=x~|~R] = {n\choose x}R^{x}(1-R)^{n-x} \\
\end{equation}

\begin{figure}[h]
    \centering
    \includegraphics[width=0.7\linewidth]{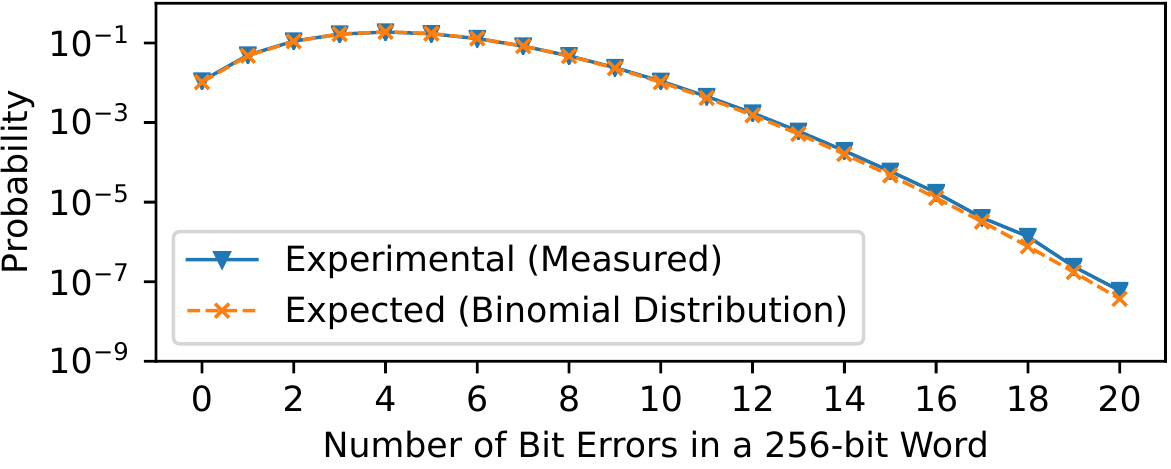}
    \caption{Expected and experimental probabilities of observing an $X$-bit
    error in a 256-bit word for a representative DRAM device without on-die
    ECC at $\bm{t_{REFW} =}$ 20s and 60$^\circ$C.}
    \label{fig:uni_rand_assumption} 
\end{figure}

Before demonstrating the use of \mechanism/, we first validate that the
independent, uniform-random \mpvnine{data-retention error} model holds for the
devices that we test by comparing experimentally-measured error distributions
to the \emph{expected} distributions. Figure~\ref{fig:uni_rand_assumption}
shows both the expected and experimental probabilities of observing an $X$-bit
error in a single 256-bit word throughout DRAM at fixed operating conditions
of $t_{REFW} = 20s$ and 60$^\circ$C for a single representative DRAM device
\emph{without} on-die ECC.

The experimental data is well predicted by the binomial distribution and
diverges only at extreme error counts \mpvsix{that have} few experimental
samples. This validates modeling retention errors using a uniform-random
distribution for the devices without on-die ECC. We repeat this experiment
across all of our devices without on-die ECC for various word sizes, refresh
windows, and temperatures, and we find that the uniform-random model holds
across all experiments.
\subsection{\mpvseven{Determining the True-/Anti-Cell Layout}}
\label{subsec:true_anti_cell_rev_engr}

\begin{figure}[h]
    \centering
    \includegraphics[width=0.7\linewidth]{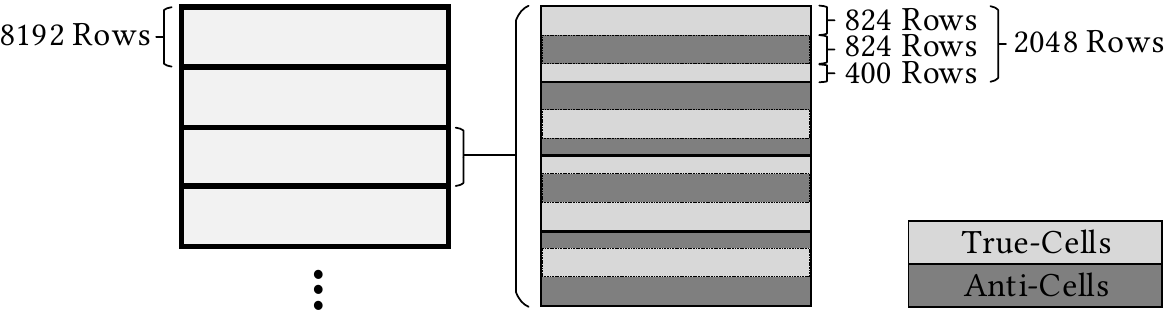}

    \caption{\mpvsix{A DRAM bank comprises groups of 824 or 400 rows with
    alternating true- and anti-cells per group.}}
    \label{fig:anticell_subarray_layout_cartoon} 
\end{figure}

We reverse-engineer the true-/anti-cell layout in the devices with on-die ECC
\mpvone{to ensure that we can accurately model the pre-correction error
distribution in simulation (as described in
Section~\ref{subsec:high_level_sim_arch}, we only inject errors in cells
programmed to \mpvfour{the} ``charged'' state)}. We do this by studying the
locations of data-retention errors after disabling refresh for a long time
(e.g., \textgreater30 minutes), which causes most cells to leak to their
discharged state.\footnote{\mpvsix{A small number of cells do not follow the overall
pattern due to either \emph{i}) extraordinarily long retention times or
\emph{ii}) ECC correction.}} Figure~\ref{fig:anticell_subarray_layout_cartoon}
illustrates the resulting pattern, showing how individual rows comprise
entirely true- or anti-cells, and contiguous groups of either 824 or 400 rows
alternate throughout a bank. In simulation, we model each DRAM burst to be
entirely composed of either true- or anti-cells with a 50\% probability. This
accurately models sampling an arbitrary burst from the entire memory address
space.

\begin{figure}[h]
    \centering
    \includegraphics[width=0.7\linewidth]{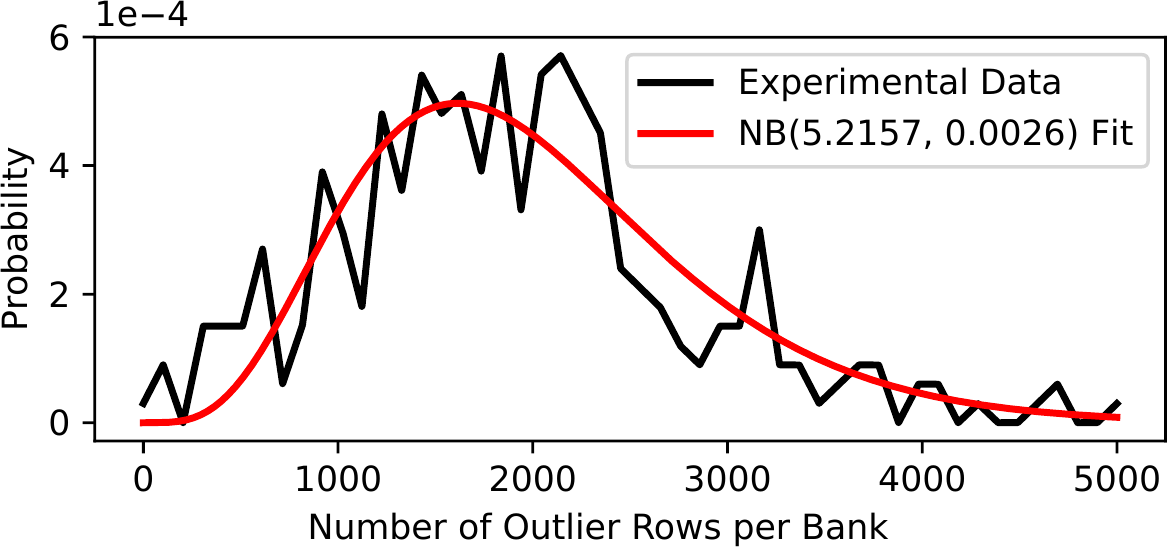}
    \caption{Histogram of the number of rows with outlier
    \mbox{true-/anti-cell} layouts per bank across all banks of all
    \mpvnine{DRAM devices with on-die ECC} (NB: negative-binomial).}
    \label{fig:remapped_rows}
\end{figure}

Despite the observed true-/anti-cell pattern, we find that a small amount of
uniquely randomly-distributed rows in each bank \emph{do not} follow the
pattern shown in Figure~\ref{fig:anticell_subarray_layout_cartoon}. Instead,
these rows alternate true- and anti-cells \emph{every byte} and are often
found in clusters of two or more. A histogram of the number of such rows,
\mpvseven{called \emph{\mpveight{outlier} rows}}, per bank across all 232 devices with
on-die ECC is shown in Figure~\ref{fig:remapped_rows} alongside a best-fit
negative-binomial distribution curve. Both the shape of the frequency
distribution and the observed clustering are consistent with
post-manufacturing repair row remapping
techniques~\cite{horiguchi2011nanoscale}. Since these rows have a different
true- and anti-cell composition, they add unwanted noise to our
reverse-engineering analysis. While we could account for them in our
simulations, we simply \mpvseven{skip testing these rows in our} experimental
analysis to avoid unnecessary complexity.

\subsection{Applying \mechanism/ to \mpvnine{DRAM} with On-Die ECC}

\label{subsec:rev_engr_ecc_scheme}

We demonstrate applying \mechanism/ to the DRAM \mpvnine{devices} with on-die
ECC \mpvseven{using} the experimental configuration shown in
Table~\ref{tab:rev_engr}. The \mpvseven{error distribution} resulting from a
single experiment provides the PMF given by Equation~\ref{eqn:observables},
which forms the observations $\mathcal{O}$ in the overall optimization problem
(Equation~\ref{eqn:final_opt_problem}).

\begin{table}[h]
\footnotesize 
\centering

\begin{tabular}{|c|c|c|}
\hline
\textbf{Model Parameter} & \textbf{Experiment} & \textbf{Simulation} \\
\hline
\hline
Word Size & \multicolumn{2}{|c|}{256-bits} \\ \hline
True-/Anti-Cell Layout & \multicolumn{2}{|c|}{50\%/50\% at word-granularity} \\ \hline
Data Pattern & \texttt{RANDOM} & \texttt{RANDOM} and \mpvsix{\texttt{0xFF}} \\ \hline
\mpveight{Outlier} Rows & \mpvseven{Skipped} & Ignored \\ \hline
Temperature & 70$^\circ$C & Encompassed in the RBER \\ \hline
$t_{REFW}$ & 5 minutes & Encompassed in the RBER \\ \hline
\end{tabular}
\caption{Experimental and simulation setup for reverse-engineering the ECC
scheme used in \mpveight{the} tested devices.}
\label{tab:rev_engr}
\end{table}
\mpvfour{Using a representative device, we perform a single experiment at the
conditions shown in Table~\ref{tab:rev_engr}, measuring} a
\emph{post-correction} BER of 0.041578. \mpveight{Then, configuring
\simulator/ with} the parameters listed under ``Simulation'' in
Table~\ref{tab:rev_engr}, we evaluate the full optimization problem of
Equation~\ref{eqn:final_opt_problem} using the grid-search approach described
in Section~\ref{subsec:determining_map_params}.

\begin{figure}[h]
    \centering
    \includegraphics[width=0.7\linewidth]{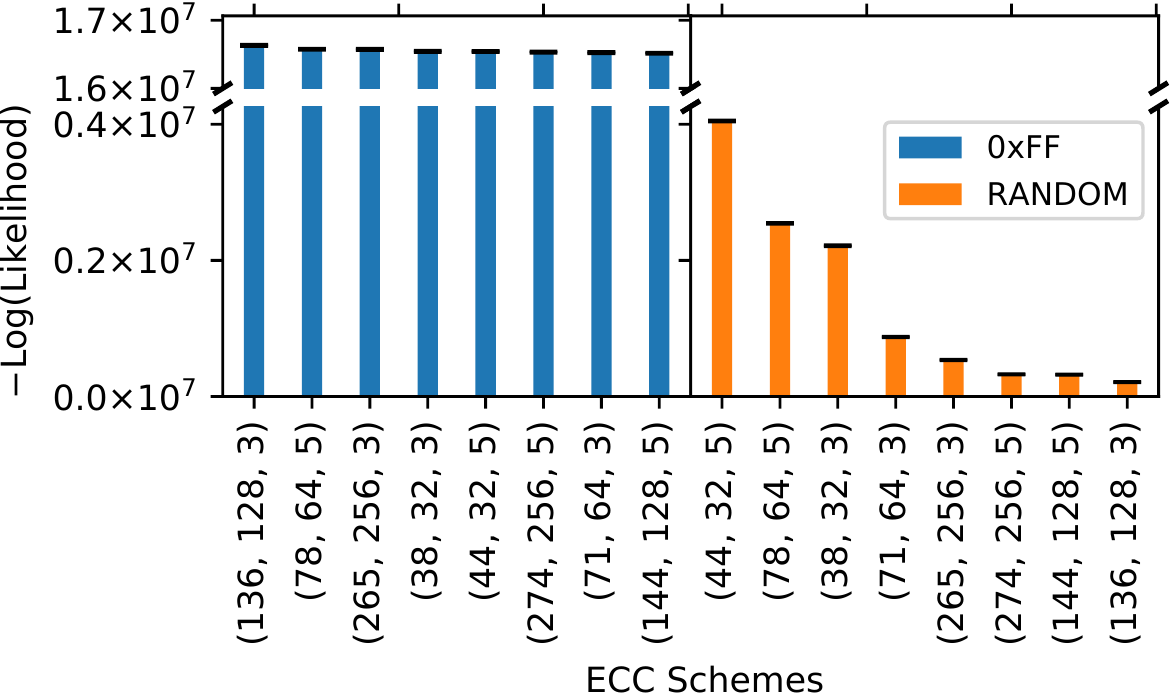}
    \caption{Likelihoods of eight different ECC schemes across two different
    data patterns, where each likelihood is individually maximized over
    the model parameter $\theta$.}
    \label{fig:posterior} 
\end{figure}

Figure~\ref{fig:posterior} presents the negative log-likelihoods
(Equation~\ref{eqn:best_fit}) of the eight highest-likelihood ECC schemes for
each of the \mpvnine{\texttt{0xFF} and \texttt{RANDOM}} data patterns. Models
are sorted in order of increasing likelihood (i.e., \emph{decreasing} negative
log-likelihood) from left to right for each data pattern. Bars show black
confidence intervals \mpvnine{spanning} the \mbox{min/max} values when
bootstrapping the observed data $10^5$ times (described in
Section~\ref{subsec:accuracy}). The confidence intervals are tight enough
\mpvnine{to appear as a single line atop each bar}. Note that the
\texttt{0xFF} models have low likelihoods \mpvnine{(i.e., higher bars)}, which
agrees with the fact that our experimental data is obtained \mpvnine{using} a
\texttt{RANDOM} data pattern.

The smallest (i.e., rightmost) bar represents the \emph{most likely model},
which provides the final reverse-engineered model parameters, including the
ECC scheme and the pre-correction error rate. For greater insight into the
results, Table~\ref{tab:best_fits} describes the five highest-likelihood
models in detail.

\setlength\tabcolsep{3pt} 
\begin{table}[h]
\footnotesize 
\centering

\begin{tabular}{|c|c|c|c|c|}
\hline
\multicolumn{2}{|c|}{$-$Log-Likelihood} & $f_i$ & \multicolumn{2}{|c|}{$\mathbf{\theta}$} \\
\hline
\textbf{Value} & \textbf{Bootstrap (Min, Max)} & \textbf{ECC Code} & \textbf{RBER} & \textbf{Data
Pattern} \\
\hline
\hline
2.12e-5 & \mpvone{(2.09e-5, 2.15e-5)} & (136, 128, 3) & 0.038326 & \texttt{RANDOM} \\ \hline
3.21e-5 & \mpvone{(3.18e-5, 3.24e-5)} & (144, 128, 5) & 0.039113 & \texttt{RANDOM} \\ \hline
3.26e-5 & \mpvone{(3.22e-5, 3.29e-5)} & (274, 256, 5) & 0.039995 & \texttt{RANDOM} \\ \hline
5.38e-5 & \mpvone{(5.32e-5, 5.43e-5)} & (265, 256, 3) & 0.039956 & \texttt{RANDOM} \\ \hline
8.74e-5 & \mpvone{(8.69e-5, 8.79e-5)} & (71, 64, 3) & 0.038472 & \texttt{RANDOM} \\ \hline
\end{tabular}
\caption{Details of the five highest-likelihood models (shown in
Figure~\ref{fig:posterior}) and their raw likelihood values.}
\label{tab:best_fits}
\end{table}
\setlength\tabcolsep{6pt} 

The data indicates that \mpvseven{a Hamming single-error correction code with ($n=136$, $k=128$,
$d=3$)} is the most likely ECC scheme out of all models considered. This
result is consistent with several industrial prior
works~\cite{micron2017whitepaper, kwak2017a, kwon2017an,
chun201816gb}.\footnote{We obtained this result \emph{without} informing the
prior distribution about the existence of prior works. If instead we had done
so as mentioned in Section~\ref{sec:bayesian_theory}, (136, 128,
3)~\cite{micron2017whitepaper, kwak2017a, kwon2017an, chun201816gb} and (71,
64, 3)~\cite{im2016im, issi2018ddr3} ECC code would have been more
overwhelmingly likely.} Compared to most of the other codes we consider, (136,
128, 3) code has a relatively low error-correction capability (i.e., 1 bit per
136 codeword bits), which is reasonable for a first-generation on-die ECC
mechanism and requires a relatively simple, low-overhead circuit
implementation.

The MAP estimate of $\theta$ provides the most likely pre-correction error
rate (i.e., RBER) and data pattern to explain the observed data.
\mpveight{Note that on-die ECC actually \emph{increases} the error rate at
these testing conditions, likely due to a high incidence of miscorrections as
described in Section~\ref{subsec:estimating_relative_freqs}.} \mechanism/
correctly infers that our experiment uses the \texttt{RANDOM} data pattern,
which is indicated by the relatively low likelihoods of the models that assume
a \texttt{0xFF} data pattern. 

\begin{figure}[h]
    \centering
    \includegraphics[width=0.7\linewidth]{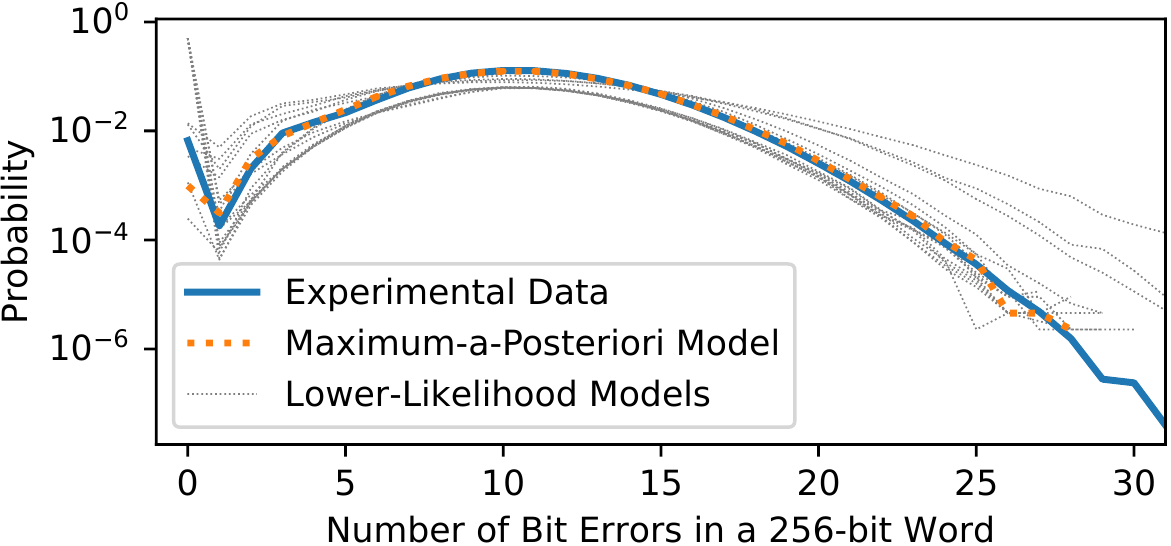}
    \caption{Full \mpvsix{PMF} for each model considered in
    Figure~\ref{fig:posterior}.}
    \label{fig:best_fit_ecc} 
\end{figure}

Figure~\ref{fig:best_fit_ecc} shows the full PMF of
Equation~\ref{eqn:observables_parameterized} for all sixteen models considered
in Figure~\ref{fig:posterior}. The maximum-a-posteriori model (dashed) and the
experimental data (solid) are shown alongside all other models (dotted). When
shown graphically, it is clear that \mechanism/ effectively performs a
rigorous best-fit analysis over several models to the experimental
observations.

We repeat this analysis across different devices, temperatures, refresh
windows, and data patterns and consistently find the (136, 128, 3) ECC code to
be the \mpveight{maximum-a-posteriori} model. Thus, we conclude that the (136,
128, 3) ECC code is the ECC scheme used in the tested devices.

\mpveight{We draw three key conclusions from this application of \mechanism/.}

\noindent
\textbf{First}, \mechanism/ infers the on-die ECC scheme with \emph{no}
visibility into the encoded data or error-detection/-correction information,
\emph{without} disabling the ECC mechanism, and \emph{without} tampering with
the hardware in any way.

\noindent
\textbf{Second}, \mechanism/ can simultaneously infer several components of
$\theta$ that might not be known. While we demonstrate a simple inference over
only two data patterns in addition to the pre-correction error rate, we could
also infer other characteristics (e.g., true-/anti-cell composition, refresh
window, temperature). In general, $\theta$ is extensible to any model
parameter that can be implemented in simulation (i.e., in \simulator/).

\noindent
\textbf{Third}, Figure~\ref{fig:best_fit_ecc} shows that the
\mpvnine{maximum-a-posteriori} model is a good fit for the empirical data,
which supports our assumption that \mpveight{data-retention} errors can be
modeled as uniformly-random events (Section~\ref{subsec:prob_model}) even for
devices with on-die ECC.

\section{Data-Retention Error Characterization \\ of DRAM Devices with
On-Die ECC}
\label{sec:proof_of_concept}

\mpvfour{Having reverse-engineered the on-die ECC scheme, we characterize
data-retention error rates with respect to both $t_{REFW}$ and temperature to
demonstrate how \mechanism/ enables studying \emph{pre-correction} errors in
practice. To our knowledge, this is the \emph{first} work to provide a
system-level error characterization study of DRAM devices with on-die ECC in
open literature in an effort to understand the pre-correction error
characteristics.}
 
\subsection{Data-Retention Error Rate vs. Refresh Window}

\begin{figure}[h]
    \centering
    \includegraphics[width=0.7\linewidth]{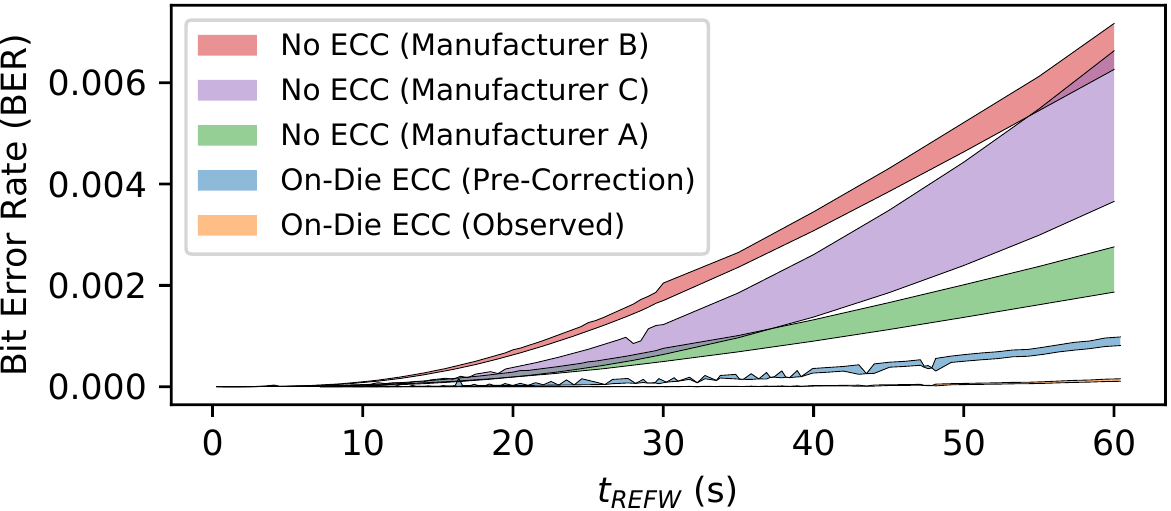}
    \caption{\mpvthree{Comparison of data-retention error rates measured using
    devices with and without on-die ECC, including the \mpvthree{inferred}
    pre-correction error rates for devices with on-die ECC.}}
    \label{fig:with_and_without_on_die_ecc} 
\end{figure}

Figure~\ref{fig:with_and_without_on_die_ecc} shows the \mpvthree{measured}
data-retention error rates for DRAM devices with and without on-die ECC using
different $t_{REFW}$ values at a fixed temperature of 50$^\circ$C using a
\texttt{0xFF} data pattern. Each of the five distributions shows the minimum
and maximum error rates observed for different groups of devices organized by
manufacturer. We also show the \emph{pre-correction} error rates
\mpvthree{inferred using \mechanism/}.

The data shows that the observed error rates for devices \emph{with} on-die
ECC lie far below those of devices \emph{without} on-die ECC. This is
consistent with observations from prior works~\cite{micron2017whitepaper,
kwon2017an, kwak2017a}, which find that on-die ECC is a strong enough
error-mitigation mechanism to allow for refresh rate reduction. Unfortunately,
the \emph{observed} error rates do not provide insight into how the core DRAM
technology has changed because it is unclear how much of the error margin
improvement is simply a result of ECC.

\mechanism/ solves this problem. By inferring the pre-correction error rates,
we observe considerable error margin for \emph{even the pre-correction error
rates}, implying that on-die ECC \mpvseven{may be} \emph{unnecessary} at these
testing conditions. This \mpvseven{may seem surprising at first sight}, since
error rates are believed to be increasing with technology
generation~\cite{kang2014co, meza2015revisiting, micron2017whitepaper,
nair2016xed}. However, on-die ECC's goal is to combat single-cell errors at
\emph{worst-case operating specifications}~\cite{micron2017whitepaper} (i.e.,
$85^\circ$C, $t_{REFW} = 32$ms~\cite{jedec2014lpddr4}, \mpvseven{worst-case
usage characteristics}). Unfortunately, our testing infrastructure
\mpvseven{currently} cannot achieve such conditions, and \mpvseven{even if it
could, the pathological access- and data-patterns depend on the proprietary
internal circuit design known only to the manufacturer}. Therefore, our
observations do not contradict expectations, and we conclude that
\mpvseven{for devices with on-die ECC: \emph{i}) on-die ECC effectively reduces the
observed error rate and \emph{ii}) both pre- and post-correction error rates are
considerably lower than those of devices without on-die ECC \emph{at our
testing conditions}.}

This example \mpvseven{demonstrates} \mechanism/'s strengths: \mechanism/
\emph{separates} the effects of a device's particular ECC mechanism from the
raw error rates of the DRAM technology and \emph{enables} a meaningful
comparison of error characteristics between devices with (or without)
different ECC schemes. \mechanism/ enables this analysis for \emph{any} error
mechanism \mpveight{that \mechanism/ is applicable to
(Section~\ref{subsec:application_to_other_tech})}.
\subsection{Data-Retention Error Rate vs. Temperature}
\label{subsec:ret_err_vs_temp}

\mpvthree{Data-retention error rates are well-known to follow an exponential
relationship with respect to temperature~\cite{hamamoto1998retention,
liu2013experimental, bacchini2014characterization}, and prior works~\cite{liu2013experimental,
kim2010high, kong2008analysis} exploit this relationship to extrapolate error
rates beyond experimentally feasible testing conditions. We show that on-die
ECC distorts this exponential relationship such that observed error rates
cannot be reliably extrapolated, and \mechanism/ recovers the underlying
exponential relationship.}

\definecolor{plotorange}{RGB}{254, 129, 18} 
\definecolor{plotblue}{RGB}{32, 119, 180} 
\definecolor{plotgreen}{RGB}{45, 160, 45} 

\begin{figure}[h]
    \centering
    \includegraphics[width=0.7\linewidth]{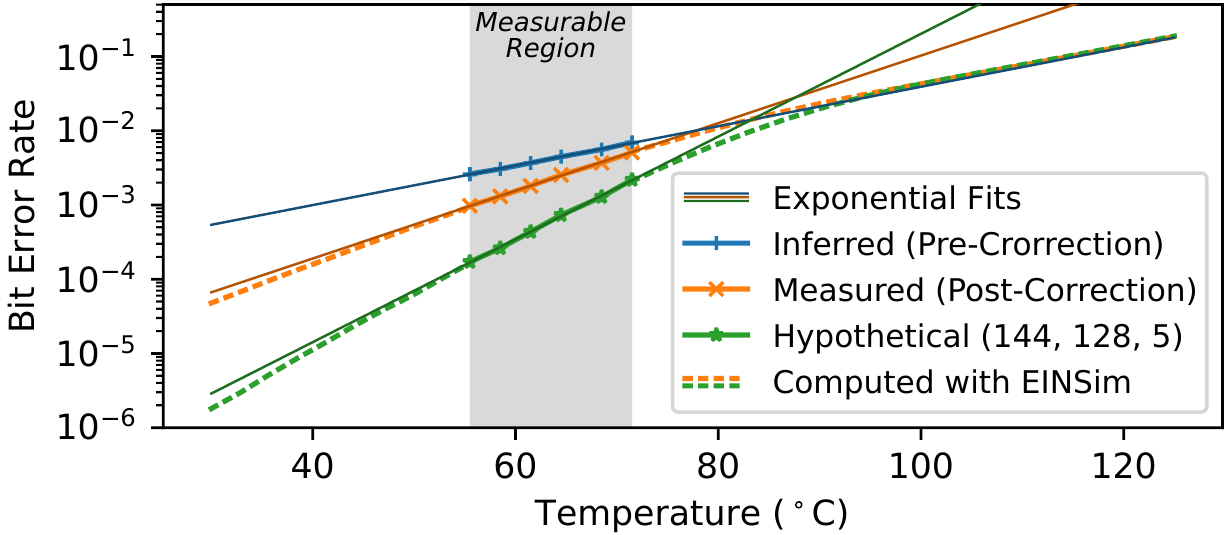}
    \caption{\mpvseven{Data-retention error rates of a single representative
    device with $\bm{t_{REFW} =}$ 10s across different temperatures, showing error
    rates: i) measured (post-correction), ii) inferred (pre-correction), and
    iii) hypothetical post-correction assuming a (144, 128, 5) ECC scheme.}}
    \label{fig:temp_fit_with_and_without_ecc} 
\end{figure}

Figure~\ref{fig:temp_fit_with_and_without_ecc} shows the exponential
relationship for a single representative device with on-die ECC at a fixed
refresh window of 10s on a semilog scale. Measurements (orange,
\textcolor{plotorange}{$\bm{\times}$}) are taken between the temperature
limits of our infrastructure (55$^\circ$C - 70$^\circ$C, illustrated with a
grey background), and the inferred pre-correction error rates (blue,
\textcolor{plotblue}{$\bm{+}$}) and the hypothetical error rates if the on-die
ECC scheme were a stronger double-error correction (144, 128, 5) code (green,
\textcolor{plotgreen}{$\bm{*}$}) are shown. \mpvseven{We show exponential fits
to data within the measurable region for all three curves. Outside of the
measurable region (i.e., \textless55$^\circ$C and \textgreater70$^\circ$C), we
use \simulator/ to extrapolate the two post-correction curves beyond the
measurable region (dashed) based on the exponential fit for the pre-correction
curve.}

\mpvthree{While all three curves \emph{appear} to fit an exponential
curve within the measurable temperature range, this is a misleading artifact
of sampling only a small fraction of the overall error distribution. Across
the full range, only the pre-correction curve follows the exponential
relationship: both post-correction curves diverge from the exponential fit on
both sides of the \mpvseven{measurable region} and follow an
\emph{ECC-specific} shape. This means that post-correction error rates cannot
be directly fitted to an exponential curve, and extrapolating along the known
exponential relationship of the data-retention error mechanism requires
knowing the pre-correction error rates.}
	
This example demonstrates how \mechanism/ \mpvseven{recovers the statistical
characteristics of the pre-correction error rates that on-die ECC
\mpvseven{obfuscates}. In general, \mechanism/ enables this for any error
mechanism that \mechanism/ \mpveight{is applicable} to (discussed in
Section~\ref{subsec:prob_model}), allowing future works to make use of
well-studied error characteristics for devices with ECC.}

\section{Summary}
\mpvtwo{We develop \mechanism/, the first statistical inference methodology
capable of determining the ECC scheme and pre-correction error rates of a DRAM
device with on-die ECC. We provide \simulator/~\cite{eccsimgithub}, a flexible
open-source simulator that \mpveight{can apply} \mechanism/ across different
DRAM devices and error models. We evaluate \mechanism/ with the \emph{first}
experimental study of 232 (82) LPDDR4 DRAM devices with (without) on-die ECC.
Using \mechanism/, we: \emph{i}) find \mpvfour{that the ECC scheme employed in
the devices we test is} \mpvseven{a single-error correction Hamming code with
($n$ = 136, $k$ = 128, $d$ = 3)}, \emph{ii}) infer \emph{pre-correction} error
rates from \emph{post-correction} errors, and \emph{iii}) recover well-known
pre-correction error distributions that on-die ECC obfuscates. With this, we
demonstrate that \mechanism/ enables DRAM error characterization studies for
devices with on-die ECC. We believe and hope that future work will use
\mechanism/ to develop new understanding and mechanisms to tackle the DRAM
scaling challenges that lie ahead.}

\chapter[BEER: Identifying the On-Die ECC Function]{Understanding Exactly How\\On-Die ECC Obfuscates Errors}
\label{chap:beer}

The previous chapter develops a statistical inference technique for
understanding high-level properties of the on-die ECC mechanism and the
pre-correction errors. This chapter takes the next step and introduces two new
testing methodologies: (1) BEER, capable of identifying the exact way in which
on-die ECC obfuscates errors; and (2) BEEP, capable of inferring the number and
bit-exact locations of the pre-correction errors responsible for an observed
post-correction error pattern.

\section{Background and Motivation}
\label{beer:sec:intro}

Dynamic random access memory (DRAM) is the predominant choice for system main
memory across a wide variety of computing platforms due to its favorable
cost-per-bit relative to other memory technologies. DRAM manufacturers maintain
a competitive advantage by improving raw storage densities across device
generations. Unfortunately, \mpi{these improvements largely rely on process
technology scaling, which} causes serious reliability issues that reduce factory
yield. DRAM manufacturers traditionally mitigate yield loss using
post-manufacturing repair techniques such as row/column
sparing~\cite{horiguchi2011nanoscale}. However, continued \mpi{technology
scaling} in modern DRAM chips requires stronger error-mitigation mechanisms to
remain viable because of random single-bit errors that are increasingly frequent
at smaller process technology \jkx{nodes~\cite{micron2017whitepaper, kang2014co,
nair2016xed, gong2017dram, kwon2014understanding, meza2015revisiting, oh20153,
kim2007low, son2015cidra, liu2013experimental, mutlu2014research,
mutlu2013memory}}. Therefore, DRAM manufacturers \mpi{have begun to} use
\emph{on-die error correction \mpi{coding}} \mpi{(\emph{on-die ECC}), which
silently corrects} single-bit errors entirely within the DRAM
chip~\cite{micron2017whitepaper, kang2014co, gong2017dram, nair2016xed,
patel2019understanding}. On-die ECC is \emph{completely invisible} outside of
the DRAM chip, \cro{so} ECC metadata (i.e., parity-check bits, error syndromes)
\mpi{that is used to correct errors} is hidden from the rest of the system.

Prior works~\cite{im2016im, nair2016xed, micron2017whitepaper,
micron2019whitepaper, oh20153, kwak2017a, kwon2017an, patel2019understanding}
indicate that \mpf{existing} on-die ECC codes are 64- or 128-bit single-error
correction (SEC) Hamming codes~\cite{hamming1950error}. However, each DRAM
manufacturer considers their on-die ECC mechanism's design and implementation to
be highly proprietary and ensures not to reveal its details in any public
documentation, including DRAM standards~\cite{jedec2012ddr4, jedec2014lpddr4},
\mpiii{DRAM} datasheets~\mpii{\cite{issi2020lpddr4, samsung2018mobile,
micron2018mobile, hynix2015366ball}}, \jkx{publications~\cite{oh20153,
kwak2017a, kwon2017an, kang2014co}}, and industry
whitepapers~\cite{micron2017whitepaper, micron2019whitepaper}.

\subsection{Motivation: On-Die ECC Obfuscates Errors}

Because the unknown on-die ECC function is encapsulated within the DRAM chip,
\mpi{it obfuscates \emph{raw bit} errors (i.e., \emph{pre-correction}
errors)\footnote{\mpi{We use the term ``error'' to refer to \emph{any} bit-flip
event, whether observed (e.g., uncorrectable bit-flips) or unobserved (e.g.,
corrected by ECC).}} in an ECC-function-specific manner. Therefore, the
locations of \mpii{software-visible} \emph{uncorrectable} errors (i.e.,
\emph{post-correction} errors) often no longer match those of the pre-correction
errors} that were caused by physical DRAM error mechanisms. While this behavior
appears desirable from a black-box perspective, it poses serious problems for
third-party DRAM customers who study, test and validate, and/or design systems
based on the reliability characteristics of the DRAM chips that they buy and
use. Section~\ref{subsec:impls_for_third_parties} describes these customers and
the problems they face in detail, including, but not limited to, three important
groups: (1) system designers who need to ensure that supplementary
error-mitigation mechanisms (e.g., rank-level ECC within the DRAM controller)
are carefully designed to cooperate with the on-die ECC
function~\cite{son2015cidra, nair2016xed, gong2018duo}, (2) large-scale
industries \mpii{(e.g., computing system providers such as
Microsoft~\cite{field2015microsoft}, HP~\cite{hp2011whitepaper}, and
Intel~\cite{intel2020platform},} DRAM module
manufacturers~\cite{kingston2012whitepaper, smart2017smart, adata2017adata}) or
government entities (e.g., national labs~\cite{nasa2016nasa,
sandia2020fabrication}) who must understand DRAM reliability characteristics
when validating DRAM chips they buy and use, and (3) researchers who need
\mpf{full visibility into physical device characteristics to study and model}
DRAM \jkx{reliability~\cite{hamamoto1995well, hamamoto1998retention,
jin2004modeling, weis2015retention, edri2016silicon, yaney1987meta,
kim2015avert, patel2019understanding, kim2020revisiting, liu2013experimental,
hassan2017softmc, kim2018dram, kim2018solar, kim2019d, khan2014efficacy,
khan2016case, khan2016parbor, khan2017detecting, chang2016understanding,
chang2017understanding, patel2017reach}}.

\begin{figure}[h]
    \centering
    \includegraphics[width=0.7\linewidth]{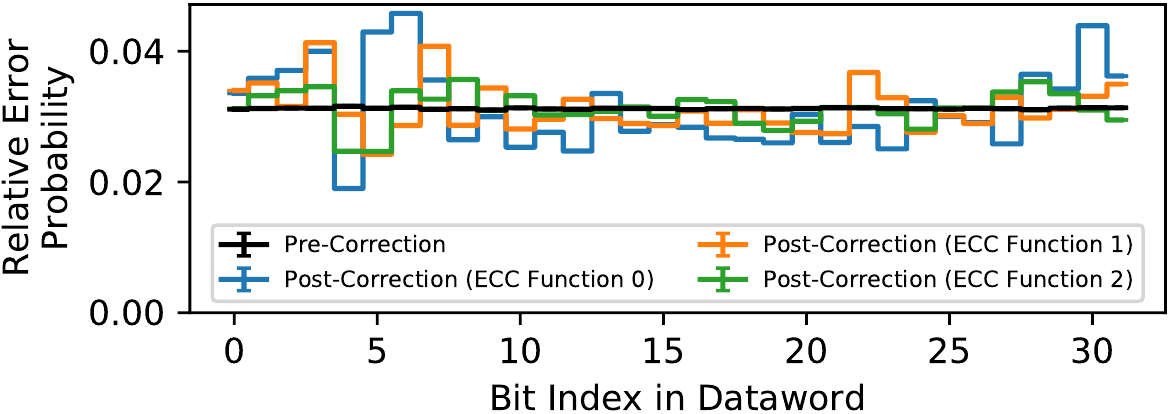}
    \caption{Relative error probabilities in different bit positions for
    different ECC functions with uniform-randomly distributed
    \mpii{pre-correction (i.e., raw)} bit errors.}
    \label{beer:fig:motivation}
\end{figure}

For each of these third parties, merely knowing or reverse-engineering the type
of ECC code (e.g., $n$-bit Hamming code) based on existing
industry~\cite{oh20153, kwak2017a, kwon2017an, micron2017whitepaper,
micron2019whitepaper, im2016im} and academic~\cite{nair2016xed,
patel2019understanding} publications is not enough to \mpi{determine} exactly
how the ECC mechanism obfuscates specific error patterns. This is because an ECC
code of a given type can have many different implementations based on how its
ECC function (i.e., its parity-check matrix) is designed, and different designs
lead to different reliability characteristics. For example,
Figure~\ref{beer:fig:motivation} shows the relative probability of observing errors
in different bit positions for three different ECC codes of the same type (i.e.,
\mpvi{single-error correction} Hamming code \mpii{with 32 data bits and 6
parity-check bits}) but that use different ECC functions. We obtain this data by
simulating $10^9$ ECC words using the EINSim
simulator~\cite{patel2019understanding, eccsimgithub} and show medians and 95\%
confidence intervals calculated via statistical
bootstrapping~\cite{efron1992bootstrap} over 1000 samples. \mpi{We simulate a
\texttt{0xFF} test pattern\footnote{Other patterns show similar behavior,
including \texttt{RANDOM} data.} with uniform-random pre-correction errors at} a
raw bit error rate of $10^{-4}$ (e.g., \mps{as often seen in} experimental
studies\hht{~\cite{hamamoto1998retention, kang2014co, liu2013experimental,
lee2015adaptive, shirley2014copula, chang2017understanding, hassan2017softmc,
chang2016understanding, patel2017reach}}).

\mpf{The data demonstrates that ECC codes of the same type can have vastly
different post-correction error characteristics. This is because each ECC
mechanism acts differently when faced with more errors than it can correct
(i.e., uncorrectable errors), causing it to mistakenly perform
ECC-function-specific ``corrections'' to bits that did not experience errors
(i.e., \mpvi{\emph{miscorrections}}, which Section~\ref{subsec:ecc_bg} expands upon).
Therefore, a researcher or engineer who studies two DRAM chips that use the same
type of ECC code but different ECC functions may find that the chips'
\mpiii{software-visible} reliability characteristics are quite different even if
the \mpiii{physical} DRAM cells' reliability characteristics are identical. On
the other hand, if we know the full ECC function (i.e., its parity-check
matrix), we can calculate exactly which pre-correction error pattern(s) result
in a set of observed errors. Figure~\ref{beer:fig:motivation} is a result of
aggregating such calculations across \mpii{$10^9$} error
patterns\footnote{\mpvi{Capturing approximately $10^9$ of the $2^{38} \approx
2.7\times 10^{11}$ unique patterns.}}, and Section~\ref{subsec:beep}
demonstrates how we can use the ECC function to infer pre-correction
error counts and locations using \mps{only} observed post-correction errors.}

\mpf{Knowing the precise transformation between pre- and post-correction errors}
benefits all of the \mpi{aforementioned} third-party use cases \mpf{because it
provides system designers, test engineers, and researchers \mpi{with} a way to
isolate the error characteristics of the memory itself from the effects of a
particular ECC function.} Section~\ref{subsec:impls_for_third_parties} provides
\mpf{several example use cases and describes the benefits of knowing the ECC
function in detail. \mpi{While} specialized, possibly intrusive methods (e.g.,
chip teardown~\mpvi{\cite{james2010silicon, torrance2009state}}, advanced
imaging techniques~\cite{ho2003method, torrance2009state}) can theoretically
extract the ECC function, such techniques are typically inaccessible to or
infeasible for many third-party users.}

\subsection{Mechanism: Bit-Exact ECC Recovery (BEER)}

To enable third parties to \mpf{reconstruct pre-correction DRAM reliability
characteristics}, \textbf{our goal} is to develop a methodology that can
reliably and accurately determine the \mpi{full on-die ECC function without
requiring hardware tools, prerequisite knowledge about the DRAM chip or on-die
ECC mechanism, or access to ECC metadata (e.g., error syndromes, parity
information)}. To this end, we develop \underline{B}it-\underline{E}xact
\underline{E}CC \underline{R}ecovery (BEER), a new methodology for determining
\cro{a DRAM chip's full on-die ECC function simply by studying the
software-visible post-correction error patterns that it generates}. Thus, BEER
requires no hardware support, hardware intrusion, or access to internal ECC
metadata (e.g., error syndromes, parity information). BEER exploits the key
insight that forcing the ECC function to act upon carefully-crafted
uncorrectable error patterns reveals ECC-function-specific behavior that
disambiguates different ECC functions. BEER comprises three key steps: (1)
\cro{deliberately} inducing uncorrectable data-retention errors by \cro{pausing
DRAM refresh} while using carefully-crafted test patterns to control \mpf{the
errors'} bit-locations, which is done by leveraging data-retention errors'
intrinsic \mpf{data-pattern asymmetry} (discussed in
Section~\ref{bg:subsec:data_retention}), (2) enumerating the bit positions where the
ECC mechanism causes miscorrections, and (3) using a SAT solver~\cite{de2008z3}
to solve for the unique parity-check matrix \mpf{that causes} the observed set
of miscorrections.

We experimentally apply BEER to 80 real LPDDR4 DRAM chips with on-die ECC from
three major DRAM manufacturers to determine \mpii{the chips'} on-die ECC
functions. We describe the experimental steps required to apply BEER to any DRAM
chip with on-die ECC and show that BEER tolerates observed experimental noise.
We show that different manufacturers appear to use different on-die ECC
functions while chips from the same manufacturer and model number appear to use
the same on-die ECC function
\mpi{(Section~\ref{subsubsec:testing_charged_patterns})}. \cro{Unfortunately},
our experimental studies with real DRAM chips have two limitations against
further validation: (1) because the on-die ECC function is considered trade
secret for each manufacturer, we are unable to obtain a groundtruth to compare
BEER's results against, even when considering non-disclosure agreements with
DRAM manufacturers and (2) we are unable to publish the final ECC functions that
we uncover using BEER for confidentiality reasons (discussed in
Section~\ref{subsec:motivation_secrecy}).

To overcome the limitations of experimental studies with real DRAM chips, we
rigorously evaluate BEER's correctness in simulation
(Section~\ref{sec:beer1eval}). We show that BEER correctly recovers the on-die
ECC function for 115300 single-error correction Hamming codes\footnote{This
irregular number arises from evaluating a \mpii{different} number of ECC
functions for different code lengths because longer codes require exponentially
more simulation time (discussed in
Section~\ref{subsec:beer1eval_correctness}).}, which are representative of
on-die ECC, with ECC word lengths ranging from 4 to 247 bits. We evaluate our
BEER implementation's runtime and memory consumption using a real system to
demonstrate that \mpii{BEER is practical and} the SAT problem that BEER requires
is realistically solvable.

\cro{To demonstrate how BEER \cro{is} useful in practice, we propose and discuss
several ways that third parties can leverage the ECC function that BEER reveals
to \mpi{more effectively design, study, and test systems that use DRAM chips
with on-die ECC} (Section~\ref{section:usecases}). As a concrete example, we
introduce and} evaluate \underline{B}it-\underline{E}xact \underline{E}rror
\underline{P}rofiling (BEEP), a new DRAM data-retention error profiling
methodology that reconstructs pre-correction error counts and locations purely
from observed post-correction errors. Using the ECC function revealed by BEER,
BEEP infers precisely which \emph{unobservable} raw bit errors correspond to
\emph{observed} post-correction errors at a given set of testing conditions. We
show that BEEP enables characterizing pre-correction errors across a wide range
of ECC functions, \mpi{ECC word} lengths, error patterns, and error rates.
\mpii{We publicly release our tools as open-source \mpii{software}: (1) a new
tool~\cite{beergithub} for applying BEER to experimental data from real DRAM
chips and (2) enhancements to EINSim~\cite{eccsimgithub} for evaluating BEER
\mpii{and BEEP} in simulation.}
\section{Challenges of Unknown On-Die ECCs}
\label{sec:motivation}

This section discusses why on-die ECC is considered proprietary, how its secrecy
causes difficulties for third-party consumers, and how \mpi{the BEER methodology
helps overcome these difficulties \mpii{by} identifying} the \mpii{full} on-die ECC
function.

\subsection{Secrecy Concerning On-Die ECC}
\label{subsec:motivation_secrecy}

On-die ECC silently mitigates increasing single-bit errors that reduce factory
\jkx{yield~\cite{micron2017whitepaper, kang2014co, nair2016xed, gong2017dram,
kwon2014understanding, meza2015revisiting, oh20153, kim2007low, son2015cidra,
liu2013experimental, mutlu2014research, mutlu2013memory}}. Because on-die ECC is
invisible to the external DRAM chip interface, older DRAM
standards~\cite{jedec2014lpddr4, jedec2012ddr4} place no restrictions \mpvi{on
the on-die ECC mechanism} while newer standards~\cite{jedec2020ddr5} specify
only a high-level description for on-die ECC to support new (albeit limited)
DDR5 features, e.g., on-die ECC scrubbing. In particular, there are no
restrictions on the \mpvi{design or implementation of the on-die ECC function
itself.}

This means that knowing \mpi{an on-die ECC mechanism's details could reveal
information about its} manufacturer's factory yield rates, which are highly
proprietary~\cite{cost1997yield, childers2015achieving} due to their direct
connection with business interests, potential legal concerns, and
competitiveness in a USD 45+ billion DRAM market~\cite{qy2019global,
verified2019global}. Therefore, manufacturers consider their on-die ECC designs
and implementations to be trade secrets that they are unwilling to disclose.
\revmp{In our experience, DRAM manufacturers will not reveal on-die ECC details
under confidentiality agreements, even for large-scale industry board vendors
for whom knowing the details stands to be mutually
beneficial.}\footnote{\revmp{Even if such agreements were possible, industry
teams and academics without major business relations with DRAM manufacturers
(i.e., an overwhelming majority of the potentially interested scientists and
engineers) will likely be unable to secure disclosure.}}

This raises two challenges for our experiments with real DRAM chips: (1) we do
not have access to ``groundtruth'' ECC functions to validate BEER's results
against and (2) we cannot publish the final ECC functions that we determine
using BEER for confidentiality reasons based on our relationships with the DRAM
manufacturers. However, this does not prevent third-party consumers from
applying BEER to their own devices, and we hope that our work encourages DRAM
manufacturers to be more open with their designs going forward.\footnote{While
full disclosure would be ideal, a more realistic scenario could be more flexible
on-die ECC confidentiality agreements. \mpi{As recent
work~\cite{frigo2020trrespass} shows, security or protection by obscurity is
likely a poor strategy in practice.}}

\subsection{On-Die ECC's Impact on Third Parties}
\label{subsec:impls_for_third_parties}

On-die ECC alters a DRAM chip's \mpii{software-visible} reliability
characteristics so that \mpi{they are no longer determined solely by how errors
physically occur within \mpi{the DRAM chip}. Figure~\ref{beer:fig:motivation}
illustrates this by showing how \mpi{using} different \mpi{on-die} ECC functions
changes how the \emph{same} underlying DRAM errors appear to the end user.
Instead of following the pre-correction error distribution (i.e., uniform-random
errors), the post-correction errors exhibit ECC-function-specific shapes that
\mpiii{are difficult to predict} without knowing precisely which ECC function
\mpiii{is} used in each case. This means that two commodity DRAM chips with
different on-die ECC functions may show similar or different reliability
characteristics irrespective of how the underlying DRAM technology and error
mechanisms behave. Therefore, the \mpii{physical} error mechanisms' behavior
alone can no longer explain a DRAM chip's  post-correction error
characteristics.}

\mpi{Unfortunately, this poses a serious problem for third-party DRAM consumers
(e.g., system designers, testers, and researchers), who can no longer
\mpii{accurately} understand a DRAM chip's reliability characteristics by
studying its \mpii{software-visible} errors. This lack of understanding prevents
third parties from \mpii{both (1)} making informed design decisions, e.g., when
building \mpii{memory-controller based} error-mitigation mechanisms to
complement on-die ECC and \mpii{(2)} developing new ideas \mpii{that rely on} on
leveraging predictable aspects of a DRAM chip' reliability characteristics,
e.g., physical error mechanisms that are fundamental to all DRAM technology.}
As error rates worsen with continued technology
\jkx{scaling~\cite{micron2017whitepaper, kang2014co, nair2016xed, gong2017dram,
kwon2014understanding, meza2015revisiting, oh20153, kim2007low,
kim2020revisiting, kim2014flipping, mutlu2013memory, mutlu2014research}},
manufacturers will likely resort to stronger codes that further distort the
post-correction reliability characteristics. The remainder of this section
describes three key ways in which an unknown on-die ECC function hinders
third-parties, and determining the function helps mitigate the problem.

\noindent
\textbf{Designing \mpt{High-Reliability} Systems.} System designers often seek
to improve memory reliability beyond that which the DRAM provides alone (e.g.,
\mpi{by including rank-level ECC within} the memory controllers of server-class
machines or ECC within on-chip caches).
In particular, rank-level ECCs are carefully designed to mitigate common DRAM
failure modes~\cite{chen2018configurable} (e.g., chip
failure~\cite{nair2016xed}, burst errors~\cite{maiz2003characterization,
dell1997white}) \mpi{in order to correct as many errors as possible. However,
designing for key failure modes requires knowing a DRAM chip's reliability
characteristics, including the effects of any underlying ECC function (e.g.,
on-die ECC)~\cite{son2015cidra, gong2018duo}. For example, Son et
al.~\cite{son2015cidra} show that if on-die ECC suffers an uncorrectable error
and mistakenly ``corrects'' a non-erroneous bit (i.e., introduces a
\emph{miscorrection}), the stronger rank-level ECC may no longer be able to even
detect what would otherwise be a detectable (possibly correctable) error. To
prevent this scenario, both levels of ECC must be carefully co-designed to
complement each others' weaknesses.}
In general,
high-reliability systems can be more effectively built around DRAM chips with
on-die ECC if its ECC function and its effects on typical DRAM failure modes are
known.

\noindent
\textbf{\mpf{Testing}, Validation, and Quality Assurance.} \mpf{Large-scale
\mpi{computing system providers} \mpf{(e.g.,
Microsoft~\cite{field2015microsoft}, HP~\cite{hp2011whitepaper},
Intel~\cite{intel2020platform}), DRAM module
manufacturers~\cite{kingston2012whitepaper, smart2017smart, adata2017adata}, and
government entities (e.g., national labs~\cite{sandia2020fabrication,
nasa2016nasa}) typically} perform extensive third-party testing of the DRAM
chips they purchase in order to ensure that the chips meet internal
performance/energy/reliability targets. These tests validate} that DRAM chips
operate as expected \mpi{and that there are \mpii{well-understood, convincing}
root-causes (e.g., fundamental DRAM error mechanisms) for any observed errors}.
Unfortunately, on-die ECC interferes with two key components of such testing.
First, it obfuscates the number and bit-exact locations \mpi{of pre-correction
errors, so diagnosing the root cause for any observed \mpii{error} becomes
challenging.} Second, on-die ECC encodes all written data into \mpi{ECC
codewords, so the values written into the physical cells likely do not match the
values observed at the DRAM chip interface. The encoding process defeats
carefully-constructed test patterns that target specific circuit-level phenomena
(e.g., exacerbating interference between bitlines~\cite{adams2002high,
mrozek2019multi, khan2016parbor}) because the encoded data may no longer have
the intended effect. Unfortunately, constructing such patterns is crucial for
efficient testing since it minimizes the testing time required to achieve high
error coverage}~\cite{horiguchi2011nanoscale, adams2002high}. In both cases, the
full on-die ECC function determined by BEER describes exactly how on-die ECC
transforms pre-correction error patterns into post-correction ones. This enables
users to infer pre-correction error locations (demonstrated in
Section~\ref{subsec:beep}) and design test patterns that result in codewords
with desired properties (discussed in Section~\ref{subsection:use_cases}).

\noindent
\textbf{Scientific \mpvi{Error-Characterization} Studies.} 
Scientific error-characterization studies explore physical DRAM error mechanisms
(e.g., data retention~\hht{\cite{hamamoto1995well, hamamoto1998retention,
shirley2014copula, jung2015omitting, weis2015thermal, weis2015retention,
jung2014optimized, khan2016parbor, khan2016case, khan2014efficacy,
liu2013experimental, patel2017reach, hassan2017softmc, khan2017detecting}},
reduced access-latency~\cite{chang2016understanding, lee2015adaptive,
lee2017design, kim2018solar, kim2019d, kim2018dram, gao2019computedram,
chang2017understanding, chandrasekar2014exploiting}, circuit
disturbance~\hht{\cite{kim2014flipping, khan2016parbor, park2016statistical,
park2016experiments, kim2020revisiting, frigo2020trrespass, khan2017detecting}})
by deliberately exacerbating the error mechanism and analyzing the resulting
errors' statistical properties (e.g., frequency, spatial distribution). These
studies help build error models~\cite{yaney1987meta, hamamoto1998retention,
shirley2014copula, edri2016silicon, lee2017design, chang2017understanding,
kim2018solar, koppula2019eden}, leading to new DRAM designs and operating points
that improve upon the state-of-the-art. Unfortunately, on-die ECC complicates
error analysis and modeling by (1) \mpi{obscuring the physical pre-correction
errors that are the object of study} and (2) \mpi{preventing direct access to
parity-check bits, thereby precluding comprehensive testing of all DRAM cells in
a given chip.} Although prior work~\cite{patel2019understanding} enables
inferring high-level statistical characteristics of the pre-correction errors,
it does not provide a precise mapping between pre-correction and post-correction
errors, which is only possible knowing the full ECC function. Knowing the full
ECC function\mpi{, via our new BEER methodology,} enables recovering the
bit-exact locations of pre-correction errors throughout the entire ECC word
(\mpi{as we demonstrate} in Section~\ref{subsec:beep}) so that
error-characterization studies \mpi{can separate the effects of DRAM error
mechanisms from those of on-die ECC}. Section~\ref{section:usecases} provides a
detailed discussion of several key characterization studies that BEER enables.

\section{Determining the ECC Function}
\label{sec:determining_ecc_func}

BEER identifies an \mpi{unknown} ECC function by systematically reconstructing
its parity-check matrix based on the error syndromes that the ECC logic
generates while correcting errors. Different ECC functions compute different
error syndromes for a given error pattern, and by \mpi{constructing and
analyzing} carefully-crafted test cases, BEER uniquely identifies which ECC
function a particular implementation uses. This section describes how and why
this process works. Section~\ref{sec:beeri} describes how BEER accomplishes this
in practice for on-die ECC.

\subsection{\mpt{Disambiguating} Linear Block Codes}
\label{subsubsec:systematically_identifying}

DRAM ECCs are linear block codes, e.g., Hamming codes~\cite{hamming1950error}
for on-die ECC~\cite{im2016im, nair2016xed, micron2017whitepaper,
micron2019whitepaper, oh20153, kwak2017a, kwon2017an, patel2019understanding},
BCH~\cite{bose1960class, hocquenghem1959codes} or
Reed-Solomon~\cite{reed1960polynomial} codes for rank-level
ECC~\cite{cojocar2019exploiting, kim2016all}, whose encoding and decoding
operations are described by \emph{linear transformations} of their respective
inputs (i.e., $\mathbf{G}$ and $\mathbf{H}$ \mpi{matrices}, respectively).
\cro{We can therefore determine the full ECC function by independently
determining each of its linear components.}

\mpt{We can isolate each linear component of the ECC function} by injecting
errors in each codeword bit position and observing the resulting error
syndromes. For example, an $n$-bit Hamming code's parity-check matrix can be
systematically determined by injecting a single-bit error in each of the $n$ bit
positions: the error syndrome that the ECC decoder computes for each pattern is
exactly equal to the column of the parity-check matrix that corresponds to
\mpi{the} position of the injected error. As an example, Equation~\ref{eqn:syn}
shows how injecting an error at position 2 (i.e., adding error pattern
$\mathbf{e_2}$ to codeword $\mathbf{c}$) extracts the corresponding column of
the parity-check matrix $\mathbf{H}$ in the error syndrome $\mathbf{s}$. \mpi{By
the definition of a block code, $\mathbf{H}\cdot\mathbf{c}=\mathbf{0}$ for all
codewords~\cite{costello1982error, huffman2010fundamentals}, so} $\mathbf{e_2}$
isolates column 2 of $\mathbf{H}$ (i.e., $\mathbf{H}_{\ast,2}$).
\vspace{-0.15\baselineskip} 
\begin{equation}
\vspace{-0.25\baselineskip} 
\arraycolsep=1.6pt
\def\arraystretch{0.2}
\footnotesize
\mathbf{s}=\mathbf{H}\cdot\mathbf{c'}=\mathbf{H}\cdot(\mathbf{c} + \mathbf{e_2})=\mathbf{H}\cdot\left(\mathbf{c}+\left[
  \begin{array}{c}
    0 \\ 0 \\ 1 \\ 0 \\ 0 \\ 0 \\ 0
  \end{array}
\right]\right)=\mathbf{0} + \mathbf{H}_{\ast,2} = \mathbf{H}_{\ast,2}
\label{eqn:syn}
\end{equation}
Thus, the entire parity-check matrix can be fully determined by testing across
all 1-hot error patterns. \jk{Cojocar et al.~\cite{cojocar2019exploiting} use
this approach on DRAM rank-level ECC, injecting errors into codewords on the
DDR bus and reading the resulting error syndromes provided by the memory controller.} 

\subsection{Determining the On-Die ECC Function}
\label{subsubsec:determining_func_from_mcp}

\jk{Unfortunately, systematically determining \mpf{an ECC function} as described
in Section~\ref{subsubsec:systematically_identifying} is not possible with
on-die ECC for two key reasons.} First, \mpf{on-die ECC's parity-check bits
\mpi{cannot be accessed directly}, so we have no easy way to inject an error
within them.} Second, on-die ECC does not signal an error-correction event
\emph{or} report error syndromes (i.e., $\mathbf{s}$). \mpf{Therefore, even if
\mpvi{specialized methods} \mpi{(e.g., chip teardown~\cite{james2010silicon,
torrance2009state}, \mpvi{advanced imaging techniques}~\cite{ho2003method,
torrance2009state})} could inject errors within a DRAM chip package where the
on-die ECC mechanism resides,}\footnote{\mpi{Such methods may reveal the exact
on-die ECC circuitry. However, they are typically inaccessible to or infeasible
for many third-party consumers.}} the error syndromes would remain invisible, so
the approach taken by Cojocar et al.~\cite{cojocar2019exploiting} cannot be
applied to on-die ECC. To \mpf{determine the on-die ECC function using the}
approach of Section~\ref{subsubsec:systematically_identifying}, we first
formalize the unknown on-die ECC function and then determine how we can infer
error syndromes within the constraints of the formalized problem.

\subsubsection{Formalizing the Unknown ECC Function}
\label{subsubsec:formalizing_ecc_func}

We assume that on-die ECC uses a systematic encoding, which means that the ECC
function stores data bits unmodified. This is a reasonable assumption for real
hardware since it greatly simplifies data access~\cite{zhang2015vlsi} and is
consistent with our experimental results in
Section~\ref{subsubsec:ecc_word_layout}. Furthermore, \mpi{because the DRAM chip
interface exposes only data bits, the relative ordering of parity-check bits
within the codeword is irrelevant from the system's perspective.}
Mathematically, the different choices of bit positions represent
\emph{equivalent codes} that all have identical error-correction properties and
differ only in their internal
representations~\cite{richardson2008modern,roth2006introduction}, which on-die
ECC does not expose. Therefore, we are free to arbitrarily choose the
\mpi{parity-check bit} positions within the codeword without loss of generality.
If it becomes necessary to identify the exact ordering of bits within the
codeword \mpiii{(e.g., to infer circuit-level implementation details)},
reverse-engineering techniques based on physical DRAM error
mechanisms~\cite{lee2017design, jung2016reverse} can potentially be used.

A systematic encoding and the freedom to choose parity-check bit positions
mean that we can assume that the ECC function is in \emph{standard form},
where we express the parity-check matrix for an $(n, k)$ code as a partitioned
matrix $\mathbf{H}_{n - k\times n} = [\mathbf{P}_{n-k \times
k}|\mathbf{I}_{n-k \times n-k}]$. $\mathbf{P}$ is a conventional notation for
the sub-matrix that corresponds to information bit positions and $\mathbf{I}$
is an identity matrix that corresponds to parity-check bit positions. Note
that the example ECC code of Equation~\ref{eqn:hg} is in standard form. With
this representation, \mpf{all} codewords take the form $\mathbf{c}_{1\times n}
= [d_0 d_1 ... d_{k-1}|p_0 p_1 ... p_{n-k-1}]$, where $d$ and $p$ are data and
parity-check symbols, respectively.

\subsubsection{Identifying Syndromes Using Miscorrections}
\label{subsubsec:using_miscorrections}

Given that \mpf{on-die ECC conceals error syndromes}, we develop a new
approach for determining the on-die ECC function that \emph{indirectly}
determines error syndromes based on how the ECC mechanism responds when faced
with uncorrectable errors. To induce uncorrectable errors, we deliberately
pause normal DRAM refresh operations \revmp{long enough (e.g., several minutes
at 80$^\circ$C) to cause a large number of data-retention errors (e.g., BER
$>10^{-4}$) throughout a chip. These errors expose a significant number of
miscorrections in different ECC words, and the sheer number of data-retention
errors dominates any unwanted interference from other possible error
mechanisms (e.g., particle strikes~\cite{may1979alpha}).}

\revmp{To control which data-retention errors occur}, we write carefully-crafted
test patterns that restrict \revmp{the} errors to specific bit locations. This
is possible because only cells programmed to the \texttt{CHARGED} state can
experience data-retention errors as discussed in
Section~\ref{bg:subsec:data_retention}. By restricting pre-correction errors to
certain cells, if a post-correction error is observed in an unexpected location,
it \emph{must} be an artifact of error correction, i.e., a \emph{miscorrection}.
Such a miscorrection is significant since it: (1) signals an error-correction
event, (2) is \emph{purely} a function of the ECC decoding logic, and (3)
indirectly reveals the error syndrome generated by the pre-correction error
pattern. \cro{The indirection occurs because, although the miscorrection does
not expose the raw error syndrome, it \emph{does} reveal that whichever error
syndrome is generated internally by the ECC logic exactly matches the
parity-check matrix column that corresponds to the position of the miscorrected
bit.}

These three properties mean that miscorrections \mpiii{are a reliable tool for
analyzing} ECC functions: for a given pre-correction error pattern, different
ECC functions will generate different error syndromes, and therefore
miscorrections, depending on how the functions' parity-check matrices are
organized. This means that a given ECC function causes miscorrections
\emph{only} within certain bits, and the locations of miscorrection-susceptible
bits differ between functions. Therefore, we can differentiate ECC functions by
\mpiii{identifying which miscorrections are possible for different test patterns.}

\subsubsection{Identifying Useful Test Patterns}
\label{subsubsec:test_patterns}

To construct a set of test patterns that suffice to uniquely identify an ECC
function, we observe that \mpiii{a miscorrection is possible in a
\texttt{DISCHARGED} data bit only if the bit's error syndrome can be produced by
some linear combination of the parity-check matrix columns that correspond to
\texttt{CHARGED} bit locations}. For example, consider the 1-\texttt{CHARGED}
patterns that each set one data bit to the \texttt{CHARGED} state and all others
to the \texttt{DISCHARGED} state. In these patterns, data-retention errors may
\emph{only} occur in either (1) the \texttt{CHARGED} bit or (2) any parity-check
bits that the ECC function also sets to the \texttt{CHARGED} state. With these
restrictions, observable miscorrections may only occur within data bits whose
error syndromes can be created by some linear combination of the parity-check
matrix columns that correspond to the \texttt{CHARGED} cells within the
codeword. 

\cro{As a concrete example, consider the codeword of
Equation~\ref{eqn:codeword}. \texttt{\textbf{C}} and \texttt{\textbf{D}} represent that the
corresponding cell is programmed to the \texttt{CHARGED} and \texttt{DISCHARGED}
states, respectively.}
\vspace{-0.15\baselineskip}
\begin{equation}
	\vspace{-0.25\baselineskip} 
	\arraycolsep=1.1pt
	\def\arraystretch{0.8}
	\footnotesize
	\mathbf{c} = \left[
	\begin{array}{cccccccc}	
		\texttt{\textbf{D}}
		& \texttt{\textbf{D}}
		& \textcolor{red}{\texttt{\textbf{C}}}
		& \texttt{\textbf{D}}
		& |
		& \texttt{\textbf{D}}
		& \textcolor{red}{\texttt{\textbf{C}}}
		& \textcolor{red}{\texttt{\textbf{C}}}
	\end{array}
	\right]
\label{eqn:codeword}
\end{equation}
\cro{Because only \texttt{CHARGED} cells \mpi{can} experience data-retention
errors, there are $2^3=8$ possible error syndromes that correspond to the
unique combinations of \texttt{CHARGED} cells failing.
Table~\ref{tab:lincomb} illustrates these eight possibilities. }

\begin{table}[H]
\centering
\arraycolsep=5pt
\def\arraystretch{1}
\footnotesize
$
\begin{array}{cll}
	\textbf{Pre-Correction} & \multirow{2}{*}{~~~~\textbf{Error~Syndrome}} & \textbf{Post-Correction} \\
	\textbf{Error Pattern} &  &  \multicolumn{1}{c}{\textbf{Outcome}} \\
	\hline
	\left[ \arraycolsep=1.5pt \begin{array}{cccccccc} 0 & 0 & 0 & 0 & | & 0 & 0 & 0 \end{array} \right] & \quad\mathbf{0} & \text{No error} \\
	\left[ \arraycolsep=1.5pt \begin{array}{cccccccc} 0 & 0 & 0 & 0 & | & 0 & 0 & \textcolor{red}{1} \end{array} \right] & \quad{\mathbf{H}_{\ast,6}} & \text{Correctable} \\
	\left[ \arraycolsep=1.5pt \begin{array}{cccccccc} 0 & 0 & 0 & 0 & | & 0 & \textcolor{red}{1} & 0 \end{array} \right] & \quad{\mathbf{H}_{\ast,5}} & \text{Correctable} \\
	\left[ \arraycolsep=1.5pt \begin{array}{cccccccc} 0 & 0 & 0 & 0 & | & 0 & \textcolor{red}{1} & \textcolor{red}{1} \end{array} \right] & \quad{\mathbf{H}_{\ast,5} +\mathbf{H}_{\ast,6}} & \textcolor{red}{\text{Uncorrectable}} \\
	\left[ \arraycolsep=1.5pt \begin{array}{cccccccc} 0 & 0 & \textcolor{red}{1} & 0 & | & 0 & 0 & 0 \end{array} \right] & \quad{\mathbf{H}_{\ast,2}} & \text{Correctable} \\
	\left[ \arraycolsep=1.5pt \begin{array}{cccccccc} 0 & 0 & \textcolor{red}{1} & 0 & | & 0 & 0 & \textcolor{red}{1} \end{array} \right] & \quad{\mathbf{H}_{\ast,2} +\mathbf{H}_{\ast,5}} & \textcolor{red}{\text{Uncorrectable}} \\
	\left[ \arraycolsep=1.5pt \begin{array}{cccccccc} 0 & 0 & \textcolor{red}{1} & 0 & | & 0 & \textcolor{red}{1} & 0 \end{array} \right] & \quad{\mathbf{H}_{\ast,2} +\mathbf{H}_{\ast,6}} & \textcolor{red}{\text{Uncorrectable}} \\
	\left[ \arraycolsep=1.5pt \begin{array}{cccccccc} 0 & 0 & \textcolor{red}{1} & 0 & | & 0 & \textcolor{red}{1} & \textcolor{red}{1} \end{array} \right] & \quad{\mathbf{H}_{\ast,2} +\mathbf{H}_{\ast,5} +\mathbf{H}_{\ast,6}} & \textcolor{red}{\text{Uncorrectable}}
\end{array}
$
\caption{\cro{Possible data-retention error patterns, their syndromes, and their
outcomes for the codeword of Equation~\ref{eqn:codeword}.}}
\label{tab:lincomb}
\end{table}

\mpiii{A miscorrection occurs whenever \mpvi{the error syndrome of an
uncorrectable error pattern} matches the parity-check matrix column of a
\emph{non-erroneous} data bit}. In this case, the column's location would then
correspond to the bit position of the miscorrection. However, \mpiii{a
miscorrection only reveals information if it occurs within one of the
\texttt{DISCHARGED} data bits, for only then are we certain that the observed
bit flip is unambiguously a miscorrection rather than an uncorrected
data-retention error}. Therefore, the test patterns we use should maximize the
number of \texttt{DISCHARGED} bits so as to increase the \mpi{number} of
miscorrections \mpvi{that yield information about the ECC function}.

To determine which test patterns to use, we expand upon the \mpvi{approach of
injecting 1-hot errors} described in
Section~\ref{subsubsec:systematically_identifying}. Although we would need to
write data to all codeword bits in order to test every 1-hot error pattern,
on-die ECC does not allow writing directly to the parity-check bits. This leads
to two challenges. First, we cannot test 1-hot error patterns for which the
1-hot error is within the parity-check bits, which means that we cannot
differentiate ECC functions that differ only within their parity-check bit
positions. Fortunately, this is not a problem because, as
Section~\ref{subsubsec:formalizing_ecc_func} discusses in detail, all such
functions are equivalent codes with identical externally-visible
error-correction properties. Therefore, we are free to assume that the
parity-check matrix is in standard form, which \mpvi{specifies parity-check
bits' error syndromes (i.e., $\mathbf{I}_{n-k \times n-k}$) and obviates the
need to experimentally determine them.} 

Second, \mpi{writing the $k$ bits of the dataword with a single \texttt{CHARGED}
cell results in a codeword with an \emph{unknown} number of \texttt{CHARGED}
cells because the ECC function independently determines the values of remaining
$n-k$ parity-check bits. As a result, the final codeword may contain anywhere
from $1$ to $n-k+1$ \texttt{CHARGED} cells, and the number of \texttt{CHARGED}
cells will vary for different test patterns. Because we cannot directly access
the parity-check bits' values, we do not know which cells are \texttt{CHARGED}
for a given test pattern, and therefore, we cannot tie post-correction errors
back to particular pre-correction error patterns. Fortunately, we can work
around this problem by considering \emph{all possible} error patterns that a
given codeword can experience, which amounts to examining all combinations of
errors that the \texttt{CHARGED} cells can experience. Table~\ref{tab:lincomb}
illustrates this for when the dataword is programmed with a 1-\texttt{CHARGED}
test pattern (as shown in Equation~\ref{eqn:codeword}). In this example, the
encoded codeword contains three \texttt{CHARGED} cells, which may experience any
of $2^3$ possible error patterns.
Section~\ref{subsubsec:testing_charged_patterns} discusses how we can accomplish
testing all possible error patterns in practice by exploiting the fact that
data-retention errors occur uniform-randomly, so testing across many different
codewords provides samples from many different error patterns \mpvi{at once}.}

\subsubsection{\mpi{Shortened Codes}}

\mpi{Linear block codes can be either of} \emph{full-length} if all possible
error syndromes are present within the parity-check matrix (e.g., all $2^p - 1$
error syndromes for a Hamming code with $p$ parity-check bits\mpi{, as is the
case for the code shown in Equation~\ref{eqn:hg}}) or \emph{shortened} if one or
more information symbols are truncated while retaining the same number of
parity-check symbols~\cite{costello1982error, huffman2010fundamentals}.
\mpi{This distinction is crucial for determining appropriate test patterns
because, for full-length codes,} the 1-\texttt{CHARGED} patterns identify the
miscorrection-susceptible bits for all possible error syndromes. In this case,
testing additional patterns that have more than one \texttt{CHARGED} bit
provides no new information because any resulting \mpi{error syndromes are
already tested} using the 1-\texttt{CHARGED} patterns.

However, for \emph{shortened codes}, the 1-\texttt{CHARGED} patterns may not
provide enough information to uniquely identify the ECC function because the
1-\texttt{CHARGED} patterns can no longer test for the missing error syndromes.
Fortunately, we can recover the \mpt{missing information by reconstructing the
truncated error syndromes using pairwise \emph{combinations} of the
1-\texttt{CHARGED} patterns. For example, asserting two \texttt{CHARGED} bits
effectively tests an error syndrome that is the linear combination of the bits'
corresponding parity-check matrix columns. Therefore, by supplementing the
1-\texttt{CHARGED} patterns with the 2-\texttt{CHARGED} patterns, we effectively
encompass the error syndromes that were shortened.}
Section~\ref{subsec:beer1eval_correctness} evaluates BEER's sensitivity to code
length, showing that the 1-\texttt{CHARGED} patterns are indeed sufficient for
full-length codes and the \{1,2\}-\texttt{CHARGED} patterns for shortened codes
\mpi{that we evaluate \mpii{with} dataword lengths \mpii{between} 4 and 247}.

\section{Bit-Exact Error Recovery (BEER)}
\label{sec:beeri}

Our goal in this work is to develop a methodology that reliably and accurately
determines the full ECC function (i.e., its parity-check matrix) for any DRAM
on-die ECC implementation without requiring hardware \mpi{tools, prerequisite
knowledge about the DRAM chip or on-die ECC mechanism,} or access to \mpf{ECC
metadata (e.g., error syndromes, parity information)}. To this end, we present
BEER, which systematically determines the ECC function by observing how it
reacts when subjected to carefully-crafted uncorrectable error patterns. BEER
implements the ideas developed throughout Section~\ref{sec:determining_ecc_func}
and consists of three key steps: (1) experimentally inducing miscorrections, (2)
analyzing observed post-correction errors, and (3) solving for the ECC function.

\mpt{This section describes each of these steps in detail in the context of
experiments using 32, 20, and 28 real LPDDR4 DRAM chips from three major
manufacturers, whom we anonymize for confidentiality reasons as A, B, and C,
respectively. We perform all tests using a temperature-controlled
infrastructure with precise control over the timings of refresh and other DRAM
bus commands.}

\subsection{Step 1: Inducing Miscorrections}

\cro{To induce miscorrections as discussed in
Section~\ref{subsubsec:test_patterns}, we must first identify the (1)
\texttt{CHARGED} and \texttt{DISCHARGED} encodings of each cell and (2) layout
of individual datawords within the address space. This section describes how we
determine these in a way that is applicable to any DRAM chip.}

\subsubsection{Determining \texttt{CHARGED} and \texttt{DISCHARGED} States} 
\label{subsubsec:tacd_layout}

We determine the encodings of the \texttt{CHARGED} and \texttt{DISCHARGED}
states by experimentally measuring the layout of true- and anti-cells throughout
the address space as done in prior works~\cite{kim2014flipping,
kraft2018improving, patel2019understanding}. \cro{We write data `0' and data `1'
test patterns to the entire chip while pausing DRAM refresh for 30 minutes at
temperatures between $30-80^\circ$C.} The resulting data-retention error
patterns reveal the true- and anti-cell layout since each test pattern isolates
one of the cell types. We find that chips from manufacturers A and B use
exclusively true-cells, and chips from manufacturer C use 50\%/50\%
true-/anti-cells organized in alternating \cro{blocks of rows with block lengths
of 800, 824, and 1224 rows. These observations are consistent with the results
of similar experiments performed by prior work~\cite{patel2019understanding}.}

\subsubsection{Determining the Layout of Datawords} 
\label{subsubsec:ecc_word_layout}

To determine which \mpt{addresses} correspond to individual ECC datawords, we
program \mpt{one cell per row\footnote{\mpt{We assume that ECC words do not
straddle row boundaries since accesses would then require reading two rows
simultaneously. However, one cell per \emph{bank} can be tested to accommodate
this case if required.}} to the \texttt{CHARGED} state with all other cells
\texttt{DISCHARGED}.} We then sweep the refresh window $t_{REFw}$ from 10
seconds to 10 minutes at $80^\circ$C \mpi{to induce uncorrectable errors.
Because \emph{only} \texttt{CHARGED} cells can fail, post-correction errors may
\emph{only} occur in bit positions corresponding to either} (1) the
\texttt{CHARGED} cell itself or (2) \mpi{\texttt{DISCHARGED} cells due to a
miscorrection.} By \mpi{sweeping the bit position of the \texttt{CHARGED} cell
within the dataword}, we observe miscorrections that are restricted exclusively
to \emph{within the same ECC dataword.} We find that chips from all three
manufacturers use identical ECC word layouts: each contiguous 32B region of DRAM
comprises two 16B ECC words that are interleaved at byte granularity. A 128-bit
dataword is consistent with prior industry and academic works on on-die
ECC~\cite{kwak2017a, kwon2017an, micron2017whitepaper, patel2019understanding}.

\subsubsection{Testing With {1,2}-\texttt{CHARGED} Patterns} 
\label{subsubsec:testing_charged_patterns}

To test each of the 1- or 2-\texttt{CHARGED} patterns, we program an equal
number of datawords with each test pattern. For example, a 128-bit dataword
yields ${128\choose1} = 128$ and ${128\choose2} = 8128$ 1- and
2-\texttt{CHARGED} test patterns, respectively. \mpi{As
Section~\ref{subsubsec:test_patterns} discusses, BEER must identify all possible
miscorrections for each test pattern. To do so, BEER must exercise all possible
error patterns that a codeword programmed with a given test pattern can
experience (e.g., up to $2^{10} = 1024$ unique error patterns for a (136, 128)
Hamming code using a 2-\texttt{CHARGED} pattern).}

\mpi{Fortunately, although BEER must test a large number of error patterns, even
a single DRAM chip typically contains millions of ECC words (e.g., $2^{24}$
128-bit words for a 16 Gib chip), and we simultaneously test them all when we
reduce the refresh window across the entire chip. Because data-retention errors
occur uniform-randomly (discussed in Section~\ref{bg:subsec:data_retention}), every
ECC word tested provides an independent sample of errors. Therefore, even one
experiment provides millions of samples of different error patterns within the
\texttt{CHARGED} cells, and running multiple experiments at different operating
conditions (e.g., changing temperature or the refresh window) across multiple
DRAM chips\footnote{\mpi{Assuming chips of the same model use the same on-die
ECC mechanism, which our experimental results in
Section~\ref{subsubsec:testing_charged_patterns} support.}} dramatically
increases the sample size, making the probability of not observing a given error
pattern exceedingly low.. We analyze experimental runtime in
Section~\ref{subsec:experiment_runtime}.}

Table~\ref{tab:misc_profiles} \mpii{illustrates testing the 1-\texttt{CHARGED}
patterns using the ECC function given by Equation~\ref{eqn:hg}. There are four
test patterns, and Table~\ref{tab:misc_profiles} shows the miscorrections that
are possible for each one} assuming that all cells are true cells. For this ECC
function, miscorrections are possible \mpiii{\emph{only}} for test pattern 0,
\mpiii{and} no pre-correction error pattern exists that can cause miscorrections
for the other test patterns. Note that, for errors in the \texttt{CHARGED}-bit
positions, we cannot be certain whether a post-correction error is a
miscorrection or simply a data-retention error, so we label it using `?'. We
refer to the cumulative pattern-miscorrection pairs as a \emph{miscorrection
profile}\mpiii{. Thus, Table~\ref{tab:misc_profiles} shows the miscorrection
profile of} the ECC function given by Equation~\ref{eqn:hg}.

\begin{table}[h]
    \scriptsize
    \centering
    \setlength\tabcolsep{3pt} 
    \def\arraystretch{1}
    \begin{tabular}{c|c|c}
        \textbf{1-\texttt{CHARGED} Pattern ID} & \multirow{2}{*}{\textbf{1-\texttt{CHARGED} Pattern}} & \multirow{2}{*}{\textbf{Possible Miscorrections}} \\ 
        \textbf{(Bit-Index of \texttt{CHARGED} Cell)} & & \\ \hline
        3 & $\left[\texttt{\textbf{D}}~\texttt{\textbf{D}}~\texttt{\textbf{D}}~\textcolor{red}{\texttt{\textbf{C}}}\right]$
        & $\left[{-}~{-}~{-}~?\right]$ \\
        2 & $\left[\texttt{\textbf{D}}~\texttt{\textbf{D}}~\textcolor{red}{\texttt{\textbf{C}}}~\texttt{\textbf{D}}\right]$
        & $\left[{-}~{-}~?~{-}\right]$ \\
        1 & $\left[\texttt{\textbf{D}}~\textcolor{red}{\texttt{\textbf{C}}}~\texttt{\textbf{D}}~\texttt{\textbf{D}}\right]$
        & $\left[{-}~?~{-}~{-}\right]$ \\
        0 & $\left[\textcolor{red}{\texttt{\textbf{C}}}~\texttt{\textbf{D}}~\texttt{\textbf{D}}~\texttt{\textbf{D}}\right]$
        & $\left[?~\textcolor{red}{\texttt{\textbf{1}}}~\textcolor{red}{\texttt{\textbf{1}}}~\textcolor{red}{\texttt{\textbf{1}}}\right]$ \\
    \end{tabular}
    \caption{\mpi{Example miscorrection profile for the ECC function given in Equation~\ref{eqn:hg}.}}
    \label{tab:misc_profiles}
\end{table}

\begin{figure}[h]
    \centering
    \includegraphics[width=0.8\linewidth]{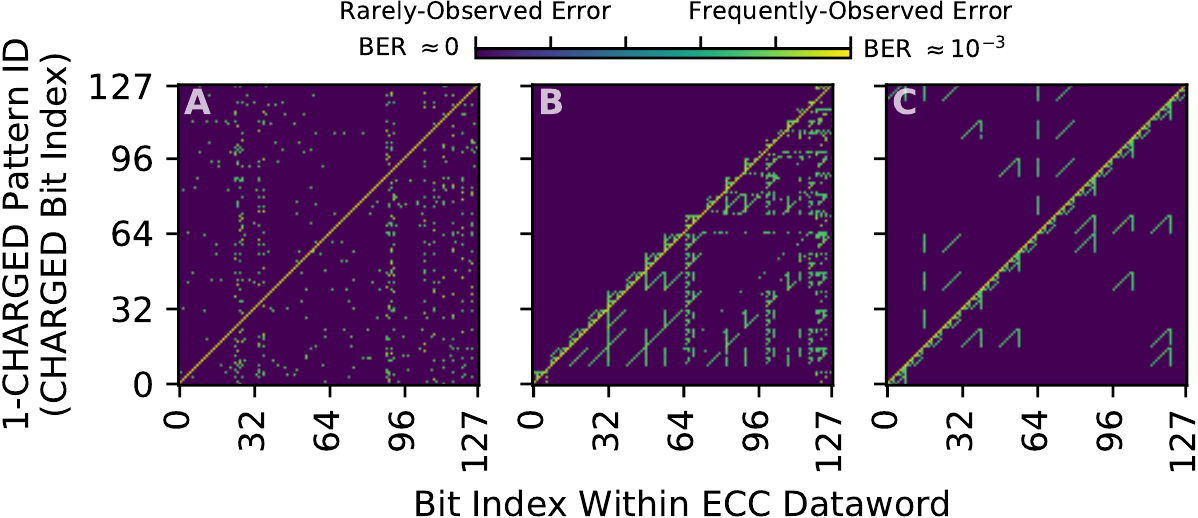}
    \caption{\mpi{Errors observed in a single representative chip from each
    manufacturer using the 1-\texttt{CHARGED} test patterns, showing that
    manufacturers appear to use different ECC functions.}}
    \label{fig:misc_profiles}
\end{figure}

To obtain the miscorrection \mpiii{profile of the on-die ECC function within
each DRAM chip that we test}, we lengthen the refresh window $t_{REFw}$ to
between 2 minutes, where uncorrectable errors begin to occur \mpi{frequently
(BER $\approx 10^{-7}$),} and 22 minutes, where nearly all ECC words exhibit
uncorrectable errors \mpi{(BER $\approx 10^{-3}$)}, in 1 minute intervals at
80$^\circ$C. \mpii{During each experiment, we record which bits are susceptible
to miscorrections for each test pattern} \mpi{(analogous to
Table~\ref{tab:misc_profiles})}. \mpii{Figure~\ref{fig:misc_profiles} shows this
information graphically, giving} the logarithm of the number of errors observed
in each bit position ($X$-axis) for each 1-\texttt{CHARGED} test pattern
($Y$-axis). \mpi{The data is taken from the true-cell regions of a single
representative chip from each manufacturer. \mpii{Errors} in the
\texttt{CHARGED} bit positions (i.e., \mpi{where} $Y=X$) stand out clearly
because \mpiii{they occur alongside all miscorrections as uncorrectable errors.}}

The data shows that miscorrection profiles vary significantly between different
manufacturers. \mpiii{This is likely because each manufacturer uses} a different
parity-check matrix: the possible miscorrections for a given test pattern depend
on which parity-check matrix columns are used to construct error syndromes. With
different matrices, different columns combine to form different error syndromes.
The miscorrection profiles of manufacturers B and C exhibit repeating patterns,
which \mpi{likely} occur due to regularities in how syndromes are organized in
the parity-check matrix, whereas the matrix of manufacturer A appears to be
relatively unstructured. We suspect that manufacturers use different ECC
functions because each manufacturer employs their own circuit design, and
specific parity-check matrix organizations lead to more favorable circuit-level
tradeoffs (e.g., layout area, critical path lengths).

We find that chips of the same model number from the same manufacturer yield
identical miscorrection profiles, which (1) validates that we are observing
design-dependent data and (2) confirms that chips from the same manufacturer and
product generation appear to use the same ECC functions. To sanity-check
our results, we use EINSim~\cite{patel2019understanding, eccsimgithub} to
simulate the miscorrection profiles of the final parity-check matrices we obtain
from our experiments with real chips, and we \mpi{observe} that \mpiii{the
miscorrection profiles obtained via simulation match those measured via real
chip experiments.}

\subsection{Step 2: Analyzing Post-Correction Errors}
\label{subsec:analyzing_postecc_errors}

In practice, BEER may either (1) fail to observe a possible miscorrection or
(2) misidentify a miscorrection due to \mpi{unpredictable} transient errors
(e.g., soft errors from particle strikes, variable-retention time errors,
voltage fluctuations). These events can theoretically pollute the miscorrection
profile with \mpi{incorrect data}, potentially resulting in an \emph{illegal}
miscorrection profile, \mpi{i.e., one} that does not match \emph{any} ECC
function.

Fortunately, case (1) is unlikely given the sheer number of ECC words even a
single chip provides for testing (discussed in
Section~\ref{subsubsec:testing_charged_patterns}). While it is possible that
different ECC words throughout a chip use different ECC functions, we believe
that this is unlikely because it \mpi{complicates the design with no clear
benefits}. Even if a chip does use more than one ECC function, the different
functions will likely follow patterns aligning with DRAM substructures (e.g.,
alternating between DRAM rows or subarrays~\cite{kim2012case, kim2018solar}),
and we can test each region individually. 

\begin{figure}[h]
    \centering
    \includegraphics[width=0.7\linewidth]{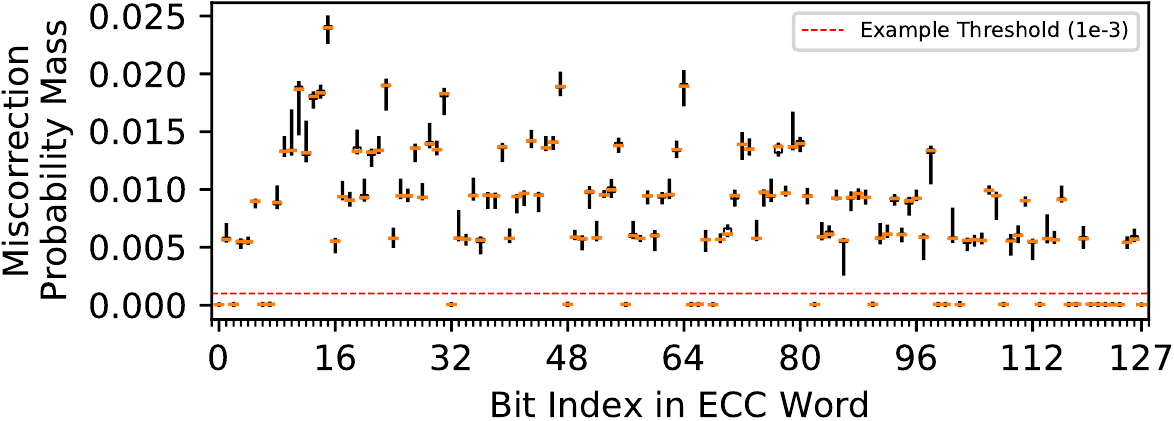}
    \caption{\mpi{Relative probability of observing a miscorrection in each bit
position aggregated across all 1-\texttt{CHARGED} test patterns for a
representative chip of manufacturer B. \mpi{The dashed line shows a threshold
filter separating zero and nonzero values.}}}
    \label{fig:misc_probs_B}
\end{figure}

\mpi{Similarly, case (2) is unlikely because transient errors occur randomly and
rarely~\cite{qureshi2015avatar} as compared with the data-retention error rates
that we induce for BEER \mpi{($>10^{-7}$)}, so transient error
occurrence counts are far lower than those of real miscorrections that are
observed frequently in miscorrection-susceptible bit positions. Therefore, we
apply a simple threshold filter to remove rarely-observed post-correction errors
from the miscorrection profile.} Figure~\ref{fig:misc_probs_B} shows the
\mpi{relative} probability of observing a miscorrection in each bit position
aggregated across all 1-\texttt{CHARGED} test patterns for a representative
chip from manufacturer B. Each data point is a boxplot that shows the full
distribution \mpi{of probability values, i.e., min, median, max, and
interquartile-range (IQR), observed when sweeping the refresh window from 2 to
22 minutes (i.e., the same experiments described in
Section~\ref{subsubsec:testing_charged_patterns}).}

We see that zero and nonzero probabilities are distinctly separated, \mpt{so we
can} robustly resolve miscorrections for each bit. Furthermore, each
distribution is extremely tight, meaning that any of the individual experiments
(i.e., any single component of the distributions) is suitable for identifying
miscorrections. Therefore, a simple threshold filter \mpi{(illustrated in
Figure~\ref{fig:misc_probs_B})} distinctly separates \mpvi{post-correction
errors} that occur near-zero times \mpvi{from miscorrections} that occur
significantly more often.

\subsection{Step 3: Solving for the ECC Function}
\label{subsubsec:rev_engr_ecc_function}

We use the Z3 SAT solver~\cite{de2008z3} (described in
Section~\ref{bg:sec:sat_solver}) to identify the exact ECC function given a
miscorrection profile. To determine the encoding ($F_{encode}$) and decoding
($F_{decode}$) functions, we express them as unknown generator ($\textbf{G}$)
and parity-check ($\textbf{H}$) matrices, respectively. We then add the
following constraints to the SAT solver for $\textbf{G}$ and $\textbf{H}$:
\begin{enumerate}
\item Basic linear code properties (e.g., unique $\textbf{H}$ columns).
\item Standard form matrices, as described in Section~\ref{subsubsec:formalizing_ecc_func}.
\item Information contained within the miscorrection profile
(i.e., pattern $i$ can(not) yield a miscorrection in bit $j$).
\end{enumerate}

\noindent
Upon evaluating the SAT solver with these constraints, the resulting
$\textbf{G}$ and $\textbf{H}$ matrices represent the ECC encoding and decoding
functions, respectively, that cause the observed miscorrection profile. To
verify that no other ECC function may result in the same miscorrection
profile, we simply repeat the SAT solver evaluation with the additional
constraint that the already discovered $\textbf{G}$ and $\textbf{H}$ matrices
are invalid. If the SAT solver finds another ECC function that satisfies the
new constraints, the solution is not unique. 

\cro{To \mpii{seamlessly} apply BEER to the DRAM chips that we test, we develop an
open-source C++ application~\cite{beergithub} that incorporates the SAT solver
and determines the ECC function corresponding to an arbitrary miscorrection
profile. The tool exhaustively searches for all possible ECC functions that
satisfy the aforementioned constraints and therefore will generate the input
miscorrection profile.} Using this tool, we apply BEER to miscorrection profiles
that we experimentally measure across all chips using refresh windows up to 30
minutes and temperatures up to 80$^\circ$C. We find that BEER uniquely
identifies the ECC function for all manufacturers. Unfortunately, we are unable
to publish the resulting ECC functions for confidentiality reasons as set out in
Section~\ref{subsec:motivation_secrecy}. \cro{Although we are confident in our
results because our SAT solver tool identifies a unique ECC function that
explains the observed miscorrection profiles for each chip, we have no way to
validate BEER's results against a groundtruth. To overcome this limitation, we
demonstrate BEER's correctness using simulation in
Section~\ref{subsec:beer1eval_correctness}.}

\subsection{Requirements and Limitations}
\label{subsec:beeri_limitations}

\mpii{Although we demonstrate BEER's effectiveness using both experiment and
simulation, BEER has several testing requirements and limitations that we review
in this section.}

\noindent
\textbf{\mpii{Testing Requirements}}

\begin{itemize}
\item \revmpf{\emph{Single-level ECC}: BEER assumes \mpvi{that there is no
second level of ECC (e.g., rank-level ECC in the DRAM controller) present}
during testing.\footnote{\revmpf{We can potentially extend BEER to multiple
levels of ECC by extending the SAT problem to the concatenated code formed by
the combined ECCs and constructing test patterns that target each level
sequentially, but \mpi{we leave this direction to future work}.}} This is
reasonable since system-level ECCs can typically be bypassed (e.g., \mpi{via}
FPGA-based testing \mpi{or disabling through the BIOS) or
reverse-engineered~\cite{cojocar2019exploiting}, even in the presence of on-die
ECC, before applying BEER}.}

\item \mpii{\emph{Inducing data-retention errors:}} BEER requires finding a
refresh window (i.e., $t_{REFw}$) for each chip that is long enough to
\mpii{induce data-retention errors and} expose miscorrections. Fortunately, we
find that \mpii{refresh windows} between 1-30 minutes at 80$^\circ$C
\mpii{reveal} more than enough miscorrections to apply BEER. In general, the
refresh window \mpi{can be} easily modified (discussed in
Section~\ref{bg:subsec:data_retention}), \mpii{and because data-retention errors
are fundamental to DRAM technology,} BEER applies to all DDRx DRAM families
regardless of their data access protocols and will likely hold for future DRAM
chips\mpiii{, whose data-retention error rates will likely be even more
prominent~\cite{micron2017whitepaper, kang2014co, nair2016xed, gong2017dram,
kwon2014understanding, meza2015revisiting, oh20153, kim2007low, son2015cidra,
liu2013experimental}.}

\end{itemize}

\noindent
\textbf{\revmp{Limitations}}

\begin{itemize}
\item \emph{ECC code type}: BEER works on systematic linear block codes, which
are \mpi{commonly employed} for latency-sensitive main memory chips since: (i) they
allow the data to be directly accessed without additional
operations~\cite{zhang2015vlsi} and (ii) stronger codes (e.g.,
LDPC~\cite{gallager1963low}, concatenated codes~\cite{forney1965concatenated})
cost considerably more area and latency~\cite{cai2017error, nishi2019advances}.

\item \emph{No groundtruth}: BEER alone cannot confirm whether the ECC function
that it identifies is the correct answer. However, if BEER finds exactly one ECC
function that explains the experimentally observed miscorrection profile, it is
\mpi{very} likely that the ECC function is correct.

\item \emph{\mpi{Disambiguating equivalent codes}}: On-die ECC does not expose
the parity-check bits, so BEER can only determine the ECC function to an
equivalent code \mpi{(discussed in Sections~\ref{subsubsec:formalizing_ecc_func}
and~\ref{subsubsec:test_patterns}). Fortunately, equivalent codes differ only in
their internal metadata representations, so this limitation should not hinder
most third-party studies. In general,} we are unaware of any way to disambiguate
equivalent codes without accessing the ECC mechanism's internals.

\end{itemize}

\section{BEER Evaluation}
\label{sec:beer1eval}

\cro{We evaluate BEER's correctness in simulation, \mpi{SAT solver} performance
on a real system, and experimental runtime analytically. Our evaluations both
(1) show that BEER is practical and correctly identifies the ECC function within
our simulation-based analyses, and (2) provide intuition for how the SAT
problem's complexity scales for longer ECC codewords.}

\subsection{Simulation-Based Correctness Evaluation}
\label{subsec:beer1eval_correctness}

We simulate applying BEER to DRAM chips with on-die ECC using a modified version
of the EINSim~\cite{patel2019understanding, eccsimgithub} open-source DRAM
error-correction simulator that we also publicly release~\cite{eccsimgithub}. We
simulate \mpi{115300} single-error correction Hamming code functions that are
representative of those used for on-die ECC~\cite{im2016im, nair2016xed,
micron2017whitepaper, micron2019whitepaper, oh20153, kwak2017a, kwon2017an,
patel2019understanding}: 2000 each for dataword \mpiii{lengths} between 4 and 57 bits,
\mpii{100 each between 58 and 120 bits, and 100 each for selected values}
between 121 and 247 bits because longer codes require significantly more
simulation time. For each ECC function, we simulate inducing data-retention
errors within the 1-, 2-, and 3-\texttt{CHARGED}\footnote{We include the
3-\texttt{CHARGED} patterns to show that \mpvii{they fail to uniquely identify
all ECC functions despite comprising combinatorially more test patterns than the
combined 1- and 2-\texttt{CHARGED} patterns.}} test patterns according to the
data-retention error properties outlined in Section~\ref{bg:subsec:data_retention}.
For each test pattern, we \mpiv{model a real experiment by simulating} $10^9$
ECC words and \mpiii{data-retention error rates} ranging from $10^{-5}$ to
$10^{-2}$ to obtain a miscorrection profile. \cro{Then, we apply BEER to the
miscorrection profiles and show that BEER correctly recovers the original ECC
functions.}

\begin{figure}[h]
    \centering
    \includegraphics[width=0.7\linewidth]{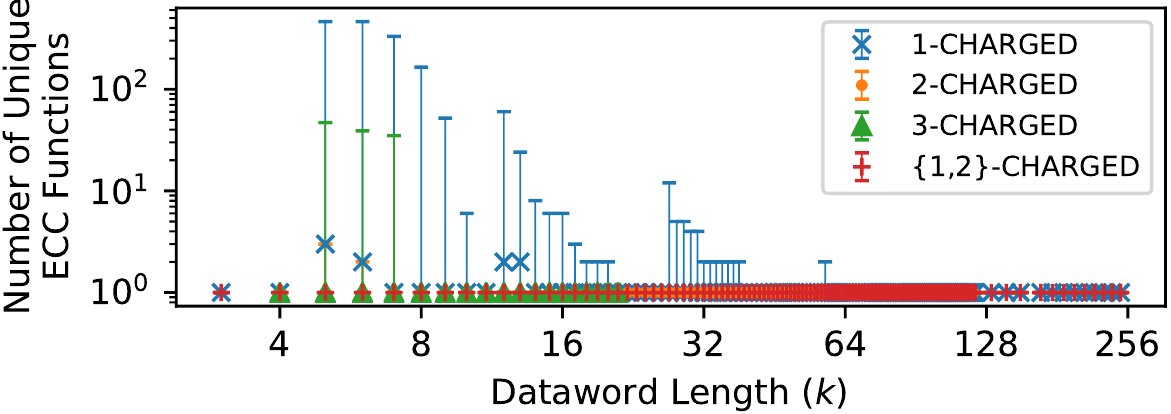}
    \caption{Number of ECC functions that match miscorrection profiles created
    using different test patterns.}
    \label{fig:beer_correctness}
\end{figure}

\mpo{Figure~\ref{fig:beer_correctness} shows how many unique ECC functions BEER
finds when using different test patterns to generate miscorrection profiles. For
each dataword length tested, we show the minimum, median, and maximum number of
solutions identified across all miscorrection profiles. The data shows that BEER
is always able to recover the original \mpi{unique} ECC function using the
\{1,2\}-\texttt{CHARGED} configuration that \mpt{uses both the
1-\texttt{CHARGED} and 2-\texttt{CHARGED} test patterns}. \mpf{For full-length
codes (i.e., \mpi{with dataword lengths} $k\in{4, 11, 26, 57, 120, 247,...}$)
that contain all possible error syndromes within the parity-check matrix by
construction, all test patterns uniquely determine the ECC function, including
the 1-\texttt{CHARGED} patterns alone.}}

On the other hand, the individual 1-, 2-, and 3-\texttt{CHARGED} patterns
sometimes identify multiple ECC functions for shortened codes, with more
solutions identified both for (1) shorter codes and (2) codes with
more aggressive shortening. However, the data shows that BEER often still
uniquely identifies the ECC function even using \mpi{only} the
1-\texttt{CHARGED} patterns \mpii{(i.e., for 87.7\% of all codes simulated)} and
\emph{always} does so with the \{1,2\}-\texttt{CHARGED} patterns. This is
consistent with the fact that shortened codes expose fewer error syndromes to
test (discussed in Section~\ref{subsubsec:test_patterns}). It is important to
note that, \mpv{even if BEER identifies} multiple solutions, it still
narrows a combinatorial-sized search space to a tractable number of ECC
functions that are \mpii{well suited} to more expensive analyses (e.g.,
intrusive error-injection, die imaging techniques, or manual inspection).

While our simulations do not model interference from transient errors, such
errors are rare events~\cite{qureshi2015avatar} when compared with the amount of
\mpiv{uncorrectable data-retention errors that BEER induces}. Even if sporadic
transient errors were to occur, Section~\ref{subsec:analyzing_postecc_errors}
discusses in detail how BEER mitigates their impact on the miscorrection profile
using a simple thresholding filter.

\subsection{Real-System Performance Evaluation}
\label{subsec:beer1eval_perf}

\begin{figure}[h]
    \centering
    \includegraphics[width=0.7\linewidth]{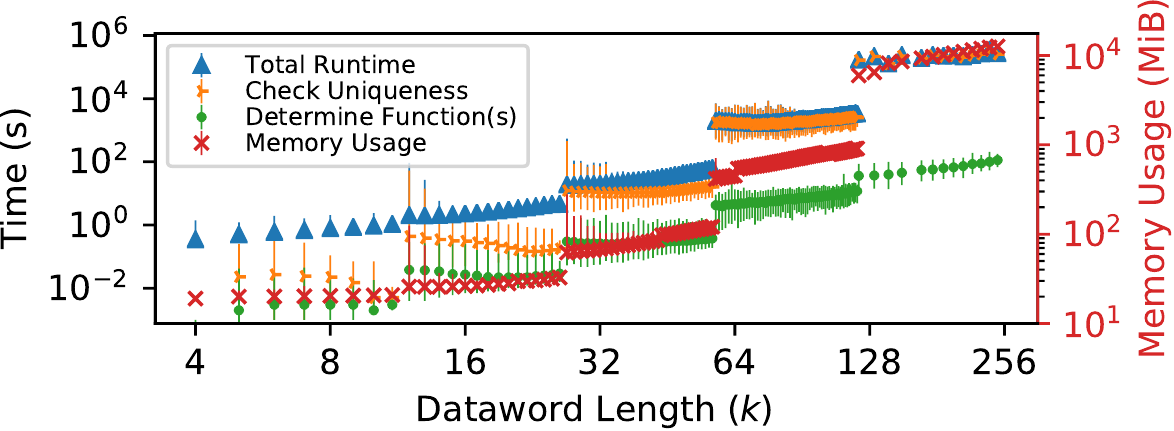}
    \caption{Measured BEER runtime (left y-axis) and memory usage (right
    y-axis) for different ECC codeword lengths.}
    \label{fig:beer1_perf}
\end{figure}

We evaluate BEER's performance and memory usage using ten servers with 24-core
2.30 GHz Intel Xeon(R) Gold 5118 CPUs~\cite{intel2020xeongold5118} and 192 GiB
2666 MHz DDR4 DRAM~\cite{jedec2012ddr4} \mpt{each}. All measurements are taken
\mpi{with Hyper-Threading~\cite{intel2020xeongold5118} enabled and all cores
fully occupied}. Figure~\ref{fig:beer1_perf} shows \mpii{overall runtime} and
memory usage when running BEER with the 1-\texttt{CHARGED} patterns for
different ECC code lengths on a log-log plot \mpii{along with} the time required
to (1) solve for the ECC function (``Determine Function'') and (2) verify the
uniqueness of the solution (``Check Uniqueness''). Each data point gives the
minimum, median, and maximum values observed across our simulated ECC
functions (described in Section~\ref{subsec:beer1eval_correctness}). We see
that the total runtime and memory usage are negligible for short codes and grow
as large as 62 hours and 11.4 GiB of memory for large codes. \mpii{For a
representative dataword length of 128 bits, the median total runtime and memory
usage are 57.1 hours and 6.3 GiB, respectively.} At each code length where we
add an additional parity-check bit, the runtime and memory usage jump
accordingly since the complexity of the SAT evaluation problem increases by an
extra dimension.

The total runtime is quickly dominated by the SAT solver checking uniqueness,
which requires exhaustively exploring the entire search space of a given ECC
function. However, simply determining the solution ECC function(s) is much
faster, requiring less than 2.7 minutes even for the longest codes evaluated and
for shortened codes that potentially have multiple solutions \mpi{using only}
the 1-\texttt{CHARGED} patterns. From this data, we conclude that BEER is
practical for reasonable-length codes used for on-die ECC (e.g., $k=64, 128$).
However, our BEER implementation has room for optimization, e.g., using
dedicated GF(2) BLAS libraries (e.g., LELA~\cite{hovinen2011lela}) or advanced
SAT solver theories (e.g., SMT bitvectors~\cite{brummayer2009boolector}), and an
optimized implementation would likely improve performance, enabling BEER's
application to an even greater range of on-die ECC functions.
Section~\ref{future:subsec:extending_proposed} discusses such optimizations in greater detail.
Nevertheless, BEER is a one-time \emph{offline} process, so it need not be
aggressively performant in most use-cases.

\subsection{Analytical Experiment Runtime Analysis}
\label{subsec:experiment_runtime}

Our experimental runtime \mpi{is overwhelmingly bound by waiting for
data-retention errors to occur during a lengthened refresh window (e.g., 10
minutes) while} interfacing with the DRAM chip requires \mpi{only on the order
of milliseconds} (e.g., 168 ms to read an entire 2 GiB LPDDR4-3200
chip~\cite{jedec2014lpddr4}). Therefore, we estimate \mpi{total} experimental
runtime \mpi{as the sum of the refresh windows that we individually test}. For
the data we present in Section~\ref{subsubsec:testing_charged_patterns},
\mpi{testing each refresh window between 2 to 22 minutes in 1 minute increments
requires a combined 4.2 hours of testing \mpii{for a single chip}. However, if
chips of the same model number use the same ECC functions (as \mpii{our data
supports in} Section~\ref{subsubsec:testing_charged_patterns}), \mpii{we can
reduce overall testing latency by parallelizing individual tests} across
different chips.} Furthermore, because BEER is \mpi{likely} a one-time exercise
for a given DRAM chip, it is sufficient that BEER is practical \mpii{offline}.

\begin{figure*}[b]
    \centering
    \includegraphics[width=\linewidth]{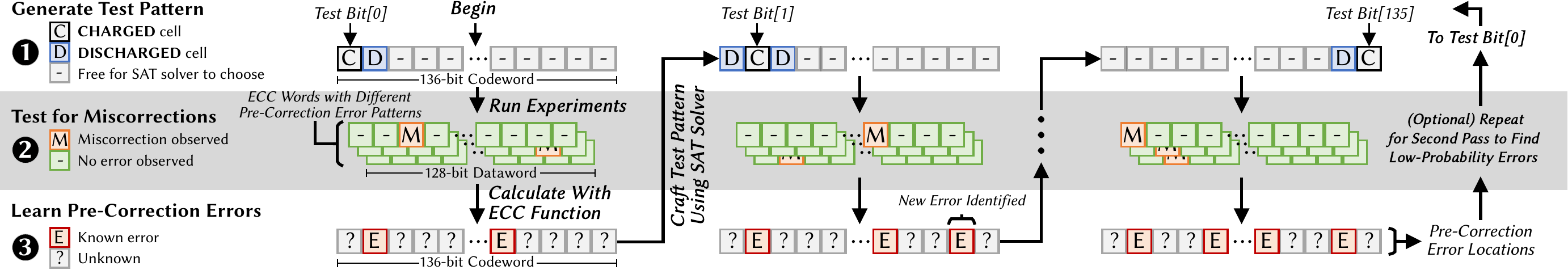}
    \caption{\mpi{Example of running BEEP on a single \mpii{136-bit ECC
    codeword} to identify locations of pre-correction errors.}}
    \label{fig:beep_steps}
\end{figure*}

\section{Example Practical Use-Cases}
\label{section:usecases}

BEER empowers third-party DRAM users to decouple the reliability
characteristics of modern DRAM chips from any particular on-die ECC function
that a chip implements. This section discusses five concrete analyses that
BEER enables. To our knowledge, BEER is the first work capable of
inferring this information without bypassing the on-die ECC mechanism. We hope
that end users and future works find more ways to extend and apply BEER
in practice.

\subsection{BEEP: Profiling for \mpt{Raw Bit} Errors}
\label{subsec:beep}

We introduce \underline{B}it-\underline{E}xact \underline{E}rror
\underline{P}rofiling (BEEP), a new data-retention error profiling algorithm
enabled by BEER that infers the number and bit-exact locations of
\mpt{pre-correction} error-prone cells when given a set of operating conditions that
cause uncorrectable errors in an ECC word. To our knowledge, BEEP is the first
DRAM error profiling methodology capable of identifying bit-exact error
locations throughout the entire on-die ECC codeword, including within the
parity bits.

\subsubsection{BEEP: Inference Based on Miscorrections}

Because miscorrections are purely a function of the ECC logic (discussed in
Section~\ref{subsubsec:using_miscorrections}), an observed miscorrection
indicates that a specific pre-correction error pattern has occurred. Although
several such patterns can map to the same miscorrection, BEEP narrows down the
possible pre-correction error locations by using the known parity-check matrix
\jkz{(after applying BEER)} to construct test patterns for additional
experiments that disambiguate the possibilities. At a high level, BEEP crafts
test patterns to reveal errors as it incrementally traverses each codeword bit,
\mpiv{possibly using multiple passes to capture low-probability errors}. As BEEP
iterates over the codeword, it builds up a list of suspected error-prone cells. 

\mpiii{BEEP comprises three} phases: \circled{1} \jkz{crafting suitable test
patterns}, \circled{2} experimental testing \jkz{with crafted patterns}, and
\circled{3} calculating pre-correction error locations from observed
miscorrections. Figure~\ref{fig:beep_steps} illustrates these three phases in an
example where BEEP profiles for pre-correction errors in a 128-bit ECC dataword.
The following sections explain each of the three phases and refer to
Figure~\ref{fig:beep_steps} as a running example.

\subsubsection{Crafting Suitable Test Patterns}
\label{subsubsec:beep_test_patterns}

Conventional DRAM error profilers (e.g.,~\mpiv{\cite{venkatesan2006retention,
liu2012raidr, liu2013experimental, khan2016parbor, lee2017design,
patel2017reach, khan2017detecting, cheng2002neighborhood, van1991testing,
kraft2018improving, hassan2017softmc, jha2003testing}}) use carefully designed
test patterns that induce worst-case circuit conditions in order to maximize
their coverage of potential errors~\cite{adams2002high, mrozek2019multi}.
\cro{Unfortunately, on-die ECC encodes all data into codewords, so the intended
software-level test patterns likely do not maintain their carefully-designed
properties when written to the physical DRAM cells.} BEEP circumvents these
ECC-imposed restrictions by using a SAT solver along with the known ECC function
\jkz{(via BEER)} to craft test patterns that \mpvi{both} (1) \emph{locally} induce the
worst-case circuit conditions and (2) result in \emph{observable miscorrections}
if suspected error-prone cells do indeed fail.

Without loss of generality, we assume that the worst-case conditions for a given
bit occur when its neighbors are programmed with the opposite charge states,
which prior work shows to exacerbate circuit-level coupling effects and increase
error rates~\mpiv{\cite{adams2002high, liu2013experimental, khan2016parbor,
mrozek2019multi, van2002address, redeker2002investigation, al2004effects,
seyedzadeh2017mitigating, konishi1989analysis, li2011dram, nakagome1988impact}}.
If the design of a worst-case pattern is not known, or if it has a different
structure than we assume, BEEP \mpi{can be adapted} by simply modifying the
relevant SAT solver constraints (described below). To ensure that BEEP observes
a miscorrection when a given error occurs, BEEP crafts a pattern that will
suffer a miscorrection if the error occurs \emph{alongside an
already-discovered} error. We express these conditions to the SAT solver using
the following constraints:

\begin{enumerate}
\item Bits adjacent to the target bit \mpt{have opposing charge states}.
\item One or more miscorrections is possible using some combination of the
already-identified data-retention errors.
\end{enumerate}

\noindent
\cro{Several such patterns typically exist, and BEEP simply uses the first one
that the SAT solver returns (although a different BEEP implementation could test
multiple patterns to help identify low-probability errors).
Figure~\ref{fig:beep_steps} \circled{1} illustrates how such a test pattern
appears physically \mpv{within the cells of a codeword}: the target cell is
\texttt{CHARGED}, its neighbors are \texttt{DISCHARGED}, and the SAT solver
freely determines the \mpv{states of the remaining cells} to increase the
likelihood of a miscorrection if the target cell fails. If the SAT solver
\mpv{fails to find such a test pattern, BEEP attempts to craft a pattern using
constraint 2 alone, which, unlike constraint 1, is} essential to observing
miscorrections. Failing that, BEEP simply skips the bit until more error-prone
cells are identified that could facilitate causing miscorrections. We evaluate
\mpi{how successfully BEEP identifies errors} in
Section~\ref{subsubsec:evaluating_beep}, finding that a second pass over the
codeword helps in cases of few or low-probability errors.}

\subsubsection{\jkz{Experimental Testing with Crafted Patterns}} 
\label{subsubsec:running_beep_test_patterns}

\mpiii{BEEP tests a pattern by writing \mpf{it} to the target ECC word, inducing
errors by lengthening the refresh window, and reading out the post-correction
data. Figure~\ref{fig:beep_steps} \circled{2} shows examples of post-correction
error patterns that might be observed during an experiment. Each miscorrection
indicates that an uncorrectable number of pre-correction errors exists, and BEEP
uses the parity-check matrix $\textbf{H}$ to calculate their precise locations.
This is possible because each miscorrection reveals an error syndrome
$\mathbf{s}$ for the (unknown) erroneous pre-correction codeword $\mathbf{c'}$
that caused the miscorrection}. Therefore, we can directly solve for
$\mathbf{c'}$ as shown in Equation~\ref{eqn:system}.
\begin{equation}
\mathbf{s} = \mathbf{H} * \mathbf{c'} = 
c'_0 \cdot \mathbf{H_{\ast,0}} + 
c'_1 \cdot \mathbf{H_{\ast,1}} + 
... + 
c'_n \cdot \mathbf{H_{\ast,n}} \label{eqn:system}
\end{equation}
This is a system of equations with \jkz{one equation for each of $n-k$
unknowns,} \mpi{i.e., one each for the} $n-k$ inaccessible parity bits. There is
guaranteed to be exactly one solution for $\mathbf{c'}$ since the parity-check
matrix always has full rank (i.e., $\mathrm{rank}(\mathbf{H}) = n - k$). Since
we also know the original codeword
($\mathbf{c}=F_{encode}(\mathbf{d})=\mathbf{G}\cdot\mathbf{d}$), we can simply
compare the two (i.e., $\mathbf{c} \oplus \mathbf{c'}$) to determine the
\emph{bit-exact error pattern} that led to the observed miscorrection.
\mpi{Figure~\ref{fig:beep_steps} \circled{3} shows how BEEP updates a list of
learned pre-correction error locations, which the SAT solver then uses to
construct test patterns for subsequent bits. Once all bits are tested, the list
of pre-correction errors yields the number and bit-locations of all identified
error-prone cells.}

\subsubsection{Evaluating BEEP's Success Rate}
\label{subsubsec:evaluating_beep}

To understand how BEEP performs in practice, we evaluate its \emph{success
rate}, i.e., the likelihood that BEEP correctly identifies errors within a
codeword. We use a modified version of EINSim~\cite{eccsimgithub} to perform
Monte-Carlo simulation across 100 codewords per measurement. To keep our
analysis independent of any particular bit-error rate model, we subdivide
experiments by the number of errors ($N$) injected per codeword. In this way, we
can flexibly evaluate the success rate for a specific error distribution using
the law of total probability over the $N$s.

\begin{figure}[h]
    \centering
    \includegraphics[width=0.7\linewidth]{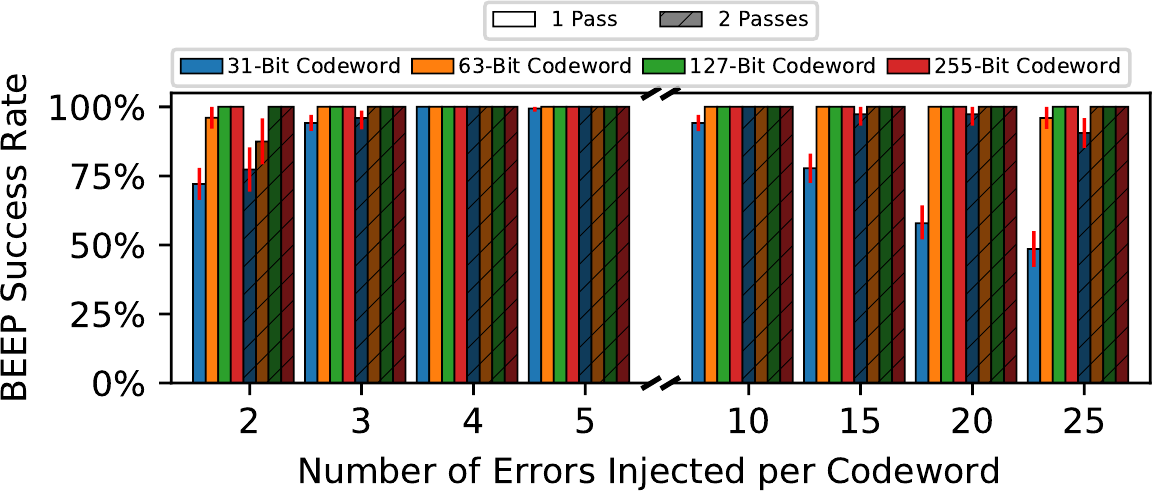}
    \caption{BEEP success rate for 1 vs. 2 passes and
    different codeword lengths and numbers of errors injected.}
    \label{fig:beer2_success_rate_npasses}
\end{figure}

\textbf{Number of Passes.} Figure~\ref{fig:beer2_success_rate_npasses} shows
BEEP's success rate when using one and two passes over the codeword for
different codeword lengths. \mpii{Each bar shows the median value over the 100
codewords with an error bar showing the 5th and 95th percentiles.} The data
shows that BEEP is highly successful across all tested error counts, especially
for longer 127- and 255-bit codewords that show a 100\% success rate \emph{even
with a single pass}. Longer codewords perform better in part because BEEP uses
one test pattern per bit, which means that longer codes \jkz{lead to} more
patterns. However, longer codewords perform better even with comparable
test-pattern counts (e.g., 2 passes with 31-bit vs 1 pass with 63-bit codewords)
because longer codewords \mpi{simply have more bits (and therefore, error
syndromes) for the SAT solver to consider when crafting a miscorrection-prone
test pattern. On the other hand, miscorrection-prone test patterns are more
difficult to construct for shorter codes that provide fewer bits to work with,
so BEEP fails more often when testing shorter codes.}

\begin{figure}[h]
    \centering
    \includegraphics[width=0.7\linewidth]{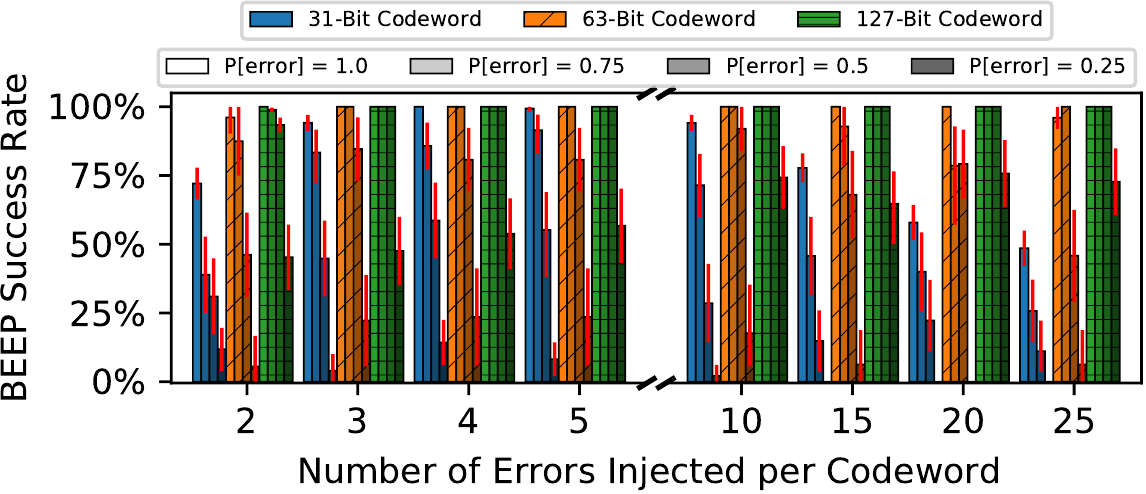}
    \caption{BEEP success rate for different single-bit error
    probabilities using different ECC codeword lengths for different numbers of
    errors injected in the codeword.}
    \label{fig:beer2_success_rate_perror}
\end{figure}

\textbf{Per-Bit Error Probabilities.}
Figure~\ref{fig:beer2_success_rate_perror} shows how \mpii{BEEP's success rate
changes using a single pass} when the injected errors have different per-bit
probabilities of error (P[error]). This experiment represents a more realistic
scenario where some DRAM cells probabilistically experience data-retention
errors. We see that BEEP remains effective (i.e., \mpii{has a near-100\% success
rate}) for realistic 63- and 127-bit codeword lengths, especially at higher
bit-error probabilities and error counts. \jkz{BEEP generally has a higher
success rate with longer codes compared to shorter ones,} and for shorter
codewords at low error probabilities, the data shows that \jkz{BEEP may require}
more test patterns (e.g., multiple passes) to reliably identify all errors.

It is important to note that, while evaluating low error probabilities is
demonstrative, it \mpiii{represents a pessimistic scenario since a real DRAM
chip exhibits} a mix of low and high per-bit error probabilities.\footnote{Patel
et al.~\cite{patel2017reach} provide a preliminary exploration of \mpii{how
per-bit error probabilities are distributed throughout a DRAM chip}, but
\mpii{formulating a detailed error model for accurate simulation} is beyond the
scope of \mpiii{our} work.} \mpi{Although \emph{any} error-profiling mechanism
that identifies errors based on when they manifest might} miss low-probability
errors,\footnote{Patel et al.~\cite{patel2017reach} increase error coverage by
exacerbating the bit-error probability, and their approach \mpii{(REAPER)} can
be used alongside BEEP to help identify low-probability errors.} \jkz{the data
shows that BEEP is resilient} to low error probabilities, especially for longer,
more realistic codewords. Therefore, our evaluations demonstrate that BEEP
effectively enables a new profiling methodology that \mpiii{uses the ECC
function determined by BEER to infer pre-correction errors from observed
post-correction error patterns}.

\subsubsection{\revmp{Other DRAM Error Mechanisms}}
\label{subsubsec:beep_other_error_mechs}

\cro{Although we demonstrate BEEP solely for data-retention errors, BEEP can
potentially be extended to identify errors that occur due to other DRAM error
mechanisms (e.g., stuck-at faults, circuit timing failures). However,
simultaneously diagnosing multiple error models is a very difficult problem
since different types of faults can be nearly indistinguishable (e.g.,
data-retention errors and stuck-at-\texttt{DISCHARGED} errors). Profiling for
arbitrary error types is a separate problem from what we tackle in this work,
and we intend BEEP as a simple, intuitive demonstration of how knowing the ECC
function is practically useful. Therefore, we leave extending BEEP to
alternative DRAM error mechanisms to future work.}

\subsection{Other Use-Cases that Benefit from BEER} 
\label{subsection:use_cases}

\mpii{We identify four additional use cases for which BEER mitigates on-die
ECC's interference with third-party studies by revealing the full ECC function
(i.e., its parity-check matrix).}

\subsubsection{\mpii{Combining Error Mitigation Mechanisms}}

\revmp{If the on-die ECC function is known, a system architect can design a
\mpii{second level of error mitigation} (e.g., rank-level ECC) that better suits
the error characteristics of a DRAM chip with on-die ECC. Figure 1 provides a
simple example of how different ECC functions cause different data bits to be
more error-prone even though the pre-correction errors are uniformly
distributed. This means that on-die ECC \emph{changes} the DRAM chip's
\mpii{software-visible} error characteristics in a way that depends on the
particular ECC function it employs. If the on-die ECC function is known, we can
calculate the expected post-correction error
characteristics\footnote{\revmp{\mpii{By assuming} a given data value
distribution, e.g., \mpii{fixed} values for a \mpii{predictable software}
application, uniform-random data for a general system.}} and build an error
model that accounts for the transformative effects of on-die ECC. Using this
error model, the system architect can make an informed decision when selecting a
secondary mitigation mechanism to complement on-die ECC. For example, architects
could modify a traditional rank-level ECC scheme to asymmetrically protect
certain data bits that are more prone to errors than others as a result of
on-die ECC's behavior~\cite{kraft2018improving, wen2013cd}. In general, BEER
enables system designers to better design secondary error-mitigation mechanisms
to suit the expected DRAM reliability characteristics, thereby improving overall
system reliability.}

\subsubsection{Crafting Targeted Test Patterns}

Several DRAM error mechanisms are highly pattern sensitive, including
RowHammer~\cite{kim2014flipping, mutlu2017rowhammer, mutlu2019rowhammer,
kim2020revisiting}, data-retention~\cite{patel2017reach, liu2013experimental,
liu2012raidr, khan2014efficacy, khan2016parbor, khan2017detecting,
hamamoto1998retention, kim2009new}, and
reduced-access-latency~\cite{kim2018solar, lee2017design, lee2015adaptive,
chang2016understanding, chang2017understanding}. Different test patterns affect
error rates by orders of magnitude~\hht{\cite{lanteigne2016how,
liu2013experimental, patel2017reach, kim2020revisiting, khan2017detecting,
khan2016parbor, khan2016case}} because each pattern
exercises different static and dynamic circuit-level effects.  \cro{Therefore,
test patterns are typically designed carefully to induce the worst-case circuit
conditions for the error mechanism under test (e.g., marching
`1's~\cite{adams2002high, liu2013experimental, patel2017reach,
hassan2017softmc, mrozek2019multi}).} As
Section~\ref{subsubsec:beep_test_patterns} discusses in greater detail, on-die
ECC restricts the possible test patterns to only the ECC function's codewords.
Fortunately, the SAT-solver-based approach that BEEP uses to craft test
patterns generalizes to crafting targeted test patterns for these error
mechanisms also.

\subsubsection{\mpiv{Studying Spatial Error Distributions}}

Numerous prior works~\hht{\cite{shirley2014copula, kim2014flipping,
chang2016understanding, park2016statistical, lee2017design, kim2018solar,
chang2017understanding}}
experimentally study the spatial distributions of errors throughout the DRAM
chip in order to gain insight into how the chip operates and how its
performance, energy, and/or reliability can be improved. These studies rely on
inducing errors at relatively high error rates so that many errors occur that
can leak information about a device's underlying structure. With on-die ECC,
studying spatial error distributions requires identifying
\emph{pre-correction} errors throughout the codeword, including within the
inaccessible parity bits. BEEP demonstrates one possible concrete way by which
BEER \mpf{enables these studies for chips with on-die ECC}.

\subsubsection{\mpiv{Diagnosing Post-Correction Errors}}

A third-party tester may want to determine the physical reason(s) behind an
observed error. For example, a system integrator who is validating a DRAM
chip's worst-case operating conditions may observe unexpected errors due to an
unforeseen defect (e.g., at a precise DQ-pin position). Unfortunately, on-die
ECC obscures both the number and locations of pre-correction errors, so the
observed errors no longer provide insight into the underlying physical error
mechanism responsible. Using BEEP, such errors can be more easily diagnosed
because the revealed pre-correction errors directly result from the error
mechanism.
\section{Summary}

We introduce Bit-Exact Error Recovery (BEER), a new methodology for determining
the full \mpii{DRAM} on-die ECC function (i.e., its parity-check matrix) without
requiring hardware support, \mpi{prerequisite knowledge about the DRAM chip or
on-die ECC mechanism,} or access to \mpf{ECC metadata (e.g., parity-check bits,
error syndromes)}. We use BEER to determine the on-die ECC functions of 80 real
LPDDR4 DRAM chips and show that \mpi{BEER} is both effective and practical using
rigorous simulations. We discuss five concrete use-cases for BEER, including
BEEP, a new DRAM error profiling methodology capable of inferring \mpi{exact}
pre-correction error counts and locations. We believe that BEER takes an
important step towards enabling effective third-party design and testing around
DRAM chips with on-die ECC and are hopeful that BEER will enable many new
studies \mpiii{going forward}.

\chapter[HARP: Profiling Memory Chips with On-Die ECC]{Practical and Effective Error Profiling\\for Memory Chips with On-Die ECC}
\label{chap:harp}

Throughout the previous three chapters, we built a detailed understanding of how
and why on-die ECC obfuscates pre-correction errors. Then, we developed new
testing techniques that provide insight into the on-die ECC mechanism and the
\emph{pre-correction errors}. In this chapter, we closely study the
characteristics of the \emph{post-correction} errors. We examine how and why
on-die ECC makes error profiling difficult from outside of the memory chip,
regardless of whether a profiler knows the details of the on-die ECC
implementation. Then, we develop a new error profiling algorithm, Hybrid
Active-Reactive Profiling (HARP) to overcome the challenges that on-die ECC
introduces for practical and effective error profiling.

\section{Background and Motivation}
\label{sec:introduction}

Modern memory technologies that are suitable for main memory \mpm{(e.g., Dynamic
Random Access Memory (DRAM)~\cite{dennard1968field, mandelman2002challenges},
Phase-Change Memory (PCM)~\gfii{\cite{lee2009architecting, lee2010phasecacm, wong2010phase,
burr2010phase, lee2010phase,qureshi2009scalable}}, Spin-Transfer Torque RAM
(STT-RAM)~\cite{huai2008spin, kultursay2013evaluating})} all suffer from various
error mechanisms that play a key role in determining reliability, manufacturing
yield, and operating characteristics such as performance and energy
efficiency~\cite{lee2016technology, zhao2014improving, itrs2015more,
burr2010phase, wang2017architecting, lee2009architecting, lee2010phasecacm,
kultursay2013evaluating, kang2014co, park2015technology, cha2017defect,
kim2020revisiting, mandelman2002challenges, lee2010phase, guo2017sanitizer}.
Unfortunately, as memory designers shrink \mpm{(i.e., scale)} memory process
\mpm{technology node sizes} to meet \mpm{ambitious capacity, cost, performance,
and energy efficiency} targets, worsening reliability becomes an increasingly
significant challenge to surmount~\cite{lee2016technology, awasthi2012efficient,
mutlu2013memory, vatajelu2018state, cha2017defect, park2015technology,
kang2014co, nair2013archshield, qureshi2015avatar, kim2014flipping,
mutlu2019rowhammer, kim2020revisiting, hong2010memory, mandelman2002challenges,
kim1998dram}. For example, DRAM process technology \mpm{scaling} exacerbates
cell-to-cell variation and noise margins, severely impacting error mechanisms
that constrain yield, including cell data-retention~\cite{liu2013experimental,
nair2013archshield, kline2020flower, khan2016parbor, kang2014co,
venkatesan2006retention, hamamoto1995well, hamamoto1998retention,
park2015technology, cha2017defect, hong2010memory, patel2017reach,
khan2014efficacy, shirley2014copula} and
read-disturb~\mpv{\cite{kim2014flipping, kim2020revisiting, mutlu2019rowhammer,
mutlu2017rowhammer, frigo2020trrespass, orosa2021deeper, hassan2021uncovering}}
phenomena. Similarly, emerging main memory technologies suffer from various
error mechanisms that can lead to high error rates if left unchecked, such as
limited endurance, resistance drift, and write disturbance in
PCM~\cite{lee2009architecting, lee2010phasecacm, lee2009study, kim2005reliability, itrs2015more,
kang2006a, awasthi2012efficient} and data retention, endurance, and read
disturbance in STT-RAM~\cite{vatajelu2018state, chen2010advances,
apalkov2013spin, naeimi2013sttram, chun2012scaling, raychowdhury2009design}.
Therefore, enabling reliable system operation in the presence of scaling-related
memory errors is a critical research \mpq{challenge} for allowing continued main
memory scaling.

\textbf{Error mitigation \gfii{mechanisms} and on-die ECC.} Modern systems
\mpm{tolerate errors} using \emph{error-mitigation mechanisms}, which prevent
errors that occur within the memory \mpq{chip} from manifesting as
software-visible bit flips. Different error-mitigation mechanisms target
different types of errors, ranging from fine- to coarse-grained mitigation using
hardware and/or software techniques. \cref{sec:bg_error_mitigations} reviews main
memory \mpu{error-mitigation mechanisms.}

To address increasing memory error rates, memory chip manufacturers \mpq{have
started} to incorporate \emph{on-die error-correcting codes} (on-die
ECC).\footnote{Our work applies to \emph{any} memory \mpq{chip} that is packaged
with a proprietary ECC mechanism; on-die ECC is one embodiment of such a chip.}
On-die ECC is already prevalent among modern DRAM (e.g.,
LPDDR4~\cite{micron2017whitepaper, oh2014a, kwak2017a, kwon2017an,
patel2019understanding, patel2020bit}, DDR5~\cite{jedec2020ddr5}) and
STT-RAM~\cite{everspin2021sttmram} chips because it enables memory manufacturers
to aggressively scale their technologies while maintaining the appearance of a
reliable memory chip. Unfortunately, on-die ECC changes how memory errors appear
outside the memory chip (e.g., to the memory controller \mpq{or the system
software}). This introduces new challenges for \mpu{designing} additional
error-mitigation mechanisms at the system level~\mpv{\cite{son2015cidra,
gong2018duo, nair2016xed, jeong2020pair, cha2017defect, pae2021minimal,
criss2020improving, luo2014characterizing, gurumurthi2021hbm3}} or test a memory
chip's reliability characteristics~\cite{patel2019understanding, patel2020bit,
gold2014providing, gorman2015memory}.

In this work, we focus on enabling a class of state-of-the-art hardware-based
error-mitigation mechanisms known as \emph{repair mechanisms}\footnote{We
discuss repair mechanisms in detail in
\cref{bg:subsec:repair_mech}.}~\cite{yoon2011free, nair2013archshield,
ipek2010dynamically, schechter2010use, lin2012secret, tavana2017remap,
kline2017sustainable, longofono2021predicting, kline2020flower, nair2019sudoku,
seong2010safer, wilkerson2008trading, zhang2017dynamic, liu2012raidr,
qureshi2015avatar, venkatesan2006retention, son2015cidra} when used alongside
memory chips with on-die ECC. These repair mechanisms \mpm{operate from outside
the memory chip (e.g., from the memory controller)} to identify and repair
memory locations that are at risk of error (i.e., are known or predicted to
experience errors). In particular, prior works~\cite{nair2013archshield,
kline2017sustainable, tavana2017remap, schechter2010use, seong2010safer} show
that \emph{bit-granularity} repair mechanisms efficiently tackle high error
rates (e.g., $>10^{-4}$) resulting from aggressive technology scaling by
focusing error-mitigation resources on \mpq{bits} that are known to be
susceptible to errors.

\begin{figure}[H]
    \centering
    \includegraphics[width=0.7\linewidth]{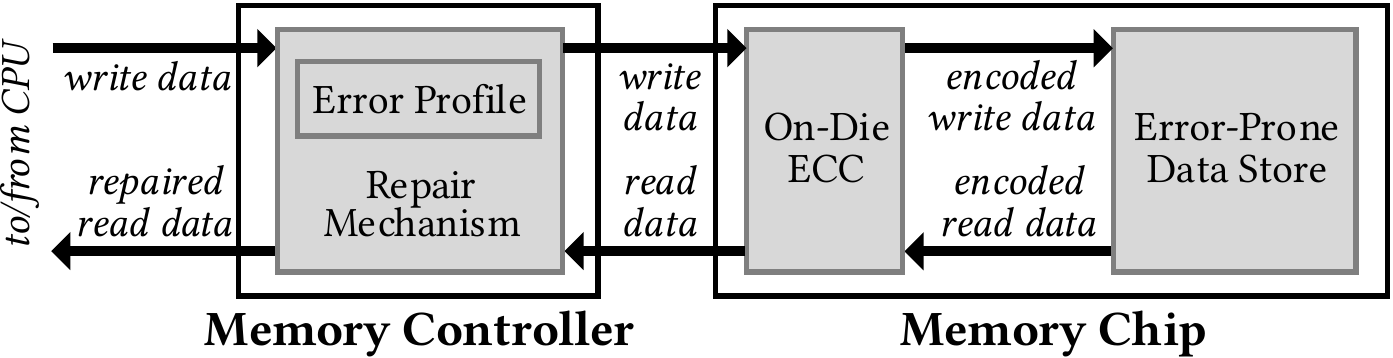}
    \caption{\mpm{High-level block diagram of a system that uses a repair mechanism with a memory chip that uses on-die ECC.}}
    \label{fig:sys_diag_simple}
  \end{figure}

\mpm{Fig.~\ref{fig:sys_diag_simple} illustrates a system that uses both a repair
mechanism (within the memory controller) and on-die ECC (within the memory
chip). \mpu{On-die ECC encodes all data provided by the CPU before writing it}
to the memory. On a read operation, \mpu{on-die ECC first decodes the stored
data, correcting any correctable errors. The repair mechanism then repairs the
data before returning it} to the CPU. The repair mechanism performs repair
operations \mpu{using} a list of bits known to be at risk of error, \mpu{called}
an \emph{error profile}.}

\subsection{Motivation: Understanding and Overcoming Profiling Challenges Introduced by On-Die ECC}

\mpq{Repair} mechanisms depend on having a practical algorithm for identifying
at-risk memory locations to repair. We refer to this as an \emph{error profiling
algorithm}. \mpm{We classify a profiling algorithm as either} \emph{active} or
\emph{reactive} depending on whether it takes action to search for at-risk
memory locations or passively monitors memory to identify errors as they occur
during normal system operation. Prior works~\cite{venkatesan2006retention,
liu2012raidr, khan2014efficacy, khan2016case, khan2016parbor, khan2017detecting,
patel2017reach, qureshi2015avatar, choi2020reducing, lee2017design,
bacchini2014characterization, sharifi2017online, kim2020revisiting,
frigo2020trrespass, cojocar2019exploiting, zhang2012memory, hamdioui2017test,
tavana2017remap, qureshi2011pay, patel2020bit, liu2013experimental,
cojocar2020are, mutlu2019rowhammer, chang2017understanding, lee2015adaptive,
kim2018solar, kim2019d, baek2014refresh, kim2003block, lin2012secret} propose a
variety of algorithms for \emph{active} profiling. In general, these algorithms
all search for errors using multiple \emph{rounds} of tests that each attempt to
induce and identify errors (e.g., by taking exclusive control of the memory
chip~\cite{tavana2017remap, liu2013experimental, patel2017reach,
patel2020bit}). \mpm{The algorithms maximize the chance of observing errors to
identify as many at-risk bits as possible by testing under \emph{worst-case
conditions} (e.g., data and access patterns, operating conditions)}. Only
BEEP~\cite{patel2020bit} accounts for the effects of on-die ECC by first
reverse-engineering the on-die ECC implementation, so we refer to all other
active profiling algorithms as \emph{Naive} in the context of this work. In
contrast, proposals for \emph{reactive} profiling passively monitor an
error-detection mechanism (typically an ECC) to identify errors as they occur
during normal system operation~\cite{jacob2010memory, mukherjee2004cache,
qureshi2015avatar, sharifi2017online, qureshi2011pay, meza2015large,
bajura2007models, saleh1990reliability}.

Regardless of the profiling algorithm, any at-risk bits that the profiler misses
will not be repaired by the repair mechanism that the profiler supports.
Therefore, achieving practical and effective repair requires a profiling
algorithm that \mpq{quickly and efficiently} achieves high coverage of at-risk
memory locations.

\textbf{On-die ECC's impact on error profiling.}
Unfortunately, on-die ECC \emph{fundamentally changes} how memory errors appear
outside of the memory chip: instead of \mpu{errors} that follow well-understood
semiconductor error models~\cite{jedec2016failure}, the system observes
\mpu{obfuscated error patterns that vary with the particular on-die ECC
implementation}~\cite{patel2019understanding, patel2020bit}. This is a serious
challenge for existing profiling algorithms because, as
\cref{sec:impl_on_error_prof} \mpu{shows}, on-die ECC both (1) \emph{increases}
the number of at-risk bits that need to be identified and (2) makes those bits
\mpu{harder} to identify. Even a profiler that knows the on-die ECC
implementation (e.g., BEEP~\cite{patel2020bit}) cannot easily identify all
at-risk bits because it \mpu{lacks} visibility into the error-correction
process.

\textbf{Our goal} is to study and address the challenges that on-die ECC
introduces for \mpm{bit-granularity error profiling}. To this end, we
\mpu{perform} the first analytical study of how \mpu{on-die ECC affects error
profiling}. We \mpu{find} that on-die ECC introduces statistical dependence
between errors in different bit positions such that, even if raw bit errors
(i.e., \emph{pre-correction} errors) occur independently, errors observed by the
system (i.e., \emph{post-correction} errors) do not. This raises three \mpm{new}
challenges for practical and effective bit-granularity error profiling
(discussed in detail in \cref{sec:impl_on_error_prof}).

First, on-die ECC transforms a small set of bits at risk of pre-correction
errors into a \emph{combinatorially larger} set of bits at risk of
post-correction errors. \cref{subsec:combinatorial_explosion} shows how this
exponentially increases the number of bits the profiler must identify. Second,
on-die ECC ties post-correction errors to specific combinations of
pre-correction errors. Only when those specific pre-correction error
combinations occur, can the profiler identify the corresponding bits at risk of
post-correction errors. \cref{subsec:bootstrapping} shows how this significantly
slows down profiling. Third, on-die ECC interferes with commonly-used memory
data patterns that profilers use to maximize the chance of observing errors.
This is because post-correction errors appear only when \emph{multiple}
pre-correction errors occur concurrently, which the data patterns must account
for. \cref{subsec:mutli_bit_test_pattern} \mpu{discusses the difficulty of
defining new data patterns for use with on-die ECC.}

\subsection{Mechanism: Hybrid Active-Reactive Profiling (HARP)}

To address these three challenges, we introduce Hybrid Active-Reactive Profiling
(HARP), a new bit-granularity error profiling algorithm that operates within the
memory controller to support a repair mechanism \mpq{for} memory chips with
on-die ECC. HARP is based on two key insights. First, given that on-die ECC uses
systematic encoding (i.e., data bits are preserved one-to-one during ECC
encoding\footnote{Nonsystematic designs require additional decoding effort
(i.e., more logic operations) because data \emph{cannot} be read \emph{directly}
from stored values~\cite{zhang2015vlsichapter41}. This increases the energy
consumption of read operations and either reduces the overall read timing
margins available for other memory operations or increases the memory read
latency.}), there are only two possible types of post-correction errors: (1)
\emph{direct} errors, corresponding to pre-correction errors within the
systematically encoded data bits; and (2) \emph{indirect} errors, resulting from
\emph{mistaken} correction operations (i.e., \emph{miscorrections}) on-die ECC
performs for \emph{uncorrectable} errors. Second, because on-die ECC corrects a
fixed number $N$ of errors, \emph{at most} $N$ indirect errors can occur
\mpq{concurrently} (e.g., $N=1$ for a Hamming code~\cite{hamming1950error}).

Based on these insights, the key idea of HARP is to \mpr{reduce the problem of
profiling a chip \emph{with} on-die ECC into that of a chip \emph{without}
on-die ECC} by separately identifying direct and indirect errors. HARP consists
of two phases. First, an active profiling phase that \mpr{uses existing
profiling techniques to identify} bits at risk of direct errors \mpr{with the
help of} a simple modification to the on-die ECC read operation that allows the
memory controller to read raw data (but not \mpq{the on-die ECC metadata})
values. Second, a reactive profiling phase that safely identifies bits at risk
of indirect errors using a secondary $N$-error-correcting ECC within the memory
controller. The secondary ECC is used \emph{only} for reactive profiling:
\mpr{upon identifying an error}, the corresponding bit is marked as at-risk,
which signals the repair mechanism to repair the bit thereafter.

Prior work~\cite{patel2020bit} shows that knowing the on-die ECC's internal
implementation (i.e., its parity-check matrix)\footnote{Potentially provided
through manufacturer support, datasheet information, or previously-proposed
reverse engineering techniques~\cite{patel2020bit}.} enables \emph{calculating}
which post-correction errors a given set \mpu{of pre-correction errors can
cause.} Therefore, we introduce two HARP variants: HARP-Aware (HARP-A) and
HARP-Unaware (HARP-U), which do and do not know the parity-check matrix,
respectively. HARP-A does \emph{not} improve coverage \mpu{over} HARP-U because
it has no additional visibility into the pre-correction errors. However, HARP-A
demonstrates that, although knowing the parity-check matrix alone does
\emph{not} overcome the three profiling challenges, it does provide a head-start
for reactive profiling based on the results of active profiling.

We evaluate HARP in simulation relative to two state-of-the-art baseline error
profiling algorithms: \emph{Naive} (which represents the vast majority of
previous-proposed profiling algorithms~\cite{venkatesan2006retention,
liu2012raidr, khan2014efficacy, khan2016case, khan2016parbor, khan2017detecting,
patel2017reach, qureshi2015avatar, choi2020reducing, lee2017design,
bacchini2014characterization, sharifi2017online, kim2020revisiting,
frigo2020trrespass, cojocar2019exploiting, zhang2012memory, hamdioui2017test,
tavana2017remap, qureshi2011pay, patel2020bit, liu2013experimental,
cojocar2020are, mutlu2019rowhammer, chang2017understanding, lee2015adaptive,
kim2018solar, kim2019d, baek2014refresh, kim2003block, lin2012secret}) and
\emph{BEEP}~\cite{patel2020bit}. We show that HARP quickly achieves coverage of
\emph{all bits at risk of direct errors} while Naive and BEEP \mpu{are either
slower or unable} to achieve full coverage. For example, when there are 2/3/4/5
bits at risk of pre-correction error that each fail with probability 0.5,
HARP\footnote{HARP-U and HARP-A have identical coverage of bits at risk of
direct error.} achieves 99th-percentile\footnote{We report 99th percentile
coverage to compare against baseline configurations that do not achieve full
coverage within the maximum simulated number of profiling rounds (due to
simulation time constraints, as discussed in \cref{subsubsec:sim_strat}).}
coverage in 20.6\%/36.4\%/52.9\%/62.1\% of the profiling rounds required by the
best baseline algorithm. Based on our evaluations, we conclude that HARP
effectively overcomes the three profiling challenges. We publicly release our
simulation tools as open-source software on Zenodo~\cite{harp-artifacts} and
Github~\cite{harpgithub}.

\mpq{To demonstrate the end-to-end importance of having an effective profiling
mechanism, we also} perform a case study of how HARP, Naive, and BEEP profiling
can impact the overall \mpi{bit error rate} of a system equipped with \mpm{an
ideal bit-repair mechanism that perfectly repairs all identified at-risk bits}.
We show that, because HARP achieves full coverage of bits at risk of direct
errors, it \mpq{enables} the bit-repair mechanism to repair all errors.
\mpq{Although Naive eventually achieves full coverage, it takes substantially
longer to do so (by $3.7\times$ for a raw per-bit error probability of 0.75). In
contrast, BEEP does \emph{not} achieve full coverage,} so the bit-repair
mechanism is unable to repair all errors that occur \mpq{during system
operation}.
\section{\mpu{Formalizing Error Profiling}}
\label{sec:motivation2}
\label{subsubsec:expressing_error_models}

We express error profiling as a statistical process to understand \mpq{the
effects of} on-die ECC. To this end, we first formalize \mpq{the concepts of}
errors and error profiling. Then, we \mpq{examine how on-die ECC changes the way
that errors appear outside of the memory chip.}

\subsection{Practical and Effective Error Profiling}

\mpj{Any repair mechanism's effectiveness strongly depends on the effectiveness
of the error profiling algorithm that it uses because the repair mechanism can
only repair memory locations that it knows are at risk of error. In this
section, we define the key properties of practical and effective active and
reactive profilers.}

\subsubsection{\mpj{Active Profiling.}}

Active profiling algorithms take exclusive control of a memory \mpp{chip} in
order to (possibly destructively) test worst-case data and access
patterns~\cite{venkatesan2006retention, liu2012raidr, khan2016case,
khan2016parbor, khan2017detecting, patel2017reach, lee2017design,
bacchini2014characterization, kim2020revisiting, frigo2020trrespass,
cojocar2019exploiting, zhang2012memory, hamdioui2017test}, \mpu{so} the system
cannot perform useful work \mpu{while profiling}. Therefore, an active profiler
\mpu{must} identify at-risk bits as quickly and comprehensively as possible. We
quantify this by measuring the \mpu{fraction} of all at-risk bits that a given
profiler \mpu{identifies (i.e., its \emph{coverage})} across a fixed number of
testing rounds.

\subsubsection{\mpj{Reactive Profiling.}}

Reactive profiling algorithms (e.g., ECC scrubbing~\cite{han2014data,
qureshi2015avatar, choi2020reducing, sharifi2017online, alameldeen2011energy,
naeimi2013sttram}) \mpp{passively monitor error-detection mechanisms during
normal system operation, so their performance and energy impact is relatively
low and can be amortized across runtime~\cite{han2014data, qureshi2015avatar}.}
However, because \mpu{reactive profilers operate during \mpv{runtime}, they}
must ensure that they can not only detect, but also correct any errors that
occur. \mpp{Any errors that are not detected and corrected by the reactive
profiler risk introducing failures to the rest of the system.} In our work, we
quantify \mpq{the error-mitigation capability of a reactive profiler in terms of
ECC correction capability}, which is well-defined for ECCs used in memory
hardware design (i.e., linear block codes~\cite{roth2006introduction,
richardson2008modern, horiguchi2011nanoscale}).

\subsection{Errors and Error Models}
\label{subsec:errors_and_models}

Our work assumes uncorrelated single-bit errors because recent experimental
studies and repair efforts from academia~\cite{nair2013archshield,
bautista2016unprotected, longofono2021predicting, nair2019sudoku, son2015cidra},
industry~\cite{kang2014co}, and memory manufacturers
themselves~\cite{micron2017whitepaper, im2020im, ipek2010dynamically,
cha2017defect, park2015technology} focus on single-bit errors as the primary
reliability challenge with increasing storage density. In particular, DRAM and
STT-RAM manufacturers \mpr{use} on-die ECC specifically to combat these errors
in recent high-density chips~\cite{gong2018duo, micron2017whitepaper, im2020im,
kang2014co, oh2014a, everspin2021sttmram, cha2017defect}. Therefore, we assume
that errors exhibit the following properties:

\begin{enumerate}
    \item \emph{Bernoulli process:} independent of previous errors.
    \item \emph{Isolated:} independent of errors in other bits.
    \item \emph{Data-dependent:} dependent on the stored data pattern.
\end{enumerate}

To first order, this \mpj{error model suits} a broad range of error mechanisms
that relate to technology scaling and motivate the use of bit-granularity
repair, including DRAM data-retention~\cite{venkatesan2006retention,
hamamoto1995well, hamamoto1998retention, liu2013experimental,
nair2013archshield, kline2017sustainable, kline2020flower, khan2016parbor,
baek2014refresh, patel2017reach, weis2015retention, jung2017platform,
jin2005prediction, kim2009new, kong2008analysis, lieneweg1998assesment} and read
disturbance~\cite{kim2014flipping, kim2020revisiting, park2016statistical}; PCM
endurance, resistance drift, and write disturbance~\cite{schechter2010use,
lee2009architecting, lee2010phasecacm, lee2009study, kim2005reliability, itrs2015more, kang2006a};
and STT-RAM data retention, endurance, and read
disturbance~\cite{vatajelu2018state, chen2010advances, apalkov2013spin,
naeimi2013sttram, chun2012scaling, raychowdhury2009design}. 
We use DRAM data-retention errors in our evaluations as a well-studied and
relevant example. However, a profiler is fundamentally agnostic to the
underlying error mechanism; \mpq{it identifies} at-risk bits based on whether or
not they \mpj{are observed to fail during} profiling. Therefore, our analysis
applies directly to any error mechanism that can be described using \mpq{the
aforementioned three properties.}

\noindent
\textbf{Correlated Errors.} Prior DRAM studies show evidence of correlated
errors~\mpv{\cite{alars2005dram, meza2015revisiting, synopsys2015whitepaper,
sridharan2012study, sridharan2013feng, sridharan2015memory}}. However, we are
not aware of evidence that such errors are a first-order concern of modern DRAM
technology scaling. Correlated errors often result from faults outside the
memory array~\cite{meza2015revisiting} and are mitigated using fault-specific
error-mitigation mechanisms (e.g., write CRC~\cite{kwon2014understanding,
jedec2012ddr4}, chipkill~\cite{dell1997white, amd2009bkdg, nair2016xed} or even
stronger rank-level ECC~\cite{synopsys2015whitepaper, jedec2012ddr4,
kim2015bamboo}).

\noindent
\textbf{Low-Probability Errors.} Other main memory error mechanisms exist that
do not conform to our model, including time-dependent errors such as DRAM
variable retention time~\cite{yaney1987meta, restle1992dram, mori2005origin,
shirley2014copula, liu2013experimental, khan2014efficacy, qureshi2015avatar} and
single-event upsets such as particle strikes~\cite{may1979alpha}. In general,
these errors are either (1) inappropriate to address using a profile-guided
repair mechanism, e.g., single-event upsets that do not repeat; or (2) rare or
unpredictable enough that no realistic amount of \mpp{active} profiling is
likely to identify them, so they are left to \mpp{reactive profiling for
detection and/or mitigation}~\cite{qureshi2015avatar}. Identifying
low-probability errors is a general challenge for \emph{any} \mpq{error}
profiler and is an orthogonal problem to our work. Prior approaches to
identifying low-probability errors during \mpp{active} (e.g., increasing the
probability of error~\cite{patel2017reach}) or \mpp{reactive (e.g., periodic
ECC scrubbing~\cite{awasthi2012efficient, qureshi2015avatar}) profiling} are
complementary to our proposed techniques and can be combined with HARP (e.g.,
during the active profiling phase \mpp{described in
\cref{subsec:active_prof_phase}, or by strengthening the secondary ECC as
described in \cref{subsubsection:incomplete_coverage}) to help identify
low-probability errors.}

\subsection{Representing the Probability of Error}
\label{subsec:rep_prob_error}

We model memory as a one-dimensional bit-addressable array with address space
$\mathcal{A}$. To describe \mpu{errors} within this address space, we define two
Boolean random variables $D_i$ and $E_i$ that represent the data stored in bit
$i \in \mathcal{A}$ and whether or not the bit experiences an error,
respectively. Based on our discussion in \cref{subsec:errors_and_models}, we
model $E_i$ as a Bernoulli random variable that is independent of $E_{j\neq i}$
but dependent on the data $D_i$. In general, $E_i$ can depend on the data stored
in other cells $D_{j \neq i}$, which expresses how \mpu{a bit's} probability of
error \mpu{changes with the data stored in nearby cells}.
Equation~\ref{eqn:bernoulli} shows the resulting probability mass function,
parameterized by $p$, the probability that the bit will experience an error.
\begin{equation}
  P(E_i=x|D_i, D_j, \cdots) =
  \begin{cases}
    p(D_i, D_j, \cdots)  & \text{if $x=1$} \\
    1 - p(D_i, D_j, \cdots) & \text{if $x = 0$}
  \end{cases}
  \label{eqn:bernoulli}
\end{equation}
In general, \mpk{each bit has its own value of $p$ depending on its intrinsic
error characteristics. For example, prior work~\cite{patel2017reach}
experimentally demonstrates that $p$ values are normally distributed across
different bits for DRAM data-retention errors}, i.e., $p\sim N(\mu,
{\sigma}^2)$, with \mpk{some} normal distribution parameters $\{\mu, \sigma\}$
\mpk{that depend on the particular} memory chip and operating conditions such as
temperature.

\subsection{Incorporating On-Die ECC}
\label{subsec:incorporating_ecc_in_prof}

With on-die ECC, we adjust our address space representation to include both
\emph{logical} bit addresses $\mathcal{A}$ as observed by the memory controller
and \emph{physical} bit addresses $\mathcal{B}$ within the memory storage array.
In general, $|\mathcal{B}|>|\mathcal{A}|$ because $\mathcal{B}$ includes
addresses for parity-check bits that are \emph{not} visible outside of the
memory chip. Next, we introduce two additional \mpq{Boolean random variables:
$D_a$ \mpp{(for \underline{d}ataword)} and $C_b$ \mpp{(for
\underline{c}odeword)} that refer to the data values of logical bit
$a\in\mathcal{A}$ and physical bit $b\in\mathcal{B}$ (i.e., before and after ECC
encoding)}, respectively. Boolean variables $E_a$ and $R_b$ represent whether
logical bit $a$ and physical bit $b$ experience post- and pre-correction errors,
respectively. Note that $E$ and $D$ represent the same information from
\cref{subsec:rep_prob_error}. We use $C'$ and $D'$ to refer to codeword
and dataword values, respectively, that may contain errors. 

On-die ECC determines $D'$ from $C'$ through \mpu{syndrome decoding} (described
in \cref{bg:subsec:ondieecc_hamming_code}) \mpu{using} the ECC parity-check matrix $H$
\gfii{comprised of} columns $H[k+p:0]$. The error syndrome is computed as
$s=H[k+p:0] \cdot R[k+p:0]$ (referred to as $H \cdot R$ to simplify notation).
Then, if $s$ matches the $i$'th column $H[i]$, the ECC decoder flips the bit at
position $i$. Given that $H$ is systematically encoded (discussed in
\cref{bg:subsec:ondieecc_hamming_code}), $c[k:0]$ is equal to $d[k:0]$.
Therefore, a post-correction error $E_i$ (i.e., a mismatch between $d_i$ and
$d'_{i}$ ) can only occur in two cases: (1) an uncorrected raw bit error at
position $i$ (i.e., $R_i \wedge s\neq H[i]$); or (2) a miscorrection at position
$i$ (i.e., $\neg R_i \wedge s=H[i]$.) We refer to these two cases as
\emph{direct} and \emph{indirect} errors, respectively.
Equation~\ref{eqn:dr_error_post_ecc} summarizes both cases that lead to a
post-correction error.
\begin{align}
    P(E_i) &= P(R_i \veebar H \cdot R=H[i])
  \label{eqn:dr_error_post_ecc}
\end{align}
Equation~\ref{eqn:dr_error_post_ecc} \mpu{shows that bit $i$'s probability of
error depends \emph{not only} on that of its encoded counterpart $R_i$, but
\emph{also} on those of} \emph{all other codeword bits} $R[k+p:0]$. This means
that on-die ECC introduces statistical dependence between \emph{all bits} in a
given ECC word through their mutual dependence on $R$. \mpq{Furthermore, just as
described in \cref{subsec:rep_prob_error}, $R_i$ itself depends on the data
value stored in cell $i$ (i.e., $C_i$). As a result, \mpu{bit $i$'s probability
of error} depends on \emph{both} the data values and the pre-correction errors
\mpu{present} throughout the codeword.}

\mpq{Consequently, a given post-correction error $E_i$ may \emph{only} occur
when a particular combination of pre-correction errors occurs. This is different
from} the case without on-die ECC, where $E_i$ does \emph{not} depend on the
data or error state of any other bit \mpq{$j \neq i$}. We conclude that on-die
ECC transforms statistically independent pre-correction errors into
ECC-dependent, correlated post-correction errors.
\section{\mpu{On-Die ECC's Impact on Profiling}}
\label{sec:impl_on_error_prof}

On-die ECC breaks the simple and intuitive assumption that profiling for errors
is the same as profiling each bit individually. In this section, we identify
three key challenges that on-die ECC introduces for bit-granularity profiling.

\subsection{Challenge 1: Combinatorial Explosion}
\label{subsec:combinatorial_explosion}

\mpq{\cref{subsec:incorporating_ecc_in_prof} shows that the position of
an indirect error depends on the locations of all pre-correction errors $R$.
This means that different uncorrectable patterns of pre-correction errors can
cause indirect errors in different bits. In the worst case, \emph{every unique
combination} of pre-correction errors within a set of at-risk bits can lead to
different indirect errors. This means that the set of bits that are at risk of
post-correction errors is \emph{combinatorially larger} than the set of bits at
risk of pre-correction error.}

\begin{figure}[H]
  \centering
  \includegraphics[width=0.7\linewidth]{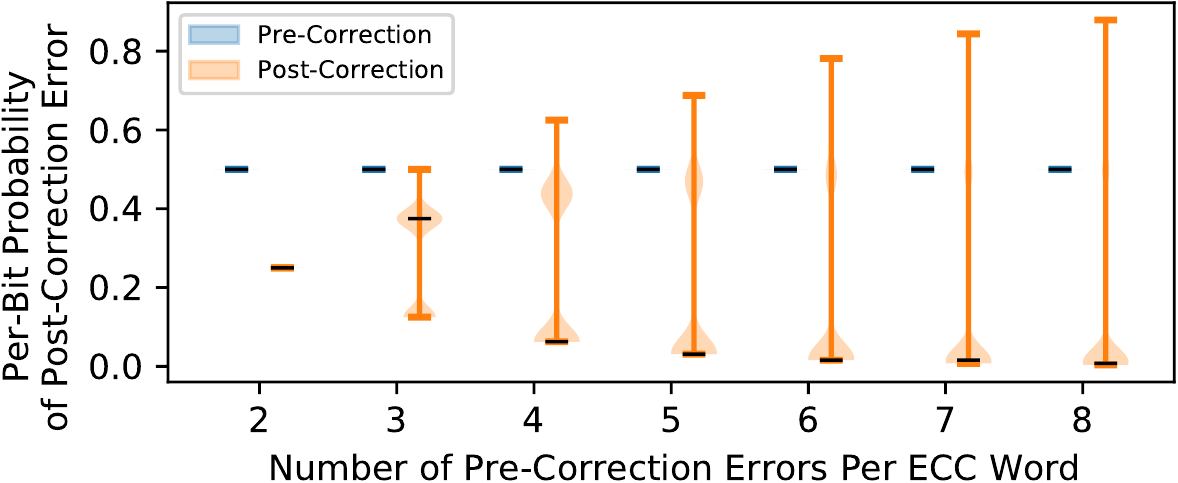}
  \caption{Distribution of \mpq{each at-risk bit's error probability} before and
  after application of on-die ECC.}
  \label{fig:pre_and_post_errors}
\end{figure}

As a concrete example, Fig.~\ref{fig:pre_and_post_errors} shows a violin plot of
each at-risk bit's per-bit probability of error (y-axis) when the codeword
contains a fixed number of bits at risk of pre-correction errors (x-axis) that
each fail with probability $0.5$. Each violin shows the distribution
\mpu{(median marked in black)} of per-bit error probabilities when simulating
70,000 ECC words for each of 1600 randomly-generated (71, 64) Hamming code
parity-check matrices assuming a data pattern of 0xFF. We make two observations.
First, the pre-correction error probabilities are all 0.5 (by design). This
means that, without on-die ECC, all bits at risk of pre-correction error are
easy to identify, i.e., each bit will be identified with probability
$p=1-0.5^{N}$ given $N$ profiling rounds. For large $N$ (e.g., $N>10$, where $p>
0.999$), the vast majority of bits will be identified. 

\mpq{Second, in contrast, the per-bit probabilities of post-correction error
exhibit a wide range. However, the probability density for each violin is
tightly concentrated at $Y\approx 0.4$ for $X=3$ and shifts towards $Y=0$ as the
number of pre-correction errors increases. This means that the bits at risk of
post-correction errors become \emph{harder to identify} because they fail less
often.}

\mpq{Table~\ref{tab:amplification} shows the maximum number of bits at risk of
post-correction errors that can be caused by a fixed number of bits at risk of
pre-correction errors. This illustrates the worst-case scenario, where
\emph{every \mpr{uncorrectable} combination} of pre-correction errors (i.e.,
\emph{pre-correction error pattern}) causes a unique indirect error. We see that
$n$ bits at risk of pre-correction errors can cause $2^{n}-1$ unique
pre-correction error patterns. Of these, $n$ are correctable error patterns,
leaving $2^{n}-n-1$ \emph{uncorrectable pre-correction error patterns}. Assuming
that each of these patterns introduces a unique indirect error, the combination
of bits at risk of direct and indirect error leads to $2^n-1$ bits at risk of
post-correction errors. Therefore, we conclude that on-die ECC
\mpr{exponentially} increases the number of at-risk bits that the profiler must
identify.}

\begin{table}[h]
    \centering
    \small
    \setlength\tabcolsep{2.5pt}
    \def\arraystretch{1}
    \begin{tabular}{lllllll}
        \textbf{Bits at risk of pre-correction errors}
          & \boldmath{$n$} & \textbf{1} & \textbf{2} & \textbf{3} & \textbf{4} & \textbf{8}\\\hline
        Unique pre-correction error patterns
          & $2^n - 1$ & 1 & 3 & 7 & 15 & 255 \\
        Uncorrectable pre-correction error patterns
          & $2^n - n - 1$ & 0 & 2 & 4 & 11 & 247\\\hline
        \textbf{Bits at risk of post-correction errors}
          & \boldmath{$2^n - 1$} & \textbf{1} & \textbf{3} & \textbf{7} & \textbf{15} & \textbf{255}\\
    \end{tabular}
    \caption{On-die ECC amplifies a few bits at risk of pre-correction errors into exponentially many bits at risk of post-correction errors.}
    \label{tab:amplification}
  \end{table}

\subsection{Challenge 2: Profiling without Feedback}
\label{subsec:bootstrapping}

\mpq{Without on-die ECC, an at-risk bit is identified when the bit fails. This
means} that every profiling round provides useful feedback about which bits
\emph{are} and \emph{are not} at risk of error. \mpq{Unfortunately, with on-die
ECC, a bit at risk of post-correction errors can \emph{only} be identified} when
\emph{particular combination(s)} of pre-correction errors occur. \mpq{This has
two negative consequences.}
 
\mpq{First, because the profiler cannot observe pre-correction errors, it} does
not know whether a particular combination of pre-correction errors has been
tested \mps{yet}. Therefore, the profiler \emph{cannot} draw meaningful
conclusions from observing a bit \emph{not} to fail. \mpq{Instead, the profiler
must pessimistically suspect every bit to be at risk of post-correction errors,
even after many \mps{profiling} rounds have elapsed without observing a given
bit fail.} \mps{Second, each ECC word can only exhibit \emph{one} pre-correction
error pattern at a time (i.e., during any given profiling round). This}
serializes the process of identifying any two bits at risk of post-correction
errors that fail under different pre-correction error patterns.

As a result, \mpu{no} profiler that identifies at-risk bits based on observing
post-correction errors \mpu{can} quickly identify all bits at risk of
post-correction errors. We refer to this problem as \emph{bootstrapping} because
the profiler must explore different pre-correction error patterns without
knowing which patterns it is exploring. In \cref{subsubsec:bootstrapping_anal},
we find that bootstrapping \gfii{limits the profiler's coverage of \mpu{at-risk
bits} to incremental improvements across profiling rounds.} 

\subsection{Challenge 3: Multi-Bit Data Patterns}
\label{subsec:mutli_bit_test_pattern}

Designing data patterns that induce worst-case circuit conditions is a difficult
problem that depends heavily on the particular circuit design of a given memory
chip and the error mechanisms it is susceptible to~\cite{dekker1990realistic,
cheng2002neighborhood, khan2018test, mrozek2019multi, mrozek2010analysis,
cui2016snake}. Without on-die ECC, a bit can fail \mpq{only} in one way, i.e.,
when it exhibits an error. Therefore, the worst-case pattern \mpq{needs to} only
consider factors that affect the bit itself (e.g., data values stored in the bit
and its neighbors).

Unfortunately, with on-die ECC, a given post-correction error can potentially
occur with multiple different pre-correction error patterns. Therefore, the
worst-case data pattern must both (1) account for different ways in which the
post-correction error can occur; and (2) for each way, consider the worst-case
conditions \mpq{for the individual pre-correction errors to occur
simultaneously}. This is a far more complex problem than without on-die
ECC~\cite{gold2014providing, gorman2015memory}, and in general, there may not
even be a single worst-case data pattern that exercises all possible cases in
which a given post-correction error might occur. To our knowledge, no prior work
has \mpq{developed a general solution to this problem,} and we identify this as
a key direction for future research.
\section{\mpu{Addressing the Three Challenges}}
\label{subsec:practical_prof}
\label{sec:addressing_challenges}

We observe that all three profiling challenges stem from the \emph{lack of
access} that the profiler has into pre-correction errors. Therefore, we conclude
that \emph{some} transparency into the on-die ECC mechanism is necessary to
enable practical error profiling in the presence of on-die ECC. \mpu{This
section discusses options for enabling} access to pre-correction errors and
\mpv{describes} our design choices for HARP.

\subsection{Necessary Amount of Transparency}

\mpn{To reduce the number of changes we require from the memory chip, we
consider the minimum amount of information that the profiler needs to make error
profiling \emph{as easy as} if there were no interference from on-die ECC. To
achieve this goal, we examine the following two insights that are derived in \cref{subsec:incorporating_ecc_in_prof}.}

\begin{enumerate}
    \item \mpn{Post-correction errors arise from either direct or indirect
    errors.}
    
    \item \mpn{The number of concurrent indirect errors is limited to the correction
    capability of on-die ECC.}
\end{enumerate} 

\mpn{First, we observe that it is \emph{not} necessary for the profiler to have
full transparency into the on-die ECC mechanism or pre-correction errors. If all
bits at risk of direct errors can be identified, all remaining indirect errors
are upper-bounded by on-die ECC's correction capability. Therefore, the indirect
errors can be safely identified from within the memory controller, e.g., using a
reactive profiler.}

\mpn{Second, we observe that the profiler can determine exactly which
pre-correction errors occurred within the data bits (though not the parity-check
bits) simply by knowing at which bit position(s) on-die ECC performed a
correction operation. This is because the data bits are systematically encoded
(\mpq{as} explained in \cref{bg:subsec:ondieecc_hamming_code}), so their
programmed values must match their encoded values. By observing which bits
experienced direct error(s), the profiler knows which pre-correction errors
occurred within the encoded data bits.}

Based on these observations, we require that the profiler be able to identify
which direct errors occur on every access, including \mpu{those that on-die ECC
corrects}. Equivalently, on-die ECC may expose the error-correction operation
that it performs so that the profiler can infer the direct \mpu{errors from the
post-correction data.}

\subsection{Exposing Direct Errors to the Profiler}

We consider two different ways to inform the profiler about pre-correction
errors within the data bits.

\begin{enumerate}

\item \emph{Syndrome on Correction.} On-die ECC reports the error syndrome
calculated on all error correction events, which corresponds to the bit
position(s) that on-die ECC corrects.

\item \emph{Decode Bypass.} On-die ECC provides a read access path that bypasses
error correction and returns the raw values stored in the data portion of the codeword. 
\end{enumerate}

\noindent 
We choose to build upon decode bypass because we believe it to be the easiest to
adopt for three key reasons. First, there exists precedent for similar on-die
ECC decode bypass paths from both academia~\cite{gong2018duo} and
industry~\cite{bains2020read} with trivial modifications to internal DRAM
hardware, and an on-die ECC disable configuration register is readily exposed in
certain DRAM datasheets~\cite{alliance2020lpddr4}. Second, prior works already
reverse-engineer both the on-die ECC algorithm~\cite{patel2020bit} and raw bit
error rate~\cite{patel2019understanding} without access to raw data bits or
insight into the on-die ECC mechanism, so we do not believe exposing a decode
bypass path reveals significantly more sensitive information. Third, we strongly
suspect that such a bypass path \emph{already exists} for post-manufacturing
testing~\cite{thun2020qualification}. This is reasonable because
systematically-encoded data bits can be read out directly without \mpv{requiring
further transformation.} If so, exposing this capability as a feature would
likely require minimal engineering effort for the potential gains of new
functionality. However, we recognize that the details of \mpn{the on-die ECC
implementation} depend on the particular memory chip design, and it is
ultimately up to the system designer to choose the most suitable option for
their system. 

\subsection{Applicability to Other Systems}

Any bit-granularity profiler operating without visibility into \mpn{the
pre-correction errors} suffers form the three profiling challenges \mpn{we
identify in this work}. Even a hypothetical future main memory system whose
memory chips and controllers are designed by the same (or two trusted) parties
will need to account for \mpq{and overcome} these profiling challenges when
incorporating a repair mechanism \mpq{that relies on practical and effective
profiling}.
\section{\mpu{Hybrid Active-Reactive Profiling}}

We introduce the Hybrid Active-Reactive Profiling (HARP) algorithm, which
overcomes the three profiling challenges introduced by on-die ECC \mpq{discussed
in \cref{sec:addressing_challenges}. HARP separates profiling into \emph{active}
and \emph{reactive} phases that independently identify bits at risk of
\emph{direct} and \emph{indirect} errors, respectively.}

\subsection{HARP Design Overview}
\label{subsec:harp_design_overview}

\begin{figure}[hb]
    \centering
    \includegraphics[width=\textwidth]{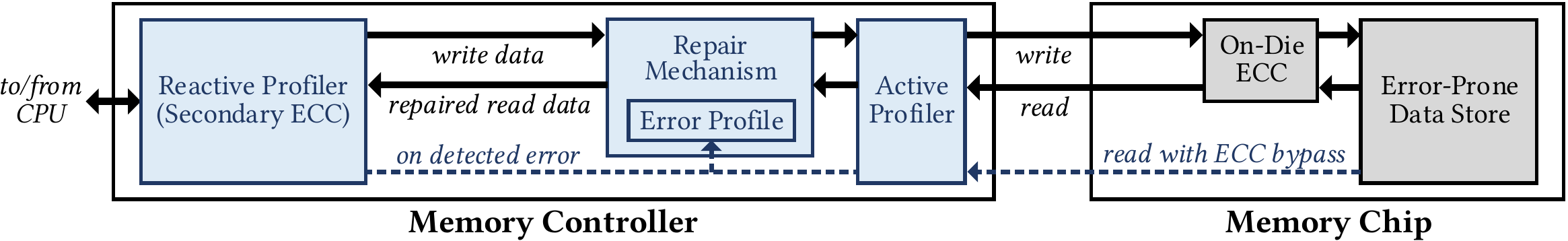}
    \caption{Block diagram summarizing the error-mitigation resources (in blue) of a HARP-enabled system.}
    \label{fig:harp_sys_diagram}
\end{figure}

\mpq{Fig.~\ref{fig:harp_sys_diagram} illustrates the high-level architecture
of a HARP-enabled system, with the required error-mitigation resources shown in
blue. The memory chip exposes a read operation with the ability to bypass on-die
ECC and return the raw data (though not parity-check) bits. The memory
controller contains a repair mechanism with an associated error profile
alongside both an \emph{active} and \emph{reactive} profiler. During active
profiling, the active profiler uses the ECC bypass path to search for bits at
risk of direct errors. Because the active profiler interfaces
\gfii{\emph{directly}} with the raw data bits, its profiling process is
equivalent to profiling a memory chip without on-die ECC. If and when direct
errors are observed, the active profiler communicates their locations to the
repair mechanism's error profile.}

\mpq{After active profiling is complete, the reactive profiler (i.e., a
secondary ECC code with correction capability \gfii{at least as strong as} that
of on-die ECC) continuously monitors for bits at risk of indirect errors. The
reactive profiler is responsible \emph{only} for identifying bits at risk of
indirect errors the first time that they fail. If and when the reactive profiler
identifies an indirect error, the location of the error is recorded to the error
profile for subsequent repair.}

\subsection{Active Profiling Implementation}
\label{subsec:active_prof_phase}

\mpq{The active profiler follows the general round-based algorithm employed by
state-of-the-art error profilers, as} discussed in
\cref{sec:introduction}. Each round of testing first programs memory
cells with a standard memory data pattern that is designed to maximize the
chance of observing errors (e.g., $0xFF$, $0x00$, random
data)~\gfii{\cite{adams2002high, mrozek2019multi, liu2013experimental, khan2016parbor, kim2020revisiting}}. Patterns may
or may not change across testing rounds depending on the requirements of the
particular data pattern. Once sufficiently many rounds are complete, the set of
at-risk bits identified comprises the union of all bits identified across all
testing rounds.

We assume that the active profiler achieves full coverage of bits at risk of
direct errors by leveraging any or all of the worst-case testing techniques
developed throughout prior works~\cite{patel2017reach, khan2016parbor,
khan2017detecting, mukhanov2020dstress, mrozek2019multi}. \mpq{This is feasible
because the active profiler can read and write to the raw data bits
\mpr{exploiting} the ECC bypass path and the systematically-encoded data bits,
respectively. Therefore, the active profiler can use techniques developed for
memory chips without on-die ECC.}

\subsection{Reactive Profiling Implementation}
\label{subsubsec:matching_ondie_and_secondary_ecc}

HARP requires that the secondary ECC \mpq{have correction capability \emph{at
least} as high as the number of indirect errors that on-die ECC can cause at one
time. This requires \mpr{the} layout of secondary ECC words to account for the
layout of on-die ECC words: the two must combine in such a way that every on-die
ECC word is protected with the necessary correction capability \gfii{by the secondary ECC}. For example,
with a single-error correcting on-die ECC that uses 128-bit words, the memory
controller must ensure that every 128-bit on-die ECC word is protected with
\emph{at least} single-error correction.}

How this is achieved \mpu{depends heavily on a given system's memory
architecture}. For example, depending on the size of an on-die ECC word and how
many memory chips the memory controller interfaces with, on-die ECC words may be
split across different data transfers. In this case, the system designer must
choose a design that \mpu{matches} their design goals, e.g., \mpu{dividing}
secondary ECC words across multiple transfers (which introduces its own
reliability challenges~\cite{gong2018duo}), or interleaving secondary ECC words
across multiple on-die ECC words (which could require stronger secondary ECC).

\mpq{Without loss of generality, we assume that the memory controller interfaces
with a single memory chip at a time (e.g., similar to \mpr{some}
LPDDR4 systems~\gfii{\cite{jedec2014lpddr4}}) and provisions a single-error correcting
code per on-die ECC word. Such a system is sufficient for demonstrating the
error profiling challenges that we address in this work.} Matching the
granularity of secondary and on-die ECC words for \mpq{arbitrary systems is
\emph{not} a problem unique to our work since \emph{any} secondary ECC 
that \gfii{is designed} to account for the effects of on-die ECC must consider how the two
interact~\cite{cha2017defect, gong2018duo}. Therefore,} we leave a general
exploration of the tradeoffs \mpr{involved} to future work.

\subsubsection{HARP-U and HARP-A\gfii{.}}
\label{subsubsection:harp_u_and_harp_a}

\mpq{We introduce two variants of HARP: HARP-A and HARP-U, which are aware and
unaware of the on-die ECC parity-check matrix $H$, respectively. HARP-A uses
this knowledge to \emph{precompute}\footnote{Using the methods described in
detail in prior work~\cite{patel2020bit}.} bits at risk of indirect error given
the bits at risk of direct error that are identified during active profiling.
HARP-A does \mpq{not} provide benefits over HARP-U during active profiling.
However, HARP-A reduces the number of bits at risk of indirect error that remain
to be identified by reactive profiling.}

\subsubsection{Increasing the Secondary ECC Strength\gfii{.}}
\label{subsubsection:incomplete_coverage}

The secondary ECC is used for reactive profiling and must \mpq{provide equal or
greater correction capability} than on-die ECC to safely identify indirect
errors. Current on-die ECC designs are limited to simple single-error correcting
codes due to area, energy, and latency constraints within the memory
die~\cite{micron2017whitepaper, gong2018duo, son2015cidra}, so the secondary ECC
can be correspondingly simple. However, if either (1) future on-die ECC designs
become significantly more complex; or (2) \mpq{the system designer wishes to
address other failure modes (e.g., component-level failures) using the secondary
ECC, the system designer will need to increase the secondary ECC strength
accordingly}. Whether profile-based repair remains a feasible error-mitigation
strategy in this case is ultimately up to the system designer and their
reliability goals, so we leave further exploration to future work.

\subsection{Limitations}
\label{subsubsec:full_coverage}

HARP relies on the active profiler to achieve full coverage of bits at risk of
direct errors so that the reactive profiler \emph{never} observes a direct error
(i.e., the reactive profiler's correction capability is never exceeded).
Consequently, if the active profiler fails to achieve full coverage, the
reactive profiler \mpq{may experience indirect errors \emph{in addition to}
direct error(s) missed by active profiling.} 

\mpq{We acknowledge this as a theoretical limitation of HARP, but we do not
believe it restricts HARP's potential impact to future designs and scientific
studies. This is because achieving full coverage of at-risk bits without on-die
ECC is a long-standing problem that is complementary to our work. Prior works
have studied this problem in detail~\cite{khan2014efficacy, khan2017detecting,
qureshi2015avatar, patel2017reach}, and any solution developed for chips
without on-die ECC can be immediately applied to HARP's active profiling phase,
effectively reducing the difficulty of profiling chips with on-die ECC to that
of chips without on-die ECC.}
\section{Evaluations}
\label{sec:evaluations}

In this section, we study how HARP's coverage of direct and indirect errors
changes with \mpn{different pre-correction error counts and per-bit error
probabilities} to both (1) demonstrate the effect of the three profiling
challenges introduced by on-die ECC and (2) show that HARP overcomes the three
challenges. 

\subsection{Evaluation Methodology}
\label{subsec:methodology}

We evaluate error coverage in simulation because, unlike with a real device, we
can accurately compute error coverage using precise knowledge of which errors
are and are not possible. This section describes our simulation methodology.

\subsubsection{Baselines for Comparison.}

We compare HARP-U and HARP-A with two baseline profiling algorithms that use
multiple rounds of testing \mpn{with different data patterns to identify at-risk
bits based only on observing post-correction errors.}

\begin{enumerate}
    \item \emph{Naive}, \mpr{which represents} the vast majority of prior
    profilers that operate without knowledge of on-die
    ECC~\cite{venkatesan2006retention, liu2012raidr, khan2014efficacy,
    khan2016case, khan2016parbor, khan2017detecting, patel2017reach,
    qureshi2015avatar, choi2020reducing, lee2017design,
    bacchini2014characterization, sharifi2017online, kim2020revisiting,
    frigo2020trrespass, cojocar2019exploiting, zhang2012memory, hamdioui2017test,
    tavana2017remap, qureshi2011pay, patel2020bit, liu2013experimental,
    cojocar2020are, mutlu2019rowhammer, chang2017understanding, lee2015adaptive,
    kim2018solar, kim2019d, baek2014refresh, kim2003block, lin2012secret}
    \mpr{(described in \cref{subsec:active_prof_phase})}.
    
    \item \emph{BEEP}, the profiling algorithm supported by the
    reverse-engineering methodology BEER~\cite{patel2020bit}. BEEP careful\gfi{ly}
    constructs data patterns intended to systematically expose post-correction
    errors based on having reverse-engineered the on-die ECC parity-check
    matrix. We follow the SAT-solver-based methodology as described by\gfi{~\cite{patel2020bit}} and use a random data pattern before the first post-correction error
    is confirmed.
\end{enumerate}
    
\subsubsection{Simulation Strategy.}
\label{subsubsec:sim_strat}

We extend the open-source DRAM on-die ECC analysis infrastructure released by
prior work~\gfi{\cite{beergithub,patel2020bit}} to perform Monte-Carlo
simulations of DRAM data retention errors. \mpn{We release our simulation tools
on Zenodo~\cite{harp-artifacts} and Github~\cite{harpgithub}.} We simulate error
injection and correction using single-error correcting Hamming
codes~\cite{hamming1950error} representative of those used in DRAM on-die ECC
\mpq{(i.e., (71, 64)~\cite{im2016im, nair2016xed} and (136, 128)~\cite{oh2014a, kwon2017an, kwak2017a, micron2017whitepaper} code configurations). All
presented data is shown for a (71, 64) code, and we verified that our
observations hold for longer (136, 128) codes.} We simulate 1,036,980 total ECC
words across 2769 randomly-generated parity-check matrices \mpq{over $\approx$20
days of simulation time. \mpq{For each profiler
configuration, we simulate 128 profiling rounds because this is enough to
understand the behavior of each configuration (e.g., the shapes of each curve in
Fig.~\ref{fig:eval_type1_coverage}), striking a good balance with simulation
time.}} 

We inject errors according to the model discussed in
\cref{subsec:errors_and_models} to simulate the effect of uniform-random,
data-dependent errors. We assume that all bits are
true-cells~\gfi{\cite{kraft2018improving, liu2013experimental}} that experience
errors \emph{only} when programmed with data `1', \mpq{which is consistent with
experimental observations made by prior work~\cite{kraft2018improving,
patel2020bit}}. To study how varying error rates impact profiling, we simulate
bit errors with Bernoulli probabilities of 1.0, 0.75, 0.5, and 0.25 and separate
our results based on the total number of pre-correction errors $n$ injected into
a given ECC word. Using this approach, one can \gfii{easily} determine the effect of
an arbitrary raw bit error rate by summing over the individual per-bit
\gfi{error} probabilities.

We define \emph{coverage} as the proportion of all at-risk bits that are
identified. We calculate coverage using the Z3 SAT solver~\cite{de2008z3},
computing the total number of post-correction errors that are possible for a
given (1) parity-check matrix; (2) set of pre-correction errors; and (3)
(possibly empty) set of already-discovered post-correction errors. Note that a
straightforward computation of coverage given on-die ECC is extremely difficult
for data-dependent errors: each data pattern programs the parity-check bits
differently, thereby provoking different pre-correction error patterns.
Therefore, using the SAT solver, we accurately measure the bit error rate
\mpr{of \emph{all possible} at-risk bits across} \emph{all possible} data
patterns.

We simulate three different \mpq{data} patterns to exercise data-dependent
behavior: \texttt{random}, \texttt{charged} \gfii{(i.e., all bits are `1's)}, and \texttt{checkered} \gfii{(i.e., consecutive bits alternate between `0' and `1')}. For the
\texttt{random} and \texttt{checkered} data patterns, we invert the data pattern
each round of profiling. For the \texttt{random} pattern, we change the random
pattern after every two profiling rounds (i.e., after both the pattern and its
inverse are tested). We ensure that each profiler is evaluated with the exact
same set of ECC words, pre-correction error patterns, and data patterns in order
to preserve fairness when comparing coverage values. Unless otherwise stated,
all data presented uses the \texttt{random} pattern, which we find performs on
par or better than the static \texttt{charged} and \texttt{checkered} patterns
that do \gfi{\emph{not}} explore different pre-correction error combinations.

\subsection{Active Phase Evaluation}

We study the number of profiling rounds required by each profiling algorithm to
achieve coverage of direct errors. We omit HARP-A because its coverage of direct
errors is equal to that of HARP-U.

\subsubsection{Direct Error Coverage.}
\label{subsubsec:direct_error_coverage}

\begin{figure}[H]
    \centering
    \includegraphics[width=0.7\linewidth]{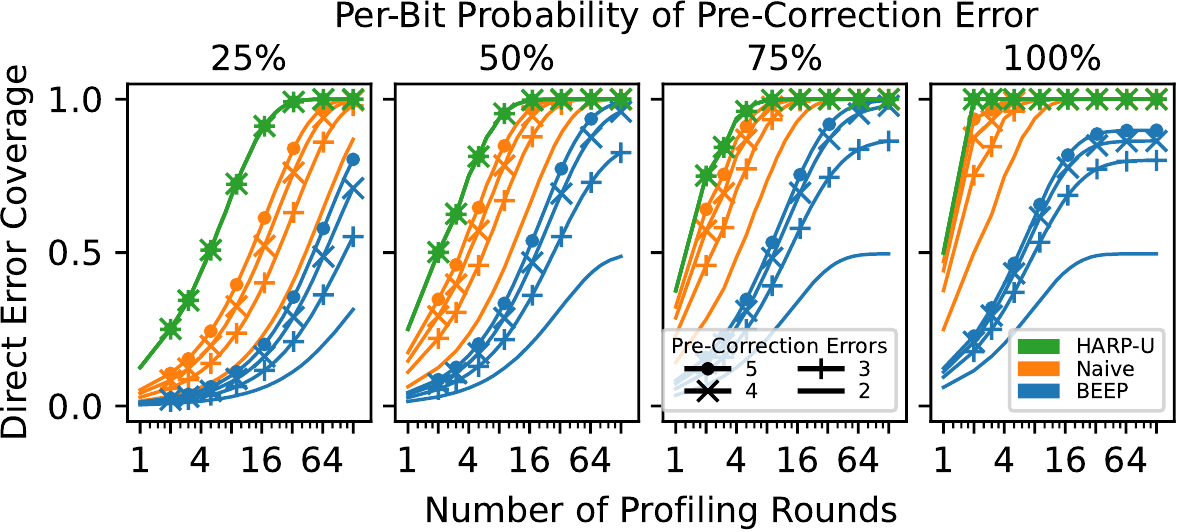}
    \caption{Coverage of bits at risk of direct errors.}
    \label{fig:eval_type1_coverage}
    \vspace{-1em}
\end{figure}

Fig.~\ref{fig:eval_type1_coverage} shows the coverage of bits at risk of
direct errors that each profiler cumulatively achieves \gfi{(y-axis)} over 128
profiling rounds \gfi{(x-axis) assuming four different values of pre-correction
errors per ECC word (2, 3, 4, and 5). We report results for four different
per-bit error probabilities of the injected pre-correction errors (25\%, 50\%,
75\%, 100\%)}. For each data point, we compute coverage as the proportion of
at-risk bits identified out of all at-risk bits across all simulated ECC words. 

We make two observations. First, HARP consistently and quickly achieves full
coverage, \emph{regardless of} the number or per-bit error probabilities of the
injected pre-correction errors. This is because HARP bypasses on-die ECC
correction, identifying each at-risk bit independently, regardless of which
error occurs in which testing round. In contrast, both Naive and BEEP \gfii{(1)}
require more testing rounds to achieve coverage parity with HARP and \gfii{(2)}
exhibit significant dependence on the number of pre-correction errors. This is a
direct result of on-die ECC: each post-correction error depends on particular
combination(s) of pre-correction errors, and achieving high coverage requires
these combinations to occur in distinct testing rounds. We conclude that that
HARP effectively overcomes the first profiling challenge by directly observing
pre-correction errors, while Naive and BEEP must both rely on uncorrectable
error patterns to incrementally improve coverage \mpv{in} each round.

Second, although, HARP and Naive both \gfi{\emph{eventually}} achieve full
coverage, BEEP \gfi{\emph{fails}} to \mpu{do so} in certain cases. This is
because BEEP does not explore all pre-correction error combinations necessary to
expose each bit at risk of direct errors. We attribute this behavior to a nuance
of the BEEP algorithm: BEEP \mpu{crafts data patterns that increase the
likelihood of indirect errors. Unfortunately, these patterns are slow to explore
different combinations of pre-correction errors, which leads to incomplete
coverage}. This is consistent with \mpu{prior work~\cite{patel2020bit}, which
finds that BEEP exhibits low coverage when pre-correction errors are sparse or
occur with low probability. We find that Naive also fails to achieve full
coverage when using static data patterns (e.g., \texttt{checkered}) for the same
reason.}

\subsubsection{Bootstrapping Analysis.}
\label{subsubsec:bootstrapping_anal}

\begin{figure}[H]
    \centering
    \includegraphics[width=0.7\linewidth]{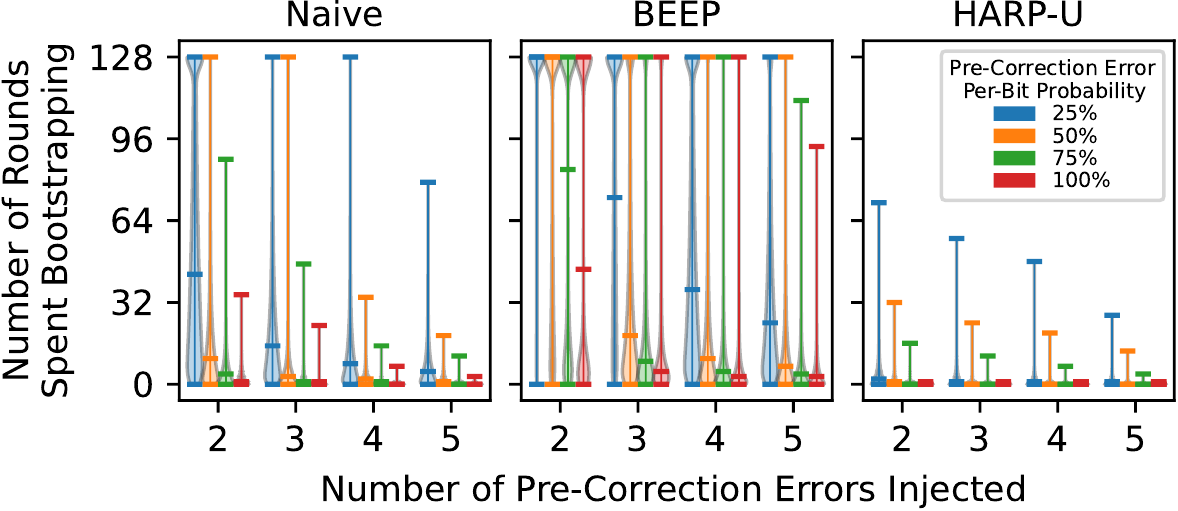}
    \caption{Distribution of the number of profiling rounds required to identify the first direct error across all simulated ECC words.}
    \label{fig:eval_bootstrapping}
\end{figure}

Fig.~\ref{fig:eval_bootstrapping} shows the distribution (median marked with a
horizontal line) of the number of profiling rounds required for each profiler
\mpu{to observe \emph{at least one} direct error in each ECC word. If} no
post-correction errors are identified, we conservatively plot the data point as
requiring 128 rounds, which is the maximum number of rounds evaluated (discussed
in \cref{subsubsec:sim_strat}). The data illustrates the difficulty of
bootstrapping because observing \mpv{any} post-correction error with on-die ECC
requires a \mpv{specific combination of} pre-correction errors to occur.

\gfi{We make three observations. First, w}e \gfi{see} that HARP identifies the
first error far more quickly than Naive or BEEP profiling across \emph{all}
configurations\gfi{. Second, HARP} never fails to identify at least one error
\gfi{within 128 rounds}. \gfi{Third, i}n contrast, BEEP sometimes \gfi{cannot}
identify an error at all due to a combination of (1) the low per-bit
pre-correction error probability and (2) the bootstrapping problem \gfi{(i.e.,}
more testing rounds does \emph{not} guarantee higher coverage unless those
rounds explore different uncorrectable patterns\gfi{)}. We conclude that HARP
effectively addresses the bootstrapping challenge by directly observing
pre-correction errors instead of relying on exploring different uncorrectable
error patterns.

\subsection{Reactive Phase Evaluation}

In this section, we study each \gfi{error} profiler's coverage of bits at risk
of indirect errors and examine the correction capability required from the
secondary ECC to safely identify the at-risk bits remaining after active
profiling. \mpq{\gfii{I}t is important to note that\gfii{, unlike HARP,} neither Naive nor BEEP
profiling achieve full coverage of bits at risk of \gfii{\emph{direct}} errors for all
configurations. In such cases, multi-bit errors can occur during reactive
profiling that are not safely identified by a single-error correcting code
(studied in \cref{subsubsec:missed_uncorr_events}), regardless of the
profiler's coverage of bits at risk of indirect errors.}

\subsubsection{Indirect Error Coverage.}

\begin{figure}[H]
    \centering
    \includegraphics[width=0.7\linewidth]{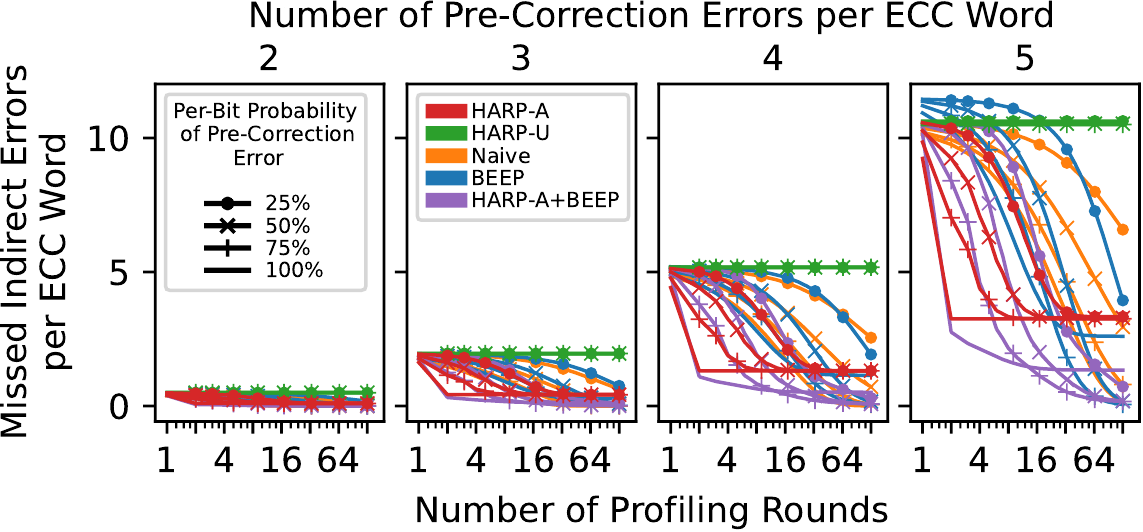}
    \caption{Coverage of bits at risk of indirect errors.}
    \label{fig:type_2_coverage}
\end{figure}

Fig.~\ref{fig:type_2_coverage} shows the proportion of all bits that are at
risk of indirect errors that each profiler \emph{has missed} per ECC word
throughout 128 rounds of profiling. This is equivalent to the number of at-risk
bits \gfi{that} reactive profiling \gfi{has} to identify. We evaluate an
additional configuration, HARP-A+BEEP, which employs BEEP to identify the
remaining at-risk bits once HARP-A has identified all bits at risk of direct
errors.

We make three observations. First, HARP-U does not identify any bits at risk of
indirect errors\footnote{Except for a small number of direct and indirect errors
that overlap.} because it \mpu{bypasses the} on-die ECC correction process that
causes indirect errors. In contrast, HARP-A quickly identifies a subset of all
bits at risk of indirect errors by predicting them from the \mps{identified}
direct errors. \mps{Note that HARP-A cannot} identify \emph{all} bits at risk of
indirect errors because \mps{doing so would require knowing which parity-check
bits are at risk of error, which the on-die ECC bypass path does not reveal.}

Second, combining HARP-A with BEEP effectively overcomes \mps{HARP-A's}
inability to identify pre-correction errors within the parity-check bits. This
is because HARP-A+BEEP synergistically combines \mpq{(1) HARP's ability to
quickly identify bits at risk of direct errors with (2) BEEP's ability to
exploit \emph{known} at-risk bits to \emph{expose} others.} The combined
configuration quickly identifies bits at risk of indirect errors, achieving
\mpq{coverage similar to Naive and BEEP profiling in less than half the number
of profiling rounds.}

Third, both Naive and BEEP achieve relatively high coverage of indirect errors
after many (i.e., $>64$) rounds compared to \mps{HARP-U and HARP-A}. This is
because \mps{both Naive and BEEP continually} explore different uncorrectable
error patterns, \mps{steadily exposing more and more indirect errors.} BEEP
achieves higher coverage because its \mpq{algorithm deliberately seeks} out
pre-correction error combinations that are more likely to cause post-correction
errors, thereby exposing more indirect errors \mps{in the long run}.

We conclude that knowing the on-die ECC parity-check matrix \mps{helps HARP-A
and BEEP identify} bits at risk of indirect errors, thereby reducing the number
of indirect errors that must be identified by \mpu{by the secondary ECC during}
reactive profiling.

\subsubsection{Secondary ECC Correction Capability.}
\label{subsubsec:missed_uncorr_events}

\begin{figure}[H]
    \centering
    \begin{subfigure}[b]{0.49\textwidth}
        \centering
        \includegraphics[width=\textwidth]{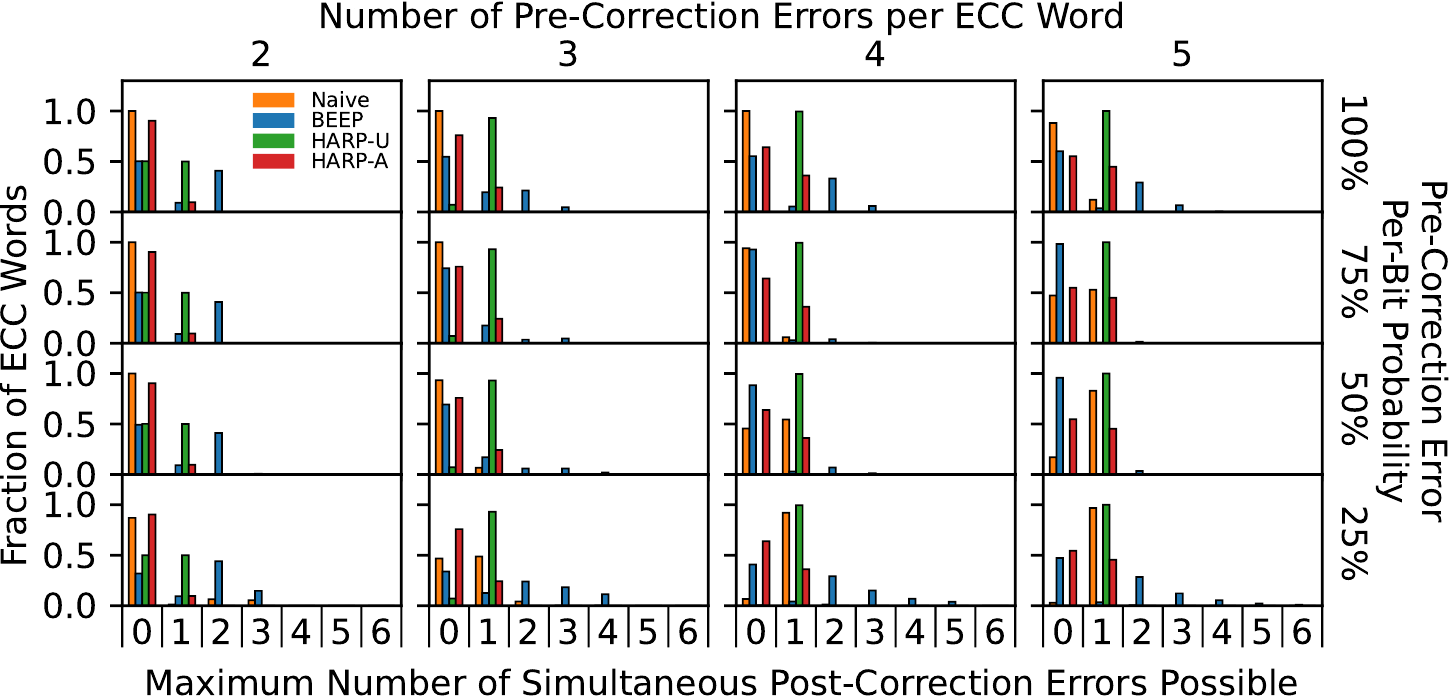}
        \caption{\mpq{Normalized histogram of the maximum number of simultaneous
        post-correction errors (x-axis) possible across all simulated ECC words
        after 128 rounds of profiling.}}
        \label{fig:max_simulteaneous_a}
    \end{subfigure}
    \hfill
    \begin{subfigure}[b]{0.49\textwidth}
        \centering
        \includegraphics[width=\textwidth]{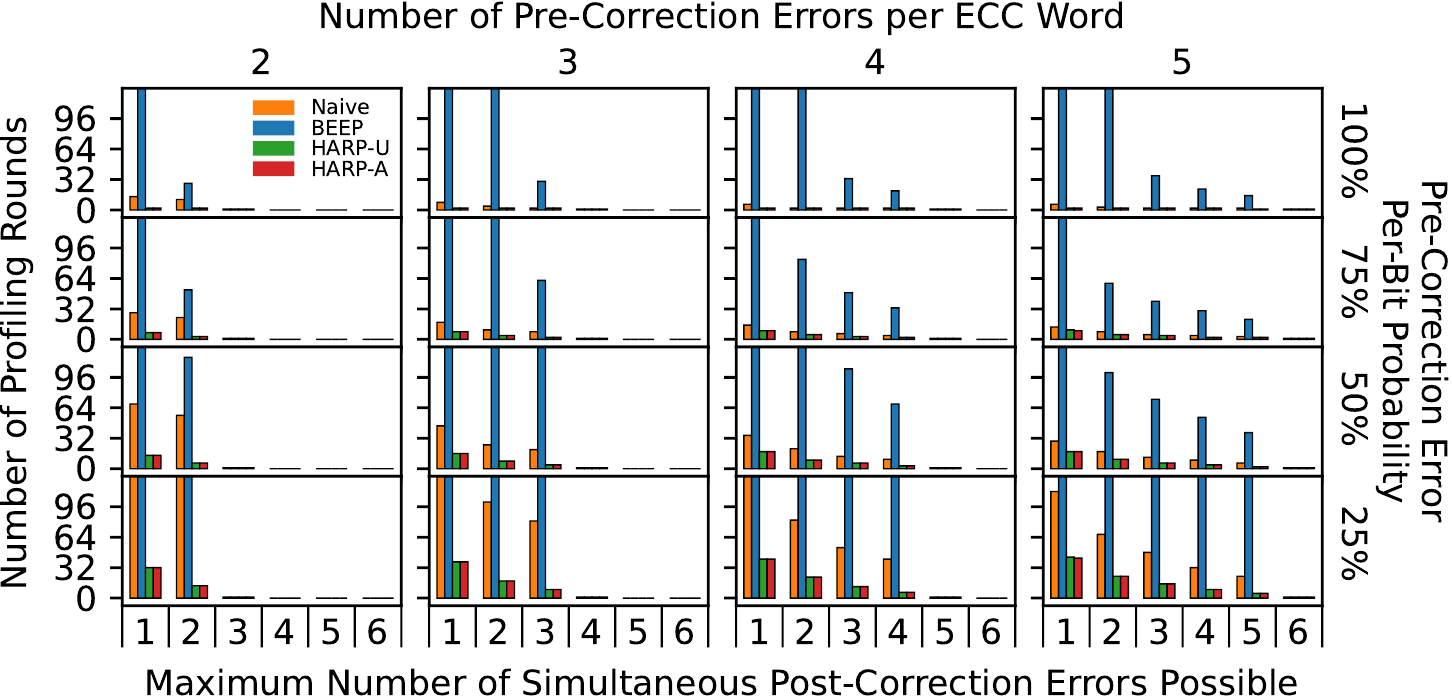}
        \caption{Number of profiling rounds (y-axis) required to achieve 99th-percentile values of the maximum number of simultaneous post-correction errors possible (x-axis).}
        \label{fig:max_simulteaneous_b}
    \end{subfigure}    
    \caption{Maximum number of simultaneous post-correction errors possible
    given all at-risk bits missed after 128 rounds of profiling.}
    \label{fig:max_simultaneous}
\end{figure}

Fig.~\ref{fig:max_simultaneous} shows the worst-case (i.e., maximum) number of
post-correction errors that can occur simultaneously within an ECC word
\mpq{after active} profiling.
\mps{This number is} the correction capability required from \mpq{secondary ECC}
to safely \mpq{perform reactive profiling}.

\noindent
\textbf{Maximum Error Count.} Fig.~\ref{fig:max_simultaneous}a shows a
normalized histogram of the maximum number of post-correction errors that can
occur simultaneously within each simulated ECC word given all at-risk bits
missed after 128 rounds of active profiling. We observe that both HARP-U and
HARP-A exhibit \emph{at most one} post-correction error across all
configurations. This is because HARP \mpq{identifies all bits at risk of direct
errors within 128 profiling rounds (shown in
Fig.~\ref{fig:eval_type1_coverage}), so only one error may occur at a time
(i.e., an indirect error).} In contrast, both Naive and BEEP are susceptible to
multi-bit errors. In particular, BEEP's relatively low coverage of bits at risk
of direct errors means that many multi-bit error patterns remain possible.
\mpq{We conclude that\gfii{, after 128 rounds of active profiling,} a
single-error correcting secondary ECC is \gfii{\emph{sufficient}} to perform
reactive profiling for \gfii{HARP \mps{but} \gfii{\emph{insufficient} \mps{to do
so} for}} Naive and BEEP.}

\noindent
\textbf{Maximum Error Count.} Fig.~\ref{fig:max_simultaneous}b shows \mpq{how
many active profiling rounds are required to ensure that \emph{no more than} an
x-axis value of post-correction errors can occur simultaneously \mpu{in a single
ECC word} during reactive profiling. We conservatively report results for the
99th percentile of all simulated ECC words because neither Naive nor BEEP
achieve full coverage of bits at risk of direct errors for all configurations
within 128 profiling rounds.} In cases where 128 profiling rounds are
insufficient to achieve 99th-percentile values, we align the bar with the top of
the plot.

We make two observations. First, \mpu{both HARP configurations perform}
\emph{significantly} faster than Naive and BEEP. For example, with a 50\%
pre-correction per-bit error probability, HARP \mpq{ensures that no more than
one post correction error can occur in 20.6\%/36.4\%/52.9\%/62.1\% of the
profiling rounds required by Naive} given 2/3/4/5 pre-correction errors. This is
because HARP quickly identifies all bits at risk of direct errors, while Naive
and BEEP \mpv{both either (1) take longer to do so; or (2) fail to do so
altogether (e.g., for the 100th percentile at a 50\% per-bit error
probability)}. Second, BEEP performs much worse than any other profiler because
it exhibits extremely low coverage of bits at risk of direct error (studied in
\cref{subsubsec:direct_error_coverage}). We conclude that achieving high
coverage of bits at risk of direct errors is essential for minimizing the
correction capability of the secondary ECC.

\subsection{\rev{Case Study: DRAM Data Retention}}
\label{subsec:harp_case_study_data_retention}

\mpq{In this section, we show how \gfii{error} profiling impacts end-to-end reliability. We
study the bit error rate of a system that uses a bit-granularity repair
mechanism (e.g., such as those discussed in \cref{bg:subsec:repair_mech}) to
reduce the DRAM refresh rate, which prior work shows can significantly improve
overall system performance and energy-efficiency~\cite{venkatesan2006retention,
liu2012raidr, qureshi2015avatar, patel2017reach} and enable continued density
scaling~\cite{nair2013archshield}. We assume that data-retention errors follow
the error model described in \cref{subsec:errors_and_models} (i.e.,
uniformly with a fixed raw bit error rate, which is consistent with prior
experimental studies~\cite{sutar2016d, patel2017reach, baek2014refresh,
hamamoto1998retention, kim2018dram, shirley2014copula, patel2019understanding})}
and that the repair mechanism perfectly repairs any at-risk bits that are
identified by either active or reactive profiling. We assume a (71, 64) SEC
on-die ECC code and a secondary ECC capable of detecting and correcting a single
error in each on-die ECC word during reactive profiling.

\mpq{Fig.~\ref{fig:case_study_ber} illustrates the fraction of all bits that
are at risk of post-correction errors (i.e., the bit error rate) before
\gfii{(Fig.~\ref{fig:case_study_ber}, left)} and after
\gfii{(Fig.~\ref{fig:case_study_ber}, right)} secondary ECC is applied (i.e.,
before and after performing reactive profiling) given an x-axis number of active
profiling rounds. \mps{Each line marker shows a} different data-retention RBER
(e.g., \gfi{due to} operating at different refresh rates).  

\begin{figure}[H]
    \centering
    \begin{subfigure}[b]{0.49\textwidth}
        \centering
        \includegraphics[width=\textwidth]{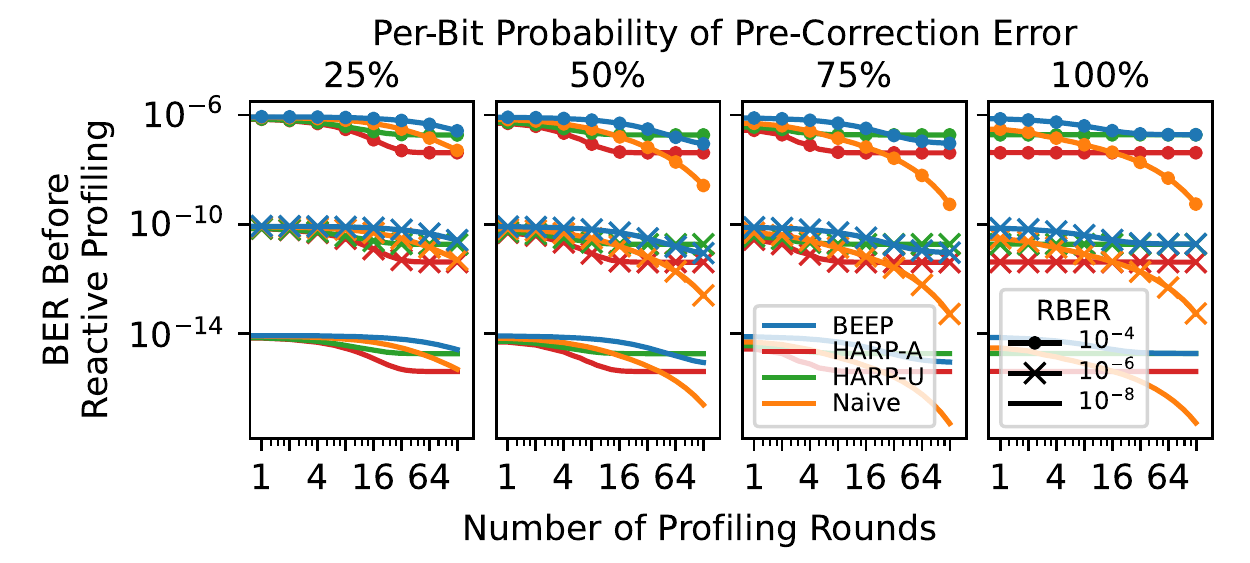}
    \end{subfigure}
    \hfill
    \begin{subfigure}[b]{0.49\textwidth}
        \centering
        \includegraphics[width=\textwidth]{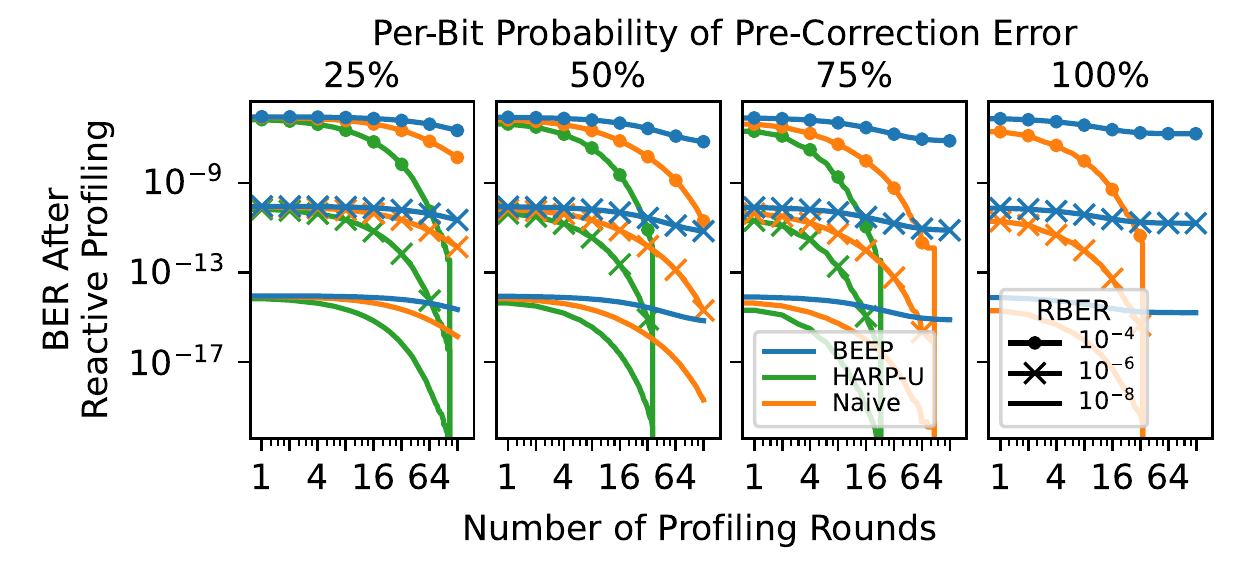}
    \end{subfigure}    
    \caption{\mps{Data-retention bit error rate (BER) using an ideal repair
    mechanism} before (left) and after (right) applying the secondary ECC.}
    \label{fig:case_study_ber}
\end{figure}

We make three observations. First, all profilers \mps{in
Fig.~\ref{fig:case_study_ber} (left)} behave consistently with the coverage
analysis of \cref{subsubsec:direct_error_coverage}. HARP quickly
identifies all bits at risk of direct error, but leaves indirect errors to be
identified by reactive profiling. Both Naive and BEEP slowly explore different
combinations of pre-correction errors, with Naive steadily reducing the BER
given more profiling rounds while BEEP fails to do so.}

Second, \mps{Fig.~\ref{fig:case_study_ber} (left) shows} the benefit of HARP-A
knowing the on-die ECC function. While both HARP-U and HARP-A quickly identify
all bits at risk of direct errors, HARP-A also identifies bits at risk of
indirect errors, thereby considerably reducing the overall BER (and therefore,
the total number of bits) that \mpr{remain to be identified by reactive
profiling.}

\mpq{Third, \mps{Fig.~\ref{fig:case_study_ber} (right) shows that} both
HARP\footnote{\mpq{HARP-A is not shown \gfii{in Fig.~\ref{fig:case_study_ber}
(right)} because it exhibits \emph{identical} BER to HARP-U after applying
secondary ECC (i.e., because both profilers have identical coverage of bits at
risk of direct errors).}} and Naive reach a BER of zero after sufficiently many
\gfi{profiling} rounds, though Naive takes significantly more profiling rounds
to do so (e.g., $3.7\times$ for a per-bit pre-correction error probability of
75\%).} This behavior is consistent with the fact that both profilers eventually
achieve full coverage of bits at risk of direct error (shown in
\cref{subsubsec:direct_error_coverage}) \gfi{In contrast}, BEEP fails to
reach a zero probability value because it fails to achieve full coverage of bits
at risk of direct error. \mpq{Note that HARP-U immediately identifies all bits
at risk of direct errors in the first profiling round with a per-bit
pre-correction error probability of 100\%, so it is not visible in the rightmost
plot.} 

\mpq{We conclude that HARP effectively identifies all bits at-risk of error
faster than the baseline profilers, thereby enabling the repair mechanism to
safely operate at the evaluated raw bit error rates.} Although this case study
\mps{uses a simple data-retention error model that does not include
low-probability errors or other failure modes (discussed in
\cref{subsec:errors_and_models}), it} demonstrates (1) the importance of
a practical and effective error profiling algorithm in enabling a repair
mechanism to mitigate errors; and (2) the advantages that HARP provides in an
end-to-end setting by overcoming the error profiling challenges introduced by
on-die ECC.
\section{Summary}

\mpq{We study how on-die ECC affects memory error profiling and identify three
key challenges that it introduces: on-die ECC (1) exponentially increases the
number of at-risk bits the profiler must identify; (2) makes individual at-risk
bits more difficult to identify; and (3) interferes with commonly-used memory
data patterns. To overcome these three challenges, we introduce Hybrid
Active-Reactive Profiling (HARP), a new bit-granularity error profiling
algorithm that enables practical and effective error profiling for memory chips
that use on-die ECC. HARP exploits the key idea that on-die ECC introduces two
different sources of post-correction errors: (1) direct errors that result from
pre-correction errors within the data portion of the ECC codeword; and (2)
indirect errors that are a result of the on-die ECC correction process. If all
bits at risk of direct error are identified, the number of concurrent indirect
errors is upper-bounded by the correction capability of on-die ECC. Therefore,
HARP uses simple modifications to the on-die ECC mechanism to quickly identify
bits at risk of direct errors and relies on a secondary ECC within the memory
controller to safely identify indirect errors. Our evaluations show that HARP
achieves full coverage of all at-risk bits in memory chips that use on-die ECC
faster than prior approaches to error profiling. We hope that the studies,
analyses, and ideas we provide in this work will enable researchers and
practitioners alike to think about \gfii{and overcome the challenge of} how to
handle error detection and correction across the hardware-software stack in the
presence of on-die ECC.}
\chapter[A Case For Transparent Reliability in DRAM Systems]{A Case for Transparent Reliability\\in DRAM Systems}
\label{chap:position}

\newcommand{\xmp}[1]{#1}
\newcommand{\xmt}[1]{#1}
\newcommand{\xmf}[1]{#1}
\newcommand{\xmg}[1]{#1}

\newcommand{\hcfirst}[0]{$\mathrm{HC}_{\mathrm{first}}$}
\newcommand{\hcfirstbold}[0]{$\mathrm{\textbf{HC}}_{\mathrm{\textbf{first}}}$}

The previous four chapters closely studied new challenges that on-die ECC
introduces because of the way that it obfuscates memory errors in the context of
error profiling, error characterization, and error mitigation. \dmpa{In this
chapter, based on the understanding and technical contributions we developed so
far, we argue for the importance of having transparency into basic commodity
DRAM reliability characteristics. We then introduce our recommendations to
facilitate such transparency in current and future commodity DRAM-based
systems.}

\section{Background and Motivation}
\label{position:bg_mot}

\mpm{Dynamic Random Access Memory (DRAM)~\cite{dennard1968field,
dennard1974design, keeth2007dram, markoff2019ibm, nature2018memory,
ibm2021dram}} is the dominant choice for main memory across a broad range of
computing systems because of its high capacity at low cost relative to other
viable main memory technologies. Building efficient DRAM chips requires
substantially different manufacturing processes relative to standard CMOS
fabrication~\cite{kim1999assessing}, so DRAM is typically designed and
manufactured separately from other system components. \mpt{In this way, system
designers who purchase, test, and/or integrate commodity DRAM chips (e.g., cloud
system designers, processor and system-on-a-chip (SoC) architects, memory module
designers, test and validation engineers) are free to focus on the particular
challenges of the systems they work on instead of dealing with the nuances of
building low-cost, high-performance DRAM.}

To ensure that system designers can integrate commodity DRAM chips from any
manufacturer, the DRAM interface and operating characteristics have long been
standardized by the JEDEC consortium~\cite{jedec2021jc42}. JEDEC maintains a
limited set of \emph{DRAM standards} for commodity DRAM chips with different
target applications, e.g., general-purpose DDR\emph{n}~\cite{jedec2008ddr3,
jedec2012ddr4, jedec2020ddr5}, \mpk{bandwidth-optimized
HBM\emph{n}~\cite{jedec2021high, jedec2022high},} mobile-oriented
LPDDR\emph{n}~\cite{jedec2014lpddr4, jedec2020lpddr5}, graphics-oriented
GDDR\emph{n}~\cite{jedec2016gddr5, jedec2016gddr6}. Given that DRAM designs are
heavily constrained by DRAM standards, manufacturers generally seek
profitability through economies of scale~\cite{kang2010study, dell1997white,
lee2013strategic, croswell2000model}: they mass produce standards-compliant DRAM
chips using highly-optimized \mpi{manufacturing} processes. High-volume
production amortizes manufacturing costs and increases per-chip profit margins.
As such, DRAM manufacturers conservatively regard design- and
manufacturing-related information as sensitive~\mpo{\cite{nair2013archshield,
gong2017dram, childers2015achieving, cost1997yield}, revealing only what DRAM
standards require.}

To maintain their competitive advantage in cost-per-capacity, DRAM manufacturers
continually improve storage densities across successive product generations
while minimizing fabrication costs (e.g., minimizing chip area, maximizing
yield). This requires a careful balance between aggressively scaling physical
feature sizes, continually optimizing circuit designs to reduce area
consumption, and mitigating reliability issues that arise with process
technology shrinkage~\mpo{\cite{kang2014co, cha2017defect, nair2013archshield,
micron2017whitepaper, park2015technology, son2015cidra, micron2022quaterly}}.
Unfortunately, focusing primarily on storage density forces DRAM manufacturers
to sacrifice potential improvements in other metrics of interest, such as
performance, energy, etc. Even if process technology shrinkage naturally
\mpo{provides} gains in these other metrics (e.g., by reducing circuit latencies
with smaller circuit elements), manufacturers typically adjust their designs to
exchange these gains for additional storage density (e.g., by building larger
array sizes that offset any reductions in access latency). As manufacturers
juggle the complex tradeoffs in chip design and manufacturing to maintain market
competitiveness, DRAM as a whole exhibits slow generational improvements in key
areas, such as access latency and power consumption~\mpo{\cite{chang2017thesis,
lee2016reducing, ghose2018your}}. 

\mpi{Figure~\ref{fig:timings_idds} provides a best-effort survey showing how
manufacturer-reported values for four key DRAM operating timings and per-chip
storage capacity \mpn{(all shown in log scale)} have evolved over time. We
extract these data values from 58 publicly-available DRAM chip datasheets from
across 19 different DRAM manufacturers with datasheet publication dates between
1970 and 2021. This data encompasses DRAM chips from both asynchronous (e.g.,
page mode, extended data out) and synchronous (e.g., SDRAM, DDR\emph{n}) DRAM
chips. Appendix~\ref{position:appendix_a} describes our data collection
methodology in further detail, and Appendix~\ref{position:appendix_b} provides
an overview of our dataset, which is publicly available on
GitHub~\cite{datasheetsurveygithub}.}

\noindent
\begin{minipage}{\linewidth}
\renewcommand{\thempfootnote}{\roman{mpfootnote}}
\begin{figure}[H]
    \centering
    \includegraphics[width=0.7\linewidth]{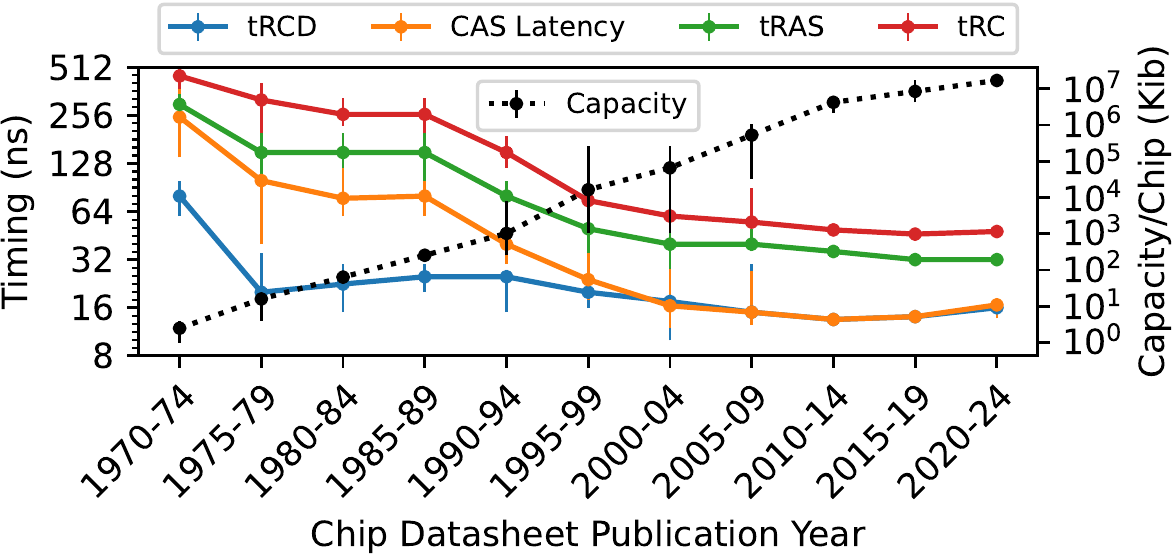}
    \caption[]{\mpo{%
    \mpn{Semi-log plot showing the evolution of key} DRAM access timings (left) and per-chip storage capacity (right)\protect\footnote{%
    \mpo{JEDEC-standardized parameters\cite{jedec2020ddr5} found in DRAM chip datasheets:}\\
    \indent\indent\indent
    \begingroup%
    \footnotesize
    \begin{tabular}{ll}%
        \textbf{Parameter} & \textbf{Definition}\\\hline
        tRCD & minimum row activation to column operation delay\\
        CAS Latency & read operation to data access latency\\
        tRAS & minimum row activation to precharge delay\\
        tRC & minimum delay between accesses to different rows
    \end{tabular}%
    \endgroup%
    }\, \mpn{across each 5-year period of time.}}}
    \label{fig:timings_idds}
\end{figure}
\end{minipage}

\subsection{Motivation: The Need for Adaptability}

\mpi{We observe a clear trend that newer DRAM chips exhibit improvements in all
four timing parameters \emph{and} storage capacity. However, \emph{none} of the
four timings have improved significantly in the last two decades. For example,
the median tRCD/CAS Latency/tRAS/tRC reduced by 2.66/3.11/2.89/2.89\% per year
on average between 1970 and 2000, but only 0.81/0.97/1.33/1.53\% between 2000
and 2015.\footnote{We report 2015 instead of 2020 because 2020 shows a
regression in CAS latency due to first-generation DDR5 chips, which we believe
is not representative because of its immature technology.} In contrast, storage
capacity improved relatively consistently with an exponential growth factor of
0.328 per year (0.341 for 1970-2000 and 0.278 for 2000-2020) across the entire
history of DRAM technology. This data is consistent with similar \mpj{studies}
done in prior work~\mpm{\cite{son2013reducing, chang2017thesis, lee2013tiered,
hennessy2011computer, chang2016understanding, lee2016reducing,
isaac2008remarkable, choi2015multiple, borkar2011future,
nguyen2018nonblocking}}, showing that commodity DRAM manufacturers have
prioritized storage capacity over access latency in recent years.}

Unfortunately, prioritizing storage density does not always align with the
increasingly diverse needs of modern computing systems. These needs change as
systems continuously evolve, so there is no single target metric (e.g., storage
capacity) that suits all DRAM-based systems. Instead, each system's design goals
differ based on factors such as cost, complexity, applications, etc. For
example, storage-focused data centers (e.g., content delivery network nodes) may
require high-reliability memory while compute-focused clusters may optimize for
performance with low-latency memory. Unfortunately, system designers today are
limited to a narrow range of commodity DRAM products,\footnote{\mpi{Custom DRAM
chips (e.g., latency-optimized~\cite{micron2021rldram, fujitsu2012fcram},
high-reliability~\cite{smart2021rugged, im2016im}) and target-specific chips
(e.g., LPDDR\emph{n}~\cite{jedec2014lpddr4, jedec2020lpddr5},
GDDR\emph{n}~\cite{jedec2016gddr5, jedec2016gddr6}) sacrifice the cost
advantages of high-volume general-purpose commodity
DRAM~\cite{lee2013strategic}.}} that effectively restrict design freedom and
limit the peak potential of DRAM-based systems. 

To address this disparity, system designers have long since developed techniques
for adapting unmodified commodity DRAM chips to varying system requirements.
Examples include: \mpl{(1)} actively \mpi{identifying and/or} mitigating errors
to improve reliability~\mpi{\cite{kim2015bamboo, cardarilli2000development,
yoon2010virtualized, udipi2012lot, jian2013low, kim2015frugal, nair2016xed,
jian2013adaptive, han2014data, chen2015ecc, chen2013e3cc, manzhosov2021muse,
patil2021dve, choi2020reducing, sharifi2017online, alameldeen2011energy,
naeimi2013sttram, awasthi2012efficient}}; \mpl{(2)} exploiting available
timing~\mpi{\cite{chandrasekar2014exploiting, chang2016understanding,
kim2018solar, lee2015adaptive, lee2017design, wang2018reducing,
zhang2016restore, koppula2019eden}} and
voltage~\mpl{\cite{chang2017understanding, david2011memory, deng2011memscale}}
margins to reduce memory access latency, power consumption, decrease refresh
overheads~\cite{venkatesan2006retention, liu2012raidr, nair2013archshield,
patel2017reach, lin2012secret, ghosh2007smart, wang2018content,
qureshi2015avatar, patel2017reach}; and \mpl{(3)} mitigating unwanted DRAM
\mpo{data persistence~\cite{halderman2008lest, gruhn2013practicability,
simmons2011security} and \mpo{read-disturb}
problems~\cite{yaglikci2020blockhammer, kim2014flipping, apple2015about,
kim2021mithril, saileshwar2022randomized}.}
Section~\ref{subsec:adaptability_benefits_consumers} discusses these proposals
in greater detail to motivate the need to adapt commodity DRAM to diverse yet
aggressive design targets.

However, these proposals are largely theoretical ideas or proofs-of-concept
based on performance and reliability characteristics that are
\mpo{\emph{assumed}, \emph{inferred}, or \emph{reverse-engineered} from a
limited set of observations and DRAM products (e.g., in-house experimental
studies)} without DRAM manufacturers' support. \mpo{Therefore, adopting such
proposals in a consumer-facing product requires a system designer to weigh the
benefits of enhancing DRAM (e.g., improving performance, security, etc.) against
both: (1) risks (e.g., failures in the field) associated with potentially
violating manufacturer-recommended operating conditions and (2) limitations due
to compatibility with only a subset of all commodity DRAM products (e.g., only
those that have been accurately reverse-engineered). These risks and limitations
are a serious barrier to adoption, especially for small-scale designers who may
have limited headroom and expertise} for exploring \mpt{unconventional designs}.

In this work, we argue that \emph{the lack of transparency} concerning DRAM
reliability characteristics is ultimately responsible for confining system
designers to conventional, specification-constrained designs. For example,
safely improving DRAM access latency by adjusting operating timings requires
understanding the possible failure modes resulting from using non-standard
timings (discussed further in Section~\ref{sec:latency_study}). This is because
selecting suitable operating timings requires the system designer to estimate
the reliability impact of the new timings, which in turn requires reliability
modeling or extensive testing under worst-case operating conditions.
\mpo{Unfortunately, obtaining the information necessary to make these estimates
(e.g., error models, worst-case testing parameters) is difficult, if not
impossible,\footnote{\mpo{For all but the largest customers capable of
independently conducting rigorous post-manufacturing testing.}} without
transparency from DRAM manufacturers. This transparency does not exist today,
even through private agreements for high-volume consumers who have significant
stake in the DRAM industry~\mpk{\cite{safari2016private, saroiu2022price,
loughlin2021stop}}.} In general, without the ability to understand how different
design choices can impact DRAM reliability (e.g., error rates), system designers
are discouraged from \mpo{deploying or even} exploring alternative designs.

To understand the source of the transparency problem, we conduct four case
studies \mpi{throughout Sections~\ref{sec:rela_study}--\ref{sec:security_study}}
that each examine a key system design concern for commodity DRAM chips: \mpi{(1)
reliability; (2) refresh overheads; (3) access latency; and (4) the RowHammer
security vulnerability. For each case study, we explain how system designers are
forced to make assumptions about DRAM reliability in order to address these
concerns without breaking design independence with DRAM manufacturers, but those
very assumptions limit the practicality and scope of the solution.} We then
argue that DRAM standards lie at the heart of the problem because they do not
adequately address the aforementioned DRAM reliability concerns. \mpo{To
overcome this reliance on assumptions, we show that incorporating specifications
for consumer-visible DRAM reliability characteristics (e.g., industry-validated
error models and testing techniques) into DRAM standards} alleviates the problem
and allows system designers to better adapt commodity DRAM to their particular
needs without requiring \mpo{changes to} how DRAM manufacturers design and build
commodity DRAM.

\mpt{We propose incorporating information transparency into DRAM standards using
a two-step approach involving all DRAM stakeholders, including consumers and
manufacturers. In Step 1, for DRAM chips already in the field, we seek the
release of basic information} about DRAM chips that consumers can use to better
understand the chips' reliability characteristics.
Section~\ref{position:subsec:what_to_release} details examples of possible
information to release, including (1) basic microarchitectural characteristics
(e.g., organization of physical rows, sizes of internal storage arrays) that can
be reverse-engineered using existing techniques \mpi{with access to appropriate
testing infrastructure}~\mpm{\cite{jung2016reverse, lee2015adaptive,
patel2019understanding, chang2016understanding, mukhanov2020dstress,
kim2018dram, kraft2018improving, kim2018solar, orosa2021deeper,
hassan2021uncovering}} and (2) industry-recommended testing best practices
(e.g., test patterns for key error mechanisms). \mpt{We believe that this
information can be released through a combination of (1) crowdsourced testing of
commodity DRAM chips on the market; and (2) DRAM chip manufacturers publishing
information (e.g., using datasheet revisions or online resources) about their
products, possibly limited to basic information that manufacturers already have
available (i.e., that requires minimal logistical effort to release). Through a
combination of these two avenues,} \mpo{\mpi{information can be provided to
\emph{all} system designers, including the majority of designers without the
ability to conduct exhaustive testing,} almost immediately without requiring
changes to existing DRAM hardware or standards (though standardizing the
information release could streamline the process). Then, armed with this
information, system designers can make more informed decisions when developing
their own solutions to system-specific design concerns while also preserving the
advantages of commodity DRAM built per general-purpose DRAM standards.}

\mpo{In Step 2, we propose extending DRAM standards with explicit DRAM
reliability standards that provide industry-standard guarantees, tools, and/or
information helpful to consumers. We envision different possibilities for these
reliability standards, including (1) reliability guarantees for how a chip is
expected to behave under certain operating conditions (e.g., predictable
behavior of faults~\cite{criss2020improving}); (2) disclosure of
industry-validated DRAM reliability models and testing strategies suitable for
commodity DRAM chips (e.g., similar to how JEDEC JEP122~\cite{jedec2016failure},
JESD218~\cite{jedec2010ssdrequirements}, and
JESD219~\cite{jedec2010ssdendurance} address \mpo{Flash-memory-specific error
mechanisms~\mpm{\cite{cai2017error, cai2012error, cai2018errors}} such as
floating-gate data retention~\mpm{\cite{cai2015data, luo2018heatwatch,
luo2018improving, cai2012flash} and \mpm{models for physical phenomena such as
threshold voltage distributions~\cite{cai2013threshold, cai2013program,
cai2015read, luo2016enabling}}}}); and (3) requirements for manufacturers to
directly provide relevant information about their DRAM chips (e.g., the
information requested in Step 1).} As the DRAM industry continues to evolve, we
anticipate closer collaboration between DRAM and system designers to efficiently
overcome the technology scaling challenges that DRAM is already
facing~\mpm{\cite{kang2014co, micron2017whitepaper, mutlu2013memory,
mutlu2015main}}. Although we hope that transparency will occur naturally as part
of this process, we believe the end result will be determined in a large part by
the direction in which DRAM standards evolve. Therefore, we believe that
ensuring transparency of reliability characteristics becomes a first-order
concern is essential for allowing innovation going forward.
\section{\xmp{The System Designer's Challenge}}
\label{sec:motivation_new}

Today's DRAM industry thrives on separation of concerns: DRAM manufacturers can
focus on designing highly-optimized DRAM chips while consumers can make use of
standardized DRAM \xmg{that conform to JEDEC standards}. This design
independence \xmt{is powerful because it allows each party to leverage their
respective expertise to build the best possible product. As a result, a system
designer who is responsible for choosing the memory substrate for a particular
system can simply select between a limited range of standardized commodity parts
that are optimized for different targets, such as general-purpose performance
(e.g., DDR\emph{n}~\cite{jedec2012ddr4, jedec2020ddr5}), high bandwidth (e.g.,
GDDR\emph{n}~\cite{jedec2016gddr5, jedec2016gddr6},
\mpo{HBM\emph{n}~\cite{jedec2021high, jedec2022high}}), and low power (e.g.,
LPDDR\emph{n}~\cite{jedec2014lpddr4, jedec2020lpddr5}).}

Unfortunately, the system designer faces a significant challenge: the designer
is unable to fully explore the memory design space \mpk{(as well as the
system-memory co-design space)} because there are only a limited number of
viable design points using commodity DRAM chips. Therefore, the limited number
of options inherently forces the designer to overlook opportunities for
customizing DRAM operation towards their system's particular design goals. As
main memory becomes an increasingly significant system
bottleneck~\mpm{\cite{mutlu2013memory, mutlu2014research, mutlu2019processing}},
we believe that enabling system designers to flexibly adapt commodity DRAM to
suit their own needs as they see fit is a promising path to reap the benefits of
adaptability while preserving the design independence between DRAM manufacturers
and system designers.

\subsection{\mpi{Benefits for DRAM Consumers}}
\label{subsec:adaptability_benefits_consumers}

Prior works~\mpk{\cite{kim2015bamboo, cardarilli2000development,
yoon2010virtualized, udipi2012lot, jian2013low, kim2015frugal, nair2016xed,
jian2013adaptive, chen2015ecc, chen2013e3cc, manzhosov2021muse, patil2021dve,
wang2018content, mcelog2021bad, nvidia2020dynamic, venkatesan2006retention,
baek2014refresh, hwang2012cosmic, meza2015revisiting, liu2012raidr,
ohsawa1998optimizing, wang2014proactivedram, lin2012secret, nair2013archshield,
ghosh2007smart, qureshi2015avatar, khan2014efficacy, khan2016case,
khan2016parbor, khan2017detecting, jafri2020refresh, kim2000dynamic,
kim2003block, katayama1999fault, patel2017reach, mathew2017using,
chandrasekar2014exploiting, chang2016understanding, kim2018solar,
lee2015adaptive, lee2017design, wang2018reducing, zhang2016restore,
hassan2016chargecache, koppula2019eden, mathew2017using, zhang2021quantifying,
gao2019computedram, kim2018dram, olgun2021pidram, olgun2021quac, kim2019d,
seshadri2013rowclone, seshadri2015fast, seshadri2017ambit, seshadri2019dram,
hajinazar2021simdram, seshadri2016buddy, yaglikci2020blockhammer,
greenfield2012throttling, mutlu2018rowhammer, kim2014flipping, konoth2018zebram,
van2018guardion, brasser2017can, saileshwar2022randomized}} demonstrate
significant system-level benefits from adapting commodity DRAM operation to
different system needs without changing the DRAM design itself. This section
reviews the benefits of four concrete examples of such customizations: \mpk{1)}
DRAM reliability improvement, \mpk{2)} DRAM refresh overhead reduction, \mpk{3)}
DRAM access latency reduction, and \mpk{4)} RowHammer security improvement. In
principle, a system designer can readily implement each customization using
existing techniques. Unfortunately, adopting these techniques in practice
requires understanding how DRAM reliability characteristics behave under
different operating conditions, which is not clearly communicated by DRAM
manufacturers or standards today. In this section, we review each example
\mpk{customization}'s potential benefits; then, our case studies throughout
Sections~\ref{sec:rela_study}-\ref{sec:security_study} explore each example
\mpk{customization} in further detail to identify the specific factors that we
believe discourage system designers from adopting the examples in practice.

\subsubsection{\mpo{DRAM Reliability Improvement}}
\label{subsubsec:mot_rela}

\mpo{DRAM is susceptible to a wide variety of error mechanisms that can impact
overall system reliability. To combat DRAM-related failures, system designers
typically incorporate reliability, availability and serviceability (RAS)
features~\cite{synopsys2015whitepaper, dell2008system, slayman2006impact} that
collectively improve system reliability beyond \mpk{what commodity DRAM chips
can provide alone}. In general, memory RAS is a broad research area with
solutions spanning the hardware-software stack, ranging from hardware-based
mechanisms within the \mpk{DRAM chip (e.g., on-die ECC
scrubbing~\cite{jedec2020ddr5, rahman2021utilizing, criss2020improving},
post-package repair~\cite{horiguchi2011nanoscale, jedec2012ddr4, jedec2020ddr5,
kim2016ecc, wada2004post}, target row refresh~\cite{hassan2021uncovering,
frigo2020trrespass})}, memory controller (e.g., rank-level
ECC~\cite{kim2015bamboo, cardarilli2000development, yoon2010virtualized,
udipi2012lot, jian2013low, kim2015frugal, nair2016xed, jian2013adaptive,
chen2015ecc, chen2013e3cc, manzhosov2021muse, patil2021dve, wang2018content},
rank-level ECC scrubbing~\mpk{\cite{han2014data, qureshi2015avatar,
choi2020reducing, sharifi2017online, alameldeen2011energy, naeimi2013sttram,
awasthi2012efficient, sridharan2015memory, rahman2021utilizing,
sharifi2017online}}, repair techniques~\cite{lin2012secret, nair2013archshield,
kline2020flower, longofono2021predicting, kline2017sustainable,
schechter2010use, nair2019sudoku, zhang2017dynamic, wang2017architecting,
kim2016relaxfault}) to software-only solutions (e.g., page
retirement~\cite{mcelog2021bad, nvidia2020dynamic, venkatesan2006retention,
baek2014refresh, hwang2012cosmic, meza2015revisiting}, failure
prediction~\cite{mukhanov2019workload, baseman2016improving,
giurgiu2017predicting, lan2010study, liang2006bluegene, boixaderas2020cost}).}

\mpo{As a specific and relevant example, an important category of hardware-based
redundancy mechanisms known as rank-level error-correcting codes (rank-level
ECC) operate within the memory controller to isolate the rest of the system from
random DRAM errors. Depending on the ECC design, rank-level ECC can protect
against random single-bit (e.g., SEC/SEC-DED Hamming
codes~\cite{hamming1950error}), multi-bit (e.g., BCH~\cite{hocquenghem1959codes,
bose1960class}, Reed-Solomon~\cite{reed1960polynomial}), and/or multi-component
(e.g., Chipkill~\cite{dell1997white, kim2015bamboo}) errors with varying
hardware and runtime overheads. The system designer must decide which ECC
mechanism is most appropriate for their particular system (e.g., which error
mechanisms are dominant and what degree of protection is required). For example,
a state-of-the-art rank-level ECC mechanism called}
\mpi{Frugal-ECC~\cite{kim2015frugal} uses data compression to provide
chipkill-correct ECC for $\times$4 non-ECC DIMMs and $\times$8 ECC DIMMs with
negligible performance (maximum of 3.8\%), energy-efficiency, and area overheads
compared with an industry-standard chipkill solution. Therefore, Frugal-ECC
enables system designers to implement chipkill reliability using commodity DRAM
chips with a fraction of the storage overheads suffered by conventional ECC DIMM
configurations.}

\subsubsection{DRAM Refresh Overhead Reduction}
\label{subsubsec:mot_dram_refresh}

DRAM stores data in volatile capacitors, which are susceptible to charge
leakage. To prevent this leakage from causing data loss, DRAM requires periodic
refresh operations that intermittently access all DRAM cells to restore their
charge levels to safe values. Unfortunately, DRAM refresh operations are well
known to waste significant system performance and
power~\mpm{\cite{ohsawa1998optimizing, kim2000dynamic, laudon2006ultrasparc,
liu2012raidr, nair2013archshield, bhati2015flexible, wang2014proactivedram,
baek2014refresh, mathew2017using, bhati2016dram}}, sacrificing almost half of
the total memory throughput and wasting \mpk{almost} half of the total DRAM
power for projected 64 Gb chips~\cite{liu2012raidr}.

To alleviate the power and performance costs of DRAM refresh, prior
works~\mpo{\cite{liu2012raidr, ohsawa1998optimizing, wang2014proactivedram,
venkatesan2006retention, lin2012secret, nair2013archshield, ghosh2007smart,
qureshi2015avatar, khan2014efficacy, khan2016case, khan2016parbor,
khan2017detecting, jafri2020refresh, kim2000dynamic, kim2003block,
katayama1999fault, patel2017reach, mathew2017using}} take advantage of the fact
that \emph{most} refresh operations are unnecessary.\footnote{Latency-hiding
techniques (e.g, prefetching, memory command scheduling, on-chip caching, etc.)
and parallelization of refresh and access
operations~\mpk{\cite{chang2014improving, nguyen2018nonblocking, pan2019hiding,
stuecheli2010elastic, mukundan2013understanding}} help mitigate performance
overheads but do not change the total number of refresh operations issued.
\mpk{As a result, such techniques cannot mitigate energy wastage due to DRAM
refresh}. These techniques are also imperfect in many cases \mpk{where
latency-hiding is impractical (e.g., row conflicts between refresh and access
commands, larger memory footprints than available caching
resources)~\mpm{\cite{nair2013case, pan2019hiding, zhang2014cream,
chang2014improving}}.}} The standard DRAM refresh algorithm refreshes all cells
frequently (i.e., at the worst-case rate) to simplify DRAM refresh and guarantee
correctness. However, each cell's data retention characteristics vary
significantly due to a combination of data-dependence~\mpk{\cite{khan2016parbor,
khan2017detecting, liu2013experimental, khan2014efficacy, patel2017reach}} and
process variation~\mpo{\cite{hamamoto1995well, hamamoto1998retention,
gong2017dram, nair2013archshield, liu2013experimental, liu2012raidr,
wang2014proactivedram}}. \mpi{As a result, eliminating unnecessary refresh
operations} can provide significant \mpk{power reduction and performance
improvement.} For example, Liu et al.~\cite{liu2012raidr} demonstrate an average
\mpl{energy-per-access and system performance improvement of 8.3\% and 4.1\%,
respectively, for 4~Gib chips (49.7\% and 107.9\% for 64~Gib chips)} when
relaxing the refresh rate at the row granularity. Therefore, reducing refresh
overheads can potentially benefit any DRAM-based system.

\subsubsection{\xmt{DRAM Access Latency Reduction}}
\label{subsubsec:mot_dram_latency}

Figure~\ref{fig:timings_idds} shows that DRAM access latency has not
significantly improved relative to storage capacity over the last two decades.
This makes DRAM an increasingly significant system performance bottleneck today,
especially for workloads with large footprints that are sensitive to DRAM access
latency~\cite{hsieh2016accelerating, ferdman2012clearing, gutierrez2011full,
hestness2014comparative, huang2014moby, zhu2015microarchitectural,
oliveira2021damov, boroumand2018google, boroumand2021google, koppula2019eden,
kanellopoulos2019smash, son2013reducing, mutlu2013memory, wilkes2001memory,
wulf1995hitting, mutlu2007stall, mutlu2003runahead, kanev2015profiling,
mutlu2014research, bera2019dspatch, bera2021pythia, liu2019binary,
ghose2019processing, shin2014nuat, ghose2019demystifying}. Although
\mpk{conventional} latency-hiding techniques (e.g., caching, prefetching,
multithreading) can potentially help mitigate many of the performance concerns,
these techniques (1) fundamentally do not change the latency of each memory
access and (2) fail to work in many cases (e.g., irregular memory access
patterns, random accesses, huge memory footprints).

To address this problem, prior works have taken two major directions. First,
many works~\mpk{\cite{chandrasekar2014exploiting, chang2016understanding,
kim2018solar, lee2015adaptive, lee2017design, wang2018reducing,
zhang2016restore, hassan2016chargecache, koppula2019eden, mathew2017using,
zhang2021quantifying}} show that the average DRAM access latency can be
shortened by reducing DRAM access timings for particular memory locations that
can tolerate faster accesses. This can be done safely because, although DRAM
standards call for constant access timings across all memory locations, the
minimum viable access timings that the hardware can support actually differ
between memory locations due to factors such as heterogeneity in the circuit
design~\mpo{\cite{lee2017design, lee2016reducing}} and manufacturing process
variation between circuit components~\mpm{\cite{chandrasekar2014exploiting,
chang2016understanding, kim2018solar, chang2017understanding, lee2015adaptive}}.

\xmt{Exploiting these variations in access timings to reduce the average memory
access latency can provide significant system performance improvement. For
example, Chang et al.~\cite{chang2016understanding} experimentally show that
exploiting access latency variations can provide an average 8-core system
performance improvement of 13.3\%/17.6\%/19.5\% for real DRAM chips from three
major DRAM manufacturers. Similarly, Kim et al.~\cite{kim2018solar} show that
exploiting access latency variations induced by DRAM sense amplifiers provides
an average (maximum) system performance improvement of 4.97\% (8.79\%) versus
using default DRAM access timings for 4-core heterogeneous workload mixes
\mpo{based on data obtained from 282 commodity LPDDR4 DRAM chips}.}

Second, other works~\mpk{\cite{gao2019computedram, kim2018dram, olgun2021pidram,
olgun2021quac, kim2019d, seshadri2013rowclone, seshadri2015fast,
seshadri2017ambit, seshadri2019dram, hajinazar2021simdram, seshadri2016buddy}}
show that commodity DRAM can perform massively-parallel computations (e.g., at
the granularity of an 8 KiB DRAM row) by exploiting \mpk{the underlying analog
behavior of DRAM operations} (e.g., charge sharing between cells). These works
show that such computations can significantly improve overall system performance
and energy-efficiency by both (1) reducing the amount of data transferred
between the processor and DRAM and (2) exploiting the relatively high throughput
of row-granularity operations. For example, Gao et al.~\cite{gao2019computedram}
show that in-DRAM 8-bit vector addition is $9.3\times$ more energy-efficient
than the same computation in the processor, primarily due to avoiding the need
for off-chip data transfers. \mpo{Similarly, Olgun et al.~\cite{olgun2021pidram}
use an end-to-end FPGA-based evaluation infrastructure to demonstrate that
in-DRAM copy and initialization techniques can improve the performance of
system-level copy and initialization by $12.6\times$ and $14.6\times$,
respectively.}

\subsubsection{\mpo{Improving Security Against RowHammer}}
\label{subsubsec:mot_dram_security}

RowHammer~\mpk{\cite{kim2014flipping, bains2014row, mutlu2017rowhammer,
mutlu2019rowhammer}} is a well-studied read-disturb phenomenon in modern DRAM
chips in which memory accesses to a given memory location can induce bit-flips
at other locations. Recent experimental studies~\cite{kim2014flipping,
kim2020revisiting} show that RowHammer is continually worsening with process
technology shrinkage. Although DRAM manufacturers incorporate internal
RowHammer-mitigation mechanisms~\cite{de2021smash, lee2014green,
frigo2020trrespass, hassan2021uncovering, cojocar2019exploiting,
kim2020revisiting, micron20208gb}, prior work~\mpk{\cite{frigo2020trrespass,
cojocar2020are, hassan2021uncovering, jattke2022blacksmith, de2021smash}} shows
that these mechanisms do not suffice. Therefore, several
works~\mpo{\cite{yaglikci2020blockhammer, park2020graphene,
kim2014flipping, yaglikci2021security, aichinger2015ddr, apple2015about}}
provide RowHammer-mitigation mechanisms that operate from outside of the DRAM
chip to provide strong security without requiring changes \mpo{to DRAM chip
hardware} or relying upon \mpo{information} from DRAM manufacturers. Such a
solution is attractive for a system designer with interest in building a secure
system because the designer can rely upon their own methods rather than relying
upon external, possibly difficult-to-verify \mpo{promises or}
guarantees~\mpk{\cite{saroiu2022price, qureshi2021rethinking}}.

\mpf{Following prior work~\cite{yaglikci2020blockhammer}, we classify
previously-proposed RowHammer defenses into four different categories as
follows.}
\begin{enumerate}
    \item \emph{Access-agnostic} mitigation hardens a DRAM chip against
    RowHammer independently of the memory access pattern. \mpk{This includes
    increasing the overall DRAM refresh rate~\cite{kim2014flipping,
    apple2015about, aichinger2015ddr} and memory-wide error correction and/or
    integrity-checking mechanisms such as strong
    ECC~\cite{qureshi2021rethinking, cojocar2019exploiting, kim2014flipping}.
    These mechanisms are algorithmically simple but can introduce significant
    system hardware, performance, and/or energy-efficiency overheads (e.g.,
    \mpm{a large number of} additional refresh operations~\cite{kim2014flipping,
    kim2020revisiting, bhati2016dram}).} 
    
    \item \mpf{\emph{Proactive} mitigations~\cite{yaglikci2020blockhammer,
    greenfield2012throttling, mutlu2018rowhammer, kim2014flipping} adjust the
    DRAM access pattern to prevent the possibility of RowHammer errors.}
    
    \item \mpf{\emph{Physically isolating} mitigations~\cite{konoth2018zebram,
    van2018guardion, brasser2017can, saileshwar2022randomized, hassan2019crow}
    physically separate data such that accesses to one portion of the data
    cannot cause RowHammer errors in another.}
    
    \item \mpf{\emph{Reactive} mitigations~\cite{kim2014flipping,
    aweke2016anvil, son2017making, seyedzadeh2018cbt, you2019mrloc,
    lee2019twice, park2020graphene, kim2014architectural, kang2020cattwo,
    bains2015row, jedec2020ddr5, bains2016distributed, bains2016row,
    devaux2021method, yaglikci2021security, marazzi2022protrr,
    kim2015architectural} identify symptoms of an ongoing RowHammer attack
    (e.g., excessive row activations) and issue additional row \mpl{activation
    or} refresh operations to prevent bit-flips from occurring.}
\end{enumerate}

\noindent
RowHammer defense is an ongoing area of research, and which mechanism type is
most effective depends on the level of security (e.g., the threat model) that
the system designer requires and the trade-offs \mpk{(e.g., performance, energy,
hardware area, complexity overheads)} they are willing to make.

\subsection{\mpi{Benefits for DRAM Manufacturers}}
\label{subsec:adaptability_benefits_manufacturers}

We believe that the ability to adapt commodity DRAM to system-specific
\mpk{design goals} also benefits DRAM manufacturers for two key reasons. First,
adaptability broadens the scope and competitive advantage of DRAM technology
relative to alternative technologies (e.g., emerging memories). Second, enabling
DRAM consumers to more easily innovate on the DRAM substrate can encourage
valuable feedback for DRAM manufacturers, including insights from customer
use-cases and \mpk{well-evaluated} suggestions for future products.

\mpi{Regardless of these benefits,} we believe making commodity DRAM adaptable
\mpi{has no significant downside for} DRAM manufacturers. The reliability
characteristics that we wish to be communicated \mpk{(as described in detail in
Section~\ref{position:subsec:what_to_release})} are either (1) already exposed
in scientific studies today; or (2) can be reverse-engineered using existing
techniques by those with access to appropriate tools (e.g., competitors,
scientific labs). We simply ask for these characteristics to be officially
provided in a trustworthy capacity. DRAM manufacturers have not previously
provided this information because there has been no pressing need to do so.
However, releasing this information makes sense today because it can enable a
broad range of benefits for DRAM consumers going forward, especially as DRAM
technology scaling continues to face increasing
difficulties~\mpk{\cite{mutlu2013memory, mutlu2021primer, mutlu2017rowhammer}}.

\subsection{\xmp{Short-Term vs. Long-Term Solutions}}
\label{subsec:mot_short_vs_long_term}

Prior works~\mpm{\cite{patterson1997case, mutlu2014research, mutlu2021primer,
mutlu2013memory, kim2014flipping, kang2014co, mutlu2017rowhammer,
mutlu2019rowhammer}} have praised the merits of cooperation between DRAM
manufacturers and system designers in order to collaboratively solve main memory
challenges across the system stack. However, this requires either (1) breaking
design independence between the two parties; (2) achieving consensus among all
DRAM stakeholders (i.e., JEDEC committee members and representatives, including
DRAM manufacturers and consumers) for every design change, followed by a lengthy
adoption period; \mpk{or (3) reducing dependence on DRAM standards and JEDEC. We
do not believe any of these options are easy to adopt} for either the (1) short
term, where we would like to quickly effect changes that enable information
transparency; or (2) long term, where breaking design independence constrains
the very freedom that we advocate system designers should have in meeting their
own design goals while preserving the cost advantages of mass-produced commodity
DRAM chips.

Instead, we argue for enabling each party to solve their own system-specific
design challenges, modifying DRAM standards only for issues that
collectively affect all DRAM stakeholders. However, regardless of how the DRAM
industry evolves over the coming years, we firmly believe that DRAM must become
more adaptable, whether that occurs through standards or
collaboration.
\section{\mpk{Quantitatively Measuring Reliability}}
\label{position:sec:formalizing}

As we will show in the following case studies
(Sections~\ref{sec:rela_study}--\ref{sec:security_study}), a system designer
exploring unconventional DRAM operating points must first understand how
reliably a chip will behave at that operating point. Given that this behavior is
not governed by DRAM standards or described by DRAM manufacturers, the system
designer must determine it themselves, e.g., through modeling and/or testing.
This section formalizes \mpk{the information that a system designer may need
(but does not necessarily have access to today) in order to quantitatively
understand DRAM reliability.} 

\subsection{Information Flow During Testing}

\begin{figure}[t]
    \centering
    \includegraphics[width=\linewidth]{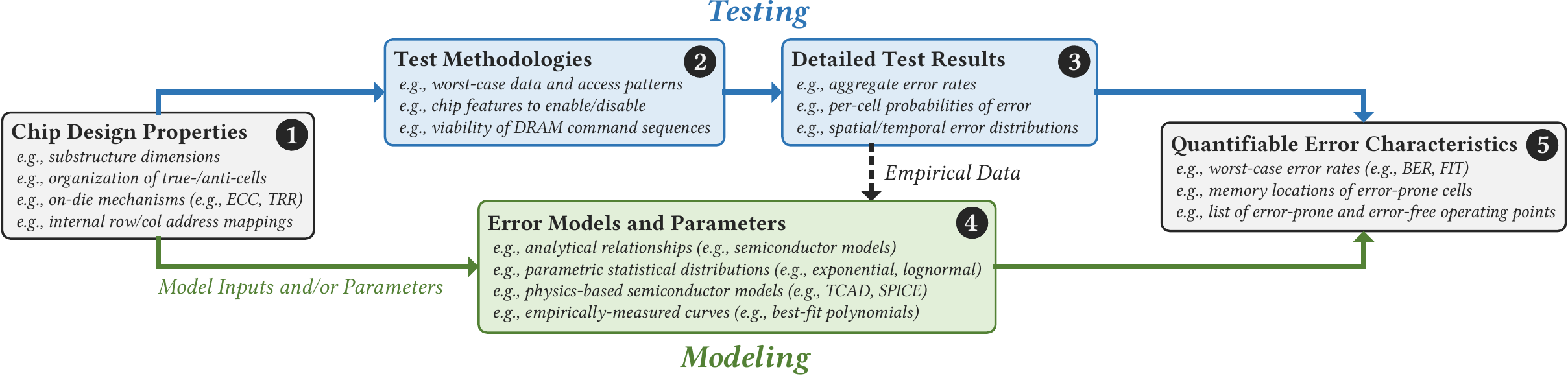}
    \caption{\mpk{Flow} of information necessary to determine key error
    characteristics for a given DRAM device.}
    \label{fig:test_flow}
\end{figure}

Figure~\ref{fig:test_flow} describes the flow of information necessary for a
\mpl{system designer to quantitatively estimate\footnote{``Estimate'' because,
in general, no \mpl{model or experiment} is likely to be perfect, including
those provided by manufacturers.} a DRAM chip's error characteristics
\circled{5} starting from basic properties of the chip \circled{1}. In
principle, these characteristics can comprise \emph{any} aspect of DRAM
reliability that a system designer wants to quantify while exploring their
system's design and/or configuration space. Examples include: (1) worst-case
error rates (e.g., bit error rate (BER) or failures in time (FIT)) across a
given set of operating points; (2) a profile of error-prone memory locations};
or (3) a list of error-free operating points \mpk{(e.g., as identified in a
shmoo analysis~\cite{baker1997shmoo})}. \mpl{The error characteristics can be
estimated in two different ways: testing or modeling.}

\subsubsection{\mpl{Determination from Testing}}

\mpl{First, a system designer may estimate error characteristics using
measurements from detailed experimental testing \circled{3} across a variety of
operating conditions. Examples of measured quantities include:} aggregate error
rates, per-cell probabilities of error, and spatial/temporal error
distributions. \mpk{These measurements can be made using testing infrastructures
ranging from industry-standard large-scale testing
equipment~\cite{advantest2022t5833, teradyne2022magnum} to home-grown tools
based on commodity FPGAs~\cite{hassan2017softmc, olgun2021pidram, hou2013fpga,
kim2014flipping, gao2019computedram, chang2017understanding, ghose2018your,
weis2015retention, chang2016understanding, khan2014efficacy, wang2018dram,
ladbury2013use} or DRAM-based computing systems~\mpl{\cite{passmark2019memtest,
cojocar2020are, veen2016drammer, francis2018raspberry, david2011memory}}.}

\mpl{To conduct accurate and rigorous testing, the system designer must use an
effective test methodology \circled{2} that suits the particular DRAM chip under
test. Prior works extensively study key aspects of effective test methodologies,
including} \mpt{appropriate data and access patterns, the effects of
enabling/disabling DRAM chip features such as \mpk{target row refresh
(TRR)~\cite{frigo2020trrespass, marazzi2022protrr, hassan2021uncovering,
jattke2022blacksmith, kim2020revisiting} and on-die error correcting codes
(on-die ECC)~\mpm{\cite{nair2016xed, micron2017whitepaper, kang2014co, oh20153,
gong2017dram, son2015cidra, oh2014a, kwak2017a, kwon2014understanding,
patel2019understanding, patel2020bit, patel2021enabling}}}, and the viability of
different DRAM command sequences (e.g., sequences that enable in-DRAM row copy
operations~\cite{seshadri2013rowclone, gao2019computedram, olgun2021pidram,
chang2016low}, \mpm{true} random-number generation~\cite{olgun2021quac,
kim2019d, talukder2018exploiting, bostanci2022dr}, and physically unclonable
functions~\cite{kim2018dram, talukder2018ldpuf}).} 

\mpl{In turn, choosing an effective test methodology requires knowledge of
basic} \mpt{properties about a DRAM chip's design and/or error mechanisms
\circled{1}. For example, DRAM manufacturer's design choices for the sizes of
internal storage arrays (i.e., mats~\mpm{\cite{lee2017design, zhang2014half,
son2013reducing, olgun2021quac}}), charge encoding conventions of each cell
(i.e., the true- and anti-cell organization~\cite{kraft2018improving,
liu2013experimental}), use of on-die reliability-improving mechanisms
\mpk{(e.g., on-die ECC, TRR)}, and organization of row and column addresses all
play key roles in determining if and how susceptible a DRAM chip is to key error
mechanisms (e.g., data retention~\cite{hamamoto1998retention,
kraft2018improving, patel2019understanding, liu2013experimental,
bacchini2014characterization, weber2005data, yamaguchi2000theoretical},
access-latency-related failures~\cite{kim2018solar, lee2013tiered,
lee2015adaptive, lee2017design, chang2016understanding, olgun2021quac,
koppula2019eden, chandrasekar2014exploiting}, and
RowHammer~\mpk{\cite{kim2014flipping, mutlu2017rowhammer, mutlu2019rowhammer,
walker2021on, yang2019trap, park2016statistical}}).
Section~\ref{position:subsubsec:design_characteristics} provides further detail
about such design properties and how knowing them is necessary to develop
effective test methodologies.}

\subsubsection{\mpl{Determination from Modeling}}

\mpl{Second, the system designer may make predictions from analytical or
empirical error models \circled{4} based on a previous understanding of DRAM
errors (e.g., from past experiments or scientific studies).} Examples of such
error models include: analytical models based on understanding DRAM failure
modes (e.g., sources of runtime faults~\mpk{\cite{croswell2000model, das2018vrl,
cardarilli2000development, hwang2012cosmic, siddiqua2013analysis,
meza2015large})}, parametric statistical models that provide useful summary
statistics (e.g., lognormal distribution of cell data-retention
times~\cite{hamamoto1995well, hamamoto1998retention, jin2005prediction,
hiraiwa1996statistical, li2011dram, hiraiwa1998local, edri2016silicon,
kim2009new, kong2008analysis}, exponential distribution of the time-in-state of
cells susceptible to variable-retention time
(VRT)~\mpm{\cite{bacchini2014characterization, kim2015avert, qureshi2015avatar,
yaney1987meta, restle1992dram, shirley2014copula, kim2011characterization,
kim2011study, kumar2014detection, mori2005origin, ohyu2006quantitative,
khan2014efficacy, kang2014co, liu2013experimental}}), physics-based simulation models (e.g.,
TCAD~\cite{yang2019trap, synopsys2018sentaurus, duan20172d, pfaffli2018tcad,
jin2005prediction} and SPICE models~\cite{luo2020clr, hassan2019crow,
lee2017design, hassan2016chargecache, lee2013tiered, shin2019dram,
wang2018reducing, zhang2016restore, wang2020figaro}), and empirically-determined
curves that predict observations well (e.g., single-bit error
rates~\mpk{\cite{patel2017reach, qureshi2015avatar, liu2013experimental,
khan2014efficacy, khan2016parbor, park2016statistical}}). \mpl{Similar to
testing, using error models to predict error characteristics ultimately relies
on understanding the DRAM chip being tested because the accuracy of the
predictions requires choosing appropriate models and model parameters (e.g.,
through testing \circled{3} or directly from fundamental chip design properties
\circled{1}).}

\subsection{\mpl{Access to Modeling and Testing Information}}
\label{position:subsec:knowing_or_assuming}

\mpl{Figure~\ref{fig:test_flow} shows that determining a DRAM chip's error
characteristics through modeling or testing ultimately relies on understanding
the chip's fundamental design properties. This reliance can be implicit (e.g.,
inherent within a pre-existing workflow designed for a specific chip) or
explicit (e.g., chosen as part of a home-grown testing methodology). Therefore,
a system designer must be vigilant of the information they (perhaps unknowingly)
rely upon at each step of their design process concerning commodity DRAM.}

\mpl{Fortunately, the system designer \emph{only} needs to be concerned with the
information flow at the children of a node whose information is already known
from a trustworthy source. For example, a system designer who wants to identify
the locations of error-prone cells (i.e., \circled{5}) using testing need not be
concerned with chip design properties (i.e., \circled{1}) if DRAM manufacturers
provide appropriate test methodologies (i.e., \circled{2}) or detailed test
results (i.e., \circled{3}).} Unfortunately, to our knowledge, neither DRAM
standards nor manufacturers provide the information in \emph{any} of the nodes
today, much less in a clear, industry-validated manner. \mpl{Therefore, the
system designer lacks a base of trustworthy information to build upon. This
creates a barrier to entry for a system designer who \mpm{wants} to explore
optimizations to commodity DRAM by compromising the designer's ability to make
well-informed \mpm{or effective} decisions.}

In general, except for the few major DRAM customers who may be able to secure
confidentiality agreements,\footnote{Even under confidentiality, DRAM
manufacturers may be unwilling to reveal certain proprietary aspects of their
designs (e.g., on-die error correction~\mpk{\cite{patel2020bit,
gurumurthi2021hbm3}}, target row refresh~\cite{saroiu2022price}) or provide
specifically requested numbers.} \mpl{system designers would need to rely on
(possibly incorrect \mpm{or incomplete}) \emph{inferences} or \emph{assumptions}
based on domain knowledge or reverse-engineering studies (e.g., similar in
spirit to~\mpm{\cite{patel2017reach, kraft2018improving, liu2013experimental,
patel2020bit, kim2020revisiting, jung2016reverse, barenghi2018software,
hassan2021uncovering, farmani2021rhat, frigo2020trrespass, wang2020dramdig,
pessl2016drama, jiang2021trrscope, kim2018solar, chang2016understanding,
lee2017design}}) that are not verified or supported by the DRAM
industry.}\footnote{DRAM manufacturers may make assumptions during their own
testing. However, they have full transparency into their own designs (i.e, the
root node in the \mpm{information} flow), so they can make the most informed
decision.} \mpk{As a result, the \emph{need} for assumptions can discourage
practitioners from exploring the full design space even when a given design
choice is otherwise beneficial. \mpl{We conclude that} the \emph{lack of
information transparency} is a serious impediment to adopting \mpm{many
promising} DRAM-related optimizations today.}
\section{Study 1: Improving Memory Reliability}
\label{sec:rela_study}

\mph{Main memory reliability is a key design concern for \emph{any} system
because \emph{when} and \emph{how} memory errors occur affects overall system
reliability. In particular, designers of reliability-critical systems such as
enterprise-class computing clusters (e.g., cloud, HPC) and systems operating in
extreme or hostile environments (e.g., military, automotive, industrial,
extraterrestrial) take additional measures \mpk{(e.g., custom
components~\mpm{\cite{agrawal1994proposed, smart2021rugged, im2016im,
infineon2022radiation, lu1989advanced, banerjee1989two, mazumder1993design,
data2022rad, 3d2022ddr4}}, redundant resources~\cite{mathew2021thermoelectric,
kobayashi2017highly, patil2021dve})} to ensure that memory errors do not
compromise their systems. Section~\ref{subsubsec:mot_rela} shows the benefits of
incorporating mechanisms to improve memory reliability.} This section explains
how the details of a DRAM chip's reliability characteristics play a major role
in determining how system designers improve overall system reliability.

\subsection{Adapting Commodity DRAM Chips}
\label{subsec:rela_study_adapting_commodity_chips}

Commodity DRAM is designed to work for a wide variety of systems at a reasonable
(albeit unspecified)\footnote{\mph{Academic works speculate that commodity DRAM
targets a bit error rate (BER) within the range of
$10^{-16}-10^{-12}$~\cite{nair2013archshield, kim2016all,
longofono2021predicting, patel2017reach}, but we are unaware of
industry-provided values.}} error rate. \mph{In general, a system designer who
needs high memory reliability must design and build their own solutions (i.e.,
outside of the DRAM chip) to tolerate memory errors.\footnote{\mph{Even
designers who adopt custom DRAM solutions that sacrifice the cost advantages of
commodity memory (e.g., high-reliability DRAM~\cite{smart2021rugged, im2016im})
may supplement the DRAM chips with additional error-mitigation mechanisms
outside of the DRAM chip}.} In doing so, the designer effectively adapts a DRAM
chip to specific system needs, enhancing DRAM reliability beyond what the DRAM
chips provide alone.}

Section~\ref{subsubsec:mot_rela} reviews examples of such memory
error-mitigation mechanisms, which span the hardware-software stack. Regardless
of where each mechanism operates from, the mechanism targets a particular
\emph{error model}, which defines the scope of the errors that it is designed to
mitigate. This is important because, while a given mechanism efficiently
mitigates errors within its target error model, it may fail to do so if errors
no longer fit the model. In such cases, a different error-mitigation mechanism
\mpk{(or possibly, a combination of multiple mechanisms)} may be more suitable.

For example, a coarse-grained approach such as page
retirement~\mpk{\cite{mcelog2021bad, nvidia2020dynamic, venkatesan2006retention,
baek2014refresh, hwang2012cosmic, meza2015revisiting}} efficiently mitigates a
small number of errors at fixed bit positions. However, page retirement exhibits
significant capacity and performance overheads at high error rates or when
mitigating errors that change positions over time~\cite{meza2015revisiting,
lee2019exploiting, mcelog2021bad}. In contrast, a fine-grained hardware-based
approach such as a block error-correcting code~\mpk{\cite{moon2005error,
richardson2008modern, roth2006introduction, clark2013error, costello1982error,
costello2004ecc}} can efficiently mitigate a limited number of
randomly-distributed errors but can fail silently (and even exacerbate the
number of errors present~\mpm{\cite{alam2021lightweight, jeong2020pair,
criss2020improving, son2015cidra, patel2019understanding, patel2020bit,
patel2021enabling}}) when its correction capability is exceeded. We conclude
that it is essential for the system designer to \mpl{know} when and how errors
occur in a given memory chip in order to make an informed choice of which
error-mitigation mechanism to use in a particular system.

\subsection{\mpm{Lack of Transparency in Commodity DRAM}}
\label{subsec:rela_study_necessary_assumptions}

Unfortunately, system designers generally do not have access to definitive error
models for commodity DRAM chips. Therefore, designers are left to rely upon
information they can gather \mpk{by themselves} (e.g., by expending testing
resources) or from external, possibly untrustworthy, sources. However, as
Section~\ref{position:sec:formalizing} discusses, obtaining the error
characteristics of a DRAM chip without input from the manufacturers requires
making a series of assumptions about the chip's design and testing
methodologies. The need for these assumptions (i.e., the lack of trustworthy
information) can easily discourage designers from pursuing custom solutions to
enhance DRAM reliability. 

To exacerbate the problem of identifying a definitive error model, DRAM
manufacturers are starting to incorporate two on-die error-mitigation mechanisms
that correct a limited number of errors from within the DRAM chip itself: (1)
\emph{on-die ECC}~\cite{micron2017whitepaper, nair2016xed, oh2014a, kwak2017a,
kwon2014understanding, oh20153, patel2019understanding, patel2020bit} for
improving reliability and yield and (2) \emph{target row
refresh}~\cite{hassan2021uncovering, frigo2020trrespass, jattke2022blacksmith,
marazzi2022protrr} for partially mitigating the RowHammer vulnerability. Prior
works on ECC~\mpm{\cite{son2015cidra, gong2018duo, nair2016xed, jeong2020pair,
cha2017defect, pae2021minimal, criss2020improving, luo2014characterizing,
gurumurthi2021hbm3, patel2019understanding, patel2020bit, patel2021harp,
patel2021enabling}} and RowHammer~\mpm{\cite{saroiu2022price,
qureshi2021rethinking, frigo2020trrespass, hassan2021uncovering}} show that both
on-die ECC and TRR change how errors appear outside of the DRAM chip, thereby
changing the DRAM error model seen by the memory controller (and therefore, to
the rest of the system). \mpk{Unfortunately, both mechanisms are opaque to the
memory controller and are considered trade secrets that DRAM manufacturers will
not officially disclose~\cite{gong2017dram, nair2013archshield,
patel2019understanding, patel2020bit, saroiu2022price, qureshi2021rethinking,
loughlin2021stop, farmani2021rhat}. As a result, both on-die ECC and TRR make it
difficult for a system designer to reason about the DRAM error model and error
rates. For example, to account for on-die ECC's and TRR's effects when designing
a system-level error-mitigation mechanism, the system designer must spend
additional time and resources using reverse-engineering techniques (e.g., for
on-die ECC~\cite{patel2019understanding, patel2020bit} or
TRR~\mpm{\cite{hassan2021uncovering, frigo2020trrespass}}) or otherwise
\mpm{find} a trustworthy source to acquire the \mpm{necessary information in
reliable manner.}}
\section{Study 2: DRAM Refresh Overheads}
\label{sec:refresh_study}

DRAM refresh is a key design concern \mpf{in modern systems.
Section~\ref{subsubsec:mot_dram_refresh} reviews evidence that reducing the
total number of refresh operations significantly benefits overall system
performance and energy efficiency}. In this section, we examine how mitigating
refresh overheads in commodity DRAM requires making assumptions about DRAM
reliability characteristics. Based on our analysis, we argue that these
assumptions limit the techniques' potential for adoption, discouraging system
designers from using these solutions in practice.

\subsection{Adapting Commodity DRAM Chips}
\label{subsec:adapting_commodity_dram_ref}

Reducing unnecessary refresh operations in commodity DRAM chips generally
requires two key steps. First, the memory controller must reduce the frequency
of periodic refresh operations. This is \mpk{achievable (though not necessarily
supported to arbitrary values) using commodity DRAM chips because the memory
controller manages DRAM refresh timings. For example, the memory controller
might} relax the rate at which it issues refresh operations \mpk{to half of the
DDR\emph{n} standard of 3.9 or 7.8 $\mathrm{\mu}$s, which is supported by
standards at extended temperature ranges~\cite{jedec2020ddr5,
jedec2020lpddr5,jedec2014lpddr4, jedec2012ddr4, jedec2008ddr3}, or even to over}
an order of magnitude less often~\cite{venkatesan2006retention, liu2012raidr,
nair2013archshield, katayama1999fault}.

Second, the system must mitigate any errors that may occur within the small
number of DRAM cells that require frequent refreshing. Doing so requires either
using additional refresh operations (e.g., by issuing extra row
activations~\cite{liu2012raidr}) or using error-mitigation mechanisms within
processor (e.g., ECC~\cite{qureshi2015avatar} and/or bit-repair
techniques~\cite{venkatesan2006retention, lin2012secret, nair2013archshield}).
Although both strategies introduce new performance and energy overheads,
\mpl{the benefits of reducing unnecessary refresh operations outweigh the
overheads introduced~\cite{liu2012raidr, ohsawa1998optimizing,
wang2014proactivedram, venkatesan2006retention, lin2012secret,
nair2013archshield, ghosh2007smart, qureshi2015avatar, patel2017reach,
nguyen2021zem}. For example, Liu et al.~\cite{liu2012raidr} project that DRAM
refresh overheads cause a 187.6\% increase in the energy-per access and a 63.7\%
system performance degradation for 64~Gib chips. By reducing the overall number
of DRAM refresh operations, the authors show that their mechanism, RAIDR, can
mitigate these overheads by 49.7\% and 107.9\%, respectively.} 

\subsection{\mpm{Lack of Transparency in Commodity DRAM}}
\label{subsec:limitng_assumptions_dram_ref}

Knowing, predicting, or identifying cells that cannot safely withstand
infrequent refreshing (i.e., retention-weak cells) is a difficult reliability
\mpk{problem because the cells' likelihood of error changes with how a DRAM chip
is used (i.e., operating conditions such as the refresh rate, voltage,
temperature) and the particular DRAM chip circuit design (e.g., random
cell-to-cell variations, locations of true and
anti-cells~\cite{liu2013experimental, kraft2018improving,
patel2019understanding}). Prior works propose two practical ways of identifying
retention-weak cells: (1) \emph{active profiling}, which uses comprehensive
tests to search for error-prone cells offline~\cite{liu2012raidr,
patel2017reach, khan2016parbor, khan2014efficacy, lin2012secret,
mathew2017using}, and (2) \emph{reactive profiling}, which constantly monitors
memory to identify errors as they manifest during runtime, e.g., ECC
scrubbing~\cite{qureshi2015avatar, han2014data, choi2020reducing}. Both
approaches require the profiler to understand the \emph{worst-case} behavior of
data-retention errors for a given DRAM chip~\cite{khan2014efficacy,
lin2012secret}: an active profiler must use the worst-case conditions to
maximize the proportion of retention-weak cells it identifies during
profiling~\cite{patel2017reach} and a reactive profiler must be provisioned to
identify (and possibly also mitigate) the worst-case error pattern(s) that might
be observed at runtime, e.g., to choose an appropriate ECC detection and
correction capability~\cite{patel2021harp, qureshi2021rethinking,
khan2014efficacy}.}

\mpk{The fact that an effective error profiling mechanism relies on
understanding the underlying error characteristics reinforces the argument
presented in Section~\ref{position:sec:formalizing}. Even though there exist
techniques for mitigating refresh overheads in commodity DRAM, practically
adopting them relies on prerequisite knowledge about a DRAM chip and its
reliability characteristics that is not provided by the DRAM industry
today.}
\section{Study 3: Long DRAM Access Latency}
\label{sec:latency_study}

Slow generational improvements in the DRAM access latency \mpf{(shown in
Section~\ref{position:bg_mot}) contrast with the growing prevalence of
latency-sensitive workloads today~\mpm{\cite{hsieh2016accelerating,
ferdman2012clearing, gutierrez2011full, hestness2014comparative, huang2014moby,
zhu2015microarchitectural, oliveira2021damov, boroumand2018google,
boroumand2021google, koppula2019eden, kanellopoulos2019smash, son2013reducing,
mutlu2013memory, wilkes2001memory, wulf1995hitting, mutlu2007stall,
mutlu2003runahead, kanev2015profiling, mutlu2014research, bera2019dspatch,
bera2021pythia, liu2019binary, ghose2019processing, shin2014nuat,
ghose2019demystifying, gomez2021benchmarking, gomez2021benchmarkingmemory,
giannoula2022towards}}. Therefore, as Section~\ref{subsubsec:mot_dram_latency}
discusses, there is significant opportunity for improving overall system
performance by reducing the memory access
latency~\mpk{\cite{chandrasekar2014exploiting, chang2016understanding,
kim2018solar, lee2015adaptive, lee2017design, wang2018reducing,
zhang2016restore, hassan2016chargecache, koppula2019eden, mathew2017using,
zhang2021quantifying, kim2020improving, lee2016simultaneous}}.} In this section,
we study how \mpk{techniques} for reducing the access latency of commodity DRAM
chips rely on making assumptions about DRAM reliability characteristics. Then,
we argue that \mpk{the need for these assumptions \mpm{(and the lack of
transparency in DRAM to allow them)} discourages system designers from adopting
the latency reduction techniques}.

\subsection{Adapting Commodity DRAM Chips}
\label{subsec:adapting_commodity_dram_lat}

Strategies for improving the access latency of commodity DRAM chips rely on
manipulating DRAM commands and/or access timings to either (1) \mpk{eliminate}
conservative timing margins that DRAM manufacturers use to account for
worst-case operation~\mpm{\cite{chang2016understanding, lee2015adaptive,
chandrasekar2014exploiting, mathew2017using, shin2014nuat, kim2018solar,
hassan2016chargecache, david2011memory, deng2011memscale,
zhang2021quantifying}}; or (2) exploit undefined DRAM chip behavior to perform
beneficial operations (e.g., performing massively-parallel computations within
DRAM rows~\mpm{\cite{hajinazar2021simdram, gao2019computedram,
seshadri2013rowclone, olgun2021pidram, seshadri2017ambit, seshadri2017simple,
seshadri2015fast, seshadri2016buddy, seshadri2020indram}}, generating random
values~\cite{kim2019d, talukder2018exploiting, olgun2021quac} or unique chip
identifiers~\cite{kim2018dram, talukder2018ldpuf, yue2020dram,
hashemian2015robust, schaller2018decay}). 

In both cases, \emph{new} DRAM access timings must be \mpk{determined that}
ensure the desired operation can be performed predictably and reliably
\mpk{under all conditions}. To identify \mpk{these access} timings, prior
works~\mpk{\cite{kim2018dram, kim2019d, talukder2019prelatpuf,
gao2019computedram, olgun2021quac, chang2016understanding,
chang2017understanding, lee2015adaptive, chandrasekar2014exploiting,
ghose2018your, hassan2017softmc, chang2017thesis, lee2016reducing,
david2011memory} perform extensive experimental characterization studies across
many DRAM chips.} These studies account for three primary sources of variation
that affect the access timings of a given memory location. First, process
variation introduces random variations between DRAM chip components (e.g.,
cells, rows, columns). Second, a manufacturer's particular circuit design
introduces structural variation (called design-induced
variation~\cite{lee2017design}) that deterministically affects access timings
based on a component's location in the overall DRAM design (e.g., cells along
the same bitline~\cite{kim2018solar}, cells at the borders of internal storage
arrays~\cite{lee2017design}). Third, the charge level of a DRAM cell varies over
time due to leakage and the effects of DRAM accesses~\cite{shin2014nuat,
hassan2016chargecache}. Experimentally determining the \mpk{new predictable and
reliable} access timings requires \mpk{properly} accounting for all three
sources of variation \mpk{under all operating conditions}.

\subsection{\mpm{Lack of Transparency in Commodity DRAM}}
\label{subsec:limitng_assumptions_dram_lat}

Unfortunately, determining new viable access timings requires developing and
executing a reliable testing methodology, which in turn requires making similar
assumptions to those discussed for \mpk{data-retention error profiling in
Section~\ref{subsec:limitng_assumptions_dram_ref}. Choosing runtime (e.g., data
and access patterns) and environmental (e.g., temperature, voltage) testing
conditions \emph{in a meaningful way} requires some understanding of the error
mechanisms involved in timing-related errors~\cite{khan2017detecting}, including
(but not limited to)} aspects of the circuit design, such as internal
substructure dimensions (e.g., subarray sizing)~\cite{kim2018solar,
lee2017design}, the correspondence between logical DRAM bus addresses and
physical cell locations~\mpm{\cite{chang2016understanding, lee2015adaptive,
khan2016parbor}}, and the order of rows refreshed by each auto-refresh
operation~\cite{shin2014nuat}. \mpk{A system designer is discouraged from
exploring improvements to the commodity DRAM access latency without trustworthy
access to this information.}
\section{Study 4: RowHammer Mitigation}
\label{sec:security_study}

\mpf{Many promising proposals exist for adding RowHammer defenses to commodity
DRAM chips} (discussed in Section~\ref{subsubsec:mot_dram_security}), but their
potential for adoption is hampered by system designers' lack of visibility
\mpf{into how the underlying error mechanism behaves.} In this section, we
examine the various assumptions that RowHammer defense proposals rely upon and
argue that these assumptions pose serious barriers for practical adoption.

\subsection{Adapting Commodity DRAM Chips}

\mpf{To effectively mitigate RowHammer bit flips, a mitigation mechanism must be
configured based on the vulnerability level of a given DRAM chip. This requires
estimating the chip's RowHammer error characteristics for different operating
conditions and access patterns. Each of the four mechanism types introduced in
Section~\ref{subsubsec:mot_dram_security} requires estimating different
characteristics.} Table~\ref{tab:rh_mitigation_info} summarizes the different
pieces of information required for each mitigation type. The first is known as
\hcfirst{}~\mpk{\cite{kim2020revisiting, orosa2021deeper} or RowHammer
Threshold~\cite{kim2014flipping, yaglikci2020blockhammer,
bennett2021panopticon}}, which describes the worst-case number of RowHammer
memory accesses required to induce a bit-flip. The second is known as the blast
radius~\cite{kim2014flipping, kim2020revisiting}, which describes how many rows
\mpf{are affected by hammering a single row}. The third is the DRAM's internal
physical row address mapping~\mpm{\cite{kim2014flipping, kim2012case}}, which is
necessary to identify the locations of victim rows.

\begin{table}[h]
    \centering
    \small
    \begin{tabular}{l|ccc}
                           & \multicolumn{3}{c}{\textbf{Required Information}} \\
        \textbf{Strategy}  & \hcfirstbold{}  &  \textbf{Blast Radius} &  \textbf{Row Mapping } \\\hline 
        Access-Agnostic       & \cmark & & \\               
        Proactive             & \cmark & \cmark & \\         
        Physically Isolating  & \cmark & \cmark & \cmark \\                  
        Reactive              & \cmark & \cmark & \cmark
    \end{tabular}
    \caption{Information needed by each of the four RowHammer-mitigation strategies.}
    \label{tab:rh_mitigation_info}
\end{table}

All three RowHammer error characteristics vary between DRAM manufacturers,
chips, \mpk{and cells} based on a combination of random process variation, a
manufacturers' particular circuit design (including yield-management techniques
such as post-manufacturing repair, target row refresh, and error correcting
codes), \mpk{and operating conditions such as temperature and
voltage~\cite{kim2014flipping, park2016statistical, park2016experiments,
kim2020revisiting, orosa2021deeper, yaglikci2021security, yun2018study,
lim2016active, farmani2021rhat}. Therefore, as with estimating DRAM refresh and
access timings (discussed in Sections~\ref{subsec:limitng_assumptions_dram_ref}
and~\ref{subsec:limitng_assumptions_dram_lat}), these studies rely on extensive
experimental testing to estimate RowHammer error characteristics that are needed
to design and/or configure the RowHammer defenses discussed in
Section~\ref{subsubsec:mot_dram_security}.}

\subsection{\mpm{Lack of Transparency in Commodity DRAM}}
\label{position:subsec:rh:necessary}

We observe that \mpk{all} previously-proposed RowHammer mitigation mechanisms
require accurately estimating RowHammer error characteristics \mpk{throughout
all valid operating conditions}. In particular, \mpm{every mechanism} must be
tuned against \mpg{at least} \hcfirst{} in order to effectively prevent
RowHammer. \mpm{Prior works~\mpk{\cite{qureshi2021rethinking, saroiu2022price,
loughlin2021stop}} make the same observation, discussing the difficulty in
practically determining and relying on this information without support from
DRAM manufacturers.} 

\mpm{Therefore, a security-focused system designer who wants to implement or
build upon one of the many previously-proposed system-level RowHammer defense
mechanisms (discussed in Section~\ref{subsubsec:mot_dram_security}) is limited
by the same information access challenges as discussed in
Section~\ref{position:subsec:knowing_or_assuming}: because neither the error
characteristics they need nor the methods to obtain them are provided by
official sources, the system designer must rely on other means to obtain the
necessary information. As a result, the system designer is likely discouraged
from exploring designs that} \mpn{address RowHammer errors in commodity DRAM
chips altogether.}
\section{\mpk{Current} DRAM Standards as the Problem}
\label{sec:spec_as_problem}

Based on our case studies, we conclude that reliance on information about DRAM
reliability characteristics poses a serious challenge for optimizing how
commodity DRAM is used. In this section, we hypothesize that the unavailability
of \mpk{information related to DRAM reliability} is caused by a \emph{lack of
transparency} within DRAM standards which provide \emph{control over}, but not
\emph{insight into}, DRAM operations. We identify DRAM standards as both (1) the
root cause of having to make assumptions about DRAM reliability \mpk{(as
standards are currently defined)} and (2) the pathway to a solution for
alleviating the need for such assumptions \mpk{(by incorporating DRAM
reliability as a key concern)}.

\subsection{\xmt{The Problem of Information Unavailability}}
\label{subsec:problem_of_unavailability}

\mpk{In each case study} throughout
Sections~\ref{sec:rela_study}--\ref{sec:security_study}, we observe that
optimizing commodity DRAM chips for key system design concerns requires knowing
information about DRAM reliability. This is unsurprising because
\mpk{reliability is central to each case study's approach: each study improves
system-level metrics (e.g., reliability, energy-efficiency, performance,
security) by leveraging key properties of one or more error mechanisms (e.g.,
spatiotemporal dependence of errors due to circuit timing
violations~\cite{kim2018solar, chang2016understanding, lee2017design}, the
localized nature of RowHammer errors~\mpm{\cite{kim2014flipping,
konoth2018zebram, van2018guardion, brasser2017can, kim2020revisiting}}).}
Therefore, identifying the best operating point requires \mpk{at least a basic
understanding} of how the error mechanisms themselves behave under
\mpk{representative} operating conditions.

Recent works~\cite{qureshi2021rethinking, saroiu2022price} discuss the pitfalls
of \mpk{designing defense mechanisms that rely on knowledge of how RowHammer
errors behave (e.g., \hcfirst{}, dependence on a chip's internal cell
organization)}, calling into question the practicality of accurately determining
\mpk{these details given an arbitrary DRAM chip.} Knowing or determining this
information is essential to guarantee protection against RowHammer. However,
determining it \mpk{without guidance from DRAM manufacturers} requires per-chip
testing and/or reverse-engineering that relies on the accuracy of the underlying
testing methodology used, which itself relies on knowledge of \mpk{DRAM chip
details} that likely needs to be assumed or inferred (as discussed in
Sections~\ref{position:sec:formalizing} and~\ref{position:subsec:rh:necessary}). 

As a result, a system designer who wants to adapt commodity DRAM for their
design requirements today is forced to make design and/or mechanism
configuration decisions based upon assumptions or inferences from unofficial
sources (e.g., self-designed experimental studies~\mpm{\cite{lee2015adaptive,
lee2017design, kim2018solar, kim2018dram, kim2019d, talukder2018exploiting,
talukder2019prelatpuf, talukder2018ldpuf, chang2016understanding,
kim2014flipping, liu2013experimental, hassan2017softmc,gao2019computedram,
olgun2021quac, chang2017understanding, chandrasekar2014exploiting,
ghose2018your, chang2017thesis, lee2016reducing, david2011memory}}).
Unfortunately, even a system designer willing to spend significant resources on
such adaptations (e.g., \mpk{to enhance system} reliability, performance,
security, etc.) may be discouraged by the underlying dependence on untrustworthy
information. In the worst case, \mpk{the designer may judge} \emph{all}
adaptations \mpk{to be} impractical without a trustworthy understanding of a
DRAM chip. \mpk{We conclude that the lack of information transparency today
discourages system designers} from exploring alternative designs that have been
shown to provide tangible benefits.

\subsection{Limitations of DRAM Standards}

Current DRAM standards do not address general reliability characteristics
because commodity DRAM is designed for a fixed, high-reliability operating point
such that the typical consumer can largely ignore errors. This follows directly
from the separation-of-concerns between system and DRAM designers: current DRAM
standards place \mpg{most of the burden of addressing DRAM} reliability
challenges (e.g., worsening error rates with continued technology
scaling~\mpk{\cite{kang2014co, micron2017whitepaper, mutlu2013memory}) on DRAM
manufacturers} alone.\footnote{\mpg{High-reliability systems may supplement DRAM
chips' base reliability with additional error-mitigation mechanisms, as
discussed in Section~\ref{subsubsec:mot_rela}.}}

We believe that this state of affairs arises naturally because establishing a
strict separation of concerns requires a clear and explicit interface between
manufacturers and customers. Consequently, ensuring that the standards leave
enough flexibility for diverse customer use-cases requires careful and explicit
attention. This is because the standards are susceptible to \emph{abstraction
inversion}~\cite{baker1990opening}, a design anti-pattern in which a previously
agreed-upon interface becomes an \emph{obstacle}, forcing \mpk{system designers}
to re-implement basic functionality in terms of the outdated abstraction. A
rigid interface limits what is and is not possible, potentially requiring
unproductive reverse-engineering to work around.

We argue that \mpk{needing to make assumptions in order to adapt} commodity DRAM
to system-specific goals \mpk{clearly indicates} abstraction inversion today.
This implies that DRAM standards have aged without sufficient attention to
flexibility. Although a fixed operating point defines a clear interface, we
believe that leaving room for (and potentially even encouraging) different
operating points \mpk{is essential today.}

\section{DRAM \mpg{Standards} as the Solution}
\label{sec:two_part_change_to_specs}

We believe that the separation of concerns provided by DRAM standards is
necessary for practicality because it enables DRAM manufacturers and system
designers to focus on designing the best possible products within their
respective areas of expertise. However, we argue that the separation must be
crafted in a way that not only does not impede progress, but ideally encourages
\mpk{and aids} it. To achieve both goals, we propose extending DRAM standards in
a way that enables system designers to make informed decisions about how their
design choices will affect DRAM operation. In other words, instead of modifying
DRAM \emph{designs}, we advocate modifying \emph{standards} to facilitate
transparency of DRAM reliability characteristics. Armed with this information,
system designers can freely explore how to best use commodity DRAM chips to
solve their own design challenges while preserving the separation of concerns
that allows DRAM designers to focus on building the best possible
standards-compliant DRAM chips.

\subsection{\mpl{Choosing Information to Release}}
\label{position:subsec:what_to_release}

\mpg{We identify what information to release using our analysis of information
flow in Section~\ref{position:sec:formalizing}. We observe that, given the
information at \emph{any} node in Figure~\ref{fig:test_flow}, system designers
can work to determine the information at each of its child nodes. As a result,
\mpi{access to trustworthy} information at \emph{any} node \mpi{provides system
designers with a foundation to make informed design decisions.} Therefore, we
recommend that the DRAM industry be free to release information at \emph{at
least one} node of their choice that they are willing and capable of doing so.
This} section examines realistic possibilities for communicating information at
each \mpg{node} of the flowchart.

\subsubsection{\xmt{Basic Design Characteristics}}
\label{position:subsubsec:design_characteristics}

At the lowest level, DRAM manufacturers could provide basic chip design
characteristics that allow system designers to develop their own test
methodologies and error models. This is the most general and flexible approach
because it places no limitations on what types of studies system designers may
pursue (e.g., in contrast to providing information \mpk{that is useful for
reasoning about only one} particular error mechanism).
Table~\ref{tab:design_characteristics} gives examples of key design
characteristics that prior works often make assumptions about in their own
efforts to optimize commodity DRAM usage. \mpk{For each design characteristic,
we list prior works that reverse-engineer the characteristic and describe
use-cases that rely on knowledge of the characteristics.}

\begin{table*}[t]
    \centering
    \scriptsize
    \setlength\tabcolsep{3pt}
    \begin{tabular}{L{3.8cm}|L{3cm}L{8.5cm}}
        \textbf{Design Characteristic} & \textbf{Reverse-Engineered By} & \textbf{Use-Case(s) Relying on Knowing the Characteristic} \\\hline\hline
        \begin{tabular}[c]{@{}l@{}}Cell charge encoding convention\\(i.e., true- and anti-cell layout)\end{tabular} & Testing~\cite{patel2019understanding, kraft2018improving, patel2017reach, liu2013experimental} & Data-retention error modeling and testing \mpk{for mitigating refresh overheads} (e.g., designing worst-case test patterns)~\cite{khan2017detecting, kraft2018improving, liu2013experimental} \\\hline
        On-die ECC details & Modeling and testing~\cite{patel2019understanding, patel2020bit} & Improving reliability (e.g., designing ECC within the memory controller)~\cite{son2015cidra, cha2017defect, criss2020improving, gong2018duo}, \mpk{mitigating RowHammer}~\cite{kim2020revisiting, hassan2021uncovering, cojocar2019exploiting, jattke2022blacksmith}  \\\hline
        Target row refresh (TRR) details & Testing~\cite{hassan2021uncovering, frigo2020trrespass} & Modeling and mitigating RowHammer~\cite{frigo2020trrespass, hassan2021uncovering, jattke2022blacksmith} \\\hline
        \mpk{Mapping between internal and external row addresses} & Testing~\cite{jung2016reverse, kim2020revisiting, tatar2018defeating, barenghi2018software, wang2020dramdig, lee2017design} & Mitigating RowHammer~\cite{kim2014flipping, barenghi2018software, jung2016reverse, kim2020revisiting} \\\hline
        \mpk{Row addresses refreshed by each refresh operation} & Testing~\cite{hassan2021uncovering} & Mitigating RowHammer~\cite{hassan2021uncovering}, improving access timings~\cite{shin2014nuat, wang2018reducing} \\\hline
        Substructure organization (e.g., cell array dimensions) & Modeling~\cite{lee2017design} and testing~\cite{chang2016understanding, lee2017design, kim2018solar} & Improving DRAM access timings~\cite{lee2017design, chang2016understanding, kim2018solar} \\\hline
        \begin{tabular}[c]{@{}l@{}}Analytical model parameters\\(e.g., bitline capacitance)\end{tabular} & Modeling and testing~\cite{hamamoto1998retention, liu2013experimental} & \mpk{Error models for reliability~\cite{li2011dram}, data-retention~\cite{das2018vrl, hamamoto1998retention, shirley2014copula, restle1992dram}, latency-related errors~\cite{lee2017design}, and RowHammer~\cite{walker2021dram, park2016statistical}} \\
    \end{tabular}
    \caption{Basic DRAM chip design characteristics that are typically assumed or inferred for experimental studies.}
    \label{tab:design_characteristics}
\end{table*}

We believe that releasing \mpg{these characteristics will minimally (if at all)}
impact DRAM manufacturer's business interests given that each of the
characteristics can be reverse-engineered with existing methods \mpk{(as shown
by Table~\ref{tab:design_characteristics}, Column 2)} and access to appropriate
tools, as demonstrated by prior studies~\mpm{\cite{patel2019understanding,
kraft2018improving, patel2017reach, liu2013experimental, patel2020bit,
hassan2021uncovering, frigo2020trrespass, jung2016reverse, kim2020revisiting,
lee2017design, chang2016understanding, kim2018solar, hamamoto1998retention,
barenghi2018software, wang2020dramdig, mukhanov2020dstress, kim2014flipping,
khan2014efficacy, farmani2021rhat}}. Releasing this information in an official capacity simply
confirms what is already suspected, providing a competitor with no more
information about a given DRAM chip than they already had available. On the
other hand, knowing this information empowers system designers \mpk{and enables
them to confidently design and implement system-level optimizations}, benefiting
both designers and manufacturers in the long run (as discussed in
Section~\ref{sec:motivation_new}).

\subsubsection{Test Methodologies}

\mpk{At a level of abstraction beyond chip design details}, DRAM manufacturers
could describe effective test methodologies that system designers can use to
study the particular aspects of DRAM reliability they are interested in.
Compared with providing chip design characteristics, directly providing test
methodologies absolves (1) manufacturers from needing to reveal chip design
information; and (2) system designers from needing the DRAM-related expertise to
determine the test methodologies from chip design characteristics.\footnote{We
believe that interested parties already have such expertise, as shown by the
fact that many studies~\mpm{\cite{patel2019understanding, kraft2018improving,
patel2017reach, liu2013experimental, patel2020bit, hassan2021uncovering,
frigo2020trrespass, jung2016reverse, kim2020revisiting, lee2017design,
chang2016understanding, kim2018solar, hamamoto1998retention,
barenghi2018software, wang2020dramdig, mukhanov2020dstress, kim2014flipping,
khan2014efficacy, farmani2021rhat}} determine the necessary test methodologies through extensive
experimentation.} \mpk{As a drawback}, providing test methodologies alone limits
system designers to working with only the particular error mechanisms that the
methodologies are designed for (e.g., data-retention, RowHammer).
Table~\ref{tab:test_params} summarizes key aspects of testing methodologies that
prior works generally need to assume throughout the course of their testing.

\begin{table}[h]
    \centering
    \small
    \begin{tabular}{L{3cm}|L{12cm}}
        \textbf{Test Parameter} & \textbf{Description} \\\hline\hline
        Data pattern 
            & Data pattern that maximizes the chance of errors occurring~\mpk{\cite{duganapalli2016modelling, liu2013experimental, khan2014efficacy, kim2014flipping, mukhanov2020dstress, patel2017reach, kim2020revisiting, orosa2021deeper,hassan2021uncovering, cojocar2020are,tatar2018defeating, cojocar2021mfit, jattke2022blacksmith, borucki2008comparison, weis2015retention, kraft2018improving, mukhanov2020dstress,kim2018solar}} \\\hline
        Environmental conditions 
            & Temperature and voltage that lead to worst-case behavior~\mpk{\cite{park2016experiments, kim2019d, liu2013experimental, yaglikci2022understanding, orosa2021deeper, schroeder2009dram, weis2015retention,wang2018dram, yaney1987meta,hamamoto1998retention}} \\\hline
        Test algorithm 
            & Sequence of representative and/or worst-case DRAM operations to test~\mpk{\cite{cojocar2020are, kim2014flipping, liu2013experimental, lee2015adaptive, kim2018dram, kim2019d, hassan2021uncovering, jattke2022blacksmith, salman2021half}}
    \end{tabular}
    \caption{Testing parameters that are typically assumed or inferred during experimental studies.}
    \label{tab:test_params}
\end{table}

\subsubsection{Test Results and/or Error Models}
\label{position:subsubsection:test_results}

\mpk{At the highest level of abstraction}, DRAM manufacturers can directly
provide test results and/or error models related to specific studies needed by
system designers. This could take the form of parametric error models (e.g., the
statistical relationship between operating timings and error rates) along with
parameter values for each chip, fine-granularity error characteristics (e.g.,
per-column minimum viable access timings) and/or summary statistics of interest
(e.g., \hcfirst{} in studies pertaining to RowHammer). In this way, system
designers can constrain (or entirely bypass) testing when developing mechanisms
using the provided information. \mpk{As a drawback}, directly releasing test
results and/or error models constrains system designers to developing solutions
only for those design concerns that pertain to the released information.
Table~\ref{tab:test_result_error_model} provides examples of key test results
and error models that prior works \mpk{leverage in order to implement
optimizations to commodity DRAM.}

\begin{table}[h]
    \centering
    \small
    \begin{tabular}{L{5cm}L{10cm}}
        \textbf{Test Result or Error Model} & \textbf{Description} \\\hline\hline
        Data-retention times
            & Minimum refresh rate required for different DRAM regions (e.g., rows, cells)~\mpk{\cite{liu2012raidr, lin2012secret, liu2013experimental,khan2014efficacy, khan2016case, kim2001block, nair2013archshield}} \\\hline
        Error profile
            & \mpk{List of cells susceptible to errors (e.g., VRT~\cite{qureshi2015avatar, liu2013experimental, khan2014efficacy}, latency-related~\cite{kim2019d, kim2018dram, kim2018solar, chang2016understanding, chandrasekar2014exploiting})} \\\hline
        Error rate summary statistics
            & \mpk{Aggregate error rates (e.g., BER~\cite{liu2013experimental, patel2017reach, patel2019understanding, weis2015retention,kang2014co}, FIT~\cite{schroeder2009dram, levy2018lessons, wang2009soft}), distribution parameters (e.g., copula~\cite{shirley2014copula}, lognormal~\cite{hamamoto1995well, hamamoto1998retention, li2011dram}, exponential~\cite{liu2012raidr,kumar2014detection})} \\\hline
        RowHammer blast radius
            & Maximum number of rows affected by hammering \mpm{one or more row(s)}~\mpk{\cite{yaglikci2020blockhammer, kim2020revisiting, yaglikci2021security, walker2021dram, loughlin2021stop, kim2014flipping}} \\\hline
        \hcfirst{} or RowHammer Threshold
            & \mpk{Minimum number of RowHammer accesses required to induce bit-flips~\cite{kim2020revisiting, orosa2021deeper, kim2014flipping, yaglikci2020blockhammer, bennett2021panopticon}}
    \end{tabular}
    \caption{\mpg{Examples of key test results and error models from prior works that study and/or optimize commodity DRAM.}}
    \label{tab:test_result_error_model}
\end{table}

\subsection{\mpl{Choosing} When to Release the Information}

\mpi{We expect that releasing information by changing DRAM standards will be a
slow process due to the need for consensus between DRAM stakeholders. Instead,
we propose decoupling the \emph{release} of information from the
\emph{requirement} to do so. To this end, we recommend a practical two-step
process with different approaches in the short- and long-term.}

\subsubsection{Step 1: \mpg{Immediate Disclosure of Information}}

\mpg{We \mpi{recommend} two independent approaches to quickly release
information in the short-term. First, we recommend a public crowdsourced database
\mpi{that aggregates already-known information, e.g., inferred through}
reverse-engineering studies. We believe this is practical given the significant
research and industry interest in optimizing how commodity DRAM chips are used.
Such a database would provide an opportunity for peer review of posted
information, increasing the likelihood that the information is trustworthy. In
the long run, we believe such a database would facilitate information release
from DRAM manufacturers themselves because the manufacturers could simply
validate database information, if not contribute directly.}

\mpg{Second, we recommend} that commodity DRAM manufacturers individually
release one or more of the aforementioned categories of information for current
DRAM chips and those already in the field. For example, manufacturers may
\mpi{update} chip datasheets to incorporate relevant design characteristics or
make more extensive information available online (e.g., similar to how some
manufacturers already provide compliance documents and functional simulation
\mpk{models} through their websites~\mpk{\cite{micron2021dram, issi2022ddr4,
nanya2022NT5AD256M16E4}}). Releasing \mpk{any of the information described
throughout Section~\ref{position:subsec:what_to_release}} requires no changes to
DRAM designs or standards, though modifying DRAM standards (e.g., via an
addendum, as we suggest in Step 2) would help unify the information release
across all manufacturers. However, in the short term, we believe it is more
important to release the information, even if \mpk{not} standardized, so that it
is available as soon as possible.

\subsubsection{Step 2: Explicit DRAM Reliability Standards}

In the long term, we recommend DRAM standards be modified to \mpk{promote (or
even require)} DRAM manufacturers to disclose any information that impacts DRAM
reliability as relevant to a system designer. This information may include any
or all of the information discussed throughout this work; we believe that the
DRAM stakeholders themselves (i.e., DRAM manufacturers and system designers) are
in \mpk{a good} position to determine and standardize which information is the
most relevant and useful to regulate.

As a concrete example of how such changes to standards may occur, we reference
test methodologies~\cite{jedec2010ssdrequirements, jedec2010ssdendurance} and
error models~\cite{jedec2016failure} that JEDEC provides for NAND \mpm{flash
memory endurance~\cite{cai2017error, cai2018errors, cai2012error}, including
floating-gate data retention~\cite{cai2015data, luo2018heatwatch,
luo2018improving, cai2012flash} and threshold voltage
distributions~\cite{cai2013threshold, cai2013program, cai2015read,
luo2016enabling}}. These documents outline standardized best practices for
studying and characterizing endurance properties of SSD devices. We envision
analogous documents released for key DRAM error mechanisms (e.g.,
data-retention, access-timing-related, RowHammer), providing a standardized and
reliable alternative to inferring the same information through unofficial
channels.

\subsection{Alternative Futures}

We anticipate consumer use-cases \mpk{to continue diversifying}, making
affordable-yet-flexible DRAM increasingly important. Ambitious initiatives such
as DRAM-system co-design~\mpk{\cite{patterson1997case, mutlu2014research,
mutlu2021primer, kim2014flipping, mutlu2015main}} and emerging, non-traditional
DRAM architectures~\mpm{\cite{devaux2019true, kwon202125, mutlu2021primer,
oliveira2021damov, he2020newton, niu2022184qps, ahn2016scalable, lee20221ynm,
mutlu2019processing, gomez2021benchmarking, gomez2021benchmarkingmemory}} will naturally engender transparency by tightening the
relationship between DRAM manufacturers and system designers. Regardless of the
underlying motivation, we believe that increased transparency of DRAM
reliability characteristics will remain crucial to allowing system designers to
make the best use of commodity DRAM chips \mpk{by enabling them to customize
DRAM chips for system-level goals}.
\section{Conclusion}
\label{sec:conclusion}

We \mpm{contend} that system designers lack the necessary transparency into DRAM
reliability to make informed decisions about how their design choices will
affect DRAM operation. Without this transparency, system designers are
discouraged from exploring \mpk{the full design space around commodity DRAM,
wasting considerable potential for system-level optimization in meeting the
particular needs of their systems. We support our argument with} four case
studies that each examine an important design concern in modern DRAM-based
systems: (1) improving DRAM reliability; (2) mitigating DRAM refresh overheads;
(3) decreasing the DRAM access latency; and (4) defending against RowHammer. For
each case study, we argue that developing an effective system-level solution
requires making restrictive, potentially incorrect assumptions about DRAM
reliability characteristics. Based on our studies, we identify DRAM standards as
the source of the problem: current standards enforce a fixed operating point
without providing the context necessary to enable safe operation outside that
point. To overcome this problem, we introduce a two-step approach that modifies
DRAM standards to incorporate transparency of key reliability characteristics.
We believe that our work paves the way for a more open and flexible DRAM
standard that \mpk{enables \mpm{DRAM} consumers to better adapt and build upon
commodity DRAM technology while allowing \mpm{DRAM} manufacturers to preserve
their competitive edge. As a result, our work enables better innovation of
customized DRAM systems to fully harness the advantages of DRAM technology into
the future.}

\begin{subappendices}
    \section{DRAM Trends Survey}
\label{position:appendix_a}

We survey manufacturer-recommended DRAM operating parameters as specified in
commodity DRAM chip datasheets in order to understand how the parameters have
evolved over time. We extract values from 58 independent DRAM chip datasheets
from across 19 different DRAM manufacturers with datasheet publishing dates
between 1970 and 2021. \mpk{Appendix~\ref{position:appendix_b} lists each
datasheet and the details of the DRAM chip that it corresponds to. We openly
release our full dataset on GitHub~\cite{datasheetsurveygithub}, \mpm{which
provides a spreadsheet with all of the raw data used in this paper, including
each timing and current parameter value, and additional fields (e.g., clock
frequencies, package pin counts, remaining IDD values) that are not presented
here}.}

\subsection{\mpl{DRAM Access Timing Trends}}
\label{position:subsec:timing_trends}

We survey the evolution of the following four DRAM timing parameters that are
directly related to DRAM chip performance.

\begin{itemize}
    \item \emph{tRCD}: time between issuing a row command (i.e., row activation) and a column command (e.g., read) to the row.
    \item \emph{CAS Latency (or tAA)}: time between issuing an access to a given column address and the data being ready to access.  
    \item \emph{tRAS}: time between issuing a row command (i.e., row activation) and a precharge command.
    \item \emph{tRC}: time between accessing two different rows.
\end{itemize}

\begin{figure}[h]
    \centering
    \includegraphics[width=0.7\linewidth]{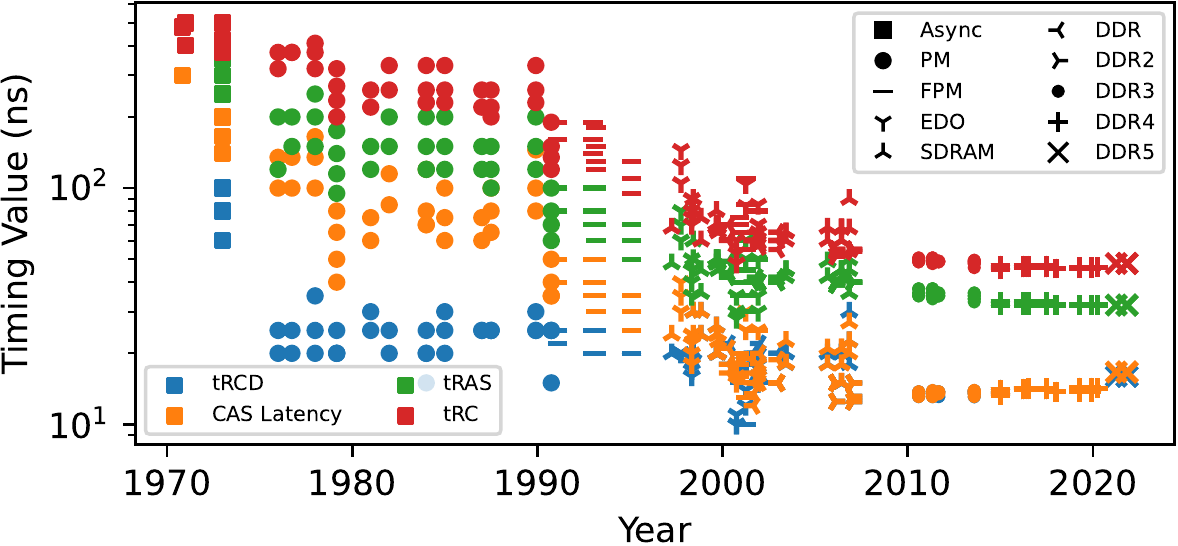}\\
    \includegraphics[width=0.7\linewidth]{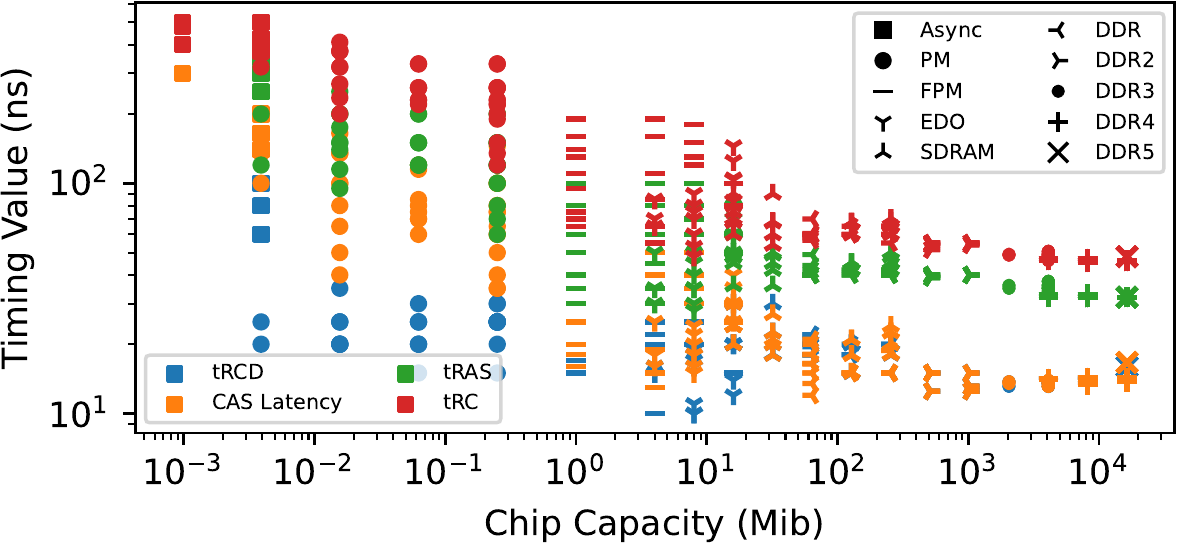}
    \caption[]{Evolution of \mpk{four} key DRAM timing parameters \mpn{(shown in
    log scale)} across years (top) and chip capacities (bottom) separated by
    DRAM standard.}
    \label{fig:da57}
\end{figure}

\noindent
Figure~\ref{fig:da57} shows how key DRAM timing parameters have evolved across
DRAM chips of different years (top) and capacities (bottom). \mpn{Timing values
are shown in log scale to better distinguish small values in newer DRAM chips.}
\mpm{Each type of marker illustrates DRAM chips of different DRAM standards.}

We make three qualitative observations. First, while all four DRAM timing values
have roughly decreased over time, improvements have been relatively stagnant for
the last two decades (note the logarithmic Y-axis). The bulk of the improvement
in timing parameter values occurred during the period of asynchronous DRAM, and
following the introduction of SDRAM and DDR\emph{n} DRAM chips, little to no
improvements have been made despite, \mpm{or possibly as a result of, continual}
increases in overall chip storage density. Second, CAS latency and tRCD
converged to roughly the same values following the introduction of synchronous
DRAM. We hypothesize that this is because \mpi{similar factors affect the
latency of these operations, including a long command and data communication
latency between the external DRAM bus and the internal storage
array~\cite{keeth2007dram}.} Third, the DDR5 data points appear to worsen
relative to previous DDR\emph{n} points. However, we believe this \mpk{might be}
because DDR5 chips are new at the time of writing this article and have not yet
been fully optimized (e.g., through die revisions and other process
improvements).

\begin{figure}[h]
    \centering
    \includegraphics[width=0.7\linewidth]{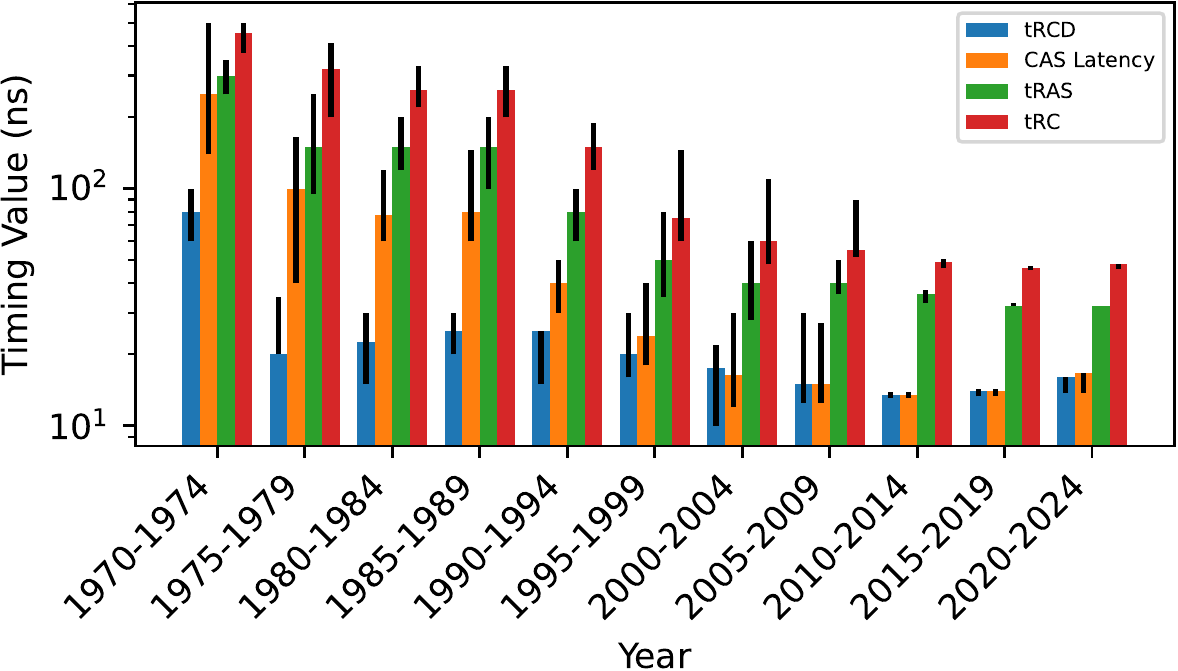}
    \includegraphics[width=0.7\linewidth]{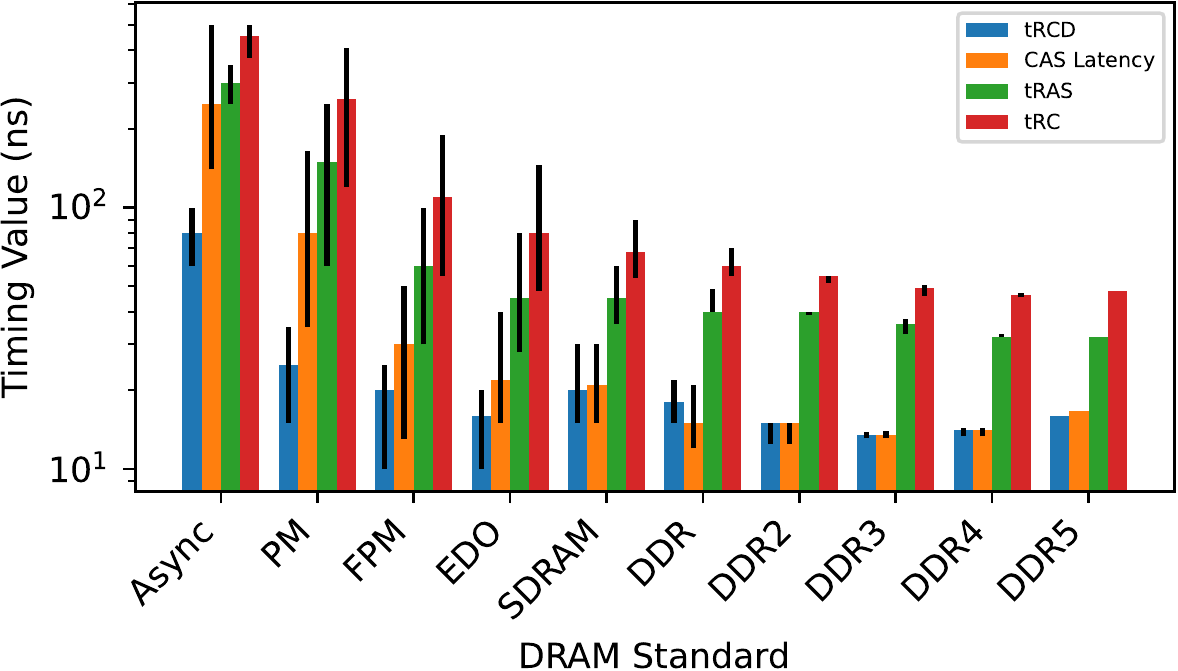}
    \caption[]{Evolution of the minimum, median, and maximum values of key DRAM timing parameters \mpn{(shown in log scale)} for each \mpk{5-year period (top) and DRAM standard (bottom)}.}
    \label{fig:da01}
\end{figure}

To quantify the changes in \mpn{access timings}, we aggregate the data points
from Figure~\ref{fig:da57} \mpm{by three different categories: time, DRAM
standard, and chip capacity. Figure~\ref{fig:da01}, shows} the minimum, median,
and maximum of the timing parameter values \mpn{(in log scale)} for
\mpk{each 5-year period (top) and DRAM standard (bottom)}. \mpk{The data shows
that the median tRCD/CAS Latency/tRAS/tRC reduced by 2.66/3.11/2.89/2.89\% per
year on average between 1970 and 2000 but only 0.81/0.97/1.33/1.53\% between
2000 and 2015\footnote{We omit the 2020 data point because 2020 shows a
regression in CAS latency due to first-generation DDR5 chips, which we believe
is not representative because of its immature technology.} for an overall
decrease of 1.83/2.10/1.99/2.00\% between 1970 and 2015.}

\begin{figure}[h]
    \centering
    \includegraphics[width=\linewidth]{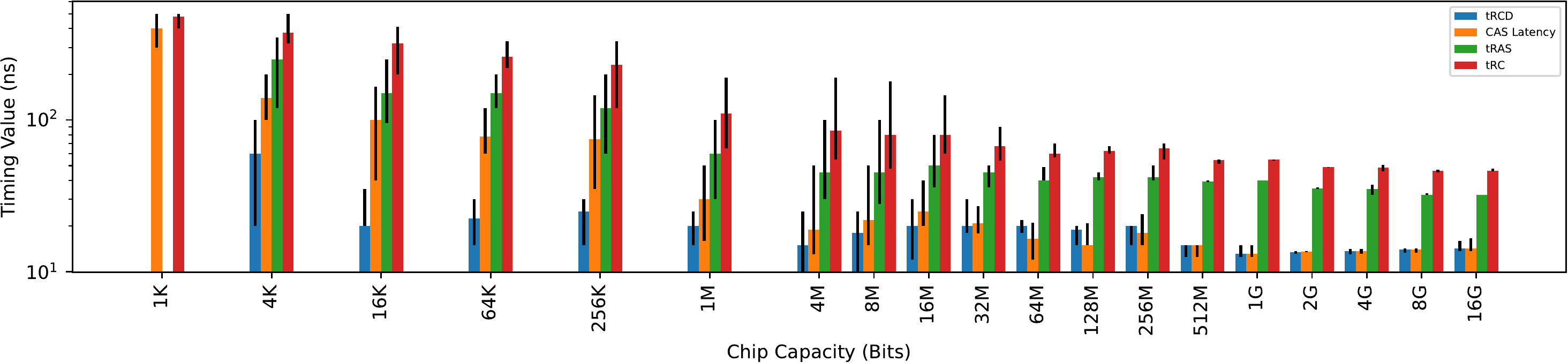}
    \caption[]{\mpm{Evolution of the minimum, median, and maximum values of key DRAM timing parameters \mpn{(shown in log scale) grouped by} DRAM chip \mpn{storage} capacity.}}
    \label{fig:da5}
\end{figure}

\mpm{Figure~\ref{fig:da5} shows the minimum, median, and maximum of the timing
parameter values \mpn{(in log scale) grouped by DRAM chip storage
capacity.}\footnote{\mpm{We omit tRCD and tRAS for the 1~Kib chips because they
do not use a row address strobe (RAS) signal.\vspace{-1ex}}} We find that the \mpn{timings
follow similar trends as in Figure~\ref{fig:da01} because higher-capacity DRAM
chips are typically introduced more recently and follow newer} DRAM standards.}

\subsection{Current Consumption Trends}

We review the evolution of the following key DRAM current consumption
measurements, which are standardized by JEDEC and are provided by manufacturers
in their datasheets.
\begin{itemize}
    \item \emph{IDD0}: current consumption with continuous row activation and
    precharge commands issued to only one bank.
    \item \emph{IDD4R}: current consumption when issuing back-to-back read operations to all banks.
    \item \emph{IDD5B}: current consumption when issuing continuous burst refresh operations.
\end{itemize}

\begin{figure}[h]
    \centering
    \includegraphics[width=0.7\linewidth]{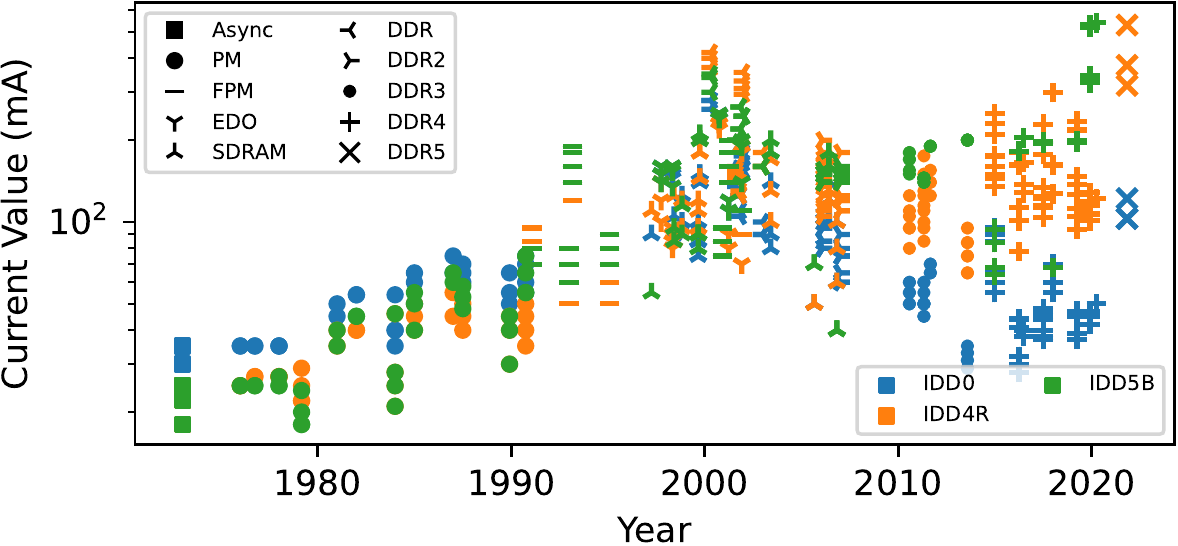}
    \includegraphics[width=0.7\linewidth]{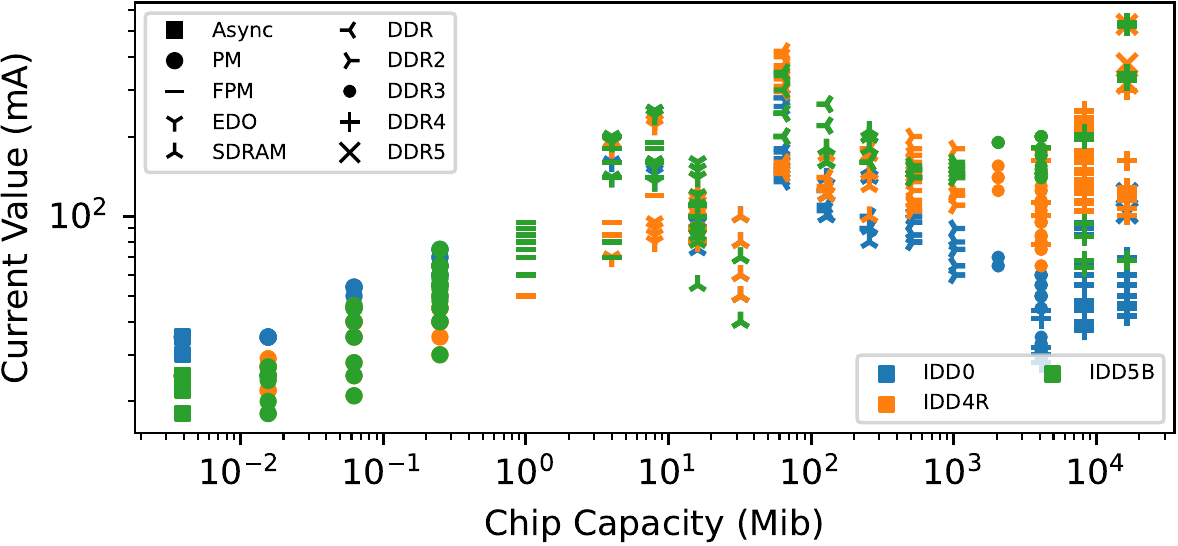}
    \caption[]{Evolution of key DRAM current consumption values \mpn{(shown in log scale)} across years (top) and chip capacities (bottom) separated by DRAM standard.}
    \label{fig:da46}
\end{figure}

\noindent
Figure~\ref{fig:da46} shows how key DRAM current consumption values \mpn{(in log
scale)} have evolved across DRAM chips of different years (top) and capacities
(bottom). We use different markers to show data points from chips of different
DRAM standards. \mpn{We qualitatively observe that current consumption increased
exponentially up until approximately the year 2000, which is about the time at
which improvements in access timings slowed down (as seen in
Figure~\ref{fig:da57}). After this point, different current consumption
measurements diverged as IDD0 values decreased while IDD4R and IDD5B stabilized
or increased. We explain this behavior by a change in the way DRAM chips were
refreshed as DRAM capacities continued to increase. Earlier DRAM chips refreshed
rows using individual row accesses (e.g., RAS-only refresh), which result in
comparable behavior for access and refresh operations. In contrast, newer DRAM
chips aggressively refresh \emph{multiple} rows per refresh operation (e.g.,
burst refresh), which differentiates refresh operations from normal row
accesses~\cite{mukundan2013understanding,balasubramonian2019innovations,
utah2013dram}.}

\begin{figure}[h]
    \centering
    \includegraphics[width=0.7\linewidth]{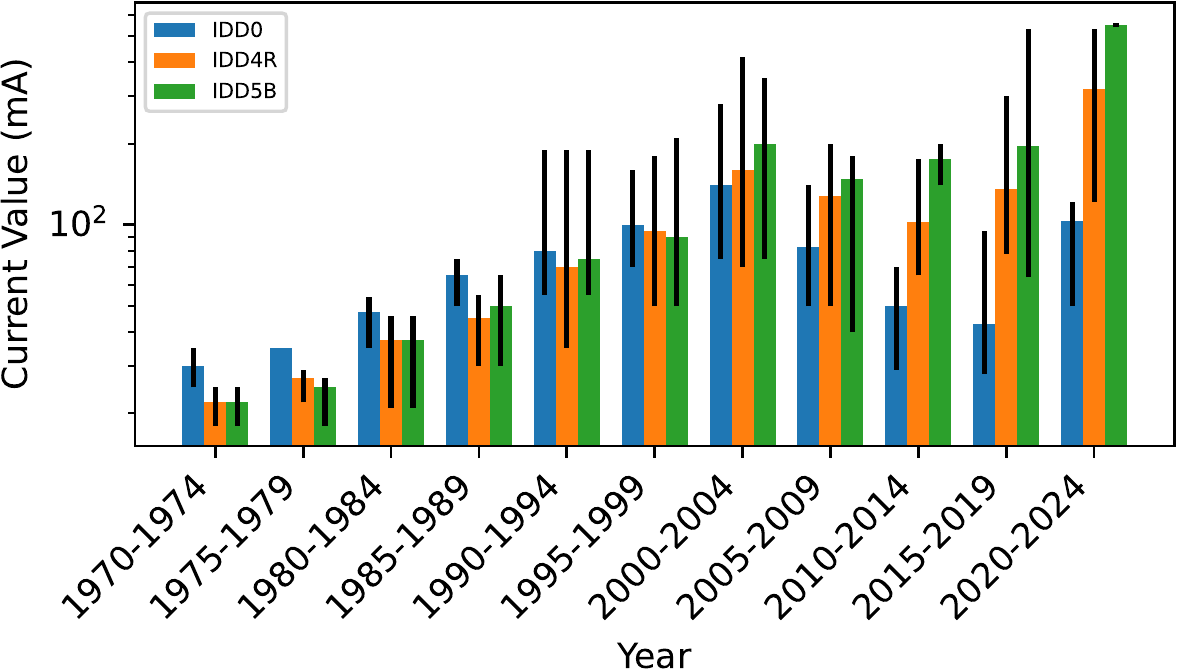}
    \includegraphics[width=0.7\linewidth]{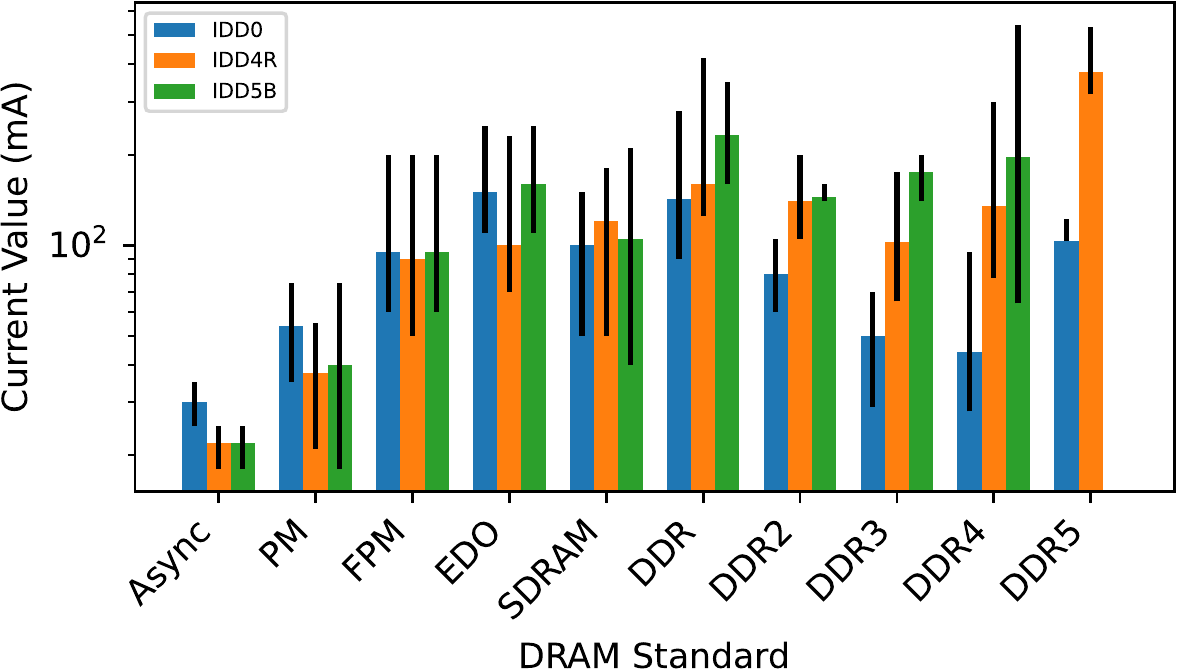}
    \caption[]{Evolution of the minimum, median, and maximum of key DRAM current consumption value \mpn{(shown in log scale)} for each \mpk{5-year period (top) and DRAM standard (bottom).}}
    \label{fig:da23}
\end{figure}

We quantify the current \mpn{consumption} values by aggregating the data points
from Figure~\ref{fig:da46} \mpm{by time and DRAM standard.}
Figure~\ref{fig:da23} shows \mpm{the minimum, median, and maximum values
\mpn{(in log scale)} across each} \mpk{5-year period (top) and DRAM standard
(bottom).} \mpk{The data shows that the median IDD0/IDD4R/IDD5B increased by
12.22/20.91/26.97\% per year on average between 1970 and 2000 but
\emph{decreased} by 4.62/1.00/0.13\% between 2000 and 2015\footnote{Similar to
Section~\ref{position:subsec:timing_trends}, we omit the 2020 data point because
the first-generation DDR5 chips exhibit outlying data values (e.g., no data
reported for IDD5B in the datasheets).} for an overall increase of
0.96/11.5/17.5\% between 1970 and 2015.}

\subsection{Relationship Between Timings and Currents}

\begin{figure}[h]
    \centering
    \includegraphics[width=0.7\linewidth]{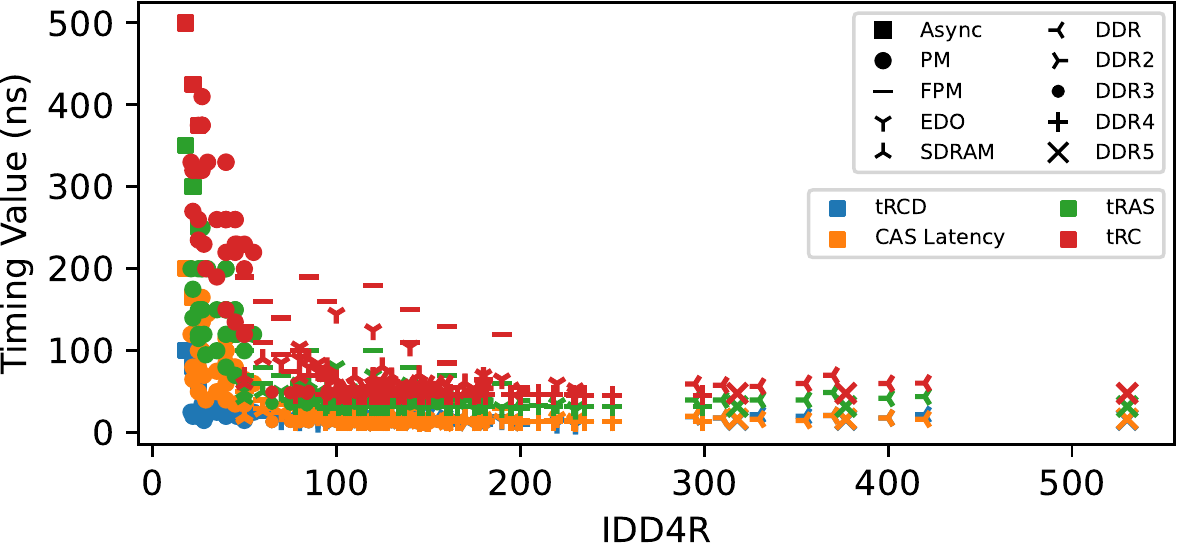}
    \caption[]{\mpk{Relationship between the four timing parameters and IDD4R separated by DRAM standard.}}
    \label{fig:da9}
\end{figure}

Finally, we examine the high-level relationship between the timing parameter and
current consumption values. We find that the two are generally inversely
related, which follows from the general principle that faster DRAM chips (i.e.,
lower timing parameters) require more power (i.e., increased current consumption
values). Figure~\ref{fig:da9} illustrates this relationship for the four timing
parameters studied in Section~\ref{position:subsec:timing_trends} relative to
IDD4R (i.e., the current consumption of read operations).

\subsection{\mpl{DRAM Refresh Timing Trends}}

\mpl{DRAM refresh is governed by two key timing parameters:}

\begin{itemize}
    \item \mpl{\emph{tREFI} (refresh interval): time between consecutive refresh commands sent by the memory controller.}
    \item \mpl{\emph{tRFC}: duration of a single refresh command.}
\end{itemize}

\begin{figure}[h]
    \centering
    \includegraphics[width=0.7\linewidth]{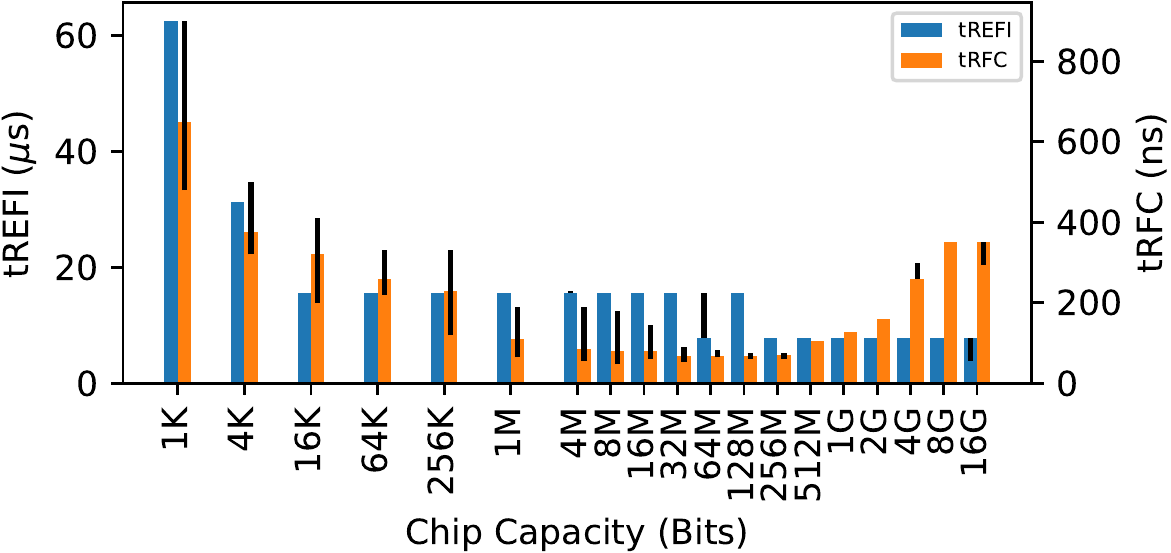}
    \caption[]{Evolution of tREFI (left y-axis) and tRFC (right y-axis) across DRAM chips of increasing storage capacity.}
    \label{fig:da12}
\end{figure}

\noindent
\mpl{Figure~\ref{fig:da12} shows how tREFI (left y-axis) and tRFC (right y-axis)
evolved across the DRAM chips in our study. We group chips by storage capacity
because DRAM refresh timings are closely related to capacity:
\mpm{higher-capacity} chips using the same technology require more time or more
refresh operations to fully refresh. \mpm{The error bars show the minimum and
maximum values observed across all chips for any given chip capacity.}}

\mpl{We make three observations. First, tREFI is shorter for
\mpm{higher-capacity} DRAM chips (e.g., 62.5~$\mu$s for an asynchronous 1~Kib
chip versus 3.9~$\mu$s for a 16~Gib DDR5 chip). This is consistent with the fact
that \mpm{higher-capacity} chips require more frequent refreshing. Second, tRFC
first decreases with chip capacity (e.g., 900 ns for an asynchronous 1~Kib chip
versus 54~ns for a 32~Mib SDRAM chip) but then increases (e.g., to 350~ns for a
16~Gib DDR4 chip). This is because rapid improvements in row access times (and
therefore refresh timings) initially outpaced the increase in storage capacity.
However, starting around 512~Mib chip sizes, row access times improved much more
slowly (as observed in Section~\ref{position:subsec:timing_trends}) while
storage capacity continued to increase. \mpm{Third, the variation in tRFC across
chips of each capacity (illustrated using the error bars) decreased for
\mpm{higher-capacity} chips. This is because higher-capacity chips follow more
recent DRAM standards (i.e., DDR\emph{n}), which standardize DRAM auto refresh
timings. In contrast, older DRAM chips were simply refreshed as quickly as their
rows could be accessed (e.g., every tRC using RAS-only refresh).}}

\begin{figure}[h]
    \centering
    \includegraphics[width=0.7\linewidth]{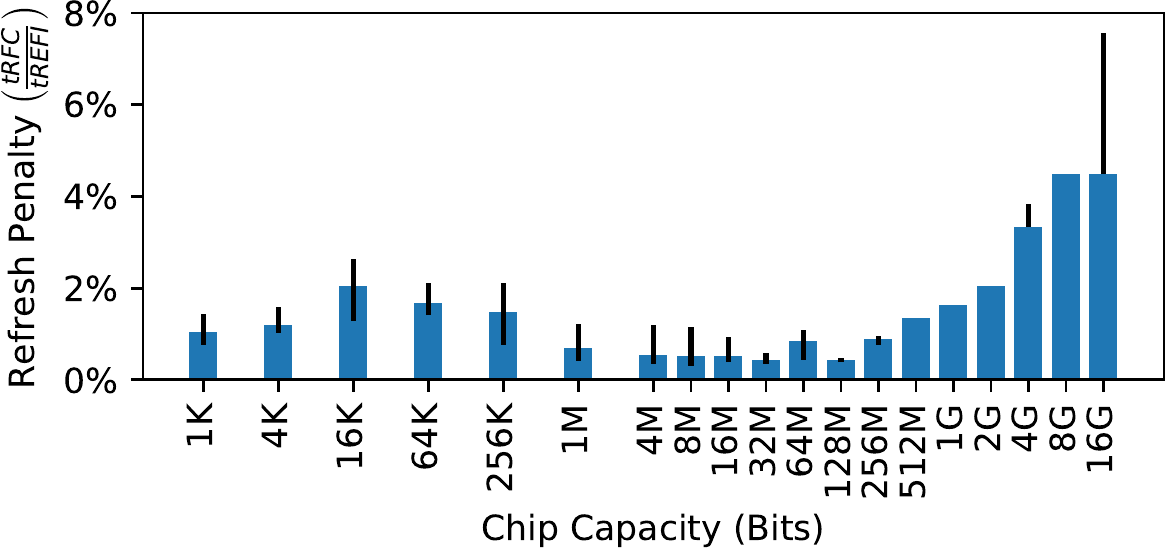}
    \caption[]{\mpm{Refresh penalty} (computed as the ratio between tRFC and tREFI) for DRAM chips of different storage capacities.}
    \label{fig:da_ref_overhead}
\end{figure}

\mpm{Figure~\ref{fig:da_ref_overhead} shows the \emph{refresh
penalty}~\cite{utah2013dram, balasubramonian2019innovations}, which is defined
as the ratio between tRFC and tREFI, for DRAM chips of different storage
capacities. The refresh penalty represents the average time that a DRAM rank (or
bank) is unavailable for access due to refresh operations~\cite{utah2013dram,
stuecheli2010elastic, balasubramonian2019innovations, cheng2019retention,
zhang2014cream}. We observe that the refresh penalty exhibits a similar trend to
tRFC: the refresh penalty worsens from a median of 1.04\% for 1~Kib chips to
2.05\% for 16~Kib chips, then improves to 0.43\% for 128~Mib chips, and finally
worsens to a median of 4.48\% (worst-case of 7.56\% for DDR5 chips) for 16~Gib
chips.}

\mpl{This non-monotonic trend is due to the relative improvements in DRAM access
times and storage capacities: DRAM capacities consistently improved while DRAM
access times did so only for older, lower-capacity chips (e.g., $\le 128$~Mib
chips). This is consistent with trends observed in prior
work~\mpm{\cite{nguyen2018nonblocking, liu2012raidr, bhati2015flexible,
nair2013case, chang2014improving, qureshi2015avatar}}, which expect that future,
higher-capacity DRAM chips will spend an even larger proportion of time
refreshing unless the DRAM refresh algorithm \mpm{and techniques} can be
improved.}
    \clearpage

\section{\mpk{Survey Data Sources}}
\label{position:appendix_b}
\mpk{Table~\ref{position:tab:sources} itemizes the 58 \mpm{DRAM} datasheets used
for our survey in Appendix~\ref{position:appendix_a}. For each datasheet, we
show the DRAM chip manufacturer, model number, DRAM standard, year, and
capacity. Our full dataset is available online~\cite{datasheetsurveygithub}.}

\begin{center}
    \scriptsize
    \renewcommand{\arraystretch}{0.75}
    \begin{tabular}{llllll}
        \textbf{Year} & \textbf{Manufacturer} & \textbf{Model Number} & \textbf{Datasheet Source} & \textbf{DRAM Standard} & \textbf{Capacity per Chip (Kib)} \\\hline
        1970 & Intel             & 1103         & \citeS{ds_intel19701103}            & Asynchronous & 1 \\
        1971 & Mostek            & MK4006       & \citeS{ds_mostek1971mk4006}         & Asynchronous & 1 \\
        1973 & Mostek            & MK4096       & \citeS{ds_mostek1973mk4096}         & Asynchronous & 4 \\
        1976 & Mostek            & MK4027       & \citeS{ds_mostek1976mk4027}         & PM           & 4 \\
        1976 & Mostek            & MK4116P      & \citeS{ds_mostek1976mk4116p}        & PM           & 16 \\
        1978 & Fairchild         & F4116        & \citeS{ds_fairchild1978f4116}       & PM           & 16 \\
        1979 & Intel             & 2118         & \citeS{ds_intel19792118}            & PM           & 16 \\
        1981 & Mitsubishi        & M5K4164ANP   & \citeS{ds_mitsubishi1981m5k4164anp} & PM           & 64 \\
        1982 & Mostek            & MK4564       & \citeS{ds_mostek1982mk4564}         & PM           & 64 \\
        1984 & NTE               & NTE4164      & \citeS{ds_nte1984nte4164}           & PM           & 64 \\
        1984 & Texas Instruments & TMS4416      & \citeS{ds_texas1984tms4416}         & PM           & 64 \\
        1985 & Mitsubishi        & M5M4256P     & \citeS{ds_mitsubishi1985m5m4256p}   & PM           & 256 \\
        1987 & Samsung           & KM41464A     & \citeS{ds_samsung1987km41464a}      & PM           & 256 \\
        1987 & Texas Instruments & TMS4464      & \citeS{ds_texas1987tms4464}         & PM           & 256 \\
        1989 & Texas Instruments & SMJ4464      & \citeS{ds_texas1989smj4464}         & PM           & 256 \\
        1990 & Intel             & 21256        & \citeS{ds_intel199021256}           & PM           & 256 \\
        1991 & Mitsubishi        & M5M44100     & \citeS{ds_mitsubishi1991m5m44100}   & FPM          & 4096 \\
        1993 & Mitsubishi        & M5M44256B    & \citeS{ds_mitsubishi1993m5m44256b}  & FPM          & 1024 \\
        1993 & Mosel Vitelic     & V404J8       & \citeS{ds_mosel1993v404j8}          & FPM          & 8192 \\
        1995 & Siemens           & HYB511000BJ  & \citeS{ds_siemens1995hyb511000bj}   & FPM          & 1024 \\
        1997 & Hyundai           & HY5118164B   & \citeS{ds_hyundai1997hy5118164b}    & EDO          & 16384 \\
        1997 & Samsung           & KM48S2020CT  & \citeS{ds_samsung1997km48s2020ct}   & SDRAM        & 16384 \\
        1998 & Micron            & MT48LC4M4A1  & \citeS{ds_micron1998mt48lc4m4a1}    & SDRAM        & 16384 \\
        1998 & Mosel Vitelic     & V53C808H     & \citeS{ds_mosel1998v53c808h}        & EDO          & 8192 \\
        1998 & Siemens           & HYB39S16400  & \citeS{ds_siemens1998hyb39s16400}   & SDRAM        & 16384 \\
        1999 & Samsung           & K4S160822D   & \citeS{ds_samsung1999k4s160822d}    & SDRAM        & 16384 \\
        1999 & Samsung           & K4S561632A   & \citeS{ds_samsung1999k4s561632a}    & SDRAM        & 262144 \\
        2000 & Amic              & A416316B     & \citeS{ds_amic2000a416316b}         & FPM          & 1024 \\
        2000 & ISSI              & IS41LV32256  & \citeS{ds_issi2000is41lv32256}      & EDO          & 8192 \\
        2000 & Samsung           & K4D623237A5  & \citeS{ds_samsung2000k4d623237a5}   & DDR          & 65536 \\
        2001 & Alliance          & AS4C256K16E0 & \citeS{ds_alliance2001as4c256k16e0} & EDO          & 4096 \\
        2001 & Alliance          & AS4C4M4FOQ   & \citeS{ds_alliance2001as4c4m4foq}   & FPM          & 16384 \\
        2001 & ISSI              & IS41C4400X   & \citeS{ds_issi2001is41c4400x}       & EDO          & 16384 \\
        2001 & Micron            & MT46V2M32    & \citeS{ds_micron2001mt46v2m32}      & DDR          & 65536 \\
        2001 & Micron            & MT46V32M4    & \citeS{ds_micron2001mt46v32m4}      & DDR          & 131072 \\
        2001 & Mosel Vitelic     & V58C265164S  & \citeS{ds_mosel2001v58c265164s}     & DDR          & 65536 \\
        2001 & TM Tech           & T224160B     & \citeS{ds_tm2001t224160b}           & FPM          & 4096 \\
        2003 & Micron            & MT46V64M4    & \citeS{ds_micron2003mt46v64m4}      & DDR          & 262144 \\
        2003 & Samsung           & K4S560432E   & \citeS{ds_samsung2003k4s560432e}    & SDRAM        & 262144 \\
        2005 & Amic              & A43L0632     & \citeS{ds_amic2005a43l0632}         & SDRAM        & 32768 \\
        2006 & Elite             & M52S32321A   & \citeS{ds_elite2006m52s32321a}      & SDRAM        & 32768 \\
        2006 & ISSI              & IS42S81600B  & \citeS{ds_issi2006is42s81600b}      & SDRAM        & 131072 \\
        2006 & Sasmung           & K4T51043QC   & \citeS{ds_sasmung2006k4t51043qc}    & DDR2         & 524288 \\
        2007 & Micron            & MT47H256M4   & \citeS{ds_micron2007mt47h256m4}     & DDR2         & 1048576 \\
        2010 & Samsung           & K4B4G0446A   & \citeS{ds_samsung2010k4b4g0446a}    & DDR3         & 4194304 \\
        2011 & Hynix             & H5TQ4G43MFR  & \citeS{ds_hynix2011h5tq4g43mfr}     & DDR3         & 4194304 \\
        2011 & Nanya             & NT5CB512M    & \citeS{ds_nanya2011nt5cb512m}       & DDR3         & 2097152 \\
        2013 & Samsung           & K4B4G0446A   & \citeS{ds_samsung2013k4b4g0446a}    & DDR3         & 4194304 \\
        2015 & Micron            & MT40A2G      & \citeS{ds_micron2015mt40a2g}        & DDR4         & 8388608 \\
        2016 & Hynix             & H5AN4G4NAFR  & \citeS{ds_hynix2016h5an4g4nafr}     & DDR4         & 4194304 \\
        2016 & Samsung           & K4A8G165WC   & \citeS{ds_samsung2016k4a8g165wc}    & DDR4         & 8388608 \\
        2017 & Hynix             & H5AN8G4NAFR  & \citeS{ds_hynix2017h5an8g4nafr}     & DDR4         & 8388608 \\
        2018 & Micron            & MT40A        & \citeS{ds_micron2018mt40a}          & DDR4         & 16777216 \\
        2019 & Hynix             & H5AN8G4NCJR  & \citeS{ds_hynix2019h5an8g4ncjr}     & DDR4         & 8388608 \\
        2019 & Samsung           & K4AAG045WA   & \citeS{ds_samsung2019k4aag045wa}    & DDR4         & 16777216 \\
        2020 & Samsung           & K4AAG085WA   & \citeS{ds_samsung2020k4aag085wa}    & DDR4         & 16777216 \\
        2021 & Hynix             & HMCG66MEB    & \citeS{ds_hynix2021hmcg66meb}       & DDR5         & 16777216 \\
        2021 & Micron            & MT60B1G16    & \citeS{ds_micron2021mt60b1g16}      & DDR5         & 16777216
    \end{tabular}
\end{center}
\captionof{table}{List of DRAM chip datasheets used in our DRAM trends survey.}
\label{position:tab:sources}

\clearpage
\balance
\bibliographystyleS{IEEEtran}
\bibliographyS{references}
\end{subappendices}

\chapter{Conclusions and Future Directions}
\label{chap:conc}

In summary, the goal of this dissertation is twofold: (1) understand how on-die
ECC obfuscates the raw memory error characteristics; and (2) develop new testing
techniques that enable scientists and engineers to overcome the obfuscation. To
this end, we use a combination of analytical and experimental studies to develop
a series of new testing techniques that allow for faster, more effective error
profiling and recovery of the error characteristics that on-die ECC obfuscates. 

First, we perform the first extensive error characterization study of
data-retention errors in LPDDR4 DRAM chips in order to understand how the errors
occur and develop a fast and efficient error profiling algorithm. We find that
there is a complex tradeoff space surrounding data-retention error profiling,
where profiling under different operating conditions can improve profiling
runtime and coverage at the cost of incurring false positives. Based on this, we
introduce reach profiling, a new data-retention error profiling algorithm that
searches for errors at a longer refresh interval and/or higher temperature than
the desired operating conditions in order to improve profiling runtime and
coverage. Our evaluations show that reach profiling can improve runtime by an
average of 2.5$\times$ relative to the best prior profiling algorithm while
achieving 99\% coverage with less than a 50\% false positive rate. 

Second, we extend our studies to DRAM chips that use on-die ECC. To understand
how on-die ECC affects DRAM data-retention error characteristics, we develop
Error Inference (EIN), a new statistical inference methodology capable of
inferring details of both the on-die ECC mechanism and the underlying raw bit
errors using only the processor-visible post-correction errors. We evaluate EIN
through the first data-retention error characterization study of DRAM chips that
use on-die ECC. We find that EIN is able to (1) reverse-engineer the type and
strength of the on-die ECC implementation; (2) infer pre-correction
data-retention error rates given only the post-correction errors; and (3)
recover the well-studied raw bit error distributions that on-die ECC obfuscates.

Third, to understand exactly how on-die ECC acts upon a given raw bit error
pattern, we develop a new testing methodology, Bit-Exact Error Recovery (BEER).
BEER systematically determines the full on-die ECC function (i.e., its
parity-check matrix) without requiring access to hardware tools, prerequisite
knowledge about the DRAM chip or the on-die ECC mechanism, or ECC metadata
(e.g., error syndromes, parity information). We demonstrate BEER using 80 real
LPDDR4 DRAM and evaluate its correctness and performance in simulation. We then
introduce Bit-Exact Error Profiling (BEEP), the first error profiling algorithm
that uses the known on-die ECC function (e.g., via BEER) to recover the number
and bit-exact locations of the pre-correction errors that are responsible for a
given post-correction error pattern.

Fourth, we study how on-die ECC operations alter the memory controller's
perspective of memory errors in order to understand how on-die ECC impacts error
profiling. Based on our study, we identify three key challenges that on-die ECC
introduces, each stemming from the fact that on-die ECC introduces statistical
dependence between otherwise independent errors. To overcome these challenges,
we introduce Hybrid Active-Reactive Profiling (HARP), a new bit-granularity
error profiling algorithm. HARP uses small modifications to the on-die ECC read
logic to enable fast and efficient error profiling in the presence of on-die
ECC. Our evaluations show that HARP significantly improves profiling speed
relative to baseline profiling algorithms that rely only on post-correction
errors.

Finally, \dmpa{based on the understanding and insights we developed through the
course of our work, we argue the importance of having transparency into basic
DRAM reliability characteristics so that system designers can more easily
implement a broad range of optimizations for better adapting commodity DRAM
chips to the particular needs of their systems.} We substantiate our arguments
through three case studies: (1) reducing DRAM refresh overhead; (2) improving
DRAM access latency; and (3) designing RowHammer defenses. In each study, we
highlight the lack of transparency that discourages system designers from
adopting the optimization. We conclude by \dmpa{providing our recommendations
for improving transparency in current and future commodity DRAM-based systems.}

\section{Future Research Directions}

Although this dissertation focuses on testing techniques to overcome on-die ECC
primarily in the context of DRAM errors, we believe that this work is applicable
in a more general sense and opens up new research directions. This section
reviews promising directions for future work.

\subsection{Extending the Proposed Techniques}
\label{future:subsec:extending_proposed}

We believe that the general principles behind each of our techniques are
applicable to other ECC designs (e.g., stronger linear block codes such as
Bose-Chaudhuri-Hocquenghem (BCH)~\cite{bose1960class, hocquenghem1959codes} and
Reed-Solomon (RS)~\cite{reed1960polynomial} codes) and architectures (e.g.,
rank-level ECC), failure modes (e.g., RowHammer), and memory technologies and
(e.g., NAND Flash memory~\cite{cai2012error}, phase-change
memory~\cite{wong2010phase, lee2009architecting, lee2010phasecacm}, magnetoresistive
memory~\cite{apalkov2013spin, zhao2014improving}, Racetrack
memory~\cite{parkin2015memory}). In this section, we review several directions
in which we believe our work can be adapted to identify and address reliability
challenges in other systems.

\subsubsection{To Other ECC Schemes} Although we demonstrate EIN, BEER, and HARP
in the context of DRAM on-die ECC, the techniques we rely upon are not
restricted to single-error correcting Hamming code. EIN is broadly applicable
because it only requires understanding how a given ECC mechanism will transform
the statistical characteristics of the pre-correction errors. As long as this
transformation can be understood for a given ECC scheme (e.g., using
simulations, as we do in our work), EIN can be extended to deduce the nature of
the ECC scheme and the pre-correction errors it obfuscates. Similarly, BEER and
HARP rely upon how the syndrome decoding process treats uncorrectable errors.
This process is not specific to Hamming codes, and in fact, can be used by all
linear block codes, including BCH and RS codes that are useful for stronger,
multi-bit error correction outside of the memory die. Therefore, we believe
future work can take inspiration from and extend our work to develop analogous
techniques for systems using different, possibly stronger ECC schemes.

\subsubsection{To Other DRAM Failure Modes} We demonstrate our work in the
context of DRAM data-retention errors because they are a relevant and
interesting problem today. However, we believe that future work can apply
similar principles to those underlying our techniques (e.g., profiling under
exacerbated operating conditions, exploiting data-dependence of the error
mechanisms) to solve error profiling and characterization challenges for other
DRAM failure modes. For example, several works~\cite{yang2019trap,
schaller2017intrinsic, park2016experiments} analytically and/or experimentally
demonstrate that RowHammer exhibits sensitivity to device temperature. Based on
this observation, reach profiling can potentially be extended to quickly and
efficiently identify those cells that are more susceptible to RowHammer errors.
In general, any failure mode that is sensitive one or more parameters that are
under the operator's control (e.g., data patterns, timings, temperature,
voltage) can theoretically be exploited in such a manner.

Similarly, a BEER-like approach can be taken with any failure mode that exhibits
strong dependence on the data value stored in a cell. This data dependence
allows control over which cells can and cannot fail, enabling the extraction of
information about the ECC mechanism and raw bit error characteristics. Although
other DRAM failure modes have not been shown to exhibit as strong a data
dependence as data-retention, we believe that BEER can be extended to work with
a failure mode that exhibits \emph{some amount of} data dependence through
creative use of statistical measurements.

\subsubsection{To Unpredictable Failure Modes} Our studies and techniques are
centered on observing large quantities of uncorrectable errors in order to infer
information about the on-die ECC mechanism and raw bit errors. However,
addressing difficult-to-predict errors (e.g., sporadic events such as VRT
errors) is already a significant challenge in DRAM
today~\cite{micron2017whitepaper} and will continue to worsen with technology
scaling. Unfortunately, these errors are difficult to anticipate beforehand
(e.g., with offline profiling), so understanding their characteristics and
developing efficient techniques to combat them is an important problem. 

\subsubsection{To Emerging Memory Technologies} Our work provides researchers
with the tools to understand and explore DRAM on-die ECC and the raw bit error
characteristics. However, the analysis and techniques we contribute are not
limited to DRAM and can theoretically be extended to work with other memory
technologies, such as PCM, MRAM, and RM. These technologies are known to suffer
from their own reliability challenges~\cite{lee2009architecting, lee2010phasecacm, lee2009study,
kim2005reliability, itrs2015more, kang2006a, awasthi2012efficient,
vatajelu2018state, chen2010advances, apalkov2013spin, naeimi2013sttram,
chun2012scaling, raychowdhury2009design}, potentially requiring approaches
similar to on-die ECC to make manufacturing feasible.\footnote{In fact, certain
commodity STT-MRAM chips~\cite{everspin2021sttmram} are already provisioned with
on-die ECC.} Although most of these technologies are yet to be productized,
identifying feasible designs and architectures is an important research problem
that will continue to become more relevant as the technologies mature. We believe
that our work will be relevant and helpful in addressing challenges that arise
from on-die error-mitigation techniques that these memories incorporate.

\subsubsection{To Other ECC Architectures} At an abstract level, our work
applies to any communication channel that suffers from statistically
well-defined errors and only exposes post-correction data. DRAM on-die ECC is an
ideal example because all error-mitigation operations occur within the DRAM die
and are therefore invisible at the chip's interface. Furthermore, DRAM failure
modes are closely tied to the technology itself and have been studied
extensively across decades of research. However, any memory architecture that
provides no visibility into the ECC mechanism likely suffers from analogous
challenges to those we identify in our work. Possible examples include (1) a
memory architecture that places ECC within an in-package memory controller
(e.g., as does NAND Flash), thereby making the ECC invisible to the rest of the
system; and (2) a DRAM-based system that uses rank-level ECC (i.e., within the
memory controller). Although the ECC's internal workings are more accessible
than on-die ECC in both of these examples~\cite{cojocar2019exploiting}, a
scientist or engineer who does not have access to the interfaces that allow
visibility into the ECC mechanism (e.g., special processor registers, hardware
debugging interfaces) can extend the techniques that we provide in our work to
conduct their studies.

\subsection{Leveraging the New-Found Visibility into Error Characteristics}

We believe that future work can use the techniques that we develop in this work
to enable a broad range of more robust systems in the context of memory chips
that use on-die ECC. We review several possibilities in this section, but
believe that other use cases exist that we have not yet foreseen.

\subsubsection{Improved System-Level Error Mitigations} Armed with the tools to
better understand on-die ECC, system designers can adapt their own
error-mitigation mechanisms to the on-die ECC mechanisms that memory
manufacturers deploy. This is especially important for systems that integrate
modern commodity DRAM because on-die ECC is already prevalent among LPDDR4 and
DDR5 chips~\cite{oh2014a, kwon202125, kwak2017a, jedec2020ddr5}. Several
works~\cite{son2015cidra, cha2017defect, gong2018duo} argue the value of being
aware of on-die ECC when designing system-level error-mitigation mechanisms, and
techniques such as EIN and BEER enable this visibility for already-deployed
memory chips that use on-die ECC.

\subsubsection{Error-Tolerant Computing} Prior works~\cite{tu2018rana,
koppula2019eden, nguyen2020approximate} have used their understanding of DRAM
error characteristics to make neural network applications tolerant to DRAM
errors. For such applications, we believe this is a promising approach to
circumvent the need for stronger, more expensive hardware error-mitigation
mechanisms. As with designing robust systems, making an application resilient to
errors requires understanding the errors' characteristics. We believe that the
techniques we contribute will allow researchers to extend existing techniques
and develop new ones that target systems equipped with on-die ECC.

\subsection{Alternative Error Mitigation Designs}

Our work details key challenges that on-die ECC's obfuscation of error
characteristics entails. We believe that modifying or replacing on-die ECC with
techniques that \emph{avoid} obfuscating the underlying error characteristics is
a crucial future research direction. Although these approaches cannot help for
DRAM on-die ECC that pervades DRAM chips today, we believe that they are
promising alternatives for future designs.

\subsubsection{Alternative On-Die ECC Architectures} Recent
work~\cite{nair2016xed, gong2018duo, jeong2020pair} has considered
rearchitecting on-die ECC in ways that allow system designers to better
complement the reliability benefits that on-die ECC already provides. Similarly,
recent DDR5 DRAM standards~\cite{jedec2020ddr5} explicitly discuss on-die ECC
within specifications and add limited functionality to expose the underlying
reliability characteristics (e.g., a limited form of ECC scrubbing). Each of
these approaches take important steps towards enabling DRAM consumers to extract
the maximum benefit from on-die ECC while minimizing the downsides (e.g., that
we identify and work to address in this dissertation). However, we believe that
significant future work remains to identify the best possible on-die ECC designs
for different system configurations, design goals, and target error rates.

\subsubsection{System-Level Error Mitigations} As memory errors worsen with
continued technology scaling, we expect memory manufacturers to consider even
stronger on-die ECC mechanisms. As an alternative to on-die ECC, we believe it
is important to explore other ways to mitigate high error rates. Different
system components have different advantages with respect to error mitigation.
For example, the software is well-equipped to determine which errors are and are
not significant at the application-level~\cite{messer2004susceptibility,
li2010realistic, luo2014characterizing} and address them using flexible
software-level techniques (e.g., using redundancy~\cite{shye2007using,
taylor1980redundancy, reis2005swift, shye2008plr} and/or failure recovery
mechanisms~\cite{fang2017letgo, chen2019care, qin2005rx}). In contrast, the
hardware can quickly and efficiently respond to errors that occur during
runtime~\cite{martinez2011compiler}. We believe that exploring the system-level
design space for error mitigation is a promising direction, not only to tackle
high error rates that are difficult to address using on-die ECC or hardware
repair mechanisms, but also to address the difficulties that proprietary on-die
ECC solutions cause for system design and test.

\subsection{Improving Transparency into DRAM Reliability and Operation}

Our work explains and addresses the consequences of using on-die ECC that is
completely invisible to the rest of the system. To prevent this problem in the
future, we believe that an in-depth reevaluation of DRAM specifications is
warranted. Although we identify certain information that can help DRAM consumers
if released (discussed in Section~\ref{sec:two_part_change_to_specs}), a broader
study of areas in which design transparency can be helpful, possibly performed
by system designers themselves, would be immensely helpful to guide future
work.

\section{Concluding Remarks}

In this dissertation, we explore memory error characterization and profiling
both with and without on-die ECC. We builds a detailed understanding of the
challenges that on-die ECC introduces for when studying memory error
characteristics. We show that on-die ECC obfuscates errors, making it difficult
for scientists and engineers to make sense of observed errors in the context of
the underlying physical memory technology. To address these challenges, we
propose four new sets of testing techniques that enable more effectively study
memory error characteristics: (1) Reach profiling, which quickly identifies bits
at risk of data-retention error with high coverage so that refresh overhead
mitigation techniques may safely eliminate unnecessary refresh operations; (2)
EIN, which allows inferring details of both the on-die ECC mechanism and the raw
bit error characteristics in the context of an error-characterization study; (3)
BEER and BEEP, which expose the precise details of how on-die ECC is affecting
raw bit error patterns, allowing for well informed design and test practices;
and (4) HARP, which enables quickly identifying bits at risk of error in memory
devices that use on-die ECC, thereby enabling bit repair mechanisms to
accurately mitigate errors. We hope that the data and methods we contribute
enable new studies and research directions that embrace the benefits of
techniques like on-die ECC without sacrificing the transparency that we believe
is crucial for continued innovation.

\appendix
\cleardoublepage%
\chapter{Other Works of the Author}

\dmpd{I led four successful projects that led towards this dissertation during
my time as a Ph.D. student. First, I developed REAPER~\cite{patel2017reach}
with significant and foundational help from Jeremie Kim.
Chapter~\ref{chap:reaper} describes this project in detail, which was my first
foray into the DRAM space. Second, I delved further into DRAM reliability with
EIN~\cite{patel2019understanding}, releasing the open-source,
experimentally-validated EINSim~\cite{eccsimgithub} tool as part of the project.
To our knowledge, EIN is the first work to delve into the real-world
implementation of DRAM on-die ECC and received the Best Paper Award at DSN 2019.
Chapter~\ref{chap:ein} summarizes this project and our findings. Next, I
continued exploring the adverse consequences of on-die ECC with
BEER~\cite{patel2020bit}, culminating in an open-source tool~\cite{beergithub}
that enables the community to replicate our study and freely apply the BEER
methodology to their own devices. BEER received the best paper award at MICRO
2020 in recognition of its contributions to the state-of-the-art.
Chapter~\ref{chap:beer} describes BEER and our open-source tool. Most recently,
I developed HARP~\cite{patel2021harp}, whose artifacts are peer-evaluated and
publicly available as an open-source tool~\cite{harp-artifacts, harpgithub}.
Chapter~\ref{chap:harp} discusses our study and findings from this project.
Finally, I used the combined learnings from these projects (and other
collaborations, described below) to develop a set of recommendations for the
DRAM community going forward. Chapter~\ref{chap:position} summarizes these
recommendations, which we believe will improve the state of the industry in both
the short and long term.}

Throughout my graduate studies, I also participated in research projects in
several distinct areas with my fellow graduate students in SAFARI. \dmpd{I
acknowledge these works in the remainder of this} chapter.

I spent the first year of my Ph.D. working in close collaboration with Amirali
Boroumand. We devolved an efficient coherence protocol for processing-in-memory
systems, culminating in two works: LazyPIM~\cite{boroumand2016lazypim} and
CoNDA~\cite{boroumand2019conda}. In these works, we show that a speculative,
optimistic approach to coherence traffic can be a good alternative to fine- or
coarse-grained accounting given workloads that access the same data from both
the host processor and the processing-in-memory cores.

Next, I delved into the DRAM reliability space and became involved with several
works that I co-authored with Jeremie Kim alongside our work with data-retention
errors. We performed extensive experimental DRAM error characterization studies,
in which we exploited DRAM access timing reductions to (1) develop reliable
device fingerprints in the form of physically unclonable functions (DRAM latency
PUF~\cite{kim2018dram}); (2) reduce memory access latency for memory locations
that are tolerant to reduced access timings (Solar DRAM~\cite{kim2018solar});
(3) and generate true random numbers (D-RaNGe~\cite{kim2019d}). Later, I was
involved with Yaohua Wang in two projects. First, we improved the DRAM access
latency by exploiting opportunities for reducing memory access timings based on
an application's memory access pattern (CAL~\cite{wang2018reducing}). Second, we
developed a fine-grained in-DRAM data movement mechanism
(FIGARO~\cite{wang2020figaro}) that can be exploited to act as a cache that
improves the average DRAM access latency, in addition to potentially helping
defend against security attacks that make use of RowHammer and DRAM timing side
channels.

Continuing on the path of DRAM access latency reduction, I worked with Hasan
Hassan to develop CROW~\cite{hassan2019crow}, a mechanism that enables quickly
duplicating data between rows in-DRAM. Using the copy operation, we show that it
is possible to use the copied rows to reduce both access latency and refresh
overheads. Next, I worked with Haocong Luo to develop
CLR-DRAM~\cite{luo2020clr}, a technique that enables a run time memory
capacity-latency tradeoff at the granularity of individual DRAM rows. CLR-DRAM
allows switching between the two states on the order of nanoseconds, providing a
new and highly reconfigurable substrate for future research to build upon. Next,
I worked with Lois Orosa to develop CODIC~\cite{orosa2021codic} a substrate that
exposes fine-grained internal DRAM access timings to the programmer. Using these
access timings, the programmer can implement a wide range of features, such as a
physically unclonable function and a cold boot attack prevention mechanism.
Finally, I worked with Ataberk Olgun to design QUAC-TRNG~\cite{olgun2021quac}, a
high-throughput true random number generator based on the error characteristics
of commodity DRAM chips.

I also contributed on the topic of RowHammer through several distinct projects.
First, I worked with Lucian Cojocar to study the effectiveness of different
RowHammer error testing strategies and develop a methodology usable by
server-class systems to test for RowHammer~\cite{cojocar2020are}. Next, I worked
with Jeremie Kim to perform a large-scale RowHammer error-characterization
study~\cite{kim2020revisiting}, in which we examined how RowHammer scales across
different DRAM standards and technology generations. Then, I worked with Giray
Ya{\u{g}}l{\i}k{\c{c}}{\i} to develop
BlockHammer~\cite{yaglikci2021blockhammer}, a new RowHammer defense mechanism
that provides strong protection while improving scalability and compatibility
with commodity DRAM chips relative to prior RowHammer defenses. \dmpd{Following
BlockHammer, I worked with} Lois Orosa and Giray Ya{\u{g}}l{\i}k{\c{c}}{\i} to
perform a large-scale RowHammer error-characterization
study~\cite{orosa2021deeper}. \dmpd{Finally, I worked with Giray
Ya{\u{g}}l{\i}k{\c{c}}{\i} to experimentally evaluate whether manipulating the
DRAM supply voltage can be an effective defense against
RowHammer~\cite{yaglikci2022understanding}.}

Lastly, I worked with Nastaran Hajinazar on two different projects. In the
first, we developed a new virtual memory framework called the Virtual Block
Interface (VBI)~\cite{hajinazar2020virtual}. VBI shifts the responsibility for
most virtual memory management tasks to the hardware, thereby relieving the
system software of significant complexity associated with virtual memory
management responsibilities. In doing so, VBI improves overall system
performance and enables several important architectural optimizations with
respect to memory management. In the second project, we develop an end-to-end
system framework for in-DRAM data processing called
SIMDRAM~\cite{hajinazar2021simdram}. SIMDRAM extends in-DRAM processing to a
much more general set of computation primitives, facilitating the use of in-DRAM
processing for a broader range of workloads.

\cleardoublepage\balance
\begin{singlespace}
\bibliographystyle{IEEEtranS}
\bibliography{references}

\begin{thebibliography}{10}
\providecommand{\url}[1]{#1}
\csname url@samestyle\endcsname
\providecommand{\newblock}{\relax}
\providecommand{\bibinfo}[2]{#2}
\providecommand{\BIBentrySTDinterwordspacing}{\spaceskip=0pt\relax}
\providecommand{\BIBentryALTinterwordstretchfactor}{4}
\providecommand{\BIBentryALTinterwordspacing}{\spaceskip=\fontdimen2\font plus
\BIBentryALTinterwordstretchfactor\fontdimen3\font minus
  \fontdimen4\font\relax}
\providecommand{\BIBforeignlanguage}[2]{{%
\expandafter\ifx\csname l@#1\endcsname\relax
\typeout{** WARNING: IEEEtran.bst: No hyphenation pattern has been}%
\typeout{** loaded for the language `#1'. Using the pattern for}%
\typeout{** the default language instead.}%
\else
\language=\csname l@#1\endcsname
\fi
#2}}
\providecommand{\BIBdecl}{\relax}
\BIBdecl

\bibitem{ds_intel19701103}
{Intel}, ``{1103},''
  \url{http://www.decadecounter.com/vta/pdf/Intel%20Memory%20Design%20Handbook%20[1973-08].pdf},
  1970.

\bibitem{ds_mostek1971mk4006}
{Mostek}, ``{MK4006},''
  \url{https://usermanual.wiki/Pdf/1974MostekIntegratedCircuitGuide.1468557856/view},
  1971.

\bibitem{ds_mostek1973mk4096}
{Mostek}, ``{MK4096},''
  \url{https://console5.com/techwiki/images/0/04/MK4096.pdf}, 1973.

\bibitem{ds_mostek1976mk4027}
{Mostek}, ``{MK4027},''
  \url{https://console5.com/techwiki/images/d/df/MK4027.pdf}, 1976.

\bibitem{ds_mostek1976mk4116p}
{Mostek}, ``{MK4116P},''
  \url{https://console5.com/techwiki/images/8/85/MK4116.pdf}, 1976.

\bibitem{ds_fairchild1978f4116}
{Fairchild}, ``{F4116},''
  \url{http://minuszerodegrees.net/memory/4116/datasheet_F4116.pdf}, 1978.

\bibitem{ds_intel19792118}
{Intel}, ``{2118},''
  \url{https://drive.google.com/file/d/0B9rh9tVI0J5mNDkwZGEwM2QtMzYzNC00YjQ4LTg4NjYtOGY2ZGRkMDMxYjFm/view?resourcekey=0-vyWj--_z6lp7BjZ-6epTng},
  1979.

\bibitem{ds_mitsubishi1981m5k4164anp}
{Mitsubishi}, ``{M5K4164ANP},''
  \url{https://datasheetspdf.com/pdf-file/1110696/Mitsubishi/M5K4164ANP-15/1},
  1981.

\bibitem{ds_mostek1982mk4564}
{Mostek}, ``{MK4564},''
  \url{http://www.minuszerodegrees.net/memory/4164/datasheet_MK4564-15_and_MK4564-20.pdf},
  1982.

\bibitem{ds_nte1984nte4164}
{NTE}, ``{NTE4164},'' \url{http://www.farnell.com/datasheets/1905614.pdf},
  1984.

\bibitem{ds_texas1984tms4416}
{Texas Instruments}, ``{TMS4416},''
  \url{http://pdf.datasheetcatalog.com/datasheets2/81/817426_1.pdf}, 1984.

\bibitem{ds_mitsubishi1985m5m4256p}
{Mitsubishi}, ``{M5M4256P},''
  \url{http://bitsavers.trailing-edge.com/components/mitsubishi/_dataBooks/1985_Mitsubishi_IC_Memories.pdf},
  1985.

\bibitem{ds_samsung1987km41464a}
{Samsung}, ``{KM41464A},''
  \url{https://console5.com/techwiki/images/2/24/KM41464A.pdf}, 1987.

\bibitem{ds_texas1987tms4464}
{Texas Instruments}, ``{TMS4464},''
  \url{https://www.silicon-ark.co.uk/datasheets/tms4464-datasheet-texas-instruments.pdf},
  1987.

\bibitem{ds_texas1989smj4464}
{Texas Instruments}, ``{SMJ4464},'' \url{http://65xx.unet.bz/ds/TMS4464.pdf},
  1989.

\bibitem{ds_intel199021256}
{Intel}, ``{21256},''
  \url{https://drive.google.com/file/d/0B9rh9tVI0J5mMjU2MDJlNzItNWVkYy00NWM0LThmZjEtYTkyYjE5MTQxOGI2/view?resourcekey=0-Q0K9JcVvNlgRngkBon8vAw},
  1990.

\bibitem{ds_mitsubishi1991m5m44100}
{Mitsubishi}, ``{M5M44100},''
  \url{https://www.datasheetarchive.com/pdf/download.php?id=74e4e0a53cd85e765cc396504a798082be9621&type=O&term=M5M44100},
  1991.

\bibitem{ds_mitsubishi1993m5m44256b}
{Mitsubishi}, ``{M5M44256B},''
  \url{https://datasheetspdf.com/pdf-file/1111257/Mitsubishi/M5M44256BP-10/1},
  1993.

\bibitem{ds_mosel1993v404j8}
{Mosel Vitelic}, ``{V404J8},''
  \url{https://www.datasheetarchive.com/pdf/download.php?id=d5e7f23416e86a5950d91ea69b37003889d50e&type=M&term=V404J8SU70},
  1993.

\bibitem{ds_siemens1995hyb511000bj}
{Siemens}, ``{HYB511000BJ},''
  \url{https://datasheetspdf.com/pdf-file/381513/Siemens/HYB511000BJ-/1}, 1995.

\bibitem{ds_hyundai1997hy5118164b}
{Hyundai}, ``{HY5118164B},''
  \url{https://www.datasheetarchive.com/pdf/download.php?id=7413fece3e0e5dee9b73ec0bb7f6c53afd4c99&type=P&term=HY5118164B},
  1997.

\bibitem{ds_samsung1997km48s2020ct}
{Samsung}, ``{KM48S2020CT},''
  \url{http://www.maxim4u.com/view_online.php?id=1777838&file=0390\km48s2020ct-fl_3917068.pdf},
  1997.

\bibitem{ds_micron1998mt48lc4m4a1}
{Micron}, ``{MT48LC4M4A1},''
  \url{https://www.digchip.com/datasheets/download_datasheet.php?id=688351&part-number=MT48LC2M8A1},
  1998.

\bibitem{ds_mosel1998v53c808h}
{Mosel Vitelic}, ``{V53C808H},''
  \url{https://www.digchip.com/datasheets/download_datasheet.php?id=1031590&part-number=V53C808H},
  1998.

\bibitem{ds_siemens1998hyb39s16400}
{Siemens}, ``{HYB39S16400},''
  \url{https://www.digchip.com/datasheets/download_datasheet.php?id=390213&part-number=HYB39S16160AT-10},
  1998.

\bibitem{ds_samsung1999k4s160822d}
{Samsung}, ``{K4S160822D},''
  \url{http://pdf.datasheetcatalog.com/datasheet/SamsungElectronic/mXtvtzs.pdf},
  1999.

\bibitem{ds_samsung1999k4s561632a}
{Samsung}, ``{K4S561632A},''
  \url{https://www.datasheetarchive.com/pdf/download.php?id=fd48625bbd5e92da34308233eb404f7635e593&type=M&term=K4S561632A},
  1999.

\bibitem{ds_amic2000a416316b}
{Amic}, ``{A416316B},''
  \url{https://pdf1.alldatasheet.com/datasheet-pdf/view/55599/AMICC/A416316BS-35.html},
  2000.

\bibitem{ds_issi2000is41lv32256}
{ISSI}, ``{IS41LV32256},''
  \url{https://www.digchip.com/datasheets/download_datasheet.php?id=442395&part-number=IS41LV32256},
  2000.

\bibitem{ds_samsung2000k4d623237a5}
{Samsung}, ``{K4D623237A5},''
  \url{https://www.datasheetarchive.com/pdf/download.php?id=1ddb8613b8da9b267a1e546d9442e36e6a9d62&type=M&term=K4D623237A},
  2000.

\bibitem{ds_alliance2001as4c256k16e0}
{Alliance}, ``{AS4C256K16E0},''
  \url{http://www.dexsilicium.com/Alliance_AS4C256K16E0.pdf}, 2001.

\bibitem{ds_alliance2001as4c4m4foq}
{Alliance}, ``{AS4C4M4FOQ},''
  \url{https://www.datasheetarchive.com/pdf/download.php?id=c12832d7fa384ccc466fa8f0d33d169415452d&type=P&term=409--1%252Ftds%252B0541},
  2001.

\bibitem{ds_issi2001is41c4400x}
{ISSI}, ``{IS41C4400X},''
  \url{https://datasheetspdf.com/pdf-file/1237264/IntegratedSiliconSolution/IS41LV44002/1},
  2001.

\bibitem{ds_micron2001mt46v2m32}
{Micron}, ``{MT46V2M32},''
  \url{https://datasheetspdf.com/pdf-file/534262/MicronTechnology/MT46V2M32/1},
  2001.

\bibitem{ds_micron2001mt46v32m4}
{Micron}, ``{MT46V32M4},''
  \url{https://www.compel.ru/item-pdf/b6f0ed7c2d40f9dc96e3fa571607bc09/ps/micron~mt46v8m16.pdf},
  2001.

\bibitem{ds_mosel2001v58c265164s}
{Mosel Vitelic}, ``{V58C265164S},''
  \url{https://datasheetspdf.com/pdf-file/295988/MoselVitelicCorp/V58C265164S/1},
  2001.

\bibitem{ds_tm2001t224160b}
{TM Tech}, ``{T224160B},''
  \url{https://www.digchip.com/datasheets/download_datasheet.php?id=945886&part-number=T224160B},
  2001.

\bibitem{ds_micron2003mt46v64m4}
{Micron}, ``{MT46V64M4},''
  \url{https://media-www.micron.com/-/media/client/global/documents/products/data-sheet/dram/ddr1/256mb_ddr.pdf?rev=7d969af24d6d4b74a34e427f350b1c77},
  2003.

\bibitem{ds_samsung2003k4s560432e}
{Samsung}, ``{K4S560432E},''
  \url{https://ru.datasheetbank.com/datasheet-download/429400/1/Samsung/K4S560432E-UC75},
  2003.

\bibitem{ds_amic2005a43l0632}
{Amic}, ``{A43L0632},''
  \url{https://datasheetspdf.com/pdf-file/672656/AMICTechnology/A43L0632/1},
  2005.

\bibitem{ds_elite2006m52s32321a}
{Elite}, ``{M52S32321A},'' \url{http://www.farnell.com/datasheets/62304.pdf},
  2006.

\bibitem{ds_issi2006is42s81600b}
{ISSI}, ``{IS42S81600B},''
  \url{https://datasheetspdf.com/pdf-file/591012/ISSI/IS42S81600B/1}, 2006.

\bibitem{ds_sasmung2006k4t51043qc}
{Sasmung}, ``{K4T51043QC},''
  \url{https://www.digchip.com/datasheets/download_datasheet.php?id=1088989&part-number=K4T51083QC},
  2006.

\bibitem{ds_micron2007mt47h256m4}
{Micron}, ``{MT47H256M4},''
  \url{https://media-www.micron.com/-/media/client/global/documents/products/data-sheet/dram/ddr2/1gb_ddr2.pdf?rev=854b480189b84d558d466bc18efe270c},
  2007.

\bibitem{ds_samsung2010k4b4g0446a}
{Samsung}, ``{K4B4G0446A},''
  \url{https://www.samsung.com/semiconductor/global.semi/file/resource/2017/11/DS_K4B4G0846D-BC_Rev123-0.pdf},
  2010.

\bibitem{ds_hynix2011h5tq4g43mfr}
{Hynix}, ``{H5TQ4G43MFR},''
  \url{https://pdf1.alldatasheet.com/datasheet-pdf/view/533445/HYNIX/H5TQ4G63MFR-H9C.html},
  2011.

\bibitem{ds_nanya2011nt5cb512m}
{Nanya}, ``{NT5CB512M},''
  \url{http://www.sunnyqi.com/upLoad/product/month_1308/NT5CB256M8GN.pdf},
  2011.

\bibitem{ds_samsung2013k4b4g0446a}
{Samsung}, ``{K4B4G0446A},''
  \url{https://www.samsung.com/semiconductor/global.semi/file/resource/2017/11/DS_K4B4G0846D-BC_Rev123-0.pdf},
  2013.

\bibitem{ds_micron2015mt40a2g}
{Micron}, ``{MT40A2G},''
  \url{https://www.micron.com/-/media/client/global/documents/products/data-sheet/dram/ddr4/8gb_ddr4_sdram.pdf},
  2015.

\bibitem{ds_hynix2016h5an4g4nafr}
{Hynix}, ``{H5AN4G4NAFR},''
  \url{https://datasheetspdf.com/pdf-file/1309166/HynixSemiconductor/H5AN4G8NAFR-xxC/1},
  2016.

\bibitem{ds_samsung2016k4a8g165wc}
{Samsung}, ``{K4A8G165WC},''
  \url{https://www.samsung.com/semiconductor/global.semi/file/resource/2017/12/x16%20only_8G_C_DDR4_Samsung_Spec_Rev1.5_Apr.17.pdf},
  2016.

\bibitem{ds_hynix2017h5an8g4nafr}
{Hynix}, ``{H5AN8G4NAFR},''
  \url{https://www.digchip.com/datasheets/download_datasheet.php?id=217237&part-number=H5AN8G8NAFR&type=pn2},
  2017.

\bibitem{ds_micron2018mt40a}
{Micron}, ``{MT40A},''
  \url{https://www.micron.com/-/media/client/global/documents/products/data-sheet/dram/ddr4/16gb_ddr4_sdram.pdf},
  2018.

\bibitem{ds_hynix2019h5an8g4ncjr}
{Hynix}, ``{H5AN8G4NCJR},''
  \url{http://www.hytic.net/upload/files/2019/10/SK%20Hynix%20%20%20-H5AN8G4NCJR.pdf},
  2019.

\bibitem{ds_samsung2019k4aag045wa}
{Samsung}, ``{K4AAG045WA},''
  \url{https://www.memory-distributor.com/pub/media/downloads/datasheets/K4AAG085WA_BIxx.pdf},
  2019.

\bibitem{ds_samsung2020k4aag085wa}
{Samsung}, ``{K4AAG085WA},''
  \url{https://www.memory-distributor.com/pub/media/downloads/datasheets/K4AAG085WA_BIxx.pdf},
  2020.

\bibitem{ds_hynix2021hmcg66meb}
{Hynix}, ``{HMCG66MEB},''
  \url{https://gzhls.at/blob/ldb/b/e/5/8/5bc212f7c92604fd3737505ee4c96014733c.pdf},
  2021.

\bibitem{ds_micron2021mt60b1g16}
{Micron}, ``{MT60B1G16},''
  \url{https://media-www.micron.com/-/media/client/global/documents/products/data-sheet/dram/ddr5/16gb_ddr5_sdram_diereva.pdf?rev=c95e4a49184145f18e105cc41e0ee643},
  2021.

\end{thebibliography}


\begin{thebibliography}{100}
\providecommand{\url}[1]{#1}
\csname url@samestyle\endcsname
\providecommand{\newblock}{\relax}
\providecommand{\bibinfo}[2]{#2}
\providecommand{\BIBentrySTDinterwordspacing}{\spaceskip=0pt\relax}
\providecommand{\BIBentryALTinterwordstretchfactor}{4}
\providecommand{\BIBentryALTinterwordspacing}{\spaceskip=\fontdimen2\font plus
\BIBentryALTinterwordstretchfactor\fontdimen3\font minus
  \fontdimen4\font\relax}
\providecommand{\BIBforeignlanguage}[2]{{%
\expandafter\ifx\csname l@#1\endcsname\relax
\typeout{** WARNING: IEEEtranS.bst: No hyphenation pattern has been}%
\typeout{** loaded for the language `#1'. Using the pattern for}%
\typeout{** the default language instead.}%
\else
\language=\csname l@#1\endcsname
\fi
#2}}
\providecommand{\BIBdecl}{\relax}
\BIBdecl

\bibitem{beergithub}
``{BEER Source Code},'' \url{https://github.com/CMU-SAFARI/BEER}.

\bibitem{datasheetsurveygithub}
``{DRAM Datasheet Survey},''
  \url{https://github.com/CMU-SAFARI/DRAM-Datasheet-Survey}.

\bibitem{drampowergithub}
``{DRAMPower Source Code},'' \url{https://github.com/ravenrd/DRAMPower}.

\bibitem{eccsimgithub}
``{EINSim Source Code},'' \url{https://github.com/CMU-SAFARI/EINSim}.

\bibitem{harpgithub}
``{HARP Source Code},'' \url{https://github.com/CMU-SAFARI/HARP}.

\bibitem{ramulatorgithub}
``{Ramulator Source Code},'' \url{https://github.com/CMU-SAFARI/ramulator}.

\bibitem{spec2006}
``{Standard Performance Evaluation Corporation},''
  \url{http://www.spec.org/cpu2006}.

\bibitem{safari2016private}
{Private communication}, 2016--2021.

\bibitem{ibm2021dram}
``{DRAM: The Invention of On-Demand Data},''
  \url{https://www.ibm.com/ibm/history/ibm100/us/en/icons/dram/transform/},
  2021.

\bibitem{micron2022quaterly}
``{Quarterly Report on Form 10-Q},'' Micron Technologies, Inc., Tech. Rep.,
  2022.

\bibitem{3d2022ddr4}
{3D PLUS}, ``{DDR4 SDRAM},'' 2022.

\bibitem{adams2002high}
R.~D. Adams, \emph{{High Performance Memory Testing: Design Principles, Fault
  Modeling and Self-Test}}.\hskip 1em plus 0.5em minus 0.4em\relax Springer
  SBM, 2002.

\bibitem{adata2017adata}
{ADATA}, ``{ADATA XPG DDR4 Officially Validated by AMD as AM4/Ryzen
  Compatible},'' ADATA, Tech. Rep., 2017.

\bibitem{avl2019testingfaq}
I.~Advanced Validation~Labs, ``{AVL Testing and Validation FAQ}.''

\bibitem{advantest2022t5833}
{Advantest}, \emph{{T5833/T5833ES Memory Test System}},
  \url{https://www.advantest.com/products/memory/t5833.html}, 2022.

\bibitem{agrawal1994proposed}
G.~Agrawal, L.~Massengill, and K.~Gulati, ``{A Proposed SEU Tolerant Dynamic
  Random Access Memory (DRAM) Cell},'' \emph{IEEE Trans. Nucl. Sci.}, 1994.

\bibitem{ahn2016scalable}
J.~Ahn, S.~Hong, S.~Yoo, O.~Mutlu, and K.~Choi, ``{A Scalable
  Processing-In-Memory Accelerator for Parallel Graph Processing},''
  \emph{ISCA}, 2016.

\bibitem{ahn2015pim}
J.~Ahn, S.~Yoo, O.~Mutlu, and K.~Choi, ``{{PIM}-Enabled Instructions: A
  Low-Overhead, Locality-Aware Processing-in-Memory Architecture},'' in
  \emph{ISCA}, 2015.

\bibitem{aichinger2015ddr}
B.~Aichinger, ``{DDR Memory Errors Caused by Row Hammer},'' in \emph{HPEC},
  2015.

\bibitem{alars2005dram}
Z.~Al-Ars, ``{DRAM Fault Analysis and Test Generation},'' Ph.D. dissertation,
  Technische Universiteit Delft, 2005.

\bibitem{al2004effects}
Z.~Al-Ars, S.~Hamdioui, and A.~J. van~de Goor, ``{Effects of Bit Line Coupling
  on the Faulty Behavior of DRAMs},'' in \emph{VTS}, 2004.

\bibitem{alam2021lightweight}
I.~Alam, ``{Lightweight Opportunistic Memory Resilience},'' Ph.D. dissertation,
  University of California, Los Angeles, 2021.

\bibitem{alameldeen2011energy}
A.~R. Alameldeen, I.~Wagner, Z.~Chishti, W.~Wu, C.~Wilkerson, and S.-L. Lu,
  ``{Energy-Efficient Cache Design Using Variable-Strength Error-Correcting
  Codes},'' \emph{ISCA}, 2011.

\bibitem{alliance2014reliability}
{Alliance Memory}, ``{Reliability Qualification Report},'' Alliance Memory,
  Tech. Rep., 2014.

\bibitem{alliance2020lpddr4}
\emph{{2Gb/4Gb/8Gb LPDDR4}}, {Alliance Memory}, 2020, rev. 1.0.

\bibitem{amd2009bkdg}
AMD, ``{BKDG for AMD NPT Family 0Fh Processors},'' 2009.

\bibitem{amd2013bkdg}
AMD, ``{BKDG for AMD Family 15h Models 00h-0Fh Processors},'' 2013.

\bibitem{amd2018amd}
AMD, ``{AMD Opteron 4300 Series Processors},'' 2018.

\bibitem{apalkov2013spin}
D.~Apalkov, A.~Khvalkovskiy, S.~Watts, V.~Nikitin, X.~Tang, D.~Lottis, K.~Moon,
  X.~Luo, E.~Chen, A.~Ong, A.~Driskill-Smith, and M.~Krounbi, ``{Spin-Transfer
  Torque Magnetic Random Access Memory (STT-MRAM)},'' \emph{JETC}, 2013.

\bibitem{apple2015about}
{Apple Inc.}, ``{About the Security Content of Mac EFI Security Update
  2015-001},'' \url{https://support.apple.com/en-us/HT204934}, 2015.

\bibitem{atp2018stringent}
{ATP Electronics}, ``{Stringent Tests from ICs to Modules Ensure DRAM
  Reliability},''
  \url{https://www.atpinc.com/blog/dram-testing-module-chips-ic-burn-in-quality-characteristics},
  2018.

\bibitem{autran2009altitude}
J.-L. Autran, P.~Roche, S.~Sauze, G.~Gasiot, D.~Munteanu, P.~Loaiza,
  M.~Zampaolo, and J.~Borel, ``{Altitude And Underground Real-Time SER
  Characterization Of CMOS 65 Nm SRAM},'' \emph{IEEE Trans. Nucl. Sci.}, 2009.

\bibitem{awasthi2012efficient}
M.~Awasthi, M.~Shevgoor, K.~Sudan, B.~Rajendran, R.~Balasubramonian, and
  V.~Srinivasan, ``{Efficient Scrub Mechanisms for Error-Prone Emerging
  Memories},'' in \emph{HPCA}, 2012.

\bibitem{aweke2016anvil}
Z.~B. Aweke, S.~F. Yitbarek, R.~Qiao, R.~Das, M.~Hicks, Y.~Oren, and T.~Austin,
  ``{ANVIL: Software-Based Protection Against Next-Generation Rowhammer
  Attacks},'' in \emph{ASPLOS}, 2016.

\bibitem{bacchini2014characterization}
A.~Bacchini, M.~Rovatti, G.~Furano, and M.~Ottavi, ``{Characterization of Data
  Retention Faults in DRAM Devices},'' in \emph{DFT}, 2014.

\bibitem{baeg2022estimation}
S.~Baeg, D.~Yun, M.~Chun, and S.-J. Wen, ``{Estimation of the Trap Energy
  Characteristics of Row Hammer-Affected Cells in Gamma-Irradiated DDR4
  DRAM},'' \emph{IEEE Trans. Nucl. Sci.}, 2022.

\bibitem{baek2014refresh}
S.~Baek, S.~Cho, and R.~Melhem, ``{Refresh Now and Then},'' in \emph{TC}, 2014.

\bibitem{bains2014row}
K.~Bains, J.~Halbert, C.~Mozak, T.~Schoenborn, and Z.~Greenfield, ``{Row Hammer
  Refresh Command},'' 2014, {US Patent App. 13/539,415}.

\bibitem{bains2015row}
K.~Bains, J.~Halbert, C.~Mozak, T.~Schoenborn, and Z.~Greenfield, ``{Row Hammer
  Refresh Command},'' 2015, {U.S.}\ Patent 9,117,544.

\bibitem{bains2020read}
K.~S. Bains, R.~Agarwal, and J.~Lee, ``{Read Retry To Selectively Disable
  On-Die ECC},'' 2020, {US} Patent App. 16/875,642.

\bibitem{bains2016distributed}
K.~S. Bains and J.~B. Halbert, ``{Distributed Row Hammer Tracking},'' 2016,
  {U.S.}\ Patent 9,299,400.

\bibitem{bains2016row}
K.~S. Bains and J.~B. Halbert, ``{Row Hammer Monitoring Based on Stored Row
  Hammer Threshold Value},'' 2016, {U.S.}\ Patent 9,384,821.

\bibitem{bairavasundaram2008analysis}
L.~N. Bairavasundaram, A.~C. Arpaci-Dusseau, R.~H. Arpaci-Dusseau, G.~R.
  Goodson, and B.~Schroeder, ``{An Analysis Of Data Corruption In The Storage
  Stack},'' \emph{TOS}, 2008.

\bibitem{bairavasundaram2007analysis}
L.~N. Bairavasundaram, G.~R. Goodson, S.~Pasupathy, and J.~Schindler, ``{An
  Analysis Of Latent Sector Errors In Disk Drives},'' in \emph{SIGMETRICS},
  2007.

\bibitem{bajura2007models}
M.~A. Bajura, Y.~Boulghassoul, R.~Naseer, S.~DasGupta, A.~F. Witulski,
  J.~Sondeen, S.~D. Stansberry, J.~Draper, L.~W. Massengill, and J.~N.
  Damoulakis, ``{Models and Algorithmic Limits for an ECC-Based Approach To
  Hardening Sub-100-nm SRAMs},'' \emph{Trans. on Nucl. Sci.}, 2007.

\bibitem{baker1997shmoo}
K.~Baker and J.~Van~Beers, ``{Shmoo Plotting: The Black Art of IC Testing},''
  \emph{IEEE Des Test}, 1997.

\bibitem{baker1990opening}
T.~Baker, ``{Opening Up Ada-Tasking},'' \emph{ACM SIGAda Ada Letters}, 1990.

\bibitem{utah2013dram}
R.~Balasubramonian, ``{A DRAM Refresh Tutorial},''
  \url{http://utaharch.blogspot.com/2013/11/a-dram-refresh-tutorial.html},
  2013.

\bibitem{balasubramonian2019innovations}
R.~Balasubramonian, ``{Innovations in the Memory System},'' \emph{Synthesis
  Lectures on Computer Architecture}, 2019.

\bibitem{banerjee1989two}
S.~K. Banerjee, ``{Two-Transistor DRAM Cell with High Alpha Particle
  Immunity},'' 1989, {US Patent 4,864,374}.

\bibitem{barenghi2018software}
A.~Barenghi, L.~Breveglieri, N.~Izzo, and G.~Pelosi, ``{Software-Only Reverse
  Engineering of Physical DRAM Mappings For RowHammer Attacks},'' in
  \emph{IVSW}, 2018.

\bibitem{baseman2016improving}
E.~Baseman, N.~DeBardeleben, K.~Ferreira, S.~Levy, S.~Raasch, V.~Sridharan,
  T.~Siddiqua, and Q.~Guan, ``{Improving DRAM Fault Characterization Through
  Machine Learning},'' in \emph{DSN-W}, 2016.

\bibitem{bauer2016lest}
J.~Bauer, M.~Gruhn, and F.~C. Freiling, ``{Lest We Forget: Cold-Boot Attacks on
  Scrambled DDR3 Memory},'' \emph{Digital Investigation}, 2016.

\bibitem{bautista2016unprotected}
L.~Bautista-Gomez, F.~Zyulkyarov, O.~Unsal, and S.~McIntosh-Smith,
  ``{Unprotected Computing: A Large-Scale Study Of DRAM Raw Error Rate On A
  Supercomputer},'' in \emph{SC}, 2016.

\bibitem{bennett2021panopticon}
T.~Bennett, S.~Saroiu, A.~Wolman, and L.~Cojocar, ``{Panopticon: A Complete
  In-DRAM Rowhammer Mitigation},'' in \emph{DRAMSec}, 2021.

\bibitem{bera2021pythia}
R.~Bera, K.~Kanellopoulos, A.~Nori, T.~Shahroodi, S.~Subramoney, and O.~Mutlu,
  ``{Pythia: A Customizable Hardware Prefetching Framework using Online
  Reinforcement Learning},'' in \emph{MICRO}, 2021.

\bibitem{bera2019dspatch}
R.~Bera, A.~V. Nori, O.~Mutlu, and S.~Subramoney, ``{DSPatch: Dual Spatial
  Pattern Prefetcher},'' in \emph{MICRO}, 2019.

\bibitem{bhati2016dram}
I.~Bhati, M.-T. Chang, Z.~Chishti, S.-L. Lu, and B.~Jacob, ``{DRAM Refresh
  Mechanisms, Penalties, and Trade-Offs},'' in \emph{TC}, 2016.

\bibitem{bhati2015flexible}
I.~Bhati, Z.~Chishti, S.-L. Lu, and B.~Jacob, ``{Flexible Auto-Refresh:
  Enabling Scalable and Energy-Efficient DRAM Refresh Reductions},'' in
  \emph{ISCA}, 2015.

\bibitem{bjorner2015nuz}
N.~Bj{\o}rner, A.-D. Phan, and L.~Fleckenstein, ``{nu-Z: An Optimizing SMT
  Solver},'' in \emph{TACAS}, 2015.

\bibitem{boixaderas2020cost}
I.~Boixaderas, D.~Zivanovic, S.~Mor{\'e}, J.~Bartolome, D.~Vicente, M.~Casas,
  P.~M. Carpenter, P.~Radojkovi{\'c}, and E.~Ayguad{\'e}, ``{Cost-Aware
  Prediction of Uncorrected DRAM Errors in the Field},'' in \emph{SC}, 2020.

\bibitem{borkar2011future}
S.~Borkar and A.~A. Chien, ``{The Future of Microprocessors},'' \emph{CACM},
  2011.

\bibitem{boroumand2021google}
A.~Boroumand, S.~Ghose, B.~Akin, R.~Narayanaswami, G.~F. Oliveira, X.~Ma,
  E.~Shiu, and O.~Mutlu, ``{Google Neural Network Models for Edge Devices:
  Analyzing and Mitigating Machine Learning Inference Bottlenecks},'' in
  \emph{PACT}, 2021.

\bibitem{boroumand2021mitigating}
A.~Boroumand, S.~Ghose, B.~Akin, R.~Narayanaswami, G.~F. Oliveira, X.~Ma,
  E.~Shiu, and O.~Mutlu, ``{Mitigating Edge Machine Learning Inference
  Bottlenecks: An Empirical Study on Accelerating {Google} Edge Models},''
  arXiv:2103.00768, 2021.

\bibitem{boroumand2018google}
A.~Boroumand, S.~Ghose, Y.~Kim, R.~Ausavarungnirun, E.~Shiu, R.~Thakur, D.~Kim,
  A.~Kuusela, A.~Knies, P.~Ranganathan, and O.~Mutlu, ``{Google Workloads for
  Consumer Devices: Mitigating Data Movement Bottlenecks},'' in \emph{ASPLOS},
  2018.

\bibitem{boroumand2019conda}
A.~Boroumand, S.~Ghose, M.~Patel, H.~Hassan, B.~Lucia, R.~Ausavarungnirun,
  K.~Hsieh, N.~Hajinazar, K.~T. Malladi, H.~Zheng, and O.~Mutlu, ``{CoNDA:
  Efficient Cache Coherence Support For Near-Data Accelerators},'' in
  \emph{ISCA}, 2019.

\bibitem{boroumand2016lazypim}
A.~Boroumand, S.~Ghose, M.~Patel, H.~Hassan, B.~Lucia, K.~Hsieh, K.~T. Malladi,
  H.~Zheng, and O.~Mutlu, ``{LazyPIM: An Efficient Cache Coherence Mechanism
  For Processing-In-Memory},'' \emph{CAL}, 2016.

\bibitem{borucki2008comparison}
L.~Borucki, G.~Schindlbeck, and C.~Slayman, ``{Comparison of Accelerated DRAM
  Soft Error Rates Measured at Component and System Level},'' in \emph{IEEE
  IRPS}, 2008.

\bibitem{bose1960class}
R.~C. Bose and D.~K. Ray-Chaudhuri, ``{On a Class of Error Correcting Binary
  Group Codes},'' \emph{{Information and Control}}, 1960.

\bibitem{bostanci2022dr}
F.~Bostanc{\i}, A.~Olgun, L.~Orosa, A.~G. Ya{\u{g}}l{\i}k{\c{c}}{\i}, J.~S.
  Kim, H.~Hassan, O.~Ergin, and O.~Mutlu, ``{DR-STRaNGe: End-to-End System
  Design for DRAM-based True Random Number Generators},'' \emph{HPCA}, 2022.

\bibitem{brasser2017can}
F.~Brasser, L.~Davi, D.~Gens, C.~Liebchen, and A.-R. Sadeghi, ``{CAn't Touch
  This: Software-Only Mitigation Against Rowhammer Attacks Targeting Kernel
  Memory},'' in \emph{USENIX Security}, 2017.

\bibitem{brummayer2009boolector}
R.~Brummayer and A.~Biere, ``{Boolector: An Efficient SMT Solver for
  Bit-Vectors and Arrays},'' in \emph{International Conference on Tools and
  Algorithms for the Construction and Analysis of Systems}, 2009.

\bibitem{burr2010phase}
G.~W. Burr, M.~J. Breitwisch, M.~Franceschini, D.~Garetto, K.~Gopalakrishnan,
  B.~Jackson, B.~Kurdi, C.~Lam, L.~A. Lastras, A.~Padilla, B.~Rajendran,
  S.~Raoux, and R.~S. Shenoy, ``{Phase Change Memory Technology},'' \emph{J Vac
  Sci Technol B}, 2010.

\bibitem{cai2017error}
Y.~Cai, S.~Ghose, E.~F. Haratsch, Y.~Luo, and O.~Mutlu, ``{Error
  Characterization, Mitigation, and Recovery In Flash-Memory-Based Solid-State
  Drives},'' \emph{Proc. IEEE}, 2017.

\bibitem{cai2018errors}
Y.~Cai, S.~Ghose, E.~F. Haratsch, Y.~Luo, and O.~Mutlu, ``{Errors in
  Flash-Memory-Based Solid-State Drives: Analysis, Mitigation, and Recovery},''
  \emph{Inside Solid State Drives}, 2018.

\bibitem{cai2017vulnerabilities}
Y.~Cai, S.~Ghose, Y.~Luo, K.~Mai, O.~Mutlu, and E.~F. Haratsch,
  ``{Vulnerabilities in MLC NAND Flash Memory Programming: Experimental
  Analysis, Exploits, and Mitigation Techniques},'' in \emph{HPCA}, 2017.

\bibitem{cai2011fpga}
Y.~Cai, E.~F. Haratsch, M.~McCartney, and K.~Mai, ``{{FPGA}-Based Solid-State
  Drive Prototyping Platform},'' in \emph{FCCM}, 2011.

\bibitem{cai2012error}
Y.~Cai, E.~F. Haratsch, O.~Mutlu, and K.~Mai, ``{Error Patterns in {MLC NAND}
  Flash Memory: Measurement, Characterization, and Analysis},'' in \emph{DATE},
  2012.

\bibitem{cai2013threshold}
Y.~Cai, E.~F. Haratsch, O.~Mutlu, and K.~Mai, ``{Threshold Voltage Distribution
  in {MLC NAND} Flash Memory: Characterization, Analysis, and Modeling},'' in
  \emph{DATE}, 2013.

\bibitem{cai2015read}
Y.~Cai, Y.~Luo, S.~Ghose, and O.~Mutlu, ``{Read Disturb Errors in MLC NAND
  Flash Memory: Characterization, Mitigation, and Recovery},'' in \emph{DSN},
  2015.

\bibitem{cai2015data}
Y.~Cai, Y.~Luo, E.~F. Haratsch, K.~Mai, and O.~Mutlu, ``{Data Retention in MLC
  NAND Flash Memory: Characterization, Optimization, and Recovery},'' in
  \emph{HPCA}, 2015.

\bibitem{cai2013program}
Y.~Cai, O.~Mutlu, E.~F. Haratsch, and K.~Mai, ``{Program Interference in {MLC
  NAND} Flash Memory: Characterization, Modeling, and Mitigation},'' in
  \emph{ICCD}, 2013.

\bibitem{cai2012flash}
Y.~Cai, G.~Yalcin, O.~Mutlu, E.~F. Haratsch, A.~Cristal, O.~S. Unsal, and
  K.~Mai, ``{Flash Correct-And-Refresh: Retention-Aware Error Management for
  Increased Flash Memory Lifetime},'' in \emph{ICCD}, 2012.

\bibitem{cai2013error}
Y.~Cai, G.~Yalcin, O.~Mutlu, E.~F. Haratsch, A.~Cristal, O.~S. Unsal, and
  K.~Mai, ``{Error Analysis and Retention-Aware Error Management for {NAND}
  Flash Memory},'' in \emph{ITJ}, 2013.

\bibitem{cai2014neighbor}
Y.~Cai, G.~Yalcin, O.~Mutlu, E.~F. Haratsch, O.~Unsal, A.~Cristal, and K.~Mai,
  ``{Neighbor-Cell Assisted Error Correction for MLC NAND Flash Memories},'' in
  \emph{SIGMETRICS}, 2014.

\bibitem{cardarilli2000development}
G.~C. Cardarilli, P.~Marinucci, and A.~Salsano, ``{Development of an Evaluation
  Model for the Design of Fault-Tolerant Solid State Mass Memory},'' in
  \emph{ISCAS}, 2000.

\bibitem{cavicchi2010patent}
P.~Cavicchi and R.~Schellingerhout, ``{Patent Ambush in Standard-Setting: the
  Comission Accepts Commitments from Rambus to Lower Memory Chip Royalty
  Rates},'' \emph{Competition Policy Newsletter}, 2010.

\bibitem{cha2017defect}
S.~Cha, O.~Seongil, H.~Shin, S.~Hwang, K.~Park, S.~J. Jang, J.~S. Choi, G.~Y.
  Jin, Y.~H. Son, H.~Cho, J.~H. Ahn, and N.~S. Kim, ``{Defect Analysis and
  Cost-Effective Resilience Architecture for Future DRAM Devices},'' in
  \emph{HPCA}, 2017.

\bibitem{chandrasekar2014exploiting}
K.~Chandrasekar, S.~Goossens, C.~Weis, M.~Koedam, B.~Akesson, N.~Wehn, and
  K.~Goossens, ``{Exploiting Expendable Process-Margins in DRAMs for Run-Time
  Performance Optimization},'' in \emph{DATE}, 2014.

\bibitem{chang2017thesis}
K.~K. Chang, ``{Understanding and Improving Latency of DRAM-Based Memory
  Systems},'' Ph.D. dissertation, Carnegie Mellon University, 2017.

\bibitem{chang2016understanding}
K.~K. Chang, A.~Kashyap, H.~Hassan, S.~Ghose, K.~Hsieh, D.~Lee, T.~Li,
  G.~Pekhimenko, S.~Khan, and O.~Mutlu, ``{Understanding Latency Variation in
  Modern DRAM Chips: Experimental Characterization, Analysis, and
  Optimization},'' in \emph{SIGMETRICS}, 2016.

\bibitem{chang2014improving}
K.~K. Chang, D.~Lee, Z.~Chishti, A.~R. Alameldeen, C.~Wilkerson, Y.~Kim, and
  O.~Mutlu, ``{Improving DRAM Performance by Parallelizing Refreshes with
  Accesses},'' in \emph{HPCA}, 2014.

\bibitem{chang2016low}
K.~K. Chang, P.~J. Nair, D.~Lee, S.~Ghose, M.~K. Qureshi, and O.~Mutlu,
  ``{Low-Cost Inter-Linked Subarrays (LISA): Enabling Fast Inter-Subarray Data
  Movement in DRAM},'' in \emph{HPCA}, 2016.

\bibitem{chang2017understanding}
K.~K. Chang, A.~G. Ya{\u{g}}l{\i}k{\c{c}}{\i}, S.~Ghose, A.~Agrawal,
  N.~Chatterjee, A.~Kashyap, D.~Lee, M.~O'Connor, H.~Hassan, and O.~Mutlu,
  ``{Understanding Reduced-Voltage Operation in Modern DRAM Devices:
  Experimental Characterization, Analysis, and Mechanisms},'' in
  \emph{SIGMETRICS}, 2017.

\bibitem{chatterjee1979leakage}
P.~K. Chatterjee, G.~W. Taylor, A.~F. Tasch, and H.-S. Fu, ``{Leakage Studies
  in High-Density Dynamic MOS Memory Devices},'' \emph{TED}, 1979.

\bibitem{chen2019care}
C.~Chen, G.~Eisenhauer, S.~Pande, and Q.~Guan, ``{Care: Compiler-Assisted
  Recovery From Soft Failures},'' in \emph{SC}, 2019.

\bibitem{chen2010advances}
E.~Chen, D.~Apalkov, Z.~Diao, A.~Driskill-Smith, D.~Druist, D.~Lottis,
  V.~Nikitin, X.~Tang, S.~Watts, S.~Wang, S.~A. Wolf, A.~W. Ghosh, J.~W. Lu,
  S.~J. Poon, M.~Stan, W.~H. Butler, S.~Gupta, C.~K.~A. Mewes, T.~Mewes, and
  P.~B. Visscher, ``{Advances and Future Prospects of Spin-Transfer Torque
  Random Access Memory},'' \emph{TOM}, 2010.

\bibitem{chen2015ecc}
H.-M. Chen, A.~Arunkumar, C.-J. Wu, T.~Mudge, and C.~Chakrabarti, ``{E-ECC: Low
  Power Erasure And Error Correction Schemes For Increasing Reliability Of
  Commodity DRAM Systems},'' in \emph{Proceedings of the 2015 International
  Symposium on Memory Systems}, 2015, pp. 60--70.

\bibitem{chen2018configurable}
H.-M. Chen, S.-Y. Lee, T.~Mudge, C.-J. Wu, and C.~Chakrabarti,
  ``{Configurable-ECC: Architecting a Flexible ECC Scheme to Support Different
  Sized Accesses in High Bandwidth Memory Systems},'' \emph{TC}, 2018.

\bibitem{chen2016ratt}
H.-M. Chen, C.-J. Wu, T.~Mudge, and C.~Chakrabarti, ``{RATT-ECC: Rate Adaptive
  Two-Tiered Error Correction Codes for Reliable 3D Die-Stacked Memory},''
  \emph{ACM Transactions on Architecture and Code Optimization (TACO)},
  vol.~13, no.~3, pp. 1--24, 2016.

\bibitem{chen2021care}
J.~Chen, X.~Jiang, Y.~Zhang, L.~Liu, H.~Xu, and Q.~Liu, ``{CARE: Coordinated
  Augmentation for Elastic Resilience on DRAM Errors in Data Centers},'' in
  \emph{HPCA}, 2021.

\bibitem{chen2013e3cc}
L.~Chen, Y.~Cao, and Z.~Zhang, ``{E3CC: A Memory Error Protection Scheme With
  Novel Address Mapping for Subranked And Low-Power Memories},'' \emph{TACO},
  2013.

\bibitem{cheng2002neighborhood}
K.-L. Cheng, M.-F. Tsai, and C.-W. Wu, ``{Neighborhood Pattern-Sensitive Fault
  Testing and Diagnostics for Random-Access Memories},'' \emph{TCAD}, 2002.

\bibitem{cheng2019retention}
W.-K. Cheng, P.-Y. Shen, and X.-L. Li, ``{Retention-Aware DRAM Auto-Refresh
  Scheme for Energy and Performance Efficiency},'' \emph{Micromachines}, 2019.

\bibitem{childers2015achieving}
B.~R. Childers, J.~Yang, and Y.~Zhang, ``{Achieving Yield, Density and
  Performance Effective DRAM at Extreme Technology Sizes},'' in \emph{MEMSYS},
  2015.

\bibitem{cho2017novel}
M.~H. Cho, N.~Jeon, M.~Jeong, S.~Lee, S.~Yamada, and H.~Hong, ``{A Novel Method
  to Characterize DRAM Process Variation by the Analyzing Stochastic Properties
  of Retention Time Distribution},'' in \emph{EDTM}, 2017.

\bibitem{choi2020reducing}
H.~Choi, D.~Hong, J.~Lee, and S.~Yoo, ``{Reducing DRAM Refresh Power
  Consumption by Runtime Profiling of Retention Time and Dual-Row
  Activation},'' \emph{Microprocessors and Microsystems}, 2020.

\bibitem{choi2015multiple}
J.~Choi, W.~Shin, J.~Jang, J.~Suh, Y.~Kwon, Y.~Moon, and L.-S. Kim, ``{Multiple
  Clone Row DRAM: A Low Latency and Area Optimized DRAM},'' in \emph{ISCA}.

\bibitem{chun201816gb}
K.~C. Chun, Y.-G. Chu, J.-S. Heo, T.-S. Kim, S.~Kim, H.-K. Yang, M.-J. Kim,
  C.-K. Lee, J.~Kim, H.~Yoon \emph{et~al.}, ``{A 16Gb LPDDR4X SDRAM with an
  NBTI-Tolerant Circuit Solution, an SWD PMOS GIDL Reduction Technique, an
  Adaptive Gear-Down Scheme and a Metastable-Free DQS Aligner In a 10nm Class
  DRAM Process},'' in \emph{ISSCC}, 2018.

\bibitem{chun2012scaling}
K.~C. Chun, H.~Zhao, J.~D. Harms, T.-H. Kim, J.-P. Wang, and C.~H. Kim, ``{A
  Scaling Roadmap and Performance Evaluation of In-Plane and Perpendicular MTJ
  Based STT-MRAMs for High-Density Cache Memory},'' \emph{JSSC}, 2012.

\bibitem{cimatti2010satisfiability}
A.~Cimatti, A.~Franz{\'e}n, A.~Griggio, R.~Sebastiani, and C.~Stenico,
  ``{Satisfiability Modulo The Theory of Costs: Foundations and
  Applications},'' in \emph{TACAS}, 2010.

\bibitem{clark2013error}
G.~C. Clark~Jr and J.~B. Cain, \emph{{Error-Correction Coding for Digital
  Communications}}.\hskip 1em plus 0.5em minus 0.4em\relax Springer SBM, 2013.

\bibitem{cojocar2020are}
L.~Cojocar, J.~Kim, M.~Patel, L.~Tsai, S.~Saroiu, A.~Wolman, and O.~Mutlu,
  ``{Are We Susceptible to Rowhammer? An End-to-End Methodology for Cloud
  Providers},'' in \emph{IEEE S\&P}, 2020.

\bibitem{cojocar2021mfit}
L.~Cojocar, K.~Loughlin, S.~Saroiu, B.~Kasikci, and A.~Wolman, ``{mFIT: A
  Bump-in-the-Wire Tool for Plug-and-Play Analysis of Rowhammer Susceptibility
  Factors},'' 2021.

\bibitem{cojocar2019exploiting}
L.~Cojocar, K.~Razavi, C.~Giuffrida, and H.~Bos, ``{Exploiting Correcting
  Codes: On The Effectiveness Of ECC Memory Against Rowhammer Attacks},'' in
  \emph{S\&P}, 2019.

\bibitem{costello1982error}
D.~J. Costello and S.~Lin, \emph{{Error Control Coding: Fundamentals and
  Applications}}.\hskip 1em plus 0.5em minus 0.4em\relax Prentice Hall, 1982.

\bibitem{courbon2016reverse}
F.~Courbon, S.~Skorobogatov, and C.~Woods, ``{Reverse Engineering Flash EEPROM
  Memories Using Scanning Electron Microscopy},'' in \emph{CARDIS}, 2016.

\bibitem{cover1994elements}
T.~M. Cover, J.~A. Thomas, and J.~Kieffer, ``{Elements of Information
  Theory},'' \emph{SIAM Review}, 1994.

\bibitem{criss2020improving}
K.~Criss, K.~Bains, R.~Agarwal, T.~Bennett, T.~Grunzke, J.~K. Kim, H.~Chung,
  and M.~Jang, ``{Improving Memory Reliability by Bounding DRAM Faults: DDR5
  Improved Reliability Features},'' in \emph{MEMSYS}, 2020.

\bibitem{croswell2000model}
J.~A. Croswell, ``{A Model for Analysis of the Effects of Redundancy and Error
  Correction on DRAM Memory Yield and Reliability},'' Master's thesis, MIT,
  2000.

\bibitem{cui2016snake}
X.~Cui, Z.~Cheng, C.~Lee, X.~Lin, Y.~Wei, X.~Chen, and Z.~Song, ``{A Snake
  Addressing Scheme for Phase Change Memory Testing},'' \emph{SCIS}, 2016.

\bibitem{cui2014dtail}
Z.~Cui, S.~A. McKee, Z.~Zha, Y.~Bao, and M.~Chen, ``{DTail: A Flexible Approach
  to DRAM Refresh Management},'' in \emph{SC}, 2014.

\bibitem{das2018vrl}
A.~Das, H.~Hassan, and O.~Mutlu, ``{VRL-DRAM: Improving DRAM Performance Via
  Variable Refresh Latency},'' in \emph{DAC}, 2018.

\bibitem{data2022rad}
{Data Device Corporation}, ``{Rad Hard Memories},'' 2022.

\bibitem{david2011memory}
H.~David, C.~Fallin, E.~Gorbatov, U.~R. Hanebutte, and O.~Mutlu, ``{Memory
  Power Management via Dynamic Voltage/Frequency Scaling},'' in \emph{ICAC},
  2011.

\bibitem{de2008z3}
L.~De~Moura and N.~Bj{\o}rner, ``{Z3: An Efficient SMT Solver},'' in
  \emph{TACAS}, 2008.

\bibitem{de2021smash}
F.~de~Ridder, P.~Frigo, E.~Vannacci, H.~Bos, C.~Giuffrida, and K.~Razavi,
  ``{SMASH: Synchronized Many-sided Rowhammer Attacks from JavaScript},'' in
  \emph{USENIX Security}, 2021.

\bibitem{dekker1990realistic}
R.~Dekker, F.~Beenker, and L.~Thijssen, ``{A Realistic Fault Model and Test
  Algorithms for Static Random Access Memories},'' \emph{TCAD}, 1990.

\bibitem{dell1997white}
T.~J. Dell, ``{A White Paper on the Benefits of Chipkill-Correct ECC for PC
  Server Main Memory},'' \emph{{IBM Microelectronics Division}}, 1997.

\bibitem{dell2008system}
T.~J. Dell, ``{System RAS Implications of DRAM Soft Errors},'' \emph{IBM JRD},
  2008.

\bibitem{deng2011memscale}
Q.~Deng, D.~Meisner, L.~Ramos, T.~F. Wenisch, and R.~Bianchini, ``{MemScale:
  Active Low-Power Modes for Main Memory},'' in \emph{ASPLOS}, 2011.

\bibitem{dennard1968field}
R.~H. Dennard, ``{Field-Effect Transistor Memory},'' 1968, {US} Patent
  3,387,286.

\bibitem{dennard1974design}
R.~H. Dennard, F.~H. Gaensslen, H.-N. Yu, V.~L. Rideout, E.~Bassous, and A.~R.
  LeBlanc, ``{Design of Ion-Implanted MOSFET's with Very Small Physical
  Dimensions},'' \emph{JSSC}, 1974.

\bibitem{desai2012process}
S.~Desai, ``{Process Variation Aware DRAM (Dynamic Random Access Memory) Design
  Using Block-Based Adaptive Body Biasing Algorithm},'' Ph.D. dissertation,
  Utah State University, 2012.

\bibitem{devaux2019true}
F.~Devaux, ``{The True Processing in Memory Accelerator},'' in \emph{HCS},
  2019.

\bibitem{devaux2021method}
F.~Devaux and R.~Ayrignac, ``{Method and Circuit for Protecting a DRAM Memory
  Device from the Row Hammer Effect},'' 2021, 10,885,966.

\bibitem{dillig2012minimum}
I.~Dillig, T.~Dillig, K.~L. McMillan, and A.~Aiken, ``{Minimum Satisfying
  Assignments for SMT},'' in \emph{CAV}, 2012.

\bibitem{ditali2007x}
A.~Ditali, M.~K. Ma, and M.~Johnston, ``{X-Ray Radiation Effect in DRAM
  Retention Time},'' \emph{T-DMR}, 2007.

\bibitem{duan20172d}
M.~Duan, F.~Adam-Lema, B.~Cheng, C.~Navarro, X.~Wang, V.~Georgiev, F.~Gamiz,
  C.~Millar, and A.~Asenov, ``{2D-TCAD Simulation on Retention Time of Z2FET
  for DRAM Application},'' in \emph{SISPAD}, 2017.

\bibitem{duganapalli2016modelling}
K.~Duganapalli, ``{Modelling and Test Generation for Crosstalk Faults in DSM
  Chips},'' Ph.D. dissertation, Universit{\"a}t Bremen, 2016.

\bibitem{edri2016silicon}
N.~Edri, P.~Meinerzhagen, A.~Teman, A.~Burg, and A.~Fish, ``{Silicon-Proven,
  Per-Cell Retention Time Distribution Model for Gain-Cell Based eDRAMs},''
  \emph{IEEE TOCS}, 2016.

\bibitem{efron1992bootstrap}
B.~Efron, ``{Bootstrap Methods: Another Look at the Jackknife},'' in
  \emph{{Breakthroughs in Statistics}}, 1992.

\bibitem{european2010competition}
{European Commission}, ``{Competition Policy Newsletter},'' \emph{{Publications
  Office of the EU}}, 2010.

\bibitem{everspin2021sttmram}
{Everspin Technologies}, ``{16Mb MRAM MR4A16B}.''

\bibitem{eyerman2008system}
S.~Eyerman and L.~Eeckhout, ``{System-Level Performance Metrics for
  Multiprogram Workloads},'' in \emph{IEEE Micro}, 2008.

\bibitem{fang2017letgo}
B.~Fang, Q.~Guan, N.~Debardeleben, K.~Pattabiraman, and M.~Ripeanu, ``{Letgo: A
  Lightweight Continuous Framework For HPC Applications Under Failures},'' in
  \emph{HPDC}, 2017.

\bibitem{farmani2021rhat}
M.~Farmani, M.~Tehranipoor, and F.~Rahman, ``{RHAT: Efficient RowHammer-Aware
  Test for Modern DRAM Modules},'' in \emph{ETS}, 2021.

\bibitem{ferdman2012clearing}
M.~Ferdman, A.~Adileh, O.~Kocberber, S.~Volos, M.~Alisafaee, D.~Jevdjic,
  C.~Kaynak, A.~D. Popescu, A.~Ailamaki, and B.~Falsafi, ``{Clearing The
  Clouds: A Study Of Emerging Scale-Out Workloads On Modern Hardware},''
  \emph{ASPLOS}, 2012.

\bibitem{field2015microsoft}
S.~Field, ``{Microsoft Azure uses Error-Correcting Code Memory for Enhanced
  Reliability and Security},''
  \url{https://azure.microsoft.com/en-us/blog/microsoft-azure-uses-error-correcting-code-memory-for-enhanced-reliability-and-security},
  2015.

\bibitem{flamm1993measurement}
K.~Flamm, ``{Measurement of DRAM Prices: Technology and Market Structure},'' in
  \emph{{Price Measurements and Their Uses}}.\hskip 1em plus 0.5em minus
  0.4em\relax University of Chicago Press, 1993.

\bibitem{flamm2010impact}
K.~Flamm, ``{The Impact Of DRAM Design Innovation On Manufacturing
  Profitability},'' \emph{Future Fab International}, 2010.

\bibitem{forney1965concatenated}
G.~D. Forney, ``{Concatenated Codes},'' \emph{MIT Press}, 1965.

\bibitem{francis2018raspberry}
P.~Francis-Mezger and V.~M. Weaver, ``{A Raspberry Pi Operating System for
  Exploring Advanced Memory System Concepts},'' in \emph{MEMSYS}, 2018.

\bibitem{frigo2020trrespass}
P.~Frigo, E.~Vannacci, H.~Hassan, V.~van~der Veen, O.~Mutlu, C.~Giuffrida,
  H.~Bos, and K.~Razavi, ``{TRRespass: Exploiting the Many Sides of Target Row
  Refresh},'' in \emph{IEEE S\&P}, 2020.

\bibitem{fromm1997energy}
R.~Fromm, S.~Perissakis, N.~Cardwell, C.~Kozyrakis, B.~McGaughy, D.~Patterson,
  T.~Anderson, and K.~Yelick, ``{The Energy Efficiency of IRAM
  Architectures},'' \emph{ISCA}, 1997.

\bibitem{fujitsu2012fcram}
\emph{{FCRAM (Fast Cycle RAM)}}, {Fujitsu Semiconductor Limited}, 2012,
  \url{https://www.fujitsu.com/cn/Images/FCRAM_catalog_2012.2.pdf}.

\bibitem{fukami2017improving}
A.~Fukami, S.~Ghose, Y.~Luo, Y.~Cai, and O.~Mutlu, ``{Improving the Reliability
  of Chip-Off Forensic Analysis of NAND Flash Memory Devices},'' in
  \emph{Digital Investigation}, 2017.

\bibitem{gallager1963low}
R.~G. Gallager, ``{Low Density Parity Check Codes},'' Ph.D. dissertation,
  Massachusetts Institute of Technology, 1963.

\bibitem{gao2019computedram}
F.~Gao, G.~Tziantzioulis, and D.~Wentzlaff, ``{ComputeDRAM: In-Memory Compute
  using Off-the-Shelf DRAMs},'' in \emph{MICRO}, 2019.

\bibitem{gautam2019row}
S.~Gautam, S.~Manhas, A.~Kumar, M.~Pakala, and E.~Yieh, ``{Row Hammering
  Mitigation Using Metal Nanowire in Saddle Fin DRAM},'' \emph{IEEE TED}, 2019.

\bibitem{ghose2019processing}
S.~Ghose, A.~Boroumand, J.~S. Kim, J.~G{\'o}mez-Luna, and O.~Mutlu,
  ``{Processing-in-Memory: A Workload-driven Perspective},'' \emph{IBM JRD},
  2019.

\bibitem{ghose2018enabling}
S.~Ghose, K.~Hsieh, A.~Boroumand, R.~Ausavarungnirun, and O.~Mutlu, ``{Enabling
  the Adoption of Processing-in-Memory: Challenges, Mechanisms, Future Research
  Directions},'' arXiv:1802.00320, 2018.

\bibitem{ghose2019demystifying}
S.~Ghose, T.~Li, N.~Hajinazar, D.~S. Cali, and O.~Mutlu, ``{Demystifying
  Complex Workload-DRAM Interactions: An Experimental Study},''
  \emph{SIGMETRICS}, 2019.

\bibitem{ghose2018your}
S.~Ghose, A.~G. Ya{\u{g}}l{\i}k{\c{c}}{\i}, R.~Gupta, D.~Lee, K.~Kudrolli,
  W.~X. Liu, H.~Hassan, K.~K. Chang, N.~Chatterjee, A.~Agrawal, M.~O'Connor,
  and O.~Mutlu, ``{What Your DRAM Power Models Are Not Telling You: Lessons
  from a Detailed Experimental Study},'' \emph{SIGMETRICS}, 2018.

\bibitem{ghosh2007smart}
M.~Ghosh and H.-H.~S. Lee, ``{Smart Refresh: An Enhanced Memory Controller
  Design for Reducing Energy in Conventional and 3D Die-Stacked DRAMs},'' in
  \emph{MICRO}, 2007.

\bibitem{giannoula2022towards}
C.~Giannoula, I.~Fernandez, J.~G{\'o}mez-Luna, N.~Koziris, G.~Goumas, and
  O.~Mutlu, ``{Towards Efficient Sparse Matrix Vector Multiplication on Real
  Processing-in-Memory Architectures},'' \emph{SIGMETRICS}, 2022.

\bibitem{giurgiu2017predicting}
I.~Giurgiu, J.~Szabo, D.~Wiesmann, and J.~Bird, ``{Predicting DRAM Reliability
  in the Field with Machine Learning},'' in \emph{Middleware}, 2017.

\bibitem{goiffon2019radiation}
V.~Goiffon, T.~Bilba, T.~Deladerrière, G.~Beaugendre, A.~Le~Roch, A.~Dion,
  C.~Virmontois, J.-M. Belloir, M.~Gaillardin, A.~Jay, and P.~Paillet,
  ``{Radiation-Induced Variable Retention Time in Dynamic Random Access
  Memories},'' \emph{IEEE Trans. Nucl. Sci.}, 2019.

\bibitem{gokhale1995processing}
M.~Gokhale, B.~Holmes, and K.~Iobst, ``{Processing in Memory: The Terasys
  Massively Parallel PIM Array},'' \emph{Computer}, 1995.

\bibitem{gold2014providing}
S.~M. Gold and A.~B. Hegde, ``{Providing Test Coverage of Integrated ECC Logic
  in Embedded Memory},'' 2014, {US Patent 8,914,687}.

\bibitem{gomes2008satisfiability}
C.~P. Gomes, H.~Kautz, A.~Sabharwal, and B.~Selman, ``{Satisfiability
  Solvers},'' \emph{Foundations of Artificial Intelligence}, 2008.

\bibitem{gomez2016dummy}
H.~{Gomez}, A.~{Amaya}, and E.~{Roa}, ``{{DRAM} Row-Hammer Attack Reduction
  using Dummy Cells},'' in \emph{NORCAS}, 2016.

\bibitem{gomez2021benchmarkingmemory}
J.~G{\'o}mez-Luna, I.~El~Hajj, I.~Fernandez, C.~Giannoula, G.~F. Oliveira, and
  O.~Mutlu, ``{Benchmarking Memory-Centric Computing Systems: Analysis of Real
  Processing-in-Memory Hardware},'' in \emph{CUT}, 2021.

\bibitem{gomez2021benchmarking}
J.~G{\'o}mez-Luna, J.~El~Hajj, I.~Fernandez, and C.~Giannoula, ``{Benchmarking
  a New Paradigm: Understanding a Modern Processing-in-Memory Architecture},''
  in \emph{{arXiv:2105.03814}}, 2021.

\bibitem{juan2021benchmarking}
J.~G{\'o}mez-Luna, I.~E. Hajj, I.~Fernández, C.~Giannoula, G.~F. Oliveira, and
  O.~Mutlu, ``{Benchmarking a New Paradigm: Understanding a Modern
  Processing-in-Memory Architecture},'' {arXiv:2105.03814}, 2021.

\bibitem{gong2017dram}
S.-L. Gong, J.~Kim, and M.~Erez, ``{DRAM Scaling Error Evaluation Model Using
  Various Retention Time},'' in \emph{DSN-W}, 2017.

\bibitem{gong2018duo}
S.-L. Gong, J.~Kim, S.~Lym, M.~Sullivan, H.~David, and M.~Erez, ``{DUO:
  Exposing On-Chip Redundancy to Rank-Level ECC for High Reliability},'' in
  \emph{HPCA}, 2018.

\bibitem{gong2015clean}
S.-L. Gong, M.~Rhu, J.~Kim, J.~Chung, and M.~Erez, ``{Clean-ECC: High
  Reliability ECC for Adaptive Granularity Memory System},'' in
  \emph{Proceedings of the 48th International Symposium on Microarchitecture},
  2015, pp. 611--622.

\bibitem{gorman2015memory}
K.~W. Gorman, M.~R. Ouellette, and P.~E. Perry, ``{Memory Test With In-Line
  Error Correction Code Logic},'' 2015, {US Patent 9,224,503}.

\bibitem{gottscho2016measuring}
M.~Gottscho, M.~Shoaib, S.~Govindan, B.~Sharma, D.~Wang, and P.~Gupta,
  ``{Measuring The Impact Of Memory Errors On Application Performance},''
  \emph{CAL}, 2016.

\bibitem{greenfield2012throttling}
Z.~Greenfield and T.~Levy, ``{Throttling Support for Row-Hammer Counters},''
  2016, {U.S.\ Patent 9,251,885}.

\bibitem{gruhn2013practicability}
M.~Gruhn and T.~M{\"u}ller, ``{On the Practicability of Cold Boot Attacks},''
  in \emph{ARES}, 2013.

\bibitem{gu2003challenges}
B.~Gu, T.~Coughlin, B.~Maxwell, J.~Griffith, J.~Lee, J.~Cordingley, S.~Johnson,
  E.~Karaginiannis, and J.~Ehmann, ``{Challenges and Future Directions of Laser
  Fuse Processing in Memory Repair},'' \emph{Proc. Semicon China}, 2003.

\bibitem{guo2017sanitizer}
X.~Guo, M.~N. Bojnordi, Q.~Guo, and E.~Ipek, ``{Sanitizer: Mitigating the
  Impact of Expensive ECC Checks on STT-MRAM Based Main Memories},''
  \emph{TOC}, 2017.

\bibitem{gurumurthi2021hbm3}
S.~Gurumurthi, K.~Lee, M.~Jang, V.~Sridharan, A.~Nygren, Y.~Ryu, K.~Sohn,
  T.~Kim, and H.~Chung, ``{HBM3: Enabling Memory Resilience at Scale},''
  \emph{CAL}, 2021.

\bibitem{gutierrez2011full}
A.~Gutierrez, R.~G. Dreslinski, T.~F. Wenisch, T.~Mudge, A.~Saidi, C.~Emmons,
  and N.~Paver, ``{Full-System Analysis And Characterization Of Interactive
  Smartphone Applications},'' in \emph{IISWC}, 2011.

\bibitem{hajinazar2021simdram}
N.~Hajinazar, G.~F. Oliveira, S.~Gregorio, J.~Ferreira, N.~M. Ghiasi, M.~Patel,
  M.~Alser, S.~Ghose, J.~G. Luna, and O.~Mutlu, ``{SIMDRAM: An End-to-End
  Framework for Bit-Serial SIMD Computing in DRAM},'' \emph{ASPLOS}, 2021.

\bibitem{hajinazar2020virtual}
N.~Hajinazar, P.~Patel, M.~Patel, K.~Kanellopoulos, S.~Ghose,
  R.~Ausavarungnirun, G.~F. Oliveira, J.~Appavoo, V.~Seshadri, and O.~Mutlu,
  ``{The Virtual Block Interface: A Flexible Alternative To The Conventional
  Virtual Memory Framework},'' in \emph{ISCA}, 2020.

\bibitem{halderman2008lest}
J.~A. Halderman, S.~D. Schoen, N.~Heninger, W.~Clarkson, W.~Paul, J.~A.
  Calandrino, A.~J. Feldman, J.~Appelbaum, and E.~W. Felten, ``{Lest We
  Remember: Cold-Boot Attacks on Encryption Keys},'' \emph{{USENIX Security}},
  2008.

\bibitem{hamamoto1995well}
T.~Hamamoto, S.~Sugiura, and S.~Sawada, ``{Well Concentration: A Novel Scaling
  Limitation Factor Derived From DRAM Retention Time and Its Modeling},'' in
  \emph{IEDM}, 1995.

\bibitem{hamamoto1998retention}
T.~Hamamoto, S.~Sugiura, and S.~Sawada, ``{On the Retention Time Distribution
  of Dynamic Random Access Memory (DRAM)},'' in \emph{TED}, 1998.

\bibitem{hamdioui2017test}
S.~Hamdioui, P.~Pouyan, H.~Li, Y.~Wang, A.~Raychowdhur, and I.~Yoon, ``{Test
  and Reliability of Emerging Non-Volatile Memories},'' in \emph{ATS}, 2017.

\bibitem{hamming1950error}
R.~W. Hamming, ``{Error Detecting and Error Correcting Codes},'' in \emph{Bell
  Labs Technical Journal}, 1950.

\bibitem{han2014data}
Y.~Han, Y.~Wang, H.~Li, and X.~Li, ``{Data-Aware DRAM Refresh to Squeeze the
  Margin of Retention Time in Hybrid Memory Cube},'' in \emph{ICCAD}, 2014.

\bibitem{hashemian2015robust}
M.~S. Hashemian, B.~Singh, F.~Wolff, D.~Weyer, S.~Clay, and C.~Papachristou,
  ``{A Robust Authentication Methodology Using Physically Unclonable Functions
  in DRAM Arrays},'' in \emph{DATE}, 2015.

\bibitem{hassan2019crow}
H.~Hassan, M.~Patel, J.~S. Kim, A.~G. Ya{\u{g}}l{\i}k{\c{c}}{\i},
  N.~Vijaykumar, N.~M. Ghiasi, S.~Ghose, and O.~Mutlu, ``{CROW: A Low-Cost
  Substrate for Improving DRAM Performance, Energy Efficiency, and
  Reliability},'' in \emph{ISCA}, 2019.

\bibitem{hassan2016chargecache}
H.~Hassan, G.~Pekhimenko, N.~Vijaykumar, V.~Seshadri, D.~Lee, O.~Ergin, and
  O.~Mutlu, ``{ChargeCache: Reducing DRAM Latency by Exploiting Row Access
  Locality},'' in \emph{HPCA}, 2016.

\bibitem{hassan2021uncovering}
H.~Hassan, Y.~C. Tugrul, J.~S. Kim, V.~Van~der Veen, K.~Razavi, and O.~Mutlu,
  ``{Uncovering In-DRAM RowHammer Protection Mechanisms: A New Methodology,
  Custom RowHammer Patterns, and Implications},'' in \emph{MICRO}, 2021.

\bibitem{hassan2017softmc}
H.~Hassan, N.~Vijaykumar, S.~Khan, S.~Ghose, K.~Chang, G.~Pekhimenko, D.~Lee,
  O.~Ergin, and O.~Mutlu, ``{SoftMC: A Flexible and Practical Open-Source
  Infrastructure for Enabling Experimental DRAM Studies},'' in \emph{HPCA},
  2017.

\bibitem{he2020newton}
M.~He, C.~Song, I.~Kim, C.~Jeong, S.~Kim, I.~Park, M.~Thottethodi, and
  T.~Vijaykumar, ``{Newton: A DRAM-Maker's Accelerator-In-Memory (AiM)
  Architecture for Machine Learning},'' in \emph{MICRO}, 2020.

\bibitem{henkels19914}
W.~Henkels, D.-S. Wen, R.~Mohler, R.~Franch, T.~Bucelot, C.~Long,
  J.~Bracchitta, W.~Cote, G.~Bronner, Y.~Taur, and R.~Dennard, ``{A 4-Mb
  Low-Temperature DRAM},'' \emph{JSSC}, 1991.

\bibitem{hennessy2011computer}
J.~L. Hennessy and D.~A. Patterson, \emph{{Computer Architecture: A
  Quantitative Approach}}.\hskip 1em plus 0.5em minus 0.4em\relax Elsevier,
  2011.

\bibitem{hestness2014comparative}
J.~Hestness, S.~W. Keckler, and D.~A. Wood, ``{A Comparative Analysis Of
  Microarchitecture Effects On CPU and GPU Memory System Behavior},'' in
  \emph{IISWC}, 2014.

\bibitem{hp2011whitepaper}
{Hewlett-Packard Development Company, L.P.}, ``{Why Buy HP Qualified Memory?}''
  {Hewlett-Packard Development Company, L.P.}, Tech. Rep., 2011.

\bibitem{hidaka1990cache}
H.~Hidaka, Y.~Matsuda, M.~Asakura, and K.~Fujishima, ``{The Cache DRAM
  Architecture: A DRAM with an on-Chip Cache Memory},'' \emph{MICRO}, 1990.

\bibitem{hilgers2014post}
C.~Hilgers, H.~Macht, T.~Müller, and M.~Spreitzenbarth, ``{Post-Mortem Memory
  Analysis of Cold-Booted Android Devices},'' in \emph{IMF}, 2014.

\bibitem{hiraiwa1996statistical}
A.~Hiraiwa, M.~Ogasawara, N.~Natsuaki, Y.~Itoh, and H.~Iwai, ``{Statistical
  Modeling of Dynamic Random Access Memory Data Retention Characteristics},''
  \emph{JAP}, 1996.

\bibitem{hiraiwa1998local}
A.~Hiraiwa, M.~Ogasawara, N.~Natsuaki, Y.~Itoh, and H.~Iwai,
  ``{Local-Field-Enhancement Model of DRAM Retention Failure},'' in
  \emph{IEDM}, 1998.

\bibitem{ho2003method}
M.-J. Ho, ``{Method of Analyzing DRAM Redundancy Repair},'' 2003, {US Patent
  6,573,524}.

\bibitem{hocquenghem1959codes}
A.~Hocquenghem, ``{Codes Correcteurs D'erreurs},'' \emph{Chiffres}, 1959.

\bibitem{hong2010memory}
S.~Hong, ``{Memory Technology Trend and Future Challenges},'' in \emph{IEDM},
  2010.

\bibitem{horiguchi2011nanoscale}
M.~Horiguchi and K.~Itoh, \emph{{Nanoscale Memory Repair}}.\hskip 1em plus
  0.5em minus 0.4em\relax Springer SBM, 2011.

\bibitem{hou2013fpga}
C.-S. Hou, J.-F. Li, C.-Y. Lo, D.-M. Kwai, Y.-F. Chou, and C.-W. Wu, ``{An
  FPGA-Based Test Platform for Analyzing Data Retention Time Distribution of
  DRAMs},'' in \emph{VLSI-DAT}, 2013.

\bibitem{hovinen2011lela}
B.~Hovinen, ``{Getting Started with LELA},''
  \url{http://www.singular.uni-kl.de/lela/tutorial.html}, 2011.

\bibitem{hsieh2016accelerating}
K.~Hsieh, S.~Khan, N.~Vijaykumar, K.~K. Chang, A.~Boroumand, S.~Ghose, and
  O.~Mutlu, ``{Accelerating Pointer Chasing In 3D-Stacked Memory: Challenges,
  Mechanisms, Evaluation},'' in \emph{ICCD}, 2016.

\bibitem{hsu2001intertwin}
L.~L. Hsu, R.~V. Joshi, J.~A. Fifield, and W.~F. Ellis, ``{DRAM Array
  Interchangeable Between Single-cell and Twin-cell Array Operation},'' 2001,
  {US} Patent 6452855B1.

\bibitem{huai2008spin}
Y.~Huai, ``{Spin-Transfer Torque MRAM (STT-MRAM): Challenges and Prospects},''
  \emph{AAPPS bulletin}, 2008.

\bibitem{huang2014moby}
Y.~Huang, Z.~Zha, M.~Chen, and L.~Zhang, ``{Moby: A Mobile Benchmark Suite For
  Architectural Simulators},'' in \emph{ISPASS}, 2014.

\bibitem{huffman2010fundamentals}
W.~C. Huffman and V.~Pless, \emph{{Fundamentals of Error-Correcting
  Codes}}.\hskip 1em plus 0.5em minus 0.4em\relax {Cambridge University Press},
  2003.

\bibitem{hwang2012cosmic}
A.~A. Hwang, I.~A. Stefanovici, and B.~Schroeder, ``{Cosmic Rays Don't Strike
  Twice: Understanding the Nature of DRAM Errors and the Implications for
  System Design},'' in \emph{ASPLOS}, 2012.

\bibitem{ieee2010iso}
IEEE, ``{ISO/IEC/IEEE 24765: 2010 Systems and Software
  Engineering-Vocabulary},'' \emph{ISO/IEC/IEEE}, vol. 24765, 2010.

\bibitem{infineon2022radiation}
{Infineon}, ``{Radiation Hardened \& High Reliability Memories},'' 2022.

\bibitem{iniewski2011nano}
K.~Iniewski, \emph{{Nano-Semiconductors: Devices and Technology}}.\hskip 1em
  plus 0.5em minus 0.4em\relax {CRC Press}, 2011.

\bibitem{cost1997yield}
{Integrated Circuit Engineering Corporation}, \emph{{Cost Effective IC
  Manufacturing}}, 1997.

\bibitem{intel1998pc}
{Intel}, ``{PC SDRAM UNBUFFERED DIMM SPECIFICATION},'' 1998.

\bibitem{intel1975data}
{Intel Corporation}, ``{Intel Data Catalog},'' 1975.

\bibitem{intel2020xeongold5118}
{Intel Corporation}, ``{Intel Xeon Gold 5118 Processor},'' 2020,
  \url{https://ark.intel.com/content/www/us/en/ark/products/120473/intel-xeon-gold-5118-processor-16-5m-cache-2-30-ghz.html}.

\bibitem{intel2020platform}
{Intel Corporation}, ``{Platform Memory Validation},''
  \url{https://www.intel.com/content/www/us/en/platform-memory/platform-memory.html},
  2020.

\bibitem{im2016im}
{Intelligent Memory}, ``{IM ECC DRAM with Integrated Error Correcting Code},''
  2016, {Product Brief}.

\bibitem{im2020im}
{Intelligent Memory}, ``{I'M ECC DRAM with Integrated Error Correcting Code},''
  2020, product Brief.

\bibitem{ipek2010dynamically}
E.~Ipek, J.~Condit, E.~B. Nightingale, D.~Burger, and T.~Moscibroda,
  ``{Dynamically Replicated Memory: Building Reliable Systems from Nanoscale
  Resistive Memories},'' in \emph{ASPLOS}, 2010.

\bibitem{isaac2008remarkable}
R.~Isaac, ``{The Remarkable Story of the DRAM Industry},'' \emph{IEEE SSCS
  News}, 2008.

\bibitem{ishigaki2010multi}
T.~Ishigaki, T.~Kawahara, R.~Takemura, K.~Ono, K.~Ito, H.~Matsuoka, and
  H.~Ohno, ``{A Multi-Level-Cell Spin-Transfer Torque Memory with
  Series-Stacked Magnetotunnel Junctions},'' in \emph{VLSI}, 2010.

\bibitem{issi2022ddr4}
{ISSI}, ``{DDR4 SDRAM},''
  \url{https://www.issi.com/US/product-dram-ddr4.shtml}.

\bibitem{nanya2022NT5AD256M16E4}
{ISSI}, ``{NT5AD256M16E4-JR},''
  \url{https://www.nanya.com/en/Product/4596/NT5AD256M16E4-JR}.

\bibitem{issi2018ddr3}
\emph{{128Mx8, 64Mx16 1Gb DDR3 SDRAM with ECC}}, {ISSI}, 2018, rev. B1.

\bibitem{issi2020lpddr4}
{ISSI}, ``{8Gb (x16 x 2 Channel) Mobile LPDDR4/LPDDR4X},'' 2020.

\bibitem{itrs}
ITRS, ``{International Technology Roadmap for Semiconductors Executive
  Summary},'' 2013, http://www.itrs2.net/2013-itrs.html.

\bibitem{itrs2015more}
ITRS, ``{More Moore},'' 2015, \url{www.itrs2.net}.

\bibitem{jacob2010memory}
B.~Jacob, S.~Ng, and D.~Wang, \emph{{Memory Systems: Cache, DRAM, Disk}}.\hskip
  1em plus 0.5em minus 0.4em\relax Morgan Kaufmann, 2010.

\bibitem{jafri2020refresh}
S.~M. Jafri, H.~Hassan, A.~Hemani, and O.~Mutlu, ``{Refresh Triggered
  Computation: Improving the Energy Efficiency of Convolutional Neural Network
  Accelerators},'' \emph{TACO}, 2020.

\bibitem{james2010silicon}
D.~James, ``{Silicon Chip Teardown to the Atomic Scale--Challenges Facing the
  Reverse Engineering of Semiconductors},'' \emph{Microscopy and
  Microanalysis}, 2010.

\bibitem{jattke2022blacksmith}
P.~Jattke, V.~van~der Veen, P.~Frigo, S.~Gunter, and K.~Razavi, ``{Blacksmith:
  Scalable Rowhammering in the Frequency Domain},'' in \emph{SP}, 2022.

\bibitem{jedec2021jc42}
JEDEC, ``{JC-42 Solid State Memories},''
  \url{https://www.jedec.org/committees/jc-42}.

\bibitem{jedec1994synchronous}
{JEDEC}, ``{Synchronous Dynamic Random Access Memory (SDRAM)},'' 1994.

\bibitem{jedec2008ddr3}
JEDEC, \emph{{DDR3 SDRAM Specification}}, 2008.

\bibitem{jedec2010ssdrequirements}
JEDEC, \emph{{JESD218: Solid-State Drive (SSD) Requirements and Endurance Test
  Method}}, 2010.

\bibitem{jedec2010ssdendurance}
JEDEC, \emph{{JESD219: Solid-State Drive (SSD) Endurance Workloads}}, 2010.

\bibitem{jedec2012ddr4}
JEDEC, \emph{{DDR4 SDRAM Specification}}, 2012.

\bibitem{jedec2014lpddr4}
JEDEC, ``{Low Power Double Data Rate 4 (LPDDR4) SDRAM Specification},''
  \emph{JEDEC Standard JESD209--4B}, 2014.

\bibitem{jedec2016gddr5}
JEDEC, ``{Graphics Double Data Rate (GDDR5) SGRAM Standard},'' \emph{JEDEC
  Standard JESD212C}, 2016.

\bibitem{jedec2016failure}
JEDEC, \emph{{JEP122H: Failure Mechanisms and Models for Semiconductor
  Devices}}, 2016.

\bibitem{jedec2020ddr5}
JEDEC, \emph{{DDR5 SDRAM Specification}}, 2020.

\bibitem{jedec2020lpddr5}
JEDEC, ``{Low Power Double Data Rate 5 (LPDDR5) SDRAM Specification},''
  \emph{JEDEC Standard JESD209--5A}, 2020.

\bibitem{jedec2016gddr6}
JEDEC, ``{Graphics Double Data Rate (GDDR6) SGRAM Standard},'' \emph{JEDEC
  Standard JESD250C}, 2021.

\bibitem{jedec2021high}
JEDEC, ``{High Bandwidth Memory (HBM) DRAM},'' \emph{JEDEC Standard JESD235D},
  2021.

\bibitem{jedec2022high}
JEDEC, ``{High Bandwidth Memory DRAM (HBM3)},'' \emph{JEDEC Standard JESD238},
  2022.

\bibitem{jeong2020pair}
S.~Jeong, S.~Kang, and J.-S. Yang, ``{PAIR: Pin-aligned In-DRAM ECC
  architecture using expandability of Reed-Solomon code},'' in \emph{DAC},
  2020.

\bibitem{jha2003testing}
N.~K. Jha and S.~Gupta, \emph{{Testing of Digital Systems}}.\hskip 1em plus
  0.5em minus 0.4em\relax {Cambridge University Press}, 2003.

\bibitem{jian2013low}
X.~Jian, H.~Duwe, J.~Sartori, V.~Sridharan, and R.~Kumar, ``{Low-Power,
  Low-Storage-Overhead Chipkill Correct via Multi-Line Error Correction},'' in
  \emph{SC}, 2013.

\bibitem{jian2013adaptive}
X.~Jian and R.~Kumar, ``{Adaptive Reliability Chipkill Correct (ARCC)},'' in
  \emph{HPCA}, 2013.

\bibitem{jiang2021trrscope}
Y.~Jiang, H.~Zhu, H.~Shan, X.~Guo, X.~Zhang, and Y.~Jin, ``{TRRScope:
  Understanding Target Row Refresh Mechanism for Modern DDR Protection},'' in
  \emph{HOST}, 2021.

\bibitem{jiang2021quantifying}
Y.~Jiang, H.~Zhu, D.~Sullivan, X.~Guo, X.~Zhang, and Y.~Jin, ``{Quantifying
  Rowhammer Vulnerability for DRAM Security},'' in \emph{DAC}, 2021.

\bibitem{jin2005prediction}
S.~Jin, J.-H. Yi, J.~H. Choi, D.~G. Kang, Y.~J. Park, and H.~S. Min,
  ``{Prediction of Data Retention Time Distribution of DRAM by Physics-Based
  Statistical Simulation},'' \emph{TED}, 2005.

\bibitem{jin2004modeling}
S.~Jin, J.-H. Yi, Y.~J. Park, H.~S. Min, J.~H. Choi, and D.~G. Kang,
  ``{Modeling of Retention Time Distribution of DRAM Cell Using a Monte-Carlo
  Method},'' in \emph{IEDM}, 2004.

\bibitem{jung2017platform}
M.~Jung, D.~M. Mathew, C.~C. Rheinl{\"a}nder, C.~Weis, and N.~Wehn, ``{A
  Platform to Analyze DDR3 DRAM's Power and Retention Time},'' \emph{IEEE
  Design \& Test}, 2017.

\bibitem{jung2016reverse}
M.~Jung, C.~C. Rheinl{\"a}nder, C.~Weis, and N.~Wehn, ``{Reverse Engineering of
  DRAMs: Row Hammer with Crosshair},'' in \emph{MEMSYS}, 2016.

\bibitem{jung2014optimized}
M.~Jung, C.~Weis, N.~Wehn, M.~Sadri, and L.~Benini, ``{Optimized Active and
  Power-Down Mode Refresh Control in 3D-DRAMs},'' in \emph{VLSI-SoC}, 2014.

\bibitem{jung2015omitting}
M.~Jung, {\'E}.~Zulian, D.~M. Mathew, M.~Herrmann, C.~Brugger, C.~Weis, and
  N.~Wehn, ``{Omitting Refresh: A Case Study for Commodity and Wide I/O
  DRAMs},'' in \emph{MEMSYS}, 2015.

\bibitem{kalter199050}
H.~L. Kalter, C.~H. Stapper, J.~E. Barth, J.~DiLorenzo, C.~E. Drake, J.~A.
  Fifield, G.~Kelley, S.~C. Lewis, W.~B. Van Der~Hoeven, and J.~A. Yankosky,
  ``{A 50-Ns 16-Mb DRAM with a 10-Ns Data Rate and On-Chip ECC},'' \emph{JSSC},
  1990.

\bibitem{kanellopoulos2019smash}
K.~Kanellopoulos, N.~Vijaykumar, C.~Giannoula, R.~Azizi, S.~Koppula, N.~M.
  Ghiasi, T.~Shahroodi, J.~G. Luna, and O.~Mutlu, ``{SMASH: Co-Designing
  Software Compression and Hardware-Accelerated Indexing for Efficient Sparse
  Matrix Operations},'' in \emph{MICRO}, 2019.

\bibitem{kanev2015profiling}
S.~Kanev, J.~P. Darago, K.~Hazelwood, P.~Ranganathan, T.~Moseley, G.-Y. Wei,
  and D.~Brooks, ``{Profiling a Warehouse-scale Computer},'' in \emph{ISCA},
  2015.

\bibitem{kang2020cattwo}
I.~Kang, E.~Lee, and J.~H. Ahn, ``{CAT-TWO: Counter-Based Adaptive Tree, Time
  Window Optimized for {DRAM} Row-Hammer Prevention},'' \emph{{IEEE} Access},
  2020.

\bibitem{kang2010study}
J.~Kang, ``{A Study of the DRAM Industry},'' Master's thesis, Massachusetts
  Institute of Technology, 2010.

\bibitem{kang2006a}
S.~Kang, W.~Y. Cho, B.-H. Cho, K.-J. Lee, C.-S. Lee, H.-R. Oh, B.-G. Choi,
  Q.~Wang, H.-J. Kim, M.-H. Park, Y.~H. Ro, S.~Kim, C.-D. Ha, K.-S. Kim, Y.-R.
  Kim, D.-E. Kim, C.-K. Kwak, H.-G. Byun, G.~Jeong, H.~Jeong, K.~Kim, and
  Y.~Shin, ``{A 0.1-um 1.8-V 256-Mb Phase-Change Random Access Memory (PRAM)
  With 66-MHz Synchronous Burst-Read Operation},'' \emph{JSSC}, 2006.

\bibitem{kang2014co}
U.~Kang, H.-s. Yu, C.~Park, H.~Zheng, J.~Halbert, K.~Bains, S.~Jang, and J.~S.
  Choi, ``{Co-Architecting Controllers and DRAM to Enhance DRAM Process
  Scaling},'' in \emph{The Memory Forum}, 2014.

\bibitem{kang1999flexram}
Y.~Kang, W.~Huang, S.-M. Yoo, D.~Keen, Z.~Ge, V.~Lam, P.~Pattnaik, and
  J.~Torrellas, ``{FlexRAM: Toward an Advanced Intelligent Memory System},'' in
  \emph{ICCD}, 1999.

\bibitem{katayama1999fault}
Y.~Katayama, E.~J. Stuckey, S.~Morioka, and Z.~Wu, ``{Fault-Tolerant Refresh
  Power Reduction of DRAMs for Quasi-Nonvolatile Data Retention},'' in
  \emph{EFT}, 1999.

\bibitem{keeth2007dram}
B.~Keeth, R.~J. Baker, B.~Johnson, and F.~Lin, \emph{{DRAM Circuit Design:
  Fundamental and High-Speed Topics}}.\hskip 1em plus 0.5em minus 0.4em\relax
  {John Wiley \& Sons}, 2007.

\bibitem{keller2014dynamic}
C.~Keller, F.~Gurkaynak, H.~Kaeslin, and N.~Felber, ``{Dynamic Memory-Based
  Physically Unclonable Function for the Generation of Unique Identifiers and
  True Random Numbers},'' in \emph{ISCAS}, 2014.

\bibitem{khan2018test}
M.~N.~I. Khan and S.~Ghosh, ``{Test Challenges and Solutions for Emerging
  Non-Volatile Memories},'' in \emph{VTS}.\hskip 1em plus 0.5em minus
  0.4em\relax IEEE, 2018.

\bibitem{khan2014efficacy}
S.~Khan, D.~Lee, Y.~Kim, A.~R. Alameldeen, C.~Wilkerson, and O.~Mutlu, ``{The
  Efficacy of Error Mitigation Techniques for DRAM Retention Failures: A
  Comparative Experimental Study},'' in \emph{SIGMETRICS}, 2014.

\bibitem{khan2016parbor}
S.~Khan, D.~Lee, and O.~Mutlu, ``{PARBOR: An Efficient System-Level Technique
  to Detect Data-Dependent Failures in DRAM},'' in \emph{DSN}, 2016.

\bibitem{khan2016case}
S.~Khan, C.~Wilkerson, D.~Lee, A.~R. Alameldeen, and O.~Mutlu, ``{A Case for
  Memory Content-Based Detection and Mitigation of Data-Dependent Failures in
  DRAM},'' in \emph{IEEE CAL}, 2016.

\bibitem{khan2017detecting}
S.~Khan, C.~Wilkerson, Z.~Wang, A.~R. Alameldeen, D.~Lee, and O.~Mutlu,
  ``{Detecting and Mitigating Data-Dependent DRAM Failures by Exploiting
  Current Memory Content},'' in \emph{MICRO}, 2017.

\bibitem{kim2015avert}
D.-H. Kim, S.~Cha, and L.~S. Milor, ``{AVERT: An Elaborate Model for Simulating
  Variable Retention Time in DRAMs},'' \emph{Microelectronics Reliability},
  2015.

\bibitem{kim2016ecc}
D.-H. Kim and L.~S. Milor, ``{ECC-ASPIRIN: An ECC-assisted Post-Package Repair
  Scheme for Aging Errors in DRAMs},'' in \emph{VTS}, 2016.

\bibitem{kim2014architectural}
D.-H. Kim, P.~J. Nair, and M.~K. Qureshi, ``{Architectural Support for
  Mitigating Row Hammering in DRAM Memories},'' \emph{CAL}, 2014.

\bibitem{kim2016relaxfault}
D.~W. Kim and M.~Erez, ``{RelaxFault Memory Repair},'' in \emph{ISCA}, 2016.

\bibitem{kim2011characterization}
H.~Kim, B.~Oh, Y.~Son, K.~Kim, S.-Y. Cha, J.-G. Jeong, S.-J. Hong, and H.~Shin,
  ``{Characterization of the Variable Retention Time in Dynamic Random Access
  Memory},'' \emph{TED}, 2011.

\bibitem{kim2011study}
H.~Kim, B.~Oh, Y.~Son, K.~Kim, S.-Y. Cha, J.-G. Jeong, S.-J. Hong, and H.~Shin,
  ``{Study of Trap Models Related to the Variable Retention Time Phenomenon in
  DRAM},'' \emph{TED}, 2011.

\bibitem{kim2010high}
I.~Kim, S.~Cho, D.~Im, E.~Cho, D.~Kim, G.~Oh, D.~Ahn, S.~Park, S.~Nam, J.~Moon,
  and C.~Chung, ``{High Performance PRAM Cell Scalable to Sub-20nm Technology
  with Below 4F2 Cell Size, Extendable to DRAM Applications},'' in
  \emph{VLSIT}, 2010.

\bibitem{kim2020improving}
J.~S. Kim, ``{Improving DRAM Performance, Security, and Reliability by
  Understanding and Exploiting DRAM Timing Parameter Margins},'' Ph.D.
  dissertation, Carnegie Mellon University, 2020.

\bibitem{kim2018solar}
J.~S. Kim, M.~Patel, H.~Hassan, and O.~Mutlu, ``{Solar-DRAM: Reducing DRAM
  Access Latency by Exploiting the Variation in Local Bitlines},'' in
  \emph{ICCD}, 2018.

\bibitem{kim2018dram}
J.~S. Kim, M.~Patel, H.~Hassan, and O.~Mutlu, ``{The DRAM Latency PUF: Quickly
  Evaluating Physical Unclonable Functions by Exploiting the
  Latency-Reliability Tradeoff in Modern Commodity DRAM Devices},'' in
  \emph{HPCA}, 2018.

\bibitem{kim2019d}
J.~S. Kim, M.~Patel, H.~Hassan, L.~Orosa, and O.~Mutlu, ``{D-RaNGe: Using
  Commodity DRAM Devices to Generate True Random Numbers With Low Latency And
  High Throughput},'' in \emph{HPCA}, 2019.

\bibitem{kim2020revisiting}
J.~S. Kim, M.~Patel, A.~G. Ya{\u{g}}l{\i}k{\c{c}}{\i}, H.~Hassan, R.~Azizi,
  L.~Orosa, and O.~Mutlu, ``{Revisiting RowHammer: An Experimental Analysis of
  Modern Devices and Mitigation Techniques},'' in \emph{ISCA}, 2020.

\bibitem{kim2000dynamic}
J.~Kim and M.~C. Papaefthymiou, ``{Dynamic Memory Design for Low Data-Retention
  Power},'' in \emph{PATMOS}, 2000.

\bibitem{kim2001block}
J.~Kim and M.~C. Papaefthymiou, ``{Block-Based Multi-Period Refresh For Energy
  Efficient Dynamic Memory},'' in \emph{IEEE International ASIC/SOC
  Conference}, 2001.

\bibitem{kim2003block}
J.~Kim and M.~C. Papaefthymiou, ``{Block-Based Multiperiod Dynamic Memory
  Design for Low Data-Retention Power},'' in \emph{TVLSI}, 2003.

\bibitem{kim2015bamboo}
J.~Kim, M.~Sullivan, and M.~Erez, ``{Bamboo ECC: Strong, Safe, and Flexible
  Codes For Reliable Computer Memory},'' in \emph{HPCA}, 2015.

\bibitem{kim2015frugal}
J.~Kim, M.~Sullivan, S.-L. Gong, and M.~Erez, ``{Frugal ECC: Efficient And
  Versatile Memory Error Protection Through Fine-Grained Compression},'' in
  \emph{SC}, 2015.

\bibitem{kim2016all}
J.~Kim, M.~Sullivan, S.~Lym, and M.~Erez, ``{All-Inclusive ECC: Thorough
  End-to-End Protection for Reliable Computer Memory},'' in \emph{ISCA}, 2016.

\bibitem{kim1998dram}
K.~Kim, C.-G. Hwang, and J.~G. Lee, ``{DRAM Technology Perspective for Gigabit
  Era},'' \emph{TED}, 1998.

\bibitem{kim2009new}
K.~Kim and J.~Lee, ``{A New Investigation of Data Retention Time in Truly
  Nanoscaled DRAMs},'' in \emph{EDL}, 2009.

\bibitem{kim2005reliability}
K.~Kim and S.~J. Ahn, ``{Reliability Investigations for Manufacturable High
  Density PRAM},'' in \emph{IRPS}, 2005.

\bibitem{kim2021mithril}
M.~J. Kim, J.~Park, Y.~Park, W.~Doh, N.~Kim, T.~J. Ham, J.~W. Lee, and J.~H.
  Ahn, ``{Mithril: Cooperative Row Hammer Protection on Commodity DRAM
  Leveraging Managed Refresh},'' \emph{arXiv:2108.06703}, 2021.

\bibitem{kim2020evanesco}
M.~Kim, J.~Park, G.~Cho, Y.~Kim, L.~Orosa, O.~Mutlu, and J.~Kim, ``{Evanesco:
  Architectural Support for Efficient Data Sanitization in Modern Flash-Based
  Storage Systems},'' in \emph{ASPLOS}, 2020.

\bibitem{kim2007low}
S.-H. Kim, W.-O. Lee, J.-H. Kim, S.-S. Lee, S.-Y. Hwang, C.-I. Kim, T.-W. Kwon,
  B.-S. Han, S.-K. Cho, D.-H. Kim, J.-K. Hong, M.-Y. Lee, S.-W. Yin, H.-G. Kim,
  J.-H. Ahn, Y.-T. Kim, Y.-H. Koh, and J.-S. Kih, ``{A Low Power and Highly
  Reliable 400Mbps Mobile DDR SDRAM With On-Chip Distributed ECC},'' in
  \emph{ASSCC}, 2007.

\bibitem{kim2020charge}
S.~Kim, W.~Kwak, C.~Kim, D.~Baek, and J.~Huh, ``{Charge-Aware DRAM Refresh
  Reduction with Value Transformation},'' in \emph{HPCA}, 2020.

\bibitem{kim2015architectural}
Y.~Kim, ``{Architectural Techniques to Enhance DRAM Scaling},'' Ph.D.
  dissertation, Carnegie Mellon University, 2015.

\bibitem{kim2014flipping}
Y.~Kim, R.~Daly, J.~Kim, C.~Fallin, J.~H. Lee, D.~Lee, C.~Wilkerson, K.~Lai,
  and O.~Mutlu, ``{Flipping Bits in Memory Without Accessing Them: An
  Experimental Study of DRAM Disturbance Errors},'' in \emph{ISCA}, 2014.

\bibitem{kim2010atlas}
Y.~Kim, D.~Han, O.~Mutlu, and M.~Harchol-Balter, ``{ATLAS: A Scalable and
  High-Performance Scheduling Algorithm for Multiple Memory Controllers},'' in
  \emph{HPCA}, 2010.

\bibitem{kim2010thread}
Y.~Kim, M.~Papamichael, O.~Mutlu, and M.~Harchol-Balter, ``{Thread Cluster
  Memory Scheduling: Exploiting Differences in Memory Access Behavior},'' in
  \emph{MICRO}, 2010.

\bibitem{kim2012case}
Y.~Kim, V.~Seshadri, D.~Lee, J.~Liu, and O.~Mutlu, ``{A Case for Exploiting
  Subarray-Level Parallelism (SALP) in DRAM},'' in \emph{ISCA}, 2012.

\bibitem{kim2016ramulator}
Y.~Kim, W.~Yang, and O.~Mutlu, ``{Ramulator: A Fast and Extensible DRAM
  Simulator},'' in \emph{IEEE CAL}, 2016.

\bibitem{kim1999assessing}
Y.-B. Kim and T.~W. Chen, ``{Assessing Merged DRAM/Logic Technology},''
  \emph{Integration}, 1999.

\bibitem{kingston2012whitepaper}
{Kingston Technology Corporation}, ``{Kingston Testing Overview},'' Tech. Rep.,
  2012.

\bibitem{kingston2021kingston}
{Kingston Technology Corporation}, ``{Kingston's Memory Module Testing is Our
  Key to Quality Control},''
  \url{https://www.kingston.com/us/company/testing-process}, 2021.

\bibitem{kline2017sustainable}
D.~Kline, R.~Melhem, and A.~K. Jones, ``{Sustainable Fault Management and Error
  Correction for Next-Generation Main Memories},'' in \emph{IGSC}, 2017.

\bibitem{kline2020flower}
D.~Kline, J.~Zhang, R.~Melhem, and A.~K. Jones, ``{Flower and Fame: A Low
  Overhead Bit-Level Fault-Map and Fault-Tolerance Approach for Deeply Scaled
  Memories},'' in \emph{HPCA}, 2020.

\bibitem{kobayashi2017highly}
K.~Kobayashi, ``{Highly-reliable Integrated Circuits for Ground and Space
  Applications},'' \emph{ASICON}, 2017.

\bibitem{kong2008analysis}
W.~Kong, P.~C. Parries, G.~Wang, and S.~S. Iyer, ``{Analysis of Retention Time
  Distribution of Embedded DRAM-A New Method to Characterize Across-Chip
  Threshold Voltage Variation},'' in \emph{ITC}, 2008.

\bibitem{konishi1989analysis}
Y.~Konishi, M.~Kumanoya, H.~Yamasaki, K.~Dosaka, and T.~Yoshihara, ``{Analysis
  of Coupling Noise Between Adjacent Bit Lines in Megabit DRAMs},''
  \emph{JSSC}, 1989.

\bibitem{konoth2018zebram}
R.~K. Konoth, M.~Oliverio, A.~Tatar, D.~Andriesse, H.~Bos, C.~Giuffrida, and
  K.~Razavi, ``{ZebRAM: Comprehensive and Compatible Software Protection
  Against Rowhammer Attacks},'' in \emph{OSDI}, 2018.

\bibitem{koppula2019eden}
S.~Koppula, L.~Orosa, A.~G. Ya{\u{g}}l{\i}k{\c{c}}{\i}, R.~Azizi, T.~Shahroodi,
  K.~Kanellopoulos, and O.~Mutlu, ``{EDEN: Enabling Energy-Efficient,
  High-Performance Deep Neural Network Inference Using Approximate DRAM},'' in
  \emph{MICRO}, 2019.

\bibitem{kraft2018improving}
K.~Kraft, C.~Sudarshan, D.~M. Mathew, C.~Weis, N.~Wehn, and M.~Jung,
  ``{Improving the Error Behavior of DRAM by Exploiting its Z-Channel
  Property},'' in \emph{DATE}, 2018.

\bibitem{kultursay2013evaluating}
E.~K{\"u}lt{\"u}rsay, M.~Kandemir, A.~Sivasubramaniam, and O.~Mutlu,
  ``{Evaluating STT-RAM as an Energy-Efficient Main Memory Alternative},'' in
  \emph{ISPASS}, 2013.

\bibitem{kumar2014detection}
N.~Kumar, ``{Detection of Variable Retention Time in DRAM},'' Master's thesis,
  Portland State University, Portland, Oregon, 2014.

\bibitem{kwak2017a}
N.~Kwak, S.-H. Kim, K.~H. Lee, C.-K. Baek, M.~S. Jang, Y.~Joo, S.-H. Lee, W.~Y.
  Lee, E.~Lee, D.~Han, J.~Kang, J.~H. Lim, J.-B. Park, K.-T. Kim, S.~Cho, S.~W.
  Han, J.~Y. Keh, J.~H. Chun, J.~Oh, and S.~H. Lee, ``{A 4.8 Gb/s/pin 2Gb
  LPDDR4 SDRAM with Sub-100$\mu$A Self-Refresh Current for IoT Applications},''
  in \emph{ISSCC}, 2017.

\bibitem{kwon2017an}
H.-J. Kwon, E.~Seo, C.-Y. Lee, Y.-H. Seo, G.-H. Han, H.-R. Kim, J.-H. Lee,
  M.-S. Jang, S.-G. Do, S.-H. Cho, J.-K. Park, S.-Y. Doo, J.-B. Shin, S.-H.
  Jung, H.-J. Kim, I.-H. Im, B.-R. Cho, J.-W. Lee, J.-Y. Lee, K.-H. Yu, H.-K.
  Kim, C.-H. Jeon, H.-S. Park, S.-S. Kim, S.-H. Lee, J.-W. Park, S.-S. Lee,
  B.-T. Lim, J.-y. Park, Y.-S. Park, H.-J. Kwon, S.-J. Bae, J.-H. Choi, K.-I.
  Park, S.-J. Jang, and G.-Y. Jin, ``{An Extremely Low-Standby-Power 3.733
  Gb/s/pin 2Gb LPDDR4 SDRAM for Wearable Devices},'' in \emph{ISSCC}, 2017.

\bibitem{kwon2014understanding}
S.~Kwon, Y.~H. Son, and J.~H. Ahn, ``{Understanding DDR4 in Pursuit of In-DRAM
  ECC},'' in \emph{ISOCC}, 2014.

\bibitem{kwon202125}
Y.-C. Kwon, S.~H. Lee, J.~Lee, S.-H. Kwon, J.~M. Ryu, J.-P. Son, O.~Seongil,
  H.-S. Yu, H.~Lee, S.~Y. Kim, Y.~Cho, J.~G. Kim, J.~Choi, H.-S. Shin, J.~Kim,
  B.~Phuah, H.~Kim, M.~J. Song, A.~Choi, D.~Kim, S.~Kim, E.-B. Kim, D.~Wang,
  S.~Kang, Y.~Ro, S.~Seo, J.~Song, J.~Youn, K.~Sohn, and N.~S. Kim, ``{25.4 A
  20nm 6GB Function-In-Memory DRAM, Based on HBM2 with a 1.2 TFLOPS
  Programmable Computing Unit Using Bank-Level Parallelism, for Machine
  Learning Applications},'' in \emph{ISSCC}, 2021.

\bibitem{kim2006an}
{Kyu-hyoun Kim}, {Uksong Kang}, {Hoe-Ju Chung}, {Duk-Ha Park}, {Woo-Seop Kim},
  {Young-Chan Jang}, {Moonsook Park}, {Hoon Lee}, {Jin-Young Kim}, {Jung
  Sunwoo}, {Hwan-Wook Park}, {Hyun-Kyung Kim}, {Su-Jin Chung}, {Jae-Kwan Kim},
  {Hyung-Seuk Kim}, {Kee-Won Kwon}, {Young-Taek Lee}, {Joo Sun Choi}, and
  {Changhyun Kim}, ``{An 8Gb/s/pin 9.6ns Row-Cycle 288Mb Deca-Data Rate SDRAM
  with an I/O Error-Detection Scheme},'' in \emph{ISSCC}, 2006.

\bibitem{ladbury2013use}
R.~Ladbury, M.~Berg, E.~Wilcox, K.~LaBel, H.~Kim, A.~Phan, and C.~Seidleck,
  ``{Use of Commercial FPGA-Based Evaluation Boards for Single-Event Testing of
  DDR2 and DDR3 SDRAMS},'' \emph{IEEE Trans. Nucl. Sci.}, 2013.

\bibitem{lan2010study}
Z.~Lan, J.~Gu, Z.~Zheng, R.~Thakur, and S.~Coghlan, ``{A Study of Dynamic
  Meta-Learning for Failure Prediction in Large-Scale Systems},'' \emph{PDC},
  2010.

\bibitem{lanteigne2016how}
M.~Lanteigne, ``{How Rowhammer Could Be Used to Exploit Weaknesses in Computer
  Hardware},'' Tech. Rep., 2016.

\bibitem{laudon2006ultrasparc}
J.~Laudon, ``{UltraSPARC T1: Architecture and Physical Design of a 32-threaded
  General Purpose CPU},'' in \emph{ISSCC}, 2006.

\bibitem{lee2009architecting}
B.~C. Lee, E.~Ipek, O.~Mutlu, and D.~Burger, ``{Architecting Phase Change
  Memory as a Scalable DRAM Alternative},'' in \emph{ISCA}, 2009.

\bibitem{lee2010phasecacm}
B.~C. Lee, E.~Ipek, O.~Mutlu, and D.~Burger, ``{Phase Change Memory
  Architecture and the Quest for Scalability},'' \emph{CACM}, 2010.

\bibitem{lee2010phase}
B.~C. Lee, P.~Zhou, J.~Yang, Y.~Zhang, B.~Zhao, E.~Ipek, O.~Mutlu, and
  D.~Burger, ``{Phase-Change Technology and the Future of Main Memory},''
  \emph{IEEE Micro}, 2010.

\bibitem{lee2012dram}
C.-Y. Lee, C.-S. Lai, C.-M. Yang, and D.~H.-L. Wang, ``{DRAM Data Retention And
  Cell Transistor Threshold Voltage Reliability Improved By Passivation
  Annealing Prior To The Deposition Of Plasma Nitride Layer},'' \emph{TDMR},
  2012.

\bibitem{lee2015adaptive}
D.~Lee, Y.~Kim, G.~Pekhimenko, S.~Khan, V.~Seshadri, K.~Chang, and O.~Mutlu,
  ``{Adaptive-Latency DRAM: Optimizing DRAM Timing for the Common-Case},'' in
  \emph{HPCA}, 2015.

\bibitem{lee2016reducing}
D.~Lee, ``{Reducing DRAM Latency at Low Cost by Exploiting Heterogeneity},''
  Ph.D. dissertation, Carnegie Mellon University, 2016.

\bibitem{lee2016simultaneous}
D.~Lee, S.~Ghose, G.~Pekhimenko, S.~Khan, and O.~Mutlu, ``{Simultaneous
  Multi-Layer Access: Improving 3D-Stacked Memory Bandwidth at Low Cost},''
  \emph{TACO}, 2016.

\bibitem{lee2017design}
D.~Lee, S.~Khan, L.~Subramanian, S.~Ghose, R.~Ausavarungnirun, G.~Pekhimenko,
  V.~Seshadri, and O.~Mutlu, ``{Design-Induced Latency Variation in Modern DRAM
  Chips: Characterization, Analysis, and Latency Reduction Mechanisms},'' in
  \emph{SIGMETRICS}, 2017.

\bibitem{lee2013tiered}
D.~Lee, Y.~Kim, V.~Seshadri, J.~Liu, L.~Subramanian, and O.~Mutlu,
  ``{Tiered-Latency DRAM: A Low Latency and Low Cost DRAM Architecture},'' in
  \emph{HPCA}, 2013.

\bibitem{lee2019twice}
E.~Lee, I.~Kang, S.~Lee, G.~{Edward Suh}, and J.~{Ho Ahn}, ``{TWiCe: Preventing
  Row-Hammering by Exploiting Time Window Counters},'' in \emph{ISCA}, 2019.

\bibitem{lee2011correcting}
H.~T. Lee, H.~Kim, Y.-J. Baek, and J.~H. Cheon, ``{Correcting Errors in Private
  Keys Obtained from Cold Boot Attacks},'' in \emph{ICISC}, 2011.

\bibitem{lee2014green}
J.~Lee, ``{Green Memory Solution},'' {Investor’s Forum}, {Samsung
  Electronics}, 2014.

\bibitem{lee2001new}
J.-G. Lee, Y.-H. Jun, K.-H. Kyung, C.~Yoo, Y.-H. Cho, and S.-I. Cho, ``{A New
  Column Redundancy Scheme For Yield Improvement Of High Speed DRAMs With
  Multiple Bit Pre-Fetch Structure},'' in \emph{VLSI}, 2001.

\bibitem{lee2013strategic}
K.~H. Lee, ``{A Strategic Analysis of the DRAM Industry After the Year 2000},''
  Master's thesis, Massachusetts Institute of Technology, 2013.

\bibitem{lee2019evaluation}
K.~Lee, C.-H. Yun, H.~Seo, T.~Kang, Y.~Lee, and K.~Cho, ``{An Evaluation of
  X-Ray Irradiation Induced Dynamic Refresh Characterization in DRAM},'' in
  \emph{IRPS}, 2019.

\bibitem{lee2010mechanism}
M.~J. Lee and K.~W. Park, ``{A Mechanism for Dependence of Refresh Time on Data
  Pattern in DRAM},'' in \emph{EDL}, 2010.

\bibitem{lee2016technology}
S.-H. Lee, ``{Technology Scaling Challenges and Opportunities of Memory
  Devices},'' in \emph{IEDM}, 2016.

\bibitem{lee20221ynm}
S.~Lee, K.~Kim, S.~Oh, J.~Park, G.~Hong, D.~Ka, K.~Hwang, J.~Park, K.~Kang,
  J.~Kim \emph{et~al.}, ``{A 1ynm 1.25 V 8Gb, 16Gb/s/pin GDDR6-based
  Accelerator-in-Memory supporting 1TFLOPS MAC Operation and Various Activation
  Functions for Deep-Learning Applications},'' in \emph{ISSCC}, 2022.

\bibitem{lee2019exploiting}
S.~Lee, N.~S. Kim, and D.~Kim, ``{Exploiting OS-Level Memory Offlining for DRAM
  Power Management},'' \emph{CAL}, 2019.

\bibitem{lee2009study}
S.~Lee, J.-h. Jeong, T.~S. Lee, W.~M. Kim, and B.-k. Cheong, ``{A Study on the
  Failure Mechanism of a Phase-Change Memory in Write/Erase Cycling},''
  \emph{EDL}, 2009.

\bibitem{levy2018lessons}
S.~Levy, K.~B. Ferreira, N.~DeBardeleben, T.~Siddiqua, V.~Sridharan, and
  E.~Baseman, ``{Lessons Learned from Memory Errors Observed Over the Lifetime
  of Cielo},'' in \emph{SC}, 2018.

\bibitem{li2016pinatubo}
S.~Li, C.~Xu, Q.~Zou, J.~Zhao, Y.~Lu, and Y.~Xie, ``{Pinatubo: A
  Processing-in-Memory Architecture for Bulk Bitwise Operations in Emerging
  Non-Volatile Memories},'' in \emph{DAC}, 2016.

\bibitem{li2010realistic}
X.~Li, M.~C. Huang, K.~Shen, and L.~Chu, ``{A Realistic Evaluation Of Memory
  Hardware Errors And Software System Susceptibility},'' in \emph{ATC}, 2010.

\bibitem{li2011dram}
Y.~Li, H.~Schneider, F.~Schnabel, R.~Thewes, and D.~Schmitt-Landsiedel, ``{DRAM
  Yield Analysis and Optimization by a Statistical Design Approach},'' in
  \emph{CSI}, 2011.

\bibitem{liang2006bluegene}
Y.~Liang, Y.~Zhang, A.~Sivasubramaniam, M.~Jette, and R.~Sahoo, ``{Bluegene/L
  Failure Analysis and Prediction Models},'' in \emph{DSN}, 2006.

\bibitem{lieneweg1998assesment}
U.~Lieneweg, D.~Nguyen, and B.~Blaes, ``{Assessment of DRAM Reliability from
  Retention Time Measurements},'' \emph{Flight Readiness Technol. Assessment
  NASA EEE Parts Prog.}, 1998.

\bibitem{lim2016active}
C.~Lim, K.~Park, and S.~Baeg, ``{Active Precharge Hammering to Monitor
  Displacement Damage using High-Energy Protons in 3x-nm SDRAM},'' \emph{IEEE
  Trans. Nucl. Sci.}, 2016.

\bibitem{lin2012secret}
C.~H. Lin, D.-Y. Shen, Y.-J. Chen, C.-L. Yang, and M.~Wang, ``{SECRET:
  Selective Error Correction for Refresh Energy Reduction in DRAMs},'' in
  \emph{ICCD}, 2012.

\bibitem{costello2004ecc}
S.~Lin and D.~J. Costello, \emph{{Error Control Coding: Fundamentals and
  Applications}}, 2004.

\bibitem{lindenlauf2015cold}
S.~Lindenlauf, H.~H{\"o}fken, and M.~Schuba, ``{Cold Boot Attacks on DDR2 and
  DDR3 SDRAM},'' in \emph{ARES}, 2015.

\bibitem{liu2013experimental}
J.~Liu, B.~Jaiyen, Y.~Kim, C.~Wilkerson, and O.~Mutlu, ``{An Experimental Study
  of Data Retention Behavior in Modern DRAM Devices: Implications for Retention
  Time Profiling Mechanisms},'' in \emph{ISCA}, 2013.

\bibitem{liu2012raidr}
J.~Liu, B.~Jaiyen, R.~Veras, and O.~Mutlu, ``{RAIDR: Retention-Aware
  Intelligent DRAM Refresh},'' in \emph{ISCA}, 2012.

\bibitem{liu2014trustworthy}
W.~Liu, Z.~Zhang, M.~Li, and Z.~Liu, ``{A Trustworthy Key Generation Prototype
  Based on DDR3 PUF for Wireless Sensor Networks},'' in \emph{Sensors}, 2014.

\bibitem{liu2019binary}
X.~Liu, D.~Roberts, R.~Ausavarungnirun, O.~Mutlu, and J.~Zhao, ``{Binary Star:
  Coordinated Reliability in Heterogeneous Memory Systems for High Performance
  and Scalability},'' in \emph{MICRO}, 2019.

\bibitem{longofono2021predicting}
S.~Longofono, D.~Kline~Jr, R.~Melhem, and A.~K. Jones, ``{Predicting and
  Mitigating Single-Event Upsets in DRAM using HOTH},'' \emph{Microelectronics
  Reliability}, 2021.

\bibitem{loughlin2021stop}
K.~Loughlin, S.~Saroiu, A.~Wolman, and B.~Kasikci, ``{Stop! Hammer Time:
  Rethinking Our Approach to Rowhammer Mitigations},'' in \emph{HotOS}, 2021.

\bibitem{lu1989advanced}
N.~C. Lu, ``{Advanced Cell Structures for Dynamic RAMs},'' \emph{IEEE Circuits
  and Devices Magazine}, 1989.

\bibitem{luo2020clr}
H.~Luo, T.~Shahroodi, H.~Hassan, M.~Patel, A.~Giray~Ya{\u{g}}l{\i}k{\c{c}}{\i},
  L.~Orosa, J.~Park, and O.~Mutlu, ``{CLR-DRAM: A Low-Cost DRAM Architecture
  Enabling Dynamic Capacity-Latency Trade-Off},'' in \emph{ISCA}, 2020.

\bibitem{luo2015warm}
Y.~Luo, Y.~Cai, S.~Ghose, J.~Choi, and O.~Mutlu, ``{{WARM}: Improving NAND
  Flash Memory Lifetime with Write-Hotness Aware Retention Management},'' in
  \emph{MSST}, 2015.

\bibitem{luo2016enabling}
Y.~Luo, S.~Ghose, Y.~Cai, E.~F. Haratsch, and O.~Mutlu, ``{Enabling Accurate
  and Practical Online Flash Channel Modeling for Modern {MLC NAND} Flash
  Memory},'' in \emph{JSAC}, 2016.

\bibitem{luo2018heatwatch}
Y.~Luo, S.~Ghose, Y.~Cai, E.~F. Haratsch, and O.~Mutlu, ``{HeatWatch: Improving
  3D NAND Flash Memory Device Reliability by Exploiting Self-Recovery and
  Temperature Awareness},'' in \emph{HPCA}, 2018.

\bibitem{luo2018improving}
Y.~Luo, S.~Ghose, Y.~Cai, E.~F. Haratsch, and O.~Mutlu, ``{Improving 3D NAND
  Flash Memory Lifetime by Tolerating Early Retention Loss and Process
  Variation},'' \emph{SIGMETRICS}, 2018.

\bibitem{luo2014characterizing}
Y.~Luo, S.~Govindan, B.~Sharma, M.~Santaniello, J.~Meza, A.~Kansal, J.~Liu,
  B.~Khessib, K.~Vaid, and O.~Mutlu, ``{Characterizing Application Memory Error
  Vulnerability to Optimize Datacenter Cost via Heterogeneous-Reliability
  Memory},'' in \emph{DSN}, 2014.

\bibitem{maiz2003characterization}
J.~Maiz, S.~Hareland, K.~Zhang, and P.~Armstrong, ``{Characterization of
  Multi-Bit Soft Error Events in Advanced SRAMs},'' in \emph{IEDM}, 2003.

\bibitem{malek2017odd}
A.~Malek, E.~Vasilakis, V.~Papaefstathiou, P.~Trancoso, and I.~Sourdis,
  ``{Odd-ECC: On-Demand DRAM Error Correcting Codes},'' in \emph{Proceedings of
  the International Symposium on Memory Systems}, 2017, pp. 96--111.

\bibitem{mandelman2002challenges}
J.~A. Mandelman, R.~H. Dennard, G.~B. Bronner, J.~K. DeBrosse, R.~Divakaruni,
  Y.~Li, and C.~J. Radens, ``{Challenges and Future Directions for the Scaling
  of Dynamic Random-Access Memory (DRAM)},'' in \emph{IBM JRD}, 2002.

\bibitem{manzhosov2021muse}
E.~Manzhosov, A.~Hastings, M.~Pancholi, R.~Piersma, M.~T.~I. Ziad, and
  S.~Sethumadhavan, ``{MUSE: Multi-Use Error Correcting Codes},''
  \emph{arXiv:2107.09245}, 2021.

\bibitem{marazzi2022protrr}
M.~Marazzi, P.~Jattke, S.~Flavien, and K.~Razavi, ``{ProTRR: Principled yet
  Optimal In-DRAM Target Row Refresh},'' in \emph{SP}, 2022.

\bibitem{markoff2019ibm}
J.~Markoff, ``{IBM's Robert H. Dennard and the Chip That Changed the World},''
  2019,
  \url{https://www.ibm.com/blogs/think/2019/11/ibms-robert-h-dennard-and-the-chip-that-changed-the-world/}.

\bibitem{martinez2011compiler}
A.~Martinez-Alvarez, S.~Cuenca-Asensi, F.~Restrepo-Calle, F.~R.~P. Pinto,
  H.~Guzman-Miranda, and M.~A. Aguirre, ``{Compiler-Directed Soft Error
  Mitigation For Embedded Systems},'' \emph{TDSC}, 2011.

\bibitem{mathew2021thermoelectric}
D.~M. Mathew, H.~Kattan, C.~Weis, J.~Henkel, N.~Wehn, and H.~Amrouch,
  ``{Thermoelectric Cooling to Survive Commodity DRAMs in Harsh Environment
  Automotive Electronics},'' \emph{IEEE Access}, 2021.

\bibitem{mathew2017using}
D.~M. Mathew, {\'E}.~F. Zulian, M.~Jung, K.~Kraft, C.~Weis, B.~Jacob, and
  N.~Wehn, ``{Using Run-Time Reverse-Engineering to Optimize DRAM Refresh},''
  in \emph{MEMSYS}, 2017.

\bibitem{may1979alpha}
T.~C. May and M.~H. Woods, ``{Alpha-Particle-Induced Soft Errors in Dynamic
  Memories},'' \emph{TED}, 1979.

\bibitem{mazumder1993design}
P.~Mazumder, ``{Design of a Fault-Tolerant Three-Dimensional Dynamic
  Random-Access Memory with On-Chip Error-Correcting Circuit},'' \emph{TOC},
  1993.

\bibitem{mcelog2021bad}
\emph{{Bad Page Offlining}}, {mcelog}, 2021,
  \url{https://mcelog.org/badpageofflining.html}.

\bibitem{mcgregor2008braving}
P.~McGregor, T.~Hollebeek, A.~Volynkin, and M.~White, ``{Braving the Cold: New
  Methods for Preventing Cold Boot Attacks on Encryption Keys},'' in
  \emph{Black Hat}, 2008.

\bibitem{meena2014overview}
J.~S. Meena, S.~M. Sze, U.~Chand, and T.-Y. Tseng, ``{Overview Of Emerging
  Nonvolatile Memory Technologies},'' \emph{Nanoscale Research Letters}, 2014.

\bibitem{messer2004susceptibility}
A.~Messer, P.~Bernadat, G.~Fu, D.~Chen, Z.~Dimitrijevic, D.~Lie, D.~D. Mannaru,
  A.~Riska, and D.~Milojicic, ``{Susceptibility Of Commodity Systems And
  Software To Memory Soft Errors},'' \emph{TOC}, 2004.

\bibitem{meza2015revisiting}
J.~Meza, Q.~Wu, S.~Kumar, and O.~Mutlu, ``{Revisiting Memory Errors in
  Large-Scale Production Data Centers: Analysis and Modeling of New Trends from
  the Field},'' in \emph{DSN}, 2015.

\bibitem{meza2015large}
J.~Meza, Q.~Wu, S.~Kumar, and O.~Mutlu, ``{A Large-Scale Study of Flash Memory
  Errors in the Field},'' in \emph{SIGMETRICS}, 2015.

\bibitem{meza2018large}
J.~J. Meza, ``{Large Scale Studies of Memory, Storage, and Network Failures in
  a Modern Data Center},'' Ph.D. dissertation, Carnegie Mellon University,
  2018.

\bibitem{micheloni2015apparatus}
R.~Micheloni, P.~Onufryk, A.~Marelli, C.~Norrie, and I.~Jaser, ``{Apparatus and
  Method Based on LDPC Codes for Adjusting a Correctable Raw Bit Error Rate
  Limit in a Memory System},'' 2015, {US Patent 9,092,353}.

\bibitem{micheloni2010inside}
R.~Micheloni, L.~Crippa, and A.~Marelli, \emph{{Inside NAND Flash
  Memories}}.\hskip 1em plus 0.5em minus 0.4em\relax Springer Science \&
  Business Media, 2010.

\bibitem{micron2021dram}
{Micron Technology}, ``{DRAM},'' \url{https://www.micron.com/products/dram/}.

\bibitem{micron2021rldram}
\emph{{RLDRAM Memory}}, {Micron Technology}, 2021,
  \url{https://www.micron.com/products/dram/rldram-memory}.

\bibitem{micron1999tn}
{Micron Technology Inc.}, ``{TN-04-30: Various Methods of DRAM Refresh},''
  \emph{Technical Report}, 1999.

\bibitem{micron2017whitepaper}
{Micron Technology Inc.}, ``{ECC Brings Reliability and Power Efficiency to
  Mobile Devices},'' {Micron Technology Inc.}, Tech. Rep., 2017.

\bibitem{micron2018mobile}
{Micron Technology, Inc.}, ``{Mobile LPDDR4 SDRAM},'' 2018.

\bibitem{micron20208gb}
{Micron Technology Inc.}, ``{8Gb: x4, x8, x16 DDR4 SDRAM Features - Excessive
  Row Activation},'' Micron Technology Inc., Tech. Rep., 2020.

\bibitem{micron2020tn}
{Micron Technology Inc.}, ``{TN-40-40: DDR4 Point-to-Point Design Guide},''
  Micron Technology Inc., Tech. Rep., 2020.

\bibitem{moon2005error}
T.~K. Moon, \emph{{Error Correction Coding: Mathematical Methods and
  Algorithms}}.\hskip 1em plus 0.5em minus 0.4em\relax John Wiley \& Sons,
  2005.

\bibitem{mori2005origin}
Y.~Mori, K.~Ohyu, K.~Okonogi, and R.~i.~Yamada, ``{The Origin of Variable
  Retention Time in DRAM},'' in \emph{IEDM}, 2005.

\bibitem{mori2001new}
Y.~Mori, R.-I. Yamada, S.~Kamohara, M.~Moniwa, K.~Ohyu, and O.~Yamanaka, ``{A
  New Method for Predicting Distribution of DRAM Retention Time},'' in
  \emph{IRPS}, 2001.

\bibitem{mostek1973mk4096}
{MOSTEK}, ``{MK4096(K/N)-6/16/11},'' 1973.

\bibitem{mrozek2010analysis}
I.~Mrozek, ``{Analysis of Multibackground Memory Testing Techniques},'' 2010.

\bibitem{mrozek2019multi}
I.~Mrozek, \emph{{Multi-Run Memory Tests for Pattern Sensitive Faults}}.\hskip
  1em plus 0.5em minus 0.4em\relax Springer, 2019.

\bibitem{mukhanov2020dstress}
L.~Mukhanov, D.~S. Nikolopoulos, and G.~Karakonstantis, ``{DStress: Automatic
  Synthesis of DRAM Reliability Stress Viruses using Genetic Algorithms},'' in
  \emph{MICRO}, 2020.

\bibitem{mukhanov2019workload}
L.~Mukhanov, K.~Tovletoglou, H.~Vandierendonck, D.~S. Nikolopoulos, and
  G.~Karakonstantis, ``{Workload-Aware DRAM Error Prediction Using Machine
  Learning},'' in \emph{IISWC}, 2019.

\bibitem{mukherjee2004cache}
S.~S. Mukherjee, J.~Emer, T.~Fossum, and S.~K. Reinhardt, ``{Cache Scrubbing in
  Microprocessors: Myth or Necessity?}'' in \emph{SDC}, 2004.

\bibitem{mukundan2013understanding}
J.~Mukundan, H.~Hunter, K.-h. Kim, J.~Stuecheli, and J.~F. Mart{\'\i}nez,
  ``{Understanding and Mitigating Refresh Overheads in High-Density DDR4 DRAM
  Systems},'' in \emph{ISCA}, 2013.

\bibitem{muller2010aesse}
T.~M{\"u}ller, A.~Dewald, and F.~C. Freiling, ``{AESSE: A Cold-Boot Resistant
  Implementation of AES},'' in \emph{{EuroSec}}, 2010.

\bibitem{mutlu2013memory}
O.~Mutlu, ``{Memory Scaling: A Systems Architecture Perspective},'' in
  \emph{IMW}, 2013.

\bibitem{mutlu2015main}
O.~Mutlu, ``{Main Memory Scaling: Challenges and Solution Directions},'' in
  \emph{{More Than Moore Technologies for Next Generation Computer
  Design}}.\hskip 1em plus 0.5em minus 0.4em\relax Springer, 2015, pp.
  127--153.

\bibitem{mutlu2017rowhammer}
O.~Mutlu, ``{The RowHammer Problem and Other Issues we may Face as Memory
  Becomes Denser},'' in \emph{DATE}, 2017.

\bibitem{mutlu2018rowhammer}
O.~Mutlu, ``{RowHammer},''
  \url{https://people.inf.ethz.ch/omutlu/pub/onur-Rowhammer-TopPicksinHardwareEmbeddedSecurity-November-8-2018.pdf},
  2018, {Top Picks in Hardware and Embedded Security}.

\bibitem{mutlu2019processing}
O.~Mutlu, S.~Ghose, J.~G{\'o}mez-Luna, and R.~Ausavarungnirun, ``{Processing
  Data Where It Makes Sense: Enabling In-Memory Computation},''
  \emph{Microprocessors and Microsystems}, 2019.

\bibitem{mutlu2021primer}
O.~Mutlu, S.~Ghose, J.~Gomez-Luna, and R.~Ausavarungnirun, ``{A Modern Primer
  on Processing in Memory},'' in \emph{{arXiv}}, 2020.

\bibitem{mutlu2019rowhammer}
O.~Mutlu and J.~Kim, ``{RowHammer: A Retrospective},'' in \emph{TCAD}, 2019.

\bibitem{mutlu2007stall}
O.~Mutlu and T.~Moscibroda, ``{Stall-Time Fair Memory Access Scheduling for
  Chip Multiprocessors},'' in \emph{MICRO}, 2007.

\bibitem{mutlu2003runahead}
O.~Mutlu, J.~Stark, C.~Wilkerson, and Y.~N. Patt, ``{Runahead Execution: An
  Alternative to Very Large Instruction Windows for Out-of-Order Processors},''
  in \emph{HPCA}, 2003.

\bibitem{mutlu2014research}
O.~Mutlu and L.~Subramanian, ``{Research Problems and Opportunities in Memory
  Systems},'' in \emph{SUPERFRI}, 2014.

\bibitem{naeimi2013sttram}
H.~Naeimi, C.~Augustine, A.~Raychowdhury, S.-L. Lu, and J.~Tschanz, ``{STTRAM
  Scaling and Retention Failure},'' \emph{Intel Technology Journal}, 2013.

\bibitem{nair2013case}
P.~Nair, C.-C. Chou, and M.~K. Qureshi, ``{A Case for Refresh Pausing in DRAM
  Memory Systems},'' in \emph{HPCA}, 2013.

\bibitem{nair2019sudoku}
P.~J. Nair, B.~Asgari, and M.~K. Qureshi, ``{SuDoku: Tolerating High-Rate of
  Transient Failures for Enabling Scalable STTRAM},'' in \emph{DSN}, 2019.

\bibitem{nair2014refresh}
P.~J. Nair, C.-C. Chou, and M.~K. Qureshi, ``{Refresh Pausing in DRAM Memory
  Systems},'' in \emph{TACO}, 2014.

\bibitem{nair2013archshield}
P.~J. Nair, D.-H. Kim, and M.~K. Qureshi, ``{ArchShield: Architectural
  Framework for Assisting DRAM Scaling by Tolerating High Error Rates},'' in
  \emph{ISCA}, 2013.

\bibitem{nair2016xed}
P.~J. Nair, V.~Sridharan, and M.~K. Qureshi, ``{XED: Exposing On-Die Error
  Detection Information for Strong Memory Reliability},'' in \emph{ISCA}, 2016.

\bibitem{nakagome1988impact}
Y.~Nakagome, M.~Aoki, S.~Ikenaga, M.~Horiguchi, S.~Kimura, Y.~Kawamoto, and
  K.~Itoh, ``{The Impact of Data-Line Interference Noise on DRAM Scaling},'' in
  \emph{JSSC}, 1988.

\bibitem{vaid2016ssd}
I.~Narayanan, D.~Wang, M.~Jeon, B.~Sharma, L.~Caulfield, A.~Sivasubramaniam,
  B.~Cutler, J.~Liu, B.~M. Khessib, and V.~Kushagra, ``{SSD Failures in
  Datacenters: What, When and Why?}'' in \emph{SIGMETRICS}, 2016.

\bibitem{nasa2016nasa}
{NASA}, ``{NASA NEPP Program Memory Technology - Testing, Analysis, and
  Roadmap},''
  \url{https://radhome.gsfc.nasa.gov/radhome/papers/radecs05\_sc.pdf}, 2016.

\bibitem{nature2018memory}
{Nature Electronics}, ``{Memory Lane},'' 2018.

\bibitem{nguyen2021zem}
D.-T. Nguyen, N.-M. Ho, M.-S. Le, W.-F. Wong, and I.-J. Chang, ``{ZEM:
  Zero-Cycle Bit-Masking Module for Deep Learning Refresh-Less DRAM},''
  \emph{IEEE Access}, 2021.

\bibitem{nguyen2021obet}
D.-T. Nguyen, N.-M. Ho, W.-F. Wong, and I.-J. Chang, ``{OBET: On-the-Fly
  Byte-Level Error Tracking for Correcting and Detecting Faults in Unreliable
  DRAM Systems},'' \emph{Sensors}, 2021.

\bibitem{nguyen2020approximate}
D.~T. Nguyen, N.~H. Hung, H.~Kim, and H.-J. Lee, ``{An Approximate Memory
  Architecture For Energy Saving In Deep Learning Applications},'' \emph{TOCS},
  2020.

\bibitem{nguyen2018nonblocking}
K.~Nguyen, K.~Lyu, X.~Meng, V.~Sridharan, and X.~Jian, ``{Nonblocking Memory
  Refresh},'' in \emph{ISCA}, 2018.

\bibitem{nishi2019advances}
Y.~Nishi and B.~Magyari-Kope, \emph{{Advances in Non-Volatile Memory and
  Storage Technology}}.\hskip 1em plus 0.5em minus 0.4em\relax Woodhead
  Publishing, 2019.

\bibitem{niu2022184qps}
D.~Niu, S.~Li, Y.~Wang, W.~Han, Z.~Zhang, Y.~Guan, T.~Guan, F.~Sun, F.~Xue,
  L.~Duan, Y.~Fang, H.~Zheng, X.~Jiang, S.~Wang, F.~Zuo, Y.~Wang, B.~Yu,
  Q.~Ren, and Y.~Xie, ``{184QPS/W 64Mb/mm 2 3D Logic-to-DRAM Hybrid Bonding
  with Process-Near-Memory Engine for Recommendation System},'' in
  \emph{ISSCC}, 2022.

\bibitem{nvidia2020dynamic}
\emph{{Dynamic Page Retirement}}, {NVIDIA}, 2020,
  \url{https://docs.nvidia.com/deploy/dynamic-page-retirement/index.html}.

\bibitem{oh2014a}
T.-Y. Oh, H.~Chung, J.-Y. Park, K.-W. Lee, S.~Oh, S.-Y. Doo, H.-J. Kim, C.~Lee,
  H.-R. Kim, J.-H. Lee, J.-I. Lee, K.-S. Ha, Y.~Choi, Y.-C. Cho, Y.-C. Bae,
  T.~Jang, C.~Park, K.~Park, S.~Jang, and J.~S. Choi, ``{A 3.2 Gbps/Pin 8 Gbit
  1.0 V LPDDR4 SDRAM with Integrated ECC Engine for Sub-1 V DRAM Core
  Operation},'' \emph{JSSC}, 2014.

\bibitem{oh20153}
T.-Y. Oh, H.~Chung, J.-Y. Park, K.-W. Lee, S.~oh, S.-Y. Doo, H.-J. Kim, C.~Lee,
  H.-R. Kim, J.-H. Lee, J.-I. Lee, K.-S. Ha, Y.~Choi, Y.-C. Cho, Y.-C. Bae,
  T.~Jang, C.~Park, K.~Park, S.~Jang, and J.~Choi, ``{A 3.2Gbps/pin 8Gb 1.0V
  LPDDR4 SDRAM with Integrated ECC Engine for Sub-1V DRAM Core Operation},'' in
  \emph{ISSCC}, 2014.

\bibitem{ohsawa1998optimizing}
T.~Ohsawa, K.~Kai, and K.~Murakami, ``{Optimizing the DRAM Refresh Count for
  Merged DRAM/logic LSIs},'' in \emph{ISLPED}, 1998.

\bibitem{ohyu2006quantitative}
K.~Ohyu, T.~Umeda, K.~Okonogi, S.~Tsukada, M.~Hidaka, S.~Fujieda, and
  Y.~Mochizuki, ``{Quantitative Identification for the Physical Origin of
  Variable Retention Time: A Vacancy-Oxygen Complex Defect Model},'' in
  \emph{IEDM}, 2006.

\bibitem{olgun2021pidram}
A.~Olgun, J.~G. Luna, K.~Kanellopoulos, B.~Salami, H.~Hassan, O.~Ergin, and
  O.~Mutlu, ``{PiDRAM: A Holistic End-to-end FPGA-based Framework for
  Processing-in-DRAM},'' \emph{arXiv:2111.00082}, 2021.

\bibitem{olgun2021quac}
A.~Olgun, M.~Patel, A.~G. Ya{\u{g}}l{\i}k{\c{c}}{\i}, H.~Luo, J.~S. Kim,
  N.~Bostanc{\i}, N.~Vijaykumar, O.~Ergin, and O.~Mutlu, ``{QUAC-TRNG:
  High-Throughput True Random Number Generation Using Quadruple Row Activation
  in Commodity DRAM Chips},'' in \emph{ISCA}, 2021.

\bibitem{oliveira2021damov}
G.~F. Oliveira, J.~G{\'o}mez-Luna, S.~Ghose, L.~Orosa, N.~Vijaykumar,
  I.~Fernandez, M.~Sadrosadati, and O.~Mutlu, ``{DAMOV: A New Methodology and
  Benchmark Suite for Evaluating Data Movement Bottlenecks},'' in \emph{IEEE
  Access}, 2021.

\bibitem{orosa2021codic}
L.~Orosa, Y.~Wang, M.~Sadrosadati, J.~S. Kim, M.~Patel, I.~Puddu, H.~Luo,
  K.~Razavi, J.~Gómez-Luna, H.~Hassan, N.~Mansouri-Ghiasi, S.~Ghose, and
  O.~Mutlu, ``{CODIC: A Low-Cost Substrate for Enabling Custom In-DRAM
  Functionalities and Optimizations},'' \emph{ISCA}, 2021.

\bibitem{orosa2021deeper}
L.~Orosa, A.~G. Ya{\u{g}}l{\i}k{\c{c}}{\i}, H.~Luo, A.~Olgun, J.~Park,
  H.~Hassan, M.~Patel, J.~S. Kim, and O.~Mutlu, ``{A Deeper Look into
  RowHammer's Sensitivities: Experiemental Analysis of Real DRAM Chips and
  Implications on Future Attacks and Defenses},'' in \emph{MICRO}, 2021.

\bibitem{pae2021minimal}
S.-I. Pae, V.~Kozhikkottu, D.~Somasekar, W.~Wu, S.~G. Ramasubramanian,
  M.~Dadual, H.~Cho, and K.-W. Kwon, ``{Minimal Aliasing
  Single-Error-Correction Codes for DRAM Reliability Improvement},'' \emph{IEEE
  Access}, 2021.

\bibitem{pan2019hiding}
X.~Pan and F.~Mueller, ``{Hiding DRAM Refresh Overhead in Real-Time Cyclic
  Executives},'' in \emph{RTSS}, 2019.

\bibitem{park2021reducing}
J.~Park, M.~Kim, M.~Chun, L.~Orosa, J.~Kim, and O.~Mutlu, ``{Reducing
  Solid-State Drive Read Latency by Optimizing Read-Retry},'' in \emph{ASPLOS},
  2021.

\bibitem{park2016experiments}
K.~Park, C.~Lim, D.~Yun, and S.~Baeg, ``{Experiments and Root Cause Analysis
  for Active-Precharge Hammering Fault In DDR3 SDRAM Under 3$\times$ Nm
  Technology},'' \emph{Microelectronics Reliability}, 2016.

\bibitem{park2016statistical}
K.~Park, D.~Yun, and S.~Baeg, ``{Statistical Distributions of Row-Hammering
  Induced Failures in DDR3 Components},'' \emph{Microelectronics Reliability},
  2016.

\bibitem{park2015technology}
S.-K. Park, ``{Technology Scaling Challenge and Future Prospects of DRAM and
  NAND Flash Memory},'' in \emph{IMW}, 2015.

\bibitem{park2020graphene}
Y.~Park, W.~Kwon, E.~Lee, T.~J. Ham, J.~H. Ahn, and J.~W. Lee, ``{Graphene:
  Strong yet Lightweight Row Hammer Protection},'' in \emph{MICRO}, 2020.

\bibitem{parkin2015memory}
S.~Parkin and S.-H. Yang, ``{Memory on the Racetrack},'' \emph{{Nature
  Nanotechnology}}, 2015.

\bibitem{patel2021enabling}
M.~Patel, ``{Enabling Effective Error Mitigation in Memory Chips That Use
  On-Die Error-Correcting Codes},'' Ph.D. dissertation, ETH Z{\"u}rich, 2021.

\bibitem{patel2021harp}
M.~Patel, G.~F. de~Oliveira, and O.~Mutlu, ``{HARP: Practically and Effectively
  Identifying Uncorrectable Errors in Memory Chips That Use On-Die
  Error-Correcting Codes},'' in \emph{MICRO}, 2021.

\bibitem{patel2020bit}
M.~Patel, J.~Kim, T.~Shahroodi, H.~Hassan, and O.~Mutlu, ``{Bit-Exact ECC
  Recovery (BEER): Determining DRAM On-Die ECC Functions by Exploiting DRAM
  Data Retention Characteristics},'' in \emph{MICRO}, 2020.

\bibitem{patel2019understanding}
M.~Patel, J.~S. Kim, H.~Hassan, and O.~Mutlu, ``{Understanding and Modeling
  On-Die Error Correction in Modern DRAM: An Experimental Study Using Real
  Devices},'' in \emph{DSN}, 2019.

\bibitem{patel2017reach}
M.~Patel, J.~S. Kim, and O.~Mutlu, ``{The Reach Profiler (REAPER): Enabling the
  Mitigation of DRAM Retention Failures via Profiling at Aggressive
  Conditions},'' in \emph{ISCA}, 2017.

\bibitem{harp-artifacts}
M.~Patel, G.~F. Oliveira, and O.~Mutlu, ``{HARP Artifacts},'' \emph{ZENODO},
  2021, doi:10.5281/zenodo.5148592.

\bibitem{patil2021dve}
A.~Patil, V.~Nagarajan, R.~Balasubramonian, and N.~Oswald, ``{Dv{\'e}:
  Improving DRAM Reliability and Performance On-Demand via Coherent
  Replication},'' in \emph{ISCA}, 2021.

\bibitem{patterson1997case}
D.~Patterson, T.~Anderson, N.~Cardwell, R.~Fromm, K.~Keeton, C.~Kozyrakis,
  R.~Thomas, and K.~Yelick, ``{A Case for Intelligent RAM},'' \emph{{IEEE
  Micro}}, 1997.

\bibitem{pessl2016drama}
P.~Pessl, D.~Gruss, C.~Maurice, M.~Schwarz, and S.~Mangard, ``{DRAMA:
  Exploiting DRAM Addressing for Cross-CPU Attacks},'' in \emph{USENIX
  Security}, 2016.

\bibitem{pfaffli2018tcad}
P.~Pf{\"a}ffli, H.~Wong, X.~Xu, L.~Silvestri, X.~Lin, T.~Yang, R.~Tiwari,
  S.~Mahapatra, S.~Motzny, V.~Moroz, and T.~Ma, ``{TCAD Modeling for
  Reliability},'' \emph{Microelectronics Reliability}, 2018.

\bibitem{pinheiro2007failure}
E.~Pinheiro, W.-D. Weber, and L.~A. Barroso, ``{Failure Trends In A Large Disk
  Drive Population},'' \emph{FAST}, 2007.

\bibitem{pirovano2004reliability}
A.~Pirovano, A.~Redaelli, F.~Pellizzer, F.~Ottogalli, M.~Tosi, D.~Ielmini,
  A.~L. Lacaita, and R.~Bez, ``{Reliability Study Of Phase-Change Nonvolatile
  Memories},'' \emph{TDMR}, 2004.

\bibitem{prasad2005survey}
M.~R. Prasad, A.~Biere, and A.~Gupta, ``{A Survey of Recent Advances in
  SAT-based Formal Verification},'' \emph{STTT}, 2005.

\bibitem{qin2005rx}
F.~Qin, J.~Tucek, J.~Sundaresan, and Y.~Zhou, ``{Rx: Treating Bugs As
  Allergies---A Safe Method To Survive Software Failures},'' in \emph{SOSP},
  2005.

\bibitem{qureshi2011pay}
M.~K. Qureshi, ``{Pay-As-You-Go: Low-Overhead Hard-Error Correction for Phase
  Change Memories},'' in \emph{MICRO}, 2011.

\bibitem{qureshi2021rethinking}
M.~K. Qureshi, ``{Rethinking ECC in the Era of Row-Hammer},'' in
  \emph{DRAMSec}, 2021.

\bibitem{qureshi2015avatar}
M.~K. Qureshi, D.-H. Kim, S.~Khan, P.~J. Nair, and O.~Mutlu, ``{AVATAR: A
  Variable-Retention-Time (VRT) Aware Refresh for DRAM Systems},'' in
  \emph{DSN}, 2015.

\bibitem{qureshi2009scalable}
M.~K. Qureshi, V.~Srinivasan, and J.~A. Rivers, ``{Scalable High Performance
  Main Memory System Using Phase-change Memory Technology},'' in \emph{ISCA},
  2009.

\bibitem{qy2019global}
{QY Research}, ``{Global DRAM Market Professional Survey Report},''
  \url{https://garnerinsights.com/Global-DRAM-Market-Professional-Survey-Report-2019},
  2019.

\bibitem{radaelli2005investigation}
D.~Radaelli, H.~Puchner, S.~Wong, and S.~Daniel, ``{Investigation Of Multi-Bit
  Upsets In A 150 nm Technology SRAM Device},'' \emph{IEEE Trans. Nucl. Sci.},
  2005.

\bibitem{rahman2021utilizing}
M.~J.~M. Rahman, ``{Utilizing Two Stage Scrubbing to Handle Single-Fault
  Multi-Error Cases in DRAM Systems},'' Master's thesis, Iowa State University,
  2021.

\bibitem{rahmati2015probable}
A.~Rahmati, M.~Hicks, D.~E. Holcomb, and K.~Fu, ``{Probable Cause: The
  Deanonymizing Effects of Approximate DRAM},'' in \emph{ISCA}, 2016.

\bibitem{rambus2000direct}
{Rambus Inc.}, ``{Direct Rambus RIMM Module Specification},'' 2000.

\bibitem{raychowdhury2009design}
A.~Raychowdhury, D.~Somasekhar, T.~Karnik, and V.~De, ``{Design Space and
  Scalability Exploration of 1T-1STT MTJ Memory Arrays in the Presence of
  Variability and Disturbances},'' in \emph{IEDM}, 2009.

\bibitem{redeker2002investigation}
M.~Redeker, B.~F. Cockburn, and D.~G. Elliott, ``{An Investigation Into
  Crosstalk Noise in DRAM Structures},'' in \emph{MTDT}, 2002.

\bibitem{reed1960polynomial}
I.~S. Reed and G.~Solomon, ``{Polynomial Codes Over Certain Finite Fields},''
  \emph{SIAM}, 1960.

\bibitem{reis2005swift}
G.~A. Reis, J.~Chang, N.~Vachharajani, R.~Rangan, and D.~I. August, ``{SWIFT:
  SOftware Implemented Fault Tolerance},'' in \emph{CGO}, 2005.

\bibitem{restle1992dram}
P.~J. Restle, J.~Park, and B.~F. Lloyd, ``{DRAM Variable Retention Time},'' in
  \emph{IEDM}, 1992.

\bibitem{richardson2008modern}
T.~Richardson and R.~Urbanke, \emph{{Modern Coding Theory}}.\hskip 1em plus
  0.5em minus 0.4em\relax Cambridge University Press, 2008.

\bibitem{rixner2000memory}
S.~Rixner, W.~J. Dally, U.~J. Kapasi, P.~Mattson, and J.~D. Owens, ``{Memory
  Access Scheduling},'' in \emph{ISCA}, 2000.

\bibitem{micron2019whitepaper}
R.~Rooney and N.~Koyle, ``{Micron DDR5 SDRAM: New Features},'' {Micron
  Technology Inc.}, Tech. Rep., 2019.

\bibitem{roth2006introduction}
R.~M. Roth, \emph{{Introduction to Coding Theory}}.\hskip 1em plus 0.5em minus
  0.4em\relax Cambridge University Press, 2006.

\bibitem{ryu2017overcoming}
S.-W. Ryu, K.~Min, J.~Shin, H.~Kwon, D.~Nam, T.~Oh, T.-S. Jang, M.~Yoo, Y.~Kim,
  and S.~Hong, ``{Overcoming the Reliability Limitation in the Ultimately
  Scaled DRAM using Silicon Migration Technique by Hydrogen Annealing},'' in
  \emph{IEDM}, 2017.

\bibitem{saileshwar2022randomized}
G.~Saileshwar, B.~Wang, M.~Qureshi, and P.~J. Nair, ``{Randomized Row-Swap:
  Mitigating Row Hammer by Breaking Spatial Correlation Between Aggressor and
  Victim Rows},'' in \emph{ASPLOS}, 2022.

\bibitem{saino2000impact}
K.~Saino, S.~Horiba, S.~Uchiyama, Y.~Takaishi, M.~Takenaka, T.~Uchida,
  Y.~Takada, K.~Koyama, H.~Miyake, and C.~Hu, ``{Impact of Gate-Induced Drain
  Leakage Current on the Tail Distribution of DRAM Data Retention Time},'' in
  \emph{IEDM}, 2000.

\bibitem{saleh1990reliability}
A.~M. Saleh, J.~J. Serrano, and J.~H. Patel, ``{Reliability of Scrubbing
  Recovery-Techniques for Memory Systems},'' \emph{TR}, 1990.

\bibitem{salman2021half}
Q.~Salman, K.~Yoongu, B.~Nicolas, S.~Eric, and N.~Mattias, ``{Half-Double:
  Next-Row-Over Assisted Rowhammer},'' 2021.

\bibitem{samsungbrochure}
Samsung, ``{SAS SSDs for Enterprise: High Performance with Extreme
  Reliability},''
  \url{http://www.samsung.com/global/business/semiconductor/file/media/SM1625\_Brochure-0.pdf},
  2012.

\bibitem{samsung2018mobile}
{Samsung Electronics}, ``{Mobile DRAM Stack Specification (LPDDR4)},'' 2018.

\bibitem{sandia2020fabrication}
{Sandia National Laboratories}, ``{Fabriation, Testing, and Validation
  Capabilities},''
  \url{https://www.sandia.gov/mesa/fabrication/index.html#tab-9}, 2020.

\bibitem{saroiu2022price}
S.~Saroiu, A.~Wolman, and L.~Cojocar, ``{The Price of Secrecy: How Hiding
  Internal DRAM Topologies Hurts Rowhammer Defenses},'' in \emph{IRPS}, 2022.

\bibitem{schaller2017intrinsic}
A.~Schaller, W.~Xiong, N.~A. Anagnostopoulos, M.~U. Saleem, S.~Gabmeyer,
  S.~Katzenbeisser, and J.~Szefer, ``{Intrinsic Rowhammer PUFs: Leveraging the
  Rowhammer Effect for Improved Security},'' in \emph{HOST}, 2017.

\bibitem{schaller2018decay}
A.~Schaller, W.~Xiong, N.~A. Anagnostopoulos, M.~U. Saleem, S.~Gabmeyer,
  B.~{\v{S}}kori{\'c}, S.~Katzenbeisser, and J.~Szefer, ``{Decay-Based DRAM
  PUFs in Commodity Devices},'' \emph{TDSC}, 2018.

\bibitem{schechter2010use}
S.~Schechter, G.~H. Loh, K.~Strauss, and D.~Burger, ``{Use ECP, Not ECC, for
  Hard Failures in Resistive Memories},'' \emph{ISCA}, 2010.

\bibitem{schroeder2007understanding}
B.~Schroeder and G.~A. Gibson, ``{Disk Failures in the Real World: What Does an
  MTTF of 1,000,000 Hours Mean To You?}'' 2007.

\bibitem{schroeder2016flash}
B.~Schroeder, R.~Lagisetty, and A.~Merchant, ``{Flash Reliability in
  Production: The Expected and the Unexpected.}'' in \emph{FAST}, 2016.

\bibitem{schroeder2009dram}
B.~Schroeder, E.~Pinheiro, and W.-D. Weber, ``{DRAM Errors in the Wild: a
  Large-Scale Field Study},'' in \emph{SIGMETRICS}, 2009.

\bibitem{seong2010safer}
N.~H. Seong, D.~H. Woo, V.~Srinivasan, J.~A. Rivers, and H.-H.~S. Lee,
  ``{SAFER: Stuck-at-Fault Error Recovery for Memories},'' in \emph{MICRO},
  2010.

\bibitem{seong2013tri}
N.~H. Seong, S.~Yeo, and H.-H.~S. Lee, ``{Tri-Level-Cell Phase Change Memory:
  Toward an Efficient and Reliable Memory System},'' in \emph{ISCA}, 2013.

\bibitem{seshadri2015fast}
V.~Seshadri, K.~Hsieh, A.~Boroum, D.~Lee, M.~A. Kozuch, O.~Mutlu, P.~B.
  Gibbons, and T.~C. Mowry, ``{Fast Bulk Bitwise AND and OR in DRAM},''
  \emph{IEEE CAL}, 2015.

\bibitem{seshadri2013rowclone}
V.~Seshadri, Y.~Kim, C.~Fallin, D.~Lee, R.~Ausavarungnirun, G.~Pekhimenko,
  Y.~Luo, O.~Mutlu, P.~B. Gibbons, M.~A. Kozuch, and T.~C. Mowry, ``{RowClone:
  Fast and Energy-Efficient In-DRAM Bulk Data Copy and Initialization},'' in
  \emph{MICRO}, 2013.

\bibitem{seshadri2016buddy}
V.~Seshadri, D.~Lee, T.~Mullins, H.~Hassan, A.~Boroumand, J.~Kim, M.~A. Kozuch,
  O.~Mutlu, P.~B. Gibbons, and T.~C. Mowry, ``{Buddy-RAM: Improving the
  Performance and Efficiency of Bulk Bitwise Operations Using DRAM},'' in
  \emph{arXiv}, 2016.

\bibitem{seshadri2017ambit}
V.~Seshadri, D.~Lee, T.~Mullins, H.~Hassan, A.~Boroumand, J.~Kim, M.~A. Kozuch,
  O.~Mutlu, P.~B. Gibbons, and T.~C. Mowry, ``{Ambit: In-Memory Accelerator for
  Bulk Bitwise Operations Using Commodity DRAM Technology},'' in \emph{MICRO},
  2017.

\bibitem{seshadri2017simple}
V.~Seshadri and O.~Mutlu, ``{Simple Operations in Memory to Reduce Data
  Movement},'' in \emph{Advances in Computers}, 2017.

\bibitem{seshadri2019dram}
V.~Seshadri and O.~Mutlu, ``{In-DRAM Bulk Bitwise Execution Engine},''
  \emph{arXiv:1905.09822}, 2019.

\bibitem{seshadri2020indram}
V.~Seshadri and O.~Mutlu, ``{In-DRAM Bulk Bitwise Execution Engine},''
  \emph{Advances in Computers}, 2020.

\bibitem{seyedzadeh2018cbt}
S.~M. {Seyedzadeh}, A.~K. {Jones}, and R.~{Melhem}, ``{Mitigating Wordline
  Crosstalk Using Adaptive Trees of Counters},'' in \emph{ISCA}, 2018.

\bibitem{seyedzadeh2017mitigating}
S.~M. Seyedzadeh, D.~Kline~Jr, A.~K. Jones, and R.~Melhem, ``{Mitigating
  Bitline Crosstalk Noise in DRAM Memories},'' in \emph{ISMS}, 2017.

\bibitem{sfikas2016testing}
Y.~Sfikas and Y.~Tsiatouhas, ``{Testing Neighbouring Cell Leakage And
  Transition Induced Faults in DRAMs},'' \emph{IEEE Trans. Comput.}, 2016.

\bibitem{sharifi2017online}
R.~Sharifi and Z.~Navabi, ``{Online Profiling for Cluster-Specific Variable
  Rate Refreshing in High-Density DRAM Systems},'' in \emph{ETS}, 2017.

\bibitem{shin2019dram}
H.~H. Shin and E.-Y. Chung, ``{In-DRAM Cache Management for Low Latency and Low
  Power 3D-Stacked DRAMs},'' \emph{Micromachines}, 2019.

\bibitem{shin2014nuat}
W.~Shin, J.~Yang, J.~Choi, and L.-S. Kim, ``{NUAT: A Non-Uniform Access Time
  Memory Controller},'' in \emph{HPCA}, 2014.

\bibitem{shirley2014copula}
C.~G. Shirley and W.~R. Daasch, ``{Copula Models of Correlation: A DRAM Case
  Study},'' in \emph{TC}, 2014.

\bibitem{shye2008plr}
A.~Shye, J.~Blomstedt, T.~Moseley, V.~J. Reddi, and D.~A. Connors, ``{PLR: A
  Software Approach To Transient Fault Tolerance For Multicore
  Architectures},'' \emph{TDSC}, 2008.

\bibitem{shye2007using}
A.~Shye, T.~Moseley, V.~J. Reddi, J.~Blomstedt, and D.~A. Connors, ``{Using
  Process-Level Redundancy To Exploit Multiple Cores For Transient Fault
  Tolerance},'' in \emph{DSN}, 2007.

\bibitem{siddiqua2013analysis}
T.~Siddiqua, A.~E. Papathanasiou, A.~Biswas, S.~Gurumurthi, I.~Corp, and
  T.~Aster, ``{Analysis and Modeling of Memory Errors From Large-Scale Field
  Data Collection},'' in \emph{SELSE}, 2013.

\bibitem{simmons2011security}
P.~Simmons, ``{Security Through Amnesia: A Software-Based Solution to the Cold
  Boot Attack on Disk Encryption},'' in \emph{ACSAC}, 2011.

\bibitem{hynix2015366ball}
{SK Hnyix}, ``{366ball FBGA Specification 32Gb LPDDR4 (x16, 4 Channel)},''
  2015.

\bibitem{slayman2006impact}
C.~Slayman, M.~Ma, and S.~Lindley, ``{Impact of Error Correction Code and
  Dynamic Memory Reconfiguration on High-Reliability/Low-Cost Server Memory},''
  in \emph{IRWS}, 2006.

\bibitem{smart2017smart}
{SMART Modular Technologies}, ``{SMART Press Release 415},'' SMART Modular
  Technologies, Tech. Rep., 2017.

\bibitem{smart2021rugged}
\emph{{Rugged Memory}}, {SMART Modular Technologies}, 2021,
  \url{https://www.smartm.com/product/rugged-memory}.

\bibitem{snavely2000symbiotic}
A.~Snavely and D.~M. Tullsen, ``{Symbiotic Jobscheduling for a Simultaneous
  Mutlithreading Processor},'' in \emph{ASPLOS}, 2000.

\bibitem{passmark2019memtest}
P.~Software, ``{MemTest86 Overview},''
  \url{https://www.memtest86.com/index.html}, 2019.

\bibitem{son2017making}
M.~Son, H.~Park, J.~Ahn, and S.~Yoo, ``{Making DRAM Stronger Against Row
  Hammering},'' in \emph{DAC}, 2017.

\bibitem{son2015cidra}
Y.~H. Son, S.~Lee, O.~Seongil, S.~Kwon, N.~S. Kim, and J.~H. Ahn, ``{CiDRA: A
  cache-Inspired DRAM resilience architecture},'' in \emph{HPCA}, 2015.

\bibitem{son2013reducing}
Y.~H. Son, O.~Seongil, Y.~Ro, J.~W. Lee, and J.~H. Ahn, ``{Reducing Memory
  Access Latency with Asymmetric DRAM Bank Organizations},'' in \emph{ISCA},
  2013.

\bibitem{sridharan2015memory}
V.~Sridharan, N.~DeBardeleben, S.~Blanchard, K.~B. Ferreira, J.~Stearley,
  J.~Shalf, and S.~Gurumurthi, ``{Memory Errors in Modern Systems: The Good,
  the Bad, and the Ugly},'' in \emph{ASPLOS}, 2015.

\bibitem{sridharan2012study}
V.~Sridharan and D.~Liberty, ``{A Study of DRAM Failures in the Field},'' in
  \emph{SC}, 2012.

\bibitem{sridharan2013feng}
V.~Sridharan, J.~Stearley, N.~DeBardeleben, S.~Blanchard, and S.~Gurumurthi,
  ``{Feng Shui of Supercomputer Memory: Positional Effects in DRAM and SRAM
  Faults},'' in \emph{SC}, 2013.

\bibitem{stone1970logic}
H.~S. Stone, ``{A Logic-in-Memory Computer},'' \emph{TC}, 1970.

\bibitem{stuecheli2010elastic}
J.~Stuecheli, D.~Kaseridis, H.~C. Hunter, and L.~K. John, ``{Elastic Refresh:
  Techniques to Mitigate Refresh Penalties in High Density Memory},'' in
  \emph{MICRO}, 2010.

\bibitem{sutar2018d}
S.~Sutar, A.~Raha, D.~Kulkarni, R.~Shorey, J.~Tew, and V.~Raghunathan,
  ``{D-PUF: An Intrinsically Reconfigurable DRAM PUF for Device Authentication
  and Random Number Generation},'' \emph{TECS}, 2018.

\bibitem{sutar2016d}
S.~Sutar, A.~Raha, and V.~Raghunathan, ``{D-PUF: An Intrinsically
  Reconfigurable DRAM PUF for Device Authentication in Embedded Systems},'' in
  \emph{CASES}, 2016.

\bibitem{synopsys2015whitepaper}
{Synopsys}, ``{Reliability, Availability and Serviceability (RAS) for Memory
  Interfaces},'' {Synopsys}, Tech. Rep., 2015.

\bibitem{synopsys2018sentaurus}
\emph{{Sentaurus Sdevice User's Manual}}, {Synopsys}, 2018.

\bibitem{takemura2007long}
R.~Takemura, K.~Itoh, T.~Sekiguchi, S.~Akiyama, S.~Hanzawa, K.~Kajigaya, and
  T.~Kawahara, ``{Long-Retention-Time, High-Speed DRAM Array with 12-F 2 Twin
  Cell for Sub 1-V Operation},'' \emph{TOE}, 2007.

\bibitem{talukder2018exploiting}
B.~Talukder, J.~Kerns, B.~Ray, T.~Morris, and M.~T. Rahman, ``{Exploiting DRAM
  Latency Variations for Generating True Random Numbers},'' \emph{ICCE}, 2019.

\bibitem{talukder2018ldpuf}
B.~Talukder, B.~Ray, M.~Tehranipoor, D.~Forte, and M.~T. Rahman, ``{LDPUF:
  Exploiting DRAM Latency Variations to Generate Robust Device Signatures},''
  \emph{arXiv preprint arXiv:1808.02584}, 2018.

\bibitem{talukder2019prelatpuf}
B.~B. Talukder, B.~Ray, D.~Forte, and M.~T. Rahman, ``{PreLatPUF: Exploiting
  DRAM Latency Variations For Generating Robust Device Signatures},''
  \emph{IEEE Access}, 2019.

\bibitem{tang2017dram}
Q.~Tang, C.~Zhou, W.~Choi, G.~Kang, J.~Park, K.~K. Parhi, and C.~H. Kim, ``{A
  DRAM Based Physical Unclonable Function Capable of Generating >$10^{32}$
  Challenge Response Pairs per 1Kbit Array for Secure Chip Authentication},''
  in \emph{CICC}, 2017.

\bibitem{tannu2017cryogenic}
S.~S. Tannu, D.~M. Carmean, and M.~K. Qureshi, ``{Cryogenic-DRAM Based Memory
  System for Scalable Quantum Computers: A Feasibility Study},'' in
  \emph{MEMSYS}, 2017.

\bibitem{tatar2018defeating}
A.~Tatar, C.~Giuffrida, H.~Bos, and K.~Razavi, ``{Defeating Software
  Mitigations Against Rowhammer: A Surgical Precision Hammer},'' in
  \emph{RAID}, 2018.

\bibitem{tavana2017remap}
M.~K. Tavana, A.~K. Ziabari, M.~Arjomand, M.~Kandemir, C.~Das, and D.~Kaeli,
  ``{REMAP: A Reliability/Endurance Mechanism for Advancing PCM},'' in
  \emph{MEMSYS}, 2017.

\bibitem{taylor1980redundancy}
D.~J. Taylor, D.~E. Morgan, and J.~P. Black, ``{Redundancy In Data Structures:
  Improving Software Fault Tolerance},'' \emph{TSE}, 1980.

\bibitem{tehranipoor2017investigation}
F.~Tehranipoor, N.~Karimian, W.~Yan, and J.~A. Chandy, ``{Investigation of DRAM
  PUFs Reliability Under Device Accelerated Aging Effects},'' in \emph{ISCAS},
  2017.

\bibitem{tehranipoor2016robust}
F.~Tehranipoor, W.~Yan, and J.~A. Chandy, ``{Robust Hardware True Random Number
  Generators using DRAM Remanence Effects},'' in \emph{HOST}, 2016.

\bibitem{teradyne2022magnum}
{Teradyne}, \emph{{Magnum EPIC Ultra-high Performance Solution for Memory
  Device Test}}, \url{https://www.teradyne.com/products/magnum-epic/}, 2022.

\bibitem{torrance2009state}
R.~Torrance and D.~James, ``{The State-of-the-Art in IC Reverse Engineering},''
  in \emph{CHES}, 2009.

\bibitem{tu2018rana}
F.~Tu, W.~Wu, S.~Yin, L.~Liu, and S.~Wei, ``{RANA: Towards Efficient Neural
  Acceleration with Refresh-Optimized Embedded DRAM},'' in \emph{ISCA}, 2018.

\bibitem{udipi2012lot}
A.~N. Udipi, N.~Muralimanohar, R.~Balsubramonian, A.~Davis, and N.~P. Jouppi,
  ``{LOT-ECC: Localized And Tiered Reliability Mechanisms For Commodity Memory
  Systems},'' in \emph{ISCA}, 2012.

\bibitem{van1991testing}
A.~J. Van~de Goor, \emph{{Testing Semiconductor Memories: Theory and
  Practice}}.\hskip 1em plus 0.5em minus 0.4em\relax {John Wiley \& Sons,
  Inc.}, 1991.

\bibitem{van2002address}
A.~J. Van De~Goor and I.~Schanstra, ``{Address and Data Scrambling: Causes and
  Impact on Memory Tests},'' in \emph{DELTA}, 2002.

\bibitem{veen2016drammer}
V.~van~der Veen, Y.~Fratantonio, M.~Lindorfer, D.~Gruss, C.~Maurice, G.~Vigna,
  H.~Bos, K.~Razavi, and C.~Giuffrida, ``{Drammer: Deterministic Rowhammer
  Attacks on Mobile Platforms},'' \emph{{CCS}}, 2016.

\bibitem{van2018guardion}
V.~van~der Veen, M.~Lindorfer, Y.~Fratantonio, H.~P. Pillai, G.~Vigna,
  C.~Kruegel, H.~Bos, and K.~Razavi, ``{GuardION: Practical Mitigation of
  DMA-Based Rowhammer Attacks on ARM},'' in \emph{DIMVA}, 2018.

\bibitem{van2015mathematical}
J.~P. van Zandwijk, ``{A Mathematical Approach to NAND Flash-Memory
  Descrambling and Decoding},'' \emph{Digital Investigation}, 2015.

\bibitem{van2017bit}
J.~P. van Zandwijk, ``{Bit-Errors as a Source of Forensic Information in
  NAND-Flash Memory},'' \emph{Digital Investigation}, 2017.

\bibitem{vatajelu2018state}
E.~I. Vatajelu, P.~Pouyan, and S.~Hamdioui, ``{State of the Art and Challenges
  for Test and Reliability of Emerging Nonvolatile Resistive Memories},''
  \emph{JCTA}, 2018.

\bibitem{venkatesan2006retention}
R.~K. Venkatesan, S.~Herr, and E.~Rotenberg, ``{Retention-Aware Placement in
  DRAM (RAPID): Software Methods for Quasi-Non-Volatile DRAM},'' in
  \emph{HPCA}, 2006.

\bibitem{verified2019global}
{Verified Market Research}, ``{Global DRAM Market By Application, By
  Technology, By Memory, By Competitive Landscape, By Geographic Scope And
  Forecast},''
  \url{https://www.verifiedmarketresearch.com/product/global-dram-market-size-and-forecast-to-2025},
  2019.

\bibitem{villanueva2019cold}
R.~Villanueva-Polanco, ``{Cold Boot Attacks on Bliss},'' in \emph{LATINCRYPT},
  2019.

\bibitem{thun2020qualification}
M.~Von~Thun, ``{Qualification and Reliability of MRAM Toggle Memory Designed
  for Space Applications},''
  \url{https://www.everspin.com/file/157395/download}, 2020.

\bibitem{wada2004post}
O.~Wada, T.~Namekawa, H.~Ito, A.~Nakayama, and S.~Fujii, ``{Post-Packaging Auto
  Repair Techniques for Fast Row Cycle Embedded DRAM},'' in \emph{TEST}, 2004.

\bibitem{walker2021on}
A.~J. Walker, S.~Lee, and D.~Beery, ``{On DRAM Rowhammer and the Physics of
  Insecurity},'' \emph{TED}, 2021.

\bibitem{walker2021dram}
A.~J. Walker, S.~Lee, and D.~Beery, ``{On DRAM Rowhammer and the Physics of
  Insecurity},'' \emph{TED}, 2021.

\bibitem{wang2015backward}
D.~Wang, ``{Backward Compatible Dynamic Random Access Memory Device and Method
  of Testing Therefore},'' 2015, {US Patent 9,123,441}.

\bibitem{wang2009soft}
F.~Wang and V.~D. Agrawal, ``{Soft Error Rates with Inertial and Logical
  Masking},'' in \emph{VLSI}, 2009.

\bibitem{wang2018dram}
F.~Wang, T.~Vogelsang, B.~Haukness, and S.~C. Magee, ``{DRAM Retention at
  Cryogenic Temperatures},'' in \emph{IMW}, 2018.

\bibitem{wang2017architecting}
H.~Wang, ``{Architecting Memory Systems Upon Highly Scaled Error-Prone Memory
  Technologies},'' Ph.D. dissertation, Rensselaer Polytechnic Institute, 2017.

\bibitem{wang2014proactivedram}
J.~Wang, X.~Dong, and Y.~Xie, ``{ProactiveDRAM: A DRAM-Initiated Retention
  Management Scheme},'' in \emph{ICCD}, 2014.

\bibitem{wang2020dramdig}
M.~Wang, Z.~Zhang, Y.~Cheng, and S.~Nepal, ``{Dramdig: A Knowledge-Assisted
  Tool To Uncover DRAM Address Mapping},'' in \emph{DAC}, 2020.

\bibitem{wang2018content}
S.~Wang, M.~N. Bojnordi, X.~Guo, and E.~Ipek, ``{Content Aware Refresh:
  Exploiting the Asymmetry of DRAM Retention Errors to Reduce the Refresh
  Frequency of Less Vulnerable Data},'' \emph{TOC}, 2018.

\bibitem{wang2020figaro}
Y.~Wang, L.~Orosa, X.~Peng, Y.~Guo, S.~Ghose, M.~Patel, J.~S. Kim, J.~G. Luna,
  M.~Sadrosadati, N.~M. Ghiasi, and O.~Mutlu, ``{FIGARO: Improving System
  Performance via Fine-Grained In-DRAM Data Relocation and Caching},'' in
  \emph{MICRO}, 2020.

\bibitem{wang2018reducing}
Y.~Wang, A.~Tavakkol, L.~Orosa, S.~Ghose, N.~M. Ghiasi, M.~Patel, J.~S. Kim,
  H.~Hassan, M.~Sadrosadati, and O.~Mutlu, ``{Reducing DRAM Latency Via
  Charge-Level-Aware Look-Ahead Partial Restoration},'' in \emph{MICRO}, 2018.

\bibitem{wang2015radar}
Y.~Wang, Y.~Han, C.~Wang, H.~Li, and X.~Li, ``{RADAR: A Case for
  Retention-Aware DRAM Assembly and Repair in Future FGR DRAM Memory},'' in
  \emph{DAC}, 2015.

\bibitem{weber2005data}
A.~Weber, A.~Birner, and W.~Krautschneider, ``{Data Retention Analysis on
  Individual Cells of 256Mb DRAM in 110nm Technology},'' in \emph{ESSDERC},
  2005.

\bibitem{weis2015retention}
C.~Weis, M.~Jung, P.~Ehses, C.~Santos, P.~Vivet, S.~Goossens, M.~Koedam, and
  N.~Wehn, ``{Retention Time Measurements and Modelling of Bit Error Rates of
  Wide I/O DRAM in MPSoCs},'' in \emph{DATE}, 2015.

\bibitem{weis2015thermal}
C.~Weis, M.~Jung, O.~Naji, C.~Santos, P.~Vivet, and A.~Hansson, ``{Thermal
  Aspects and High-Level Explorations of 3D Stacked DRAMs},'' in \emph{ISVLSI},
  2015.

\bibitem{weis2017dramspec}
C.~Weis, A.~Mutaal, O.~Naji, M.~Jung, A.~Hansson, and N.~Wehn, ``{DRAMSpec: A
  High-Level DRAM Timing, Power and Area Exploration Tool},'' \emph{IJPP},
  2017.

\bibitem{wen2013cd}
W.~Wen, M.~Mao, X.~Zhu, S.~H. Kang, D.~Wang, and Y.~Chen, ``{CD-ECC:
  Content-Dependent Error Correction Codes for Combating Asymmetric Nonvolatile
  Memory Operation Errors},'' in \emph{ICCAD}, 2013.

\bibitem{wicht2004yield}
B.~Wicht, T.~Nirschl, and D.~Schmitt-Landsiedel, ``{Yield and Speed
  Optimization of a Latch-Type Voltage Sense Amplifier},'' in \emph{JSSC},
  2004.

\bibitem{wilkerson2008trading}
C.~Wilkerson, H.~Gao, A.~R. Alameldeen, Z.~Chishti, M.~Khellah, and S.-L. Lu,
  ``{Trading Off Cache Capacity for Reliability to Enable Low Voltage
  Operation},'' \emph{ISCA}, 2008.

\bibitem{wilkes2001memory}
M.~V. Wilkes, ``{The Memory Gap and The Future of High Performance Memories},''
  \emph{SIGARCH Computer Architecture News}, 2001.

\bibitem{wise2018reverse}
J.~Wise, ``{Reverse Engineering a NAND Flash Device Management Algorithm},''
  \url{https://joshuawise.com/projects/ndfrecovery#ecc\_recovery}, 2014.

\bibitem{wong2012metal}
H.-S.~P. Wong, H.-Y. Lee, S.~Yu, Y.-S. Chen, Y.~Wu, P.-S. Chen, B.~Lee, F.~T.
  Chen, and M.-J. Tsai, ``{Metal--Oxide RRAM},'' \emph{Proc. IEEE}, 2012.

\bibitem{wong2010phase}
H.-S.~P. Wong, S.~Raoux, S.~Kim, J.~Liang, J.~P. Reifenberg, B.~Rajendran,
  M.~Asheghi, and K.~E. Goodson, ``{Phase Change Memory},'' \emph{Proc. IEEE},
  2010.

\bibitem{wu2019protecting}
X.-C. Wu, T.~Sherwood, F.~T. Chong, and Y.~Li, ``{Protecting Page Tables From
  Rowhammer Attacks Using Monotonic Pointers in DRAM True-Cells},'' in
  \emph{ASPLOS}, 2019.

\bibitem{wulf1995hitting}
W.~A. Wulf and S.~A. McKee, ``{Hitting the Memory Wall: Implications of the
  Obvious},'' \emph{SIGARCH Computer Architecture News}, 1995.

\bibitem{wyns1988temperature}
P.~Wyns, R.~Anderson, and W.~DesJardins, ``{Temperature Dependence of Required
  Refresh Time in Dynamic Random Access Memories},'' in \emph{Proc. Symp. Low
  Temperature Electronics and High Temperature Superconductors}, 1988.

\bibitem{xiong2016run}
W.~Xiong, A.~Schaller, N.~A. Anagnostopoulos, M.~U. Saleem, S.~Gabmeyer,
  S.~Katzenbeisser, and J.~Szefer, ``{Run-Time Accessible DRAM PUFs in
  Commodity Devices},'' in \emph{CHES}, 2016.

\bibitem{yaglikci2021security}
A.~G. Ya{\u{g}}l{\i}k{\c{c}}{\i}, J.~S. Kim, F.~Devaux, and O.~Mutlu,
  ``{Security Analysis of the Silver Bullet Technique for RowHammer
  Prevention},'' 2021.

\bibitem{yaglikci2021blockhammer}
A.~G. Ya{\u{g}}l{\i}k{\c{c}}{\i}, M.~Patel, J.~S. Kim, R.~Azizibarzoki,
  A.~Olgun, L.~Orosa, H.~Hassan, J.~Park, K.~Kanellopoullos, T.~Shahroodi,
  S.~Ghose, and O.~Mutlu, ``{BlockHammer: Preventing RowHammer at Low Cost by
  Blacklisting Rapidly-Accessed DRAM Rows},'' in \emph{HPCA}, 2021.

\bibitem{yaglikci2020blockhammer}
A.~G. Ya{\u{g}}l{\i}k{\c{c}}{\i}, M.~Patel, J.~S. Kim, R.~Azizibarzoki,
  A.~Olgun, L.~Orosa, H.~Hassan, J.~Park, K.~Kanellopoullos, T.~Shahroodi,
  S.~Ghose, and O.~Mutlu, ``{BlockHammer: Preventing RowHammer at Low Cost by
  Blacklisting Rapidly-Accessed DRAM Rows},'' in \emph{HPCA}, 2021.

\bibitem{yaglikci2022understanding}
A.~G. Ya{\u{g}}l{\i}k{\c{c}}{\i}, H.~Luo, A.~Olgun, G.~F. de~Oliveira~Junior,
  J.~Park, M.~Patel, H.~Hassan, L.~Orosa, J.~Kim, and O.~Mutlu,
  ``{Understanding the RowHammer Vulnerability Under Reduced Wordline Voltage:
  An Experimental Study Using Real Devices},'' in \emph{DSN}, 2022.

\bibitem{yamaguchi2000theoretical}
K.~Yamaguchi, ``{Theoretical Study of Deep-Trap-Assisted Anomalous Currents in
  Worst-Bit Cells of Dynamic Random-Access Memories (DRAM's)},'' \emph{TED},
  2000.

\bibitem{yaney1987meta}
D.~S. Yaney, C.-Y. Lu, R.~A. Kohler, M.~J. Kelly, and J.~T. Nelson, ``{A
  Meta-Stable Leakage Phenomenon in DRAM Charge Storage-Variable Hold Time},''
  in \emph{IEDM}, 1987.

\bibitem{yang2016suppression}
C.~M. Yang, C.~K. Wei, Y.~J. Chang, T.~C. Wu, H.~P. Chen, and C.~S. Lai,
  ``{Suppression of Row Hammer Effect by Doping Profile Modification in
  Saddle-Fin Array Devices for Sub-30-nm DRAM Technology},'' \emph{IEEE
  Transactions on Device and Materials Reliability}, 2016.

\bibitem{yang2017scanning}
C.-M. Yang, C.-K. Wei, H.-P. Chen, J.-S. Luo, Y.~J. Chang, T.-C. Wu, and C.-S.
  Lai, ``{Scanning Spreading Resistance Microscopy for Doping Profile in
  Saddle-Fin Devices},'' \emph{IEEE Transactions on Nanotechnology}, 2017.

\bibitem{yang2019trap}
T.~Yang and X.-W. Lin, ``{Trap-Assisted DRAM Row Hammer Effect},'' \emph{EDL},
  2019.

\bibitem{yeleswarapu2018sscmsd}
R.~Yeleswarapu and A.~K. Somani, ``{Sscmsd-Single-Symbol Correction
  Multi-Symbol Detection For DRAM Subsystem},'' in \emph{PRDC}, 2018.

\bibitem{yitbarek2017cold}
S.~F. Yitbarek, M.~T. Aga, R.~Das, and T.~Austin, ``{Cold Boot Attacks Are
  Still Hot: Security Analysis of Memory Scramblers in Modern Processors},'' in
  \emph{HPCA}, 2017.

\bibitem{yoon2010virtualized}
D.~H. Yoon and M.~Erez, ``{Virtualized and Flexible ECC for Main Memory},'' in
  \emph{ASPLOS}, 2010.

\bibitem{yoon2011free}
D.~H. Yoon, N.~Muralimanohar, J.~Chang, P.~Ranganathan, N.~P. Jouppi, and
  M.~Erez, ``{FREE-p: Protecting Non-Volatile Memory Against Both Hard and Soft
  Errors},'' in \emph{ISCA}, 2011.

\bibitem{you2019mrloc}
J.~M. You and J.-S. Yang, ``{MRLoc : Mitigating Row-Hammering Based on Memory
  Locality},'' in \emph{DAC}, 2019.

\bibitem{yue2020dram}
M.~Yue, N.~Karimian, W.~Yan, N.~A. Anagnostopoulos, and F.~Tehranipoor,
  ``{DRAM-Based Authentication Using Deep Convolutional Neural Networks},''
  \emph{IEEE Consumer Electronics Magazine}, 2020.

\bibitem{yun2018study}
D.~Yun, M.~Park, C.~Lim, and S.~Baeg, ``{Study of TID Effects on One Row
  Hammering using Gamma in DDR4 SDRAMs},'' in \emph{IRPS}, 2018.

\bibitem{zhang2015hi}
C.~Zhang, G.~Sun, X.~Zhang, W.~Zhang, W.~Zhao, T.~Wang, Y.~Liang, Y.~Liu,
  Y.~Wang, and J.~Shu, ``{Hi-Fi Playback: Tolerating Position Errors in Shift
  Operations of Racetrack Memory},'' in \emph{ISCA}, 2015.

\bibitem{zhang2021quantifying}
D.~Zhang, G.~Panwar, J.~B. Kotra, N.~DeBardeleben, S.~Blanchard, and X.~Jian,
  ``{Quantifying Server Memory Frequency Margin and Using it to Improve
  Performance in HPC Systems},'' in \emph{ISCA}, 2021.

\bibitem{zhang2018exploring}
D.~Zhang, V.~Sridharan, and X.~Jian, ``{Exploring And Optimizing
  Chipkill-Correct For Persistent Memory Based On High-Density Nvrams},'' in
  \emph{MICRO}, 2018.

\bibitem{zhang2017dynamic}
J.~Zhang, D.~Kline, L.~Fang, R.~Melhem, and A.~K. Jones, ``{Dynamic
  Partitioning To Mitigate Stuck-At Faults in Emerging Memories},'' in
  \emph{ICCAD}, 2017.

\bibitem{zhang2014half}
T.~Zhang, K.~Chen, C.~Xu, G.~Sun, T.~Wang, and Y.~Xie, ``{Half-DRAM: A
  High-Bandwidth and Low-Power DRAM Architecture from the Rethinking of
  Fine-Grained Activation},'' in \emph{ISCA}, 2014.

\bibitem{zhang2014cream}
T.~Zhang, M.~Poremba, C.~Xu, G.~Sun, and Y.~Xie, ``{CREAM: A
  Concurrent-Refresh-Aware DRAM Memory Architecture},'' in \emph{HPCA}, 2014.

\bibitem{zhang2019boosting}
X.~Zhang, R.~Wang, Y.~Zhang, and J.~Yang, ``{Boosting Chipkill Capability Under
  Retention-Error Induced Reliability Emergency},'' in \emph{ASP-DAC}, 2019.

\bibitem{zhang2015exploiting}
X.~Zhang, Y.~Zhang, B.~R. Childers, and J.~Yang, ``{Exploiting DRAM Restore
  Time Variations In Deep Sub-Micron Scaling},'' in \emph{DATE}, 2015.

\bibitem{zhang2016restore}
X.~Zhang, Y.~Zhang, B.~R. Childers, and J.~Yang, ``{Restore Truncation for
  Performance Improvement in Future DRAM Systems},'' in \emph{HPCA}, 2016.

\bibitem{zhang2015vlsichapter41}
X.~Zhang, in \emph{{VLSI Architectures for Modern Error-Correcting Codes}},
  2015, ch.~41, p.~53.

\bibitem{zhang2015vlsi}
X.~Zhang, \emph{{VLSI Architectures for Modern Error-Correcting Codes}}.\hskip
  1em plus 0.5em minus 0.4em\relax {CRC Press}, 2015.

\bibitem{zhang2012multi}
Y.~Zhang, L.~Zhang, W.~Wen, G.~Sun, and Y.~Chen, ``{Multi-Level Cell STT-RAM:
  Is it Realistic or Just a Dream?}'' in \emph{ICCAD}, 2012.

\bibitem{zhang2012memory}
Z.~Zhang, W.~Xiao, N.~Park, and D.~J. Lilja, ``{Memory Module-Level Testing And
  Error Behaviors For Phase Change Memory},'' in \emph{ICCD}, 2012.

\bibitem{zhao2014improving}
B.~Zhao, ``{Improving Phase Change Memory (PCM) and Spin-Torque-Transfer
  Magnetic-RAM (STT-MRAM) as Next-Generation Memories: A Circuit
  Perspective},'' Ph.D. dissertation, University of Pittsburgh, 2014.

\bibitem{zhu2015microarchitectural}
Y.~Zhu, D.~Richins, M.~Halpern, and V.~J. Reddi, ``{Microarchitectural
  Implications Of Event-Driven Server-Side Web Applications},'' in
  \emph{MICRO}, 2015.

\bibitem{zuravleff1997controller}
W.~K. Zuravleff and T.~Robinson, ``{Controller for a Synchronous DRAM that
  Maximizes Throughput by Allowing Memory Requests and Commands to be Issued
  Out of Order},'' 1997, {US} Patent 5,630,096.

\end{thebibliography}
\end{singlespace}

\bookmarksetup{startatroot}
\end{document}